\documentclass[a4paper,11pt,oneside]{report} 

\usepackage[T1]{fontenc}
\usepackage[utf8]{inputenc}
\usepackage{textcomp}
\usepackage{gensymb}
\usepackage{amsmath}
\usepackage{amssymb}
\usepackage{caption}
\usepackage[export]{adjustbox}
\usepackage{datetime}
\usepackage[a4paper,top=2.5cm,bottom=2.5cm,width=15cm]{geometry}
\usepackage{afterpage}
\usepackage{hhline}
\usepackage[hang,flushmargin,bottom]{footmisc}
\usepackage[dvipsnames]{xcolor}

\usepackage{graphicx}
\usepackage{float}
\graphicspath{ {images/} }

\usepackage{csquotes,xpatch}
\usepackage[numbers,sort&compress]{natbib}

\usepackage{setspace}

\definecolor{SussexCobaltBlue}{HTML}{1d4289}

\usepackage[colorlinks=true,
urlcolor=blue,
anchorcolor=blue,
citecolor=SussexCobaltBlue,
filecolor=blue,
linkcolor=SussexCobaltBlue,
menucolor=blue,
linktocpage=true,
pdfa=true,
pdfauthor={Alessandro Maraio},
pdftitle={Maximising information from weak lensing galaxy surveys --- Alex Maraio PhD thesis},
pdfproducer={pdfLaTeX with Hyperref},
pdfcreator={pdfLaTeX compiled on Ubuntu 20.04},
pdfsubject={Weak lensing cosmology power spectra analysis},
pdfencoding=auto,
psdextra
]{hyperref}

\makeatletter
\let\oldHyPsd@CatcodeWarning\HyPsd@CatcodeWarning
\renewcommand{\HyPsd@CatcodeWarning}[1]{
  \ifnum\pdfstrcmp{#1}{math shift}=0    
  \else                               
    \ifnum\pdfstrcmp{#1}{subscript}=0    
    \else                                 
      \oldHyPsd@CatcodeWarning{#1}
    \fi  
  \fi
  
}
\makeatother

\usepackage{titlesec}
\usepackage{fancyvrb}
\usepackage{mathrsfs}

\usepackage{lmodern}
\usepackage{ragged2e}
\newcommand{\seriffontregular}{\sffamily\selectfont}
\newcommand{\seriffont}{\sffamily\fontseries{b}\selectfont}
\newcommand{\seriffontsemi}{\sffamily\fontseries{b}\selectfont}
\newcommand{\seriffontsemiitalics}{\sffamily\itshape\selectfont}

\usepackage{tocloft}

\titleformat{\section}{\seriffontsemi\LARGE}{\thesection}{1em}{}
\titleformat{\subsection}{\seriffontsemi\Large}{\thesubsection}{1em}{}
\titleformat{\subsubsection}{\seriffontsemi\large}{\thesubsubsection}{1em}{}
\titleformat{\paragraph}{\seriffontsemiitalics\normalsize}{\theparagraph}{0em}{\vspace*{0em}}[\vspace{-0.66em}]

\usepackage[most]{tcolorbox}

\definecolor{nicecolour}{HTML}{A4278D}

\newtcolorbox{mytext}[1][]{%
    nobeforeafter,
    tcbox raise base,
    enhanced,
    colframe=nicecolour,
    boxrule=1.15pt,
    boxsep=0pt,
    left=2.5mm,
    top=2.5mm,
    bottom=2.5mm,
    right=2.5mm,
    #1}

\newtcolorbox{mypaper}[1][]{%
    nobeforeafter,
    tcbox raise base,
    enhanced,
    colframe=nicecolour,
    colback=green!20,
    boxrule=1.15pt,
    boxsep=0pt,
    left=2.5mm,
    top=2.5mm,
    bottom=2.5mm,
    right=2.5mm,
    #1}

\usepackage{fancyhdr}
\fancypagestyle{plain}{%
\fancyhf{} 
\fancyfoot[C]{\sffamily\fontsize{10pt}{10pt}\selectfont \hfill-- \thepage\ --\hfill} 

}
\usepackage{blindtext}

\definecolor{gray75}{gray}{0.75}
\definecolor{chaptertitle}{HTML}{A4278D}

\NewDocumentCommand{\insquare}{omo}{%
  \IfValueTF{#1}{\setlength{\fboxsep}{#1}}{}%
  \IfValueTF{#3}{\setlength{\fboxrule}{#3}}{}%
  \fbox{\textcolor{red}{#2}}
}

\renewcommand{\d}{\textrm{d}}

\newcommand{\fNL}{f_\textrm{NL}}

\usepackage{relsize}
\usepackage{xspace}
\newcommand\Cpp{C\nolinebreak\hspace{-.05em}\raisebox{.4ex}{\relsize{-3}{\textbf{+}}}\nolinebreak\hspace{-.10em}\raisebox{.4ex}{\relsize{-3}{\textbf{+}}}\xspace}

\newcommand{\eV}{\textrm{e\kern-0.25ex V}\xspace}

\usepackage[nottoc,notlof,notlot]{tocbibind}

\newcommand{\Camb}{{\seriffontregular CAMB}\xspace}

\newcommand{\Planck}{{\textit{Planck}}}

\newcommand{\NumPy}{{\seriffontregular NumPy}}
\newcommand{\SciPy}{{\seriffontregular SciPy}}
\newcommand{\pandas}{{\seriffontregular Pandas}\xspace}
\newcommand{\mpl}{{\seriffontregular Matplotlib}\xspace}
\newcommand{\seaborn}{{\seriffontregular Seaborn}\xspace}

\newcommand{\healpy}{{\seriffontregular healpy}\xspace}
\newcommand{\HEALPix}{{\seriffontregular HEALPix}\xspace}
\newcommand{\GetDist}{{\seriffontregular GetDist}\xspace}
\newcommand{\CCL}{{\seriffontregular CCL}\xspace}
\newcommand{\Flask}{{\seriffontregular Flask}\xspace}
\newcommand{\Glass}{{\seriffontregular Glass}\xspace}
\newcommand{\NaMaster}{{\seriffontregular NaMaster}\xspace}
\newcommand{\Cosmosis}{{\seriffontregular CosmoSiS}\xspace}
\newcommand{\Eigen}{{\seriffontregular Eigen}\xspace}

\newcommand{\HaloFit}{{\seriffontregular HaloFit}\xspace}
\newcommand{\HMCode}{{\seriffontregular HMCode}\xspace}
\newcommand{\HMCodett}{{\seriffontregular HMCode-2020}\xspace}

\newcommand{\MultiNest}{{\seriffontregular MultiNest}\xspace}
\newcommand{\PolyChord}{{\seriffontregular PolyChord}\xspace}

\newcommand{\As}{\ensuremath{A_\textrm{s}}}
\newcommand{\ns}{n_\textrm{s}}
\newcommand{\Ylm}{Y_{\ell m}}
\newcommand{\Cl}{C_{\ell}}
\newcommand{\alm}{a_{\ell m}}
\newcommand{\almprime}{a_{\ell' m'}^{*}}

\newcommand{\rhomatter}{\ensuremath{\rho_{\textsc{m}}}}
\newcommand{\rhorad}{\ensuremath{\rho_{\textsc{r}}}}
\newcommand{\rhocrit}{\ensuremath{\rho_{\textsc{crit}}}}

\newcommand{\Omegar}{\ensuremath{\Omega_{\textsc{r}}}}
\newcommand{\Omegam}{\ensuremath{\Omega_{\textsc{m}}}}
\newcommand{\Omegab}{\ensuremath{\Omega_{\textsc{b}}}}
\newcommand{\Omegac}{\ensuremath{\Omega_{\textsc{c}}}}

\newcommand{\deltam}{\delta_{\textsc{m}}}

\newcommand{\deltamdot}{\dot{\delta}_{\textsc{m}}}

\newcommand{\deltamddot}{\ddot{\delta}_{\textsc{m}}}

\newcommand{\Sigmacrit}{\ensuremath{\Sigma_{\textsc{cr}}}}
\newcommand{\fk}{\ensuremath{f_{k}}}

\newcommand{\chimax}{\ensuremath{\rchi_{\textsc{max}}}}

\newcommand{\lmax}{\ensuremath{\ell_{\textsc{max}}}}

\DeclareRobustCommand{\rchi}{{\mathpalette\irchi\relax}}
\newcommand{\irchi}[2]{\raisebox{\depth}{$#1\chi$}}

\newcommand{\Lag}{\ensuremath{\mathscr{L}}}

\newcommand{\TAGN}{\ensuremath{T_{\textrm{AGN}}}}

\newcommand{\Tr}{\ensuremath{\textrm{Tr}}\xspace}

\usepackage{blindtext}
\usepackage{quotchap}

\usepackage{soul}
\usepackage{tensor}

\usepackage{fontawesome}
\usepackage{bbold}
\usepackage{bm}

\newtcbox{\mychar}[1][]{nobeforeafter, tcbox raise base, colback=red, colframe=black, sharp corners, colupper=white, size=fbox, #1}

\newcommand{\Cltilde}{\tilde{C}_{\ell}}

\newcommand{\Clvec}{\mathbfit{C}_\ell}
\newcommand{\Cltildevec}{\tilde{\mathbfit{C}}_{\ell}}

\newcommand{\yl}{y_{\ell}}

\newcommand{\almE}{a_{\ell m}^{E}}
\newcommand{\almEtilde}{\tilde{a}_{\ell m}^{E}}
\newcommand{\almEprime}{a_{\ell' m'}^{E}}
\newcommand{\almB}{a_{\ell m}^{B}}
\newcommand{\almBtilde}{\tilde{a}_{\ell m}^{B}}
\newcommand{\almBprime}{a_{\ell' m'}^{B}}

\newcommand{\Wllmm}{\, _{\pm 2}W^{\ell \ell'}_{m m'}}
\newcommand{\Mmat}{\mathbfss{M}_{\ell \ell'}}

\newcommand{\ClEE}{C_{\ell}^{EE}}
\newcommand{\ClBB}{C_{\ell}^{BB}}
\newcommand{\ClEB}{C_{\ell}^{EB}}

\newcommand{\Nl}{N_{\ell}}

\newcommand{\Ymat}{\mathbfss{Y}}

\newcommand{\sYlm}{_{s}Y_{\ell m}}
\newcommand{\stwoYlm}{\, _{\pm 2}Y_{\ell m}}
\newcommand{\splustwoYlm}{\, _{+2}Y_{\ell m}}
\newcommand{\sminustwoYlm}{\, _{-2}Y_{\ell m}}
\newcommand{\stwoYlmprime}{\, _{\pm 2}Y_{\ell' m'}}

\newcommand{\nhatvec}{\hat{n}}

\newcommand{\Nside}{N_{\mathrm{side}}}
\newcommand{\Npix}{N_{\mathrm{pix}}}

\newcommand{\fsky}{f_{\mathrm{sky}}}

\newcommand{\cs}{c_{\mathrm{s}}}

\newcommand{\sigmaeight}{\sigma_{8}}
\newcommand{\Seight}{S_{8}}

\newcommand{\arcmin}{\mathrm{arcmin}}

\newcommand{\mathbfit}[1]{\vec{#1}}
\newcommand{\mathbfss}[1]{\mathbf{#1}}

\usepackage{nameref}
\usepackage{booktabs}  
\usepackage[absolute]{textpos}
\usepackage{fancyhdr}
\fancypagestyle{lscape}{%
\fancyhf{} 
\fancyfoot[c] {%
\sffamily\fontsize{10pt}{10pt}\selectfont \hfill-- \thepage\ --\hfill%
\begin{textblock}{1}(13,10.5){\rotatebox{90}{\sffamily\fontsize{10pt}{10pt}\selectfont \hfill-- \thepage\ --\hfill}}\end{textblock}}
 
}

\usepackage{pdflscape}
\usepackage{wrapfig}
\usepackage{lscape}
\usepackage{rotating}
\newcommand{\Mpc}{\ensuremath{\textrm{Mpc}}}

\newcommand{\Clp}{C_{\ell'}}

\newcommand{\Omegamgeom}{\ensuremath{\Omega_{\textsc{m}\textrm{ geom}}}}

\newcommand{\Mb}{M_{\textrm{b}}}
\newcommand{\Tagn}{T_{\textrm{AGN}}}

\newcommand{\Halofit}{\textsc{HaloFit}\xspace}
\newcommand{\Bacco}{\textsc{Bacco}\xspace}
\newcommand{\BCEmu}{\textsc{BCemu}\xspace}
\newcommand{\KiDS}{\textrm{KiDS-1000}\xspace}
\newcommand{\DES}{\textrm{DES-Y3}\xspace}

\newcommand{\Bahamas}{\textsc{Bahamas}\xspace}
\newcommand{\Eagle}{\textsc{Eagle}\xspace}
\newcommand{\HZAGN}{\textsc{Horizon-AGN}\xspace}
\newcommand{\Illustris}{\textsc{Illustris}\xspace}
\newcommand{\Simba}{\textsc{Simba}\xspace}
\newcommand{\TNG}{\textsc{Illustris-TNG}\xspace}
\newcommand{\Camels}{\textsc{Camels}\xspace}

\newcommand{\Phy}{\ensuremath{P^{\textsc{Hy}}(k)}}
\newcommand{\Pdm}{\ensuremath{P^{\textsc{DM}}(k)}}

\newcommand{\Phykz}{\ensuremath{P^{\textsc{Hy}}(k, z)}}
\newcommand{\Pdmkz}{\ensuremath{P^{\textsc{DM}}(k, z)}}

\usepackage{colortbl}
\usepackage{tikz}
\newcommand{\tikzcolorbox}[3]{\tikz[baseline]\node[fill=#1,draw=#2,scale=1.75]{#3 };}

\usepackage{appendix}

\usepackage{enumitem}
\newlist{galitemize}{itemize}{4}
\setlist[galitemize,1]{
  leftmargin=\dimexpr0.3cm+\labelsep\relax,
  label={\smash{\raisebox{-0.2\height}{\includegraphics[width=0.4cm]{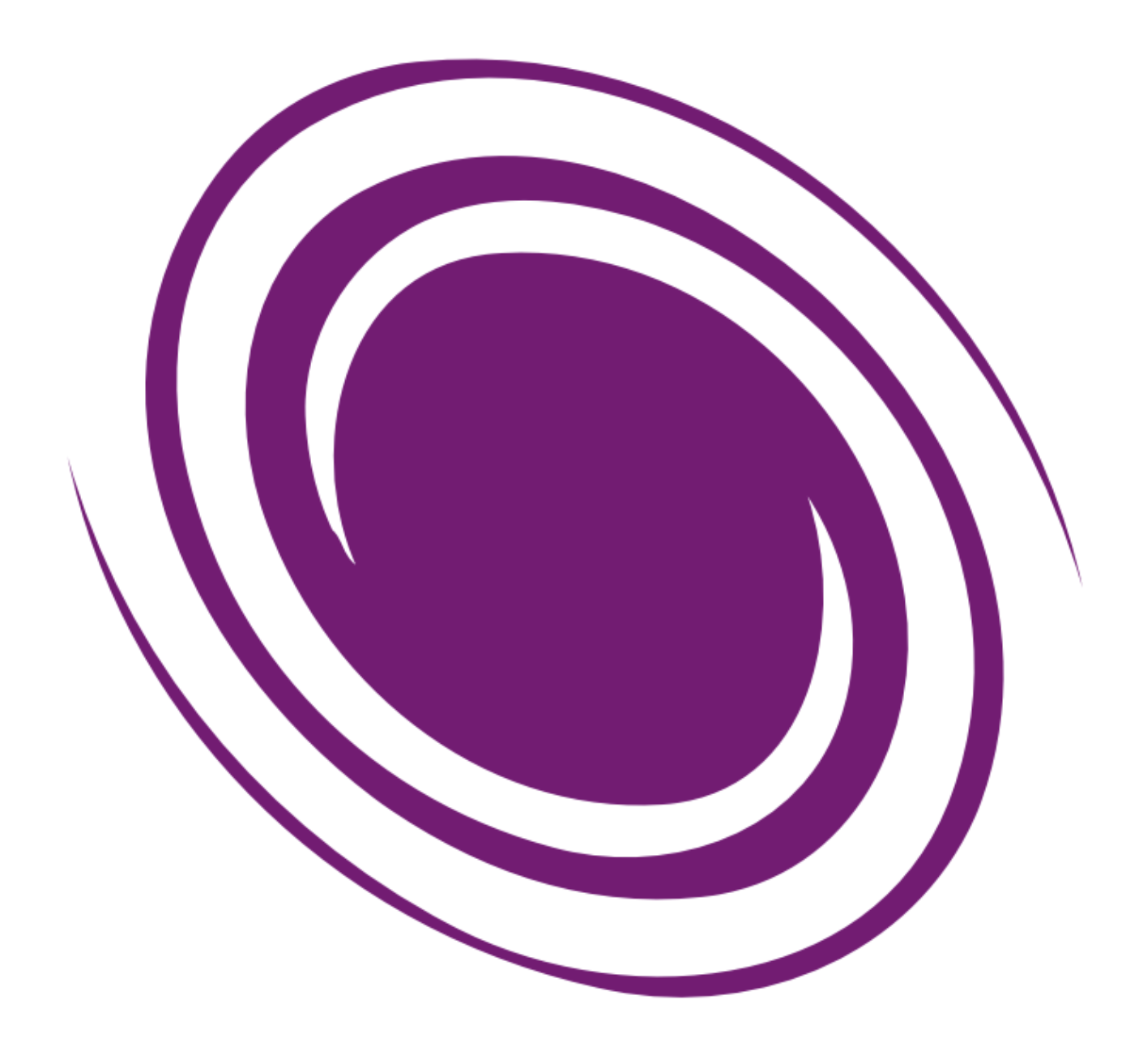}}}}
}

\newcommand{\kmax}{k_{\textrm{max}}}

\newcommand{\ngal}{\bar{n}_{\textrm{gal}}}
\newcommand{\chisqcrit}{\chi^2_{\textrm{crit}}}
\newcommand{\bins}[2]{#1 \! \times \! #2}

\newcommand{\mnu}{\Sigma m_{\nu}}
\newcommand{\Neff}{N_{\textrm{eff}}}

\makeatletter
  \let\oldr@@t\r@@t 
  \def\r@@t#1#2{%
    \setbox0=\hbox{$\oldr@@t#1{#2\,}$}\dimen0=\ht0
    \advance\dimen0-0.2\ht0
    \setbox2=\hbox{\vrule height\ht0 depth -\dimen0}%
    {\box0\lower0.4pt\box2}}
  \NewCommandCopy{\oldsqrt}{\sqrt}
  \renewcommand*{\sqrt}[2][\ ]{\oldsqrt[#1]{#2}}
\makeatother

\RequirePackage[normalem]{ulem} 
\RequirePackage{color}\definecolor{RED}{rgb}{1,0,0}\definecolor{BLUE}{rgb}{0,0,1} 

\begin{document}
\titleformat{\chapter}[hang]{\filleft\Huge\bfseries}{}{0pt}{\Huge\seriffontsemi}[\vspace{0ex}\rule{\textwidth}{1.15pt}\vspace*{1.25ex}]

\pagenumbering{roman}

\begin{titlepage}
\begin{center}


\includegraphics[width=225px]{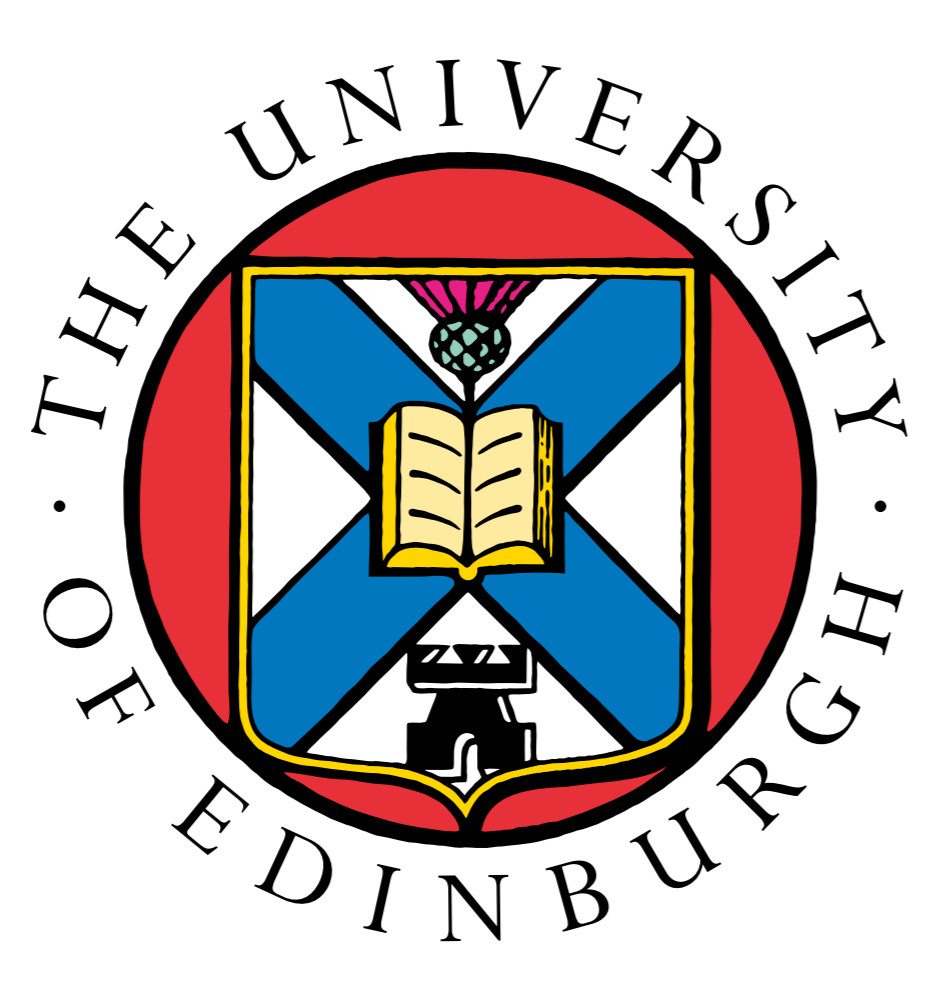}


\vspace{2cm}
        
\rule{\textwidth}{1.15pt}
\vspace{0.25cm}

{\Huge\seriffont Maximising information from\\ weak lensing galaxy surveys}

\vspace{0.75cm}
\rule{\textwidth}{1.15pt}
\vspace{1.2cm} 

{\LARGE\seriffont Alessandro Maraio }

{\large\seriffontregular\textit{Institute for Astronomy, University of Edinburgh}}
\vspace{0.5cm}

\vspace*{2cm}       
Doctor of Philosophy \\
March 2025 

\vspace{5cm} 

\end{center}
\end{titlepage}

\chapter*{Abstract}
\addcontentsline{toc}{chapter}{Abstract}
Weak lensing galaxy surveys are currently undergoing a dramatic
revolution as the dawn of the Stage-IV surveys is upon us. The quality and quantity
of data that we are to receive over the next decade dwarfs the data we have from
existing observatories and will undoubtedly lead to many cosmological discoveries.
Therefore, ensuring that our analysis methods are as accurate and precise as the
raw data is of upmost importance, and the driving force behind this thesis and
the work contained within. 

Understanding the details behind a modern cosmic shear analysis requires us
to start with the fundamentals of cosmology and go from there. 
Chapter~\ref{chp:introduction} provides us with a gentle introduction to some of
the driving ideas and results that have compounded to form our modern cosmological
model. Starting with Einstein's general relativity, Chapter~\ref{chp:background_cosmology}
derives many key results which are required for any cosmological analysis.
We then focus on the specifics of a cosmic shear analysis in Chapter~\ref{chp:grav_lensing},
with a brief look into some of the motivations behind the Stage-IV surveys
in Chapter~\ref{chp:Cosmology_2024}.

Returning to the theme of the development of accurate and precise statistical
methods for use in cosmic shear analyses, Chapter~\ref{chp:QML_estimator}
investigates a new implementation of the Quadratic Maximum Likelihood (QML)
estimator. Previous implementations of the QML estimator required the direct
evaluation of dense, high-dimensional matrices of either spherical harmonics
or Legendre polynomials. This severely limits the application of these methods
to the increased precision of Stage-IV survey data. The evaluation of each
entry in these matrices is a major bottleneck in computational speed, combined 
with the extreme limitations in the RAM usage from the high dimensionality of
the matrices makes these existing implementations of QML unsuitable for the next generation
of cosmic shear surveys. I led the development of a new, alternative implementation
of the QML estimator which side-steps the need to compute and store these
massive matrices though using the conjugate-gradient and finite-differences
methods. These allow for dramatically reduced run-times and RAM usage of
my new estimator when compared to previous implementations. Applying this
estimator to Stage-IV cosmic shear data finds that we achieve a $20\,\%$
decrement in the error bars on the largest angular scales for the $E$-modes, 
and over an order of magnitude decrement in the $B$-modes. This highlights
the usefulness of QML methods when applied to cosmic shear survey data.  

An accurate measurement of the observational power spectrum is useless without robust
modelling of all the systematic effects that contribute to the cosmic shear
signal. Of particular importance is the impact of baryonic feedback physics 
that exists on small-scales within our Universe, which can dramatically affect
the cosmic shear power spectrum on small angular scales. Mitigating the effects of
baryonic feedback bias in cosmic shear surveys has been the discussion of
many papers in the literature. Chapter~\ref{chp:baryonic_effects}
investigates the implementation of the `theoretical uncertainties' approach
applied to baryonic feedback in the matter power spectrum, the underlying
quantity that cosmic shear probes. By benchmarking several baryonic feedback
models to an ensemble of hydrodynamical simulations, we can quantify the error
on each model as a function of wavenumber and redshift. These errors then inform
us on how to smoothly down-weight the contaminating modes in the power spectrum,
allowing us to gain maximum information from our observables without being
susceptible to bias from inadequate modelling of baryonic feedback physics.
We find that using a simple, one-parameter model of baryonic feedback leads
to an unacceptable level of bias when including angular scales down to
$\lmax = 5000$, the nominal target for a Stage-IV survey. The inclusion of the
theoretical error covariance mitigates these biases, with the results preferring
a multi-parameter model of baryonic feedback to further mitigate biases.

We then turn our attention to deriving new methods for determining sets of
binary scale cuts for future cosmic shear surveys in Chapter~\ref{chp:binary_cuts}.
Binary cuts have been
well-studied in the literature, where they have been used extensively in
previous Stage-III cosmic shear surveys and thus form the default choice
for Stage-IV surveys. We find that if we apply existing
methods for deriving binary cuts to Stage-IV survey data, then the results from
our new surveys will be no better than existing results in the literature.
Using this as motivation, we present three alternative methods for how the
derivation of binary scale cuts might work for a Stage-IV survey. These aim to
keep the maximum amount of information possible from the data, without incurring
significant biases in the final results.

We present a summary of work completed in this thesis along with conclusions and
some thoughts for the future in Chapter~\ref{chp:conclusions}.

\chapter*{Universal abstract}
\addcontentsline{toc}{chapter}{Universal abstract}
While astronomy may be considered one of the oldest natural sciences, it is
remarkable that we could only begin to describe the large-scale properties of
our Universe a little over a hundred years ago. Since then, the study of
our Universe, cosmology, has evolved from us making primitive observations
that distant galaxies were moving away from us, to making unimaginably precise
measurements throughout our Universe's evolution and across the entire spectrum
of light. These observations have led to extraordinary discoveries about the
properties of our Universe: that $25\,\%$ of it is in a type of matter that 
we've never observed on Earth, and $70\,\%$ of it acts in a way that no type of
matter does on Earth, it has a negative gravitational attraction, and if we try to calculate
its value, we get a value that is just fantastically away from what we measure!

What is even more perplexing is that the $5\,\%$ of our Universe that we are familiar with in
our everyday lives --- the ordinary `baryonic' material that we, the Earth and 
planets, our Sun and distant stars are made from --- is often too challenging
to analyse or understand the physics of on cosmological scales. This motivates
us to use our advanced telescopes to look out into our night's sky and make
observations that allow us to constrain the properties of our Universe and
the physical laws that govern it.

One of the most powerful probes of our Universe comes from the effect of
weak gravitational lensing, which is akin to how a magnifying glass can
magnify and distort the image of objects behind it. However, instead of
having a magnifying glass in space, we use the distribution of matter within
our Universe to produce a similar effect. Sorry, there's no \textit{2001}-style
space magnifying monoliths!

This lensing by the large-scale structure of our Universe induces slight
changes in the apparent images of distant galaxies, which we can then try to
measure with our telescopes. Since each individual galaxy's distortion is only very slight, 
we rely on large-scale statistical analyses of galaxy images to derive
insights into the physics of our Universe from them. The first such statistical
technique that this thesis investigates is that of \textit{power spectrum 
estimation}. This involves quantifying how important each scale is in distorting
the images of galaxies. We are used to breaking down a signal into its components,
for example a bass guitar produces very low frequency (boom) sounds whereas a
high-hat cymbal can produce high frequency (tiss) sounds. We do much the same,
but instead of sound frequencies, we break our signals down into large and small
angular scales on the sky. Think of this very much like looking at a Dalmatian
and counting the number of small and large spots that it has.

The technique that allows us to go from maps of galaxy images on the sky to
a power spectrum is called a power spectrum estimator, with this thesis
presenting a new implementation of the Quadratic Maximum Likelihood (QML)
estimator. This aims to maximise the probability (or likelihood) of the values
of the amplitude of the large- and small-scale distortions given a map of galaxy
shapes. Our new implementation circumvents the need to compute and store very 
many numbers in the computer's memory, and so we can analyse far higher
quality maps with improved performance than previous method allowed for.  

Using the new implementation of our estimator, we found that we could
decrease the error bars (how certain we are of different measurements) with
respect to the nominal analysis choice by around $20 \,\%$ on the largest angular
scales for the `$E$-modes', which carry the cosmological signal, and 
with a decrease of about a factor of $10$ for the
`$B$\nobreakdash-modes', which are tests of Einstein's general relativity. 
These decrements represents a dramatic decrease in the error bars, and thus
by using our new estimator we can be much more certain of our power spectrum
estimate, and so more certain in any results using it --- such as investigating 
the properties of our Universe.  

As mentioned previously, the ordinary matter in our Universe that we are made 
from produce some remarkable physical processes within the Universe. We are already
familiar with our Sun, which provides us with essential sunlight and warmth
that's essential to life on Earth. As our Sun ages, it will gradually run out
of hydrogen to fuse, and so expand to form a red giant. After that, it will
shrink down, forming a white dwarf star. We only have one star in our Solar
System, but, like Tatooine, stars can form in pairs --- a binary star system.
If one of these stars becomes a white dwarf, and then eats material from
the other star, the white dwarf can undergo a brilliant explosion, a supernova,
which are bright enough to outshine entire galaxies. This puts a huge amount
of energy into the galaxy, decimating any objects that are trying to form
under gravity. 

We have also observed that in the centre of many galaxies exist the presence
of supermassive black holes, which can have masses from a million to billions of
times the mass of our Sun. These truly massive objects pull in huge amounts of
matter from their galaxy and emit vast quantities of radiation from the infalling
matter back into their galaxies. This extreme amount of radiation further 
disrupts the gravitational attraction of matter within galaxies. 

The physical processes that describe the properties of supernovae and supermassive
black holes (dubbed baryonic feedback) are extremely complicated, and so we cannot write down simple
formulae for their properties with pen-and-paper. Thus, we aim to capture their
properties by running hydrodynamical simulations (hydro-sims), which aim to implement `the Universe
in a computer'. However, it has been found that different collaborations who
produce their own hydro-sims come up with very different results for the 
properties of these objects within their simulations! Thus, when we use models
of baryonic feedback in our cosmological analyses of observational data, we
need to ensure that our models can recreate all predictions from every hydro-sim. 
Otherwise, if we use an inadequate model on our observational data, we could be
susceptible to bias in our results and come up with the wrong conclusions from
it. 

Thus, the process of comparing our analytic models of baryonic feedback to
hydrodynamical simulations results in the creation of the theoretical error
covariance, which quantifies how badly our models fit the data as a function
of scale and evolution in our Universe. This theoretical error covariance
can then be included in our cosmological analyses, which, this thesis shows, can
successfully mitigate biases associated from baryonic feedback. We find that
we naturally down-weight angular scales that are contaminated with baryonic
feedback, without having to impose hard cuts on our data-vectors as previous 
analyses have found. This comes at the cost of moderately increased error bars
on our parameters. However, as ensuring that our final results remain unbiased is the
most important criterion for any cosmological analysis, this is an acceptable
trade-off compromise for correct results.

We then go on to develop new methods of constructing hard cuts in the data,
again to alleviate the impact of baryonic feedback on our data-vectors, but in
a way that is much more easily extendable to include other sources of errors and
uncertainties in our modelling. We extend existing methods that previous
collaborations have used, finding that our new methods allow for much more
information to be kept from our observations in a way that still leaves the
final results unbiased. 

We have now entered an era in which the quality of observations that we
are getting from the next generation of telescopes is just amazing. It is hoped
that the vast increase in the precision on the data, along with more accurate
and precise statistical methods which this thesis has developed, will allow us
to further uncover the fundamental physical properties of much of our Universe
which has thus far eluded us.

\phantom{\cite{Maraio:2022ywi,Maraio:2024xjz}}

\begin{savequote}[65mm]
  And they were a good friend
  
  \qauthor{---Obi Wan Kenobi}
  \end{savequote}
\vspace*{5cm}
\chapter*{Acknowledgements}
\addcontentsline{toc}{chapter}{Acknowledgements}
First and foremost, I would like to thank my amazing supervisors, Alex and Andy,
for their unwavering guidance, support, and infinite knowledge throughout the
whole of my PhD. Especially since, while I may have tried to channel my
inner Kimi Räikkönen of \textit{``Leave me alone, I know what I'm doing''} ---
I am, in fact, not Kimi Räikkönen and I did \textit{not} know what I was doing most of the time! This thesis,
and the work contained within, would be nowhere near the quality that it is 
(or at least, I hope it is) without both of your help, patience, and many, many
comment boxes! A sincere \textit{thank you} to you both. \\ 
I would also like to thank other members of the IfA, in particular Joe
Zuntz, Catherine Heymans, and Britton Smith for their help, humour, and support
in research, becoming a PhD student, and undergraduate teaching.

Secondly, I would like to thank all my amazing friends that I got to share the
highs and lows of a PhD, and MPhys before that, with them:

To Ewan, thanks for putting up with this idiot for three years as your
flatmate and friend. It's fair to say that we were put in the deep end, but
it was always fun to share our hours-long rants about Cuillin and what a 
`$\rchi^2$ bathtub' is together!  

Poppy: \textit{Thank you}. Thank you for your unlimited infectious excitement
and enthusiasm that always provided me with fresh motivation whenever we met up
or Discorded. Thank you for
our endless \textit{Minecraft} nights where you were the Steve (or Shrek) to my Alex,
spending over three hundred hours across one hundred evenings since 2020
with me. I look forward to reading your thesis, soon-to-be
Dr Joshi, and for continuing our very many \textit{Minecraft} nights together!

To Zo\"e: Thank you for introducing me to \textit{Stardew Valley} and our many
farming evenings spent together! Thank you also for a very comfy sofa in
Geneva and for getting me into CERN (even if it was just for lunch!). 
I very much look forward to ABBA \textit{Voyage} together, hopefully both as doctors! 

Andrew \& Annie: I would probably not be doing a PhD had I not signed up to
a random `Doctor Who society' one fateful Wednesday morning fresher's fair. I most definitely
caught the doctoral bug from you two, so I'm blaming you for being upgraded from
The Master to The Doctor. I look forward to more Eurovision parties, fort 
building, and \texttt{<Stephan>}\,\textit{dinner}\,\texttt{</Stephan>} parties together.

Hannah \& Tom: \textit{What can I say except thank you} for so many fun afternoons spent messing about
by Kate's hole. I never laugh as hard as I do when we're together and Fordy
takes all of our pieces in Ludo, we flail around in \textsc{speed!},
get blown up in MarioKart, mess around in the Mii creator, or Tom puts a
black hole right by spawn. While Beales may be long gone, the spirit of the spirit
of \textsc{cmo} will live on with all of us.

To Stephen: My life would absolutely have been different if not
for our friendship dating back to our Broadwater years: no \textit{Age of Empires II},
no \textit{Star Wars: Battlefront II}, no Bon Jovi, no Doctor Who, and no 3am Monopoly! 
Thank you also to the rest of the Dice (\textsc{rip}) crew for many happy
board game evenings together. \textit{We're not old, just older}.

Matt: Thanks for being a great and close friend since Year 8, sorry I ignored you
the first months at Durrington --- but I hope I've redeemed myself somewhat in
the last fourteen years! I may still be slightly salty about not signing up to
the `Nerd Academy' with Debbie and missing out on a free graphical calculator :p

Sam, I also probably wouldn't have embarked on a PhD had it not been for my wonderful
experience as an MPhys student --- of which I have you to thank for a large
portion of that. I couldn't have asked for anyone better to spend many evenings
in the MPhys Lab deeply troubled over that laser with, and I still crack up at
our implementation of \texttt{FedjaFunc} and \texttt{FedjaFit}. Thankfully, I have 
not need to resort to putting pieces of BluTack on my monitor to measure graphs, 
and I've checked that this document is not called anything related to the 
Fabry-P\'erot etalon! The question of \textit{``Whom's't thou mind if I were
to subscript with an eegrick and a YOT?''} still remains unanswered to this day.

I would also like to give a big cheers to all my old and new Sussex friends
for accepting this interloper and for sharing very many Falmer Friday post-grad
pints together. And to my countless old Q-Soc and DocSoc friends; thanks for very many
evenings spent together, which always ended in a trip to Falmer Bar! After all,
\textit{who says you can't go home?}

The absolute highlight of my PhD experience has been the travel to amazing places
for conferences and summer schools. I would like to thank everyone I met, 
interacted, karaoked, or shared a pint or meal with
during my travels for making these events such welcoming and unforgettable to me.
In particular, the memories made in Oslo, Les Houches, Copenhagen, Innsbruck, and Rome with my 
\textit{Euclid} \textsc{swg-boat} buddies (Natalie, Jonathan, Lucy, Josh,
Casey, and Conor) will stay with me for a very, \textit{very} long time.

It goes without saying that I would not be completing a PhD without inspiration
and support from very many teachers and lecturers at school, college, and 
university. In particular, I would like to thank Ms Trignano, Mrs Baker,
Mr Fairbairn, Miss Holt, Miss~Schuler, Bernie Flint, Debbie Collier, 
Nicole Cozens, David Seery, and Kathy Romer.

Thank you to my paw-some companions Billy, Bonnie, Benny, Ralph, and Misty for 
very many fluffy cuddles spent together throughout the years.~\faicon{paw}

Finally, I would also like to thank all developers of free and open-source software (FOSS)
for making their codes, documentation, and examples publicly available allowing
such incredible science to be done by the entire community. Some of the FOSS
software which have been essential to my MPhys and PhD include, but not limited to:
\NumPy~\cite{Numpy:2020array}, \SciPy~\cite{SciPy:2020},
\pandas~\cite{Pandas:2010}, \mpl~\cite{Matplotlib:2007},
\seaborn~\cite{Seaborn:2021}, \Camb~\cite{Lewis:2002ah,Lewis:1999bs,Howlett:2012mh},
\GetDist~\cite{Lewis:2019xzd}, \CCL~\cite{LSSTDarkEnergyScience:2018yem}, 
\Flask~\cite{Xavier:2016elr}, \Glass~\cite{Tessore:2023zyk}, \HEALPix~\cite{Gorski:2004by},
\healpy~\cite{Zonca2019}, \NaMaster~\cite{Alonso:2018jzx}, \Eigen~\cite{eigenweb},
Linux~\faicon{linux}, Ubuntu, Python, and \Cpp; 
with particular thanks to Joe Zuntz for his amazing \Cosmosis framework~\cite{Zuntz:2014csq}. 
A large part of my second and third projects embodied the spirit of Gene Kranz
of \textit{Apollo~13}: \textit{`I don't care what anything was \textsc{designed}
to do, I care what it \textsc{can}~do'},
and \Cosmosis certainly did not let me down --- just like \textit{Aquarius} didn't.

\vspace*{1cm}

\begin{flushright}
    {\LARGE\faicon{heart}} \\
    \textit{---Alex\\
     Gauss House, Sussex\\
     15$^{\textit{th}}$ November 2024}
\end{flushright}


\vspace*{-5cm}
\begin{savequote}[65mm]
  Look at these graphs
  
  \qauthor{---Nickelback}
  \end{savequote}
\vspace*{5cm}
\chapter*{Table of contents}
\vspace*{-4.5cm}
\tableofcontents 









\clearpage
\titleformat{\chapter}[hang]{\filleft\Huge\bfseries}{}{0pt}{\seriffontsemi\textcolor{white}{\setlength{\fboxrule}{25.5pt}\tikzcolorbox{chaptertitle}{chaptertitle}{\thechapter} \\ \vspace*{2.25ex} \Huge\bfseries}}[\vspace{-1ex}\rule{\textwidth}{1.15pt}\vspace*{1.25ex}]
\pagenumbering{arabic}

\begin{savequote}[65mm]
  We choose to go to the Moon!
  We choose to go to the Moon in this decade, 
  not because it is easy,
  but because it is hard; because that goal will serve to
   organize and measure the best of our energies and skills, 
   because that challenge is one that we are willing to accept,
    one we are unwilling to postpone, 
    and one we intend to win.
  \qauthor{---John F\!. Kennedy}
\end{savequote}

\chapter{Introduction} 
\label{chp:introduction}
\textit{Gentlemen, a short view back to the past}.
I think every cosmology PhD student since the Big Bang has said something along 
the lines of `now is a very exciting time to be a cosmologist\dots', however, as much
as it is clichéd to say, \textit{it really is an exciting time to be a
weak-lensing cosmologist!} We are now in an era where the precision of the data
that we are receiving really is \textit{awesome}\footnote{And I mean this in the
traditional dictionary definition: it inspires awe, and fills someone with 
reverential fear, wonder, or respect.}. The exploration and exploitation of this
data will allow us to test the fundamental properties of our Universe with a
precision that has never been possible before. Hence, the statistical methods 
that are applied to this extremely exquisite data need to as accurate, precise,
and robust as the data. It would be a travesty to spend billions of pounds and
tens of millions of human-hours to send the finest telescopes into space,
observe galaxies whose light has taken over ten billion years to get to us, only to analyse the data 
with sub-optimal statistical methods. Thus, the development and verification of
these cosmic shear analysis methods is the fundamental driving force behind the
work presented in this thesis. But before we get bogged down in $E$-/$B$-modes,
the mixing matrix, baryonification methods, and what a `theoretical error' is,
it's good to take stock of just how rapidly cosmology, astrophysics, and 
astronomy have become precision sciences in just over a century.


While humankind may have stared at the nights sky and admired the wonders of
the heavens that was laid out before them for a hundred millennia, it wasn't until Isaac Newton's 
\textit{Principia Mathematica} of 1687 that we started to have the mathematical
framework to describe the observed motion of the heavenly bodies~\cite{Newton:1687eqk}. It is almost
remarkable that it took over two hundred years for us to advance beyond
Newtonian theories of motion, requiring us to wait for the great Albert Einstein
to develop his theory of general relativity (GR) in 1915 to describe the
large-scale properties of our Universe~\cite{Einstein:1916vd}. 

The introduction of general relativity heralded the beginnings of modern
cosmology, where now we could, for the first time, study the fundamental
properties of our Universe, ask and answer questions such as: What was the history of
our Universe? What is our Universe made of? And what will the future of our
Universe be? 

Most of these questions could be answered through the assumption
of the \textit{cosmological principle} and the subsequent development of the
Friedmann-Lemaître-Robertson-Walker metric, which predicted a global expansion
or contraction of the Universe that is governed by the contents within it.
Such an expansion was in direct contradiction with the static Universe theory,
which was a widely held belief at the turn of the twentieth century that
our Universe was infinite in space and eternal in time and was neither 
expanding nor contracting. Einstein constructed his static universe such that
the free parameter $\Lambda$, present in his field equations, was such that it
would oppose the gravitational attraction of the matter within our Universe,
and we would be left with a static Universe~\cite{Einstein:1917ce,ORaifeartaigh:2017uct}. 

While there was no \textit{a priori} theoretical reason not to include $\Lambda$
in the field equations, there was a dearth of experimental evidence which could
discriminate between a static or expanding Universe. This all changed when
Vesto Slipher made observations of the spectral lines on extragalactic nebulae
and found that they were slightly shifted to red or blue colours. These red-
or blue-shifts in the spectra were correctly interpreted as resulting from the
Doppler shift of the motion of the extragalactic nebulae with respect to 
us~\cite{Slipher:1917Obs40304S,Weinberg:1993ftmmW}. These Doppler shifts could
have just indicated that these galaxies just had large peculiar velocities,
which is any motion not resulting from the large-scale bulk flow of the Universe,
and so consistent with the static Universe model. In the next decade, 
Georges Lemaître derived that in an expanding spacetime, there is a simple
direct proportionality between a distant galaxy's recessional velocity and
its distance from Earth~\cite{Lemaitre:1931MNRAS91490L}. Edwin Hubble was
then able to get accurate distance measurements to Slipher's receding galaxies
though observations of Cepheid variable stars, building upon work by 
Henrietta Leavitt and Harlow Shapley~\cite{Weinberg:1993ftmmW,Leavitt:1907AnHar6087L,Leavitt:1912HarCi1731L}. In 1929, Hubble
announced that he found a proportionality between the redshift of galaxies and
their distance to us, thus forming Hubble's law. When combined with the cosmological 
principle, the observations that distant galaxies were receding away from us
with a velocity that was proportional to their distance, became observational
unequivocal proof that our Universe was 
expanding~\cite{Hubble168,Lemaitre:1931MNRAS483L}. These observations
signalled the end of the inclusion of $\Lambda$ in the Einstein equations, since
there was no desire to force a static universe anymore, though its death was
rather short-lived$\dots$

While the derivation of the Friedmann equations and the observational verification
of Hubble's law was a powerful test of Einstein's general relativity, perhaps
the most famous `killer' piece of evidence of Newtonian mechanics' demise, and
establishment of general relativity as the correct framework for our Universe's
dynamics, was the result from the Eddington expedition of 1919~\cite{Dyson:1920abc}. Both general
relativity and Newtonian mechanics predict that matter can deflect
the path of light-rays, though the prediction for the deflection angle in general
relativity is double that of Newtonian mechanics. Since this deflection of light
by matter is equivalent to the presence of an optical lens along the light's path,
this phenomenon is called \textit{gravitational lensing}. By measuring the slight
shift in apparent positions of distant stars by our Sun during the eclipse of 
1919, Eddington and collaborators confirmed that their deflection angles were
consistent with the predictions of general relativity, ushering in the
general relativistic age. What Eddington and his collaborators may not have been
realised at the time is that their primitive observations to test general 
relativity will be re-employed a century later, with unimaginable precision, to
yet again test the physical properties of our Universe~\cite{Bartelmann:1999yn,Kilbinger:2014cea}. 

The first cosmic mystery that general relativity could not explain were the
observations made by Fritz Zwicky who imaged the Coma cluster and, using the
assumption of the virial theorem, found that the average density in the Coma
cluster would have to be at least four hundred times larger than estimates based
on the apparent luminous matter~\cite{Zwicky:1933AcHPh6110Z,Zwicky:2009GReGr41207Z}.
This extra, non-luminous matter was dubbed `dark matter' (originally 
`dunkle Materie' in German) by Zwicky, and the name has stuck ever since.
Zwicky's findings of non-luminous dark matter were followed up by, 
among many other results, observations of galaxy rotation curves by Vera
Rubin and collaborators. Their observations suggested that stars which were
a large radial distance from their galaxy's centre were orbiting too fast for
the gravitational attraction of the apparent luminous matter~\cite{Rubin:1980ApJ471R}.
This either suggests that either the Newtonian limit of general relativity
does not hold on intra-galactic scales, or that there is extra material within
galaxies that exerts gravitational attraction without being luminous. Thanks
to Zwicky, we still call this extra non-luminous gravitational material
`dark matter', and has become an accepted part of the modern standard
cosmological model, even if we cannot explain its exact phenomenology.

The second cosmological mystery which was unravelled at the end of the twentieth
century (a mere four months after my birth!) came from observations of 
high-redshift supernovae, and suggested that they were fainter than otherwise
predicted for a matter-dominated Universe~\cite{Perlmutter:1998np,Riess:1998cb}.
This gave strong evidence that our Universe was not just expanding, as Hubble
and co found, but \textit{accelerating}. The only way to get accelerated
expansion in general relativity is by having the cosmological constant being
the dominant contribution to our Universe's energy density. 
Hence, the cosmological constant $\Lambda$ was back with a vengeance, but 
instead of forcing our Universe to be static, it was acting to accelerate its
expansion!\footnote{I was privileged to attend a talk given by Nobel laureate Brian Schmidt
at the University of Southampton in November 2015. His talk came at a time
when I considering university applications but was as yet undecided what to
study between chemistry, mathematics, and physics. After his talk, I had no
doubt that physics, and more specifically cosmology, was my natural calling.
\textit{The rest, as they say, is history.}}

The detection of both dark matter and dark energy have been corroborated by
numerous independent probes and observational groups, and form the foundations
of our modern standard cosmological model~\cite{Lahav:2022poa}. But the physical properties of
both dark matter and dark energy remain today as mysterious as their first
detection: dark matter is some form of matter that only interacts gravitationally
and contributes around $25 \, \%$ of our Universe's energy density today; dark energy
can be perfectly described by the cosmological constant~$\Lambda$ in the Einstein equation,
and contributes around $70 \, \%$ of our Universe's energy density today~\cite{PlanckCollaboration:2018eyx}.
So we really don't have much of a clue about what $95 \, \%$ of our Universe is!

While our tried and tested method of investigating physical phenomena in our
Universe of simply `looking at it' might not work for dark matter and dark
energy, it does not mean that we have no hope of constraining their behaviour.
Since both interact gravitationally, we can use any probe that is sensitive to
the total amount of matter present, not just the luminous stuff that we're made
from. Such a probe is gravitational lensing, first proved by Eddington and his
collaborators, and experienced a great revival to test dark matter and dark 
energy in the late 1990s~\cite{Schneider:1992bmb,Bartelmann:1999yn}. It is the perfect probe of the dark sector, since
lensing is sensitive to both the amount of matter and evolution of the
energy densities over time, allowing us to constrain the time-evolution of the
dark matter and dark energy fields -- a key goal of the latest cosmic shear 
surveys~\cite{Kaiser:1992ps}. 

Measurements of weak lensing by large-scale structure were first published by four independent groups 
in 2000~\cite{Kaiser:2000if,Bacon:2000sy,vanWaerbeke:2000rm,Wittman:2000tc},
which was made possible by improvements in CCD technology enabling much
wider field of views and increased sensitivity, providing the ideal conditions
for observing large quantities of distant galaxies. Further developments in
telescope and detector technology have enabled weak lensing surveys to go
from covering around $1\,\deg^2$ to nearly $15\,000 \,\deg^2$~\cite{Kilbinger:2014cea},
and has evolved into one cosmology's most precision sciences. 

Hence, we are suitably motivated to go out looking into our cosmos and try
to make measurements of gravitational lensing across the sky and across
cosmic time. However, while Eddington's observations were to measure the
simple deflection in the light from a handful of stars, a modern weak lensing 
survey aims to measure the deflection of light from well over a billion galaxies,
and fraught with intricacies and subtleties that need the correct
modelling to ensure that we reach the correct conclusions from our data. In this
thesis, we investigate the effect of power spectrum estimation techniques
and the impact of baryonic feedback for weak lensing galaxy surveys such that
we can maximise information from them.

\section*{Layout of the thesis}

In Chapter~\ref{chp:background_cosmology} we present background material on
modern cosmology, leading to a discussion of some of the current observational 
probes our Universe.
This feeds into a detailed discussion of gravitational lensing and its application
in cosmology in Chapter~\ref{chp:grav_lensing}, and culminates in a short 
discussion about the state of weak lensing cosmology in 2024 in 
Chapter~\ref{chp:Cosmology_2024}.

We then go on to present new work introducing a new cosmic shear power spectrum
estimation technique in Chapter~\ref{chp:QML_estimator}, and work on mitigating
the effects of baryonic feedback physics on cosmic shear surveys through a 
theoretical error covariance in Chapter~\ref{chp:baryonic_effects} and
binary scale cuts in Chapter~\ref{chp:binary_cuts}. 

We end on a short discussion and conclusions
about the work presented in this thesis with an outlook for the future
in Chapter~\ref{chp:conclusions}.

\section*{Notation}

In this thesis we use natural units, where $c = \hbar = k_{\textsc{b}} = 1$, and the `mostly
negative' metric signature $(+, -, -, -)$ for coordinates $(t, \,x_1, \,x_2, \,x_3)$. 
Latin indices from the start of the alphabet $(a, \, b, \, c, \,\dots)$ represent
space-time indices, while Latin indices from the middle of the alphabet 
$(i, \, j, \, k, \,\dots)$ represent spatial indices only.

\clearpage

\begin{savequote}[65mm]
  Ah, shit, \\
  the shipping and receiving!

  \qauthor{---Sips}

  Crabs are people, \\
  clams are people!

  \qauthor{---Lewis Brindley}
\end{savequote}
\chapter{Modern cosmology}
\label{chp:background_cosmology}
\vspace*{-1cm}
\begin{mytext}
	\textbf{Outline.} In this chapter, I present a brief overview of the ideas
	and motivations of modern cosmology starting from general relativity,
	deriving various well-known results along the way, and ending with 
	presentation and discussion of key results from several observational
	probes which form the foundation of modern cosmology. \textit{Okay, we do it.}
\end{mytext}

\section{Cosmology from general relativity}

The era of modern cosmology started with the introduction of general relativity
(GR) by Einstein in 1915~\cite{Einstein:1916vd}. This provided insights into the
underlying nature of our Universe, such as how energy, matter, and space-time are
related, and that gravity is nothing more than the curvature of space-time. One
of the key equations underpinning general relativity are the Einstein equations, which are a
series of ten coupled non-linear partial differential equations, and can be written
as~\cite{Hobson:2006se}
\begin{align}
	R_{ab} - \frac{1}{2} R g_{ab} + \Lambda g_{ab} = 8 \pi G \,  T_{ab},
	\label{eqn:Einstein_fild_eqn}
\end{align}
where $g_{ab}$ is the space-time metric, $R_{ab}$ is the Ricci tensor,
$R$ is the Ricci scalar (which are both built from derivatives of the
metric), $T_{ab}$ is the energy-momentum tensor, and $\Lambda$ is a constant.
This constant is the famous cosmological constant, or `fudge factor', that
Einstein introduced to make his model of the Universe static. The raw equations
are agnostic to $\Lambda$'s value; it could be zero, close to zero, or highly
non-zero. From theory alone, there is no \textit{a prior} constraint that can
be placed on the value of $\Lambda$. Einstein, realising that his model of the
Universe predicted one that expanded, refused to believe that such Universe
existed, and so ansatzed\footnote{\textit{Ansatz the answer}} the value of $\Lambda$ to be exactly that which gave
a static evolution of the Universe. This static Universe was unstable with the
dynamics of the Universe hanging in a delicate balance, with the slightest
perturbation leading to expansion or contraction. 

With Edwin Hubble's 1929 work finding that the Universe was expanding, and thus
not a steady-state solution \cite{Hubble168}, Einstein abandoned the cosmological
constant as it was no longer necessary --- dubbing his inclusion of it in the first 
place as his ``greatest mistake''. However, this was not the end of
the cosmological constant's story as we will see shortly.

\subsection{Ricci curvature}

To start with our analysis of the Einstein field equations (Equation~\ref{eqn:Einstein_fild_eqn}),
let us look at the left-hand side of the equation. This is dubbed the `geometry
side', as it encodes the geometrical properties of a system. We start out
with the Ricci scalar, $R$, which is a contraction of the Ricci tensor,
\begin{align}
	R \equiv \tensor{R}{^{a}_{a}}.
\end{align}
We can now go one step up and ask how do we define the Ricci tensor? We find
that it is a contraction over the first and third indices of the Riemann tensor\footnote{Yes, it is confusing having three different quantities all with the symbol $R$, however we can tell which one we are using by the number of indices each term is carrying.}
$\tensor{R}{^a_{bcd}}$ as
\begin{align}
	R_{ab} \equiv \tensor{R}{^c_{acb}}.
\end{align}
We now go one step higher and ask what the Riemann tensor is made out of, and we
find that it's built from products and derivatives of connection coefficients,
or Christoffel symbols, $\Gamma^{a}_{bc}$ as
\begin{align}
	\tensor{R}{^a_b_c_d} = \partial_c \Gamma^a_{bd} - \partial_d \Gamma^a_{bc} + \Gamma^a_{cf} \Gamma^f_{bd} - \Gamma^a_{df} \Gamma^f_{bc}.
\end{align}
We can \textit{finally} ask what are the connection coefficients\footnote{Despite having indices, the connection coefficients are not tensors since \textit{they do not transform like a tensor}.} made from, and
we happily find that they're computed from derivatives of the space-time metric
as
\begin{align}
	\Gamma^a_{bc} = \frac{1}{2} g^{a d} \left[ \partial_b g_{dc} + \partial_c g_{bd} - \partial_d g_{bc} \right].
\end{align}
Thus, starting from a space-time metric, we can fully derive the geometry
part of the Einstein equations. We are now lead down a new rabbit hole as we try
to work out how to construct our space-time metric\dots

\subsection{The space-time metric}

As shown above, one attempts to solve the Einstein field equations to obtain
the space-time metric $g_{ab}$. The metric, $g_{ab}$, is usually written as the
space-time element, $\d s^2$, of
\begin{align}
	\d s^2 \equiv g_{ab} \d x^a \d x^b,
\end{align}
where $\d x$ are the infinitesimal elements of our chosen coordinate system.
The space-time element is especially useful since it is an invariant and thus
the same for all observers.

\subsubsection{The Schwarzschild metric}

For an arbitrary physical system, a general solution of the Einstein equations
would be analytically intractable. To simplify the problem, one can
place certain constraints, or symmetries, on the physical problem to arrive at
solutions for the metric. The first such solution for the metric came from
Karl Schwarzschild in 1916, arriving at the Schwarzschild metric for the
gravitational field generated by a point mass of mass
$M$~\cite{Schwarzschild:1916Abh}. This solution is obtained by noting that
outside of the point mass, the region that we are interested in, the
energy-momentum tensor $T_{ab}$ is zero. Additionally, we note the spherical
symmetry of the problem, and thus spherical coordinates
$(t, r, \vartheta, \phi)$ are the natural choice. In these coordinates, the
Schwarzschild metric is given as~\cite{Hobson:2006se}
\begin{align}
	\d s^2 = \left( 1 - \frac{2 G M}{r} \right) \d t^2
	- \left( 1 - \frac{2 G M}{r} \right)^{-1} \! \d r^2
	- r^2 \left[ \d \vartheta^2 + \sin^2 \vartheta \, \d \phi^2 \right].
	\label{eqn:Schwarzschild_metric}
\end{align}

\subsubsection{The Friedmann-Robertson-Walker metric}

While the Schwarzschild metric was solved in a vacuum, where the energy-momentum
tensor vanishes ($T_{ab} = 0$), one can try and find a solution to the Einstein
equations where this condition is not true. This is motivated by the fact that
our Universe is not totally empty and does feature stuff in it, such as me and
you! As discussed above, a general solution for an arbitrary universe's energy
distribution is totally intractable, so we need to place some constraints on
our system if we are to obtain a solution.

The first such assumption is that we do not hold any special place in the
Universe, and thus any such observations that we make could have been made from
any other position in the Universe. This is the Copernican principle, or the
modern assumption of \textit{homogeneity}.

The second assumption is that when we make observations of the Universe, such as
temperature maps of the cosmic microwave background (CMB), or the distribution of
galaxies on the sky, the Universe appears to look very much the same in every
direction. For example, the temperature of the CMB radiation was measured to be
within one part in one hundred thousand by the FIRAS instrument on the
\textit{COBE} satellite~\cite{COBE:1993ij}. This extreme uniformity across the
entire sky gives rise to the assumption of \textit{isotropy}. Though, of course,
there is a significant dipole term which arises from the Solar System's
movement through the universe, which has been measured to be around $370 \, 
\textrm{km/s}$, which is a whopping $0.1\,\%$ of the speed of light,
with respect to the CMB rest frame \cite{Planck:2018nkj}. This dipole is almost
always removed in any cosmological analyses using CMB data.

It is important to note that homogeneity and isotropy are different properties,
and a system obeying one does not necessarily obey the other. For example,
consider the electric field between two infinite plates with potential difference
$\Delta V$ and separation $d$. The electric field is then given by $\vec{E} = \frac{\Delta V}
{d} \, \hat{n}$, where $\hat{n}$ points towards the negative plate~\cite{Griffiths_2023}. This is
clearly homogenous everywhere between the two plates, but clearly not isotropic
as the field gives rise to a preferential direction. Equally, we can consider
the electric field of a point-charge $q$ at the origin. As a function of radial
distance $r$, the field is $\vec{E}(r) = \frac{1}{4\pi \epsilon_0} \, \frac{q}{r^2} \, \hat{r}$.
At the origin, this is clearly an isotropic field as it looks the same in
every direction, but is not homogenous. The observations that our Universe
obeys \textit{both} homogeneity and isotropy was a cornerstone of modern
observational cosmology. 

Together, the joint assumptions of homogeneity and isotropy form the
modern \textit{cosmological principle}, from which much of modern cosmology is
built upon. Imposing this cosmological principle, with a non-zero energy density,
gives rise to the Friedmann-Robertson-Walker (FRW) metric, written in spherical
polars, of
\begin{align}
	\d s^2 = \d t^2 - a^2(t) \left( \frac{\d r^2}{1 - k r^2} +
	r^2 \! \left[ \d \vartheta^2 + \sin^2 \! \vartheta \, \d \phi^2 \right] \right)\!,
	\label{eqn:FRW_metric}
\end{align}
where $a(t)$ is the scale-factor (which describes the homogenous expansion of
space with cosmic time $t$, normalised such that it is unity at present times:
$a(t_0) = 1$), and $k$ is the curvature parameter which describes the
spatial geometry of the universe: flat ($k=0$), closed ($k > 0$), or open
($k < 0$). A subtlety in our FRW metric of Equation~\ref{eqn:FRW_metric} is that
$r$ is the dimensionful comoving coordinate. That is, objects that simply
move with the bulk Hubble flow have fixed comoving coordinates. The physical
distance at time $t$ is $a(t) r$.

Since the radial element of our FRW metric changes depending on the value of
the curvature of the universe, we can transform our FRW metric to one where the
radial comoving coordinate is given by $\rchi$ and is independent of curvature.
This gives~\cite{Hobson:2006se}
\begin{align}
	\d s^2 = \d t^2 - a^2(t) \left( \d \rchi^2 +
	f_{k}^2(\rchi) \! \left[ \d \vartheta^2 + \sin^2 \! \vartheta \, \d \phi^2 \right] \right)\!,
	\label{eqn:modified_FRW_metric}
\end{align}
where the function $r = f_{k}(\rchi)$ is given by
\begin{align}
	f_{k}(\rchi) \, = \, \begin{cases}
		         \sin \rchi     & \textrm{for } k > 0,          	\\
		         \rchi			& \textrm{for } k = 0,			\\
		         \sinh \rchi	& \textrm{for } k < 0.
	         \end{cases}
	\label{eqn:f_k_chi}
\end{align}

\subsubsection{The perturbed metric}

While the FRW metric is very useful to describe the large-scale homogenous
properties of the universe, many of which will be derived just later, it is
important to note that the Universe is \textit{not} totally homogenous. For
example, the air in the room that you are reading this thesis has an
average density of $1.2 \, \textrm{kg m}^{-3}$ so when compared to the
average density of the universe of approximately
$10^{-26} \, \textrm{kg m}^{-3}$, this represents an extraordinary perturbation
of the order $10^{26}$. While none of the cosmological perturbations will come
close to this size, we can extend the FRW metric to include perturbations
as
\begin{align}
	\d s^2 = \left(1 + 2 \Psi \right) \d t^2 - a^2(t) \left(1 - 2 \Phi\right) \d \vec{x}^2,
	\label{eqn:perturbed_FRW_metric}
\end{align}
where $\Psi$ and $\Phi$ are the Bardeen potentials, and are functions of the
four space-time coordinates, and describe the first-order perturbations to
a homogenous universe, and $\d \vec{x}^2$ is the spatial-only FRW metric.
We will revisit this perturbed metric when we consider space-time perturbations
later.

\subsection{The Friedmann equations}

When we introduced the FRW metric in Equation~\ref{eqn:FRW_metric}, we
introduced a function $a(t)$ which we invoked was the scale-factor of the
universe. We can now ask many questions about the properties of this
scale-factor: How does it behave to different matter densities? How does
curvature affect the evolution of the scale-factor? And what physical intuition
can we obtain from the scale-factor?

To answer these questions, we turn to the Einstein field equations. Since we
have the key ingredient, the metric, we can start solving these field
equations. Recalling that the right-hand side of the field equations
(Equation~\ref{eqn:Einstein_fild_eqn}) was in terms of the energy-momentum
tensor, we need to find such energy-momentum tensor that describes the Universe.
While this may seem like another intractable question, we turn to a
cosmologist's favourite activity: assumptions! Since we are only after the
general solution for a homogenous and isotropic universe (remember we are
ignoring perturbations for the time being), we can describe the Universe as
being a perfect fluid. This perfect fluid is characterised by having an
energy density $\rho$ and pressure $p$, both of which are functions of cosmic
time $t$ alone, since we are following the cosmological principle. This
perfect fluid yields an energy-momentum tensor of the form
\begin{align}
	T_{ab} = (\rho + p) u_a u_b + p g_{ab},
	\label{eqn:energy_mot_tensor_univ}
\end{align}
where $u_a$ is the four-velocity which, for a fluid at rest in our comoving
coordinate basis, is simply $u_a = (1, 0, 0, 0)$. Since we are dealing with
an FRW universe, no time or positional-dependence can be included in our 
energy-momentum tensor.

\subsubsection{The Friedmann equation}

Now that we have both the metric and the energy-momentum tensor, we arrive at
the first Friedmann equation (often called just the Friedmann equation)
derived from the $tt$-component of the Einstein field equations of
\begin{align}
	3 H^2 = 8 \pi G \, \rho -\frac{3 k}{a^2} + \Lambda,
	\label{eqn:Friedmann_eqn}
\end{align}
where we have introduced the Hubble parameter $H$ defined as
$H \equiv \dot{a} / a$, which measures the relative expansion rate of the
universe, where $\dot{a} \equiv \d a / \d t$ -- the derivative of
the scale-factor with respect to cosmic time.

\subsubsection{The acceleration equation}

Differentiating the Friedmann equation, and using the conservation of mass and 
energy, yields the second Friedmann equation, called the acceleration equation,
of
\begin{align}
	3\frac{\ddot{a}}{a} = - 4 \pi G \left( \rho + 3 p \right) + \Lambda.
	\label{eqn:acceleation_eqn}
\end{align}
One immediate result of the acceleration equation is that clear to see, is that both
the energy density $\rho$ and pressure $p$ of any matter in our Universe will
act to decelerate the expansion of the universe, whereas the cosmological
constant $\Lambda$, depending on its sign, can act to accelerate ($\Lambda > 0$)
or decelerate ($\Lambda < 0$) the universe. Thus, the sign of $\Lambda$ and its
relative importance when compared to $\rho$ and $p$ are key to understanding the
dynamics of a universe, which we will see shortly.

\subsubsection{The fluid equation}

With our energy-momentum tensor of Equation~\ref{eqn:energy_mot_tensor_univ},
we can apply the conservation of energy-momentum ($\nabla^{a} T_{ab} = 0$, where
$\nabla^a$ is the covariant derivative), to find a third Friedmann equation, 
often called the fluid or continuity equation, of
\begin{align}
	\dot{\rho} + 3 H (\rho + p) = 0.
	\label{eqn:fluid_eqn}
\end{align}
It is important to note that the three Friedmann equations are not independent
of each other, and so, in general, only two are used to obtain solutions for the
dynamics of the scale-factor.

\subsection{Solutions to the Friedmann equations}
\label{sec:Friedmann_solutions}

Now that we are armed with the Friedmann, acceleration, and continuity equations, we can investigate
solutions to the Friedmann equations --- that is, solving for the time evolution
of the scale-factor $a(t)$ --- given different dominating components of our
cosmological fluid. In our three scenarios, we will consider flat universes
($k = 0$) only. A discussion on curvature will come thereafter.

\subsubsection{Cold matter dominated}

The first solution that we will investigate is one of a universe that is
dominated by a cold, pressureless matter. Here, the `pressureless' qualifier
gives us the condition that $\rhomatter \gg p_{\textsc{m}}$, and thus we can
ignore any pressure terms in the Friedmann equations for this matter. Solving
the fluid equation, we arrive at
\begin{align}
	\rhomatter(t) = \rho_{\textsc{m,0}} \, a^{-3}(t),
\end{align}
where $\rho_{\textsc{m,0}}$ is the density of this matter today. It is important
to note the triviality of this result, if energy is conserved (which it
\textit{always} is), then the energy density evolves with time as the inverse of
volume and thus we arrive at the negative cubic power present on the
scale-factor.

With our solution for the evolution of our matter density, we can plug this
into the Friedmann equation to obtain
\begin{align}
	a(t) = \left[\frac{t}{t_0}\right]^{2/3}\!\!\!\!\!\!\!,
	\label{eqn:matter_dom_evo}
\end{align}
where $t_0$ is the age of the universe today, and that we have normalised the
scale-factor to be unity today ($a(t_0) = 1$).

\subsubsection{Radiation dominated}

Unlike our cold matter, radiation has a non-negligible pressure term which is
related to its energy density through $p_\textsc{r} = \rho_{\textsc{r}} / 3$. 
Again, solving the fluid
equation we find the evolution of our radiation of
\begin{align}
	\rho_{\textsc{r}}(t) = \rho_{\textsc{r,0}} \, a^{-4}(t),
\end{align}
and so we find that the radiation energy density decays with an additional
factor of $1/a$ compared to our matter, this corresponds to the cosmological
redshifting of the photons --- a phenomena that we will investigate shortly.

Again, now what we have the time evolution of our fluid, the Friedmann equation
gives the scale-factor evolution as
\begin{align}
	a(t) = \left[\frac{t}{t_0}\right]^{1/2}\!\!\!\!\!\!\!,
	\label{eqn:radiation_dom_evo}
\end{align}
and thus we see that a radiation-dominated universe expands slower than a
matter-dominated one.

\subsubsection{\boldmath$\Lambda$ dominated}

A fairly ominous term that's been appearing in many of our equations thus far
has been $\Lambda$, dubbed the cosmological constant\footnote{\textit{Somehow, \st{Palpatine} the cosmological constant returned}}. But what are the
properties of this cosmological constant, and what would the dynamics of a
universe that is dominated by it look like?

If $\Lambda$ is truly a constant, then $\dot{\rho}_{\Lambda}$ would be zero, since there
would be no evolution in its density $\rho_{\Lambda}$, given as
$\rho_{\Lambda} = \Lambda / 8\pi G$. Thus, when comparing to the fluid
equation of Equation~\ref{eqn:fluid_eqn}, we find that we require
$p_{\Lambda} + \rho_{\Lambda} = 0$, or equally $p_{\Lambda} = -\rho_{\Lambda}$,
for our constant. This has the odd physical property that our cosmological
constant fluid has a \textit{negative} pressure, akin to tension in a stretched 
rubber band, but nevertheless makes physical sense. For our cosmological
constant dominated universe, the Friedmann equation becomes
\begin{align}
	\left(\frac{\dot{a}}{a}\right)^2 = \frac{\Lambda}{3},
\end{align}
which has the simple exponential solution of
\begin{align}
	a(t) \propto \exp \left[ \sqrt{\frac{\Lambda}{3}} \, t \right]\!.
	\label{eqn:scale_factor_lambda}
\end{align}
This solution is also called \textit{de Sitter} space, and shows that a 
$\Lambda$-dominated universe will expand exponentially forever.

\subsubsection{Arbitrary fluids}

Looking at our three solutions for the scale-factor above, while the physical
properties of our three cosmological fluids might be quite different, if one
looks hard enough, they'll find only one common difference: the pressure term.
We can express the pressure of a fluid in terms of its energy density through
the equation of state
\begin{align}
	p_{i} = w_{i} \rho_{i}
\end{align}
for any arbitrary fluid $i$. Here, $w$ is the \textit{equation of state
	parameter}, which takes values for our three scenarios of
\begin{align*}
	w_i \, = \, \begin{cases}
		         w_\textsc{m} = 0           & \textrm{cold, pressureless matter,}           \\
		         w_\textsc{r} = \frac{1}{3} & \textrm{radiation \& relativistic particles,} \\
		         w_\Lambda = -1             & \textrm{cosmological constant.}
	         \end{cases}
\end{align*}

If we now assume that the universe is dominated by our
arbitrary fluid $i$, then we can re-write the acceleration equation
(Equation~\ref{eqn:acceleation_eqn}) in terms of its equation of state $w_i$,
to give
\begin{align}
	\frac{\ddot{a}}{a} = - \frac{4 \pi G}{3} (1 + 3 w_i) \rho_i,
	\label{eqn:acceleration_eos}
\end{align}
and so we find that if our fluid's equation of state satisfies $w_i < -\frac{1}{3}$
then $\ddot{a} > 0$ and thus we get accelerated expansion.

\begin{figure}[t]
    \centering
    \includegraphics[width=\linewidth,trim={0.0cm 0.0cm 0.0cm 0.0cm},clip]{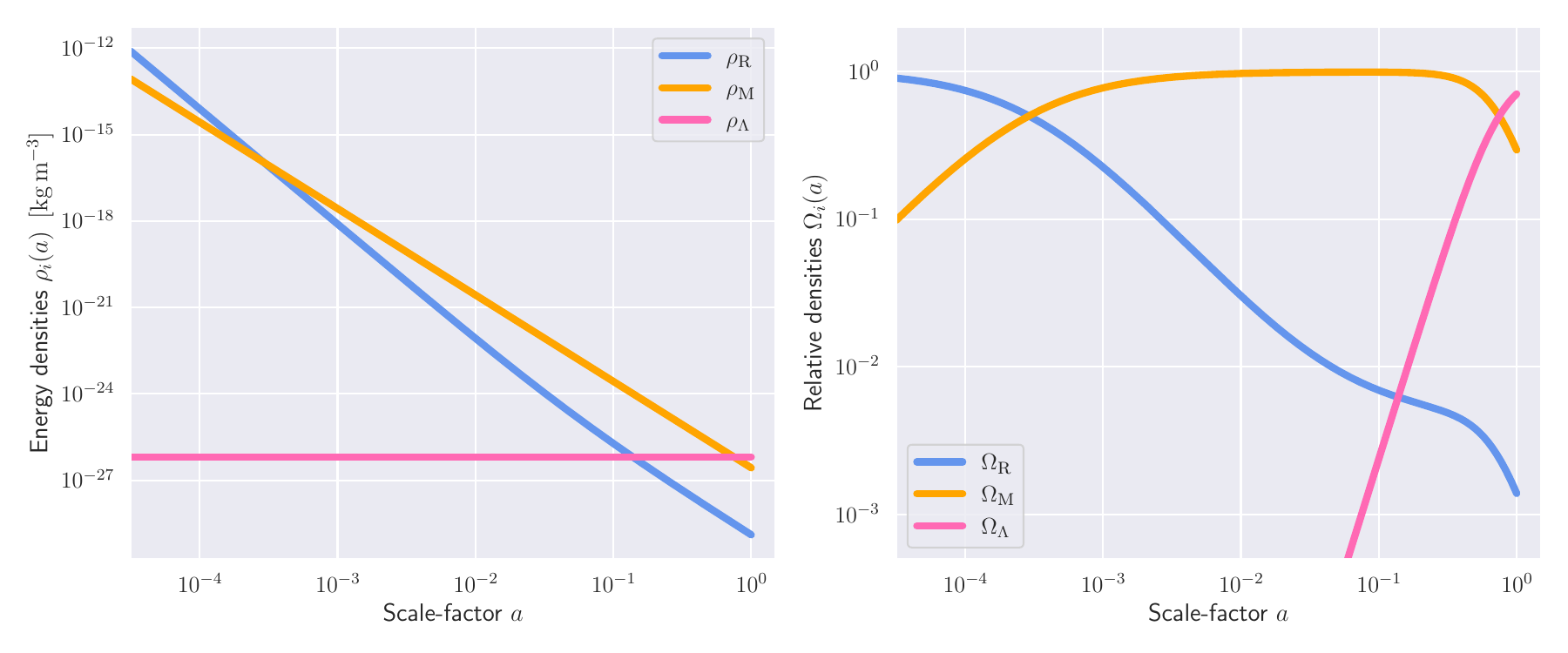}
    \caption{Evolution of the energy density (left panel) and relative density 
	(right panel) of the three main components of
	our Universe: radiation, matter, and the cosmological constant. We see that
	at early times (small $a$), the universe was radiation dominated. This was
	quickly redshifted away leading to an extended period of matter domination.
	It was only relatively recently that we entered the epoch of 
	$\Lambda$-domination, and thus hit the turning point of accelerated 
	exponential expansion.}
    \label{fig:Energy_densitites}
\end{figure}

\subsubsection{Dark energy equation of state}

\begin{figure}[t]
    \centering
    \includegraphics[width=\linewidth,trim={0.0cm 0.0cm 0.0cm 0.0cm},clip]{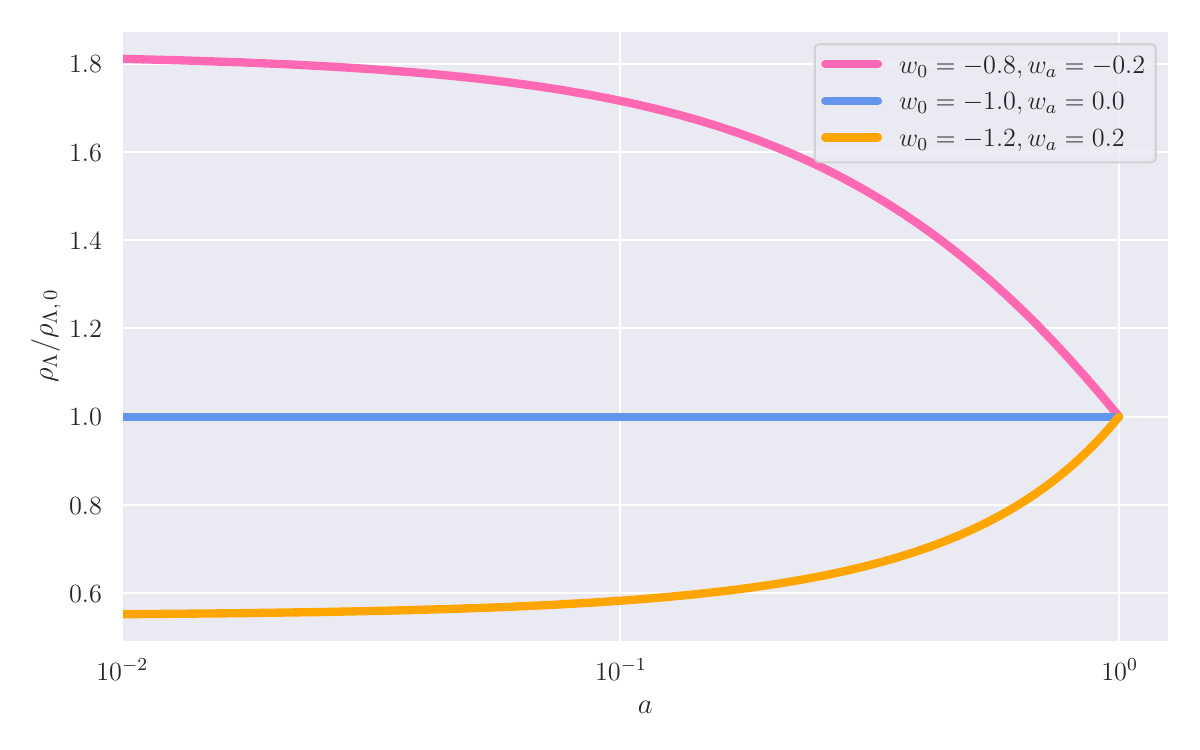}
    \caption{Evolution of the dark energy density for three different
		combinations of $w_0$ and $w_a$ values (Equation~\ref{eqn:w0_wa}).
		The blue curve corresponds to the cosmological constant~$\Lambda$,
		showing constant energy density over time. The non-constant curve's
		evolution can be explained through Equation~\ref{eqn:de_time_evo}.
		}
    \label{fig:dark_energy_density}
\end{figure}

Thus far, we have been dealing with a static cosmological constant $\Lambda$
present in the Einstein equations. Since this cosmological constant contributes
around $70 \, \%$ of our Universe's energy density, and cannot be directly 
observed, it was coined `dark energy' to mimic the equally unobservable
dark matter~\cite{Huterer:1998qv}. While $\Lambda$ may be a constant in the
Einstein equations, we can entertain the possibility that dark energy
might not be a complete constant. If we allow for dark energy to have a time
dependence, $w_{\Lambda} = w_{\Lambda}(t)$\footnote{We shall use $\Lambda$ to
represent a generic dark energy fluid which acts to accelerate our Universe,
even if it is not a complete constant.}, a simple first-order linear expansion
for its equation of state with cosmic expansion is~\cite{Linder:2002et,Linder:2005in}
\begin{align}
	w_{\Lambda}(a) = w_0 + w_a \left(1 - a\right).
	\label{eqn:w0_wa}
\end{align}
The cosmological constant then becomes a sub-class of our evolving dark energy
model, and corresponds to $w_0 = -1$ and $w_a = 0$. We plot three different
combinations of ($w_0, w_a$) in Figure~\ref{fig:dark_energy_density}. We can
explain the different evolution of our three curves by re-casting the
fluid equation (Equation~\ref{eqn:fluid_eqn}) into
\begin{align}
	\dot{\rho}_{\Lambda} = - 3 H \rho_{\Lambda} \left(1 + w_{\Lambda}\right),
	\label{eqn:de_time_evo}
\end{align} 
and so if $w=-1$, then we arrive at our cosmological constant with its constant
energy density. Likewise, we find that if $w < -1$ then the dark energy
density grows with time ($\dot{\rho}_{\Lambda} > 0$), and if $w > -1$ then
the dark energy density decreases with time ($\dot{\rho}_{\Lambda} < 0$). The dark
energy field supports accelerated expansion as long as $w_{\Lambda} < -\frac{1}{3}$.

\subsubsection{Curvature}

In our three scenarios above, we considered the case for a spatially flat
universe (one where $k = 0$). There is no \textit{a prioiri} reason that we
should consider flat universes only since the Einstein field equations are
valid for all values of $k$. However, one can consider the properties of
such a flat universe. If we impose that $k = 0$ on the Friedmann equation of
Equation~\ref{eqn:Friedmann_eqn}, then we find that the total energy density
$\rho = \rhomatter + \rhorad + \rho_{\Lambda}$ is constrained to be
that of the \textit{critical density} $\rhocrit$ of
\begin{align}
	\rhocrit(t) \equiv \frac{3 H^2(t)}{8 \pi G}.
\end{align}
Since the Hubble parameter $H$ depends on time, so will the critical density.
If we use that the Hubble parameter today is around $H_0 \simeq 70 \, \textrm{km/s} \,\,
\textrm{Mpc}^{-1}$, then we find that the Universe's critical density today is
around $9 \times 10^{-30} \, \textrm{g} \, \textrm{cm}^{-3}$, or about
five hydrogen atoms per cubic metre.

From the critical density, one can define the \textit{density parameters} of
each component of the cosmological fluid $\Omega_i$, defined as
\begin{align}
	\Omega_i(t) \equiv \frac{\rho_i(t)}{\rhocrit(t)}.
	\label{eqn:Omega_i}
\end{align}
From this, one can define a density parameter associated with the curvature of
an arbitrary universe $\Omega_k$, defined as
\begin{align}
	\Omega_k(z) = -\frac{k}{a^2 H^2},
\end{align}
such that the sum of the density parameters is defined to be unity
\begin{align}
	\Omega_{\textsc{tot}} = \Omega_{\textsc{r}}(z) + \Omega_{\textsc{m}}(z)
	+ \Omega_{\Lambda}(z) + \Omega_{k}(z) \equiv 1.
\end{align}

\begin{figure}[t]
    \centering
    \includegraphics[width=0.85\linewidth,trim={0.0cm 0.0cm 0.0cm 0.0cm},clip]{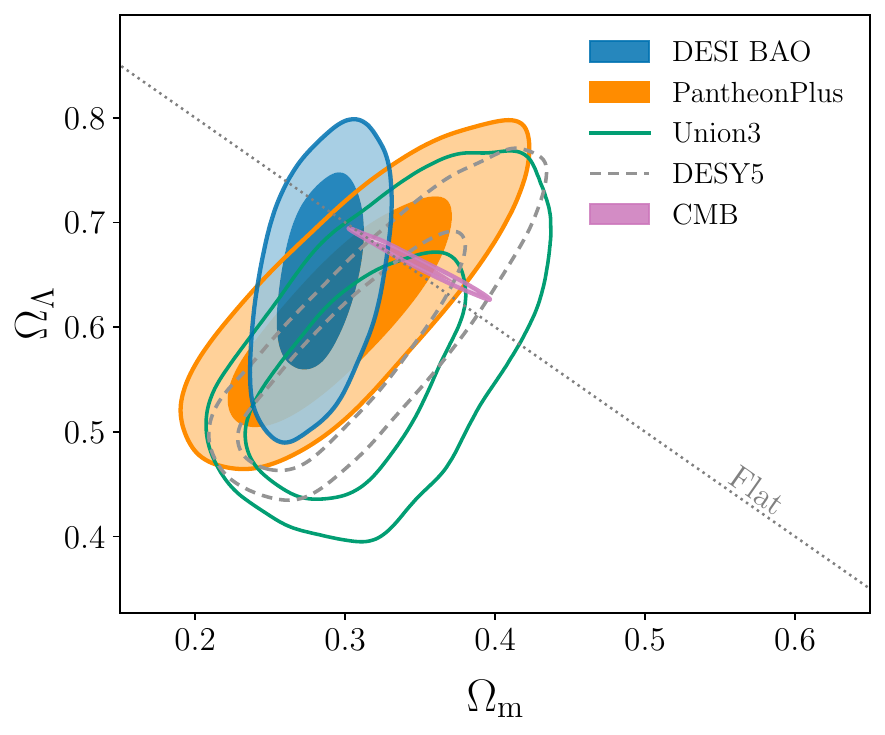}
    \caption{Parameter constraints on the matter ($\Omegam$) and dark energy
	densities ($\Omega_{\Lambda}$) from
	a number of different cosmological probes. While each individual probe's average
	value might vary from one another, all agree that $\Omega_{\Lambda} > 0$, we are living in
	an accelerating universe which is consistent with a flat universe of $\Omega_k = 0$.
	We introduce these cosmological probes in Section~\ref{sec:probe_of_lss}.
	Figure taken from the DESI 2024 release~\cite{DESI:2024mwx}.}
    \label{fig:DESI_omega}
\end{figure}

We measure values of $\Omega_k$ that are consistent with a flat universe of
$\Omega_k = 0$ (see Figure~\ref{fig:DESI_omega}),
and thus most analyses specialise to the flat-only case.
However, in general, it should be a free parameter in any cosmological model.

\subsubsection{A note on notation}

The density parameter for each component of the cosmological fluid, the $\Omega_i$'s,
are time / redshift / scale-factor dependent, and thus should be written as
$\Omega_i(t)$. Their values today, at $t = t_0$, should be written as
$\Omega_{i_0}$, however since cosmologists are lazy people, the subscript zero
is usually dropped and thus $\Omegam$ should be interpreted as the 
matter density parameter today, likewise for $\Omegar$ and $\Omega_{\Lambda}$.



\subsection{Cosmological redshift}

One of the most basic fundamentals to modern cosmology is our ability to collect
photons of all different wavelengths and from different epochs in our Universe's
history. Thus, understanding the physics of what happens to these photons
between when they were emitted and detected by us is key to ensuring that we
can deduce the correct properties of our Universe from these observations. One
such key property to these propagating photons is the phenomena of
cosmological redshift, which we will derive here.

Firstly, consider a source of photons at rest (in its reference frame) at
fixed radial comoving distance $\rchi$. Since photons travel along null
geodesics ($\d s ^2 = 0$), we find for a radially incoming photon that
\begin{align}
	\d t = -a(t) \, \d \rchi
\end{align}
Integrating the total comoving distance from us (the observer at $\rchi = 0$)
out to the source at $\rchi = \rchi$, we find
\begin{align}
	\rchi = \int_{0}^{\chi} \d \rchi' = -\int_{t_2}^{t_1} \frac{\d t}{a(t)} = \int_{t_1}^{t_2} \frac{\d t}{a(t)},
\end{align}
since the photon was emitted at time $t = t_1$ and we observed the photon
at $t = t_2$ and $\rchi = 0$. Thinking of photons not as of particles, but of
continuous electromagnetic waves, $t_1$ also corresponds to the time that a
crest of the wave is emitted. A second crest of the wave will be emitted
$\delta t = 1 / \nu_1$ later, where $\nu_1$ is the frequency in the source rest
frame. This will be detected by us at the origin at time $t_2 + \delta t_2$.
Since the comoving radial distance is invariant, we find
\begin{align}
	\rchi = \int_{t_1}^{t_2} \frac{\d t}{a(t)} =
	\int_{t_1 + \delta t_1}^{t_2 + \delta t_2} \frac{\d t}{a(t)}.
\end{align}
For this to hold true, we require that
\begin{align}
	\frac{\delta t_2}{a(t_2)} - \frac{\delta t_1}{a(t_1)} = 0.
\end{align}
This gives us the result for cosmological redshifting of
\begin{align}
	\frac{\delta t_2}{\delta t_1} = \frac{\nu_1}{\nu_2} = \frac{a(t_2)}{a(t_1)}.
\end{align}
Expressing this in terms of the wavelength of the light emitted ($\lambda_1$)
and detected ($\lambda_2$), we find that the wavelength of the detected photons
are given by
\begin{align}
	\lambda_2 = \lambda_1 \frac{a(t_2)}{a(t_1)}.
\end{align}
Thus, the properties of the detected light now depend on the evolution of the
scale-factor of the universe. If, for example, a universe was expanding
($a(t_2) > a(t_1)$), then the wavelength of the detected light will be larger
than that of when it was emitted, i.e. it was shifted to the redder part of the
electromagnetic spectrum -- and thus the term redshift was born. If both
$\lambda_1$ and $\lambda_2$ are known (say they correspond to a known
elements emission line), then one can assign \textit{a redshift} to the source
of
\begin{align}
	1 + z = \frac{\lambda_2}{\lambda_1}.
	\label{eqn:redshift_def}
\end{align}
Additionally, since we have normalised the scale-factor to be unity today
(recall $a(t_0) = 1$), then we find
\begin{align}
	a = \frac{1}{1 + z},
\end{align}
a very useful relation between redshift and the scale-factor.

\subsection{The generalised Hubble parameter}

The density parameter for each component of the cosmological
fluid introduced in Equation~\ref{eqn:Omega_i}, $\Omega_i$, becomes extremely
useful when we want to manipulate the first Friedmann equation
(Equation~\ref{eqn:Friedmann_eqn}) into the generalised Hubble parameter $H(z)$.
By noting the redshift evolution of each individual component, we can write
the Hubble parameter as
\begin{align}
	H^2(z) = H_0^2 \left( \Omega_{\textsc{r}_0} (1 + z)^4
	+ \Omega_{\textsc{m}_0} (1 + z)^3 + \Omega_{\Lambda_0} 
	+ \Omega_{k_0} (1 + z)^2 \right).
	\label{eqn:generalised_Hubble}
\end{align}
This form of the Hubble parameter becomes incredibly useful when dealing with
distances in cosmology, since it can be anchored to values observed today.

\subsection{Cosmological distances}
\label{sec:cosmo_distances}

Large amounts of observational cosmology rely on us being able to measure
accurate distances to objects, such as galaxies or galaxy clusters. While
accurate measurements are essential, a solid theoretical understanding of
these distances are also key to correctly interpret these distances. Measuring
distances in a flat, Euclidean space-time are simple; we can produce different
estimates for distance to an object using different methods and they will all
agree, to within the measurement's statistical errors. However, when we go to
an expanding space-time this no longer holds, and so different measurement
techniques yield different result for distances to the same object. Here, we
discuss different measurement techniques to astronomical and cosmological
objects and how they are related to each other.

\subsubsection{Comoving distances}

The comoving distance to an object at redshift $z$ can be given in terms of
an integral over the Hubble parameter as
\begin{align}
	\rchi = \int_{0}^{z} \frac{\d z'}{H(z')},
\end{align}
and hence if one can accurately measure redshifts and comoving distances to
objects, then one can constrain the evolution of the Hubble parameter -- a key
goal of modern cosmology.

\subsubsection{Hubble velocity distance}

Another simple distance measure is the Hubble velocity distance
($d_{\textsc{h}}$), which only works for local objects ($z \ll 1$). Here, we
can re-write Hubble's law to find
\begin{align}
	d_{\textsc{h}} = \frac{c z}{H_0}.
\end{align}

\subsubsection{Angular diameter distance}

The angular diameter distance is useful when we measure objects that have
a known physical size and want to relate it to the angular size that it appears
from at Earth. If we have an object of proper length $l$ at
distance $d$, then the angle that it subtends on the sky would be given by
$\Delta \theta = l / d$ (in the limit $d \gg l$) for flat Euclidean space. Using
this relation, we can define the angular diameter distance $d_{\textsc{a}}$ to
satisfy
\begin{align}
	d_{\textsc{a}} \equiv \frac{l}{\Delta \theta}.
\end{align}
The solution to this can be given in terms of the comoving distance as
\begin{align}
	d_\textsc{a} = \frac{f_{k}(\rchi)}{1 + z}.
\end{align}

It is often useful to calculate the angular diameter distance between a
source at redshift~$z_2$ and observer at redshift $z_1$ ($z_1 < z_2$), which
is given by~\cite{Schneider:2015eaci}
\begin{align}
	d_{\textsc{a}}(z_1, z_2) = \frac{1}{1 + z_2} f_{k}(\rchi(z_2) - \rchi(z_1)).
\end{align}
We see that $d_{\textsc{a}}(z_1, z_2) \neq d_{\textsc{a}}(z_2) - d_{\textsc{a}}(z_1)$,
which is an important result when we consider the equations of gravitational
lensing shortly.

\subsubsection{Luminosity distance}

Analogously, the luminosity distance ($d_{\textsc{l}}$) can be defined
from the classical luminosity-flux relation which follows an inverse-square law
of
\begin{align}
	F = \frac{L}{4 \pi \, d_{\textsc{l}}^2},
\end{align}
where $F$ is the flux received, and $L$ is the luminosity of an object. Again,
this can be written in terms of the comoving distance to give
\begin{align}
	d_{\textsc{l}} = (1 + z) f_k(\rchi) = (1 + z)^2 d_{\textsc{a}}.
	\label{eqn:luminosity_dist}
\end{align}
From this, we see that in an expanding universe, the distance inferred from
measuring its luminosity alone would be larger than that from measuring its size
and inferring its distance from that. This shows that objects in the sky
appear fainter in an expanding universe than they otherwise would be in a
static space-time, an important result of observational cosmology.

\begin{figure}[t]
	\centering
	\includegraphics[width=0.975\linewidth,trim={0.0cm 0.0cm 0.0cm 0.0cm},clip]{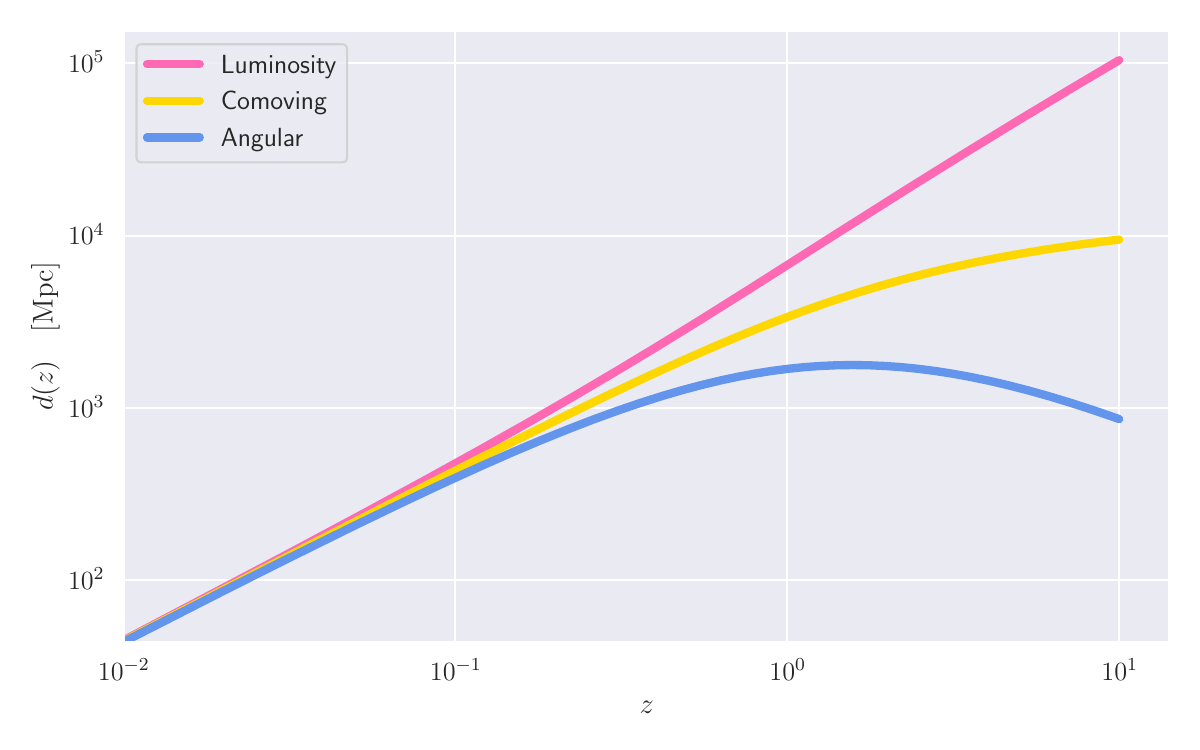}
	
	\caption{Plot of the different measurements of cosmological distances as
		a function of redshift. Here, we see that in an expanding universe 
		objects are fainter, and thus appear further away than they actually are.
		We also see the peak in the angular diameter distance which is unique 
		to an expanding space-time. For the best explanation of this effect, see 
		\href{https://xkcd.com/2622}{\texttt{xkcd~2622.}}
		}

	\label{fig:distances}
\end{figure}

Figure~\ref{fig:distances} plots the luminosity, comoving, and angular diameter
distances as a function of redshift. This demonstrates the differences between
these three measurements, and why they're important.

\section{Inflation \& the early universe}

In the beginning (queue \textit{2001: A Space Odyssey} theme) there was
nothing.... Then a wild Big Bang occurred! Much has happened to the universe
in its (approximate) 13.6 billion years of evolution, and thus it's important
to look across the universe's entire existence for us to start answering the
fundamental questions: How did we get here? What's going to happen to the
Universe? And what is the meaning of life? (Though that has a simple answer
of 42.)

\subsection{The hot Big Bang model}

With Edwin Hubble's observations that distant objects were receding from us
equally in all directions, one could wind the clock back in which one would 
predict that at the start of the universe, all matter was at one location
and thus everything started from a single `Big Bang'\footnote{It's interesting
to note that British astronomer Fred Hoyle who coined this name, was actually
trying to discredit the theory -- favouring the steady state solution instead.}.  
The Big Bang model got modified to the `hot' Big Bang model by George Gamow's
work on trying to explain the abundance of elements within the universe~\cite{Gamow:1948pob}.
By proposing that the early universe at some point had sufficient energy
to overcome the repulsive nature of the strong nuclear force, such that
nuclear fusion reactions can occur, but not too high energy which would result 
in instant destruction of any newly created atoms from rouge high-energy 
photons, one can precisely calculate the exact ratios of elements formed in
the early universe. This process of element creation is called Big Bang
nucleosynthesis (BBN), lasting from about one second to three minutes after the
initial Big Bang, and created elements D, $^{3}$He, $^{4}$He, and $^{7}$Li~\cite{ParticleDataGroup:2024cfk}.
The famous `Alpha-Bethe-Gamow', or $\alpha \beta \gamma$, paper~\cite{Alpher:PhysRev73803}
was one of the first attempts to calculate the abundances of such elements from
BBN, with subsequent calculations being in very good agreement with observational
data~\cite{ParticleDataGroup:2024cfk}. This agreement underpins our hot Big Bang
model; that our Universe started out extremely hot and dense which rapidly
expanded and cooled.

\subsection{The Big Bang break down}

While the hot Big Bang model has been very successful in providing many physical
answers to many questions that arose, such as the relative abundances of
elements (through primordial nucleosynthesis~\cite{Gamow:1948pob}) and the
prediction of the cosmic microwave background~\cite{Peebles:1965ApJ}, it is not
an exhaustive theory. Such is life, you answer some questions only to be left
with more that are unresolved. Why can't it be turtles all the way down?

\subsubsection{Flatness problem}

The flatness problem is a problem with the hot Big Bang model that when we
observe the Universe on the large scales, we observe a universe that is
consistent with being spatially flat. Some of the latest observational
constraints on spatial curvature come from the \textit{Planck} 2018 data
release~\cite{PlanckCollaboration:2018eyx}, which constrain the curvature
density $\Omega_k$ to be $\Omega_k = (0.0007 \pm 0.0019)$, which is very much
consistent with a flat universe of $k = 0$.

One can now ask: Why is this a `problem'? Surely\footnote{I am serious, and don't call me Shirley.}
we could just be living in a flat, or very close to flat universe. On the flip
side, we could have been living in a very non-flat universe that experiences
severe spatial curvature with no way for the Big Bang model to favour one of the
other. Thus, for our theoretical model to match the experimental evidence, one
needs to `fine-tune' the curvature parameter such that it has the value that it
has today, for which we have no \textit{a priori} reason for doing so.

Furthermore, it should be noted that from the Friedmann equations, the evolution
of the curvature density during matter domination (which is what the vast
majority of the universe's history has been in) scales as~\cite{Liddle:2015int}
\begin{align}
	\lvert \Omega_k(t) \rvert \propto t^{\frac{2}{3}}.
	\label{eqn:Omega_k_matter}
\end{align}
Since this is an increasing function with time, if we observe $\Omega_k$ to be
very close to zero today, some 13.6 billion years after the Big Bang, then it
must have been \textit{exceedingly} close to zero at the start of the Universe.
For example, during the period of nucleosynthesis, approximately one second
after the Big Bang, then we require an upper limit on the curvature density of
$\lvert \Omega_k \rvert \lesssim 10^{-18}$ \cite{Liddle:2015int}. Since
$\Omega_k$ should be a free parameter in any cosmological model, requiring it
to have such a tight constraint is highly unphysical. Thus, we would like
to introduce some theoretically motivated mechanism for which the value of
$\Omega_k$ is driven to very small values, regardless of its initial value.

Note that if, for some reason, $\Omega_k$ was \textit{exactly} zero at the
time of the Big Bang, then it would \textit{always} be zero and thus there
would be no flatness problem today.

\subsubsection{Horizon problem}

When we look at the Universe on the largest scales, we see that it is very
nearly isotropic (this was the motivation behind the cosmological principle,
and the FRW metric). For example, since the level of temperature anisotropies
in the cosmic microwave background are at the level of one in one hundred
thousand, this prompts us to believe that any two points on the opposite side
of the sky must have been in thermal equilibrium for them to have such similar
statistical properties. However, given that the Universe has a finite age, the
hot Big Bang model predicts that these two vastly separated regions could not
have been in causal contact for them to establish this equilibrium.

Therefore, in the framework of the hot Big Bang model, we require very nearly
uniform initial conditions across the entire universe for us to observe such
statistical isotropy. This requires additional `fine-tuning' in the initial
conditions, which again is dissatisfying and theoretically unmotivated.

\subsubsection{Cosmological perturbations}

While the horizon problem looks at the large-scale statistical isotropies that
we see on the sky, it is important to note that the Universe is \textit{not}
totally homogenous. For example, the existence of planets, stars,
and galaxies all show that there are significant overdensities in the Universe
-- the formation and evolution of which need to be explained in our cosmological
models. Thus, we require some mechanism in the early universe to `seed' these
perturbations, from which they can evolve to form structures, and you and me.

\subsection{Inflate our problems away}
\label{sec:inflate_our_probs_away}

As we've discussed, there are numerous `problems' with the hot Big Bang model,
namely that there involves much `fine-tuning' of the initial conditions if we
were to match up our experimental evidence to our theoretical models. Thus, we
wish to introduce some physical mechanism that can solve many of these problems
while also producing its own predictions which we can test to verify or nullify
the theory.

The inflationary paradigm was introduced in Refs.~\cite{Guth:1980zm,Linde:1981mu}
as a period of rapid expansion in the \textit{very} early universe, when it
was only around $10^{-34}$ seconds old. Inflation was introduced to solve many
of the problems with the hot Big Bang model (namely the flatness and horizon
problems), as noted by the title of Ref~\cite{Guth:1980zm} ``Inflationary
universe: \textit{A possible solution to the horizon and flatness problems}''.
The principle of inflation is that the universe undergoes accelerated expansion,
i.e. $\ddot{a} > 0$. As we will see, by postulating that there was this rapid
expansion very early in the Universe's evolution, it solves many of the
problems that we have seen in the hot Big Bang model.

\subsubsection{Solving the flatness problem}

Previously, we have seen that during matter domination the curvature density
grows with time (Equation~\ref{eqn:Omega_k_matter}), and so a small perturbation
from zero in the early universe would grow into a much larger value today. We
can repeat this analysis during a period of inflation and see how the curvature
density changes during this epoch.

If we assume the case of perfect exponential inflation, which is highly
motivated as we will see later, then the evolution of the scale-factor evolves 
with time has the same functional form as 
our $\Lambda$-dominated universe of Equation~\ref{eqn:scale_factor_lambda}.
Using the Friedmann equations again, it can be shown that the curvature density
changes with time as~\cite{Liddle:2015int}
\begin{align}
	\lvert \Omega_k(t) \rvert \propto \exp \left(- \sqrt{\frac{4 \Lambda}{3}} \,t \right).
	\label{eqn:Omega_k_inflation}
\end{align}
Hence, no matter what the initial value of $\Omega_k$ was from the initial
conditions of the Big Band singularity, $\Omega_k$ is exponential suppressed
and the universe is rapidly forced to being spatially flat provided that
there is sufficient inflation.

\subsubsection{Solving the horizon problem}

Inflation also naturally solves the horizon problem, as the accelerated
expansion allows for a, then small, patch of the Universe to reach thermal
equilibrium before the exponential expansion kicks in to enlarge the patch to
a size that is many times larger than our observable Universe. Hence, if we
assume that the initial path thermalised,\footnote{Now there's no particular
	reason for the initial path to be in thermal equilibrium before inflation, which
	is one of the major criticisms of the inflationary paradigm.} then the resulting
expanded volume will also be in thermal equilibrium -- matching the statistical
isotropy that we see today.

\subsubsection{Solving the monopole problem}

There also exists the `monopole problem'. This is where, as modern particle
physics theory try to unify the electromagnetic, strong, and weak forces into a
single `Grand Unified Theory' (GUT), extraordinarily heavy particles are predicted
to have numerously formed right after the Big Bang. Such particle could be a magnetic
monopole, which are analogous to electric charge carries, such as quarks and
electrons, but for the magnetic force and arise from such GUT
theories~\cite{Liddle:2015int,Milton:2006cp}. These magnetic monopoles would have had masses
around the GUT scale, $E_{\textsc{gut}} \sim 10^{16} \, \textrm{GeV}$, which
when compared to the heaviest elementary particle found thus far, the top quark
with mass around $173\, \textrm{GeV}$~\cite{ParticleDataGroup:2022pth}, shows
that even if there were only very few of these magnetic monopole particles they
would dominate the universe's energy density contributions. This would
significantly alter the previously well-understood dynamics of radiation
domination in the early universe followed by matter domination, and thus would
be incompatible with the hot Big Bang model.

Inflation solves this problem as the rapid, exponential expansions will serve
to vastly dilute the abundances of relic particles. Thus with sufficient
expansion, it could be probable that none of these particles would exist in an
observable universe despite initially dominating the energy distribution of the
pre-inflated universe~\cite{Linde:1983psb}.

\subsubsection{How much inflation}

While inflation solves many of the problems that we have discussed, by
postulating that the Universe underwent a rapid exponential expansion soon
after it formed, it does not tell us exactly how long this expansion lasted. To
quantify how long inflation went on for, we can introduce the idea of
$e$-foldings, $N$, which are simply the number of times that the universe expanded
by a factor of $e$. Mathematically, this is defined as the ratio of the
scale-factors at the end ($a_2$) to the beginning ($a_1$) of inflation
\begin{align}
	N \equiv \ln \frac{a_2}{a_1} = \int_{t_1}^{t_2} \!\! \d t \, H(t).
\end{align}
Estimates for this place the number of $e$-foldings around $50 \lesssim N \lesssim 70$
to adequately explain the flatness, horizon, and monopole problems~\cite{Liddle:2003as,Dodelson:2003vq}.
For example, if we assume inflation started at the GUT scale, with a temperature
$T_{\textsc{gut}} \sim 10^{28} \, \textrm{K}$, and lasted for $N = 65$
$e$-foldings, then it would be driven to a temperature of
$T_2 \sim e^{-65} \,T_{\textsc{gut}} \sim 0.6 \, \textrm{K}$~\cite{Ryden:1970vsj}.
Though due to the wide variety of inflationary models that have been studied in
the literature, there exist many models where inflation lasts for much longer,
such as models where $N$ can be of the order one thousand or even larger than
ten thousand~\cite{Marzouk:2021tsz,Torrado:2017qtr}.

\subsubsection{Ending inflation and reheating}

During inflation, all existing matter and radiation are rapidly expanded away
with the universe rapidly cooling to a temperature of $1\,\textrm{K}$ or less,
as shown above. The entire energy density of the universe is now held in the
inflaton field(s). Thus, to match the existing well-tested predictions of the
hot Big Bang model, such as primordial nucleosynthesis~\cite{Peacock_1998},
these inflaton fields will need to decay, creating the very many high-energy
elementary particles that are present in the Standard Model~\cite{Earnshaw:2024zbv}. This then allows
the hot Big Bang to progress normally, with a hot radiation-dominated universe,
though one with now the correct initial conditions to match the phenomena
that we observe today.

It should be noted that the energy scale of reheating, $T_{\textsc{r}}$, should
be large enough that it produces an abundance of highly energetic Standard Model
elementary particles, but much smaller the GUT scale so as to not re-form
troublesome particles, such as magnetic monopoles, that inflation was invoked
to solve~\cite{Kofman:1994rk,Liddle:2015int}.

\subsection{The physics of inflation}

While the exact physics of inflation is the subject of much lengthy discussion,
we will only present a set of various key results and derivations. Many
textbooks have been dedicated to inflation, with Refs.~\cite{Peacock_1998,Lyth:2009zz,Liddle:2000cg}
covering the topic particularly well.

\subsubsection{Inflation, or dark energy?}

While we are conceded with the primordial universe for the time being, it is
important to note that the results that we derive here hold for any scalar field
that is dominating the universe's energy density -- irrespective of if its
dynamics occur $10^{-34}$ seconds after the Big Bang or after 13 billion years.
Thus, these results also apply to dark energy domination period that our
universe is currently experiencing.

\subsubsection{Inflating with scalar fields}

Currently, the most accepted idea for what drove inflation was that there were
some scalar field(s) active during the early universe whose dynamics gave rise
to accelerated expansion. This is because, as we shall see, if the energy
density is dominated by scalar fields, then the exponential expansion is a
natural consequence. Furthermore, by quantising the scalar field to a quantum
scalar field, we produce perturbations in the late-universe that are compatible
with observations today.

Scalar fields exist in a wide range of physics, with an example of a classical
scalar field being temperature -- each point in spacetime has
a specific value for the field, which may evolve with time.
As we are dealing with relativistic quantum fields, we require that the
scalar field is invariant under Lorentz transforms.

The idea of inflating with scalar fields is physically motivated from elementary
particle physics, due to the fact that scalar fields are an established part of
the Standard Model of particle physics, as one of its constituent particles, the
Higgs boson, is a particle derived from a scalar field \cite{griffiths2008}.
In the 1960s, the Englert-Brout-Higgs mechanism \cite{Higgs:1964pj,Englert:1964et}
proposed to use scalar fields to help solve the \textit{small} problem of particles having no
mass! Weinberg then applied the mechanism to produce particle passes in the
newly developed Standard Model~\cite{Djouadi:2024has}.
With the detection of the Higgs boson in 2012 \cite{ATLAS:2012tfa,CMS:2012xdj},
we have strong experimental evidence that scalar fields exist in nature, and so
may be included in our cosmological models. Therefore, it is conceivable that
some scalar fields were active during the early universe in such a way that can
give rise to inflationary dynamics.

\subsubsection{Scalar field Lagrangian}

To determine the dynamics of the scalar field during inflation, we can make use
of Lagrangian mechanics. This introduces the concept of the Lagrangian density
$\Lag$ which encodes the physical properties of the scalar fields. In
classical mechanics, this is a function of the coordinates $x^i$ and their
derivatives $\dot{x}^i$ (giving $\Lag = \Lag(x^i, \, \dot{x}^i)$). This
translates to field theory as a function of the field $\phi$ and its
derivatives $\partial_a \phi$ (giving $\Lag = \Lag(\phi, \, \partial_a \phi)$).
Using our field theory Lagrangian, we find the action to be
\begin{align}
	S = \int \!\! \d^4 x \, \, \Lag(\phi, \, \partial_a \phi).
\end{align}
Using the calculus of variations, and applying the principle of least action --
which states that the action is extremised such that $\delta S = 0$ -- gives
rise to the Euler-Lagrange field equations of
\begin{align}
	\partial_a \!\!\left( \frac{\partial \Lag}{\partial \left(\partial_a \phi \right)} \right)
	- \frac{\partial \Lag}{\partial \phi} = 0,
\end{align}
which are analogous to the Euler-Lagrange equations for a system in classical
mechanics.

Let us consider the simplest case: a single scalar field $\phi$ (called the
inflaton field). From field theory, the Lagrangian density for a free scalar
field is~\cite{Weinberg:1995mt}
\begin{align}
	\mathscr{L} = \frac{1}{2} g^{ab} (\partial_a \phi) (\partial_b \phi) - V(\phi),
	\label{eqn:general_lagrangian}
\end{align}
where the first term is a generalised kinetic energy, and $V(\phi)$ is the
potential. Applying the Euler-Lagrangian equations, we find the field equation
of motion to be
\begin{align}
	\nabla^a \nabla_a \phi + \frac{\partial V}{\partial \phi} = 0,
\end{align}
where $\nabla_{a}$ is the covariant derivative ($\nabla_a x^b = \partial_a x^b +
\Gamma^{b}_{ca} x^c$) since we wish to include the properties of a non-trivial
metric~\cite{Hobson:2006se}.

\subsubsection{The action in curved spacetime}

So far, our discussion of scalar fields have been for the case of flat
Minkowski spacetime, where the metric is $g_{ab} = \eta_{ab}$. However, since we
know that the observable Universe can be accurately described by the FRW metric, we
wish to understand how our scalar fields act in non-Minkowski spacetimes. Since
the action should be invariant upon general coordinate transformations, the
action in GR becomes~\cite{Hobson:2006se}
\begin{align}
	S = \int \! \d^4 x \, \sqrt{ {\scriptstyle |} g {\scriptstyle |}} \, \, \mathscr{L}(\phi, \, \partial_a \phi),
\end{align}
where $g$ is the determinant of the metric. This gives an `effective Lagrangian',
which is now the correct term to use in the Euler-Lagrange equations, of
$\sqrt{ {\scriptstyle |} g {\scriptstyle |}} \, \, \mathscr{L}(\phi, \, \partial_a \phi)$.

If we now consider a FRW universe that is dominated by our single scalar field
that is homogenous, we can neglect the spatial derivatives that feature in our
field-space Lagrangian of Equation~\ref{eqn:general_lagrangian}, to give simply
\begin{align}
	\mathscr{L}(\phi, \partial_a \phi) = \frac{1}{2} \dot{\phi}^2 - V(\phi).
\end{align}

The determinant of the FRW metric (Equation~\ref{eqn:FRW_metric}) is
$\sqrt{ {\scriptstyle |} g {\scriptstyle |}} = a^3(t)$, and thus we find the
evolution of our scalar fields in FRW spacetime to be given by
\begin{align}
	\ddot{\phi} + 3H\dot{\phi} = - \partial_\phi V.
\end{align}
This can be recognised as a form of a damped harmonic oscillator, where the
damping term ($3 H \dot{\phi}$) is analogous to the friction generated by
an expanding universe -- and is called the Hubble drag~\cite{Lyth:2009zz}.

\subsubsection{The equation of state for a scalar field}

Previously, in Equation~\ref{eqn:acceleration_eos}, we saw that to get
accelerated expansion, we required that the Universe be dominated by a
fluid that has equation of state $w$ such that $w < -\frac{1}{3}$. Hence, we
need to verify that this condition on $w$ can be satisfied by a scalar field.

The pressure and density of a scalar field can be found by coupling it to
gravity through the Einstein-Hilbert action which, for a spatially
homogenous field ($\partial_i \phi = 0$), yields~\cite{Hobson:2006se,Liddle:2003as}
\begin{subequations}
	\begin{align}
		P_\phi    & = \frac{1}{2} \dot{\phi}^2 - V(\phi), \\
		\rho_\phi & = \frac{1}{2} \dot{\phi}^2 + V(\phi).
	\end{align}
\end{subequations}
Hence, the equation of state for a scalar field is
\begin{align}
	w_{\phi} = \frac{\frac{1}{2} \dot{\phi}^2 - V(\phi)}{\frac{1}{2} \dot{\phi}^2 + V(\phi)},
\end{align}
which gives accelerated expansion provided that $V(\phi) > \dot{\phi}^2$.
Additionally, we see that in the limit that the potential dominates over the
kinetic term ($V(\phi) \gg \dot{\phi}^2$), the equation of state is driven to
negative one ($w_\phi \rightarrow -1$), which is the limit of the classical
cosmological constant.

\subsection{The primordial perturbations}

While inflation was largely motivated to solve the many problems with the hot
Big Bang model (Section~\ref{sec:inflate_our_probs_away}), one problem not yet
addressed is the generation of perturbations in the early universe that are
necessary for structure to grow, and thus result in the creation of you, dear
reader, and I. This is where inflation provides perhaps its greatest success,
a physical mechanism for how these primordial perturbations were produced.
A full discussion of cosmological perturbation theory during inflation is
outside the scope of this thesis, see Refs.~\cite{Peacock_1998,Liddle:2000cg,Lyth:2009zz}
(yet again) for comprehensive descriptions, and reserve our discussion to some
key results that will be needed later.

Nevertheless, we can at least start with a brief physical insight into the basic
mechanisms that drove the primordial perturbations, which have subsequently
evolved in the large-scale structure we see today some 13.6 billion years later. 
The foundation of the generation of inflationary perturbations come from the
quantum fluctuations in the scalar field that drove inflation. Quantum field
theory tells us that even in an empty vacuum, particle-antiparticle pairs
are continually popping in and out of existence, so called `virtual particles',
whose energies and lifetimes are governed by Heisenberg's uncertainty 
principle~\cite{Hobson:2006se}. The effect of these virtual particles is that a
small region of space will have a slightly higher energy than average and one with
a correspondingly lower energy.  In a non-inflationary spacetime, this has no
macroscopic effects, as these virtual particle pairs are confined to the quantum
realm and the overall average energy is conserved. However, when we subject our
spacetime to the dramatic exponential expansion of inflation, these 
particle-antiparticle pairs are driven apart much faster than the speed of light,
and so are out of causal contact~\cite{Liddle:2000cg}. 

Hence, the virtual particle pair cannot recombine since they are out of causal
contact (outside of the horizon), and thus the under- and over-dense regions are
`locked in' by inflation. These quantum fluctuations then provide the seeds of
structure formation during the decelerating periods of radiation- and then
matter-domination of our Universe.

We can decompose our inflaton field $\phi$ into a zeroth-order homogenous
background $\bar{\phi}$, and a perturbative part, $\delta \phi$, to give~\cite{Dodelson:2021ft}
\begin{align}
	\phi(t, \vec{x}) = \bar{\phi}(t) + \delta \phi(t, \vec{x}).
\end{align}
These perturbations in our scalar field during inflation give rise to the curvature perturbation
$\zeta$, which has a general interpretation as the fractional perturbation
in the scale factor $\zeta = \delta a / a$. It is an extremely useful quantity
in inflationary cosmology since $\zeta$ is conserved when perturbation modes
move outside the horizon.

We now wish to investigate the statistical properties of $\zeta$, such that we
can hope to obtain something that can be measured experimentally, and thus test
our theories of inflation.

\subsubsection{The two-point function}

Since we have constructed the curvature perturbation field $\zeta$ to have
a mean of zero, $\langle \zeta(\vec{x}) \rangle = 0$, the first non-zero 
statistical measure of $\zeta$ will be the total variance of the perturbations.
This is called the two-point function.
If we consider the Fourier transform of $\zeta$ into momentum-space, we get
\begin{align}
	\zeta(\vec{x}) = \int \!\! \frac{\d^3 \vec{k}}{\left( 2 \pi \right)^3} \, \,
	\zeta_{\vec{k}} \,  e^{i \vec{k}\cdot \vec{x}},
	\label{eqn:Fourier_xfm_zeta}
\end{align}
where we are integrating over all wavenumbers $\vec{k}$. If we now compute the
variance of $\zeta$ at point $\vec{x}$ using the above Fourier transform, which
is the definition of the variance of the field $\zeta$, we find
\begin{align}
	\langle \zeta^2(\vec{x}) \rangle = \int \!\! \frac{\d^3 \vec{k}}{\left( 2 \pi \right)^3}
	\frac{\d^3 \vec{k}'}{\left( 2 \pi \right)^3} \, \,
	\langle \zeta_{\vec{k}} \, \zeta_{\vec{k}'} \rangle \, e^{i \vec{x} \cdot (\vec{k} + \vec{k}')}.
	\label{eqn:Pk_variance_kspace}
\end{align}
Statistical homogeneity and isotropy (the cosmological principle) requires that
this expectation value be position independent, i.e.
$\langle \zeta^2(\vec{x}) \rangle  = \langle\zeta^2\rangle$. For this to hold,
we require that $\vec{k} + \vec{k}' = 0$, and thus we find that the expectation
value of our two $\zeta$ terms in momentum-space are proportional to a
$\delta$-function of the momenta: $\langle \zeta_{\vec{k}} \, \zeta_{\vec{k}'}
	\rangle \propto \delta^{(3)}(\vec{k} + \vec{k}')$. From this, we can
\textit{define} the primordial power spectrum $P(k)$ such that
\begin{align}
	\langle \zeta_{\vec{k}} \, \zeta_{\vec{k}'} \rangle = (2 \pi)^{3} \,
	\delta^{(3)}(\vec{k} + \vec{k}') \, P_{\zeta}(k).
	\label{eqn:power_spec_def}
\end{align}
Using this definition of the power spectrum in
Equation~\ref{eqn:Pk_variance_kspace} gives
\begin{align}
	\langle \zeta^2 \rangle & = \int \!\! \frac{\d^3 \vec{k}}{\left( 2 \pi \right)^3}
	\frac{\d^3 \vec{k}'}{\left( 2 \pi \right)^3} \,
	(2 \pi)^3 \, \delta^{(3)}(\vec{k} + \vec{k}') \, P_{\zeta}(k) \, e^{i \vec{x} \cdot (\vec{k} + \vec{k}')}. \nonumber \\
	                        & = \int \!\! \frac{\d^3 \vec{k}}{\left( 2 \pi \right)^3} \,  P_\zeta(k) \nonumber           \\
	                        & = \frac{1}{2 \pi^2} \int \! \d k \,\, k^2 \,  P_\zeta(k).
\end{align}
It is often useful to use the dimensionless power spectrum $\mathcal{P}_{\zeta}(k)$ instead,
which is related to $P(k)$ through
\begin{align}
	\mathcal{P}_\zeta(k) = \frac{1}{2\pi^2} k^3 P_\zeta(k).
	\label{eqn:P_zeta_dimless}
\end{align}
This has the useful statistical property that it is the variance per
logarithmic interval in $k$-space.
The power spectrum can be parametrised as a power-law relation of \cite{Byrnes:2014pja}
\begin{align}
	\mathcal{P}_\zeta(k) = \As \! \left( \frac{k}{k_*}\right)^{\ns - 1}\!\!\!\!\!\!\!\!\!\!\!\!,
	\label{eqn:power_law_inflation}
\end{align}
where $\As$ is the scalar amplitude, $\ns$ is the scalar spectral index --
with $\ns = 1$ giving a true scale-independent spectrum -- and $k_*$ is the
`pivot' scale, which characterises the observational scale of interest.
The latest observational
constraints on the power spectrum come from the 2018 \Planck\ data release,
placing $\ln( 10^{10} \, \As) = 3.047 \pm 0.014$, evaluated at the pivot
scale $k_* = 0.05\,$Mpc$^{-1}$,
and $\ns = 0.9665 \pm 0.0038$ \cite{PlanckCollaboration:2018eyx}.

\begin{figure}[t]
	\centering
	\includegraphics[width=0.975\linewidth,trim={0.0cm 0.0cm 0.0cm 0.0cm},clip]{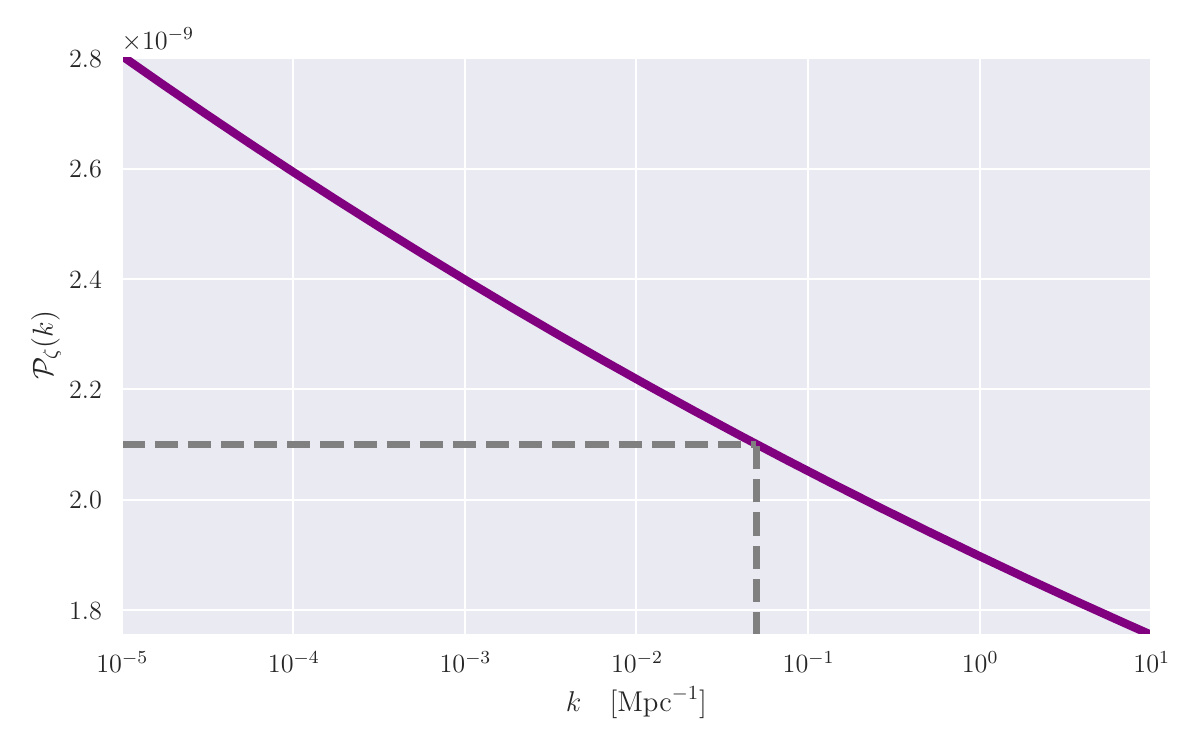}
	\caption{The dimensionless inflationary power spectrum plotted using
		the \textit{Planck} 2018 values. Here, we show the `pivot' scale
		of $0.05 \, \textrm{Mpc}^{-1}$ and the amplitude at this scale of
		$2.1 \times 10^{-9}$. Since $\ns < 1$, we have a `red' spectrum giving
		more power\protect\footnotemark\, on larger physical scales than small scales, which propagates
		into the density contrasts that we see today.
		}
	\label{fig:inflationary_pow_spec}
\end{figure}

The inflationary power spectrum is plotted in Figure~\ref{fig:inflationary_pow_spec}.

\footnotetext{\textit{More power} --- Cyber-Leader}

\subsubsection{The three-point function}

For a pure Gaussian field, the two-point function encodes all the statistical
properties about the field, since higher-order correlators are either zero
(for an odd number of terms), or can be expressed entirely out of the two-point
function (for an even number of terms). This arises from how a Gaussian
distribution can be described using only two numbers: its mean and variance.
However, while the curvature perturbations are thought to be heavily Gaussian,
they may not be exactly Gaussian and thus have small non-Gaussianities. Thus,
the first higher-order correlator up from the two-point function is the
three-point function defined through~\cite{Wands:2010af,Lyth:2005fi}
\begin{align}
	\langle \zeta_{\vec{k}_1} \, \zeta_{\vec{k}_2} \, \zeta_{\vec{k}_3} \rangle =
	\left( 2 \pi \right)^3 \, \delta^{(3)}( \vec{k}_1 + \vec{k}_2 + \vec{k}_3 ) \,
	B_\zeta (k_1, k_2, k_3),
	\label{eqn:zeta_bispec}
\end{align}
where $B_\zeta(k_1, k_2, k_3)$ is the curvature bispectrum. The relative
amplitude of the bispectrum compared to the power spectrum is parameterised
through $\fNL$, defined as~\cite{Lyth:2009zz}
\begin{align}
	\frac{6}{5} \fNL(k_1, k_2, k_3) = \frac{B_\zeta(k_1, k_2, k_3)}{
		P_\zeta(k_1)P_\zeta(k_2) +P_\zeta(k_1)P_\zeta(k_3) + P_\zeta(k_2)P_\zeta(k_3) }.
	\label{eqn:inf_red_bsp}
\end{align}
Measuring a statistically significant value for $\fNL$ has been the goal of
very many experimental cosmologists for the last couple decades as vary many
inflationary models could be ruled out by its value~\cite{Bartolo:2004if}.

\section{Evolution of structure across cosmic time}
\label{sec:perturbation_evolution}

While we have seen that inflation can generate small perturbations in the very
early universe, our next task is to find out how these small perturbations
evolve into the large-scale structure that we see today.

\subsection{Linear density evolution}

Let us start with assuming that the density of the universe can be expressed
as a homogenous background $\bar{\rho}$ and a spatially varying relative
perturbative part $\delta$, to give
\begin{align}
	\rho(t, \vec{x}) \equiv \bar{\rho}(t) \left(1 + \delta (t, \vec{x})\right).
	\label{eqn:def_density_pert_delta}
\end{align}
Since the density of the Universe is constrained to be non-negative, $\delta$
is bound to be $-1 \leq \delta < \infty$. By construction, the average of 
$\delta$ is zero, $\langle \delta(t, \vec{x})\rangle = 0$. We now want to work out how the
perturbations $\delta$ evolve during the expanding evolution of our Universe.
The first such equation that we will use is the Poisson equation of~\cite{Peacock_1998}
\begin{align}
	\vec{\nabla}^2 \Phi = 4 \pi G \, \bar{\rho} \delta,
	\label{eqn:Poisson_grav}
\end{align}
where $\Phi$ is the Newtonian gravitational potential. This gives rise to
the evolution of our perturbations $\delta$ in an expanding space-time as
\begin{align}
	\ddot{\delta} + 2 H \dot{\delta} - 4 \pi G \, \bar{\rho} \delta  = 0.
	\label{eqn:delta_evo}
\end{align}
For approximately constant $\bar{\rho}$ and small $H$, we find that the solution
for the time-evolution of $\delta$ is exponential growth. In this case, we find
that an initially small overdensity would continuously grow as its gravitational
potential continued to bring in nearby matter. However, the presence of the
$2 H \dot{\delta}$ (the `Hubble drag') term serves to slow this exponential
growth, and thus we see that in an expanding space-time, the growth of

We now wish to go about solving this equation to obtain the evolution of $\delta$
for different components of our Universe during its different epochs.

\subsection{Matter perturbations during radiation domination}
\label{sec:pert_evo_rad_dom}

First, we shall consider the perturbations in the matter density $\deltam$ 
during the epoch of radiation domination -- early on in our Universe's evolution.
Since we are considering the matter perturbations, we can transform 
Equation~\ref{eqn:delta_evo} into
\begin{align}
	\deltamddot + 2 H \deltamdot - \frac{3}{2} H^2(t) \Omegam(t) \deltam  = 0,
	\label{eqn:delta_evo_matter}
\end{align}
where we have used the Friedmann equation to write the physical matter density 
in terms of the Hubble parameter and matter density parameter. During radiation
domination, we can approximate that $\Omegam(t) \rightarrow 0$, and recalling
that the time evolution of the scale factor was $a(t) \propto \sqrt{t}$ 
(Equation~\ref{eqn:radiation_dom_evo}) and thus the Hubble rate is given by
$H(t) = 1/2t$, we find the evolution of the matter perturbations as  
\begin{align}
	\deltamddot + \frac{1}{t} \deltamdot = 0.
\end{align}
This has solutions of the form
\begin{align}
	\deltam(a) = A \ln a + B,
\end{align}
and so we see that the matter perturbations during radiation domination grow
with the logarithm of the scale factor.

\subsection{Matter perturbations during matter domination}
\label{sec:pert_evo_mat_dom}

We can now repeat the same exercise, but ask how do the matter perturbations
evolve during the matter dominated epoch of our Universe. Since we are now
in matter domination, we can approximate $\Omegam(t) \rightarrow 1$,
the time evolution of the scale factor was $a(t) \propto t^{2/3}$ 
(Equation~\ref{eqn:matter_dom_evo}) giving the Hubble rate as $H = 2/3t$.
This gives the evolution of our matter perturbations to be governed by
\begin{align}
	\deltamddot + \frac{4}{3t} \deltamdot - \frac{2}{3t^2}\deltam = 0.
\end{align}
This has a solution of the form
\begin{align}
	\deltam(a) = Aa + Ba^{-\frac{3}{2}}.
\end{align}
Hence, we see that during matter domination the matter perturbations grow as
$\deltam(a) \propto a$ which is significantly faster than their growth during
radiation domination of $\deltam(a) \propto \ln a$. This shows the suppression
in the growth of structure that arose during the radiation dominated epoch 
and is called the Mészáros effect~\cite{Meszaros:1974tb}.

\subsection[Matter perturbations during $\Lambda$ domination]{Matter perturbations during $\bm\Lambda$ domination}

Since we currently observe that our Universe is undergoing accelerated expansion
due to a non-zero cosmological constant, we can consider how the matter
perturbations evolve during $\Lambda$ domination. In this case, we can again
approximate that $\Omegam(t) \rightarrow 0$ and that the Hubble rate tends
to a constant, $H(t) \rightarrow$ const. This gives a solution as
$\deltam \rightarrow$ const, and thus during $\Lambda$ domination the matter
perturbations stop growing and become static. Hence, there will be no new 
large-scale structure in our Universe, and that everything that will ever form
would have formed by now.

\subsection{The Fourier underworld}

We have already encountered the Fourier transform when looking at the curvature
perturbation $\zeta$ (Equation~\ref{eqn:Fourier_xfm_zeta}), and so we yet again
turn to it when dealing with our density perturbation $\delta$. This is 
especially useful since we can now investigate how the perturbations evolve
on different physical scales.  We can define the Fourier components of
the density perturbation $\delta(t, \vec{k}) \equiv \delta_{\vec{k}}$ through
\begin{align}
	\delta_{\vec{k}} = \int \!\! \d^3 \vec{x} \,\, e^{-i \vec{k} \cdot \vec{x}} \,\,
	\delta(t, \vec{x}).
\end{align}
$\delta_{\vec{k}}$ are complex quantities, where the magnitude represents the
amplitude of the perturbation and the phase represents its spatial dependence.
Momentum-space is especially useful when considering that any spatial derivatives of
the density perturbations become algebraic in Fourier-space:
$\partial_i \delta(t, \vec{x}) \rightarrow i k_i \delta(t, \vec{k})$.
Larger values of the wavenumber $k$ represent smaller physical scales and
vice-versa.

\begin{figure}[t]
	\centering
	\includegraphics[width=0.975\linewidth,trim={0.0cm 0.0cm 0.0cm 0.0cm},clip]{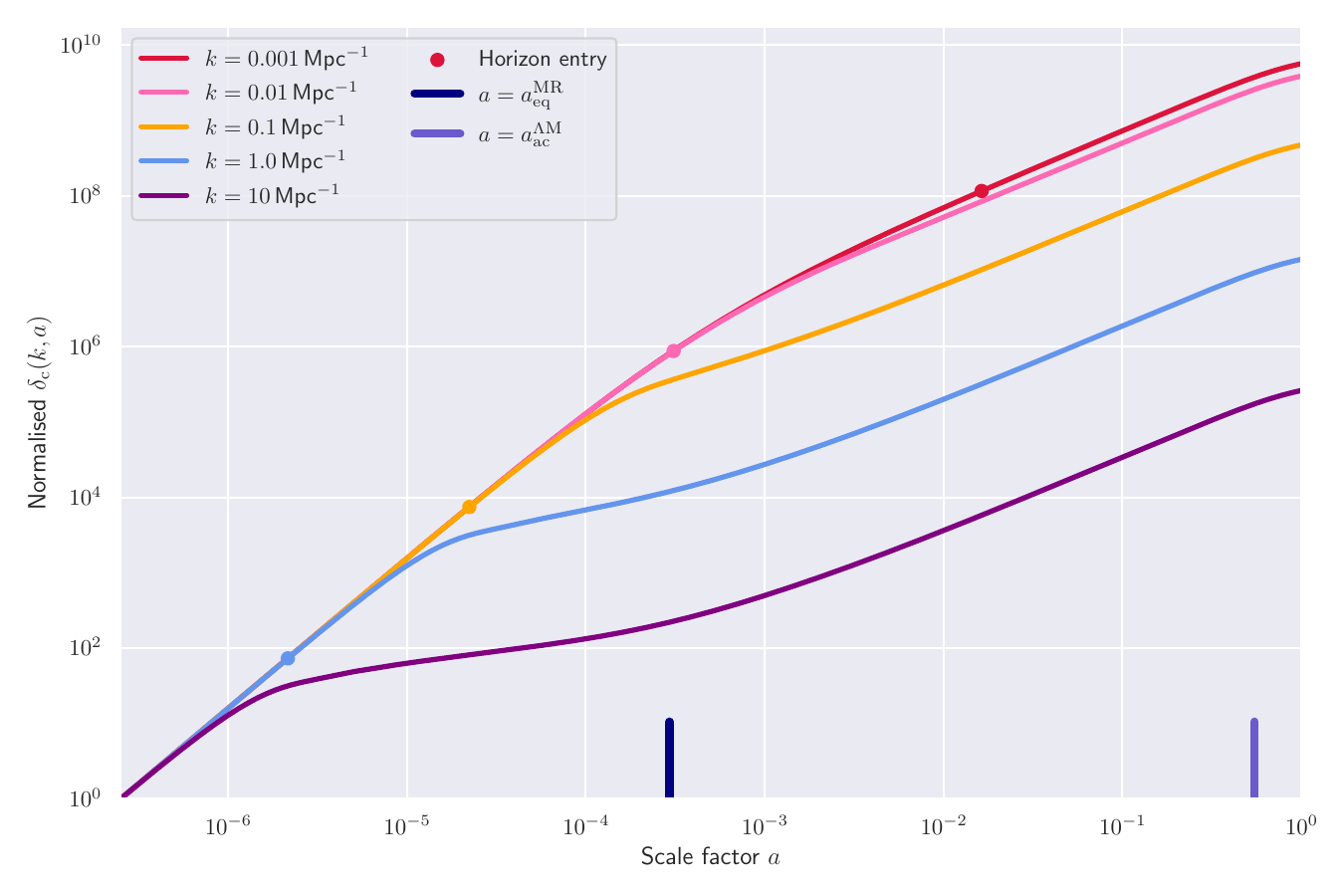}
	\caption{
		Evolution of the cold dark matter density perturbations for five 
		different Fourier modes across the evolution of the Universe, normalised
		to their value on the left-hand side. The coloured circles represent the
		scale factor at which that coloured curve's wavenumber re-enters the
		horizon (given when $k = aH(a)$), and so undergo sub-horizon evolution
		again.  We see that as the small-scale
		modes enter the horizon during radiation domination (up to 
		matter-radiation equality of $a_{\mathrm{eq}}^{\textsc{mr}} \sim 10^{-3}$,
		highlighted by the navy band), they undergo only logarithmic growth in their
		amplitudes, which suppresses power on the smallest scales today. 
		As the Universe begins its epoch of matter domination (after the navy band),
		these modes then grow with the scale factor, $\deltam \propto a$, as
		do the modes newly entering the horizon. We also see the effects of the
		current epoch of $\Lambda$ domination towards the right-hand side of the
		plot (the Universe entering accelerated expansion at 
		$a_{\mathrm{ac}}^{\Lambda\textsc{m}} \sim 0.5$, highlighted by
		the light blue band), where the growth of the perturbations slows for all
		scales, and so will tend to a constant in the far future. These curves 
		were computed using \Camb for a $\Lambda$CDM universe that is compatible
		with  observations.
	}
	\label{fig:delta_c_evo}
\end{figure}

Figure~\ref{fig:delta_c_evo} plots the evolution across the history of our 
Universe of the dark matter density perturbations for five physical scales,
ranging from the largest scales that we can observe in the observable Universe
today to the smallest scales. This clearly shows the suppression in the growth
of perturbations during the radiation dominated epoch, how the perturbations
grow with the scale factor during matter domination, and that the growth of
perturbations is already being suppressed due to the cosmological constant --
and will become static in the far future.

\subsection{The photonic and baryonic perturbations}
\label{sec:photon_baryon_pert}

While investigating the evolution of the dark matter perturbations has given us
vital insights into how these perturbations grow with the evolution of the 
Universe, which sow the seeds and form the large-scale structure that we see 
today, we do have to remember that our Universe isn't totally made of
cold dark matter and that it has vital contributions from radiation and baryons!
Working out the evolution of the dark matter perturbations were relatively easy
since they simply evolve under gravity in an expanding spacetime and are
decoupled with the rest of the goings-on in the Universe. In the early epochs
of our Universe, radiation and baryons are coupled together and have a 
non-negligible self-pressure. This dramatically reduces their ability to
clump together under gravity, and the perturbations have the ability to `bounce'
back out again due to their pressure. These physical effects making evaluating
the evolution of the baryonic and photonic perturbations significantly more
challenging, and so we leave the full derivation of these to other works (e.g.
Refs.~\cite{Peacock_1998} and~\cite{Amendola:2015ksp}).

As the photonic and baryonic fluids have non-zero self-pressure, these fluids
have a non-zero sound speed, $\cs$, defined as
\begin{align}
	\cs^2 \equiv \frac{\partial p}{\partial \rho}.
\end{align}
This modifies the perturbation evolution equation to that of a damped
harmonic oscillator,
\begin{align}
	\ddot{\delta} + 2 H \dot{\delta} - \left(4 \pi G \bar{\rho} - \frac{\cs^2 k^2}{a^2} \right) \delta = 0,
	\label{eqn:baryonic_photonic_per_evo}
\end{align}
where we are already working in Fourier space. Hence, the wavenumber determines
the evolution of that perturbation mode: long range modes (small $k$) continue
to grow, whereas short range modes (large $k$) undergo damped oscillations.
Specifically, if the physical wavelength is smaller than the Jeans length,
given by~\cite{Peacock_1998}
\begin{align}
	\lambda_{\mathrm{J}} = \cs \sqrt{\frac{\pi}{G \bar{\rho}}},
\end{align} 
then the perturbations will undergo oscillatory evolution. As the Universe
expands and the average density $\bar{\rho}$ decrease, this will increase the
Jeans length and so longer range modes will start to oscillate. We also see 
that once recombination takes place and the baryonic fluid's sound speed
drops to zero, then it oscillations will cease and ordinary gravitational
attraction can continue.

\begin{figure}[t]
	\centering
	\includegraphics[width=\linewidth,trim={0.45cm 0.0cm 0.5cm 0.0cm},clip]{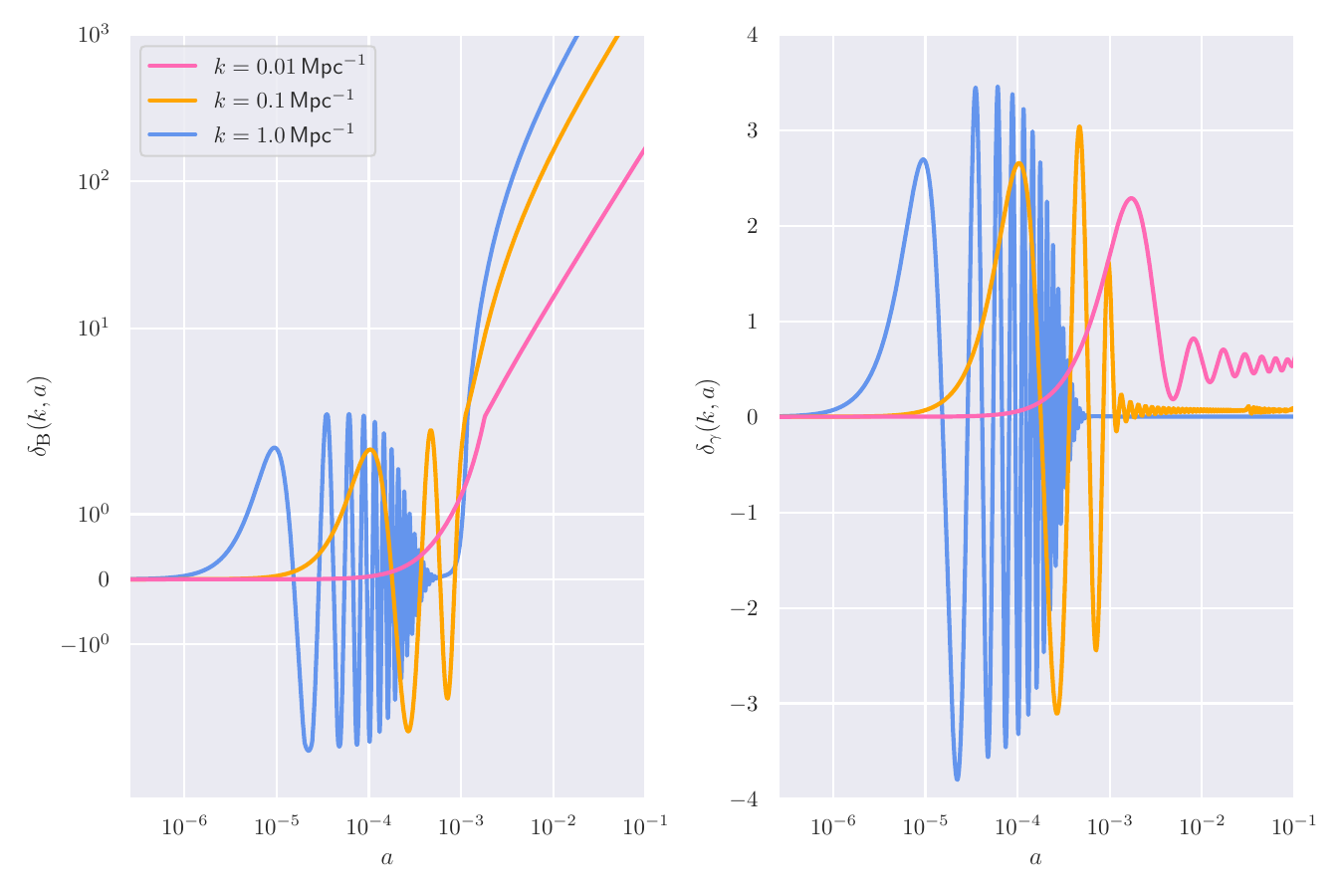}
	\caption{
		Evolution of the perturbations in the baryonic (left panel) and 
		radiation (right panel) components of our Universe for three different
		scales. We see that prior to decoupling ($a \sim 10^{-3}$), the baryons
		are highly coupled to the photonic fluid, which rapidly undergoes
		expansion and contraction due to its self-supporting pressure. This
		inhibits the gravitational collapse of the baryonic fluid, and forces
		their perturbations to oscillate. Once recombination has taken place
		and the baryons are no longer coupled to the photons, the baryonic
		perturbations can gravitationally collapse, `catching up' with the
		dark matter perturbations which have been free to grow under gravity
		since the early Universe (Figure~\ref{fig:delta_c_evo}).
		These curves were computed using \Camb for a $\Lambda$CDM universe that
		is compatible with observations.
	}
	\label{fig:delta_b_r_evo}
\end{figure}

Figure~\ref{fig:delta_b_r_evo} plots the evolution of the baryonic and photonic
perturbations for three physical scales across the evolution of our Universe.
We see that prior to recombination these two fluids are highly coupled, and that
both feature heavy oscillations in their amplitudes, as predicted from
Equation~\ref{eqn:baryonic_photonic_per_evo}. We also see that the frequency of
oscillation is highly $k$ dependent, with short-range modes exhibiting higher
frequency oscillations. It also shows that the longer range modes start 
oscillating later on in the Universe's evolution, which is predicted from the
Jeans length and the fact that these long-range modes re-entered the horizon
significantly after the shorter-range modes. After recombination, the
baryonic perturbations are free to catch up with the dark matter perturbations
which is shown by their steep increase after $a \sim 10^{-3}$, while the 
photonic fluid which still features a self-supporting pressure term 
undergoes damped harmonic motion with an amplitude and frequency related to
their physical scale.

\section{The matter power spectrum}

Now that we have an understanding of how density perturbations evolve during
the Universe's evolution, we wish to use this to produce estimates for
quantities that we hope are either directly observable, or lead to quantities
that are. Hence, we want to consider statistics of the density perturbations
$\delta$, just as how we looked into statistics of the curvature perturbation
$\zeta$ during inflation.

By construction, the spatial average of the density perturbation $\delta$ is
zero at any given time (cf. Equation~\ref{eqn:def_density_pert_delta})
\begin{align}
	\langle \delta \rangle = \left\langle \frac{\rho}{\bar{\rho}} - 1 \right\rangle
	= \frac{\langle \rho \rangle}{\bar{\rho}} - 1 = 0,
\end{align}
and thus the first non-zero statistic is the two-point correlation function
$\xi$, defined through
\begin{align}
	\xi(t, \vec{x}, \vec{y}) = \langle \delta(t, \vec{x}) \, \delta(t, \vec{y}) \rangle.
\end{align}
However, due to our assumptions of homogeneity and isotropy (the cosmological
principle), should not depend on the exact positions $\vec{x}$ and $\vec{y}$,
but of the magnitude separation vector $\vec{r}$ between the two
$r \equiv | \vec{x} - \vec{y}|$ giving $\xi$ as $\xi(t, r)$.

We can now consider the Fourier transform of our correlation function, to
find the two-point function for the momenta of
\begin{align}
	\langle \delta(t, \vec{k}_1) \, \delta(t, \vec{k}_2) \rangle =
	\iint \!\! \d^3 \vec{x} \, \d^3 \vec{y} \,\, e^{i \vec{k}_1 \cdot \vec{x}}
	e^{i \vec{k}_2 \cdot \vec{y}} \,
	\langle \delta(t, \vec{x}) \, \delta(t, \vec{y}) \rangle.
\end{align}
Since we know that this should not depend on the raw values of $\vec{x}$ and
$\vec{y}$, we can use that $\vec{y} = \vec{x} - \vec{r}$, to give
\begin{align}
	\langle \delta(t, \vec{k}_1) \, \delta(t, \vec{k}_2) \rangle & =
	\iint \!\! \d^3 \vec{x} \, \d^3 \vec{r} \,\, e^{i (\vec{k}_1 + \vec{k}_2) \cdot \vec{x}}
	e^{-i \vec{r} \cdot \vec{k_2}} \, \xi(t, r)                      \\
	                                                             & =
	\int \!\! \d^3 \vec{x} \,\, e^{i (\vec{k}_1 + \vec{k}_2) \cdot \vec{x}}
	\int \!\! \d^3 \vec{r} \,\, e^{-i \vec{r} \cdot \vec{k_2}} \, \xi(t, r)
\end{align}
Thus, we can evaluate the $\vec{x}$ integral, to give out two-point function as
\begin{align}
	\langle \delta(t, \vec{k}_1) \, \delta(t, \vec{k}_2) \rangle =
	(2 \pi)^3 \delta^{(3)}(\vec{k}_1 + \vec{k}_2) \, P(k),
	\label{eqn:matter_pow_def}
\end{align}
where we have defined the power spectrum $P(k)$ as
\begin{align}
	P(k) \equiv \int \!\! \d^3 \vec{r} \,\, e^{-i \vec{r} \cdot \vec{k_2}} \, \xi(t, r).
\end{align}
It can often be convenient to work in the dimensionless power spectrum
$\Delta(k)$ defined through
\begin{align}
	\Delta(k) \equiv \frac{k^3}{2 \pi^2 } \, P(k),
\end{align}
since it is related to the total variance of the density perturbation field,
$\sigma^2$, though
\begin{align}
	\sigma^2 & \equiv \int \!\! \d^3 \vec{x} \,\, \langle \delta^2(\vec{x}) \rangle, \label{eqn:tot_delta_var}             \\
	         & = \int \!\! \frac{\d^3 \vec{k}}{(2 \pi)^3} \, \lvert \delta_{\vec{k}} \rvert^2, \nonumber \\
	         & = \int \!\! \d \ln k  \,\, \frac{k^3}{2 \pi^2} P(k),
\end{align}
and thus $\Delta(k)$ is the contribution to the total variance per logarithmic
interval in $k$.

\subsection[Sigma-8]{\boldmath$\sigma_8$}

Equation~\ref{eqn:tot_delta_var} gives the total variance of the density
perturbation field, however it can useful to consider the variance smoothed
by a filter of comoving radius $R$. This smoothed variance is therefore given as
\begin{align}
	\sigma^2(R) = \int \!\! \d \ln k \, \, \Delta(k) W^2(kR),
\end{align}
where $W$ is a window function. This is usually chosen to be a spherical top-hat
profile given by~\cite{Schaefer:2018jwu}
\begin{align}
	W(x) = \frac{3}{x^3} \left[\sin x - x \cos x\right].
\end{align}

From our spatially smoothed variance, by choosing a fixed scale, we can define a
parameter that allows for easy comparison for the amplitude of the power
spectrum between different models and for comparison between theory and
experiment. For historical reasons, this
scale was chosen to be $8 \, h^{-1} \, \textrm{Mpc}$\footnote{The value of 
$R= 8 \, h^{-1} \, \textrm{Mpc}$ was chosen as when we look into our local
Universe, the relative fluctuations of the number density of galaxies are
order unity on this scale, and thus $\sigmaeight$ will be of order unity without
having to do any complicated cosmology. A value of order unity is nice property
for our cosmological parameter~\cite{Prat:2025ucy,Bartelmann:1999yn}.}, and thus we find the smoothed
variance to be $\sigma_8$ as
\begin{align}
	\sigma_8^2 = \int \!\! \d \ln k \, \, \Delta_{\textsc{lin}}(k) \,
	W(k \, 8 h^{-1} \, \textrm{Mpc}),
	\label{eqn:sigma_8}
\end{align}
where we are explicitly using the \textit{linear} dimensionless power spectrum.

\subsection{Evolution of the matter power spectrum}

Now that we have the tools to investigate the matter power spectrum, we can
look at how it changes during the Universe's evolution and for different scales.
Schematically, we can write the evolution of the Newtonian potential as a
$k$-dependant transfer function $T(k)$, and an overall scaling/growth function
$D_{+}(k)$ to give~\cite{Dodelson:2021ft}
\begin{align}
	\Phi(k, a) = \frac{3}{5} \zeta(k) \times \left[\textrm{Transfer function} (k)\right]
	\times \left[\textrm{Growth function}(a)\right].
\end{align}
The transfer functions are defined to `transfer' the perturbation of a given
wavenumber $k$ from an initial time $t_i$ to a later time $t_f$,
\begin{align}
	\Phi(t_f, k) = T(k) \Phi(t_i, k).
\end{align}
These transfer functions depend on their wavenumber compared to the
wavenumber of the horizon during matter-radiation equality
 ($z \sim 3400$~\cite{Tkachev:2017fdu})
\begin{align}
	k_{\textsc{eq}} \equiv (aH)|_{\textsc{eq}},
\end{align}
and thus modes smaller than this ($k > k_{\textsc{eq}}$) entered the horizon
during radiation domination, while larger modes ($k < k_{\textsc{eq}}$) entered
the horizon during matter domination. We have seen in Sections~\ref{sec:pert_evo_rad_dom}
and~\ref{sec:pert_evo_mat_dom} that the matter perturbations, $\deltam$, grow
with different rates in radiation domination and matter domination, respectively.
Hence, the evolution of these density perturbation modes will be determined by
when that mode enters the horizon (which is in turn is determined by its
wavenumber $k$). Thus, different modes will have different different transfer
functions.

The transfer functions can be computed, and are
\begin{align}
	T(k) \propto \begin{cases}
		             1 \,\,      & k < k_{\textsc{eq}}, \\
		             k^{-2} \,\, & k > k_{\textsc{eq}}.
	             \end{cases}
\end{align}

Recalling the definition of our primordial power spectrum,
Equation~\ref{eqn:power_law_inflation}, we find the linear matter power spectrum
at late times to be
\begin{align}
	P_\textsc{lin} (k, a) = \As \, T^2(k) \left(\frac{k}{k_*}\right)^{\ns},
\end{align}
and thus we find the overall $k$-scaling to be
\begin{align}
	P_{\textsc{lin}}(k) \propto \begin{cases}
		                            k^{\ns} \,\,     & k < k_{\textsc{eq}}, \\
		                            k^{\ns - 4} \,\, & k > k_{\textsc{eq}}.
	                            \end{cases}
	\label{eqn:P_k_break}
\end{align}

\subsection{Non-linear evolution}

So far, we have concerned ourselves with linear perturbations where they
have been described using perturbation theory with linearised equations.
However, this is only applicable for small fluctuations, where $\delta \ll 1$.
Thus, when these fluctuations start to become order unity, $\delta \sim 1$,
perturbation theory breaks down, and thus the evolution of perturbations are
described by solving the fully-coupled differential equations in an expanding
FRW spacetime. These calculations are extremely complex, and even non-linear
perturbation theory fails to properly describe the perturbations in the
densest and smallest-scales~\cite{Bernardeau:2001qr}.

Hence, we are forced to turn to numerical methods in order to correctly
determine the behaviour of the matter power spectrum on the smallest scales,
the scales that hold a huge amount of cosmological information. This is
because there are very many more small-scale modes than large-scales for
a given volume of a cosmological survey. Hence, as the amplitude of shot noise
is inversely proportional to the number of modes, the small-scale modes 
inherently carry more cosmological information. 

\begin{figure}[t]
	\centering
	\includegraphics[width=0.975\linewidth,trim={0.0cm 0.0cm 0.0cm 0.0cm},clip]{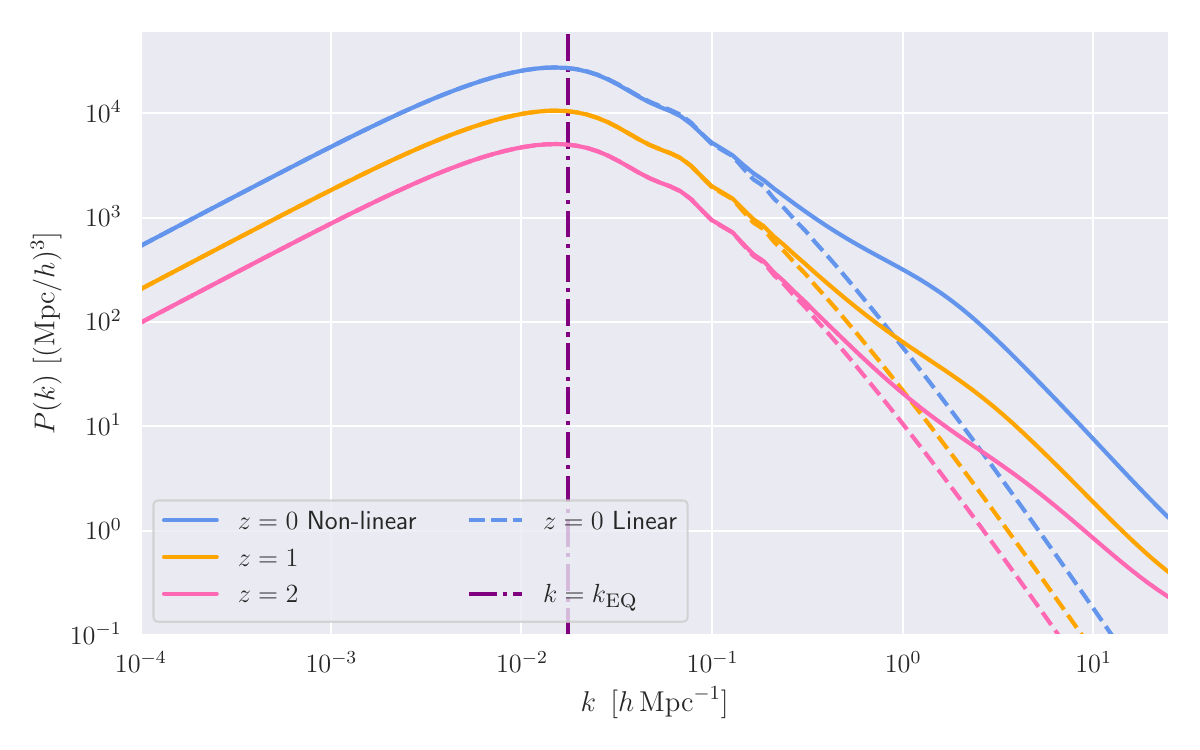}
	\caption{Linear and non-linear matter power spectrum evaluated at redshifts 
	0, 1, and 2 using \texttt{CAMB}~\cite{Lewis:1999bs,Howlett:2012mh} with
	the approximate $k$-scale of matter-radiation equality 
	($k_\textsc{eq} \simeq 1.7\times 10^{-2} \,\, h \, \textrm{Mpc}^{-1}$)~\cite{Bahr-Kalus:2023ebd}
	plotted as the dashed vertical line. This shows the extremely strong break
	in the linear matter power spectrum around $k_\textsc{eq}$
	as predicted from Equation~\ref{eqn:P_k_break}, which is a direct result of
	how perturbations grow more slowly in radiation domination (scales smaller
	than $k_{\textsc{eq}}$) and so the growth of structure is suppressed
	on small scales. This also shows the small oscillations in the power
	spectrum around $k \sim 10^{-1} \, h$Mpc$^{-1}$, which arises from the
	oscillations in the photonic and baryonic density perturbations prior to recombination
	(see Figure~\ref{eqn:baryonic_photonic_per_evo}), and are called
	baryonic acoustic oscillations (BAO). We also see the significant
	boost in the power on small scales from non-linear effects, and thus
	accurate determination of the non-linear effects are crucial for the
	correct evaluation of $P(k)$ at small-scales. }
	\label{fig:MatPowSpec}
\end{figure}

The linear and non-linear matter power spectrum is plotted in 
Figure~\ref{fig:MatPowSpec} which shows the clear break in the power spectrum
around the $k_\textsc{eq}$ scale, and the effects of the non-linear effects
on small-scale $k$ modes.

\section{The halo model}

As we have already touched upon, non-linear perturbation theory is both
extremely analytically complex and fails to describe the proper dynamics of
the scales in the matter power spectrum. Thus, to accurately obtain predictions
for the matter power spectrum in the non-linear regime, we turn to numerical
tools as our saviour. However, these numerical $N$-body simulations are
extremely computationally expensive to run, especially for a wide range of
cosmological parameters that we want to test in our analyses of data.
$N$-body simulations capture the non-linear gravitational clustering and 
dynamics of dark matter only, which can produce estimates for the non-linear 
matter power spectrum down to very small physical scales (large $k$). From these
$N$-body simulations, physically-motivated semi-analytic models for the 
non-linear spectra can be constructed, with values for free-parameters obtained
by fitting the models to the $N$-body simulations. Using a suite of such 
$N$-body simulations allows for the comparison and benchmarking of these
semi-analytic models. Note that $N$-body simulations produce predictions for the
non-linear dark matter power spectrum only, without any baryonic feedback present.

It was suggested in Refs.~\cite{Peebles:1974aaa,Peebles:1980abc} that clustering
on highly non-linear scales could be understood through assuming that regions
of large overdensities undergo virialisation and detach from the expanding FRW
background. This then evolved into the halo model~\cite{Seljak:2000abc,Peacock:2000qk,Ma:2000ik,Asgari:2023mej},
which models the total non-linear matter power spectrum as a sum of two 
components
\begin{align}
	P_{\textsc{nl}}(k) = P_{\textsc{2h}}(k) + P_{\textsc{1h}}(k),
\end{align} 
where $ P_{\textsc{2h}}(k)$ is the two-halo term, which is a quasi-linear
term that represents the power generated by the large-scale distribution of
dark matter haloes throughout the universe, and $P_{\textsc{1h}}(k)$ is the
one-halo term which describes the power contributions from the dark matter
clustering within a single halo. By choosing physically motivated functions
for the one- and two-halo terms, one can fit the free parameters of the model
to $N$-body simulations to provide accurate results for the non-linear
matter power spectrum. This approach was first taken in Ref.~\cite{Smith:2002dz}
to form the \HaloFit model, and was extended in Ref.~\cite{Takahashi:2012em}
which increased the accuracy of the fitting-functions from the increase in 
accuracy from $N$-body simulations in the intervening decade. Further
improvements were made in Refs.~\cite{Mead:2015yca,Mead:2016zqy,Mead:2020vgs}
culminating in the \HMCodett software package which claims to have an RMS error
of less than $2.5\,\%$ across a range of cosmologies for scales $k < 10 \, h\textrm{Mpc}^{-1}$
and redshifts $z < 2$.

\begin{figure}[t]
	\centering
	\includegraphics[width=0.975\linewidth]{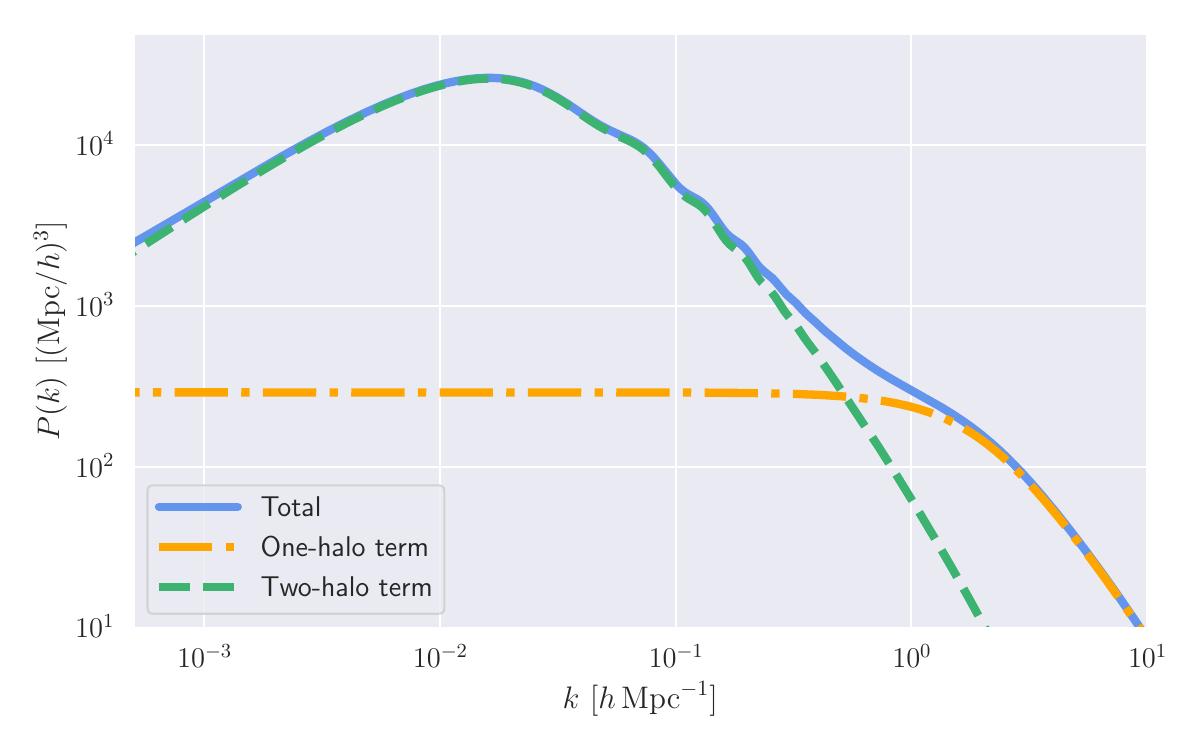}
	\caption{Total matter power spectrum with the one- and two-halo terms 
	from the halo model computed at redshift $z=0$ using 
	\texttt{PyHaloModel}~\cite{Asgari:2023mej}. This shows that the two-halo 
	term dominates the linear power spectrum, with the one-halo term
	dominating the non-linear dynamics.}
	\label{fig:One_two_halo}
\end{figure}

Figure~\ref{fig:One_two_halo} plots the total non-linear matter power spectrum
along with the one- and two-halo terms. We see how the two-halo term
dominates the linear spectrum, since these wavenumbers correspond to
distances much greater than the size of a typical halo, whereas the one-halo
term describes the non-linear dynamics of the power spectrum on the smallest
scales, since these wavenumbers are probing the dynamics of individual haloes.

\section{Baryon feedback in the matter power spectrum}

In our solutions to the background Friedmann equations (Section~\ref{sec:Friedmann_solutions}),
we had that the matter density $\Omegam$ was simply the sum of the cold dark
matter density ($\Omegac$) and the baryonic matter density ($\Omegab$), since
in the late-time Universe the baryonic fluid can be considered pressureless
and so its dynamics are governed from gravitational attraction only. Thus, 
on large scales the Universe's dynamics is driven by only the  total
matter density $\Omegam$, not the individual values of $\Omegab$ and $\Omegac$.
This also holds true for the matter power spectrum, where on scales larger than
$k_\textsc{eq}$ it is only sensitive to $\Omegam$, whereas on smaller scales
the physics of baryons becomes sufficiently different from the physics of
dark matter such that the degeneracy breaks and is sensitive to both $\Omegab$
and $\Omegac$.

We have already seen the first of these different physical processes: during the
radiation dominated era, the baryonic matter was coupled to the radiation and thus
supported from gravitational collapse through radiation pressure. However, the
dark matter has no such coupling to the radiation, and thus it is free to 
condense into gravitational potentials, growing them further. Thus, in a universe
with more dark matter, its gravitational potentials can grow much faster in the
primeval universe than one with more baryonic matter. This produces an increased
amplitude in the matter power spectrum on scales smaller than $k_\textsc{eq}$.

Baryons have many other ways in which they can affect the matter power spectrum
on small scales, which we describe below.

\subsection{Supermassive black holes and active galactic nuclei}

\begin{figure}[t]
	\centering
	\includegraphics[width=0.975\linewidth, trim={0cm 6cm 0cm 5cm}, clip]{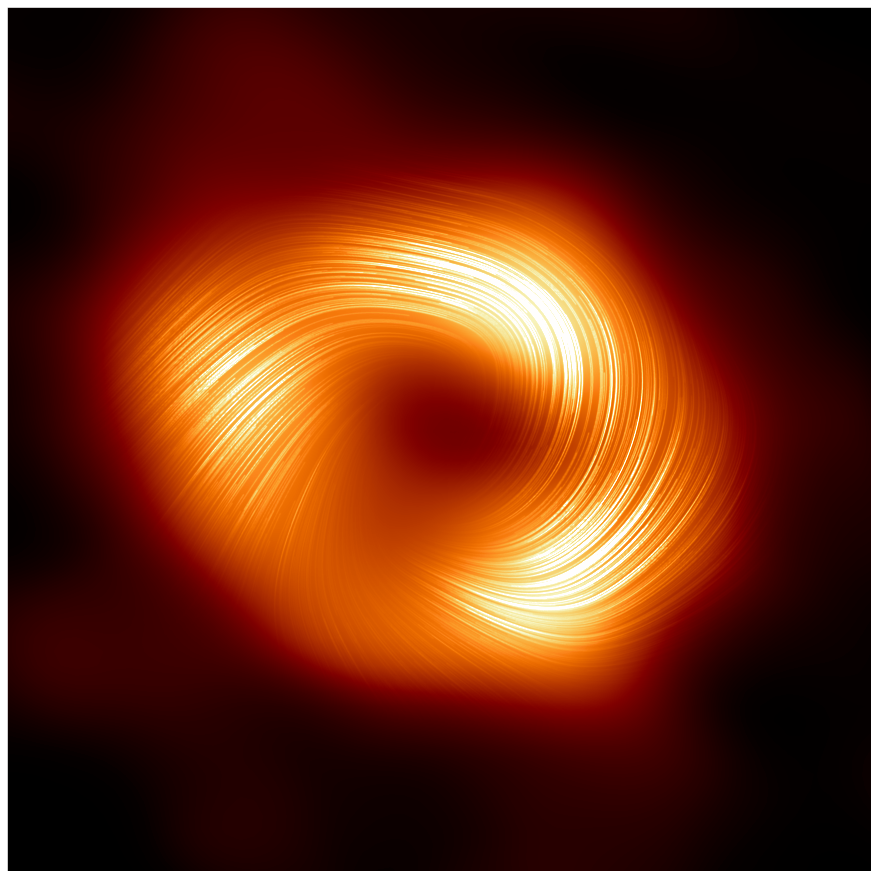}
	\caption{Image of the Milky Way's central supermassive black hole
	Sagittarius~A\!* with magnetic field lines overlaid, taken by the Event Horizon Telescope collaboration
	released in 2024~\cite{EHT:2024a,EHT:2024b}.
	 Sgr~A\!* was discovered in 1974 by Balick and Brown~\cite{Balick:1974ApJ},
	 with the 2020 Nobel physics prize being awarded to Genzel and Ghez for
	 showing that it was a single compact object.}
	\label{fig:EHT_polarisation}
\end{figure}

In order to explain supermassive black holes, let us first take a detour to
explain `ordinary' stellar black holes\footnote{Given the nature of black holes,
with how they represent an unphysical singularity within the laws of general
relativity, there is nothing ordinary about any type of black hole.}. These 
are formed by the gravitational collapse of a dying star, when the gravitational
pull of its matter is greater than what the repulsive force of the Pauli exclusion
principle can even provide, thus collapsing the star to a single point in 
spacetime~\cite{Hawking:1988qt}. While, in theory, general relativity
does not place any constraints on the masses of black holes, since stellar remanent
black holes are the result of the collapse of a dying star, their masses are
restricted to those above the Chandrasekhar limit (about $1.4 \, M_\odot$) up to
the mass limit for which stars can form (which is of the order $100 \, M_\odot$
\cite{Bambi:2019xzp}). Gravitational wave signals from inspiralling black hole
binaries have been confirmed by the LIGO and Virgo detectors~\cite{LIGOScientific:2016aoc,KAGRA:2021vkt},
confirming the GR prediction of both the existence of black holes and
the propagation of gravitational waves. 

Supermassive black holes are extraordinarily massive black holes which can be
found in the centre of galaxies, with typical masses of millions to billions of
solar masses. The black hole at the centre of our galaxy, Sagittarius~A\!*, 
has an estimated mass of $4.2 \times 10^6 \, M_{\odot}$~\cite{GRAVITY:2023avo},
with the one at the centre of Messier~87 having a mass of approximately
$6.5 \times 10^9 \, M_{\odot}$~\cite{EventHorizonTelescope:2019ggy}. Both of
these supermassive black holes have been imaged by the Event Horizon Telescope
collaboration, with Figure~\ref{fig:EHT_polarisation} showing an image of
Sagittarius~A\!* with polarisation field lines overlaid. What we see here is the
electromagnetic radiation emitted from the accretion disk that surrounds the
supermassive black hole. The formation of these supermassive black holes are not
yet completely understood. Observations of quasars with masses around
$10^{10} M_{\odot}$ at redshift $z \sim 6$ (when the universe was less than
1 Gyr old) presents a challenging puzzle for how these objects grew so large so
quickly~\cite{Wu_2015}.

New observations from the James Webb Space Telescope (JWST) further strengthen our
understanding that supermassive black holes formed during the primeval
universe ($z > 6$~\cite{Sorce:2022sgz}), at which point the Universe was only 
800 million years old~\cite{Bosman:2024NatAs81054B}. Spectroscopic observations
from JWST have confirmed the presence of such a supermassive black hole at
$z = 7.0848$ having mass $M = \left(1.52 \pm 0.17 \right) \times 10^{9}\,$M$_{\odot}$,
estimated from observations of the H$\alpha$ emission line~\cite{Bosman:2024NatAs81054B},
and eight objects at redshift $z > 6.5$ with masses $M \sim 10^{9}\,$M$_{\odot}$
using H$\beta$ and [Fe~II] emission lines~\cite{Yang:2023pnw}.
As JWST continues to perform exquisite observations of the primeval universe,
it will undoubtedly find more high-redshift objects which challenge our theoretical
understanding of how these objects got so massive so quickly. 

The accretion of the surrounding interstellar gas onto the supermassive black
hole causes the matter to emit huge amounts of energy across the entire
electromagnetic spectrum, from radio to $\gamma$-rays \cite{Padovani:2017zpf}.
These powerful emissions from the galactic centre are called active galactic
nuclei (AGN, plural AGNi~\cite{Gow:2024yba}).
These emissions energises the interstellar medium, having the effect of moving baryons from
high-density to lower-density regions~\cite{Levine:2006ApJ57L}.
Additionally, the supermassive black hole may power two perpendicular, 
highly columnated jets of relativistic matter which spew forth into the 
interstellar medium, further displacing the baryons out to very large 
distances~\cite{vanDaalen:2011xb}. 

We have detected the presence of such thermal and jet emission from AGN through
high-resolution X-ray observations of galaxy clusters~\cite{McNamara:2007ww}.
These observations have revealed the existence of giant cavities within clusters,
which disrupts the ordinary gravitational collapse of gas, and thus suppressing
or even terminating star formation and the growth of luminous galaxies~\cite{Fabian:2012xr}.
AGN feedback is also essential for reproducing the observed stellar luminosity
function. In particular, we observe a sharp decrement in the luminosity function
of massive galaxies which can only be reproduced through the inclusion of
gas heating from AGN feedback in our theoretical models~\cite{Croton:2005hbr}.

Additionally, measurements of the thermal Sunyaev-Zeldovich effect of the
cosmic microwave background allows the probing of the thermodynamics of hot gas
in massive galaxies, galaxy groups and clusters~\cite{Sunyaev:1980ARA&A18537S}.
This allows us to probe the physics of AGN feedback, and the comparison of
observational data to simulated models~\cite{LeBrun:2015sgq,Amodeo:2020mmu}.

\subsection{Supernovae}
\label{sec:supernovae_feedback}

The baryons in galaxies can condense under gravity to form stars, some of which
undergo violent supernova explosions at the end of their life. These explosions
expel huge amounts of matter at extremely high velocities into the interstellar
medium, further disrupting its original structure. Supernovae are also major
sources of heavy elements, which alters the chemical composition of the 
interstellar medium, which changes the physics of baryonic cooling~\cite{Mohammed:2014lja}.
Supernovae can also trigger star formation via compression of the interstellar
medium from their shock waves, thus increasing clustering on very small scales~\cite{Li:stu1571}.

Supernovae feedback can particularly affect small, or dwarf, galaxies by driving
out large quantities of interstellar gas from these dwarf galaxies. This
dramatically quenches their star formation rates~\cite{Gallart:2021ApJ909192G,Dashyan:2020qnw},
with JWST again providing recent high-quality data on these low-mass quiescent
galaxies~\cite{Strait:2023ApJ949L23S}.

\subsection{Modelling baryon feedback}
\label{sec:modelling_baryonic_feedback}

As we have seen, baryonic physics can have an extremely large effect on the
matter distribution within the interstellar medium, and thus a strong effect on
the matter power spectrum. Since the evaluation of the matter power spectrum
is crucial for many cosmological analyses, the accurate determination of 
baryonic effects within it is essential to ensure unbiased cosmological
constraints~\cite{vanDaalen:2011xb}. While pure analytic models, in the form of
a Taylor expansion in powers of $k$, have been attempted~\cite{Mohammed:2014lja},
more accurate methods in the form of semi-analytic models have had more success
in matching results from fully hydrodynamical simulations (hydro-sims). These
hydro-sims are much more complex than $N$-body simulations, since they aim to 
implement full prescriptions for baryonic feedback, such as AGNi and supernovae
feedback, radiative heating and cooling, chemical processes in the intergalactic
and interstellar mediums, and galaxy formation, in addition to the usual
gravitational dynamics~\cite{Jones:2024MNRAS5351293J}.  

\begin{figure}[t]
	\centering
	\vspace*{-1cm}
	\includegraphics[width=0.975\linewidth, trim={0cm 0cm 0cm 0cm}, clip]{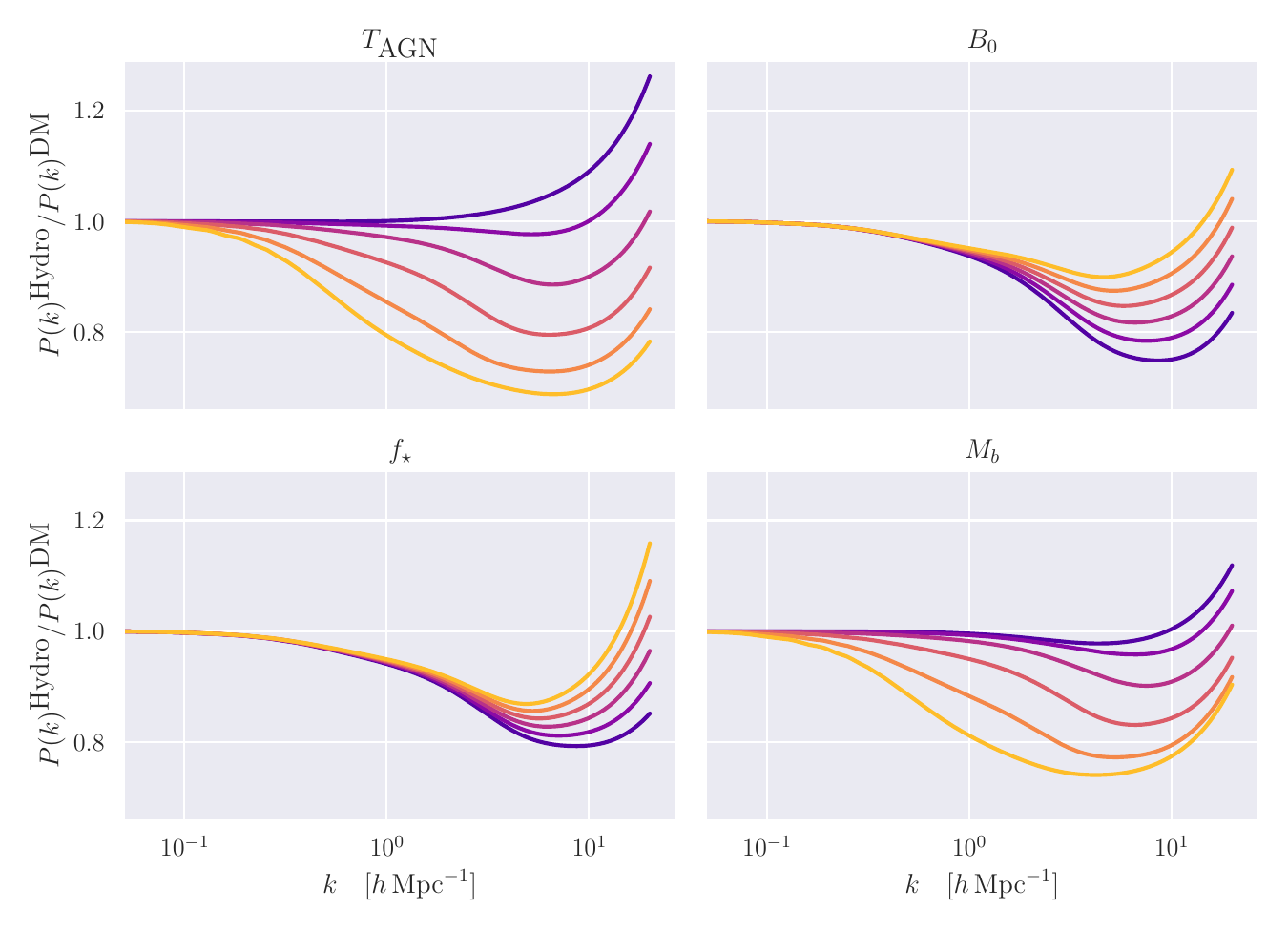}
	
	\caption{Ratios of matter power spectra for models with baryon feedback
		to no feedback at redshift $z=0$ when varying feedback parameters 
		(with smaller values in blue and larger values in yellow for
		each parameter).
		We see that in the single-parameter model, $\TAGN$ both the
		amplitude of the suppression and the scale at which the suppression
		starts to take place (with larger values shifting to larger scales).
		In the three-parameter model, we see that $B_0$, which is the halo 
		concentration parameter, mainly changes the depth of the suppression,
		with smaller values leading to more suppression since more gas has been
		expelled from haloes. $f_\star$ accounts for the presence of stars within
		haloes, and thus we see larger values corresponds to an increase in power
		on very small scales. $M_b$ controls the break-scale which modifies the 
		Navarro-Frenk-White (NFW) window function~\cite{NFW:1997ApJ493N} to
		account for gas expulsion from the halo, and changes both the amplitude
		and starting scale of the suppression. Note that we start plotting the
		ratios from $k \sim 10^{-1} \,h$Mpc$^{-1}$, and so the linear spectrum
		down to $k \sim 10^{-4} \, h$Mpc$^{-1}$ is unaffected by baryonic feedback.
		}
	
	\label{fig:HMCode_baryons}
\end{figure}

Such semi-analytic model including baryonic feedback is \HMCodett~\cite{Mead:2020vgs}, 
which can either have one or three physical parameters that quantify the 
strength and effect of baryonic feedback. The underlying halo model is modified
to change the halo concentration ($B$) and gas content ($M_b$) of haloes, which accounts for
how mass is removed from halo centres due to feedback, and the inclusion of a
central mass term ($f_\star$) which corresponds to the presence of stars in the centres of
haloes~\cite{Mead:2020vgs}. Figure~\ref{fig:HMCode_baryons}
plots the ratio of the matter power spectrum with baryonic feedback to that of
dark matter only clustering for a range of different values for each \HMCodett
astrophysical parameter, at redshift zero.

\section{Probes of the large-scale structure}
\label{sec:probe_of_lss}

Thus far, we have built up a physical picture of how quantum fluctuations
generated in the inflationary universe seeded slight regions of under- and
over-densities right at the beginning of our Universe. 
These perturbations then grow in an expanding radiation-,
matter-, and then a cosmological constant-dominated universe where the
perturbation dynamics are governed by the dominating fluid. Thus, by probing
the large-scale structure throughout the Universe's history, we can constrain
the physics and contents of our Universe. Since different probes are sensitive
to different types of physics, we find that our overall picture of the universe
today is formed from the combination of many individual probes\footnote{\textit{What\textinterrobang \,\,No, not the mind probe.}}, many of which
we will describe here.

\subsection{Supernovae and the distance-redshift relation}
\label{sec:supernovae}

As described above in Section~\ref{sec:supernovae_feedback}, supernovae explosions
can occur at the end of a massive star's life and emit enormous amounts of
electromagnetic radiation, often outshining their entire host galaxy.
Since they are only observable for a short period of time, they are classified
as transient objects. Despite the low chance of any star undergoing a
supernova explosion, we are now observing upwards of ten supernovae per day~\cite{Cappellaro:2022xen}.
Due to their extreme brightness, supernovae have been observed as far back as
185 A.D.~\cite{Zhao:2006ChJAA}. There exists many different classes of
supernova explosions, of which the `Type Ia', which originate from the 
thermonuclear explosions of white dwarfs, are most applicable to cosmological
analyses. This is because their explosions can be thought of as a `standardisable 
candle', and thus their absolute magnitude is approximately the same between
all supernovae Ia explosions -- up to empirical corrections~\cite{Branch:1992rv}.
This makes them prime candidates to test the
distance-redshift relation, which constrains the present-day energy densities
and Hubble expansion rate~\cite{Cappellaro:2022xen}.

In the low redshift limit, Hubble's law of the expanding universe can be written as\footnote{\textit{If my
calculations are correct, when an object hits 560 parsecs away, you're
gonna see some serious shit.} --- Dr Emmett Brown (probably).}
\begin{align}
	v = H_0 \, d,
\end{align}
where $H_0$ is the present expansion rate. Transforming this to a source's redshift,
we find Hubble's law to be
\begin{align}
	z = H_0 d.
\end{align}
Thus, a measurement of both the redshift and distance to supernovae will allow
the constraint on the Hubble rate, though this approximation only holds for
low redshift sources.

For sources at larger redshifts ($z \gtrsim 0.1)$, we turn to the full 
distance-redshift relation for the luminosity distance of Equation~\ref{eqn:luminosity_dist}.
Specialising for a flat universe consisting of matter and a cosmological
constant only (a suitable approximation for any late-universe study, since
any contributions from radiation and relativistic particles have long since
been redshifted away), we find the luminosity distance to be
\begin{align}
	d_{\textsc{l}}(z) = \frac{1 + z}{H_0}  \, \int_{0}^{z} \!\! 
	\frac{\d z'}{\sqrt{\Omega_{\textsc{m}} (1+z)^3 + \Omega_{\Lambda}}}.
\end{align}
If one is able to measure the apparent magnitude $m$ of a source, one
can define the \textit{distance modulus} $\mu$ of a source as $m - M$, where $M$
is its absolute magnitude. It is given in terms of the logarithm of the
luminosity distance as
\begin{align}
	\mu = 5 \log_{10} \left[\frac{d_{\textsc{l}}(z)}{1 \, \textrm{Mpc}}\right]
	+ 25.
\end{align}

\begin{figure}[tp]
	\centering
	\includegraphics[width=\linewidth]{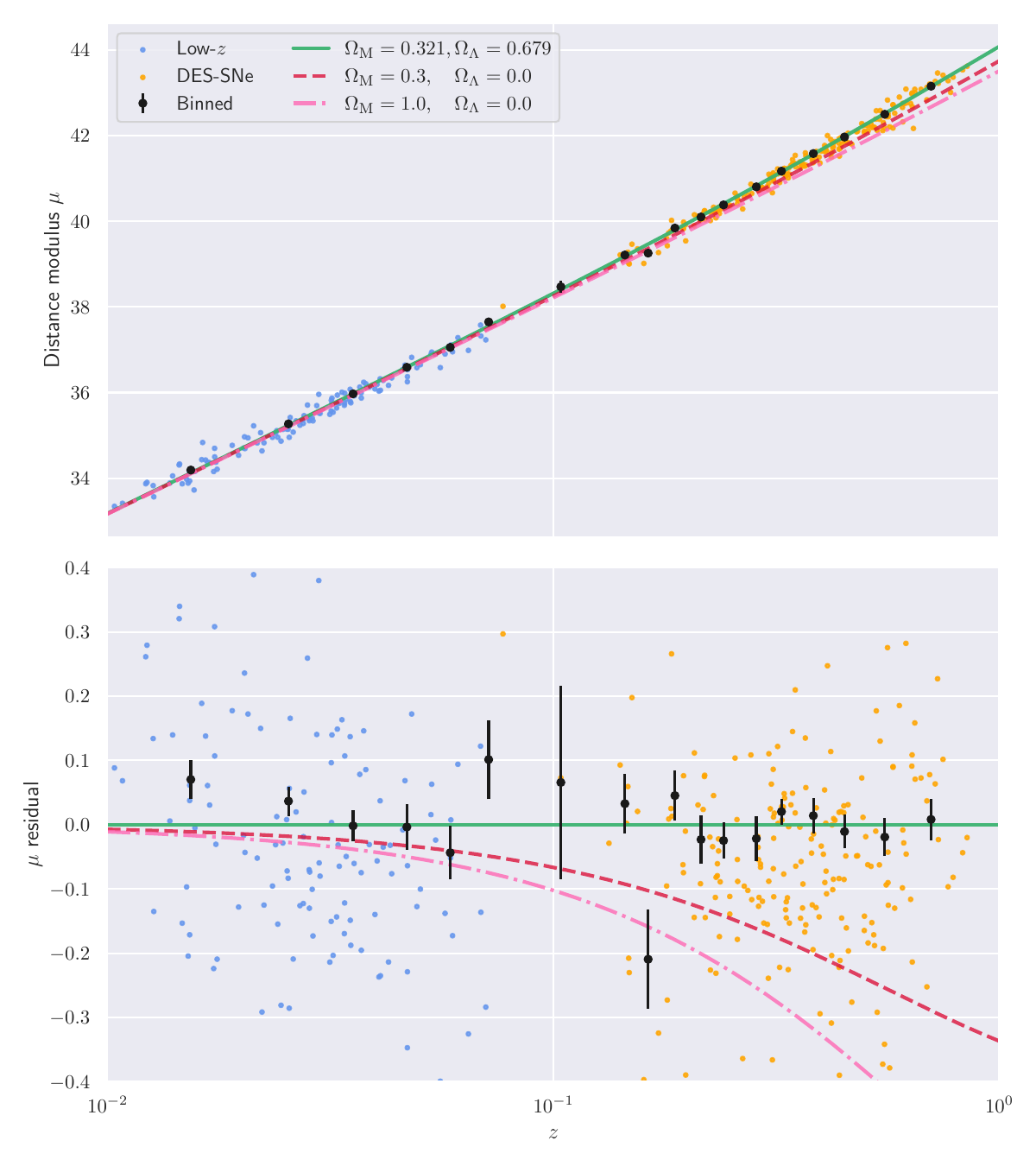}
	\caption{Hubble diagram created using high-redshift supernovae data from the
	 	Dark Energy Survey (DES)~\cite{DES:2018ioi,DES:2018paw} and low-redshift supernovae from the 
	 	Harvard-Smithsonian Center for Astrophysics surveys~\cite{Hicken:2009ApJ331H,Hicken:2012ApJS12H}
	 	and Carnegie Supernova Project~\cite{Contreras:2010AJ519C,Stritzinger:2011AJ156S}.
		We plot the distance modulus to redshift relation for the observed Type Ia supernovae,
		along with binned data-points. Plotted in the solid, dotted, and dashed-dotted
		lines are predictions for this relation for three different cosmologies --
		where the solid green line represents the best-fit cosmology from the data.
		These results are consistent with first precision observations of supernovae from
		the late 1990s~\cite{Riess:1998cb}, showing that we live in a
		universe with a non-zero cosmological constant $\Omega_\Lambda \neq 0$, which
		was a major advancement in observational cosmology.
		}
	\label{fig:DES_SNe}
\end{figure}

Thus, by making measurements of the distance modulus and redshift of many
sources, one can constrain the evolution of the Hubble parameter -- both 
determining the value of $H_0$ and the density parameters $\Omega_i$. Such
example is in Figure~\ref{fig:DES_SNe}, which clearly shows that
supernovae data supports $\Omega_{\Lambda} \neq 0$.

\clearpage
\subsection{The cosmic microwave background}
\label{sec:cmb}

\begin{figure}[t]
    \centering
    \includegraphics[width=0.9\linewidth, page=4, angle=180, trim={2.5cm 2.85cm 2.75cm 2.35cm}, clip]{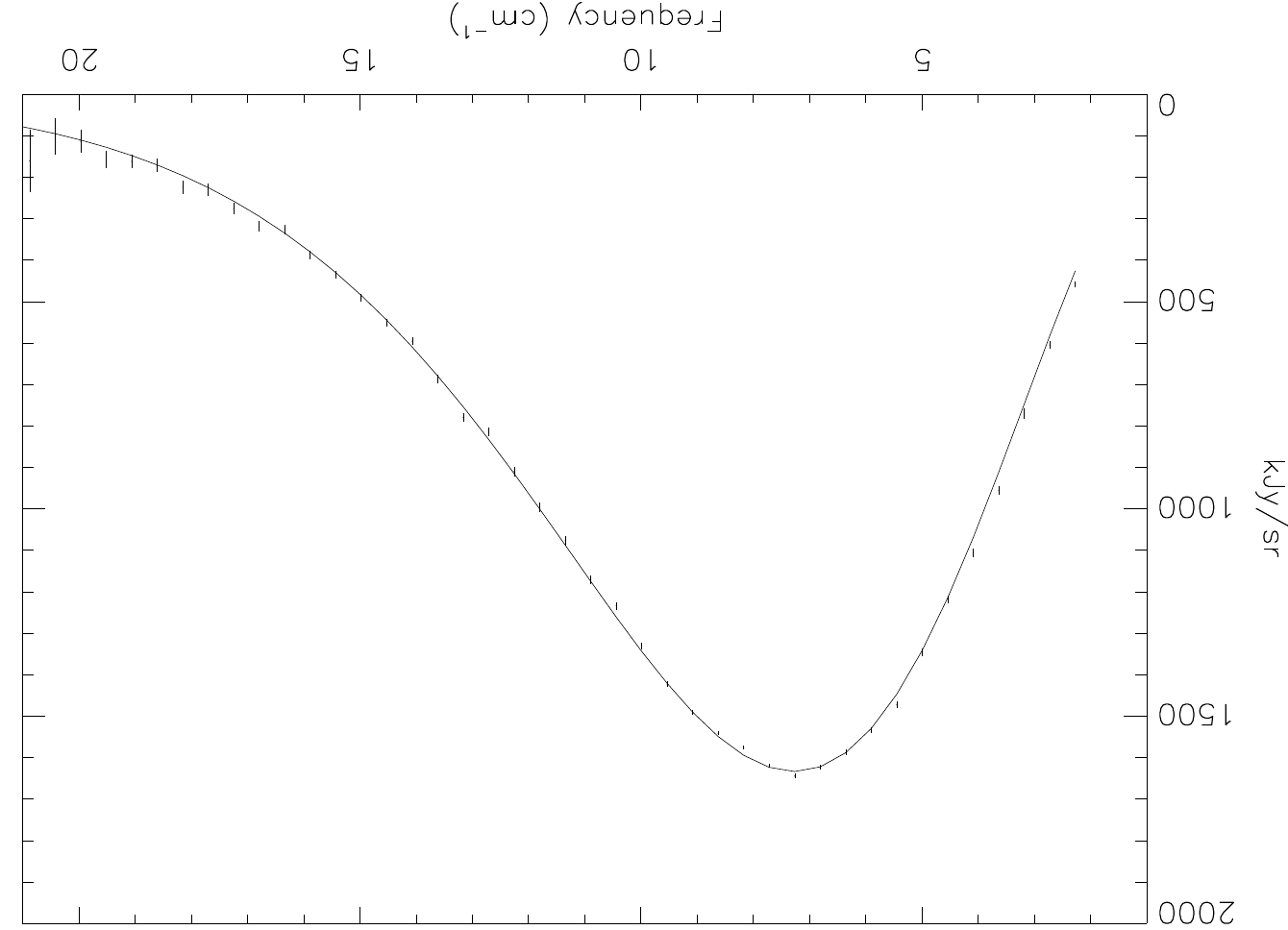}
    \caption{Plot of the CMB black-body spectrum, as measured by the 
    \textit{FIRAS} experiment on the \textit{COBE} satellite. Here, the 
    points represent the observed values, and the line is a fitted black-body 
    spectrum. We see that the black-body spectrum is an excellent fit to the
	data, providing one of the best-fits in all of cosmology and a core proponent
	of the hot Big Bang model.
	Plot taken from Ref.~\cite{COBE:1996fi}.}
    \label{fig:COBE_CMB}
\end{figure}

The cosmic microwave background (CMB) is another hugely powerful probe of 
the large-scale structure of our Universe. It has been measured by very many
experiments over many decades, starting with the initial detection by 
Penzias \& Wilson in 1965~\cite{Penzias:1965wn}. They were trying to measure the
radio emission of our galaxy, the Milky Way. However, what they found was a
persistent signal was present no matter where they pointed their radio telescope
and did not depend on the days or the seasons. They initially believed that this
low-level signal was the result of a pair of pigeons making the 6-metre reflector
their home, leaving the reflector coated in ``a white dielectric 
material''~\cite{Weinberg:1993ftmmW}. After thorough checks of their experiment,
they concluded that their radio signal, which they estimated to be between 2.5\,K 
and 4.5\,K, was of extragalactic origin. These observations were accompanied by
theoretical predictions that such microwave spectra should exist in our Universe
today, if the early Universe was hot and filled with radiation~\cite{Peebles:1965ApJ}.
These observations and accompanying theoretical derivations started the
era of cosmic microwave background cosmology, and has evolved to be one of the
most precise measurements in all of cosmology.

The first accurate, all-sky
CMB survey was performed by NASA's \textit{Cosmic Background Explorer} 
(\textit{COBE}) satellite. Figure~\ref{fig:COBE_CMB} shows the spectrum of the
CMB as measured by its \textit{FIRAS} (Far InfraRed Absolute Spectrophotometer)
instrument with a black-body spectrum at $T = 2.728 \, \textrm{K}$ plotted on
top. This shows that the CMB follows an almost exact black-body spectrum, 
which is the equilibrium distribution for photons of frequency $\nu$ at
temperature $T$ of
\begin{align}
	f_\gamma(\nu) = \frac{2}{\left(2 \pi \right)^3} \, \frac{1}{\exp \left( \nu / T \right) -1 },
\end{align}
where the factor of two arises from the two polarisation states for photons. 
Thus, the energy density distribution for photons following a black-body
distribution is given as
\begin{align}
	\epsilon_\gamma(\nu) = \frac{8 \pi \, \nu^3}{\exp \left( \nu / T \right) -1}.
\end{align}

Since we observe the CMB to be an almost perfect black-body today, we can 
`rewind' the universe backwards to find out the properties of the CMB at
much earlier times. As photons get redshifted due to the expansion of the
universe, we find $\nu \propto 1/a$. Thus, the CMB at earlier times will still
be a black-body spectrum, but one with a shifted temperature of
\begin{align}
	T(z) = (1 + z) T_0,
\end{align}
where $T_0$ is the temperature of the CMB observed today.

\subsubsection{Anisotropies in the CMB}

While the initial studies of the CMB confirmed the existence of the CMB and
that it followed a black-body spectrum, which is an extremely useful probe
of the thermal history of the Universe, the study of the background alone
cannot help us in understanding the origin and evolution of perturbations in
the Universe that we see today in galaxies and superclusters. To investigate
these perturbations, we turn to the \textit{anisotropies} in the CMB, which are
the small deviations from the average temperature over a specific patch of sky.

\begin{figure}[t]
	\centering
	\includegraphics[width=0.975\linewidth]{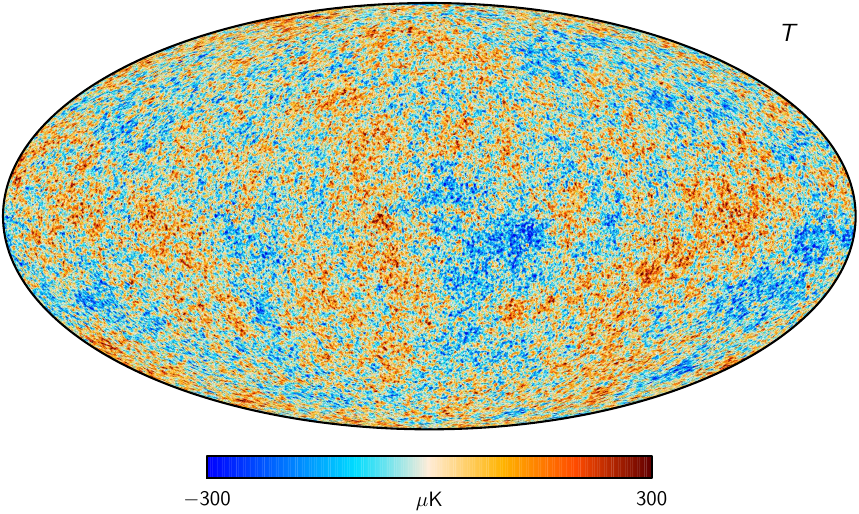}
	\caption{Map of the CMB temperature anisotropies, as measured by the \textit{Planck} 
	satellite and taken from their 2018 data release~\cite{Planck:2018yye}.
	Here, we see the exquisite precision to which \textit{Planck} observed the CMB
	over its five-year lifetime. Thus, \textit{Planck} has provided some of the
	most stringent tests of the general relativity, $\Lambda$CDM, and hot Big
	Bang model yet. The quality of the data is such that \textit{Planck} has almost saturated
	the information content from the primary temperature anisotropies, and so
	future CMB experiments, such as the Simons Observatory~\cite{Simons:2018sbj,Simons:2019BAAS147L}
	and CMB-S4~\cite{CMB-S4:2019BAAS,CMB-S4:2022ght}, aim to measure the 
	polarisation of the CMB to unprecedented accuracy.}
	\label{fig:Planck_CMB_map}
\end{figure}

In the early universe, photons were in thermal equilibrium with the matter, and
thus the study of anisotropies in the CMB will allow us to probe the 
inhomogeneities in the matter density field. The all-sky study of these
CMB anisotropies were initially led by the Wilkinson Microwave
Anisotropy Probe (\textit{WMAP}), and then followed up by the \textit{Planck}
satellite, as shown in Figure~\ref{fig:Planck_CMB_map}. Since the temperature
anisotropies are expected to closely follow a Gaussian distribution (stemming
from how the curvature perturbations from inflation should be largely Gaussian), 
the first non-zero statistical tool for studying the anisotropies is the 
two-point function.

\subsubsection{Cosmology on the sphere}

Since we observe the CMB anisotropies as a function of angular direction only, 
as we can assume that all photons originate from the surface of last-scattering, 
we can treat the anisotropies as a function of angular size on a 2D sphere 
instead of physical wavenumber size, as was done for the inflationary 
perturbations. Thus, we can perform an expansion in terms of spherical harmonics
$\Ylm$, which satisfy the eigenvalue equation for the two-dimensional Laplacian
of
\begin{align}
	r^2 \, \vec{\nabla}^2 \Ylm = - \ell(\ell + 1) \Ylm.
\end{align}
Thus, we can expand the temperature anisotropy at a specific angular point, 
$\Delta T(\hat{n})$, defined as $\Delta T(\hat{n}) \equiv T(\hat{n}) - \langle T \rangle$
where $\langle T \rangle$ is the average CMB temperature, as \cite{Fergusson:2006pr}
\begin{align}
    \Delta T(\hat{n}) = \sum_{\ell = 0}^{\infty} \, \sum_{m=-\ell}^{\ell} \!\! \alm \, \Ylm(\hat{n}),
    \label{eqn:temp_expansion}
\end{align}
where $\alm$ are the expansion coefficients and $\hat{n}$ is the unit vector on
the sky. The $\alm$'s can be found by inverting the above equation, and using 
the orthonormality of the spherical harmonics, to obtain
\begin{align}
    \alm = \int \!\! \d \Omega(\hat{n}) \,\,  \Delta T(\hat{n}) \, \Ylm^*(\hat{n}),
	\label{eqn:alm_def}
\end{align}
where $\d \Omega$ is the area element on a 2-sphere.
The $\ell$ index determines the angular size of the anisotropy on the 
sky, where small $\ell$ corresponds to large angular scales, and vice versa.
The $\ell = 0$ term corresponds to the monopole which, for a correctly normalised
$\Delta T$, should be zero. The $\ell = 1$ term is the dipole, of which the
primary contribution is due to the Earth's motion. Therefore, $\ell \geq 2$ 
modes give us information about the intrinsic CMB anisotropies.

Invariance under rotations impose that $\langle \alm \rangle = 0$, and also 
leads to the two-point function of the $\alm$ to be given as
\begin{align}
    \langle \alm \, a^*_{\ell' m'} \rangle = \delta_{\ell \ell'} \, \delta_{m m'} \, \Cl,
    \label{eqn:Cl_definition}
\end{align}
where $\Cl$ is the power spectrum of the CMB anisotropies.

\begin{figure}[t]
    \centering
    \includegraphics[width=0.975\linewidth,trim={0.25cm 0.3cm 0.7cm 0.7cm},clip]{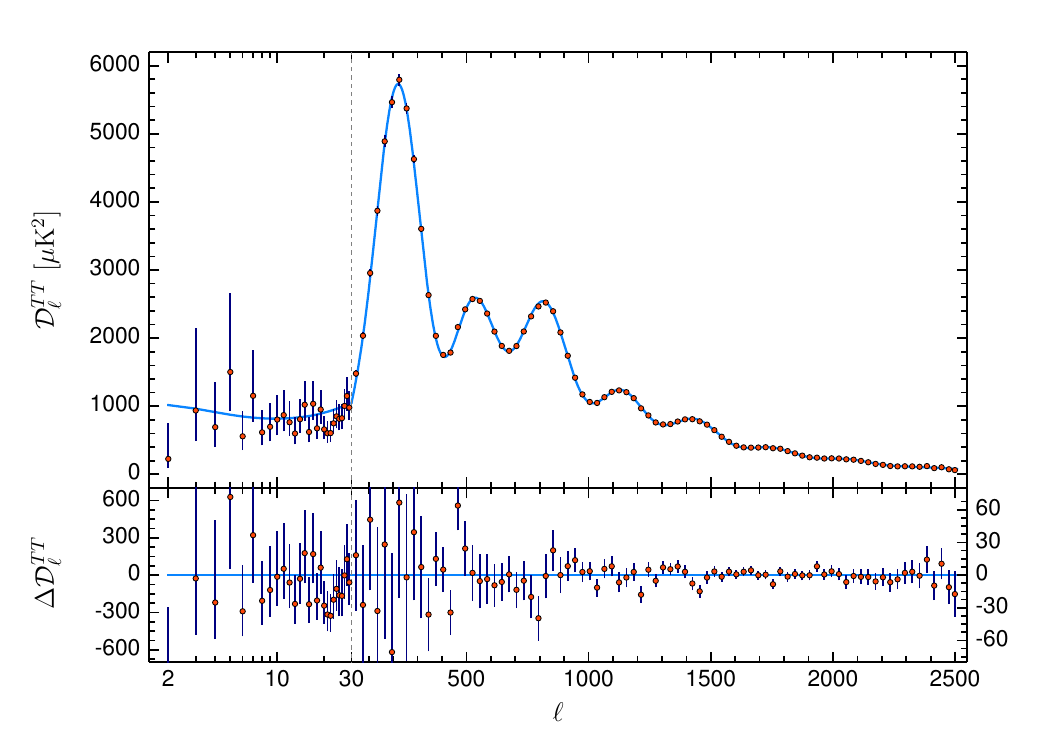}
    \caption{The measured CMB temperature power spectrum performed by the \textit{Planck} satellite,
    taken from their 2018 data release \cite{PlanckCollaboration:2018eyx}.
    The red dots are the measured values, with $\pm 1 \sigma$ errors, and the
	blue curve is the best-fit for the $\Lambda$CDM model. This demonstrates
	the remarkable success of the $\Lambda$CDM model in describing our Universe
	using just six free parameters: $\Omegab$, $\Omegac$, $H_0$, $\As$, $\ns$,
	and $\tau$. The \textit{Planck} results are a real triumph of both
	observational and theoretical cosmology since they produce such amazing
	agreement over across all scales.
    Note that what is plotted is $D_\ell \equiv \ell (\ell + 1)\, \Cl / 2 \pi$.}
    \label{fig:Planck_Cl}
\end{figure}

\subsubsection{Relating the CMB to inflation}

Since we can make accurate measurements of the anisotropies in the cosmic
microwave background, we want to form a theoretical prescription for the angular
power spectrum, such that it can be predicted from a theory of cosmology.
Using Equations~\ref{eqn:alm_def} and~\ref{eqn:Cl_definition}, we find the 
angular power spectrum as
\begin{align}
	\Cl = 4\pi \! \int \! \d \ln k \,\, T_{\ell}^2(k) \, \mathcal{P}_{\zeta}(k),
\end{align}
where $T_{\ell}(k)$ are the CMB transfer functions and $\mathcal{P}_{\zeta}$
is the dimensionless inflationary power spectrum (Equation~\ref{eqn:P_zeta_dimless}).

We can now investigate the nature of the theoretical CMB power spectrum to see 
if we can gain any physical intuition about either the properties or physics of
our Universe which can then be matched to observational data.

\paragraph{The Sachs-Wolfe effect}
\noindent If we consider very large angular scales, then the CMB anisotropies
are determined by large-scale modes that have entered the horizon very recently.
This provides a great way of directly measuring the initial conditions from the
CMB~\cite{Dodelson:2021ft}. This is because the large-scale modes have only just
undergone sub-horizon evolution, and so are mostly fixed to when they left the
horizon during inflation. 
In this large-scale limit, the transfer functions reduce to $-\frac{1}{5} j_{\ell}(k \eta_{*})$,
where $j_{\ell}$ are the spherical Bessel functions, and $\eta_{*}$ is the 
comoving distance to recombination.
Thus, we find that the CMB anisotropies on the largest-scales reduce
to a simple form, which, for a scale-invariant primordial power spectrum of
$\ns = 1$, is approximately
\begin{align}
	\frac{\ell (\ell + 1)}{2 \pi} \Cl \simeq \frac{\As}{25},
\end{align} 
where $\As$ was the amplitude of the primordial power spectrum. This holds for 
angular scales $\ell \lesssim 40$~\cite{Ryden:1970vsj} and can be verified by
comparing to Figure~\ref{fig:Planck_Cl} which measured a roughly constant
Sachs-Wolfe plateau from which we can read off a value of $D_{\ell} \sim 1000\,\mu$K$^2$,
yielding a value of $\As \sim 3\times 10^{-9}$, correct to within a factor of two.

\paragraph{Acoustic peaks}
\noindent Looking at the CMB temperature power spectrum of Figure~\ref{fig:Planck_Cl},
it is clear that there are several strong peaks in the power spectrum. These
are called the acoustic peaks and arise from the photon-baryon interactions
in the early Universe, which we discussed in Section~\ref{sec:photon_baryon_pert}.
Prior to photon decoupling, the photon-baryon fluid
energy density was only about forty percent of the dark matter energy 
density~\cite{Ryden:1970vsj}. Thus, the dynamics of the photon-baryon fluid
are mostly dominated by gravitational interactions of the dark matter. This
dark matter, interacting only gravitationally, is free to clump together and
form deep potential wells in the matter-dominated era, with growth suppressed
in the radiation-dominated era. The photon-baryon fluid will then try to condense
within these potential wells, increasing its pressure as it is compressed by
gravity. This self-interaction pressure eventually grows to a point where it is
greater than the gravitational attraction, and so causes the photon-baryon
fluid to expand outwards. As the fluid expands, the self-interaction pressure
decreases until it starts to fall back into the potential wells. The cycle of
infalling and expulsion is called acoustic oscillations and are what cause the
acoustic peaks in the CMB power spectrum.

The primary acoustic peak occurs at around $\ell \simeq 220$~\cite{Ryden:1970vsj}
and corresponds to potential wells where the photon-baryon fluid reached
maximum compression at the time of decoupling. The subsequent peaks represent
integer multiples of the compression peaks for the photon-baryon fluid.

Since the location of these acoustic peaks in multipole-space strongly depend
on the angular diameter distance to the surface of last scattering, which 
in-turn depends on the spatial curvature of the Universe, measurements of the
acoustic peaks provide strong constraints on the curvature of our Universe. 

Additionally, measurements of the relative amplitudes of the acoustic peaks
provide strong constraints on the `physical' baryon density, $\Omegab h^2$.

\subsection{Baryonic acoustic oscillations}
\label{sec:bao}

\begin{figure}[t]
    \centering
    \includegraphics[width=0.825\linewidth,trim={0.0cm 0.0cm 0.0cm 0.0cm},clip]{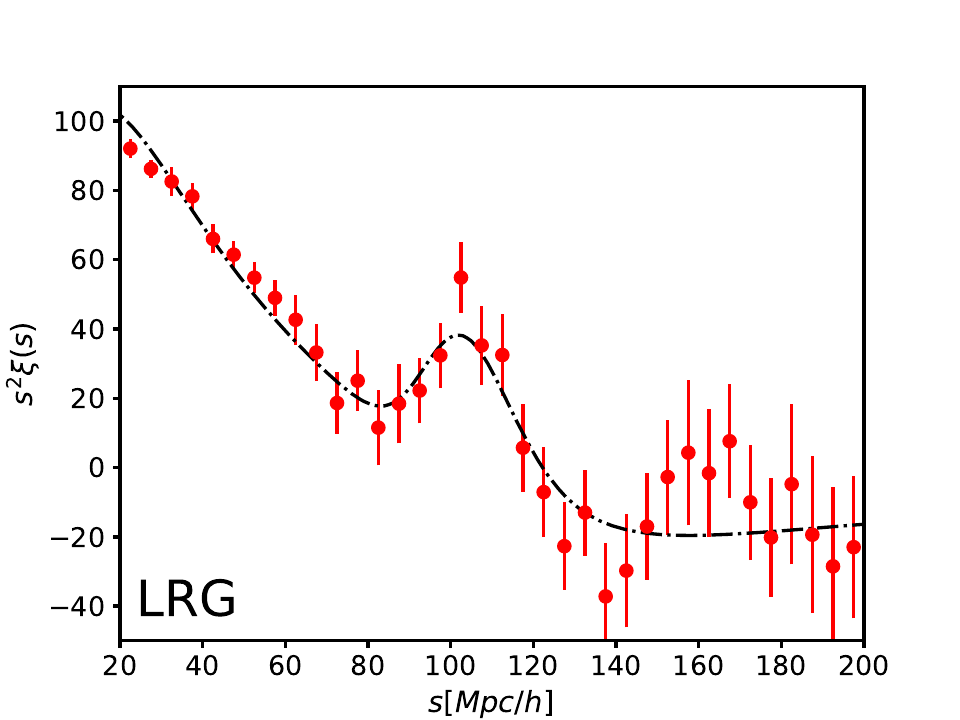}
    \caption{Measurements of the BAO signal from the DESI early data release 
	one~\cite{DESI:2023bgx}. This shows a strong and significant increase in 
	the clustering of galaxies around the $100\,\textrm{Mpc}/h$ scale, which is
	in perfect agreement with the theory (the dashed black line).}
    \label{fig:DESI_BAO}
\end{figure}

As we have seen in the cosmic microwave background above, the acoustic oscillations
in the photon-baryon fluid causes detectable acoustic peaks in the photon
temperature anisotropies. These oscillations also perturb the baryonic density
distribution, resulting in the creation of baryon acoustic oscillations (BAO).
The primary peak in the baryon fluid can be translated from the angular scale
used in the CMB to a comoving distance at the redshift of last-scattering,
which gives the acoustic scale of around $r_{\textsc{bao}} \approx 100 \, \textrm{Mpc}/h$~\cite{Ryden:1970vsj}.

To see if we can detect this peak in the baryonic fluid, we can measure the
correlation function of galaxies, $\xi(r)$. If you chose any random galaxy, the
correlation function quantifies, on average, how many galaxies are likely to be 
found at a comoving distance $r$ away from it. Thus, if these baryonic acoustic
oscillations caused an increase in the baryonic density on the $r_{\textsc{bao}}$
scale, then we would find a small increase in the galaxy density 
correlation function around the BAO scale, $r_{\textsc{bao}}$. 

This increase in the correlation function is exactly what has been measured by
various galaxy surveys, such as the \textit{Baryon Oscillation Spectroscopic 
Survey} (BOSS) from the Sloan Digital Sky Survey (SDSS), and in the
\textit{Dark Energy Spectroscopic Instrument} (DESI) data release one, as
plotted in Figure~\ref{fig:DESI_BAO}. Since this BAO scale 
acts as a `standard ruler', measuring its location provides very tight
constraints on the evolution of the angular diameter distance, and thus on the
geometry of the universe.

\subsection{Why use different probes?}

It is a natural question to ask why do we have so many different independent
probes of the structure and properties of our Universe? Wouldn't a single, 
high-quality survey provide us with all the information that we could possibly want
about our Universe? To answer this, we must consider that each cosmological
survey that we have discussed (along with many others) probe slightly different
physical processes within our Universe, and at different times within its evolution.
Thus, we find that different probes have
different degeneracy directions between cosmological parameters. For example, 
the amplitude of the high-$\ell$ CMB power spectra constrains the combination
of $\As e^{-2\tau}$, thus having a significant degeneracy between the amplitude 
$\As$ and the optical depth $\tau$.
Similarly, there is a degeneracy between the value of $H_{0}$ and the absolute
magnitude $M$ when analysing supernovae data. Furthermore, measurements of BAO
alone are only able to measure $\Omegam$ and $H_0 r_{\textrm{d}}$\footnote{$r_{\textrm{d}}$ is 
the distance travelled by sound waves between the end of inflation and the 
decoupling of baryons from photons after recombination.}, and thus 
information from Big Bang nucleosynthesis is needed to provide constraints on 
the physical baryon density $\Omegab h^2$, and thus break the degeneracy 
between $H_0$ and $r_{\textrm{d}}$.

Thus, by using and combining different probes of our Universe, we extract much
more information about the physics and properties of our Universe than any 
single experiment can. 

\begin{figure}[tp]
    \centering
    \includegraphics[width=0.975\linewidth,trim={0.0cm 0.0cm 0.0cm 0.0cm},clip]{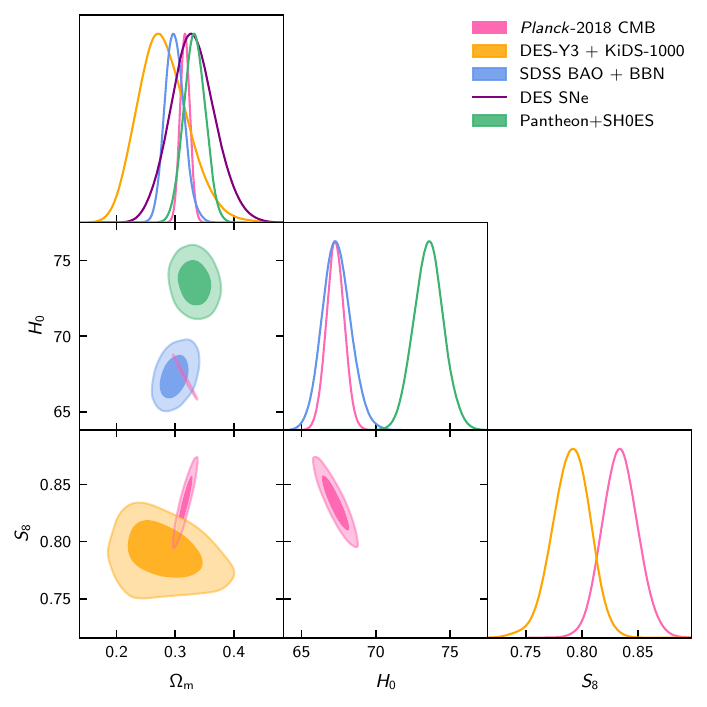}
    \caption{Triangle plot showing the results for three key cosmological
		parameters (the overall matter density today $\Omegam$, the Hubble
		parameter today $H_0$ [in units of $\textrm{km/s}\, \textrm{Mpc}^{-1}$]
		and the clustering of matter $\Seight$) that are well-measured by a
		variety of experimental probes. We see that since each probe measures a
		different physical process within our Universe, this leads to different
		degeneracy directions within the parameter-space. We also see that some
		probes show non-overlapping contours for certain cosmological parameters\dots
	}
    \label{fig:combined_probes}
\end{figure}

\begin{savequote}[65mm]
  Science knows it doesn't know everything,\\
  otherwise it'd stop.

  \qauthor{---Dara \'O Briain}
\end{savequote}

\chapter{Gravitational lensing analysis}
\label{chp:grav_lensing}
\begin{mytext}
    \textbf{Outline.} In this chapter, I review the basic formalism and derive
    the gravitational lensing effect from general relativity. I then go on to
    explain how gravitational lensing can be used in an analysis to constrain
    the properties and content of our Universe. 
\end{mytext}

\section{Gravitational lensing from a point-mass}
\label{sec:grav_lensing_point_mass}

Since we wish to gain a physical understanding of the gravitational lensing
phenomena, we can start with the simplest example in general relativity: a 
light-ray being deflected by a point mass $M$ surrounded by empty space. For
this case, the spacetime metric is the Schwarzschild geometry 
(Equation~\ref{eqn:Schwarzschild_metric}), which we had as
\begin{align}
    \d s^2 = \left( 1 - \frac{2 G M}{r} \right) \d t^2
	- \left( 1 - \frac{2 G M}{r} \right)^{-1} \! \d r^2
	- r^2 \left[ \d \vartheta^2 + \sin^2 \vartheta \, \d \phi^2 \right].
\end{align} 
To solve for how light propagates through this spacetime, we need to solve the
GR equivalent of Newton's second law of classical mechanics ($F = ma$), which
is the geodesic equation. This can be written as
\begin{align}
    \frac{\d^2 x^a}{\d \lambda^2} + \Gamma^a_{bc} \frac{\d x^b}{\d \lambda} \frac{\d x^c}{\d \lambda} = 0,
\end{align}
where $\lambda$ is an affine parameter used to parametrise the particle's path,
and $\Gamma^a_{bc}$ are the Christoffel symbols, which can be computed from 
the metric through
\begin{align}
    \Gamma^a_{bc} = \frac{1}{2} g^{a d} \left[ \partial_b g_{dc} + \partial_c g_{bd} - \partial_d g_{bc} \right].
\end{align}
This is quite an involved approach, and thus to gain additional physical insight
into the process of gravitational lensing, we can instead consider the use
of calculus of variations.

\subsection{Lagrangians}

The fundamental quantity when using the calculus of variations is the Lagrangian
$\mathscr{L}$, which is defined as
\begin{align}
    \mathscr{L}(x^a, \, \dot{x}^a, \, \lambda) = g_{ab} \dot{x}^a \dot{x}^b,
\end{align}
where 
\begin{align}
    \dot{x}^a \equiv \frac{\d x^a}{\d \lambda}.
\end{align}
Since we have a diagonal metric, computing our Lagrangian is straightforward, 
giving
\begin{align}
    \mathscr{L} =  \left(1 - \frac{2GM}{r} \right) \dot{t}^2 
    - \left( 1 - \frac{2GM}{r} \right)^{-1} \dot{r}^2 
    - r^2 \left[\dot{\vartheta}^2 + \sin^2\!\vartheta \, \dot{\phi}^2 \right].
\end{align}
To obtain the geodesic equation, we can simply apply the Euler-Lagrange
equations of 
\begin{align}
    \frac{\d}{\d \lambda} \! \left[ \frac{\partial \mathscr{L}}{\partial \dot{x}^a}\right]
    - \frac{\partial \mathscr{L}}{\partial x^a} = 0.
\end{align}
Noting that our Lagrangian is independent of $t$ and $\phi$, depending only on
their derivatives, we find their associated Euler-Lagrange equation is that of
a conserved quantity
\begin{align}
    \left(1 - \frac{2GM}{r} \right) \dot{t} &= k, \label{eqn:time_geodesic}\\
    r^2 \sin^2\! \vartheta \, \dot{\phi} &= h, \label{eqn:phi_geodesic}
\end{align}
where $k$ and $h$ are constants. Unfortunately, the Lagrangian depends explicitly
on $r$ and $\vartheta$, and thus their Euler-Lagrange equations give
\begin{align}
    \left(1 - \frac{2GM}{r} \right)^{-1} \ddot{r} + \frac{GM}{r^2} \dot{t}^2 
    -  \left(1 - \frac{2GM}{r} \right)^{-2} GM \frac{\dot{r}^2}{r^2} 
    - r\left[ \dot{\vartheta}^2 + \sin^2\!\vartheta \, \dot{\phi}^2 \right] = 0,  \label{eqn:third_geodesic}\\
    \ddot{\vartheta} + 2 \frac{\dot{r}}{r} \dot{\vartheta} - \sin \vartheta \cos \vartheta \dot{\phi}^2 = 0. \label{eqn:fourth_geodesic}
\end{align}
Equation~\ref{eqn:fourth_geodesic} provides an immediate solution for $\vartheta$
as $\vartheta = \frac{\pi}{2}$, which allows us to restrict the light-ray's path
to the equatorial plane with no loss of generality. However,
Equation~\ref{eqn:third_geodesic} provides no such trivial answer. 

To continue, we recall that photons propagate along null geodesics ($\d s^2 = 0$),
and thus we find the line element as
\begin{align}
    \left(1 - \frac{2GM}{r} \right)\dot{t}^2 
    -\left(1 - \frac{2GM}{r} \right)^{-1} \dot{r}^2 
    -r^2 \, \dot{\phi}^2 = 0, \label{eqn:new_geodesic}
\end{align}
where we have specialised to the case where $\vartheta = \frac{\pi}{2}$. To
obtain the light-ray's trajectory, we wish to solve this equation for a function
of the form $r(\phi)$. Using our two conserved quantities (Equations~\ref{eqn:time_geodesic}
and~\ref{eqn:phi_geodesic}), and implementing a coordinate transform of $u \equiv
\frac{1}{r}$, we find the light-ray's propagation to be given by
\begin{align}
    k^2 -h^2 \left[ \frac{\d u}{\d \phi} \right]^2 
    -\left(1 - 2GM u \right) h^2 u^2 = 0,
\end{align}
where now we have an equation containing only $\phi$ and $u$ (and thus $r$).

By performing a $\phi$ derivative, we find our final form to be
\begin{align}
    \frac{\d^2 u}{\d \phi^2} + u - 3GMu^2 = 0, \label{eqn:final_diff_eqn}
\end{align}
which can be solved to find a solution for $u(\phi)$.

This can be solved by considering a zeroth-order solution and then perturbing
it. To zeroth-order, there is no point mass ($M = 0$), and thus our light-ray
propagates along a straight line. In this case, our differential equation 
becomes
\begin{align}
    \frac{\d^2 u}{\d \phi^2} = -u,
\end{align}
which has a solution of
\begin{align}
    u(\phi) = \frac{\sin \phi}{b},
\end{align}
where $b$ is the \textit{impact parameter} and is the radial distance to the
point-mass at closest approach. Thus, our general solution becomes
\begin{align}
    u(\phi) = \frac{\sin \phi}{b} + \Delta u(\phi),
\end{align}
where $\Delta u(\phi)$ is our perturbation. Solving for the full solution gives
\begin{align}
    u(\phi) = \frac{\sin \phi}{b} + \frac{3 G M}{2 b^2} \left[1 + \frac{1}{3} \cos \left(2 \phi\right)\right].
    \label{eqn:u_phi}
\end{align}
To find the deflection angle of our light-ray due to our point mass, we can
consider the angle in the limit $r \rightarrow \infty$ (thus $u \rightarrow 0$)
and double it. Since we expect the angle to be small, we can apply the 
small-angle approximations ($\sin \phi \approx \phi$ and $\cos \phi \approx 1$)
to Equation~\ref{eqn:u_phi} to find $\phi$ as
\begin{align}
    \phi \simeq \frac{-2GM}{b},
\end{align}
and thus the total deflection angle is
\begin{align}
    \lvert \Delta \phi \rvert \simeq \frac{4 \hspace{.1em} G \hspace{-.1em} M}{b}.
\end{align}
This is exactly \textit{double} the expected deflection when calculated from
Newtonian mechanics, and thus an accurate measurement of the deflection angle 
would be able to discern between the Newtonian and Einsteinian theories of
gravity.

\subsection{The Eddington expedition}

With the announcement of Einstein's general theory of relativity at the
Prussian Academy of Sciences in 1915, cosmologists were eager to test this new
theory. Since many of the predictions from Einstein's general relativity
required incredible observational precision, such as it requiring over a century
of technological advances to measure the gravitational waves predicted from 
GR~\cite{LIGOScientific:2016aoc}, the additional factor of two in the deflection
angle over the Newtonian prediction, which could be measured terrestrially and
occurred within our Solar System, became the perfect test-bed for GR. 

Astronomer Royal
Frank Watson Dyson and director of the Cambridge Observatory Arthur Stanley 
Eddington realised that the total solar eclipse of 1919 would be the perfect
opportunity to test the deflection angle and thus sought to lead an expedition
to make measurements during the eclipse. Since the eclipse's path was from
northern Brazil to west Africa, measurements from the UK were not possible. 
Thanks to the recent ending of World War One, international travel was now 
possible, with
Eddington leading a group to the west African island of Príncipe, and fellow
astronomer Andrew Crommelin from the Royal Observatory Greenwich led a group
to the north-eastern Brazilian town of Sobral~\cite{Coles:2019abc}.

\begin{figure}[t]
	\centering
	\includegraphics[width=0.925\linewidth]{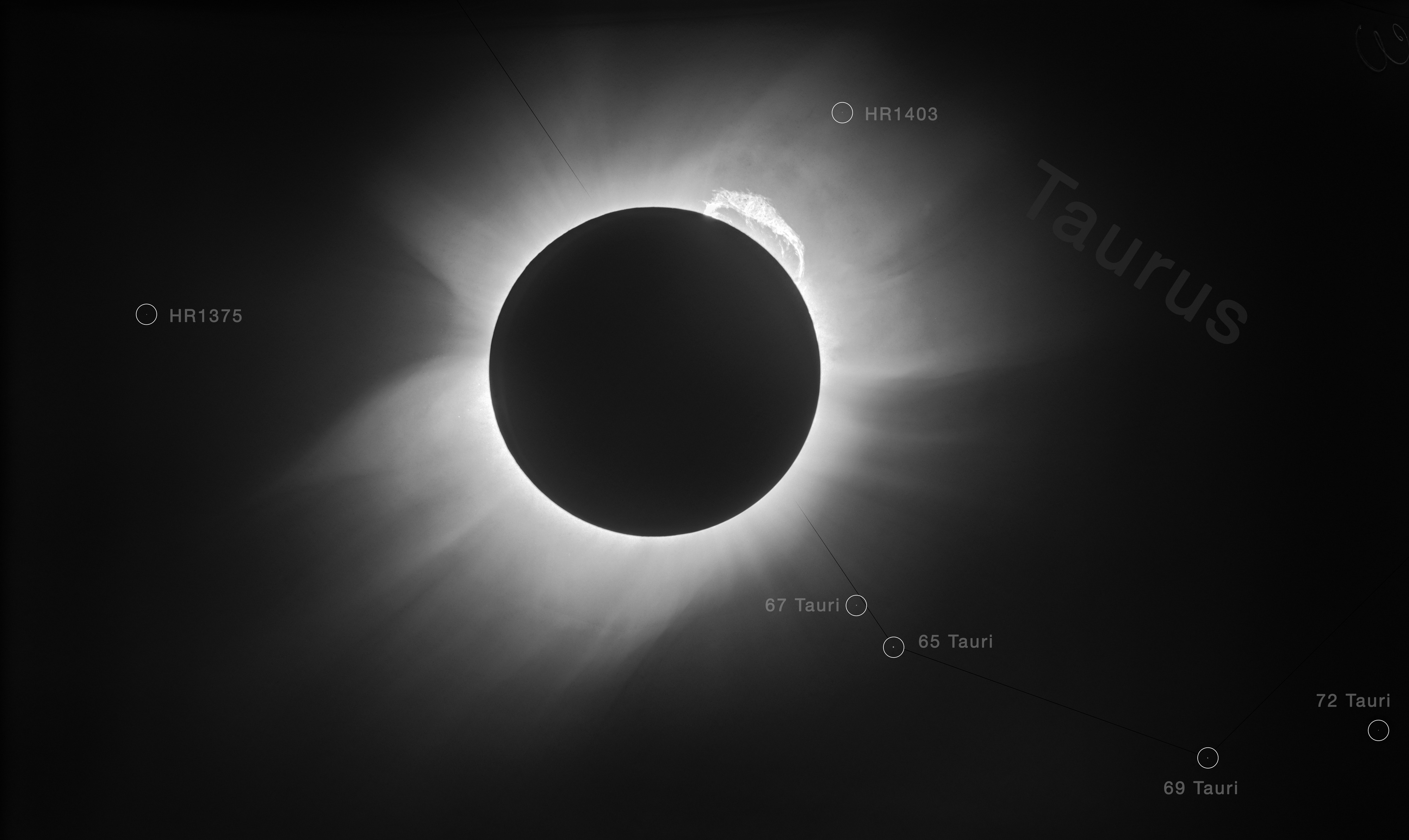}
	\caption{Modern restoration of an original plate as measured by Eddington
        and collaborators during the 1919 Solar eclipse, with the position of stars
        in the Taurus cluster labelled. It was plates like these that allowed the
        astronomers to measure the gravitational deflection angle from the sun,
        and confirm that Einsteinian gravity more accurately predicts the
        angle over Newtonian mechanics.
        Credit: European Southern Observatory / Landessternwarte Heidelberg-Königstuhl / F. W. Dyson, A. S. Eddington, \& C. Davidson.
        }
	\label{fig:Eddington1919}
\end{figure}

Figure~\ref{fig:Eddington1919} shows a modern restoration of one of Eddington's
plates that recorded the position of stars in the Taurus constellation 
during the eclipse. The results of
the expedition gave a result for the deflection angle that was consistent with
the GR prediction, and thus was one of the first major confirmations of GR over
Newtonian gravity~\cite{Dyson:1920abc}.

\section{Weak gravitational lensing}

\begin{figure}[t]
	\centering
	\includegraphics[width=0.675\linewidth]{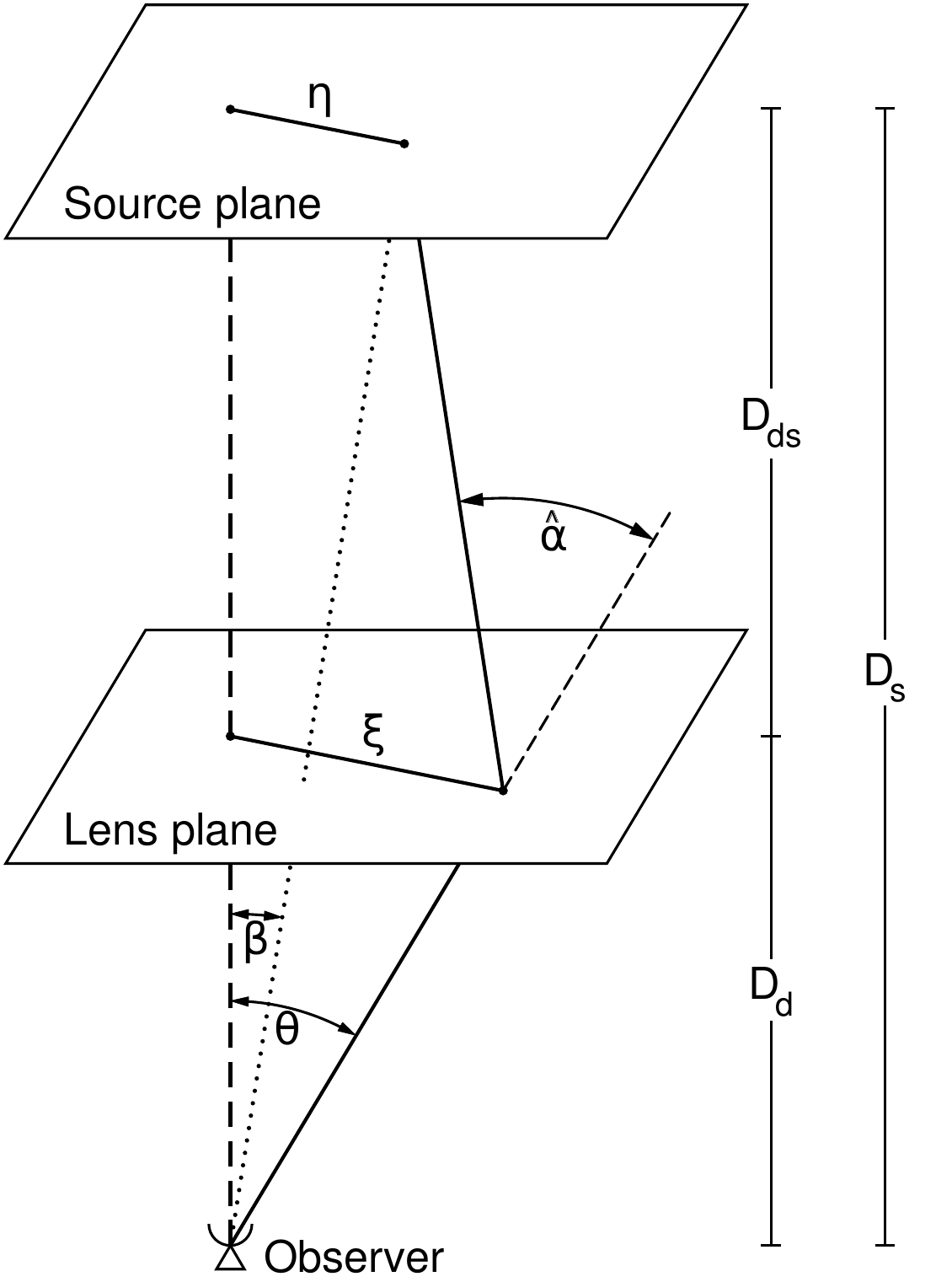}
	\caption{Simplified diagram of a typical lensing scenario: we have a
        source located on the source plane at an angular diameter distance $D_\textsc{s}$ from
        the observer, whose light gets deflected by a lens in the lens plane
        at an angular diameter distance $D_{\textsc{d}}$ from the observer. The lens acts to
        move the observed position of the source from the undeflected angle
        $\beta$ to the lensed angle $\theta$, and so $\hat{\alpha}$ becomes
        the deflection angle. 
        Figure taken from Ref.~\cite{Bartelmann:1999yn}. }
	\label{fig:Lensing_diag}
\end{figure}

In Section~\ref{sec:grav_lensing_point_mass}, we saw how point masses can
deflect light-rays propagating through spacetime. The effect of this is that
when we observe the angular positions of objects on the sky, this may not be
their true positions, since their light-rays could have been deflected by massive
objects located between us (the observer) and the source. Thus, gravitational
lensing creates a map between the observed angular positions and the true
angular positions of objects on the sky. Figure~\ref{fig:Lensing_diag} shows
a typical singular-lensed configuration, where light rays are emitted from an
object located on the source plane, which are then deflected by masses in the
lens plane, and then detected by the observer. Here, we are assuming that the
deflections occur over a much shorter distance in the lens plane than the
distance between the source and lens and lens and observer planes (the thin-lens
approximation). Using some geometry, we can relate the observed angular
separation $\vec{\theta}$, the true angular separation $\vec{\beta}$ and the
deletion angle $\hat{\vec{\alpha}}$ through
\begin{align}
    \vec{\eta} = \vec{\beta} D_{\textsc{s}} = \vec{\theta} D_{\textsc{s}} -
    \hat{\vec{\alpha}} D_{\textsc{ds}},
\end{align}
where $D_{\textsc{s}}$ and $D_{\textsc{ds}}$ are the angular diameter distance
between the observer and source, and source and lens, respectively, and 
$\vec{\eta}$ is the angular separation in the source plane. Defining the
reduced deflection angle as 
$\vec{\alpha} \equiv \left( D_\textsc{ds} / D_\textsc{s} \right) \hat{\vec{\alpha}}$,
we find that the lens equation is given as
\begin{align}
    \vec{\beta} = \vec{\theta} - \vec{\alpha}(\vec{\theta}).
    \label{eqn:lens_equation}
\end{align}
If the lens equation has more than one solution for fixed source position 
$\vec{\beta}$, then the same source can be seen by the observer at multiple
angular separations $\vec{\theta}$ and thus the source is multiply-imaged. This
can happen for the cases where the lenses are `strong'. 

The strength of lenses is characterised by the dimensionless surface mass
density $\kappa(\vec{\theta})$, which is given as
\begin{align}
    \kappa(\vec{\theta}) = \frac{\Sigma (\vec{\theta} \, D_{\textsc{d}} )}{\Sigmacrit},
\end{align} 
where $\Sigmacrit$ is the critical surface mass density and given as
\begin{align}
    \Sigmacrit = \frac{1}{4 \pi G} 
    \frac{D_{\textsc{s}}}{D_{\textsc{d}} \, D_{\textsc{ds}}}.
\end{align}
Thus, as $\Sigmacrit$ is a function of distances, it is therefore a function of
redshifts of the sources and lenses. A lens plane which has $\kappa \geq 1$
can produce multiple images for some positions $\vec{\beta}$, and thus the 
condition of $\kappa \geq 1$ distinguishes sources between `weak' and `strong'.
Though, while images where $\kappa < 1$ may be considered weak, the sources
used in real weak lensing analyses feature $\kappa \ll 1$~\cite{Bartelmann:1999yn}.

\subsection{Propagation of light through the Universe}

Now that we know how light-rays react to the presence of both point masses 
and thin lenses, we which to extend our discussions to deflections of light 
from the large-scale structure of the Universe, not singular lenses. Since
large-scale structures are scattered throughout a photon's path, we need to
consider how the photon propagates through this perturbed spacetime. In
Equation~\ref{eqn:perturbed_FRW_metric}, we introduced the perturbed FRW metric
of
\begin{align*}
    \d s^2 = \left(1 + 2 \Psi \right) \d t^2 - a^2(t) \left(1 - 2 \Phi\right) \d \vec{x}^2,
\end{align*} 
where $\Psi$ and $\Phi$ are our perturbation potentials. Under Einsteinian
gravity, and where there are no anisotropic stresses in the energy-momentum
tensor $T_{ab}$, we find that they are equal, $\Psi = \Phi$. Since photons
propagate along null geodesics ($\d s^2 = 0$), we find that the time taken
for a general photon's path through our perturbed spacetime to be
\begin{align}
    t = \int \! (1 - 2 \Phi) \, \d r,
\end{align}
where the integral is in physical coordinates, $r$. This is analogous to light
propagating through a medium with a variable refractive index given by 
$n = 1 - 2 \Phi$. We can now apply Fermat's principle, that light takes the
path of minimum time, thus $\delta t = 0$~\cite{Prat:2025ucy,Kilbinger:2014cea}.
This gives us the Euler-Lagrange equations for the light ray, and thus can be
integrated to find the deflection angle $\hat{\vec{\alpha}}$ 
(cf. Figure~\ref{fig:Lensing_diag}) can be written as
\begin{align}
    \hat{\vec{\alpha}} = - 2 \int \! \nabla_{\!\!\perp} \Phi \, \, \d r,
\end{align}
where the gradient and integral are both evaluated in physical coordinates.
Since deflections in the photon's path arise from the gradient of the 
gravitational potential perpendicular to the photon's path (parallel gradients
simply change the arrival time of a photon for a source at fixed distance),
this is why we take the two-dimensional gradient only.
Again, this factor of two exists only in Einsteinian gravity and not in
Newtonian dynamics.

To see how light-rays are deflected as they travel through spacetime on 
cosmological scales, let us consider two light rays that separated by a
comoving distance $\vec{x}$. This separation will change as the two light
rays encounter slightly different gradients in the gravitational potential as
they propagate though spacetime. In a completely homogenous universe, one
where there are no perturbations, then the comoving separation between the
two light rays are given by the zeroth-order solution $\vec{x}_0$ of
\begin{align}
    \vec{x}_0 = \fk(\rchi) \vec{\theta}.
\end{align}
However, in a non-homogenous universe where perturbations are present, we
find that these perturbations induce lensing deflections along the 
line-of-sight. For this case, the separation can be written as an integral
along the line-of-sight of the photon, where the superscript zero corresponds
to the fiducial ray from which the separation vector $\vec{x}$ is defined from,
to give
\begin{align}
    \vec{x}(\chi) = \fk(\chi) \,\vec{\theta} - 2 \int_0^\chi \!\! \d \chi' \,\,
    \fk(\chi - \chi') \left[\nabla_{\!\!\perp}\Phi(\chi', \vec{x}) - 
    \nabla_{\!\!\perp} \Phi^{0}(\chi') \right],
\end{align}
where $\vec{\theta}$ is the angle subtended by the two light rays for the
observer, and the source is at some comoving distance $\rchi$. The second, 
fiducial ray $\Phi^{(0)}$ does not require a second argument since we have
defined it to travel along $\vec{x} = 0$.

Applying the lens
equation, we find that we can write the comoving separation $\vec{x}$ in terms
of the observed angle ($\vec{x}(\rchi) = \fk(\rchi) \vec{\beta}$), and thus we
find the reduced deflection angle $\vec{\alpha}$ as
\begin{align}
    \vec{\alpha} = 2 \int_0^\chi \!\! \d \chi' \,\,
    \frac{\fk(\chi - \chi')}{\fk(\chi)}
    \left[\nabla_{\!\!\perp} \! \Phi(\chi', \vec{x}) - \nabla_{\!\!\perp} \! \Phi^{0}(\chi')\right].
    \label{eqn:deflection_angle}
\end{align}

\subsection{The amplification matrix}

The lens equation (Equation~\ref{eqn:lens_equation}) provided us with a relation
between the lensed (observed) coordinates $\vec{\theta}$ to the original
(source) coordinates $\vec{\beta}$. Alternatively, we can express this mapping
through the differential Jacobian transformation matrix $\mathbf{A}$, given as
\begin{align}
    \mathbf{A} \equiv \frac{\partial \vec{\beta}}{\partial \vec{\theta}},
\end{align}
where $\mathbf{A}$ is our amplification matrix. Taking components of this
matrix, we find that they can be written as
\begin{align}
    A_{ij} &= \frac{\partial \beta_i}{\partial \theta_j} 
    = \delta_{ij} - \frac{\partial \alpha_i}{\partial \theta_j}, \\
    &= \delta_{ij} - 2 \int_0^\chi \!\! \d \chi' \,
    \frac{\fk(\chi - \chi') \fk(\chi')}{\fk(\chi)}
    \frac{\partial^2}{\partial x_i \partial x_j} 
    \Phi(\chi', \, \fk(\chi') \vec{\theta}),
\end{align}
where we have used the zeroth-order approximation for $\vec{x}$ as
$\vec{x}_0 = \fk(\rchi) \vec{\theta}$, which in differential form gives
$\partial x_i = \fk(\rchi')\, \partial \theta_i$, and dropped the term
proportional to the gradient of $\Phi^{(0)}$ since it does not depend on the
angle $\vec{\theta}$. Here $\delta_{ij}$ is the Kronecker-$\delta$ symbol which
is either one or zero.

Using our amplification matrix, we can define a scalar lensing potential 
$\psi$ such that it satisfies
\begin{align}
    A_{ij} = \delta_{ij} - \partial_i \partial_j \psi,
    \label{eqn:lensing_pot_def}
\end{align}
where the partial derivates are with respect to $\vec{\theta}$. This definition 
gives $\psi$ as
\begin{align}
    \psi(\rchi, \vec{\theta}) =  2 \!\int_0^\chi \!\! \d \chi' \, 
    \frac{\fk(\chi - \chi')}{\fk(\chi) \, \fk(\chi')}
    \Phi(\chi', \, \fk(\chi') \, \vec{\theta}). 
\end{align}
The amplification matrix $\mathbf{A}$ can be parameterised in terms of a 
scalar field, the convergence $\kappa$, and a spin-2 field, the shear 
$\gamma = (\gamma_1, \gamma_2)$, as 
\begin{align}
    \mathbf{A} = \begin{pmatrix}
        1 - \kappa - \gamma_1 && - \gamma_2 \\
        - \gamma_2 && 1 - \kappa + \gamma_1
    \end{pmatrix}.
    \label{eqn:amplification_mat_explicit}
\end{align} 
This assumes that the amplification matrix is symmetric, which is a very good 
assumption as galaxies are only very weakly lensed, $|\gamma|, |\kappa| \ll 1$,
and thus we can ignore higher-order terms in the expansion. Comparing our
definition of $\mathbf{A}$ in terms of $\gamma$ and $\kappa$ to our definition 
of the lensing potential, as given in Equation~\ref{eqn:lensing_pot_def}, 
we find that the shear and convergence are
given in terms of derivates of the potential $\psi$ as
\begin{subequations}
    \label{eqn:gamma_kappa_psi}
    \begin{align}
        \kappa &= \frac{1}{2} \left(\partial_1 \partial_1 + \partial_2 \partial_2\right) \psi
        = \frac{1}{2} \vec{\nabla}^2 \psi, \label{eqn:kappa_nabla_psi} \\
        \gamma_1 &= \frac{1}{2} \left(\partial_1 \partial_1 - \partial_2 \partial_2\right) \psi, \label{eqn:gamma_1_psi}\\
        \gamma_2 &= \partial_1 \partial_2 \psi. \label{eqn:gamma_2_psi}
    \end{align}
\end{subequations}
Hence, both the convergence $\kappa$ and shear $\gamma$ are related to the
lensing potential $\psi$, which allows us to transform between these 
quantities, such as using the Kaiser-Squires reconstruction 
method~\cite{Kaiser:1992ps}.

\subsection{Physical interpretation of shear and convergence}

\begin{figure}[t]
    \centering
    \includegraphics[width=0.65\linewidth]{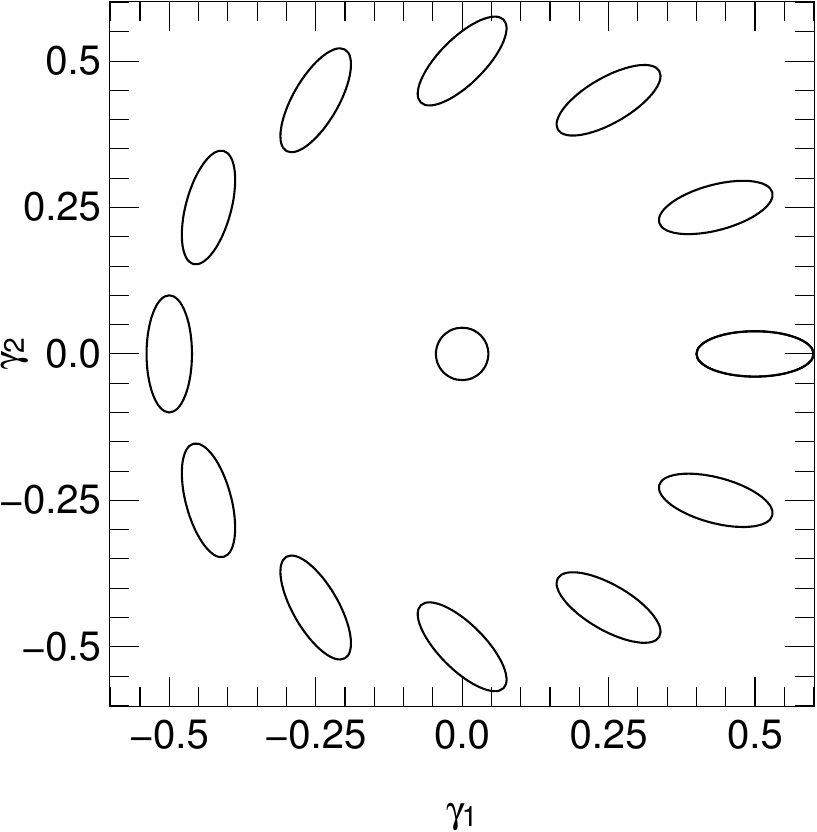}
    \caption{Figure showing that if we started out with a perfectly circular
    original image, the action of $\gamma_1$ and $\gamma_2$ on it serves to
    distort our image into an ellipse, with the orientation of the ellipse
    as a function of the shear components  $\gamma_1$ and $\gamma_2$. We note
    that the distortions from $\gamma_1$ are offset $\frac{\pi}{4}$ from 
    $\gamma_2$, such is the nature of the spin-2 field. 
    Taken from Ref.~\cite{Kilbinger:2014cea}.}
    \label{fig:Lensing_wheel}
\end{figure}

Introduced in Equation~\ref{eqn:amplification_mat_explicit}, we can characterise
the distortions in galaxy images through a spin-0 (scalar) field the convergence
$\kappa$, and a spin-2 (tensor) field the shear $\gamma$. Physically, the
convergence corresponds to an isotropic increase or decrease in the size of
observed images when compared to their undistorted original size. This isotropic
stretching is called the magnification, $\mu$, and is the determinant of the
inverse amplification matrix, $\mathbf{A}^{-1}$. Thus, it is given as
\begin{align}
    \mu &\equiv \det \mathbf{A}^{-1}, \\
        &= \left[\left(1 - \kappa\right)^2 - \lvert\gamma\rvert^2\right]^{-1}, \\
        &\simeq 1 + 2 \kappa,
\end{align}
where the last approximation is taken in the weak-field limit~\cite{Kilbinger:2014cea}.
Thus, $\kappa < 0$ is a demagnification, whereas $\kappa > 0$ is a magnification
in this limit. The shear field $\gamma$ corresponds to the anisotropic 
stretching of images on the sky, turning
circular objects into elliptical images, as shown in
Figure~\ref{fig:Lensing_wheel}.

The shear is called a spin-2 field since the rotation of an ellipse of $\pi$
radians returns the original ellipse, as demonstrated in Figure~\ref{fig:Lensing_wheel}.
With this, it is convenient to express shear in terms of a polar complex number:
$\gamma = \gamma_1 + i \gamma_2 = |\gamma| \exp(2 i \varphi)$, where $\varphi$
is the angle between the two shear components. From this definition, it is
immediately obvious that a rotation of $\pi$ in $\varphi$ returns the original
shear $\gamma$.

\subsection{Overdensity projections}

Equation~\ref{eqn:kappa_nabla_psi} relates the convergence $\kappa$ to a
Poisson equation for the lensing potential $\psi$, and thus the convergence
can be interpreted as a projected surface density. We now wish to use this
relation to find an equation for $\kappa$ in terms of the density perturbation
$\delta$ (Equation~\ref{eqn:def_density_pert_delta}). To do so, we can `add' 
a second-order derivative of the comoving distance $\rchi$, 
$\frac{\partial^2}{\partial \rchi^2}$, which, to good approximation, integrates
to zero along the line-of-sight. This then gives us a 3D Laplacian acting on the
Newtonian potential $\Phi$, $\vec{\nabla}^2 \Phi$, and so we can use the
gravitational Poisson equation (Equation~\ref{eqn:Poisson_grav}) to relate the
Laplacian to the density contrast $\delta$. Thus, we find the convergence field
out to a conformal distance $\rchi$ to be
\begin{align}
    \kappa(\rchi, \vec{\theta}) = \frac{3 H_0^2 \Omegam}{2}
    \int_{0}^{\rchi} \!\! \frac{\d \rchi'}{a(\rchi')} \, \frac{\fk(\rchi - \rchi') \, \fk(\rchi') }{\fk(\rchi)}
    \, \, \delta(\rchi', \fk(\rchi')\vec{\theta}),
    \label{eqn:kappa_chi_theta}
\end{align}
where we have used that $\bar{\rho} \propto a^{-3}$. 

Here, we see that our convergence field is a projection of the density
contrast along the line-of-slight in comoving coordinates, weighted by some
geometrical factors involving distances between the source, lenses, and observer.
For the case of a flat universe, where $\fk(\rchi) = \rchi$, then we find
the geometrical weight to be $(\rchi - \rchi')\rchi'$ which is a parabola with
a maximum at $\rchi' = \frac{\chi}{2}$, and thus lensing structures around 
halfway between the source and observer are the most efficient at generating
lensing perturbations. 

\begin{figure}[tp]
    \centering
    \includegraphics[width=\linewidth]{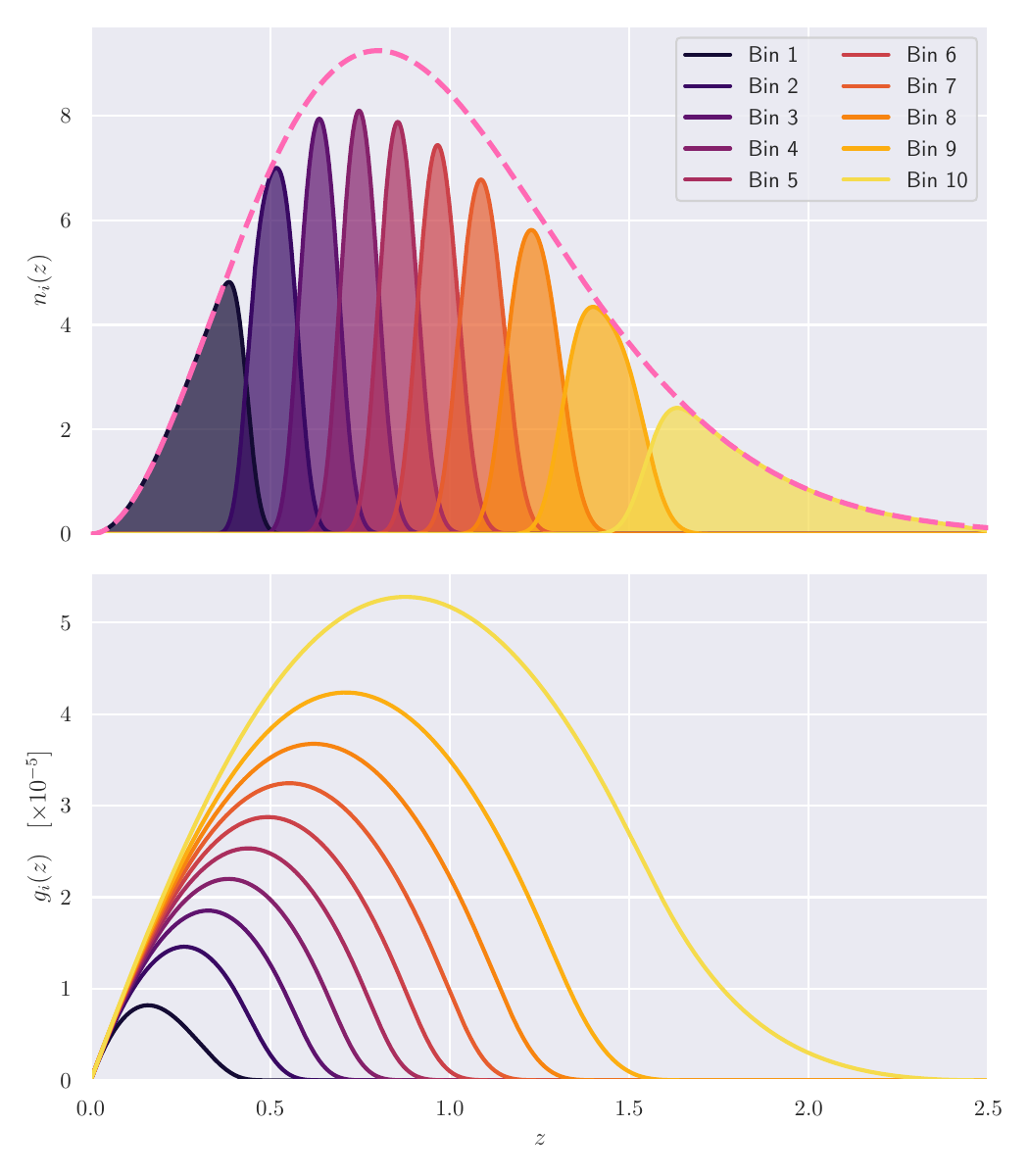}
    \caption{
        \textit{Top panel}: An example Smail distribution (Equation~\ref{eqn:smail_distribution})
        for the total source galaxy number density, in dashed pink, which is then
        split into ten equally-populated tomographic redshift bins (the coloured
        curves), computed using the \textsc{glass} public code~\cite{Tessore:2023zyk}.\\
        \textit{Bottom panel}: The lensing kernels computed for each of the
        corresponding photometric redshift bins in the top panel. We see
        that the lensing kernels peak at around half the redshift of the number
        density peak, which follows from the lens equation, and that higher
        redshift bins have large support in their kernels over a wide range
        of redshifts.
        }
    \label{fig:lensing_kernels}
\end{figure}

In practice, we have a distribution of source galaxies in redshift, $n(z)$,
and thus we wish to find the average convergence for these sources. This 
requires an additional weighting with respect to this $n(z)$ in our
line-of-sight integral. A typical distribution for $n(z)$ is the
Smail distribution~\cite{Smail:1994sx,Euclid:2019clj} of
\begin{align}
    n(z) \propto \left(\frac{z}{z_0}\right)^n \exp \left[- \left(\frac{z}{z_0}\right)^{\frac{3}{2}}\right],
    \label{eqn:smail_distribution}
\end{align}
where $z_0$ is given in terms of the median redshift $z_\textsc{med}$ by
$z_0 = z_\textsc{med} / \sqrt{2}$. For the \textit{Euclid} survey, $z_{\textsc{med}} = 0.9$~\cite{Euclid:2011zbd}.
We can relate this galaxy distribution in redshift-space to comoving 
distance-space by noting that the number of galaxies are conserved, and thus
$n(z) \d z = n(\rchi) \d \rchi$. Thus, the convergence from a distribution of
source galaxies is given by
\begin{align}
    \kappa(\vec{\theta}) = \int_{0}^{\chimax} \!\! \d \rchi 
    \,\, n(\rchi) \, \kappa(\rchi, \vec{\theta}).
\end{align}
where we integrate out to the limiting comoving distance of our galaxy 
distribution $\chimax$. Inserting our equation for $\kappa(\rchi, \vec{\theta})$
of Equation~\ref{eqn:kappa_chi_theta}, we find
\begin{align}
    \kappa(\vec{\theta}) = \frac{3 H_0^2 \Omegam}{2} \int_{0}^{\chimax} \!\!
    \d \rchi \, \, g(\rchi) \frac{\fk(\rchi)}{a(\rchi)} \, \delta(\rchi,\, \fk(\rchi) \vec{\theta}),
    \label{eqn:kappa_theta}
\end{align}
where we have defined $g(\rchi)$ as the lensing efficiency kernel, and is given
as
\begin{align}
    g(\chi) = \int_{\chi}^{\chimax} \!\! \d \chi' \, n(\chi') \, 
    \frac{\fk(\chi' - \chi)}{\fk(\chi')}. 
    \label{eqn:lensing_kernels}
\end{align} 
The lensing kernel represents the efficiency of a lens at distance $\rchi$ 
combined with the source galaxy distribution. We plot the distribution
of source number densities $n(z)$ and their associated lensing kernels $g(z)$
in Figure~\ref{fig:lensing_kernels}.

\subsection{Measuring shear}
\label{sec:measuring_shear}

\begin{figure}[t]
    \centering
    \includegraphics[width=0.65\linewidth]{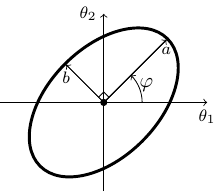}
    \caption{An image of a galaxy can be approximated by an ellipse, 
    characterised by its semi-major axis $a$, semi-minor axis $b$, and
    orientation $\varphi$.
    }
    \label{fig:ellipse_shear}
\end{figure}

To measure the shear field, we first need to measure the ellipticity of
galaxies on the sky. A galaxy can be approximated by an ellipse, which is 
characterised by its semi-major axis $a$, semi-minor axis $b$, and
orientation $\varphi$ (see Figure~\ref{fig:ellipse_shear}). This allows us to
quantify each galaxy with a (complex) ellipticity $\varepsilon$, defined as
\begin{align}
    \varepsilon = \frac{a - b}{a + b} \, e^{2 i \varphi}.
\end{align}
Now, it is highly unlikely that every galaxy is perfectly spherical in its
source plane (e.g. galaxy morphology and classifications from the Hubble
tuning fork~\cite{Hubble:1926ApJ}), and thus we can assume that every galaxy
has some intrinsic ellipticity in its source plane, $\varepsilon^{\textsc{s}}$. Thus 
the observed ellipticity will be its source ellipticity convoluted with
gravitational lensing, to give
\begin{align}
    \varepsilon = \frac{\varepsilon^{\textsc{s}} + g}{1 + g^{*} \, \varepsilon^{\textsc{s}}},
    \label{eqn:observed_ellip}
\end{align}
where $g$ is the reduced shear, defined as
\begin{align}
    g \equiv \frac{\gamma}{1 - \kappa}.
\end{align}
In the weak lensing limit, where $\kappa \ll 1$, then Equation~\ref{eqn:observed_ellip}
reduces to
\begin{align}
    \varepsilon \simeq \varepsilon^{\textsc{s}} + \gamma.
\end{align}
If the original intrinsic ellipticities of galaxies are randomly oriented on
the sky, then the average of $\varepsilon^{\textsc{s}}$ is zero,
$\langle \varepsilon^{\textsc{s}} \rangle = 0$. Thus, the average of the
observed ellipticity of galaxies over a small patch of sky is an unbiased
estimator for the shear field in that patch
\begin{align}
    \langle \varepsilon \rangle = \gamma.
\end{align}

\section{Estimating source galaxy redshift}

A major source of information from weak lensing is our ability to perform
redshift tomography on our data, that is to split our sheared, source galaxies 
into different redshift bins which allows us to measure the \textit{evolution}
of the shear signal with redshift, and increases the number of data-points
in our data-vectors~\cite{Hu:1999ek}. Thus, to perform a tomographic analysis,
we need estimates of the source galaxies redshifts. The most accurate estimates 
for the redshifts of extragalactic objects come from spectroscopy, which can
measure the wavelength of individual atomic emission and absorption lines,
such as: H$\alpha$, H$\beta$, H$\gamma$, $\left[\textrm{OII}\right]$, 
$\left[\textrm{OIII}\right]$, and $\left[\textrm{NeIII}\right]$~\cite{Baker:2025NatAs141B}.
These observations can be compared to measurements performed at-rest in a laboratory,
and so the redshift can be easily computed from the increase in wavelength 
from the Doppler shift (Equation~\ref{eqn:redshift_def}). 

Exceedingly high-quality spectroscopic redshifts of distant galaxies have
been observed using the Near-Infrared Spectrograph (NIRSpec) instrument on
the James Webb Space Telescope~\cite{Baker:2025NatAs141B,NIRSpec:2022A&A80J}.
While we would love to perform spectroscopy on our weak lensing source galaxies,
the amazingly long exposure times required for spectroscopy, which can be
of the order ten thousand seconds~\cite{Baker:2025NatAs141B}, makes spectroscopy
absolutely prohibitively expensive for over the billion galaxies whose shapes we
wish to include in our tomographic analyses. Hence, we are forced to use
alternative methods to estimate galaxy redshifts that can still deliver (somewhat)
accurate results, but with much reduced observation time. 

\begin{figure}[t]
    \centering
    \includegraphics[width=0.975\linewidth,trim={0.0cm 0.0cm 0.0cm 0.0cm},clip]{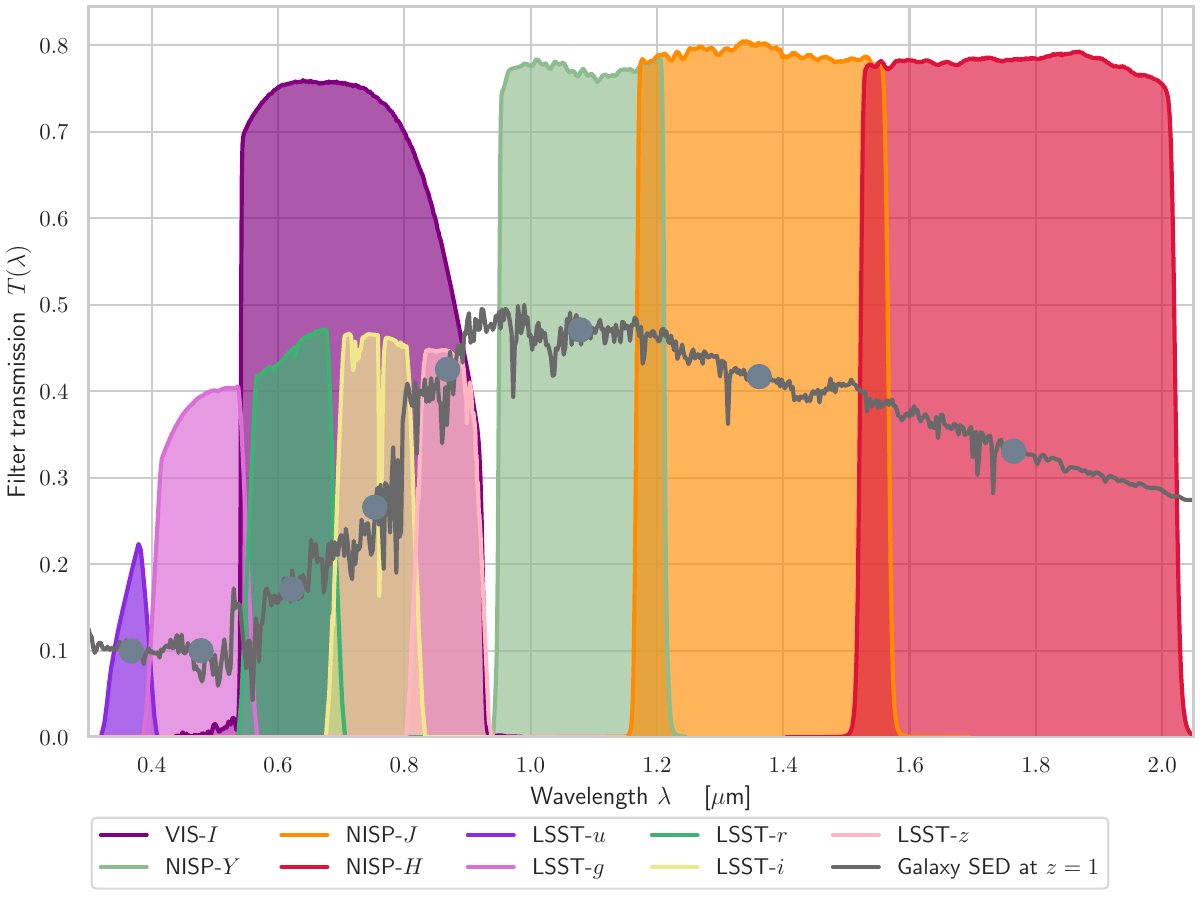}
    \caption{Plot of filter transmission profiles for the space-based 
        \textit{Euclid} VIS and NISP instruments~\cite{Euclid:2024yrr}, along
        with the ground-based LSST survey at the \textit{Rubin} 
        observatory~\cite{LSST:2008ijt}\protect\footnotemark, and an example
        galaxy spectral energy distribution created using the \textsc{Bagpipes}
        code~\cite{Carnall:2018MNRAS4804379C} where the galaxy is located at
        redshift $z=1$ (the SED is normalised to $0.5$ for easier comparison
        to the filter profiles). The photometric measurements using these nine
        filters for this SED is shown in the grey circles. 
        \textit{Euclid} is able to image in the infrared since there is no
        atmosphere in space that absorbs these wavelengths as for Earth-based
        observatories, which is seen by the lower transmission amplitude for the
        five LSST curves. By complementing space-based infrared measurements
        with ground-based optical measurements, higher accuracy photometric
        redshifts can be obtained for the source galaxies, thus optimising the
        weak lensing analysis pipeline.
        }
    \label{fig:photo_band_passes}
\end{figure}

Photometric redshifts work by imaging a galaxy in multiple colours (band-filters)
which can approximately recover the galaxy's spectral energy distribution (SED),
and by comparing this to simulated galaxies SED's at known redshifts an estimate
for the photometric redshift can be obtained (SED fitting~\cite{Bolzonella:2000js}).
Figure~\ref{fig:photo_band_passes}
plots the transmission profile curves for \textit{Euclid}'s VIS instrument, the
three filters associated with its NISP instrument~\cite{Euclid:2024yrr}, the
five filters to be used by the LSST survey at the \textit{Rubin} 
observatory~\cite{LSST:2008ijt}, and an example galaxy SED located at $z=1$ 
created using the \textsc{Bagpipes} code~\cite{Carnall:2018MNRAS4804379C}.
The full SED is hugely complicated, featuring many emission and absorption lines,
which is what allows for precision spectroscopic redshifts to be obtained. The
solid circles in Figure~\ref{fig:photo_band_passes} shows the photometric 
measurements of the SED in the nine filters (excluding the broad VIS filter),
and is what would be used to estimate this galaxies redshift photometrically.
We see that by using the combination of these
nine filters, we can obtain information about a galaxy's SED from the 
near-ultraviolet into the near-infrared which can approximate spectroscopic
information -- though is, of course, far coarser in wavelength than spectroscopy.

\textit{Euclid} alone will be unable to accurately determine photometric 
redshifts, and so needs combining with ground-based optical measurements --
these measurements will come from LSST for the southern sky and the UNIONS 
collaboration for the northern sky.

By performing a tomographic analysis of cosmic shear, that is to utilise the
auto- and cross-correlations of our galaxy shapes at different redshifts, we
can directly constrain the growth and evolution of structure across cosmic time.
This will allow us to place tight constraints on the properties of dark energy,
since it directly affects the background evolution and growth of structure in
our Universe -- with the end goal to determine if dark energy is compatible
with the cosmological constant $\Lambda$ or is some new time-evolving field. 

The uncertainties in the estimates of photometric redshifts is a leasing source
of systematic uncertainty for current weak lensing surveys~\cite{Dalal:2023olq}.
Hence, we require extremely accurate assessment of photometric redshifts for
any forthcoming weak lensing survey which aim to significantly reduce statistical
and systematic uncertainties. Thus, considerable efforts have been spent
to optimise the estimation of photometric redshifts~\cite{Euclid:2020gbk,Euclid:2022oea}.

\clearpage

\footnotetext{\textit{Euclid} data taken from \url{https://euclid.esac.esa.int/msp/refdata/nisp/NISP-PHOTO-PASSBANDS-V1} \cite{Euclid:2022vkk},
LSST data taken from \url{https://github.com/LSSTDESC/SN-PWV/tree/main/data/filters/lsst_baseline}~\cite{LSST:2008ijt}.}

\section{Estimators of weak lensing}

Now that we have built up a solid theoretical understanding of cosmic shear, 
we can look at how it is actually measured in practice.

\subsection{Lensing power spectrum}

Since galaxy images are observed on the sky, though unlike the CMB we \textit{can}
obtain radial information of galaxies through their redshift, we are motivated
to express the statistics of cosmic shear through the angular power spectrum
$\Cl$, analogous to the CMB. Here, we wish to derive the angular power spectrum
for cosmic shear. Equation~\ref{eqn:kappa_theta} gave us the angular convergence
field as a line-of-sight integral over the density perturbation $\delta$, 
weighted with some geometrical factors. To start our derivation, we consider
the Fourier transform of the density perturbation field into momentum-space of
\begin{align}
    \delta(\rchi, \rchi \vec{\theta}) = \int \!\! \frac{\d^3 \vec{k}}{(2 \pi)^3} 
    \, \delta_{\vec{k}} \,  e^{i \vec{k} \cdot \vec{x}},
\end{align}
where $\vec{x} \equiv (\rchi, \hat{\vartheta} \rchi)$. Thus, the convergence field
on the sky can be written as 
\begin{align}
    \kappa(\vec{\theta}) &= \frac{3 H_0^2 \Omegam}{2} \! \int \! \d \rchi \,
    \frac{g(\chi) \, \fk(\chi)}{a(\chi)} \,
    \int \! \frac{\d^3 \vec{k}}{(2 \pi)^3} \, 
    \delta_{\vec{k}} \,  e^{i \vec{k} \cdot \vec{x}}, \\
    &= \Upsilon \! \int \! \d \rchi \,\,
    \Gamma(\rchi )\,
    \int \! \frac{\d^3 \vec{k}}{(2 \pi)^3} \, 
    \delta_{\vec{k}} \,  e^{i \vec{k} \cdot \vec{x}},
\end{align}
where we have defined $\Upsilon$ as the cosmological prefactors, and 
$\Gamma(\rchi)$ as the collection of geometrical functions for convenience.
Now, using the Rayleigh plane-wave expansion, we can express the complex 
exponential into a decomposition of spherical harmonics and spherical Bessel
functions $j_\ell(x)$ \cite{Mehrem:2009ip}, as
\begin{align}
    e^{i \vec{k} \cdot \vec{x}} = 4\pi \sum_{\ell, \, m} 
    i^\ell \, j_\ell(k \rchi) \, \Ylm(\hat{k}) \, \Ylm^*(\vec{\theta}).
    \label{eqn:Rayleigh_plane_wave}
\end{align}
Thus, the angular convergence field becomes
\begin{align}
    \kappa(\vec{\theta}) = 4 \pi \Upsilon \sum_{\ell, \, m} i^{\ell}  \int \! \d \rchi  \,
    \, \Gamma(\rchi) \,
    \int \! \frac{\d^3 \vec{k}}{(2 \pi)^3} \, 
    j_{\ell}(k \rchi) \, Y_{\ell m}(\hat{k}) Y_{\ell m}^{*}(\vec{\theta}) \,
    \delta_{\vec{k}}.
\end{align}
We can now compare this formula with that of a spherical harmonic expansion on
a function of the sphere (Equation~\ref{eqn:temp_expansion}), to find the
expansion coefficients $\alm$ of the convergence field as
\begin{align}
    \alm = 4 \pi \Upsilon i^{\ell}  \int \! \d \rchi  \,
    \, \Gamma(\rchi) \,
    \int \! \frac{\d^3 \vec{k}}{(2 \pi)^3} \, 
    j_{\ell}(k \rchi) \, Y_{\ell m}^*(\hat{k}) \,
    \delta_{\vec{k}}.
\end{align}
Recalling that the definition of the power spectrum arose from the expectation
value of two $\alm$ values (Equation~\ref{eqn:Cl_definition}), we can take this
expectation value to find
\begin{multline}
    \langle \alm, \almprime \rangle = (4 \pi \Upsilon)^2 \, i^{\ell} (-i)^{\ell'} \times \\
    \int \! \d \rchi \d \rchi' \, 
    \! \frac{\d^3 \vec{k}}{(2 \pi)^3} \frac{\d^3 \vec{k}'}{(2 \pi)^3} \,
    \, \Gamma(\rchi) \Gamma(\rchi') \,
    j_{\ell}(k \rchi) \, j_{\ell'}(k' \rchi')
    Y_{\ell m}(\hat{k}) \, Y_{\ell' m'}^{*}(\hat{k}')
    \langle \delta_{\vec{k}} \, \delta_{\vec{k}'}^{*} \rangle.
\end{multline}
Now recalling our definition of the matter power spectrum as (Equation~\ref{eqn:matter_pow_def})
\begin{align}
	\langle \delta_{\vec{k}} \, \delta_{\vec{k}'}^{*} \rangle. =
	(2 \pi)^3 \delta^{(3)}(\vec{k}_1 - \vec{k}_2) \, P(k),
\end{align}
we find that we require $\vec{k}' = \vec{k}$, and so we can immediately perform
the $\vec{k}'$ integral, to give
\begin{multline}
    \langle \alm, \almprime \rangle = (4 \pi \Upsilon)^2 \, i^{\ell} (-i)^{\ell'} \times \\
    \int \! \d \rchi \d \rchi' \, 
    \! \frac{\d^3 \vec{k}}{(2 \pi)^3} \,
    \, \Gamma(\rchi) \Gamma(\rchi') \,
    j_{\ell}(k \rchi) \, j_{\ell'}(k \rchi')
    Y_{\ell m}(\hat{k}) \, Y_{\ell' m'}^{*}(\hat{k})
    P(k)
\end{multline}
We can now express our $\d^{3} \vec{k}$ volume element as $k^2 \d k \, \d \Omega$,
where $\d \Omega$ is the volume element on the sphere, to give
\begin{multline}
    \langle \alm, \almprime \rangle = \frac{(4 \pi \Upsilon)^2}{(2 \pi)^3} \,
     i^{\ell} (-i)^{\ell'} \, 
    \int \! \d \rchi \d \rchi' \, 
    \! \d k \,
    \, \Gamma(\rchi) \Gamma(\rchi') \,
    j_{\ell}(k \rchi) \, j_{\ell'}(k \rchi') \, k^2 P(k) \times \\
    \int \!\! \d \Omega \, Y_{\ell m}(\hat{k}) \, Y_{\ell' m'}^{*}(\hat{k}).
\end{multline}
Thus, using the orthonormality of the spherical harmonics over the complete
sphere, we can evaluate the angular integral to find
\begin{align}
    \langle \alm, \almprime \rangle = \frac{2 \Upsilon^2}{\pi} \, i^{\ell} (-i)^{\ell'}
    \delta_{\ell \ell'} \, \delta_{m m'} 
    \int \! \d \rchi \d \rchi' \, 
    \! \d k \,
    \, \Gamma(\rchi) \Gamma(\rchi') \,
    j_{\ell}(k \rchi) \, j_{\ell'}(k \rchi') \, k^2 P(k).
\end{align}
Since the Kronecker-$\delta$ forces $\ell' = \ell$, we find that 
$i^{\ell} (-i)^{\ell} = 1$ for all $\ell$. Thus, we find 
\begin{align}
    \langle \alm, \almprime \rangle = \frac{2 \Upsilon^2}{\pi} \,
    \delta_{\ell \ell'} \, \delta_{m m'} 
    \int \! \d \rchi \d \rchi' \, 
    \! \d k \,
    \, \Gamma(\rchi) \Gamma(\rchi') \,\,
    j_{\ell}(k \rchi) \, j_{\ell}(k \rchi') \,\, k^2 P(k).
    \label{eqn:exact_cl_vals}
\end{align}

\subsubsection{The Limber approximation}

Thus far, we have been performing an exact calculation with no approximations
yet. So, if we wish to keep our derivation of the angular power spectrum
coefficients to be exact, then this would be where our derivation ends with
our $\Cl$ values given by Equation~\ref{eqn:exact_cl_vals}. However, this is not
such a nice equation for the $\Cl$ values since it is a triple integral
over $\rchi$, $\rchi'$, and $k$, and involves two spherical Bessel functions
which are highly oscillatory, especially for large $\ell$ values, and thus
challenging to numerically evaluate to make theoretical predictions for the
$\Cl$ values. 

Hence, we can employ Limber's approximation of the spherical Bessel functions,
which allows us to re-write the spherical Bessel functions as Dirac-$\delta$
functions in the following way~\cite{Limber:1954zz}
\begin{align}
    j_{\ell}(x) \Rightarrow \sqrt{\frac{\pi}{2 \alpha}} \, \delta(\alpha - x),
\end{align}
where $\alpha$ is determined by the $\ell$ index. Originally, $\alpha$ was taken to be
$\alpha = \ell$, however $\alpha = \ell + \frac{1}{2}$ has been found to increase
the accuracy of the approximation for smaller values of $\ell$, where it is
accurate down to $\ell \simeq 10$~\cite{Gebhardt:2017chz}. An even further
improvement can be made by using $\alpha = \sqrt{\ell(\ell + 1)}$~\cite{LoVerde:2008re}.
For simplicity, we will be using $\alpha = \ell$ in the continuing derivation
since it is trivial to replace it with either of the two approximations in the
final result. 

Returning to Equation~\ref{eqn:exact_cl_vals}, we can employ Limber's 
approximation on the $j_{\ell}(k \rchi')$ term to eliminate the $k$ integral as
\begin{align}
    j_{\ell}(k \rchi') \Rightarrow \sqrt{\frac{\pi}{2 \ell}} \, \delta (\ell - k \rchi'),
\end{align}
and thus enforces that $k = \ell / \rchi'$. This gives our integral\footnote{It is important to remember that when evaluating 
an integral of the $\delta$-function when it is scaled by a constant $\beta$,
the integral gives
\begin{align}
    \int \d x \,\, \delta(\beta x) = \frac{1}{\beta}
\end{align}
since the integral is required to be normalised, 
and thus we pick up a $1/\rchi'$ term in our integrand.
} as
\begin{align}
    \langle \alm, \almprime \rangle \simeq \frac{2 \Upsilon^2}{\pi} \, \sqrt{\frac{\pi}{2 \ell}} \,
    \delta_{\ell \ell'} \, \delta_{m m'} 
    \int \! \d \rchi \d \rchi' \,
    \, \Gamma(\rchi) \Gamma(\rchi') \, j_{\ell} \! 
    \left(\frac{\ell}{\rchi'} \rchi \right) \, \left[\frac{\ell}{\rchi'}\right]^2
    \, \frac{1}{\rchi'} \, P(k).
\end{align}
Thus, we can use our remaining spherical Bessel function to eliminate the 
$\rchi'$ integral through
\begin{align}
    j_{\ell} \! \left(\frac{\ell}{\rchi'} \rchi \right) \Rightarrow 
    \sqrt{\frac{\pi}{2 \ell}} \, \, \delta \! \left(\ell - \frac{\ell \rchi}{\rchi'} \right),
\end{align}
which enforces that $\rchi' = \rchi$. This gives our remaining 1D integral as
\begin{align}
    \langle \alm, \almprime \rangle \simeq \frac{2 \Upsilon^2}{\pi} \, \sqrt{\frac{\pi}{2 \ell}}^2 \,
    \delta_{\ell \ell'} \, \delta_{m m'} 
    \int \! \d \rchi  \, \, \Gamma^2(\rchi) \, \left[\frac{\ell}{\rchi}\right]^2
    \frac{1}{\chi} \, \frac{\chi}{\ell} \, P(k).
\end{align}
Simplifying terms, and substituting in $\Upsilon$ and $\Gamma$, we find the
angular power spectrum of convergence to be~\cite{Kilbinger:2014cea}
\begin{align}
    _{\kappa \kappa} C_{\ell}^{ab} = \frac{9 \Omegam^2 H_0^4}{4}
    \int_{0}^{\chimax} \!\! \d \rchi \, \, 
    \frac{g_a(\rchi) \, g_b(\rchi)}{a^2(\rchi)} \, 
    P_{\delta} \left( k = \frac{\ell}{\fk(\rchi)}, \, z=z(\rchi)\right),
    \label{eqn:converg_power_spec}
\end{align}
where now $g_a$ is the lensing kernel associated with the distribution of source
galaxies, $n_a(z)$, in redshift bin $a$.

\subsubsection{\boldmath $E$- and $B$-modes}

So far, we have been concerned with the statistics of the convergence field
$\kappa$. However, this is not directly observable, what is observable is the
shapes of galaxies and thus we make measurements of the spin-2 field shear
$\kappa$. Because shear is a spin-2 field, when we decompose it into power
spectra, we actually get three unique spectra: $EE$, $EB$, and $BB$ modes
which are analogous to the same three mode decomposition from the polarisation
of the CMB (also a spin-2 field). This arises from Helmholtz's decomposition
theorem, which states that any vector field can be written as the
gradient of a scalar potential ($\vec{\nabla} \Phi$) plus the curl of
a vector potential ($\vec{\nabla} \times \vec{A}$)~\cite{Helmholtz+1858+25+55}.
The $E$-modes subsequently correspond to the gradient-like term, and the
$B$-modes correspond to the curl-like term of our shear fields. Hence,
the $EE$ and $BB$ modes are the auto-correlation of the $E$- and $B$-modes, 
respectively, with the $EB$ mode being the cross-correlation.

Since Equation~\ref{eqn:gamma_kappa_psi}
relates convergence and shear to derivates of the same lensing potential $\psi$,
we find that the shear-$EE$ power spectrum can be written as 
\begin{align}
    C_{\ell}^{EE} = \frac{(\ell - 1) (\ell + 2)}{\ell (\ell + 1)} \, C_{\ell}^{\kappa \kappa},
\end{align}
and due to parity considerations under Einsteinian gravity $C_{\ell}^{EB} = 0$
for all $\ell$. For spherically-symmetric mass distributions, the induced 
deflections are aligned tangentially and so we expect zero $B$-mode power
($C_{\ell}^{BB} = 0$) from cosmic shear~\cite{Prat:2025ucy}.

\subsubsection{Cosmic variance \& noise}
\label{sec:cosmic_variance}

In Section~\ref{sec:measuring_shear}, we argued that while every galaxy has
an intrinsic shape, if we average over enough galaxies then the observed
ellipticities are an unbiased estimator of the intrinsic shear field. These
intrinsic ellipticities contribute to a white-noise effect on the power spectrum
of the form
\begin{align}
    N_{\ell}^{ab} = \frac{\sigma_{\epsilon}^2}{\bar{n}} \, \delta_{ab},
    \label{eqn:shape_noise}
\end{align}
where $\sigma_{\epsilon}$ is the standard deviation of the intrinsic galaxy 
ellipticity dispersion per component, and $\bar{n}$ is the expected number of
observed galaxies per steradian per redshift bin. $\bar{n}$ depends on the 
specifics of the galaxy survey, for the main \textit{Euclid} survey it is
expected to observe 30 galaxies per square arcminute, divided into ten 
redshift\footnote{Generally, the more tomographic redshift bins that the
observed source galaxy distribution is divided into provides additional
information from the increased number of cross-correlations of the angular
power spectrum. However, by splitting the source galaxy distribution into
more bins decreases the number density of each bin $\bar{n}$, increasing the
noise level of each bin. This reduces the amount of information that can
be extracted at small-scales. Thus, detailed investigations were required to
establish the optimal number of bins to split the source galaxies into.
It has been suggested that the first \textit{Euclid} data release (DR1) be
split into six tomographic bins~\cite{Euclid:2025nwk}, and \textit{Euclid}'s 
final data release (DR3) using either ten or thirteen bins~\cite{Euclid:2024yrr}
for optimal constraints on the dark energy equation of state.}
bins gives $\bar{n} = 3 \, \textrm{gals / arcmin}$, with 
$\sigma_\epsilon = 0.21$~\cite{Euclid:2011zbd}. 

While shape noise primarily reduces the information content at small-scales, the
effect of cosmic variance impacts at large-scales. This is because each 
$\ell$-mode is only estimated from $2 \ell + 1$ $m$-modes, and so each
$\Cl$ mode has a variance of $2 / (2 \ell + 1)$. This significantly affects the
precision of which large-scale $\Cl$ modes can be measured, and is a result of
us only being able to make measurements from one point in the Universe.

\begin{figure}[t]
    \centering
    \includegraphics[width=0.975\linewidth,trim={0.0cm 0.0cm 0.0cm 0.0cm},clip]{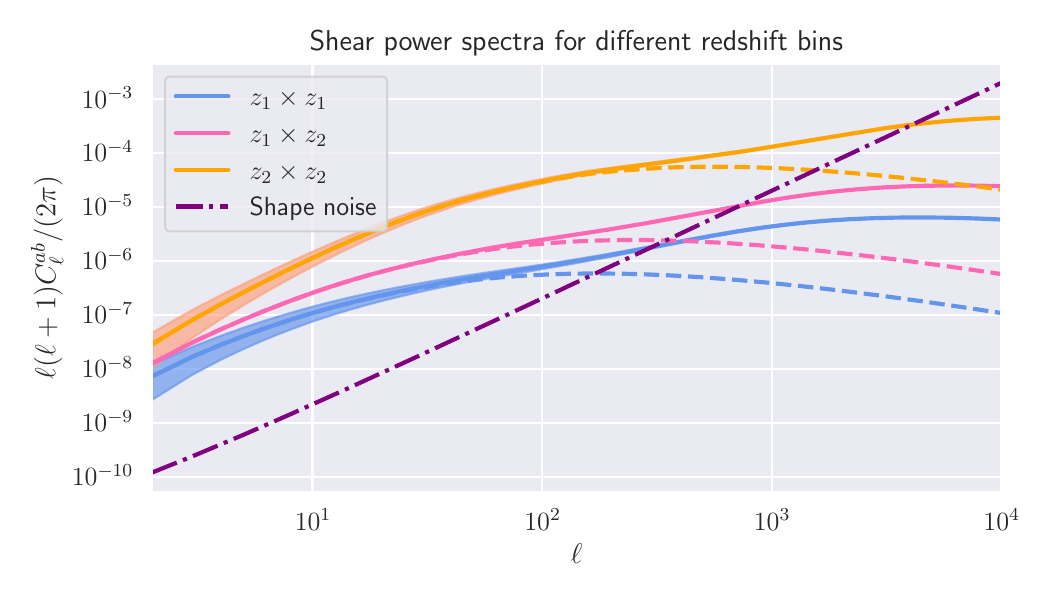}
    \caption{Plot of the auto and cross shear power spectrum for two Gaussian 
    redshift bins with centres $z_1 = 0.5$ and $z_2 = 2.0$ and widths of $\sigma_z = 0.25$.
    Dashed lines indicate where the linear matter power spectrum has been
    used instead of the full non-linear prescription. The shaded regions on
    the auto-spectra indicate the $1\sigma$-width of the cosmic variance. 
    The dashed purple line corresponds to the shape noise amplitude
    expected for \textit{Euclid}~\cite{Euclid:2011zbd}.}
    \label{fig:ShearPowSpec}
\end{figure}

\begin{figure}[p]
    \centering
    \includegraphics[width=0.975\linewidth,trim={0.0cm 0.0cm 0.0cm 0.0cm},clip]{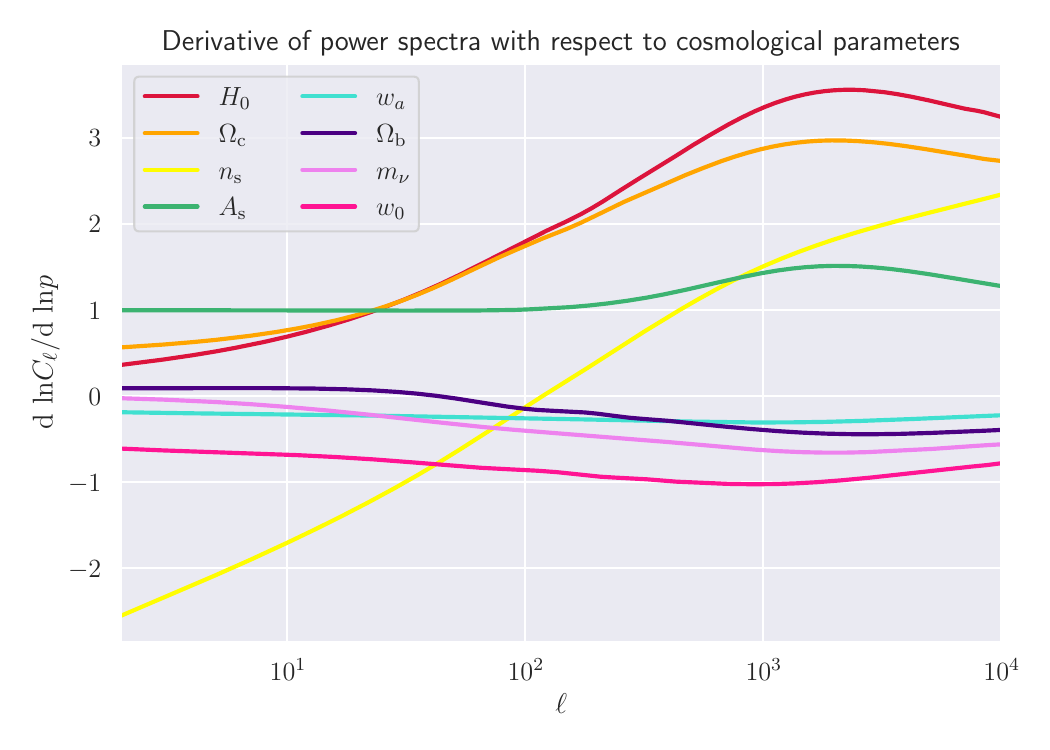}
    \caption{Derivatives of the shear power spectrum with respect to input 
    cosmological parameters, $\d \ln \Cl / \d \ln p$, except for the $m_\nu$, 
    $w_0$, and $w_a$ curves which are $\d \ln \Cl / \d p$. The curves are coloured
    and ordered by their amplitude on the right-hand side. 
    Derivates were
    evaluated for the Gaussian redshift bin located at $\bar{z} = 2.0$, and 
    a fiducial cosmology of $A_\textsc{s} = 2.1\times 10^{-9}$, $n_\textsc{s} = 0.96$, 
    $H_0 = 70$ km/s Mpc$^{-1}$, 
    $\Omega_\textsc{c} = 0.25$, $\Omega_\textsc{b} = 0.046$, $w_0 = -1$, $w_a=0$,
    and $m_\nu = 0.06\, \eV$.
    We can understand the nature of these curves by recalling that the angular
    power spectra is the Limber integral over the matter power spectrum weighted
    by the lensing kernels (Equation~\ref{eqn:converg_power_spec}). $\As$ and $\ns$
    are the amplitude and tilt of the primordial perturbation power spectrum
    (Equation~\ref{eqn:power_law_inflation}), which then directly feed into the
    amplitude and tilt of the matter power spectrum (Equation~\ref{eqn:P_k_break}).
    Thus, we see that $\As$ provides an overall scaling (save for the small
    non-linear increase) of the $\Cl$ values, and that increasing $\ns$ produces
    a bluer spectrum -- providing the tilt seen above. Increasing the cold dark
    matter density $\Omegac$ increases the gravitational clustering of matter
    in our Universe, which increases the lensing signal, particularly on
    smaller-scales. Increasing the baryon density $\Omegab$ serves to increase
    the effect of baryonic feedback processes, and thus suppresses power
    on smaller-scales. In our lensing integral, have a prefactor of $H_0^4$
    which is then modulated by a $\mathcal{O}(h^{-3.5})$ term in the matter
    power spectrum on the largest physical scales~\cite{Hall:2021qjk}, which
    combine for form the $H_0$ dependence shown. Neutrinos serve to wash out
    clustering on small-scales thanks to their free-streaming effects, with
    an increased neutrino mass increasing this suppression of small-scale 
    clustering~\cite{Copeland:2019bho}. More positive values of the dark energy 
    equation of state parameters $w_0$ and $w_a$ require that dark energy had a 
    larger physical density in the past (Equation~\ref{eqn:de_time_evo} and 
    Figure~\ref{fig:dark_energy_density}) and so suppresses the gravitational
    attraction of matter, decrease the clustering of matter and thus
    decreasing the lensing angular power spectrum amplitudes. A more thorough
    discussion about the impacts of time evolving dark energy ($w_0$ and $w_a$)
    and massive neutrinos on the weak lensing power spectrum is presented in
    Appendix~\ref{chp:appendix_B}.
    } 
    \label{fig:ShearPowSpecDeriv}
\end{figure}

The shear power spectrum is plotted in Figure~\ref{fig:ShearPowSpec} for two
Gaussian redshift bins. We plot the logarithmic derivates of the angular power
spectrum with respect to the underlying cosmological parameters in 
Figure~\ref{fig:ShearPowSpecDeriv}. We see that cosmic shear is generally sensitive
to the density perturbation amplitude $\As$ and matter density $\Omegac$ (which
combine to form the lensing amplitude parameter $S_8$~\cite{Hall:2021qjk}), with
weaker dependency on the dark energy equation of state ($w_0$ and $w_a$) and
the neutrino masses ($m_{\nu}$), and so require large quantities of accurate
and precise statistical measurements of cosmic shear to place tight constraints
on these less well constrained cosmological parameters.

\subsection{Real-space correlation functions \& COSEBIs}

An alternative statistical quantity to the power spectrum is the angular 
two-point correlation function $\hat{\xi}_{\pm}$. This can be estimated from 
the same set of galaxy images and can be written as (for 
redshift bins $a$ and $b$) \cite{Heymans:2013fya}
\begin{align}
    \hat{\xi}^{ab}_{\pm}(\theta) = \frac{\sum w_i w_j \left[
        \varepsilon^a_\textsc{t}(\vec{x}_i) \varepsilon^b_\textsc{t}(\vec{x}_j) \pm
        \varepsilon^a_\times\!(\vec{x}_i) \varepsilon^b_\times\!(\vec{x}_j) 
    \right] }{\sum w_i w_j},
\end{align}
where $w_i$ are the individual weights of each galaxy, the sum is taken over 
pairs of galaxies with angular separation $\lvert \vec{x}_i - \vec{x}_j \rvert = \theta$,
and $\varepsilon_\textsc{t}$ and $\varepsilon_\times$ are the tangential and 
cross-ellipticity parameters, respectively. Galaxy weights are used as, for 
example, one might want to down-weight galaxies that have poor or noisy shape 
measurements, but still include them in the analysis.

The two-point correlation functions can be predicted from the cosmic shear
angular power spectrum through~\cite{Lemos:2017arq}
\begin{subequations}
    \begin{align}
        \xi_{+}^{ab}(\theta) &= \sum_{\ell} \! \frac{2 \ell + 1}{4 \pi}
        \, \left[ _{EE}C_{\ell}^{ab} + _{BB\!\!}C_{\ell}^{ab} \right] 
        \, d^{\ell}_{2\, 2}(\theta) , \\ 
        \xi_{-}^{ab}(\theta) &= \sum_{\ell} \! \frac{2 \ell + 1}{4 \pi}
        \, \left[ _{EE}C_{\ell}^{ab} - _{BB\!\!}C_{\ell}^{ab} \right] 
        \, d^{\ell}_{2\, -2}(\theta) ,
    \end{align}
    \label{eqn:2PCF_def}
\end{subequations}
where $d_{mn}^{\ell}$ are the reduced Wigner $D$-matrices, and we have 
generalised to a case where $C_{\ell}^{BB} \neq 0$.

A further statistic can be derived from the correlations in the form of 
the COSEBIs (Complete Orthogonal Sets of E/B-mode Integrals) statistics which
provide a set of $E_n$ and $B_n$ values for the $E$- and $B$-mode shear 
statistics, respectively, where larger values of $n$ are typically more sensitive
to small-scale information. In terms of the two-point correlation functions, 
the COSEBI statistics are given as~\cite{Schneider:2010pm,Asgari:2012ir}
\begin{subequations}
    \begin{align}
        E_n^{ab} &= \frac{1}{2} \int_{\theta_{\textrm{min}}}^{\theta_{\textrm{max}}}
        \! \d \theta \, \, \theta \left[T_{+n}(\theta) \xi_{+}^{ab}(\theta)
        + T_{-n}(\theta) \xi_{-}^{ab}(\theta)\right], \\ 
        B_n^{ab} &= \frac{1}{2} \int_{\theta_{\textrm{min}}}^{\theta_{\textrm{max}}}
        \! \d \theta \, \, \theta \left[T_{+n}(\theta) \xi_{+}^{ab}(\theta)
        - T_{-n}(\theta) \xi_{-}^{ab}(\theta)\right],
    \end{align}
    \label{eqn:COSEBI_def}
\end{subequations}
where $T_{\pm n}(\theta)$ are the COSEBI filter functions.

Since the angular power spectrum values are related to the underlying
geometry and clustering of our Universe through a single integral 
(Equation~\ref{eqn:converg_power_spec}), rather than two or three
sums/integrals for the case of the correlation functions and COSEBIs,
the angular power spectrum can be much simpler to model -- and thus will
be our cosmic shear summary statistic of choice throughout this thesis.

\section{Parameter estimation}

Thus far, we have laid down a large amount of theoretical background for
modern cosmology. We started with some fundamental theories about the nature
of the Universe, and developed them into predictions for physical phenomena
that we can go out and measure. While all of our theories are precisely calculated and 
well-motivated, we still have numerous unknown parameters in our models that
we would like to constrain, along with the fundamental assumptions that allowed
us to develop these theoretical models. This is where we use our very expensive
telescopes and instruments to go looking into our Universe to measure physical
properties via physical processes (Section~\ref{sec:probe_of_lss}). Since
every measurement comes with it an associated uncertainty, the propagation of
uncertainties from our measurements to the parameters that we aim to measure is
just as, if not more, important than the measurement of the parameter's value
itself.

This is because a measurement without an uncertainty is no measurement at all;
we have no idea if the measurement is important or not and to which degree
the new measurement agrees or disagrees with existing values. Thus, the
determination of cosmological parameter values and their uncertainties underpin
all of modern experimental observational cosmology. 

Bayesian statistics form the heart of our parameter estimation statistical
analyses, since Bayesian statistics naturally incorporate the degree of
\textit{certainty} that we have in our measurements. The foundation of Bayesian
statistics is Bayes' theorem, which is often stated in terms of events
$A$ and $B$ as
\begin{align}
    P(A|B) = \frac{P(B|A) \, P(A)}{P(B)},
    \label{eqn:cosmo_Bayes}
\end{align}
where $P(A)$ and $P(B)$ are the probabilities of events $A$ and $B$, respectively,
and $P(A|B)$ and $P(B|A)$ are the condition probabilities of event $A$ occurring
given that event $B$ happened, and vice-versa. While this form is useful for
dealing with discrete events $A$ and $B$, and their probabilities associated
with these events, it becomes less useful when applying it to a cosmological
analysis. 

Therefore, we shall introduce the idea that we have some data vector $\vec{x}$,
which has been derived from some observations (e.g. the CMB power spectrum),
and a model with some parameters $\vec{\theta}$ that we want to estimate
from our observations. In this case, Bayes' theorem becomes
\begin{align}
    P(\vec{\theta}|\vec{x}) = \frac{P(\vec{x} | \vec{\theta}) \, P(\vec{\theta})}{P(\vec{x})}.
    \label{eqn:Bayes_thm}
\end{align}
Since each term is important in its own right, we have
\begin{galitemize}
    \item
    $P(\vec{\theta} | \vec{x})$ is the \textit{posterior} for our parameters,
        and is what we are trying to estimate.
    
    \item
    $P(\vec{x} | \vec{\theta})$ is the \textit{likelihood} and is often
        denoted by its own symbol $\mathcal{L}(\vec{x} | \vec{\theta})$ because
        of its importance. 
    
    \item
    $P(\vec{\theta})$ is the \textit{prior} and quantifies our level 
        of certainty in the values of our model, and may encode information
        from previous results.
    
    \item
    $P(\vec{x})$ is the \textit{evidence} which acts to normalise the 
        posterior, and is mostly ignored in cosmological analyses since the
        relative probabilities of parameters do not change.
\end{galitemize}

\subsection{Likelihoods}

The key to evaluating the posterior distribution of Equation~\ref{eqn:cosmo_Bayes}
is the likelihood. It turns out that the log-likelihood $\mathscr{L}$ is
often the more useful quantity, defined as
\begin{align}
    \mathscr{L} \equiv - \ln \mathcal{L}.
\end{align}
Unfortunately, for such an important quantity, the likelihood is rarely
known exactly for an arbitrary data-set and has to be approximated.

Due to the central limit theorem,
the principal likelihood is that of a Gaussian distribution. It quantified by
the mean of the observables (which are a function of the model parameters)
$\vec{\mu}(\vec{\theta})$, its covariance matrix $\mathbf{C}$, and number of
observables $N$. Here, probability density function is given as
\begin{align}
    p(\vec{x} | \vec{\mu}, \mathbf{C}) = \frac{1}{(2\pi)^{N/2} \, \sqrt{| \mathbf{C}|}}
    \, \exp \! \left(-\frac{1}{2} \, [\vec{x} - \vec{\mu} ]^{\textsc{t}} \, \mathbf{C}^{-1} \, [\vec{x} - \vec{\mu}] \right),
\end{align}
and thus we can see that the $\rchi^2$ for the Gaussian distribution is given by
\begin{align}
    \rchi^2 = \left[ \vec{x} - \vec{\mu}(\vec{\theta}) \right]^{\textrm{T}}
    \mathbf{C}^{-1} \left[\vec{x} - \vec{\mu}(\vec{\theta})\right].
\end{align}
If each observable is independent of each other, then we have a diagonal
covariance matrix where the entries are the variances of each observable
$\sigma^2_i$, and thus the $\rchi^2$ becomes
\begin{align}
    \rchi^2 = \sum_{i=1}^{N} \frac{1}{\sigma^2_i} \left[x_i - \mu_i(\vec{\theta})\right]^2.
\end{align}

\subsection[Likelihood of $\Cl$ values]{Likelihood of \boldmath$\Cl$ values}

Since having the correct form of the likelihood is essential for any 
cosmological analysis, we can look at the distribution of the $\Cl$ values,
which are the observables that we are interested in. Since the underlying $\alm$
values follow a Gaussian distribution of mean zero and variance $\Cl$, we 
have that their distribution is given by~\cite{Percival:2006ss}
\begin{align}
    p(\alm | \Cl) = \frac{1}{\sqrt{2 \pi \, \Cl}} \, \exp \! \left[ - \frac{\lvert \alm \rvert^2}{2 \, \Cl} \right],
\end{align}
which gives our previous result of $\langle \lvert \alm \rvert^2 \rangle = \Cl$
(Equation~\ref{eqn:Cl_definition}). We have that the \textit{estimator} of the
power spectrum, $\tilde{C}_{\ell}$, is the average over the $\alm$ values as
\begin{align}
    \tilde{C}_{\ell} = \frac{1}{2 \ell + 1} \sum_{m} \lvert \alm \rvert^2.
\end{align}
This gives that the $\Cl$ values are distributed as the sum of $\nu \equiv 2 \ell + 1$
squared Gaussian distributions, which is a $\Gamma$ distribution~\cite{Percival:2006ss}
of the form
\begin{align}
    p(\tilde{C}_{\ell} | \Cl) \propto \Cl \left[\frac{\tilde{C}_{\ell}}{\Cl}\right]^{\frac{\nu}{2} - 1}
    \exp \left[-\frac{\nu}{2} \frac{\tilde{C}_{\ell}}{\Cl}\right].
\end{align}
This distribution has a mean of $\Cl$ and a variance of $2 \Cl^2 / \nu$ (this
factor of $2 / \nu$ is the cosmic variance that we saw in Section~\ref{sec:cosmic_variance}).
The maximum of this distribution occurs not at the mean value, but instead
at $\tilde{C}_{\ell} = \Cl (\nu - 2)/\nu$. In the high-$\ell$ limit ($\nu \rightarrow \infty$),
we find this distribution tends to the Gaussian distribution, due to the central
limit theorem\footnote{Unless, of course, if \textit{the limit does not exist}.}.

\begin{figure}[tp]
	\centering
    \vspace*{-1cm}
	\includegraphics[width=0.975\linewidth,trim={2.0cm 1.6cm 1.7cm 0.5cm},clip]{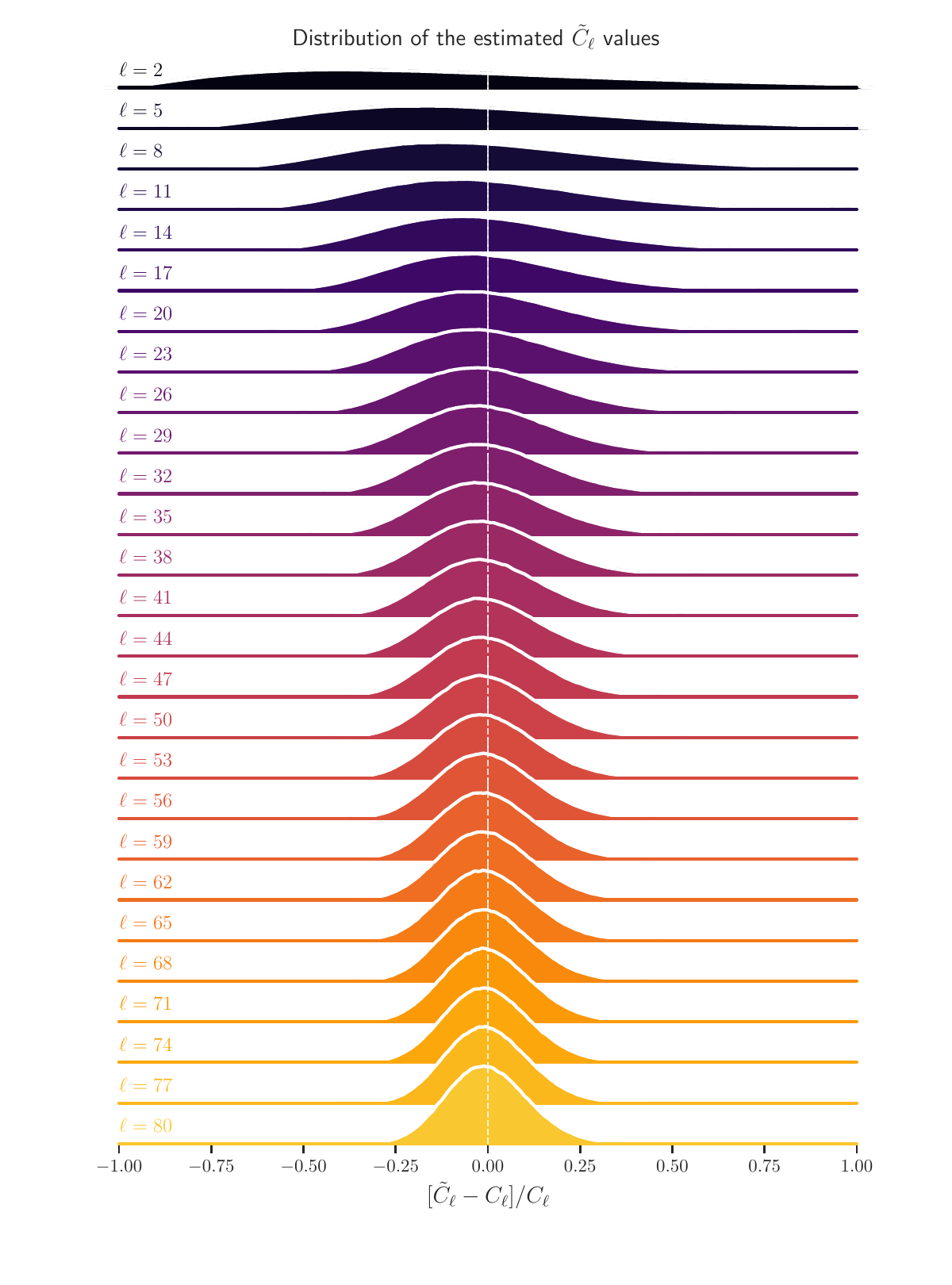}
	\caption{Relative distribution of the estimated power spectrum coefficients,
        $\tilde{C}_{\ell}$, with respect to their average value, $\Cl$, for
        an ensemble of one million realisations.
        We can see that for low $\ell$ values,
        the distribution is significantly non-Gaussian, where the $\Cl$
        distributions exhibit significant skewness and excess kurtosis, since 
        they follow the $\Gamma$ distribution. Once the $\ell$ value becomes
        significantly large ($\ell \gtrsim 50$), their distribution can
        be well-approximated by a Gaussian.}
	\label{fig:ClRidgePlot}
\end{figure}

Figure~\ref{fig:ClRidgePlot} shows this $\Gamma$ distribution of the estimated
$\Cl$ values through their relative differences to the mean $\Cl$. This clearly
shows that for low $\ell$ values, the distribution is significantly 
non-Gaussian and as $\ell$ increases the distribution tends to a Gaussian
due to the central limit theorem.

For the case where we have more than one redshift bin, we find that we have a
mix of auto- and cross-spectra for each $\ell$ value. For the case of two
redshift bins, we find the signal matrix for each $\ell$-mode,
$\mathbf{S}_{\ell}$, to be
\begin{align}
    \mathbf{S}_\ell = \begin{pmatrix}
        \tilde{C}_\ell^{11} & \tilde{C}_\ell^{12} \\
        \tilde{C}_\ell^{12} & \tilde{C}_\ell^{22}
    \end{pmatrix},
\end{align}
where we have used that $\Cl^{12} = \Cl^{21}$ from symmetry arguments. The
corresponding theory matrix, at each $\ell$-mode, is $\mathbf{W}_{\ell}$ given
as
\begin{align}
    \mathbf{W}_\ell = \frac{1}{2\ell + 1} \begin{pmatrix}
        C_\ell^{11} && C_\ell^{12} \\ 
        C_\ell^{12} && C_\ell^{2}
    \end{pmatrix}.
\end{align}
These combine to form the Wishart likelihood of~\cite{Percival:2006ss}
\begin{align}
    \mathcal{L}(\mathbf{S}_\ell \vert \mathbf{W}_\ell) = 
    \frac{\lvert \mathbf{S}_\ell \rvert ^{(\nu - p - 1) / 2} \,
    \exp\! \left[-\mathrm{Tr}\!\left(\mathbf{W}_\ell^{-1} \mathbf{S}_\ell / 2\right)\right] }{
    2^{p\nu/2} \, \lvert \mathbf{W}_\ell \rvert ^{\nu / 2} \,  \Gamma_p(\nu/2)
    },
    \label{eqn:WishartLikelihood}
\end{align}
where $\mathbf{S}_\ell$ and $\mathbf{W}_\ell$ are positive-definite symmetric 
$p \times p$ matrices, $\nu > p$, and $\Gamma_p$ is the multivariate 
Gamma function. In the high-$\ell$ limit, this Wishart distribution tends 
towards the multivariate Gaussian distribution \cite{Upham:2020klf}.

\subsection{Priors}

The fundamental philosophy of Bayesian parameter estimation is that we evaluate
the posterior distribution using Bayes' theorem (Equation~\ref{eqn:Bayes_thm}) 
to `update' our knowledge about some parameters or models as more information
is gained about the system. With zero evaluations of the posterior, the initial
knowledge about the system comes from the prior distribution $P(\vec{\theta})$.
This distribution is usually informed by existing experiments and measurements
of our parameters, or by physical constraints (e.g. the sum of neutrino masses
being non-negative).

\subsubsection{Flat priors}

A weakly-informative prior is the flat, or uniform, prior. For a single
parameter $\theta$, its flat prior is defined as
\begin{align}
    P(\theta) = \begin{cases}
        \frac{1}{\theta_{\textsc{max}} - \theta_{\textsc{min}}}           & \textrm{for} \,\, \theta_{\textsc{min}} \leq \theta \leq \theta_{\textsc{max}},           \\
        0 & \textrm{otherwise.}
    \end{cases}
\end{align}
Thus, with a flat prior, the posterior is simply directly proportional to
the likelihood.

While flat priors may appear harmless since they just provide lower and upper
bounds on the parameter's value, in high dimensional spaces the very many 1D flat priors serve to
cause samplers to concentrate on a thin shell of a $D$-dimensional 
hypercube~\cite{Trotta:2017wnx}. To obtain a truly uninformative prior, one
must use the Jeffreys prior, which partially mitigates the impact of volume
effects and is insensitive to model 
parametrisations~\cite{Jeffreys:1946RSPSA,Hadzhiyska:2023wae}.

\subsection{Markov chain Monte Carlo}

Thus far, we have seen that the posterior distribution is incredibly useful for
parameter estimation problems in cosmology. Hence, we wish to have numerical
tools which allow us to form estimates of the posterior, and thus form 
constraints on our parameters. The main numerical tool of choice are
Markov chain Monte Carlo\footnote{Name comes from the Monte Carlo Casino in
Monaco and the random nature of the gambling within it.} (MCMC) which aims to
generate a series of points whose distribution is the same as the posterior, and
thus an unbiased estimator for the posterior.

\subsubsection{Metropolis-Hastings algorithm}

The simplest MCMC algorithm is the Metropolis-Hastings algorithm which 
gives the acceptance probability of a new point in the chain with parameters
$\vec{\theta}'$ over the existing set of parameters $\vec{\theta}$ as~\cite{Heavens:2009nx}
\begin{align}
    P(\textrm{acceptance}) = \min \! \left[ \frac{P(\vec{\theta'} | \vec{x})}{P(\vec{\theta}|\vec{x}) }, \, 1 \right].
\end{align} 
This probability is then compared to a uniform random number $\alpha$ in the interval
$[0, 1)$ and accept the new parameters if $P > \alpha$, otherwise the new
parameters are rejected and the existing ones are written into the chain with
the algorithm starting again with a new set of proposed points.

\subsubsection{Nested samplers}

An alternative method to chain-based MCMC algorithms are the nested samplers, 
which directly evaluates the evidence and is advantageous in high-dimensional
parameter spaces and complex likelihood surfaces~\cite{Skilling:BA127,Ashton:2022grj}.
Popular implementations of nested sampling algorithms are \MultiNest~\cite{Feroz:2007kg,Feroz:2008xx,Feroz:2013hea} and
\PolyChord~\cite{Handley:2015fda,Handley:2015vkr}.

\subsection{Fisher matrix formalism}

The Fisher matrix formalism provides an alternative method to obtaining 
cosmological parameter covariances from MCMC simulations, thus saving
significant computational costs since MCMC analyses require significantly
many evaluations of the likelihood ($10^{5}$ to $10^{6}$ evaluations). The
Fisher matrix formalism has primarily been used to derive parameter covariance
estimates, not for parameter values which requires us to fall-back to 
MCMC analyses.  The
Fisher matrix is formally defined as the second derivative of the log-likelihood
($\mathscr{L})$ with respect to our parameters of a model ($\vec{\vartheta}$).
The Fisher matrix for parameters $\alpha$ and $\beta$ is
\begin{align}
    F_{\alpha \beta} \equiv - \left\langle \frac{\partial^2 \mathscr{L}}{\partial \vartheta_{\alpha} \, \partial \vartheta_{\beta}}  \right\rangle,
\end{align}
and thus for an $N$-parameter model, its corresponding Fisher matrix is an
$N \times N$ matrix~\cite{Tegmark:1997rp}. We can directly obtain parameter 
standard deviations from the Fisher matrix by taking the diagonal of its
inverse matrix which gives the Cram\'er-Rao inequality of
\begin{align}
    \sigma_\alpha \geq \sqrt{\left[\mathbf{F}^{-1}\right]_{\alpha \alpha}}
\end{align}  
which gives the minimum $1$-$\sigma$ marginalised error bars for any unbiased
estimator of the parameters from the data~\cite{Tegmark:1996bz}. 

If we assume that our data follows a multivariate Gaussian distribution, then
it is simply characterised by the mean vector $\vec{\mu}$ and covariance
matrix $\mathbf{C}$. In this case, we can analytically calculate the Fisher
matrix, which is a function of these two objects only, as~\cite{Tegmark:1997rp}
\begin{align}
    \mathbf{F}_{\alpha \beta} = \frac{1}{2} \Tr \! \left[ 
        \mathbf{C}^{-1} \frac{\partial \mathbf{C}}{\partial \vartheta_{\alpha}} \mathbf{C}^{-1} \frac{\partial \mathbf{C}}{\partial \vartheta_{\beta}}
         \right]
         + \frac{\partial \vec{\mu}}{\partial \vartheta_{\alpha}} \mathbf{C}^{-1} \frac{\partial \vec{\mu}}{\partial \vartheta_{\beta}}.
\end{align}

Since the mean of the cosmic shear field is zero, we are
free to transform our Fisher matrix into one for the power spectrum coefficients,
In which case the cosmological dependence of the covariance matrix drops out
and we are left with~\cite{Euclid:2019clj}
\begin{align}
    \mathbf{F}_{\alpha \beta} = 
    \frac{\partial \vec{C}_{\ell}}{\partial \vartheta_{\alpha}}
    \mathbf{C}^{-1}
    \frac{\partial \vec{C}_{\ell}}{\partial \vartheta_{\beta}},
\end{align}
which is analogous to the Gaussian likelihood, but now we're inverse covariance
weighting the derivatives of the power spectra, not the differences between
data and theory.

Since we can easily compute the partial derivatives of the power spectrum
coefficients with respect to parameters and form a fiducial covariance matrix, 
the estimation of the Fisher matrix becomes straightforward and thus
parameter constraints are readily computed.

\subsubsection{The figure of merit}

When comparing different models with different Fisher matrices, it is often
useful to see how constraining each model is for a sub-set of two parameters. This can be
done through the figure of merit (FoM), which is inversely proportional to the
area of the $2$-$\sigma$ marginalised contour for two parameters $\vartheta_{\alpha}$
and $\vartheta_{\beta}$. The figure of merit for any two parameters is given as
in terms of the sub-Fisher matrix $\tilde{\mathbf{F}}$ as~\cite{Euclid:2019clj}
\begin{align}
    \textrm{FoM}_{\alpha \beta} = \sqrt{\det \left[\tilde{\mathbf{F}}_{\alpha \beta}\right]}.
\end{align}

\subsubsection{The figure of bias}

Using the Fisher matrix, we can also compute the level of expected bias in the
cosmological parameters when using a different model for the recovery of
parameters to one that was used to generate the original data. This is useful 
since we can see how changing models affects the bias in the parameters,
hopefully optimising the models to have the smallest possible biases. The
bias on parameter $\vartheta_{\alpha}$, $\delta \vartheta_{\alpha}$, is
given by~\cite{Amara:2007as}
\begin{align}
    \delta \vartheta_{\alpha} = \left[\mathbf{F}^{-1}\right]_{\alpha \beta}
    \left( \vec{C}_{\ell}^{\textrm{true}} - \vec{C}_{\ell}^{\textrm{model}}  \right)
    \mathbf{C}^{-1} \, \frac{\partial \vec{C}_{\ell}}{\partial \vartheta_{\beta}}.
\end{align}
The figure of bias (FoB), which is an overall summary statistic quantifying how
biased the parameters are overall, can then be computed from the combination of
individual parameter biases though~\cite{Gordon:2024jaj}
\begin{align}
    \textrm{FoB} = \sqrt{\overrightarrow{\delta \vartheta}_{\alpha} \, \tilde{\mathbf{F}}_{\alpha \beta} \, \overrightarrow{ \delta \vartheta}_{\beta}}.
\end{align}
Thus, with the Fisher matrix, we can provide both constraints on both the
size and offset of parameter contours for different models, thus enabling us
to select the most optimal methods for a given data-set.

\section[$\Cl$ Covariance matrix]{\boldmath $\Cl$ Covariance matrix}

Core to the accurate and precise measurement of cosmological parameters
though MCMC methods is the data covariance matrix, $\mathbf{C}$. The incorrect
specification of this matrix could lead to parameter constraints that are
either off-set from their true values, and/or contours that are too large or
small, neither of which are desirable outcomes from the data. Thus, correct
evaluation of the covariance matrix is critical for any cosmological survey.

There exists two broad categories from which a covariance matrix can be
estimated from: analytically, and numerically. Analytic methods aim to provide
the covariance matrix in closed-form expressions, usually as a function of the
underlying power spectra. Numerical methods aim to produce very many realisations
of the underlying fields, and then take a numerical covariance of the resulting
observables. Both methods provide a useful cross-check of the other, and
we expand on these methods significantly in Chapter~\ref{chp:QML_estimator}, 
however a small discussion of analytic methods is presented here.

\subsection{Gaussian covariance}
\label{sec:Gaussian_cov}

While the proper distribution of the $\Cl$ values follow a Wishart
distribution (Equation~\ref{eqn:WishartLikelihood}), this can be accurately
approximated to the Gaussian distribution for cosmic shear~\cite{Upham:2020klf}.
In the full-sky limit, the covariance of two power spectra $C^{ab}_{\ell}$
and $C^{cd}_{\ell'}$, including the effects of shape noise, is given by the
four-point function of
\begin{align}
    \textrm{Cov}\left[C^{ab}_{\ell}, \, C^{cd}_{\ell'}\right] = \frac{\delta_{\ell \ell'}}{(2 \ell + 1)}
    \left(C^{ac}_{\ell} \, C^{bd}_{\ell} + C^{ad}_{\ell} \, C^{bc}_{\ell} \right).
\end{align}
This can be modified to account for the effects of the cut-sky though the
scaling relation approximation of 
\begin{align}
    \textrm{Cov}\left[C^{ab}_{\ell}, \, C^{cd}_{\ell'}\right] = \frac{\delta_{\ell \ell'}}{(2 \ell + 1) \, \fsky}
    \left(C^{ac}_{\ell} \, C^{bd}_{\ell} + C^{ad}_{\ell} \, C^{bc}_{\ell} \right),
    \label{eqn:gaussian_cl_cov}
\end{align}
where $\fsky$ is the fraction of sky observed. Thus, the error on a single
power spectrum is
\begin{align}
    \Delta C_{\ell} = \sqrt{\frac{2}{(2 \ell + 1) \, \fsky}} \, \left[C^{EE}_{\ell} + N_{\ell}\right], 
\end{align}
where $N_{\ell}$ is the noise spectrum, given by Equation~\ref{eqn:shape_noise}.
Thus, there are three regimes to the errors on the cosmic shear power spectrum:
\begin{enumerate}
    \item \textbf{Cosmic variance limited.} Here, the $1/\sqrt{\ell}$ term 
        dominates the errors, since for low multipoles there are only a
        limited number of $\alm$ modes that can contribute to each $\Cl$. Thus,
        cosmic shear is not particularly sensitive to the largest angular scales.
    
    \item \textbf{Signal dominated.} In this region, we have enough $\alm$
        modes per $\Cl$ such that each measurement is statistically significant,
        and that our cosmic shear signal dominates over the shape noise - and
        thus we are able to exact maximum information from our shape measurements.

    \item \textbf{Shape noise dominated.} Here, the shape noise spectrum is 
        much larger than the cosmic shear signal ($N_{\ell} \gg C^{EE}_{\ell}$),
        which holds on the smallest scales. Thus, to adequately probe the 
        smallest scales, large numbers of galaxies are needed to be observed
        over the sky, which requires deep and precise measurements. 
\end{enumerate}

\subsection{Super-sample and non-Gaussian covariances}

In addition to the simple Gaussian term, there exists the residual covariance
from the non-Gaussianity of the underlying field, which generates higher-order
correlations and forms the non-Gaussian covariance term. There also
exists the super-sample covariance, which arises from the correlation
between modes within the survey and those on larger scales than the survey
footprint. Both the non-Gaussian and super-sample covariance terms contribute
significantly to the overall covariance matrix, particularly for the
off-diagonal terms~\cite{Euclid:2021ilj,Euclid:2023ove}. If these additional
terms are not present, then this can lead to an underestimate in the
uncertainties on cosmological parameters of up to $70 \, \%$. Thus, it is
essential that any cosmic shear survey
include these terms in their covariance matrices to ensure accurate parameter
constraints. 

While these additional covariance terms are important to keep in a full
cosmic shear analysis, we are going to neglect them --- opting for a 
Gaussian-only covariance matrix for the rest of this thesis --- for the sake
of simplicity.

\section{Information gain at high redshift}

When discussing the Gaussian covariance matrix, Section~\ref{sec:Gaussian_cov},
we noted that there were three regions in which the errors on the power spectrum
fall into - with the signal dominated region driving the constraining power
of cosmic shear. Thus, we would like to have the signal-dominated region to be
as large as possible, which requires that $\Cl \gg N_{\ell}$ for a wide range
of $\ell$ values. Since the angular power spectra values increase as we
observe galaxies at higher redshift, due to their light rays passing through
more of the Universe's large-scale structure (see Figure~\ref{fig:ShearPowSpec}),
galaxies at higher redshifts have more statistical power than those at low
redshifts

We can quantify this statistical power though the signal-to-noise (S/N).
For a signal vector $\vec{d}$ with covariance matrix $\mathbf{C}$, its
signal-to-noise is defined as
\begin{align}
    \textrm{S/N} \equiv \sqrt{\vec{d} \,\, \mathbf{C}^{-1} \, \vec{d}^{\textsc{t}}}.
\end{align}
Applying this to a single cosmic shear power spectrum bin, and using
the fact that the inverse of a diagonal matrix is a matrix of the reciprocal
values along the diagonal, we find its signal-to-noise to be given by
\begin{align}
    \left[\textrm{S/N}\right]^2 &= \sum_{\ell = 2}^{\lmax} \frac{(2 \ell + 1) \, \fsky}{2} \, \frac{\left[C^{EE}_{\ell} \right]^2}{\left[C^{EE}_{\ell} + N_{\ell}\right]^2}, \label{eqn:sig_to_noise_single_bin} \\
                   &= \sum_{\ell = 2}^{\lmax} \frac{(2 \ell + 1) \, \fsky}{2} \, \left[1 + \frac{N_{\ell}}{C^{EE}_{\ell}}\right]^{-2}.
\end{align}
We can now study how this signal-to-noise changes as a function of the maximum
multipole ($\lmax$) considered. 

\begin{figure}[t]
    \centering
    \includegraphics[width=0.975\linewidth,trim={0.0cm 0.0cm 0.0cm 0.0cm},clip]{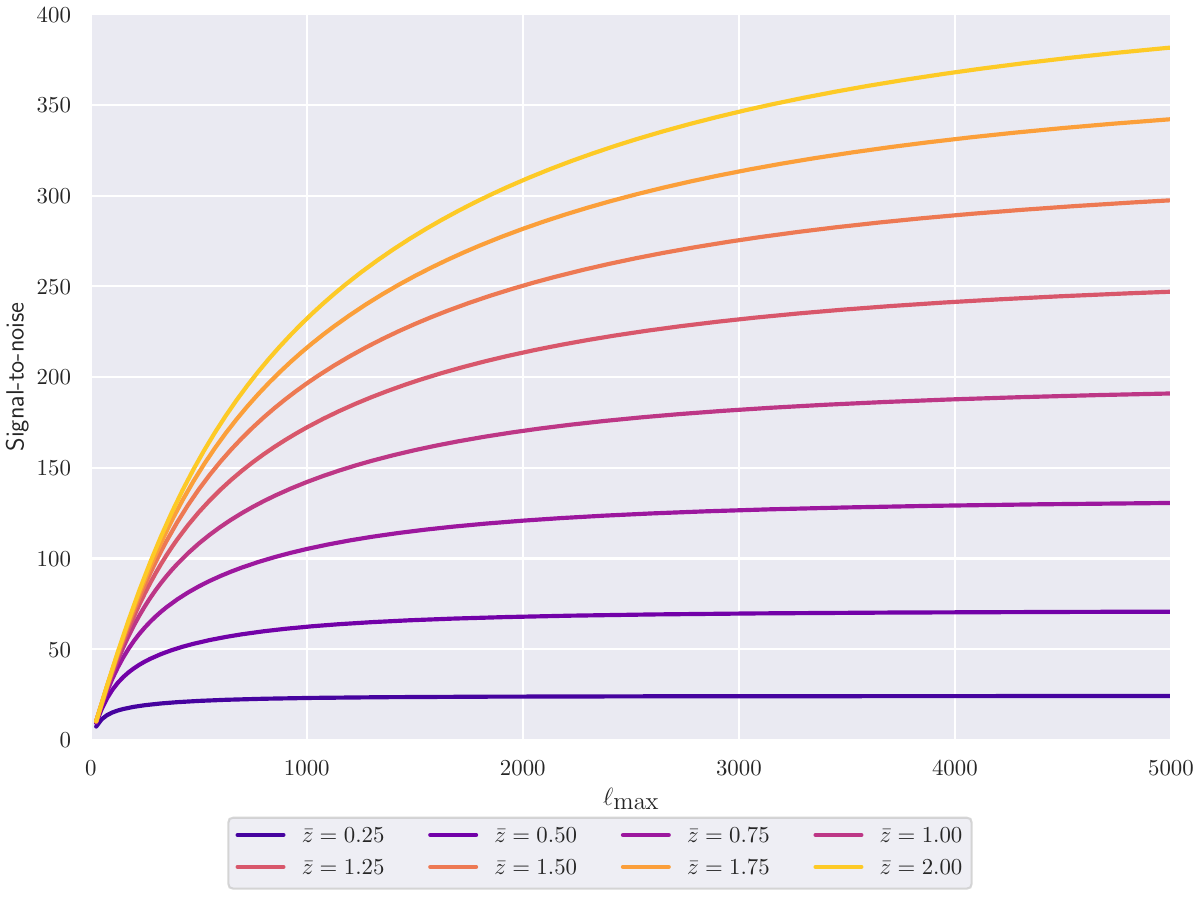}
    \caption{The signal-to-noise of the auto-correlation for eight different
        Gaussian redshift bins with centres $\bar{z}$, as a function of the
        maximum multipole considered. We see that the signal-to-noise
        amplitude and profile is highly bin dependent. The low-redshift bins
        have both an intrinsically smaller signal-to-noise and saturate
        around $\lmax \sim 1000$. In contrast, the higher redshift bins,
        which have a larger cosmic shear signal, have a larger signal-to-noise,
        and do not saturate until around $\lmax \sim 4000$.
        Thus, the optimal cosmic shear survey would probe the smallest scales at
        the highest redshifts possible.
        }
    \label{fig:signal_to_noise_ell_max}
\end{figure}

In Figure~\ref{fig:signal_to_noise_ell_max} we plot the signal-to-noise for 
the auto-correlations of eight Gaussian redshift bins centred at $\bar{z} =
\{0.25, 0.50, 0.75, 1.00, 1.25, 1.50, 1.75, 2.00\}$, all with widths $\sigma_z = 0.15$.
We see that due to the increasing amplitude of the shear power spectrum with
redshift, the further away bins have a significantly greater signal-to-noise
than the close redshift bins. Additionally, the lower redshift bins saturate
their information content around $\lmax \sim 2000$, providing very little benefit
to measuring the power spectrum beyond that, whereas the higher redshift bins 
still gain information even down to an $\lmax$ of $5000$.

\subsection{Including cross-correlations}

So far, we have been looking at the signal-to-noise of a single bin's auto
spectra as a function of redshift. In a real cosmic shear survey, not only
do we have access to the auto-spectra, but we can cross-correlate these
bins to obtain cross-spectra. This has the powerful advantages of both
dramatically increasing the number of spectra in our data-vector (the number of
spectra scales as $n_{\textrm{spectra}} = n_{\textrm{bins}} [n_{\textrm{bins}} + 1]/ 2$),
but also having the property that the variances of the cross-spectra feature a
reduced impact from contributions from shape noise (though while their
power spectrum coefficients $_{EE}\Cl^{ab}$ \, [$a \neq b$] do not feature any
shape noise, their covariances are still partially dependent on shape noise).
This can be seen by specialising our Gaussian covariance 
(Equation~\ref{eqn:gaussian_cl_cov}) to that of a cross-correlation 
(where $ab = cd$, $a \neq b$),
\begin{align}
    \textrm{Cov}\left[C^{ab}_{\ell}, \, C^{ab}_{\ell'}\right] = 
    \frac{\delta_{\ell \ell'}}{(2 \ell + 1) \, \fsky}
    \left(
        \Big[ \,\!_{EE}C^{aa}_{\ell} \, + N_{\ell} \Big] \,
        \left[ _{EE}C^{bb}_{\ell} \, + N_{\ell} \right]
        + \left[_{EE}C^{ab}_{\ell} \right]^2
    \right),
\end{align}
Thus, when considering the covariances of the cross-spectra, we find that the
impact of shape noise is reduced when compared to the auto-spectra. This reduces
the effect of the shape-noise dominated region at high $\ell$, and so
we can extract more information from this region due to the larger signal-to-noise
than from auto-spectra, which motivates us to measure the shear power spectrum 
to large $\ell$ values (small angular scales).

Now, while it is true that the signal-to-noise will be higher for the
cross-spectra, to properly model them we need to build a full covariance
matrix featuring both auto- and cross-spectra. Let us consider the case for two
photometric redshift bins, which gives us three unique spectra: $1 \times 1$, 
$1 \times 2$, and $2 \times 2$. Thus, the full covariance matrix will be made
out of nine sub-blocks, as follows
\begin{align}
    \mathbf{C} = \begin{pmatrix}
        \mathbf{C}_{11 \times 11} & \mathbf{C}_{11 \times 12} & \mathbf{C}_{11 \times 22} \\
        \mathbf{C}_{12 \times 11} & \mathbf{C}_{12 \times 12} & \mathbf{C}_{12 \times 22} \\
        \mathbf{C}_{22 \times 11} & \mathbf{C}_{22 \times 12} & \mathbf{C}_{22 \times 22}
    \end{pmatrix},
\end{align}
where each sub-block $\mathbf{C}_{ab \times cd}$ can be computed using 
Equation~\ref{eqn:gaussian_cl_cov}. Hence, when we invert this covariance
matrix, the sub-blocks get convoluted with each other, and so even though the
cross-spectra have a reduced impact from shape noise, they still get
additional contributions to their (inverse) covariances arising from the mixing
from the auto-spectra.

\begin{figure}[p]
    \centering
    \includegraphics[width=\linewidth,trim={0.0cm 0.0cm 0.0cm 0.0cm},clip]{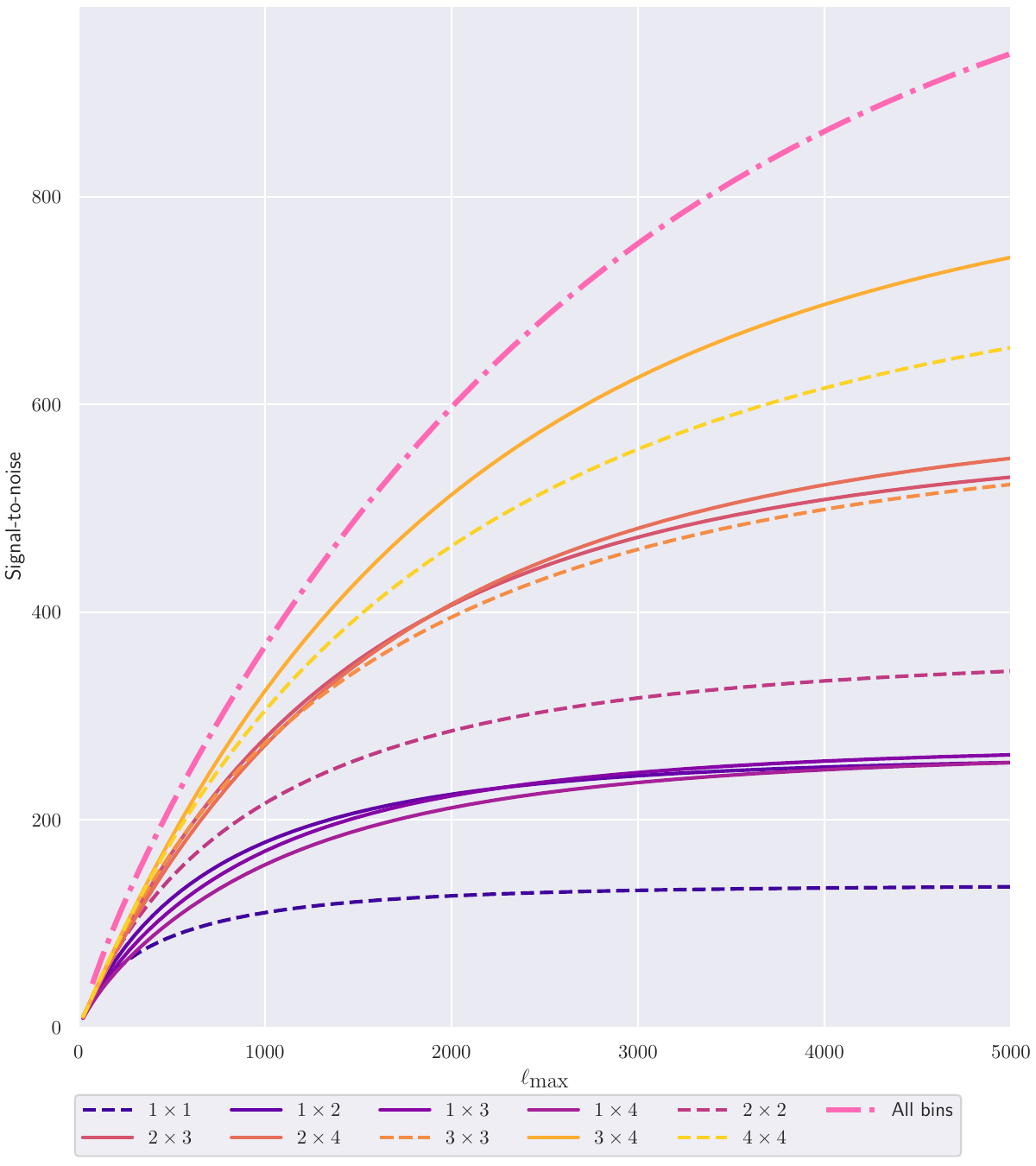}
    \caption{The signal-to-noise of the auto- and cross-correlations for
        four different Gaussian redshift bins, as a function of the
        maximum multipole considered, along with the total signal-to-noise
        of all bin combinations in dot-dashed pink. We see that the signal-to-noise
        amplitude and profile is still highly bin dependent, with low-redshift
        bins having a smaller signal-to-noise than higher redshift bins. 
        We also see that the cross-correlations (plotted in the solid curves)
        have systematically larger signal-to-noises than the auto-correlations
        (plotted in the dashed lines), which reflects the reduced impact of 
        shape noise in their covariances since the cross-correlations lack
        shape noise contributions in their data-vectors. 
        The total signal-to-noise of the entire data-vector (dot-dashed pink 
        curve) still increases even as we approach an $\lmax$ of 5000.
        Hence, while many of the individual spectra appear to saturate on these
        small-scales, we still gain considerable overall constraining power
        on the smallest scales. It is also important to note that the total
        signal-to-noise is not just the sum of the individual signal-to-noises.
        This is because there is significant correlation between the higher and
        lower redshift curves, which reduces the total overall information
        content available~\cite{Hu:1999ek}.
        }
    \label{fig:signal_to_noise_ell_max_combined}
\end{figure}

We can now repeat the same exercise as shown in Figure~\ref{fig:signal_to_noise_ell_max},
but now including the effects of cross-spectra in the signal-to-noises.
Figure~\ref{fig:signal_to_noise_ell_max_combined} plots the signal-to-noise
of the auto- and cross-correlations, along with the total signal-to-noise of
the entire data-vector. Here, we see the same trend with higher redshift bins
having larger signal-to-noise, but we also see how the suppression of shape noise
in the cross-correlations leads to higher signal-to-noise for these spectra.
Furthermore, we see the total signal-to-noise (plotted in the pink dot-dashed curve)
continues to grow even for our maximum multipole of 5000.
Thus, probing the smallest scales with cosmic shear
continues to provide valuable information content and constraining power here.

\clearpage
\section{Baryonic feedback in the angular power spectrum}
\label{sec:baryon_feedback_in_cls}

Since the cosmic shear angular power spectrum is a Limber integral of the
matter power spectrum, weighted by the lensing kernels, any baryonic feedback
effects that alter the matter power spectrum (see Section~\ref{sec:modelling_baryonic_feedback})
will also alter the observed angular cosmic shear power spectrum. As baryonic
feedback mostly impacts the small-scale clustering, that is large wavenumbers
($k \gtrsim 1 \, h$Mpc$^{-1}$), this will propagate into contaminating our
small-scale angular modes ($\ell \gtrsim 100$) -- modes provide significant
contributions to the signal-to-noise, particularly for the dark energy equation
of state parameters $w_0$ \& $w_a$ and the neutrino masses $m_\nu$.

Since we are aiming to measure the angular power spectrum down to small-scales
with high precision, in order to place tight constraints on the values of $w_0$,
$w_a$, and $m_\nu$, baryonic feedback could catastrophically bias our results for
these parameters (along with measurements of the lensing amplitude $\Seight$ 
and matter density $\Omegam$). Hence, we are driven to apply our models of
baryonic feedback in the matter power spectrum to our cosmic shear analyses.
Else, we would be forced to discard all data which is contaminated by baryonic
feedback, leaving only a tiny number of data-points from our original
observations, a highly unsatisfactory process.

\begin{figure}[p]
    \centering
    \includegraphics[width=\linewidth,trim={0.5cm 0.0cm 0.7cm 0.0cm},clip]{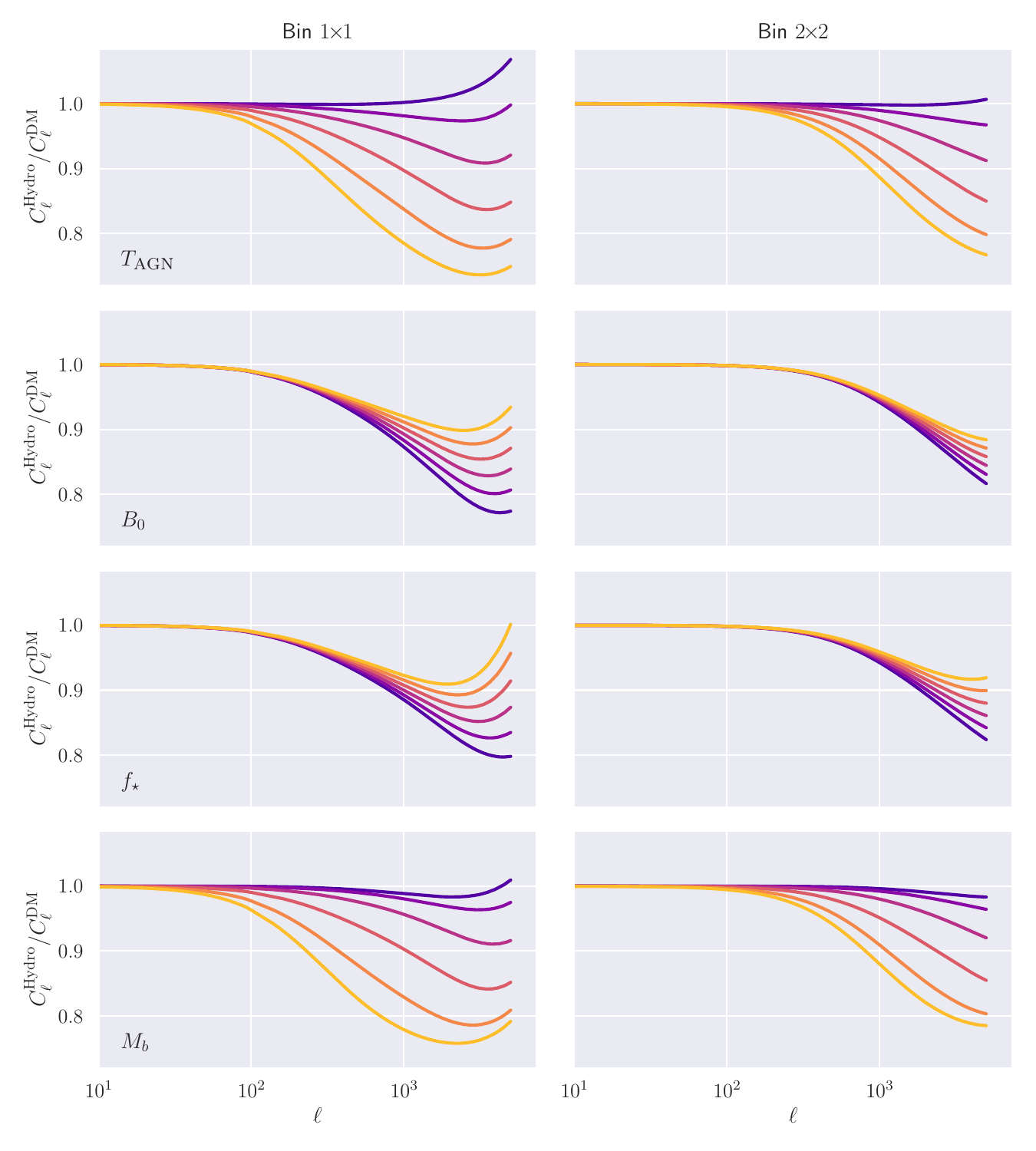}
    \caption{Ratio of the cosmic shear angular power spectrum for the case with
        baryonic feedback ($C_\ell^{\mathrm{Hydro}}$) to that without any
        feedback present ($C_\ell^{\mathrm{DM}}$), for two Gaussian redshift 
        bins, one centred at $\bar{z} = 0.5$ and one at $\bar{z} = 2.0$ (the
        columns), and for a range of the four astrophysical parameters
        present in the \HMCode-2020 model (by row, where the blue [yellow] curve
        represents a smaller [larger] value of that parameter). We plot the
        ratios our to an $\lmax$ of 5000, the optimistic target for Stage-IV
        cosmic shear surveys, such as \textit{Euclid}~\cite{Euclid:2024yrr}.
        }
    \label{fig:cl_baryon_ratio_3param}
\end{figure}

Using the \HMCode-2020 model of baryonic feedback, Figure~\ref{fig:cl_baryon_ratio_3param}
plots the ratio of the angular power spectrum values obtained using the model 
of baryonic feedback to that without baryonic feedback in, and for the four
different astrophysical parameters of the model (by row). This is broadly 
analogous to Figure~\ref{fig:HMCode_baryons}, but now we are comparing the
effects of baryonic feedback on the angular scales, which is the observable. 
Additionally, we plot the ratio for two Gaussian redshift bins, one centred at 
$\bar{z} = 0.5$ and one at $\bar{z} = 2.0$, which shows that the effects of
baryonic feedback changes with redshift.  

\HMCode-2020 is only one model of baryonic feedback and while its functional
form and values of its astrophysical parameters have been motivated from hydrodynamical
simulations, there is no guarantee that it can reproduce the exact form of
baryonic feedback that takes place within our Universe. Hence, while it is an
extremely useful model that allows our analyses to probe much smaller angular
scales than otherwise possible, we should remember that it's only a model and
thus susceptible to errors, as \textit{any} model is.

\begin{savequote}[65mm]
  I think we agree,

  the past is over
  \qauthor{---George W. Bush}
  \end{savequote}
\chapter{Concordance cosmology in 2024}
\label{chp:Cosmology_2024}
\begin{mytext}
    \textbf{Outline.} I briefly overview some of the challenges and
    outlook for cosmology in 2024\footnote{I am aware that this chapter
    will probably age horribly; \textit{Tempora mutantur, nos et mutamur in illis.}}.
\end{mytext}

\section{Current tensions in cosmology}

I'm sure that every cosmology PhD since the dawn of science have said something
along the lines of ``Now is a very exciting time to be a cosmologist'', but we
are living in an era of precision cosmology such that it's an entirely
different kind of science, altogether!\footnote{\textit{It's an entirely
        different kind of science }}
With the dramatic increase in the precision of data, and thus cosmological
constraints, any small disagreement between data sets becomes increasingly
large as the error bars on the measurements decrease. We measure the disagreement
between two values through their $\sigma$ significance. Assuming Gaussian
errors, then two measurements that have a $2$-$\sigma$ tension have a
$2.3\,\%$ one-sided probability of being a statistical fluctuation which,
while somewhat significant, is still unsatisfactory to call this a genuine
discrepancy.

\subsection[The $H_0$ tension]{The \boldmath $H_0$ tension}

The Hubble rate $H$ was defined through Equation~\ref{eqn:Friedmann_eqn} as
$H \equiv \dot{a} / a$, where $\dot{a} \equiv \d a / \d t$, and measures the
rate of expansion in the Universe. The Hubble constant $H_0$ is the present
expansion rate of our Universe today, given as $H_0 \equiv H(t = t_0)$, and is an
important observational quantity for any cosmological survey to measure,

Measurements of $H_0$ fall into two broad categories: direct and indirect
measurements, as described below.

\subsubsection{Direct measurements}

We call direct measurements anything that does not need to infer the $H_0$ value
through a cosmological model and instead measures the $H_0$ value directly.
Such example is the measurements of the recessional velocity of supernovae
through the distance-redshift relation, as seen in Section~\ref{sec:supernovae}.
Some of the tightest direct measurements on $H_0$ using Type Ia supernovae
come from the SH0ES collaboration using measurements from the \textit{Hubble}
space telescope, finding a value of
$H_0 = (73.04 \pm 1.04) \, \textrm{km/s Mpc}^{-1}$~\cite{Riess:2021jrx}.

\subsubsection{Indirect measurements}

Indirect measurements differ by requiring a model, usually the $\Lambda$CDM model,
and a number of assumptions to relate quantities that were measured in the
earlier universe to the Hubble constant today. The leading constraint from the
indirect measurements come from the CMB anisotropy measurements
(Section~\ref{sec:cmb}) from the \textit{Planck} 2018 data release finding
$H_0 = (67.36 \pm 0.54) \, \textrm{km/s Mpc}^{-1}$
using their `TT,TE,EE+lowE+lensing' dataset~\cite{PlanckCollaboration:2018eyx}.

\subsubsection{Tensions}

\begin{figure}[tp]
    \centering
    \includegraphics[width=0.9975\linewidth,trim={0.0cm 0.0cm 0.0cm 0.0cm},clip]{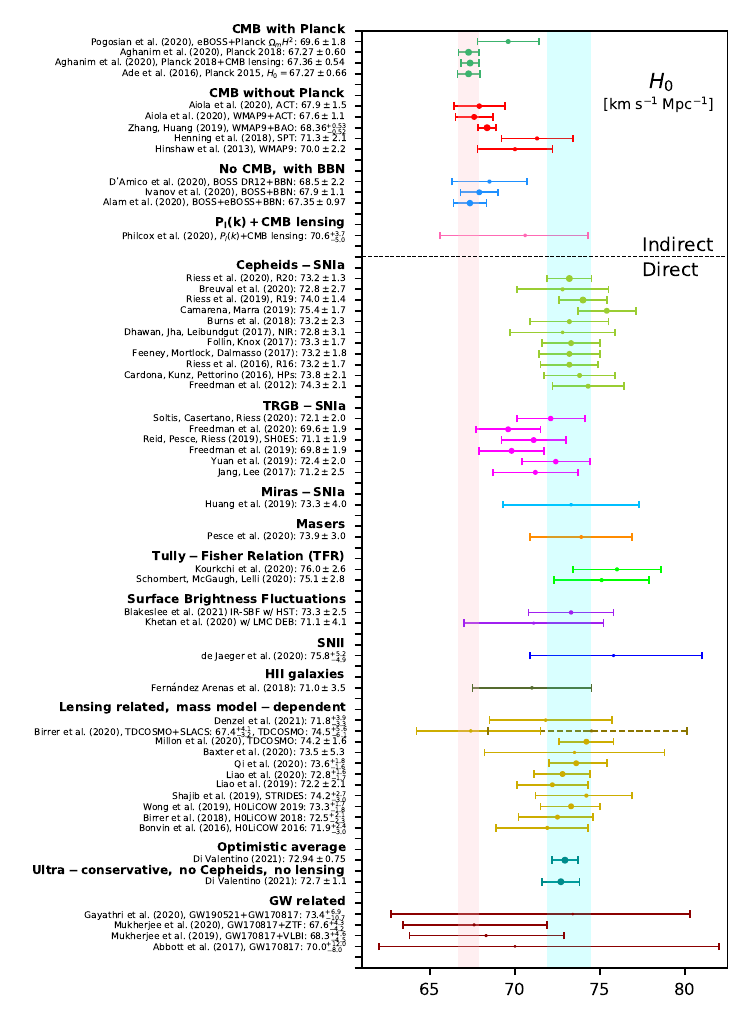}
    \caption{Whisker plot showing the different $1$-$\sigma$ ($68\,\%$)
    widths on $H_0$ results from different 
    probes and analyses choices. This shows a clear distinction between results
    obtained from direct and indirect means. 
    Figure taken from Ref.~\cite{Abdalla:2022yfr}.}
    \label{fig:H0_tension}
\end{figure}

If we compare the \textit{Planck} 2018 and latest SH0ES results, then we find 
a $5$-$\sigma$ discrepancy between their two $H_0$ values, which is highly
statistically significant. This tension extends beyond these two individual 
results, as Figure~\ref{fig:H0_tension} shows a summary of results of 
various direct and indirect measurements showing that \textit{in general} 
indirect measurements predict a low $H_0$ value whereas direct measurements
find a high $H_0$ value. The nature of this tension has been extensively
discussed in the literature but with no solid theoretical framework in place,
much of the discussion have been speculation with no concrete results thus far.

\clearpage
\subsection[The $S_8$ tension]{The \boldmath$S_8$ tension}

\begin{figure}[t]
    \centering
    \includegraphics[width=0.8\linewidth,trim={0.45cm 0.6cm 0.25cm 0.25cm},clip]{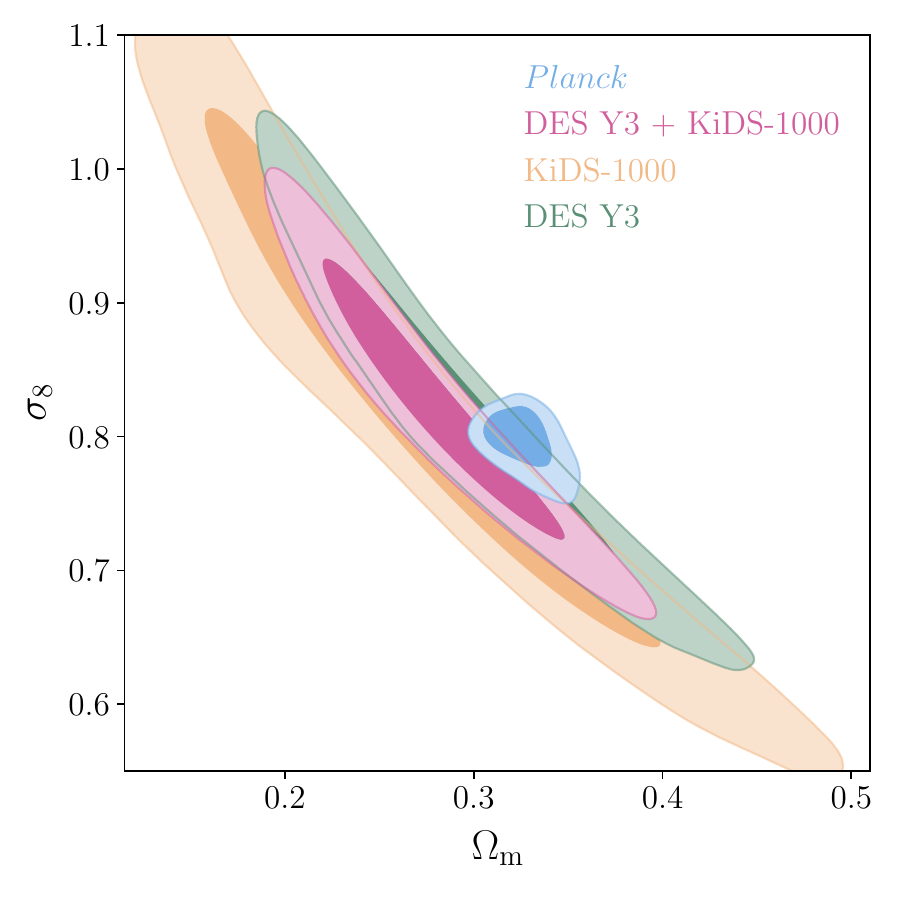}
    \caption{Joint $\sigma_8-\Omegam$ constraints from the \textit{Planck} 2018
    results~\cite{PlanckCollaboration:2018eyx}, the individual KiDS-1000 and
    DES Y3 results, and the joint KiDS-DES analysis~\cite{Kilo-DegreeSurvey:2023gfr}
    showing the distinctive `lensing bananas' from the weak lensing analyses, 
    how the CMB constraints are not subject to this degeneracy, and showing the
    approximately $1.7\sigma$ discrepancy between the cosmic shear and
    CMB constraints. Figure taken from Ref.~\cite{Kilo-DegreeSurvey:2023gfr}.}
    \label{fig:KiDS_DES_bannana}
\end{figure}

While not as statistically significant as the $H_0$ tension at present, a tension
in the value of $S_8$ has been lurking in the literature for a number of years
and is a significant target for current and future cosmological surveys.
The $S_8$ parameter roughly approximates the amplitude of
weak lensing distortions, and is defined as $S_8 \equiv \sigma_8 \, \sqrt{\Omegam / 0.3}$
where $\sigma_8$ is defined through Equation~\ref{eqn:sigma_8}. This was chosen
as $S_8$ is related to the width of the `lensing bananas' from weak lensing
studies in the $\sigma_8-\Omegam$ plane. Here, we find large degeneracies
in the joint $\sigma_8-\Omegam$ distribution, but the width of this degeneracy
is tightly constrained by weak lensing (see Figure~\ref{fig:KiDS_DES_bannana}).

Again, this tension arises between the late universe (weak lensing and galaxy clustering)
and early universe (CMB) measurements. The tightest CMB constraints again come from
the \textit{Planck} 2018 release using their `TT,TE,EE+lowE+lensing' dataset 
finding an $S_8$ value of $0.832 \pm 0.013$~\cite{PlanckCollaboration:2018eyx}.
The tightest weak lensing 
constraints come from the joint analyses of the KiDS-1000 and DES Y3 datasets 
finding an $S_8$ value of $0.790^{+0.018}_{-0.014}$~\cite{Kilo-DegreeSurvey:2023gfr}
This gives an agreement at approximately $1.7\sigma$ which, while not totally
consistent, is not large enough to say that there is a significant deviation
between the two results at the moment. 

\begin{figure}[tp]
    \centering
    \includegraphics[width=0.9975\linewidth,trim={0.0cm 0.0cm 0.0cm 0.0cm},clip]{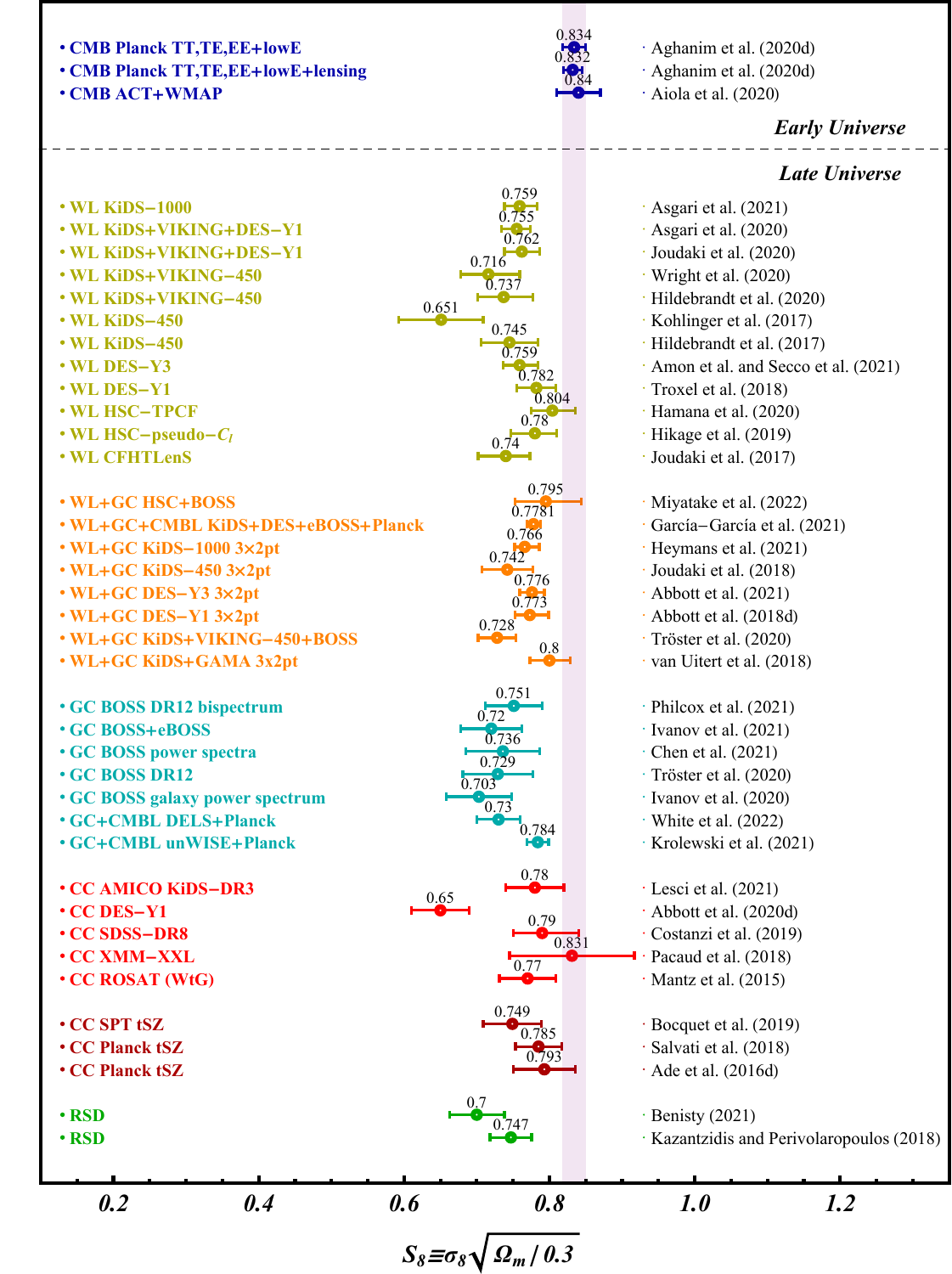}
    \caption{Whisker plot showing the different $1$-$\sigma$ ($68\,\%$)
    widths on $\Seight$ results from different 
    probes and analyses choices. We see that late-time measurements, such as
    weak lensing and galaxy clustering, are systematically lower than
    early-Universe measurements, principally \textit{Planck}'s observations of the CMB.  
    The further narrowing down of the uncertainties on $\Seight$ are one of the
    key goals of the Stage-IV cosmic shear surveys. 
    Figure taken from Ref.~\cite{Abdalla:2022yfr}.}
    \label{fig:S8_tension}
\end{figure}

Figure~\ref{fig:S8_tension} presents a summary of current $S_8$ results, again
broken down into early and late universe measurements, where the early
universe results preferring a slightly higher $S_8$ value and late universe
results preferring a lower $S_8$ value.

\clearpage
\section{The Next Generation\texttrademark}

Figure~\ref{fig:S8_tension} summarises our current best constraints on $S_8$
using existing cosmic shear surveys: the Kilo Degree Survey (KiDS), Dark Energy
Survey (DES), and Hyper Suprime-Cam (HSC) -- the so called `Stage III'\footnote{\textit{Revenge of the Sithtematics} \faicon{empire}} 
cosmic shear surveys~\cite{Albrecht:2006um}.
We are now entering the next level of cosmic shear surveys,
the `Stage IV'\footnote{\textit{A New Telesc(h)ope} \faicon{rebel}} surveys of: the \textit{Euclid}
space telescope, 
the Legacy Survey of Space and Time (LSST) at the Vera Rubin Observatory,
and the \textit{Roman} space telescope. 

\begin{figure}[t]
    \centering
    \includegraphics[width=0.9975\linewidth,trim={0.0cm 0.0cm 0.0cm 0.0cm},clip]{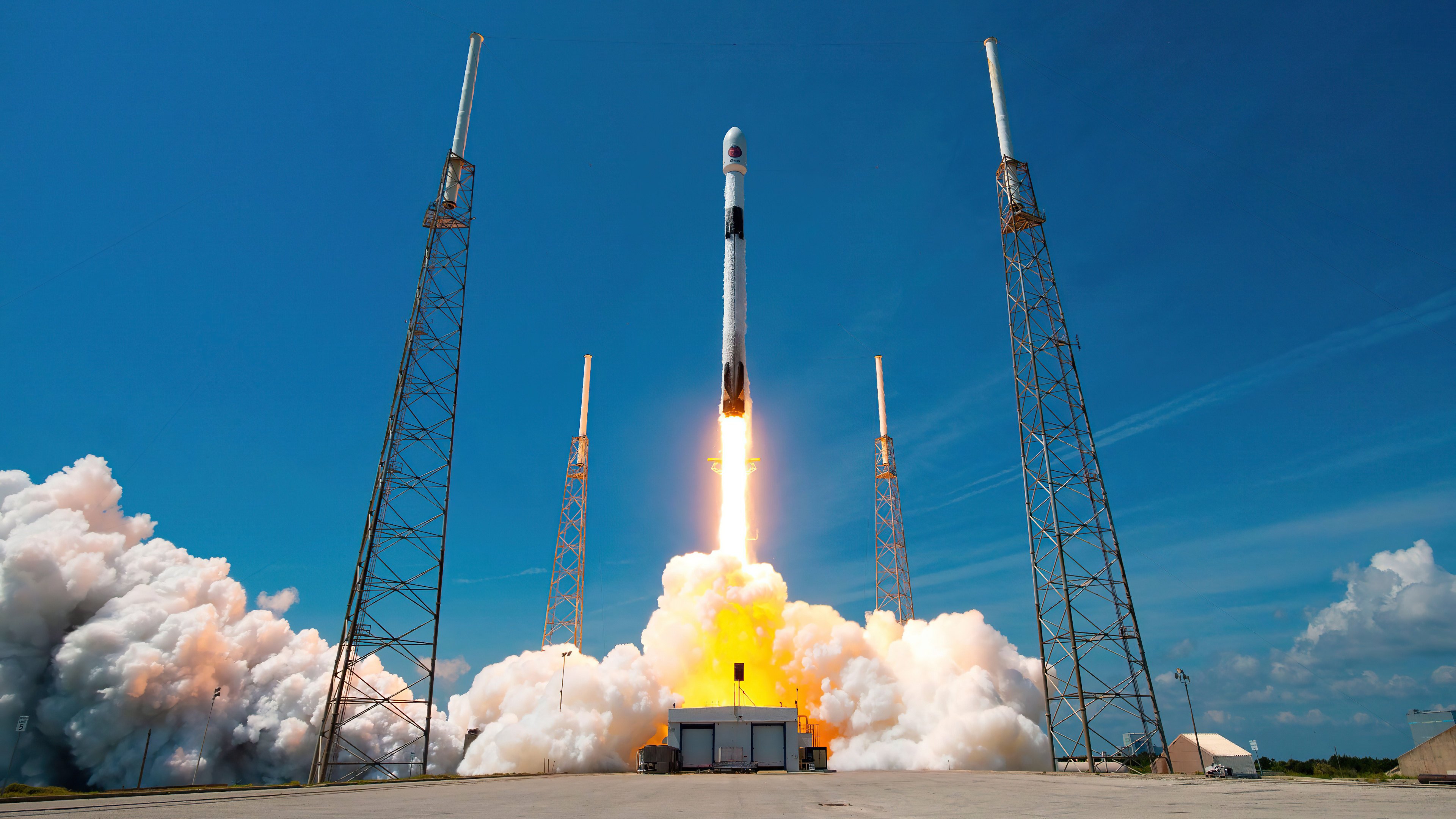}
    \caption{The \textit{Euclid} satellite atop a SpaceX Falcon 9 rocket 
        launching off from the Kennedy Space Centre on 1$^{\textrm{st}}$ July 2023\protect\footnotemark.
        Credit: ESA/SpaceX/J.-C. Cuillandre }
    \label{fig:EuclidLaunch}
\end{figure}

\begin{figure}[tp]
    \centering
    \includegraphics[width=0.9975\linewidth,trim={0.0cm 0.0cm 0.0cm 0.0cm},clip]{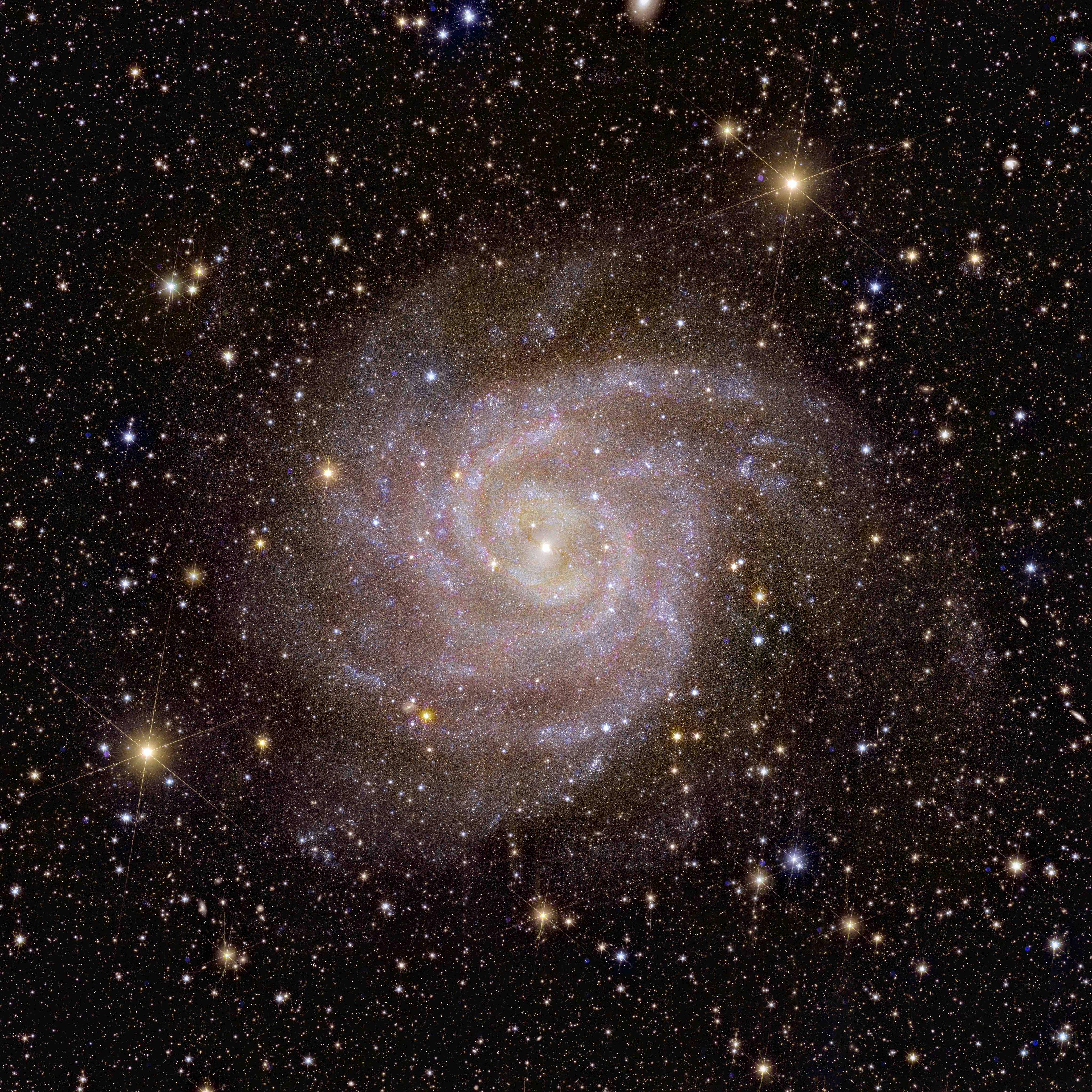}
    \caption{An image of nearby spiral galaxy IC~342 taken by the \textit{Euclid} 
        satellite as part of its early release observations scheme. 
        \textit{Euclid}'s VIS camera ordinarily takes black and white images for maximum 
        signal-to-noise, however this colour image was taken by combining
        VIS data and NISP photometry. We see spectacular sub-structure within
        this spiral galaxy, the precision shown here will allow us to
        execute cosmic shear measurements for far fainter galaxies with
        unprecedented precision.
        Credit: ESA/Euclid Consortium, data processed by J.-C.~Cuillandre}
    \label{fig:Euclid_ERO}
\end{figure}

As I write (November 2024), \textit{Euclid} has been at $L_2$ for approximately
sixteen months after launching in July 2023 on a SpaceX Falcon~9 rocket
(Figure~\ref{fig:EuclidLaunch}). While it is already taking incredible data,
for example Figure~\ref{fig:Euclid_ERO} showing a \textit{Euclid} image of
spiral galaxy IC~342, it will take time for \textit{Euclid} to complete its
planned six-year survey~\cite{Euclid:2024yrr} and provide us with its
fully complete cosmological data.

All three surveys promise to deliver
cosmic shear data that is unprecedented in terms of its quality and quantity.
This will produce a significance reduction in error bars for observed
quantities, in particular $S_8$, and so we want to develop tools that allow
us to extract the maximum amount of information possible from these new
awesome observatories. This will be the theme of the rest of my thesis: 
Chapter~\ref{chp:QML_estimator} discusses a new implementation of a power
spectrum estimator to minimise error bars, and Chapters~\ref{chp:baryonic_effects}
and~\ref{chp:binary_cuts} discussing modelling of baryonic feedback induced
scale cuts to maximise information gain.

\footnotetext{\textit{Euclid}'s launch certainly went better than any of my many attempts in \textit{Kerbal Space Program}!
    \\\textit{Bon voyage, Jeb, Val, Bill, and Bob. Bon voyage.  \faicon{rocket}
    }
}

\clearpage
\vspace*{7.5cm}
\noindent\rule{\linewidth}{0.5pt}

\vspace*{0.25cm}
\quad \noindent \textit{And now excuse me while I interrupt myself,}

\quad \noindent \textit{With half the thesis gone, there is still half the thesis to go!}

\vspace*{-0.25cm}

\begin{flushright}
    {---Murray Walker}
\end{flushright}

\quad \noindent \textit{Whoa, we're half way there}

\begin{flushright}
    {---Bon Jovi}
\end{flushright}

\vspace*{-0.5cm}
\noindent\rule{\linewidth}{0.5pt}

\begin{savequote}[75mm]
  Remember kids, the only difference
  
  between science and screwing around

  is writing it down
  
  \qauthor{---Adam Savage}
\end{savequote}
\chapter{Testing quadratic maximum likelihood estimators for \texorpdfstring{\\}{} forthcoming Stage-IV weak lensing surveys}
\label{chp:QML_estimator}
\vspace*{-1cm}
\begin{mypaper}
    This chapter was published in \textit{Monthly Notices of the Royal 
    Astronomical Society} as Maraio, Hall, and Taylor (2023)~\cite{Maraio:2022ywi}.
\end{mypaper}
\vspace*{0.5cm}

\noindent
\begin{mytext}
    \textbf{Outline.} 
    Headline constraints on cosmological parameters from current weak lensing
  surveys are derived from two-point statistics that are known to be
  statistically sub-optimal, even in the case of Gaussian fields. We study the
  performance of a new fast implementation of the Quadratic Maximum Likelihood
  (QML) estimator, optimal for Gaussian fields, to test the performance of
  Pseudo-$\Cl$ estimators for upcoming weak lensing surveys and quantify the
  gain from a more optimal method. Through the use of realistic survey
  geometries, noise levels, and power spectra, we find that there is a decrease
  in the errors in the statistics of the recovered $E$-mode spectra to the level
  of $\sim \!\! 20\,\%$ when using the optimal QML estimator over the
  Pseudo-$\Cl$ estimator on the largest angular scales, while we find
  significant decreases in the errors associated with the $B$-modes. This raises
  the prospects of being able to constrain new physics through the enhanced
  sensitivity of $B$-modes for forthcoming surveys that our implementation of
  the QML estimator provides. We test the QML method with a new implementation
  that uses conjugate-gradient and finite-differences differentiation methods
  resulting in the most efficient implementation of the full-sky QML estimator
  yet, allowing us to process maps at resolutions that are prohibitively
  expensive using existing codes. In addition, we investigate the effects of
  apodisation, $B$-mode purification, and the use of non-Gaussian maps on the
  statistical properties of the estimators. Our QML implementation is publicly
  available and can be accessed from
  \href{https://github.com/AlexMaraio/WeakLensingQML}{\texttt{GitHub}
  \faicon{github}}.
\end{mytext}

\section{Introduction}
\label{sec:Introduction}

Cosmic shear is the study of the coherent distortion in the shapes of background
galaxies due to the matter distribution of the intervening large-scale
structure~\citep{Bartelmann:1999yn,Bartelmann:2010fz,Kilbinger:2014cea}. Since
these distortions are sensitive to the total matter distribution, with
contributions from both ordinary baryonic matter and non-luminous dark matter,
cosmic shear is a powerful probe of dark matter. By measuring the cosmic shear
signal in multiple redshift bins, we can place constraints on the evolution of
structure in the Universe, ultimately placing constraints on the properties of
dark energy -- a key goal for cosmology in the current decade. See
Appendix~\ref{chp:appendix_B} for a detailed discussion on why cosmic shear
is sensitive to the evolution of dark energy, and the observational
prospects of measuring it.

Given the large quantity of high-precision cosmic shear data that forthcoming
Stage-IV weak gravitational lensing surveys, such as the \textit{Euclid} space
telescope \citep{Euclid:2011zbd}, the Legacy Survey of Space and Time (LSST) at
the \textit{Rubin} observatory \citep{LSSTDarkEnergyScience:2012kar}, and the
\textit{Roman} space telescope \citep{Spergel:2015sza}, are expected to take, it
is important to ensure that we are using the most optimal methods possible
throughout the data analysis pipeline. Given that each of these observatories
will observe and measure the shear of well over one billion galaxies, it is
unfeasible to perform data analysis on each of these individual galaxies. Hence,
some form of data compression steps are needed to make the data processable.
Here, we have investigated the process of compressing maps of the observed
ellipticities of galaxies into two-point summary statistics, namely the power
spectrum. The use of two-point statistics is well motivated because for Gaussian
fields the power spectrum contains all of the information about the field, and
two-point statistics have been extensively studied leading to the development of
robust, well-tested models. The estimation of two-point statistics from data is
an important process as it allows comparisons between observations and values
predicted from cosmological theories to be performed. These comparisons allow us
to constrain the cosmological parameters that feature in the models, and the
cosmic shear power spectrum has been used to obtain competitive
results~\citep{Heymans:2020gsg,DES:2022qpf,Hamana:2019etx}. Recent works have
suggested that there could be a possible tension to the level of $\sim\!3
\sigma$ in the value for the structure of growth parameter, $\Seight \equiv
\sigmaeight \sqrt{\Omegam / 0.3}$, measured between data from cosmic shear
surveys and those from cosmic microwave background experiments~\citep[e.g.][]{Heymans:2020gsg,Abdalla:2022yfr}. Therefore, to help determine if
this tension has physical origins or not, it is essential to ensure that our
analysis methods are as optimal as possible.

The process of compressing maps into two-point summary statistics is crucial,
and so naturally a number of competing methods to do so have been developed and
applied to cosmic shear data. Most notably are the two-point correlation
functions (2PCF) $\xi_{\pm}(\theta)$~\citep{Kaiser:1991qi,Schneider:2002jd}, the
Complete Orthogonal Sets of \mbox{E-/B-mode} Integrals 
(COSEBIs)~\citep{Schneider:2010pm,KiDS:2020suj}, and the power spectrum 
coefficients $\Cl$~\citep{Hu:2000ax,Brown:2004jn,Hikage:2010sq}. In this work, we focus on
analysing cosmic shear maps using the power spectrum. The power spectrum has the
advantage that it provides the most direct comparison between theory- and
data-vectors, without the need to perform any additional transformations when
comparing them, as is required for analyses using correlation
functions~\citep{Schneider:2002jd}\footnote{The cosmic shear angular power 
spectrum is a simple Limber integral over the matter power spectrum weighted
by the lensing kernels (Equation~\ref{eqn:cosmic_shear_powspec}). Many methods
have been developed to ensure that this integral step is as fast and accurate
as possible. If we now wish to use another estimator, say the two-point 
correlation functions or COSEBIs, then one or two additional summation steps
are necessary, respectively (Equations~\ref{eqn:2PCF_def} and~\ref{eqn:COSEBI_def}).
This also requires the angular power spectrum coefficients to be computed
to much larger $\ell$ values than when using the angular power spectrum
natively. This further evaluation of the $\Cl$ values and the additional
summation steps significantly slows down the application of 2PCF and COSEBIs to
cosmic shear analyses using Markov chain Monte Carlo methods -- which are the baseline
choice for any cosmic shear survey. Thus, we are motivated to use the $\Cl$
values as our theoretical data-vector, and so we are
tasked with finding the most-optimal experimental $\Cl$ estimator.}. 
Additionally, the power spectrum provides a
cleaner separation between linear and non-linear scales, which aids the
investigation of biases from the non-linear modelling of the matter power
spectrum and intrinsic alignments \citep{DES:2022qpf}, and the scale-dependent
signatures in the power spectrum -- for example arising from the properties of
massive neutrinos and baryonic effects. We note that none of the methods
discussed in this work employ the flat-sky approximation, with all quantities
being evaluated on the full, curved-sky.

While power spectrum estimators are a sub-set of two-point correlators, we can
further break down this category of estimators into two main methods: the
Pseudo-$\Cl$ method~\citep{Hivon:2001jp,Brown:2004jn}, and the quadratic maximum
likelihood method (QML) \citep{Tegmark:1996qt,Tegmark:2001zv}. In addition,
there is the \texttt{PolSpice} algorithm which uses the correlation functions to
produce estimates of the power spectrum, and is statistically equivalent to the
Pseudo-$\Cl$ method~\citep{Szapudi:2000xj,Chon:2003gx}. The Pseudo-$\Cl$ class of estimators work in
harmonic-space and utilise very efficient spherical harmonic transformation
algorithms which makes this class of estimator extremely numerically efficient
even for high-resolution maps \citep{Hivon:2001jp,Gorski:2004by}. Alternatively,
the QML method works in pixel-space, which results in a much larger
computational demand when compared to Pseudo-$\Cl$ method for the same
resolution maps. Traditionally, this has limited analyses using the QML method
to low-resolution maps only, and thus confined the values for the recovered
power spectrum to low multipoles. Hence, when power spectrum methods have been
applied to existing weak lensing data, the Pseudo-$\Cl$ estimator has been the
method of choice for the vast majority of weak lensing analyses~\citep{HSC:2018mrq,KiDS:2020suj,Nicola:2020lhi,
Garcia-Garcia:2019bku,DES:2022qpf}, primarily using the \texttt{NaMaster} code
which is a fast implementation of the Pseudo-$\Cl$ estimator that can be easily
applied to weak lensing analyses \citep{Alonso:2018jzx}. The Pseudo-$\Cl$
estimator will form part of the analysis pipeline for upcoming weak lensing
surveys~\citep{KiDS:2021opn}. However, it has been shown that while the QML
estimator is optimal, in the sense that it estimates a power spectrum with the
minimal possible errors \citep{Tegmark:1996qt}, it is known that the
Pseudo-$\Cl$ method is not optimal \citep{Efstathiou:2003dj}, and thus could be
introducing additional errors into the data.

Another aspect of the data analysis expected to benefit significantly from the
application of an optimal estimator is in $B$-mode measurement. In the limit of
weak gravitational lensing, the produced signal should form a curl-free field,
and thus the predicted $B$-mode signal for cosmic shear should be zero.
This is because the distortion in the shapes of galaxies due to cosmic shear
should be tangentially aligned to spherical distributions of mass, which galaxies
and galaxy clusters can be approximated as~\cite{Prat:2025ucy}. Thus, the
rotated deflections to induce a $B$-mode signal must come from non-cosmic shear
sources, such as unaccounted detector systematics~\cite{Schneider:2001af}, 
the presence of intrinsic 
alignments~\cite{Heavens:2000ad}, or  physics beyond general relativity. 
Traditionally, a statistically significant detection of $B$-modes would indicate
the presence of unaccounted systematic effects present in the data. However,
with the increased precision of forthcoming Stage-IV experiments, the $B$-mode
signal will be treated as a potential signal that could give hints of new
physical phenomena. An application of using $B$-modes to constrain novel
cosmological models was presented in Ref.~\cite{Thomas:2016xhb}. Hence, ensuring that
the $B$-mode errors are as small as possible (to help determine if any residual
$B$-mode signal is statistically significant, and to distinguish between
systematic effects and a cosmological $B$-mode signal) is another key feature
for any power spectrum estimation technique that would be applied to upcoming
experimental data. 

Additionally, in a power spectrum analysis a choice for how the survey mask
should be modelled is present. Either the effects of the mask can be deconvolved
from the observed values, giving predictions for the full-sky power
\citep{Hikage:2010sq}, or the effects of the mask can be convolved into the
theory predictions, the so-called `forward-modelling' approach
\citep{KiDS:2021opn}. Here, we focus on the full-sky predictions, which the QML
estimator naturally produces, as this allows for the most straightforward
comparisons between experimental results and theoretical predictions to be made.
This is because there is no need to convolve the theory data-vector with the
mask at every step in an analysis chain when using Monte Carlo Markov Chain
methods, and thus reducing the per-step computational requirements resulting in
faster run-times. We also note that by producing estimates of the full-sky power
spectra, our covariance matrix is less affected by the mask on large scales.

Previous attempts at applying QML methods to existing weak lensing surveys have
found little differences in results when compared to other analysis techniques.
Ref.~\cite{Kohlinger:2015tza} applied a QML implementation to estimate band powers
from the data from the Canada-France-Hawaii Telescope Lensing Survey (CFHTLenS)
finding general consistency between their QML analysis and all other studies
using CFHTLenS data. This implementation was then applied to data from the first
data release from the Kilo Degree Survey (KiDS-450) in Ref.~\cite{Kohlinger:2017sxk}.
Here, they again found broadly consistent results between their analysis and
previous works, though finding a slightly smaller value for $\Seight$ which
could be explained by their work using a slightly smaller range of $\ell$
multipole values \citep{vanUitert:2017ieu}. Quadratic and Pseudo-$\Cl$
estimators were applied to cosmic shear measurements performed by the Sloan
Digital Sky Survey in Ref.~\cite{SDSS:2011gwu}, again finding strong consistency
between the two methods. This demonstrates that analysing weak lensing data
using QML methods provides a strong consistency check between different
two-point estimators and ensuring that results are robust to the different
analysis choices that are required for the different methods. While these
previous analyses of weak lensing data using QML methods have shown strong
consistencies in the main cosmological results, though their application as a
cross-check remains an important use case, we note that the CFHTLenS and
KiDS-450 surveys covered about $154 \, \deg^2$ and $450 \, \deg^2$ of sky,
respectively. These sky areas are around two orders of magnitude smaller than
the expected sky area that forthcoming Stage-IV experiments are expected to
cover. While it has been shown that the QML and Pseudo-$\Cl$ estimators are
statistically equivalent in the high noise
regime~\citep{Efstathiou:2003dj,Efstathiou:2006eb}, the expected noise levels
for forthcoming surveys will be much lower than CFHTLenS and KiDS. Therefore,
the huge increase in statistical precision that forthcoming Stage-IV surveys
will bring means that the use of non-optimal methods (the Pseudo-$\Cl$
estimator) needs to be reassessed and their affects on cosmological constraints
quantified. 

Despite the advantages of quadratic estimators, the development of maximum
likelihood estimators, and in particular their applications to cosmic shear, has
traditionally been less explored than other techniques because of their
computational complexity and associated slowness. In general, they require the
computation and inversion of a dense pixel-space covariance matrix of the map(s)
which is a slow and inefficient process when compared to other analysis methods.
Recent theoretical developments presented in Refs.~\cite{Horowitz:2018tbe} and
\cite{Seljak:2017rmr}, along with using methods presented in Ref.~\cite{Oh:1998sr},
have provided a set of key tools that has allowed us to build a new novel QML
implementation that is highly efficient. We note that the construction of the
QML estimator is very analogous to the Wiener filtering of the data, for which
fast implementations have been recently developed and applied to CMB data-sets
\citep{Elsner:2012fe,Bunn:2016lxi,Ramanah:2018enp}. We also note that recent
works have applied quadratic estimators to galaxy clustering in Refs.
\cite{Estrada:2021hdo} and \cite{Philcox:2020vbm} applying their quadratic
estimators to data from the VIPERS and BOSS surveys, respectively.

This Chapter is structured as follows: in
Section~\ref{sec:Power_spectrum_estimators} we present a review of both the QML
and Pseudo-$\Cl$ estimators, including a detailed derivation of the QML method
in Section~\ref{sec:The_qml_estimator}. Then in
Sections~\ref{sec:Conjugate_gradient_method} and~\ref{sec:Finite_diff_fisher} we
present our new highly efficient implementation of the QML estimator.
Section~\ref{sec:Methodology} outlines our methodology for generating mock weak
lensing data, which is used for our results that we present in
Section~\ref{sec:Results}. Our conclusions are presented in
Section~\ref{sec:Conclusions}.

\section{Power spectrum estimators}
\label{sec:Power_spectrum_estimators}

As discussed in Section~\ref{sec:Introduction}, there exists two broad classes
of power spectrum estimation techniques. Here, we first derive the set of key
results for the QML method, and then present a brief review of the Pseudo-$\Cl$
method.

\subsection{The quadratic maximum likelihood estimator}
\label{sec:The_qml_estimator}

Consider a spin\footnote{\textit{Pronto, Sebastian?} S\mychar{B}innala.}-0 input map as a data-vector $\mathbfit{x}$ (which may
be complex, such as for real-space shears) that has zero mean, an example of
such a map would be convergence estimates in pixels over the sky, and covariance
$\mathbfss{C}$. This data-vector has a length equal to the number of pixels in
the map $\Npix$. We can write it as
\begin{align}
    \mathbfit{x} = \mathbfit{s} + \mathbfit{n},
\end{align}
where $\mathbfit{s}$ and $\mathbfit{n}$ are the signal and noise data-vectors,
respectively. Assuming that the signal and noise data-vectors are uncorrelated
and both follow the Gaussian distribution, then the likelihood function for the
power spectrum coefficients recovered from the map, $\Cltilde$, is given by
\begin{equation}
    \mathcal{L}(\mathbfit{x} \,|\, \Cltilde) = 
    \frac{\exp \left( -\frac{1}{2} \mathbfit{x}^{\dagger} \, \mathbfss{C}^{-1} \, \mathbfit{x} \right)}{\left(2 \pi\right)^{\Npix / 2} \lvert \mathbfss{C} \rvert^{1/2}},
    \label{eqn:Cl_likelihood}
\end{equation}
where $\mathbfss{C}$ is the total pixel-covariance matrix, given as 
\begin{equation}
    \mathbfss{C} = \langle \mathbfit{x} \, \mathbfit{x}^{\dagger} \rangle = \mathbfss{S}(\Cl) + \mathbfss{N},
    \label{eqn:total_cov_mat}
\end{equation}
where $\mathbfss{S}$ is the signal covariance matrix, $\mathbfss{N}$ is the
noise matrix, and $\Cl$ are the fiducial power spectrum coefficients. The signal
covariance matrix can be written as
\begin{equation}
    \mathbfss{S}(\Cl) = \sum_{\ell} \mathbfss{P}_\ell \, \Cl,
\end{equation}
where the $\mathbfss{P}_\ell$ matrices are defined, in pixel-space, as
\begin{equation}
    \mathbfss{P}_\ell \equiv 
    \frac{2\ell + 1}{4 \pi} P_{\ell}(\hat{r}_i \cdot \hat{r}_j),
\end{equation}
where $P_\ell$ are the Legendre polynomials, and $\hat{r}_i$ is the unit vector
for pixel $i$. This matrix can be decomposed into spherical harmonics through
the addition theorem, giving
\begin{equation}
    \mathbfss{P}_{\ell} = \sum_{m=-\ell}^{\ell} \Ylm(\hat{r}_i) \, \Ylm^*(\hat{r}_j).
\end{equation}
We note an important result of
\begin{equation}
    \frac{\partial \mathbfss{C}}{\partial \Cl} = \mathbfss{P}_\ell.
\end{equation}
In harmonic-space, these $\mathbfss{P}_{\ell}$ matrices are simply zeros with
ones along the diagonal corresponding to their $\ell$ value. This makes 
evaluating the signal matrix very easy in spherical-harmonic space.

For uncorrelated noise, the noise matrix $\mathbfss{N}$ in pixel-space is simply
given by the noise variance of the $i$-th pixel along the diagonal with zeros
elsewhere. This makes evaluating the noise matrix very easy in pixel-space. We
note that the QML method may require the manual insertion of a small level of
white noise into the covariance matrix to ensure that it is invertible, as in
some extreme cases the covariance matrix can be
singular~\citep{Bilbao-Ahedo:2017uuk}.

A minimum-variance quadratic estimator of the power spectrum can be
formed as~\citep{Tegmark:1996qt}
\begin{equation}
    y_{\ell} \equiv s_{\ell} - b_{\ell} = \mathbfit{x}^{\mathrm{T}} \, \mathbfss{E}_{\ell} \, \mathbfit{x} - b_{\ell},
    \label{eqn:y_ell_definition}
\end{equation}
where the $\mathbfss{E}_{\ell}$ matrices are given by
\begin{equation}
    \mathbfss{E}_{\ell} = \frac{1}{2} \mathbfss{C}^{-1} \frac{\partial \mathbfss{C}}{\partial \Cl} \mathbfss{C}^{-1} =
    \frac{1}{2} \mathbfss{C}^{-1} \mathbfss{P}_{\ell} \mathbfss{C}^{-1},
\end{equation}
and the noise bias terms $b_{\ell}$ are given as
\begin{equation}
    b_{\ell} = \Tr \left[ \mathbfss{N} \, \mathbfss{E}_{\ell} \right].
\end{equation}
Arranging our $\yl$ and $\Cl$ values into vectors $\mathbfit{y}$ and $\mathbfit{C}$, 
respectively, we can relate our quadratic estimator to the true power spectrum
as
\begin{equation}
    \langle \mathbfit{y} \rangle = \mathbfss{F} \, \mathbfit{C},
    \label{eqn:y_ell_avg}
\end{equation}
where $\mathbfss{F}$ is the Fisher matrix. Formally, this is defined through
the likelihood as~\citep{Bond:1998zw}
\begin{equation}
    \mathbfss{F}_{\ell_1 \ell_2} = - \left\langle \frac{\partial^2 \ln \mathcal{L}}{\partial C_{\ell_1} \, \partial C_{\ell_2}} \right\rangle.
    \label{eqn:Fisher_deriv_likelihood}
\end{equation}
When applying the likelihood of Equation~\ref{eqn:Cl_likelihood}, we find the
Fisher matrix can be written as
\begin{align}
    \mathbfss{F}_{\ell \ell'} &= \frac{1}{2} \Tr \left[ \mathbfss{C}^{-1} \frac{\partial \mathbfss{C}}{\partial C_{\ell}} \mathbfss{C}^{-1} \frac{\partial \mathbfss{C}}{\partial C_{\ell'}} \ \right] \! ,
    \label{eqn:AnalyticClFisher}
    \\
    &= \frac{1}{2} \Tr \left[ \mathbfss{C}^{-1} \, \mathbfss{P}_{\ell} \, \mathbfss{C}^{-1} \, \mathbfss{P}_{\ell'} \right] \!\! . \nonumber
\end{align}
Assuming that $\mathbfss{F}$ is regular, and thus can be inverted, we can form
an estimator for the recovered power spectrum from our map,
$\tilde{\mathbfit{C}}$, as
\begin{equation}
    \tilde{\mathbfit{C}} = \mathbfss{F}^{-1} \, \mathbfit{y}.
\end{equation}
This estimator is unbiased in the sense that its average is the true, underlying
spectrum \citep{Tegmark:1996qt},
\begin{equation}
    \langle \tilde{\mathbfit{C}} \rangle =  \mathbfit{C},
\end{equation}
and it is optimal in the sense that its covariance matrix of our estimator is 
the inverse Fisher matrix $\mathbfss{F}^{-1}$,
\begin{equation}
    \langle (\tilde{\mathbfit{C}} - \mathbfit{C}) (\tilde{\mathbfit{C}} - \mathbfit{C})^{\mathrm{T}} \rangle 
    = \mathbfss{F}^{-1}
\end{equation}
and thus satisfies the Cram\'{e}r-Rao inequality~\citep{Tegmark:1996bz}.

\subsubsection{Extension to spin-2 fields}

Above, we have derived a set of key-results of the QML estimator applied to a
scalar spin-0 field. These set of equations can be extended to cover spin-2
fields, as presented in Ref.~\cite{Tegmark:2001zv}. Such spin-2 field is cosmic
shear, of which the observed field can be decomposed into two components through
$\bm{\gamma}(\nhatvec) = \gamma_1 (\nhatvec) + i \gamma_2 (\nhatvec)$. The
data-vector will now have length $2 \Npix$, where it will be given by
$\mathbfit{x} = \left\{ \vec{\gamma}, \vec{\gamma}^{*}  \right\}$, where
$\vec{\gamma}$ is the values of the complex shears at each pixel, and the
covariance matrix (and other associated pixel-space matrices) have dimensions $2
\Npix \times 2 \Npix$, where their block structure will be given by
\begin{equation}
    \mathbfss{C} = \begin{pmatrix}
        \langle \vec{\gamma} \, \vec{\gamma}^{\dagger} \rangle & \langle \vec{\gamma} \, \vec{\gamma}^{\mathrm{T}} \rangle \\
        \langle \vec{\gamma}^{*} \, \vec{\gamma}^{\dagger} \rangle & \langle \vec{\gamma}^{*} \, \vec{\gamma}^{\mathrm{T}} \rangle
    \end{pmatrix}.
\end{equation}
Similarly, the Legendre polynomial matrix $\mathbfss{P}_{\ell}$ will have a
structure for the spin-2 case of~\citep{Tegmark:2001zv}
\begin{equation}
    \mathbfss{P}_\ell = \begin{pmatrix}
        \sum_{m} \splustwoYlm(\hat{r}_i) \, \splustwoYlm^*(\hat{r}_j) &
        \sum_{m} \splustwoYlm(\hat{r}_i) \, \sminustwoYlm^*(\hat{r}_j) \\
        \sum_{m} \splustwoYlm^*(\hat{r}_i) \, \sminustwoYlm(\hat{r}_j) &
        \sum_{m} \splustwoYlm^*(\hat{r}_i) \, \splustwoYlm(\hat{r}_j) \\
    \end{pmatrix},
\end{equation}
and the signal covariance matrix is given by
\begin{equation}
    \mathbfss{S} = \begin{pmatrix}
        \sum_{\ell} \left[\ClEE + \ClBB\right] \mathbfss{P}_{\ell}^{(1, \, 1)} &
        \sum_{\ell} \left[\ClEE - \ClBB\right] \mathbfss{P}_{\ell}^{(1, \, 2)} \vspace*{0.15cm} \\
        \sum_{\ell} \left[\ClEE - \ClBB\right] \mathbfss{P}_{\ell}^{(2, \, 1)} &
        \sum_{\ell} \left[\ClEE + \ClBB\right] \mathbfss{P}_{\ell}^{(2, \, 2)} \\
    \end{pmatrix},
\end{equation}
where $\sYlm$ are the spin-weighted spherical harmonics.

The observed spin-2 shear field can be decomposed on the full-sky into a
curl-free $E$-mode and divergence-free $B$-mode fields through
\citep{Brown:2004jn}
\begin{equation}
    (\gamma_1 \pm i \gamma_2)(\nhatvec) = \sum_{\ell} \sum_{m \,=\, - \ell}^{\ell} 
    \left[ \almE \pm i \almB \right] \stwoYlm(\nhatvec).
    \label{eqn:alm2map}
\end{equation}
This relation can be inverted to give the $\alm$ coefficients on the full-sky as
\begin{equation}
    \almE \pm i \almB = \int \!\! \d \Omega \, \left( \gamma_1 \pm i \gamma_2 \right)\!(\nhatvec)
    \stwoYlm^{*}(\nhatvec).
    \label{eqn:map2alm}
\end{equation}
These $\alm$ coefficients can then be combined to form values for the all-sky
power spectrum through
\begin{equation}
    \Cl^{XY} = \frac{1}{2 \ell + 1} \sum_{m = - \ell}^{\ell} 
    \alm^X \left[\alm^{Y}\right]^{*},
    \label{eq:alm2cl}
\end{equation}
where $X, Y$ denote either $E$ or $B$.

\subsubsection{Affect of the fiducial cosmology}

We note that to construct the covariance matrix, we have to provide our
estimator with a set of fiducial $\Cl$ values. Given that the whole point of the
estimator is to estimate the $\Cl$ values from map(s), of which their underlying
power spectrum are unknown prior to the analysis, it may appear that the
estimated power spectrum will somehow depend on the input cosmology. However,
provided that the same fiducial power spectrum is applied consistently to the
estimator, it will still produce unbiased estimates, but ones that may not
necessarily be truly optimal. An iterative scheme where the results of the
estimator are fed back into the construction of the covariance matrix, with this
process repeating for a number of times, was investigated in
Ref.~\cite{Bilbao-Ahedo:2021jhn}.

\subsection{Inverting the pixel covariance matrix}
\label{sec:Conjugate_gradient_method}

To evaluate our quadratic estimator, we need an efficient way to evaluate the
set of $y_{\ell}$ values for a given map. These in turn require efficient
evaluation of the inverse-covariance weighted map, $\mathbfss{C}^{-1}
\mathbfit{x}$. Na\"ively, one may want to compute these terms through first
evaluating the total covariance matrix $\mathbfss{C}$ and then inverting it.
However, as we have seen through Equation~\ref{eqn:total_cov_mat},
$\mathbfss{C}$ is made up of both the signal and noise covariance matrices,
resulting in there not being a single efficient basis to evaluate $\mathbfss{C}$
in without having to resort to using expensive massive matrix multiplications
using matrices of spherical harmonics, $\Ymat$. Since these operations scale as
$\mathcal{O}(\Npix^3)$, this is an important limiting factor to the resolution
that can be obtained with traditional QML estimation techniques. 

An alternative approach that negates the need to evaluate the total covariance
matrix is required to obtain competitive resolution results using QML methods.
Previous attempts at this problem have used either Newton-Raphson iteration
techniques to find the root of $\partial \mathcal{L} / \partial C_{\ell} =
0$ \citep{Bond:1998zw,Seljak:1997ep,Hu:2000ax}, or alternatively used conjugate
gradient techniques \citep{Oh:1998sr}, both of which avoid the need to directly
evaluate and invert the covariance matrix and thus offers significantly better
computational performance. Alternative techniques also include an iterative
scheme presented in Ref.~\cite{Pen:2003mu} and renormalisation-inspired methods
presented in Refs.~\cite{McDonald:2018mfm,McDonald:2019efe}. Here, we employ the
conjugate gradient approach, although minimisation approaches have also been
shown to give good results \citep{Horowitz:2018tbe}.

The conjugate gradient method \citep{Oh:1998sr} utilises a converging iterative
scheme to find a best-fitting solution vector $\mathbfit{z}$ such that it
satisfies the linear equation
\begin{equation}
    \mathbfss{C} \mathbfit{z} = \mathbfit{x}
\end{equation}
for a given covariance matrix and input maps. Using the conjugate gradient
method allows us to only compute the action of a trial-vector on the covariance
matrix instead of the traditional approach of computing the full form of the
covariance matrix and inverting it. As this is an iterative approach to finding
the best-fitting vector $\mathbfit{z}$, the value of $\mathbfss{C} \mathbfit{z}$
needs to be computed many times. Therefore, an efficient numerical approach to
computing this is required. This is achieved through the use of splitting the
pixel covariance matrix into a signal part, which is best suited to
harmonic-space, and a noise part, which is best represented in pixel-space.
Hence, this allows us to rapidly compute the action of our trial vector
$\mathbfit{z}$ on the covariance matrix through
\begin{equation}
    \mathbfss{S} \mathbfit{z} + \mathbfss{N} \mathbfit{z} = \mathbfit{x}.
\end{equation}
We can rapidly transform our trial vector $\mathbfit{z}$ between pixel- and
harmonic-space through efficient spherical harmonic transform functions
\texttt{map2alm} \& \texttt{alm2map} from the \texttt{HEALPix}
library~\citep{Gorski:2004by}. Qualitatively, the per-iteration computation of
$\mathbfss{C} \mathbfit{z}$ proceeds along the following steps:
\begin{enumerate}
    \item Convert our map-based trial-vector $\mathbfit{z}$ into a set of
    $\almE$ and $\almB$ values through the use of \texttt{map2alm} which
    implements Equation~\ref{eqn:map2alm},
    \item Re-scale the $\alm$ values with the input fiducial power spectrum
    $\ClEE$ and $\ClBB$, respectively,
    \item Generate a new set of spin-2 maps with these new $\alm$ coefficients
    to obtain the contribution from the cosmological signal using
    \texttt{alm2map} which implements Equation~\ref{eqn:alm2map},
    \item Take our original trial-vector $\mathbfit{z}$ and multiply all elements
    by the noise variance in the respective pixel to obtain the noise contribution,
    \item Finally sum the signal and noise contributions giving a final set of
    two maps in pixel space.
\end{enumerate} 

With this efficient computation of $\mathbfss{C} \mathbfit{z}$ in place, we can
use standard implementations of the conjugate gradient algorithm to find
$\mathbfit{z}$. We used the
\texttt{Eigen}\footnote{\url{https://eigen.tuxfamily.org}} \Cpp linear algebra
package to perform our conjugate gradient computations resulting in a quick and
efficient numerical implementation.

Since we are now only computing the action of the covariance matrix on our trial
vector, instead of explicitly computing the full form of the covariance matrix,
we find that our method provides much better scaling to higher map resolutions
than previous implementations. We explore the speed and memory performance of
our new estimator in Section~\ref{sec:QML_estimtors_benchmark}. In our analysis,
we used map resolutions of $\Nside = 256$, which compares to a maximum of
$\Nside = 64$ that was explored in previous QML
implementations~\citep{Bilbao-Ahedo:2021jhn}.

In general, the conjugate gradient technique can benefit greatly from an
appropriate choice of matrix preconditioner \citep{Oh:1998sr}. Given a linear
system $\mathbfss{A} \mathbfit{x} = \mathbfit{b}$, the preconditioner matrix
$\tilde{\mathbfss{A}}$ should be such that $\tilde{\mathbfss{A}}^{-1}
\mathbfss{A} = \mathbfss{I} + \mathbfss{R}$, where $\mathbfss{I}$ is the
identity matrix and $\mathbfss{R}$ is a matrix whose eigenvalues are all less
than unity. This minimises the number of iterations required for the conjugate
gradient method to converge, and thus can offer significant performance
improvements if properly set. Since our method requires a strictly diagonal
preconditioner, this placed strict constraints on the form and values of the
preconditioner. We investigated the use of different values for the diagonal of
the preconditioner finding little change in the performance of the iterative
method. Thus we used the identity matrix as our preconditioner.

We also note that we have free choice over the initial guess $\vec{z}_0$ in
the conjugate-gradient method. An educated ansatz for the choice of $\vec{z}_0$
will dramatically speed up the iterative method, since there will be less work
that the numerical method needs to perform. Approximating the Legendre polynomial
matrix ($\mathbf{P}_{\ell}$) as diagonal values of unity in pixel-space,
the signal covariance matrix will be given by 
$\mathbf{S} = \textrm{diag}(\Sigma_{\ell} \, \Cl)$. Since our noise matrix 
$\mathbf{N}$ is already diagonal in pixel-space, we can form a zeroth-order 
inverse of our map $\vec{x}$ such that it is given by $\vec{z}_0 = \vec{x} /
\left[\Sigma_{\ell} \, \Cl + N \right]$. This choice dramatically speeds up
the convergence, particularly in the noise-dominated regime.

\subsection{Forming the Fisher matrix}
\label{sec:Finite_diff_fisher}

Since the covariance matrix of our quadratic estimator is the inverse Fisher
matrix, we can get estimates for the estimator's errors through computation of
this Fisher matrix. Direct computation of the Fisher matrix through
Equation~\ref{eqn:AnalyticClFisher} requires many massive $2 \Npix \times 2
\Npix$ matrix multiplications and inversions, which has
$\mathcal{O}(\Npix^{3})$ scaling, even for efficient implementations of this
technique~\citep{Bilbao-Ahedo:2021jhn}. Thus, this direct-evaluation technique
becomes unfeasible for map resolutions above about $\Nside = 64$ for Stage-IV
cosmic shear experiments. Hence, to get power spectrum estimates for
higher-resolution maps, which increases the range of $\ell$-values that we can
estimate the power spectrum over, an alternative method to direct computation is
needed. 

We note that the Fisher matrix is related to the second-order derivative of the
likelihood function through Equation~\ref{eqn:Fisher_deriv_likelihood}. Taking a
single derivative of the likelihood yields
\begin{equation}
    \frac{\partial \ln \mathcal{L}}{\partial C_\ell} = s_\ell - b_\ell - \Tr\!\left[\mathbfss{S} \, \mathbfss{E}_{\ell}\right]
    \label{eqn:Likelihood_one_diff}
\end{equation}
where $s_{\ell}$ is our quadratic form of the map as introduced in
Equation~\ref{eqn:y_ell_definition}.
Therefore, to evaluate our Fisher matrix, we wish to take a further derivative
of the above quantity, which yields
\begin{align}
    \left\langle \frac{\partial^2 \ln \mathcal{L}}{\partial C_{\ell} \, \partial C_{\ell'}} \right\rangle &= 
    - \Tr \! \left[\mathbfss{C}^{-1} \, \mathbfss{P}_{\ell} \, \mathbfss{C}^{-1} \, \mathbfss{P}_{\ell'} \right] +
    \frac{1}{2} \Tr \! \left[\mathbfss{C}^{-1} \, \mathbfss{P}_{\ell} \, \mathbfss{C}^{-1} \, \mathbfss{P}_{\ell'}\right]\!\!, \\
    &= -\frac{1}{2} \Tr \left[\mathbfss{C}^{-1} \, \mathbfss{P}_{\ell} \, \mathbfss{C}^{-1} \, \mathbfss{P}_{\ell'} \right]\!,
\end{align}
where the first trace term comes from the differentiation of the quadratic term
$s_{\ell}$ and the second trace arises from the differentiation of the other two
terms in Equation~\ref{eqn:Likelihood_one_diff}. Since we note that the
differentiation of just the $s_{\ell}$ term alone yields twice the negative
Fisher matrix, we can form an estimator for the Fisher matrix using just this
term. Therefore, we can use the method of finite differences to differentiate
$s_{\ell}$ to give~\citep{Seljak:2017rmr,Horowitz:2018tbe}
\begin{equation}
    F_{\ell \ell'} \, \Delta C_{\ell'} = -\frac{1}{2} \left[
    \langle s_{\ell}(C_{\ell}^{\mathrm{fid}} + \Delta C_{\ell'}) \rangle 
    - \langle s_{\ell}(C_{\ell}^{\mathrm{fid}}) \rangle \right].
    \label{eqn:FisherMatrixEstimation}
\end{equation}
Here, we are manually injecting power into a specific $\ell$-mode (given as
$\Delta C_{\ell'}$), generating a map with the modified power spectrum, and
recovering the set of $s_{\ell}$ values. This gives the estimate of our Fisher
matrix associated where we are averaging over many realisations of maps
generated with the specified power spectrum, and $C_{\mathrm{fid}}$ is our
original best-guess for the power spectrum coefficients used when building the
covariance matrix $\mathbfss{C}$.

This approach of using finite-differences to estimate the Fisher matrix performs
much faster than the brute-force calculation, as described in
Equation~\ref{eqn:AnalyticClFisher}, due to our ability to estimate the Fisher
matrix directly from the $s_{\ell}$ values, which are vector quantities and for
which we already have an efficient method to compute though the
conjugate gradient method, and so we avoid having to compute the matrices and
matrix products of Equation~\ref{eqn:AnalyticClFisher}.

Here, the amount of power injected into the maps at the specific $\ell$-mode is
a free parameter of the method. We used a value of $\Delta C_{\ell} =
10^{7} \, C_{\ell}^{\mathrm{fid}}$ and verified that our results were
insensitive to the choice of this value, provided that it is sufficiently large.

Note that in our analysis presented in this paper, we are not able to recover
any of the covariances associated with any of the $EB$ modes. This is because
these modes are not linearly independent of either the $EE$ or $BB$ spectra, and
thus with our choice of fiducial spectrum containing zero $B$-modes we cannot
inject power into the $EB$ modes. $EB$-spectra can be obtained by setting the
fiducial $B$-mode power to small non-zero values, for example
\cite{Horowitz:2018tbe} use a $B$-mode spectra that has the same shape as their
$E$-mode spectra but has an amplitude that is $10^{-5}$ times smaller. Since we
used zero $B$-mode power as our fiducial model, we are unable to report on any
$EB$ results in this work.

Our new code implementing these approaches is publicly available and can be
downloaded from \texttt{GitHub}:
\href{https://github.com/AlexMaraio/WeakLensingQML}{\texttt{https://github.com/AlexMaraio/WeakLensingQML}
\faicon{github}}.

\subsection[Review of the Pseudo-$\Cl$ estimator]{Review of the Pseudo-$\bm\Cl$ estimator}
\label{sec:Review_of_pseudocl}

We refer the reader to \cite{Alonso:2018jzx,Leistedt:2013gfa}, and references
therein, for detailed reviews of the Pseudo-$\Cl$ method, but here we discuss the
key features of the estimator.

Since we cannot observe the shear field on the full-sky, our observed field is
modulated through some window function, $\mathcal{W}(\nhatvec)$, through
$\tilde{\bm{\gamma}}(\nhatvec) = \mathcal{W}(\nhatvec) \bm{\gamma} (\nhatvec)$.
Throughout this work, we consider binary masks only, and thus
$\mathcal{W}(\nhatvec)$ is either zero or one. This turns the recovered harmonic
modes into `pseudo modes', given as  
\begin{equation}
    \almEtilde \pm i \almBtilde = \int \!\! \d \Omega \,\, 
    \mathcal{W}(\nhatvec) \left( \gamma_1 \pm i \gamma_2 \right)\!(\nhatvec)
    \stwoYlm^{*}(\nhatvec),
\end{equation}
where we are denoting quantities evaluated on the cut-sky with a tilde. These
pseudo-modes are related to the underlying modes through
\begin{equation}
    \almEtilde \pm i \almBtilde = \sum_{\ell', \, m'} (\almEprime \pm i \almBprime) 
    \Wllmm,
\end{equation}
where $\Wllmm$ is the convolution kernel for our window function, which can be
written as \citep{Brown:2004jn}
\begin{align}
    \Wllmm &= \int \!\! \d \Omega \,
    \stwoYlmprime(\nhatvec) \, \mathcal{W}(\nhatvec) \, \stwoYlm^{*}(\nhatvec).
\end{align}
Expanding the window function in spherical harmonics and evaluating the
integrals yields
\begin{equation}
    \begin{aligned}
        \Wllmm = \sum_{\ell'', \, m''} (-1)^{m}
        \sqrt{\frac{(2\ell + 1)(2 \ell' + 1)(2 \ell^{''} + 1)}{4 \pi}} \,
        \mathcal{W}_{\ell'' \! m''} \\
        \times 
        \begin{pmatrix}
            \ell & \ell' & \ell'' \\
            \pm 2 & \mp 2 & 0
        \end{pmatrix} 
        \begin{pmatrix}
            \ell & \ell' & \ell'' \\
            \pm m & \mp m' & m''
        \end{pmatrix},
    \end{aligned}
\label{eqn:pcl_mixing_matrix}
\end{equation}
where $\mathcal{W}_{\ell'' m''}$ is the spin-0 spherical harmonic
transform of the mask and the terms in brackets are the Wigner-$3j$ symbols.

When forming the pseudo-multipoles, there is a freedom to add pixel weights
through the window function, for example using inverse-variance weighting
scheme, though this is typically not done in large-scale structure applications.

Combining the three shear spectra into a single vector, $\Clvec = \left( \ClEE,
\, \ClEB, \, \ClBB \right)$, we find that the cut-sky power spectrum can be
written in terms of the full-sky power spectrum through
\citep{Brown:2004jn,Hikage:2010sq}
\begin{equation}
    \Cltildevec = \sum_{\ell'} \Mmat \, \mathbfit{C}_{\ell'},
    \label{eqn:PCl_cutskycls}
\end{equation}
where $\Mmat$ is the mode coupling matrix. Provided that $\Mmat$ is invertible,
which is only the case when there is enough sky area such that the two-point
correlation function $C(\theta)$ can be evaluated on all angular scales
\citep{Mortlock:2000zw}, we can invert this relationship to give an estimate
of the full-sky spectra from the pseudo-modes,
\begin{equation}
    \Clvec = \sum_{\ell'} \Mmat^{-1} \, \tilde{\mathbfit{C}}_{\ell'}.
    \label{eqn:PCl_allskycls}
\end{equation}
This is the final expression for the estimated power spectrum of a map using the
Pseudo-$\Cl$ method. An alternative strategy that avoids this inversion is the
forward-modelling of the mask's mode-coupling matrix into the theory power
spectrum values though Equation~\ref{eqn:PCl_cutskycls}.

Since our analysis was focused on the \textit{errors} associated with the
recovered power spectrum, a detailed description of the covariance matrix
associated with the Pseudo-$\Cl$ estimator is worthy of discussion. 
In general, the exact analytic covariance of two Pseudo-$\Cl$ fields involve
terms of the form \citep{Brown:2004jn,Euclid:2021ilj}
\begin{align}
    \mathrm{Cov}\! \left[\tilde{C}_{\ell}, \, \tilde{C}_{\ell'} \right] \sim
    \sum_{\ell_1, \, \ell_2} C_{\ell_1} C_{\ell_2}
    \sum_{\substack{m, \, m', \\ m_1, \, m_2}} 
    W^{\ell \ell_1}_{m m_1} \, 
    \left( W^{\ell' \ell_1}_{m' m_1} \right)^{*} \, 
    W^{\ell' \ell_2}_{m' m_2} \, 
    \left( W^{\ell \ell_2}_{m m_2} \right)^{*}
    \label{eqn:exact_pcl_covariance}
\end{align}
Computing this involves summing $\mathcal{O}(\lmax^{6})$ terms which becomes
computationally intractable for even moderate-resolution maps. Hence, certain
assumptions are used to speed up this calculation. The principle of these is the
narrow-kernel approximation, which assumes that the power spectrum of the mask
has support only over a narrow range of multipoles when compared to the power
spectrum \citep{Efstathiou:2003dj,Garcia-Garcia:2019bku}. This involves making
substitutions of the form $\left\{ C_{\ell_1}, \, C_{\ell_2} \right\}
\rightarrow \left\{ C_{\ell}, \, C_{\ell'} \right\}$, and so the power spectrum
terms in Equation~\ref{eqn:exact_pcl_covariance} can be extracted, and then the
symmetric properties of the convolution kernels can be used to simplify the
summations. This approximation is known to be inaccurate on large scales, though
it has been shown that this has negligible impact on parameter constraints, and
for power spectra that contain $B$-modes \citep{Garcia-Garcia:2019bku}.
Alternatively, Gaussian covariances can be estimated from an ensemble of
realisations. While this produces a more accurate estimate of the covariance
matrix, especially for low multipoles, it is much more computationally demanding
due to the large number of realisations required in the ensemble - especially
for accurately determining the off-diagonal elements of the covariance matrix.

\section{Methodology}
\label{sec:Methodology}

Our aim is to investigate to what extent that QML estimators can give improved
statistical errors on the recovered shear power spectra compared to Pseudo-$\Cl$
methods. We will test the estimators on a set of mock shear maps. In this
section, we describe the fiducial setup of these mocks.

\subsection{Theory power spectrum}
\label{sec:Theory_power_spectrum}

For our analysis, we used a single redshift bin with sources following a 
Gaussian distribution centred at $z=1$ with standard deviation of 
$\sigma_z = 0.15$. We used fiducial cosmological values of 
$h = 0.7$, $\Omega_\mathrm{c} = 0.27$,
$\Omega_\textrm{b} = 0.045$, $\sigma_{8} = 0.75$, $\ns = 0.96$, and massless
neutrinos.

\begin{figure}[t]
    \includegraphics[width=\columnwidth]{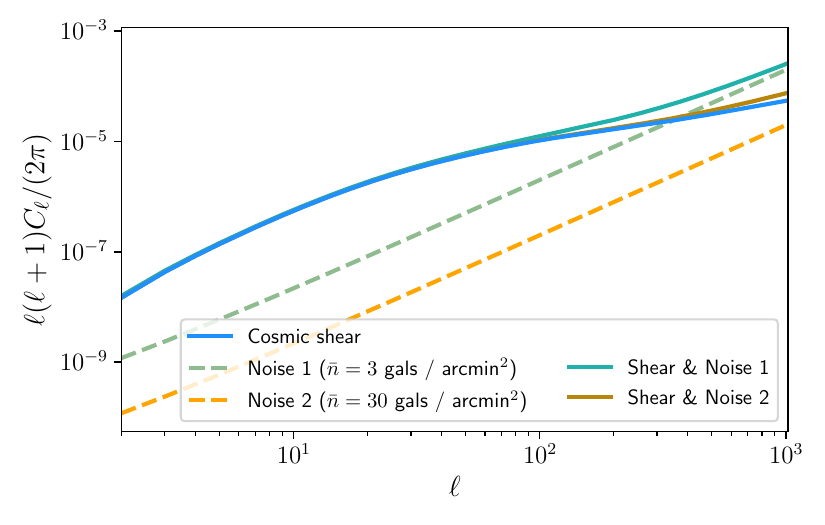}
    \vspace*{-0.75cm}
    \caption{Plot of the fiducial power spectrum values for the cosmic shear
      signal for our single bin of source galaxies (blue curve) that we model as
      following a Gaussian distribution centred at $z=1$ and width $\sigma_z =
      0.15$. We also plot the power spectrum of the shape noise corresponding to
      number densities of $\bar{n} = 3\,\textrm{gals / arcmin}^{2}$ (dashed
      green curve) and $\bar{n} = 30\,\textrm{gals / arcmin}^{2}$ (dashed orange
      curve), and the combined signal and noise spectra (solid green and orange
      curves).}
    \label{fig:FiducialPowerSpectrum}
\end{figure}

The cosmic shear theory signal for this distribution of sources was calculated
using the Core Cosmology Library (\texttt{CCL})
\citep{LSSTDarkEnergyScience:2018yem}. This implements the standard prescription
for the weak lensing power
spectrum~\citep{Bartelmann:1999yn,Bartelmann:2010fz,Kilbinger:2014cea}, where
the convergence power spectrum can be written in natural units where $c=1$ as
\begin{equation}
    C^{\kappa \kappa}_{\ell} = \frac{9}{4} \Omegam^2 \, H_{0}^{4} \! \int_{0}^{\chi_{h}} \!\!\!\!
    \d \chi \, \frac{g(\chi)^2}{a^2(\chi)} \,
    P_\delta \!\left(\! k=\frac{\ell}{f_{K}(\chi)}, \, \chi \! \right)\!,
\end{equation}
where $a(\chi)$ is the scale factor, $P_{\delta}$ is the non-linear matter power
spectrum, $f_{K}$ is the comoving angular diameter distance, and $g(\chi)$ is
the lensing kernel given as
\begin{equation}
    g(\chi) = \int_{\chi}^{\chi_{h}} \!\! \d \chi' \, n(\chi') \frac{f_{K}(\chi' - \chi)}{f_{K}(\chi')},
\end{equation}
where $n(\chi)$ is the number density of source galaxies. The convergence power
spectrum can be transformed into values for the $E$-mode power through
\citep{Hu:2000ee}
\begin{equation}
    \ClEE = \frac{(\ell - 1)(\ell + 2)}{\ell (\ell + 1)} C_{\ell}^{\kappa \kappa}.
\end{equation}
A plot of the $\ClEE$ power spectrum used, including the contribution from shape
noise (described below), is shown in Figure~\ref{fig:FiducialPowerSpectrum}.

Shape noise from the intrinsic ellipticity dispersion of galaxies is an
important factor in cosmic shear analyses. We modelled it as a flat
power-spectrum with value $\Nl$ given as 
\begin{equation}
    \Nl = \frac{\sigma_{\epsilon}^2}{\bar{n}},
    \label{eqn:Noise_power_spectrum}
\end{equation}
where $\sigma_{\epsilon}$ is the standard deviation of the intrinsic galaxy
ellipticity dispersion per component, and $\bar{n}$ is the expected number of
observed galaxies per steradian. For our main analysis, we assume
\textit{Euclid}-like values where it is expected that 30 galaxies per square
arcminute will be observed and divided into ten equally-populated photometric
redshift bins, giving $\bar{n} = 3\,\mathrm{gals /
arcmin}^{2}$~\citep{Euclid:2011zbd}. We investigate the effect of not splitting
the sources into different bins, giving rise to a much lower noise level where
$\bar{n} = 30\,\mathrm{gals / arcmin}^{2}$, in
Section~\ref{sec:Varying_noise_levels}. We take $\sigma_{\epsilon} = 0.21$. 

The shape noise spectrum produces a noise matrix with components given by
\begin{equation}
    N_{ij} = \frac{\sigma_{\epsilon}^2}{n_i} \delta_{ij},
\end{equation}
where $i, j$ are pixel indices, and $n_i$ is the expected number of galaxies in
the $i$-th pixel, which we are assuming is constant and related to $\bar{n}$
through the area of each pixel.

Figure~\ref{fig:FiducialPowerSpectrum} shows the contribution to the total
signal from cosmic shear alone and the shape noise. For the case where we
consider an observed source galaxy density of $\bar{n} = 3\,\textrm{gals /
arcmin}^{2}$, we see three distinct regions: the first is for $\ell \lesssim
200$ where the cosmic shear signal dominates, and thus the uncertainties are
dominated by cosmic variance, the second is an intermediate set of scales where
the cosmic shear and noise have roughly equal amplitude, and the third is for
scales above $\ell \gtrsim 400$ where the noise dominates. Since we consider the
statistics of our estimators up to a maximum multipole of $\lmax = 512$, this
choice of noise level allows us to test the behaviour of our estimators in these
three regions. Hence, we are sensitive to any differences in the statistics that
might arise in the different regimes. For the case where $\bar{n} =
30\,\textrm{gals / arcmin}^{2}$, we see that we are signal-dominated over our
entire multipole range.

\subsection{Survey geometry}

Since much of the comparison between our two power spectrum estimation
techniques will depend on the specific geometry of the sky mask used, we needed
to use a single, generic mask that can be applied consistently to both
estimators to highlight the effects of the estimators only. For our analysis, we
generated a custom mask that would be applicable to a space-based full-sky weak
lensing observatory. This comprises of a main cut that corresponds to the
galactic-plane combined with a slightly narrower cut that corresponds to the
ecliptic-plane. These two features alone capture the majority of features that
are expected for a \textit{Euclid}-like survey and so our simple model for the
mask will yield representative results. 

In a weak lensing analysis, stars that are present in the data need to be masked
out due to their detrimental effects on determining the shapes of the lensed
galaxies. In our analysis, we looked at the effects of our estimators with and
without stars to see if the inclusion of stars makes any meaningful difference
in either the errors or induced mode-coupling of the recovered power spectra.

\begin{figure}
    \includegraphics[width=\columnwidth]{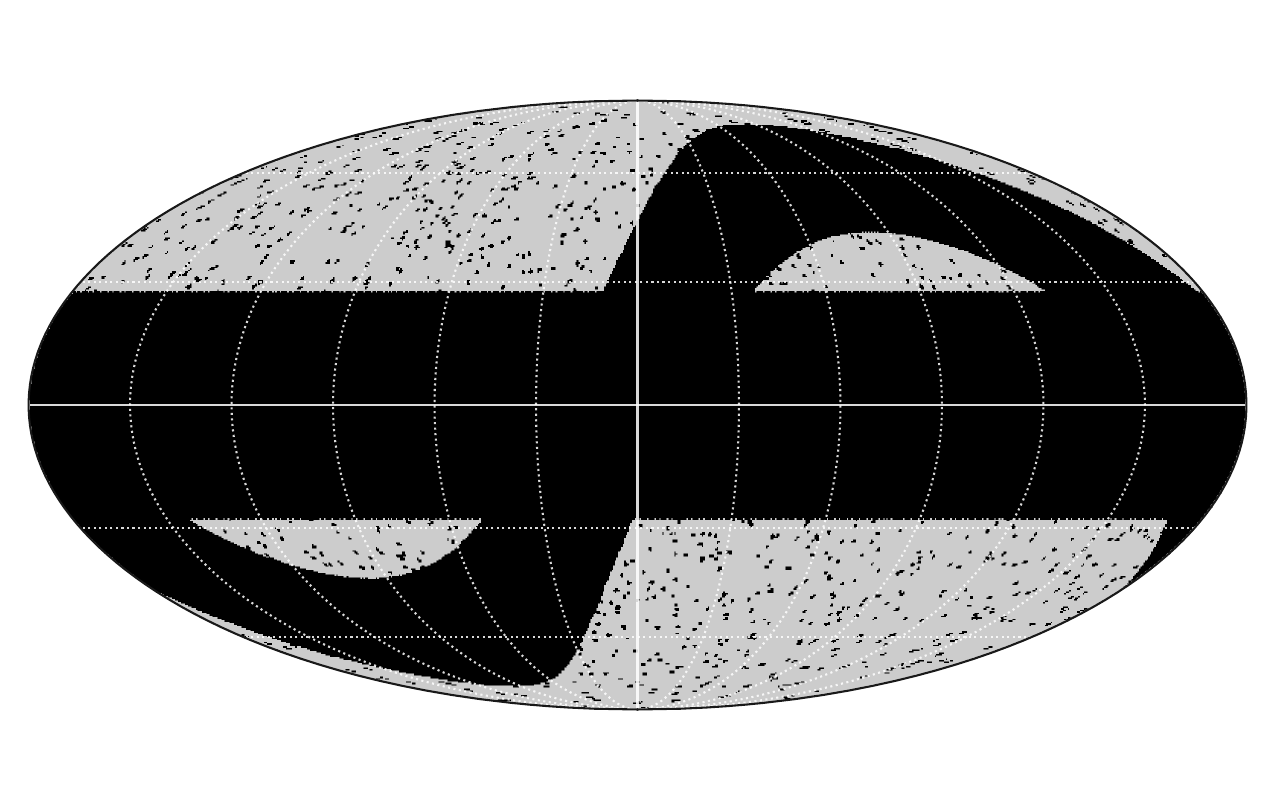}
    \vspace*{-1.0cm}
    \caption{Plot of the sky mask used in our analysis applicable for a
      space-based weak lensing experiment. Here we see the main galactic-cut as
      the thick horizontal band, the ecliptic-cut as the slightly thinner
      sinusoidal band, and our star mask consisting of random circular cut-outs
      that were generated through the prescription of
      Section~\ref{sec:Star_mask_generation}. The galactic- and ecliptic- cuts
      dominate the large-scale behaviour of the mask, whereas the star mask
      introduces strong small-scale effects.}
    \label{fig:SkyMaskWithStars}
\end{figure}

The sky masked used in our analyses is shown in
Figure~\ref{fig:SkyMaskWithStars}.

\subsubsection{Star mask generation}
\label{sec:Star_mask_generation}

We investigated the statistics of our estimators at a map resolution of $\Nside
= 256$. This corresponds to a pixel angular scale of $14 \, \arcmin$. The
prescription described in Ref.~\cite{Martinet:2020mqm} can be followed to generate a
realistic \textit{Euclid}-like star mask. This involves modelling stars as disks
that are distributed randomly on the sky and that have a radii drawn from a
random uniform distribution taking values between $0.29 \, \arcmin$ and
$8.79 \, \arcmin$. For a galaxy survey such as \textit{Euclid}, stars can be
considered point-sources, which are then broadened through the telescope's
optical system point-spread function (PSF) into disks which are then recorded
on its CCD detector, and masked over to remove their effect~\cite{Euclid:2024yrr}. 
Stars can be placed on a map until the desired sky area covered
by stars is reached. However, we note that this distribution of radii of stars
is smaller than the pixel scale for our map resolution, and so a star mask
generated using such values as presented in Ref.~\cite{Martinet:2020mqm} would give
rise to under-sampling in the star mask produced, as all stars would be single
pixels, which was found to induce errors in the Pseudo-$\Cl$ estimator.

Since we are not after an exact realistic distribution of stars in our analysis,
we can instead base our star mask on the distribution of `blinding stars', or
avoidance areas, that are expected to be encountered for a space-based
observatory~\citep{Euclid:2021icp}. These avoidance areas are expected to have
an average area of $0.785\,\mathrm{deg}^2$ and total an area of
$635\,\mathrm{deg}^2$ over the expected survey area. Assuming that these
avoidance areas can be modelled as disks, this corresponds to an average radii
of $30 \, \arcmin$, and so is large enough to cover multiple pixels in our limited
resolution maps. This approximate treatment should capture the main features
brought to the analysis by a more realistic star mask.

We used an edited version of the \texttt{GenStarMask} utility as provided with
the \texttt{Flask}\footnote{\url{https://github.com/ucl-cosmoparticles/flask}}
code \citep{Xavier:2016elr} to generate our star mask using the avoidance area
specification. Our edits were made to draw the radii from a uniform distribution
rather than a log-normal. To add some scatter to our avoidance area mask, we
generated the disks with radii between $25 \, \arcmin$ and $35 \, \arcmin$. To match the
desired total avoidance area, avoidance areas were added until they covered
$5\,\%$ of the full-sky.

\subsubsection{Power spectrum of mask}

\begin{figure}
    \includegraphics[width=\columnwidth]{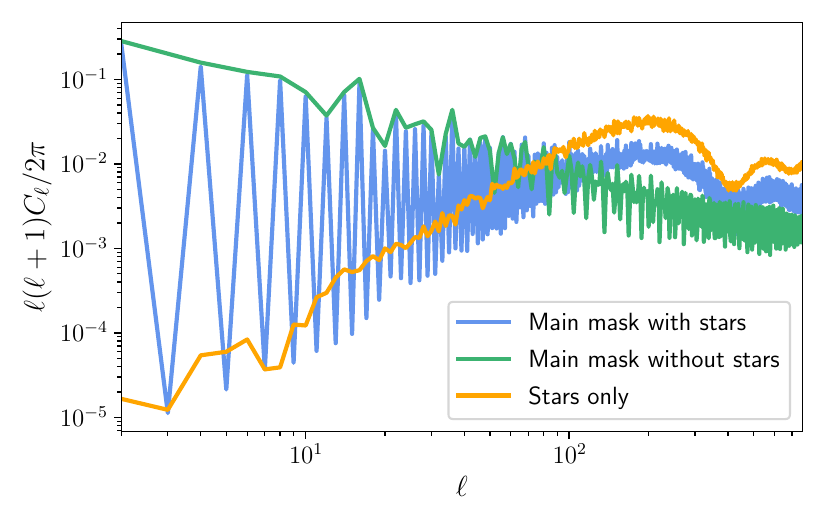}
    \vspace*{-0.75cm}
    \caption{Power spectrum of the sky masks used in our analysis. Note that for
      the `main cuts without stars' curve we plot the even $\ell$-modes only due
      to the very small values that odd $\ell$-modes take, arising from the
      parity of the mask.}
    \label{fig:Mask_power_spectrum}
\end{figure}

Since the exact form of Pseudo-$\Cl$ mixing matrix is highly sensitive to the
power spectrum of the mask used, through Equation~\ref{eqn:pcl_mixing_matrix},
we computed the spherical harmonic transform of
our generated mask. This is shown in Figure~\ref{fig:Mask_power_spectrum}.

We plot the power spectrum for our main galactic- and ecliptic- cuts only, the
two cuts with added star mask, and star mask only. For the case without stars
added, we plot the $\Cl$ values for even $\ell$-modes only. This is due to the
very small values for odd-$\ell$ modes arising from the parity of the mask.
Here, we see that the power spectrum for our main mask with stars added has two
distinct regions: dominated by the two main cuts for $\ell \lesssim 10^{2}$, and
dominated by the star mask above this threshold.

We can understand the primary behaviour of the mask's power spectrum through
computing the analytic power spectrum for a simple mask that is comprised of a
single horizontal cut ranging from $\theta = A$ to $\theta = B$. Doing so, we
find that the $\Cl$ values are given by
\begin{align}
    \Cl = \frac{\pi}{\left(2 \ell + 1\right)^{2}} 
    [P_{\ell + 1}\left(\cos A\right) - P_{\ell - 1}\left(\cos A\right)
    - P_{\ell + 1}\left(\cos B\right) + P_{\ell - 1}\left(\cos B\right)]^2.
    \label{eqn:Mask_Cl_theory_vals}
\end{align}
The full derivation of this result can be found in Appendix~\ref{chp:appendix_A}.
Enforcing that the mask is symmetric around $\theta = \pi / 2$, and using the
parity of the Legendre polynomials, we find that the analytic prediction for the
odd $\ell$-modes are zero. The addition of a second cut of equal width, for
example the ecliptic-cut, keeps the reflective symmetry. While we use a slightly
thinner ecliptic-cut, we still keep this approximate symmetry. Propagating these
suppressed odd-$\ell$ modes into the mixing matrix through
Equation~\ref{eqn:pcl_mixing_matrix} explains  why we see strong coupling in the
covariance matrices between $\Cl$ values that have even-$\ell$ offsets, and
little coupling between odd differences.

We also see that the amplitude of the power spectrum coefficients for the mask
without stars generally decreases at larger multipole values, where the $\Cl$
values roughly scale as $\ell ^2 \Cl \propto 1 / \ell$. This arises from the
large-$\ell$ behaviour of the Legendre polynomials, where they scale as
$P_{\ell} \propto 1 / \sqrt{\ell}$ \citep{Szego1975orthogonal}.

The behaviour of the power spectrum of the star mask can also be broadly split
into two distinct regions. The first is for multipoles $\ell \lesssim 200$ where
the $\Cl$ values are constant, which is a result of the random scatter of the
stars on the sphere resulting in a noise-like signal. The second is for
multipoles larger than $\ell \gtrsim 200$ where the $\Cl$ values start to
oscillate in a sinc-like behaviour, which is where the features of the
individual circular disks dominate.

\subsection[Pseudo-$\Cl$ implementation]{Pseudo-$\bm\Cl$ implementation}
\label{sec:PseudoCl_implementation}

An essential part of our work is the accurate computation of the $\Cl$ 
covariance matrices both for our new QML implementation and its comparison to
results obtained using the Pseudo-$\Cl$ method. Here, we used the 
\texttt{NaMaster}\footnote{\url{https://github.com/LSSTDESC/NaMaster}} code to
produce all estimates for the Pseudo-$\Cl$ method \citep{Alonso:2018jzx}.

In general, the computation of the exact Gaussian covariance is a difficult
problem that has been discussed extensively in previous literature. In our work,
we employed the narrow kernel approximation as presented in Ref.~\cite{Garcia-Garcia:2019bku}
to compute this Gaussian approximation. However, it
should be noted that the narrow kernel approximation overestimates the variances
for the lowest $\ell$ multipoles. Since it is these exact multipoles that we are
most interested in, we instead opt to estimate the covariance from an ensemble
of $5\,000$ maps when investigating the raw variances. However, as the
estimation of the off-diagonal elements of the covariance matrix are highly
sensitive to the number of realisations in the ensemble, when we investigate
parameter constraints that are derived from the $\Cl$ covariance matrix, we use
the `analytic' result as returned from the narrow kernel approximation. In the
limit of using large numbers of realisations in the ensemble, at the cost of
extensive run-time, Ref.~\cite{Garcia-Garcia:2019bku} demonstrated that these two
estimation techniques are consistent.

\section{Results}
\label{sec:Results}

\subsection{Benchmark against existing estimators}
\label{sec:QML_estimtors_benchmark}

\begin{figure}
  \includegraphics[width=\columnwidth]{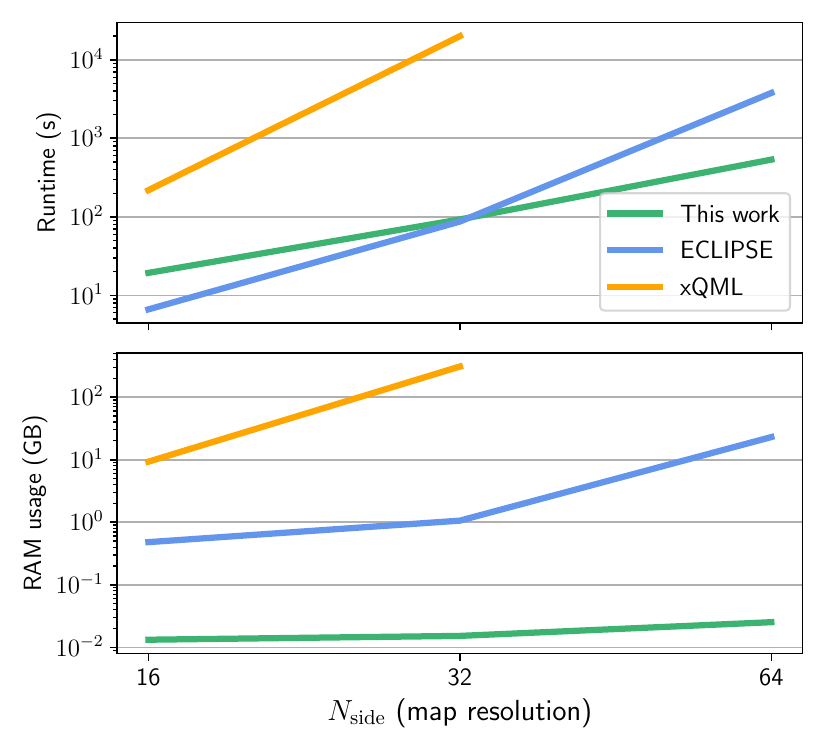}
  \vspace*{-0.5cm}
  \caption{Comparison of RAM usage and run-time for different implementations
  of the QML estimator. We show that our new method has significantly reduced
  RAM usage compared to existing estimators, which is why we can extend our
  method to increased map resolutions that the other methods cannot process.
  Results were obtained using an average over ten maps for our method, and 
  averaged over three runs for each method at each resolution. Computations
  were performed using 32 cores of an Intel Cascade Lake processor.}
  \label{fig:QML_code_comparison}
\end{figure}

In this section, we present the results of a comparative study between our new
QML implementation and the Pseudo-$\Cl$ method. First, we wish to investigate
how using the novel techniques employed by our new estimator impacts the ability
to recover the power spectrum when compared to existing QML implementations. We
compare with two leading public implementations of the QML estimator:
\begin{galitemize}
    \item \texttt{xQML}\footnote{\url{https://gitlab.in2p3.fr/xQML/xQML}} 
    as presented in Ref.~\cite{Vanneste:2018azc}. This is a
    straightforward implementation of the QML method as presented in 
    \cite{Tegmark:1996qt} and \cite{Tegmark:2001zv} that has been generalised
    to cross-correlations between maps. It is written primarily in Python with
    small parts written in C.
    
    \item \texttt{ECLIPSE}\footnote{\url{https://github.com/CosmoTool/ECLIPSE/}}
    as presented in Ref.~\cite{Bilbao-Ahedo:2021jhn}. This is a more numerically
    efficient implementation of the QML estimator compared to the original
    prescription and thus exhibits somewhat better performance scaling with
    resolution over the naive method. It is written in \textsc{Fortran}.
\end{galitemize}

We wish to compare the performance of our new code, written in \Cpp, with these
existing methods. In Figure~\ref{fig:QML_code_comparison}, we present a
comparison for the total run-time and RAM usage for the two codes described
above and our new method described in this work for a range of map resolutions
pushing to the highest $\Nside$ possible with these codes and the computational
resources available to us. Here, we see that while our new code is competitive
in total run-time when compared to \texttt{ECLIPSE}, we see many orders of
magnitude improvement in the total RAM usage for our method over the other two
methods. This is because we never have to explicitly store, invert, and compute
the product of any of the massive $\Npix \times \Npix$ matrices that the other
two methods employ. Since we are only interested in computing the action of the
covariance matrix on a trial pixel-space vector, we keep all of our
working-quantities as $\mathcal{O}(\Npix)$ which clearly have much better RAM
scaling with resolution over the pixel-space matrices. It is this vastly reduced
RAM usage requirement that allows us to push our method to resolutions that are
simply not possible on standard high performance clusters using the two
previously discussed methods. It is important to note that our new
implementation is now only run-time limited, and thus can be pushed to even
higher resolutions than have been considered in this work if more extensive
computing resources are available. The time-limiting steps to our implementation
is the transformations of the trial vector between pixel- and harmonic-space
through the use of the \texttt{HealPix} functions \texttt{alm2map} \&
\texttt{map2alm}. Since both of these functions are implemented using
\texttt{OpenMP} parallelism, faster run-times can be achieved through simply
running our code on higher core count processors.

Our code is publicly available and can be downloaded from GitHub at\\
\href{https://github.com/AlexMaraio/WeakLensingQML}{\texttt{https://github.com/AlexMaraio/WeakLensingQML}
\faicon{github}}.

\subsection{Accuracy of numerical Fisher matrix}
\label{sec:Accuracy_of_numerical_fisher}

\begin{figure}
  \includegraphics[width=\columnwidth]{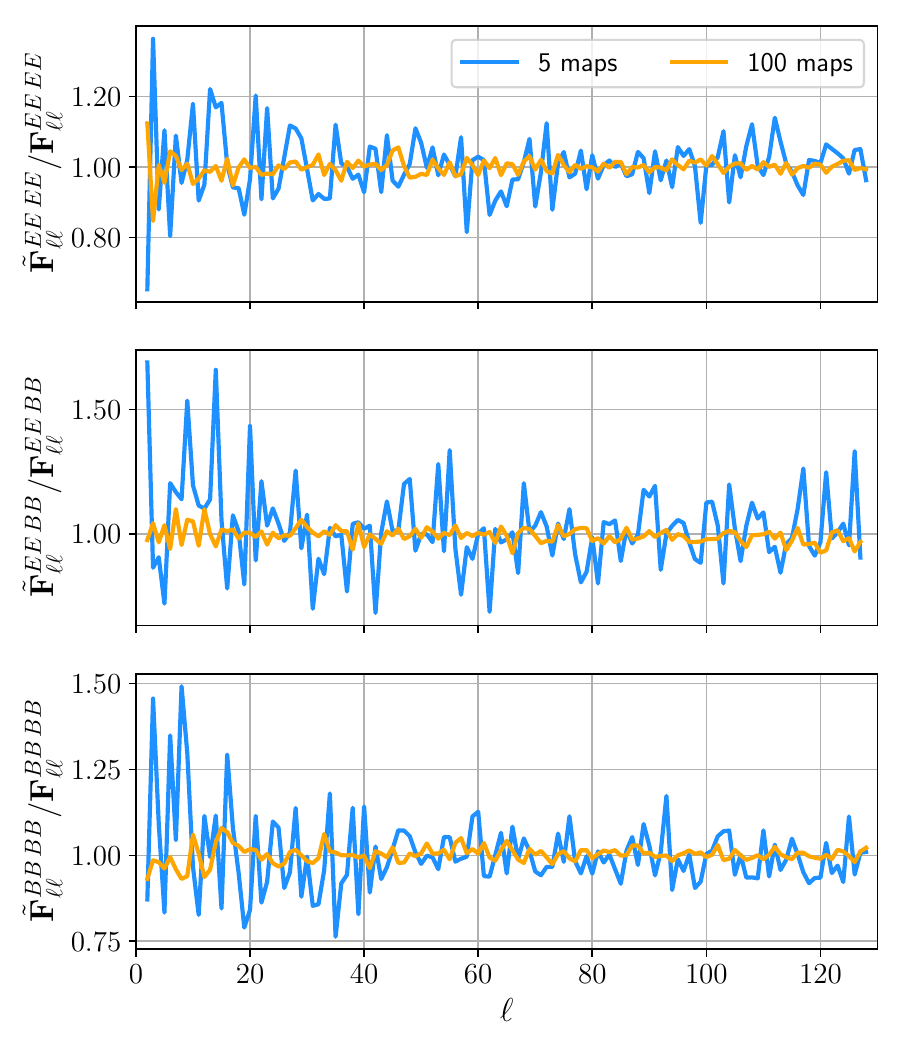}
  \vspace*{-0.75cm}
  \caption{Ratio of our numerical Fisher matrix with respect to the analytic
  result for a map resolution of $\Nside = 64$. Here, we plot the diagonal of
  the $EE$-$EE$, $EE$-$BB$, and $BB$-$BB$ components. All three components
  show good consistency with unity regardless of the number of maps averaged
  over when computing the Fisher matrix through
  Equation~\ref{eqn:FisherMatrixEstimation}, with the amplitude of the scatter
  decreasing with increasing number of maps.}
  \label{fig:Fisher_ratio}
\end{figure}

As explained in Section~\ref{sec:Finite_diff_fisher}, we use finite-differences
to compute the approximate form of the Fisher matrix from a set of $s_{\ell}$
values estimated using conjugate gradient techniques. To validate the accuracy
of these methods we compared our estimates of the Fisher matrix with the
`analytic' result as computed using the formalism presented
in~\cite{Bilbao-Ahedo:2021jhn}. While this is still a `brute-force' QML
implementation, where the covariance matrix still needs to be computed and
stored in full, their method allows many quantities to be expressed in terms of
the spherical harmonic transform matrix $\mathbfss{Y}$ and thus reduce the
computational demands of the estimator. This comparison was performed at a map
resolution of $\Nside = 64$, which was the maximum resolution possible for the
computation of the analytic result. Note that at this resolution, the typical
pixel scale is larger than the angular size of our cut-outs generated for our
star mask, and so this comparison was computed for the case without the star
mask added to the main cuts. The result of this comparison is presented in
Figure~\ref{fig:Fisher_ratio}, where we plot the ratio of the diagonal of the
$EE$-$EE$, $EE$-$BB$, and $BB$-$BB$ components of the Fisher matrix. We plot the
cases for where we average over five and one hundred maps when injecting power
into the generated maps when estimating the Fisher matrix (see
Equation~\ref{eqn:FisherMatrixEstimation}).
Figure~\ref{fig:Numeric_to_analytic_Fisher_ratio_grid} shows this ratio extended
to several of the off-diagonal strips, specifically for the cases with $\Delta
\ell = 2, 8, 32$. Both figures show that while the amplitude of the random
scatter in the ratio decreases significantly when averaging over more maps, both
cases are simply random scatter around unity - and thus our numerically obtained
Fisher matrix is a true representation of the actual Fisher matrix. Propagating
the two different numerical and analytical $\Cl$-Fisher matrices to parameter
constraints using Fisher forecasts shows negligible differences in parameter
contours which again highlights our trust in our new method to estimate the
$\Cl$-Fisher matrix at any resolution. Hence, we are free to use our validated
method to reliably increase the resolution of our implementation beyond what is
possible with current implementations. In our analysis, we averaged over twenty
five random realisations which provided a good compromise between run-time and
numerical accuracy.

\subsection[Comparing $\Cl$ variances of QML to Pseudo-$\Cl$]{Comparing $\bm \Cl$ variances of QML to Pseudo-$\bm\Cl$}

With our new implementation, we can extend the analysis of the properties of the
QML estimator to map resolutions of $\Nside = 256$, which allows us to
accurately recover the power spectrum up to a maximum multipole of $\lmax = 512$.
At this resolution, the storage of the full pixel covariance matrix alone would
require approximately $5\,\mathrm{TB}$ of RAM, which is clearly an unfeasible
requirement for any current computer and thus any analysis at this resolution is
not achievable using current QML implementations. 

We generate maps of the weak lensing shear through producing Gaussian
realisations with the power spectrum described in
Section~\ref{sec:Theory_power_spectrum} (see
Figure~\ref{fig:FiducialPowerSpectrum}). We then add shape noise to these maps
according to Equation~\ref{eqn:Noise_power_spectrum}. We generate twenty five
such maps for each power spectrum multipole that we are injecting power into
($EE$ and $BB$, from $\ell = 2$ to $\ell = 767$) and compute the QML power
spectrum ($EE$ and $BB$) from each. We estimate the QML covariance matrix using
the fact that for Gaussian maps the inverse Fisher matrix is the covariance
matrix. The methods described in Section~\ref{sec:Finite_diff_fisher} were used
to estimate the Fisher matrix, where we average over twenty five maps per
multipole when evaluating Equation~\ref{eqn:FisherMatrixEstimation}. The
Pseudo-$\Cl$ covariance matrix was computed using the methods described in
Section~\ref{sec:PseudoCl_implementation}, and was constructed from an ensemble
of $5\,000$ maps. We do not bin in $\ell$ either of our QML or Pseudo-$\Cl$
estimators, noting that our mask is small enough that the Pseudo-$\Cl$ mixing
matrix is invertible without binning as we are able to reconstruct all modes
that we have generated. We note that our results were robust to different
binning strategies that were applied though not used in our final results.
We remind the reader that we are interested in estimates of the
deconvolved full-sky Pseudo-$\Cl$ values, not forward modelling of the mask into
the original power spectrum. While this deconvolution could introduce additional
sub-optimality over forward modelling through the inversion of the mixing
matrix, for the types of mask that a forthcoming Stage-IV galaxy survey will
typically use this should not be an issue.
Appendix~\ref{app:unbiased_estimators} shows that the means of the two
estimators applied to an ensemble of maps are consistent with the input spectra.
Hence, this demonstrates that both estimators are unbiased in their means, and
thus any differences in their variances are a result of the intrinsic properties
between the two estimators.

\begin{figure}
    \includegraphics[width=\columnwidth]{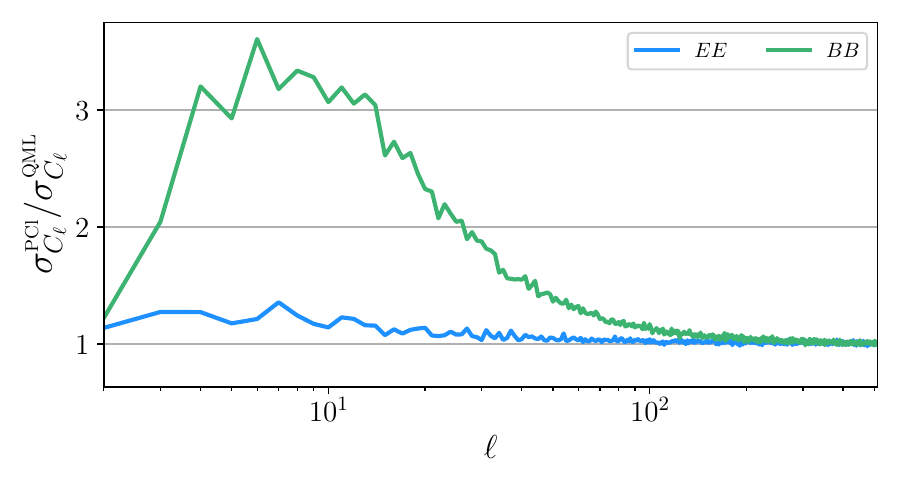}
    \vspace*{-0.75cm}
    \caption{Ratio of standard deviation of the $\Cl$ values (top curve $\ClBB$,
      bottom curve $\ClEE$) obtained using the deconvolved Pseudo-$\Cl$ method
      using \texttt{NaMaster} to those obtained using our new QML
      implementation. We see that the QML estimator provides the largest
      improvements over the Pseudo-$\Cl$ method on the largest angular scales,
      with a very significant improvement for the $B$-modes.}
    \label{fig:Cl_err_ratio}
\end{figure}

In Figure~\ref{fig:Cl_err_ratio} we plot the ratio of the standard deviations
associated with the Pseudo-$\Cl$ estimator with respect to our QML
implementation for the diagonal values associated with the $EE$-$EE$ and
$BB$-$BB$ block of the covariance matrix. Here, we see that the Pseudo-$\Cl$
estimator is sub-optimal to the level of $\sim\!\!20\%$ for $\ell \lesssim 50$
for the $EE$ spectra. This corresponds to an equivalent increase in the survey
area of around $40\%$ on these scales, which is a massive increase in equivalent
area considering that forthcoming Stage-IV surveys are expected to maximise the
possible sky area that is observable for ground- or space-based cosmic shear
surveys~\citep{Euclid:2021icp}. Hence, getting this additional area `for free'
by analysing the data through QML methods demonstrates the advantages of using
such methods and such investigations into their behaviour for cosmic shear
analyses. For the $BB$ spectra, we find that the Pseudo-$\Cl$ estimator produces
errors that are many times that of the optimal QML estimator, peaking at over
three times the standard deviation for the Pseudo-$\Cl$ estimator with respect
to our QML method. This ratio remains significantly above unity for multipoles
that are well above one hundred, which shows that there is a huge advantage to
be gained in $B$-mode precision when using QML methods over the Pseudo-$\Cl$
estimator. We find that the ratio for both sets of spectra decays to unity (with
some random scatter) for larger $\ell$ values. This matches previous QML
studies, which have principally been conducted in the context of the CMB and
ground-based galaxy clustering surveys, which found that the Pseudo-$\Cl$
estimator is close to optimal on small scales and for homogenous noise, and we
find similar results here in the weak lensing
context~\citep{Efstathiou:2003dj,Leistedt:2013gfa}. 

Figure~\ref{fig:Cl_err_ratio} also shows that the statistical precision of the
$B$-mode power spectrum is significantly higher for the QML method compared with
Pseudo-$\Cl$ method. This is relevant because cosmic shear theory predicts zero
$\ClBB$ modes and so any detection of a non-zero $\ClBB$ signal would prompt a
thorough investigation of the data \citep{Kilbinger:2014cea}. Potential sources
of $B$-mode power include residual point-spread function uncertainties,
telescope detector defects, and intrinsic alignments, all of which should be
investigated if non-zero $B$-mode power was found to be statistically
significant. The Pseudo-$\Cl$ estimator is very sub-optimal on large scales, and
so this loss of sensitivity to the $B$-modes can arise from contamination
leakage from the $E$-modes into the $B$-modes due to the nature of the cut-sky.
Since the QML estimator is derived from the likelihood for the maps, which
depends on the input fiducial power spectrum which contains zero $B$-mode power,
any $B$-mode power present in the masked maps must arises from leakage from the
$E$-modes, and thus the estimator can weight the data optimality through the
covariance matrix to minimise the variance from the $E$-modes contributing to
the $B$-modes. For the Pseudo-$\Cl$ estimator, this leakage can be mitigated
through the map-level procedure of $B$-mode purification
\citep{Lewis:2001hp,Smith:2005gi,Grain:2009wq} and has been shown to decrease
dramatically the associated $B$-mode errors, particularly at low $\ell$
multipoles~\citep{Alonso:2018jzx}. However, a requirement for purification to
work is that the mask must be differentiable along its edges. This can be
achieved through the apodisation of the mask which convolves the mask with some
smoothing window function that ensures differentiability. This has most commonly
been applied to cosmic microwave background experiments where their masks are
generally formed of a single much simpler cut applied to the sky
\citep{Planck:2018yye}. This allows apodisation to work effectively on the mask
without significant reduction to $\fsky$. However as previously discussed, a
weak lensing experiment also needs to mask out small regions corresponding to
bright stars or other objects that need removing from the data, or areas that
feature no galaxies due to limitations in depth and/or pixel size. These small
regions cause significant issues with the apodisation process as the convolution
with the smoothing function serves to dramatically increase their apparent area
- producing a significant reduction in $\fsky$. We investigate the effects of
apodising our mask in Appendix~\ref{app:Apodising_mask}. We find that while
apodisation strongly reduces widely separated mode coupling arising from the
suppression in small-scale power of the mask (as now the shape edges from our
star-like disks are smoothed out), this could not offset the significant
reduction in sky area (from $\fsky = 33\,\%$ to $\fsky = 22\,\%$) that
apodisation brought about. This resulted in apodisation providing looser
parameter constraints than for the case without apodisation. A key advantage of
the QML estimator is the natural $E$/$B$-mode separation without a loss in
sky area \citep{Bunn:2016lxi}.



\subsubsection{Varying noise levels}
\label{sec:Varying_noise_levels}

\begin{figure}
    \centering
    \includegraphics[width=\columnwidth]{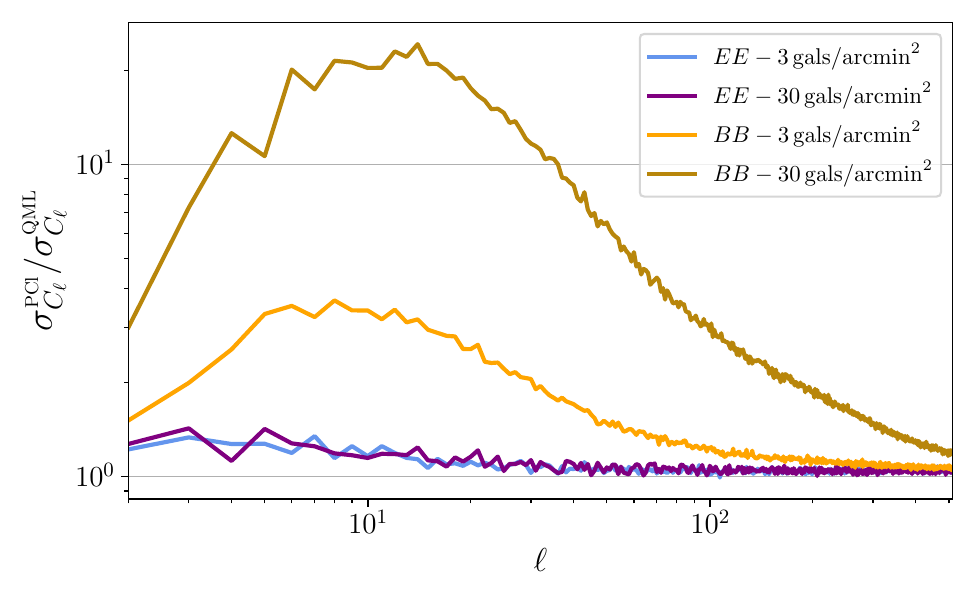}
    \vspace*{-0.75cm}
    \caption{Ratio of the errors on the power spectrum for the Pseudo-$\Cl$
      estimator with respect to the QML estimator for the case of two different
      noise levels of $\bar{n} = 3\,\textrm{gals / arcmin}^{2}$ and $\bar{n} =
      30\,\textrm{gals / arcmin}^{2}$. Here, we see the decreased noise level
      for the curves for the case of $\bar{n} = 3\,\textrm{gals / arcmin}^{2}$
      has negligible effect on the errors associated with the $EE$-spectra,
      whereas there is a large increase in the ratio for the $BB$-spectra.}
    \label{fig:PCl_to_QML_stdev_ratio_noise}
\end{figure}

\begin{figure}
    \includegraphics[width=\columnwidth, trim={0cm 0cm 0cm 0cm}, clip]{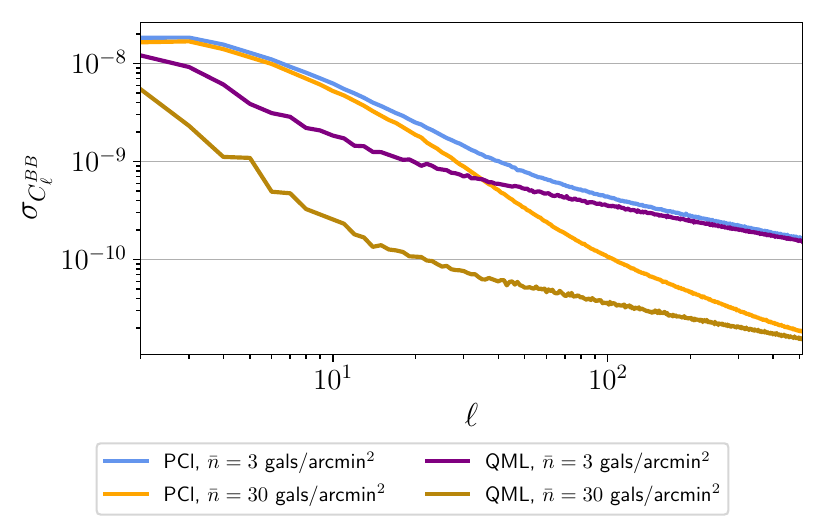}
    \vspace*{-0.6cm}
    \caption{Errors on the $BB$-spectra for the two different noise levels
      considered. Here, we see that the errors for the Pseudo-$\Cl$ estimator
      remain relatively unchanged for the lowest multipoles, whereas the errors
      for the QML estimator decrease dramatically when the amplitude of the
      noise is reduced. }
    \label{fig:Cl_std_BB_noise}
\end{figure}

Thus far, we have assumed a noise level corresponding to an experiment that has
observed thirty galaxies equally divided into ten redshift bins, giving $\bar{n}
= 3\,\mathrm{gals / arcmin}^{2}$. Here, we wish to investigate the statistics
of our estimators for the case where the observed galaxies are combined into a
single redshift bin, giving a much lower noise level of $\bar{n} =
30\,\mathrm{gals / arcmin}^{2}$. Performing this comparison produced results as
shown in Figure~\ref{fig:PCl_to_QML_stdev_ratio_noise}. Here, we see that while
there is negligible differences in the relative statistics between the
$E$-modes, there was a large increase in ratio between our two estimators for
the $B$-modes. We can investigate this large difference in the $B$-mode ratios
by plotting the raw errors of the $B$-mode spectra for our two estimators for
the two noise cases, which is shown in Figure~\ref{fig:Cl_std_BB_noise}.

Here, we see that for low multipoles there is a large difference in the errors
for the QML estimator between the two noise levels whereas the errors for the
Pseudo-$\Cl$ estimator remains relatively unchanged. At this decreased noise
level, the noise is subdominant to the cosmic variance from the $E$-modes on
large-scales. Hence, the QML estimator can vastly outperform the Pseudo-$\Cl$
method on these large-scales because the QML likelihood can efficiently minimise
the cosmic variance from the $E$-modes leaking into the $B$-modes through the
cut-sky. The increased noise level corresponds to a genuine increase in $B$-mode
power which the QML estimator cannot suppress as efficiently as the $E$-mode
cosmic variance leakage. Hence, we see an associated decrease in the ratio
between its errors and that of the Pseudo-$\Cl$ estimator.

\subsection{Cosmological parameter inference}
\label{sec:Parameter_contours}

\begin{figure}
    \includegraphics[width=\columnwidth, trim={0cm 0cm 0cm 0cm}, clip]{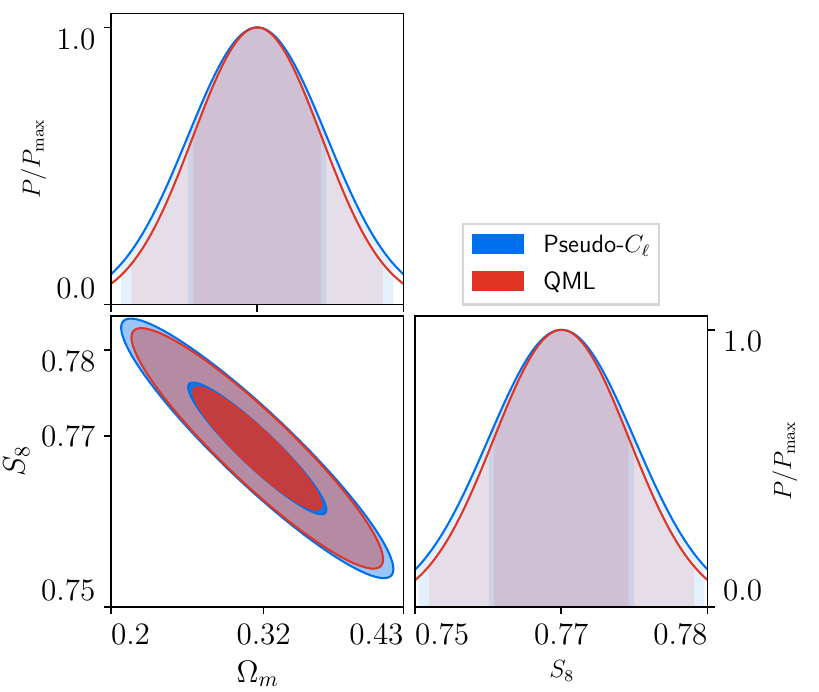}
    \vspace*{-0.5cm}
    \caption{Parameter constrains on $\Seight$ and $\Omegam$ obtained from a Fisher
      matrix analysis up to a maximum multipole of $\lmax = 512$ for our two
      estimators. Here, we see that the increased errors associated with the
      Pseudo-$\Cl$ method propagate through to slightly broadened parameter
      contours.}
    \label{fig:Param_comparison}
\end{figure}

A Fisher matrix forecast was used to propagate our estimated $\Cl$ covariance
matrices into parameter constraints. For an arbitrary set of cosmological
parameters $\vartheta_{\alpha}$ and $\vartheta_{\beta}$, the corresponding
Fisher matrix element is given by \citep{Tegmark:1996bz}
\begin{equation}
    \mathcal{F}_{\alpha \beta} = 
    \sum_{\ell, \, \ell'}
    \frac{\partial C_{\ell}}{\partial \vartheta_{\alpha}} \, 
    \mathbfss{C}^{-1}_{\ell \ell'} \, 
    \frac{\partial C_{\ell'}}{\partial \vartheta_{\beta}},
\end{equation}
where $\mathbfss{C}$ is the $\Cl$ covariance matrix. In our analysis, we focused
on the two parameters that cosmic shear places the tightest constraints on: the
clustering amplitude $\Seight$ and the total matter density $\Omegam$.
Figure~\ref{fig:Param_comparison} shows a comparison of the derived constraints
for these two parameters between our two estimators. Here, we see the effect of
the slightly increased errors associated with the Pseudo-$\Cl$ estimator have
propagated into slightly increased contours for these two parameters when
compared to the QML estimator's contours. This result could have been
anticipated from Figure~\ref{fig:Cl_err_ratio}, since most of the information on
these parameters originates from small scales where the ratio of the errors
approaches unity. In contrast, parameters affecting large angular scales, such
as local primordial non-Gaussianity, are expected to benefit substantially from
an optimal method.

\begin{figure}
    \includegraphics[width=\columnwidth]{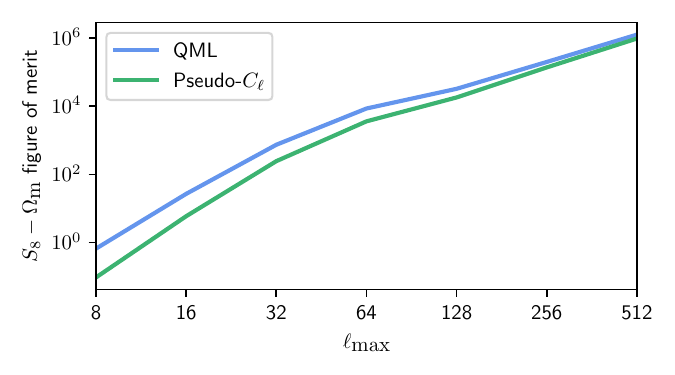}
    \vspace*{-0.75cm}
    \caption{Values for the figure of merit for the combination of $\Seight$ and
      $\Omegam$ as a function of the maximum $\ell$ multipole used in the
      analysis. We see that as the maximum multipole increases, the relative
      sub-optimality of the Pseudo-$\Cl$ estimator decreases and results from
      the two estimators converge.}
    \label{fig:figure_of_merit_vs_lmax}
\end{figure}

The figure of merit, which quantifies how well constrained parameters are, is
related to the Fisher matrix through~\citep{Euclid:2019clj} 
\begin{equation}
    \mathrm{FoM}_{\Seight \Omegam} = \sqrt{\mathrm{det}\left(\mathcal{F}\right)}.
\end{equation}
A plot of the figure of merit for the combination of $\Omegam$ and $\Seight$ as
a function of maximum multipole is shown in
Figure~\ref{fig:figure_of_merit_vs_lmax}. Here, we see that the sub-optimality
of the Pseudo-$\Cl$ method is most apparent when we are limited to low
multipoles. As the maximum multipole increases we see that the figures of merits
converge, however showing that the QML method consistently out performs the
Pseudo-$\Cl$ method. 

The application of Fisher forecasting to predict parameter constraints from the
covariance matrix is done under the assumption that the $\Cl$ values recovered
from the estimators can be described by a Gaussian likelihood. While it has been
shown that for the full sky case, an analytic calculation of the likelihood of
the power spectra can be computed \citep{Hall:2022das}, which can be accurately
modelled as a Gaussian on small scales, the exact likelihood of the recovered
power spectrum using either the QML or Pseudo-$\Cl$ estimators is still unknown.
Previous works have simply used the Gaussian approximation citing the central
limit theorem \citep{Seljak:2017rmr}, though the exact likelihood on large
scales remains unknowns for both the QML and Pseudo-$\Cl$ estimators.

\subsubsection{Inclusion of stars}

Our results presented thus far have been all for the case where we have applied
a star mask to a large-scale mask featuring ecliptic and galactic cuts, as shown
in Figure~\ref{fig:SkyMaskWithStars}. Here, we wish to investigate the detailed
effects on the covariances of our estimators when we apply the star mask to our
main two cuts. All results here are presented for the case without any mask
apodisation applied.

\begin{figure}
    \includegraphics[width=\columnwidth]{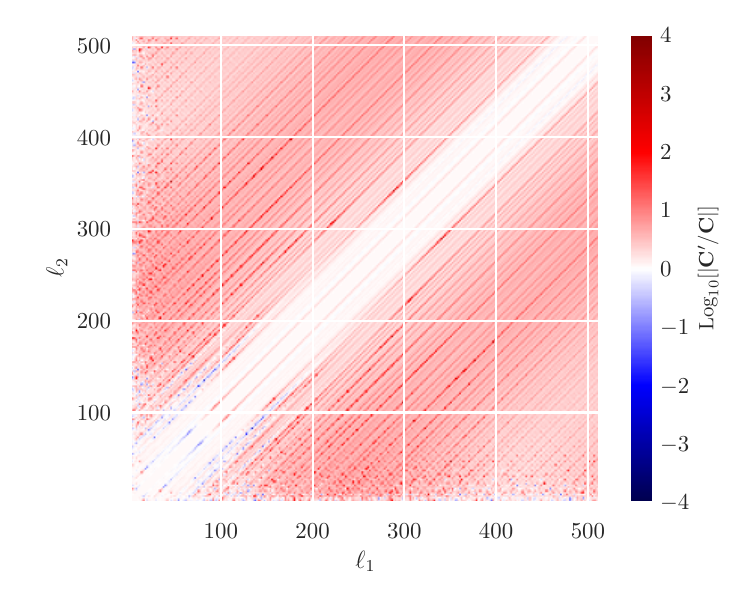}
    \vspace*{-0.5cm}
    \caption{Ratio of the analytic Pseudo-$\Cl$ covariance matrix for the
      $\ClEE$ power spectrum for the cases with ($\mathbfss{C}^{\prime}$) and
      without stars ($\mathbfss{C}$) applied to the main mask. Note that we only
      plot the covariance matrix for even-$\ell$ values only (due to the very
      small values for the odd-$\ell$ case and so their ratios are dominated
      by numerical noise).}
    \label{fig:PCl_cov_star_ratio}
\end{figure}

The ratio of the Pseudo-$\Cl$ covariance matrix for the cases with and without
stars is presented in Figure~\ref{fig:PCl_cov_star_ratio}. Here, we see that the
primary effect of the addition of stars into the mask is to increase the
correlation between widely separated $\ell$-modes, while leaving the values close to
the diagonal in the covariance matrix relatively unchanged.

\begin{figure}
    \includegraphics[width=\columnwidth, trim={0cm 0cm 0cm 0cm}, clip]{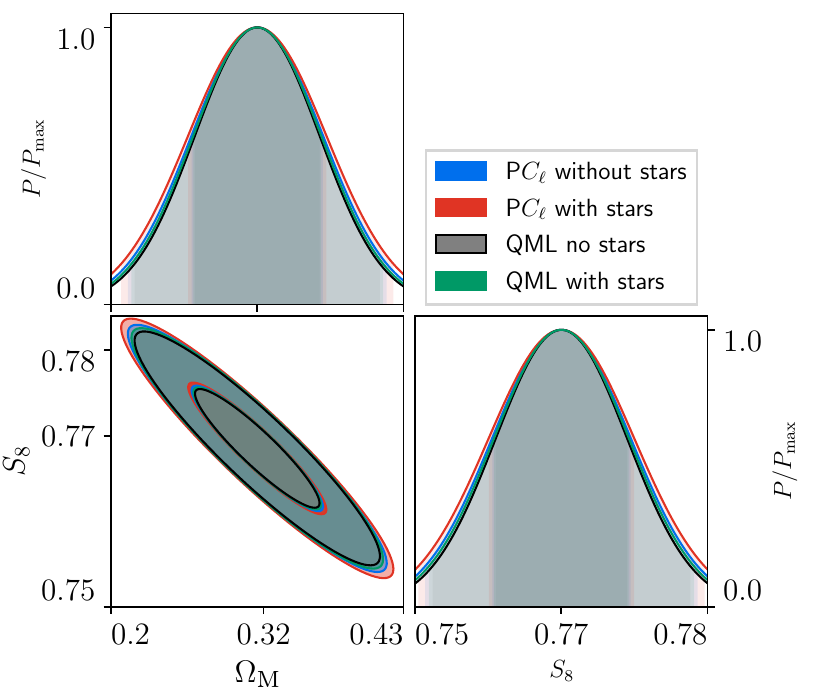}
    \vspace*{-0.5cm}
    \caption{Fisher parameter constraints comparison between QML and
      Pseudo-$\Cl$, where both estimators have a maximum multipole of $\lmax =
      512$, for the cases with and without the star mask applied to both
      estimators. Here we see that the Pseudo-$\Cl$ estimator is more sensitive
      to the inclusion of the star mask through the relative increase in
      parameter contours when compared to the QML contours.}
    \label{fig:Parameter_constraints_stars}
\end{figure}

This new covariance matrix can then be propagated into parameter contours to see
if these increased long-range correlations (which fiducially have very small
values) have any meaningful effect on cosmological parameter constraints. This
is shown in Figure~\ref{fig:Parameter_constraints_stars}. Here, we see that
there are negligible differences on the parameter contours between the two cases
for our QML estimator, however there is a slight broadening in the contours for
the Pseudo-$\Cl$ estimator which is consistent with the loss of sky area to the
star mask. This shows that the Pseudo-$\Cl$ estimator is more sensitive to the
addition of a star mask than the QML estimator, further highlighting the
benefits of the QML method. 

\subsection{Non-Gaussian maps}
\label{sec:Non_gaussian_maps}

Throughout our paper, we have been applying our estimator to Gaussian
realisations of the cosmic shear field. However, as it has been shown that the
convergence field $\kappa$ is more accurately described by a log-normal
distribution \citep{Taruya:2002vy,Hilbert:2011xq}, an investigation of how our
estimators perform when applied to these non-Gaussian maps was undertaken. This
is because as the QML estimator assumes that the underlying power spectrum
coefficients follow a Gaussian distribution, any non-Gaussianities present
in the shear field could induce non-optimality into the recovered power spectra
which would increase errors.

\begin{figure}
    \includegraphics[width=\columnwidth, trim={0cm 0cm 0cm 0cm}, clip]{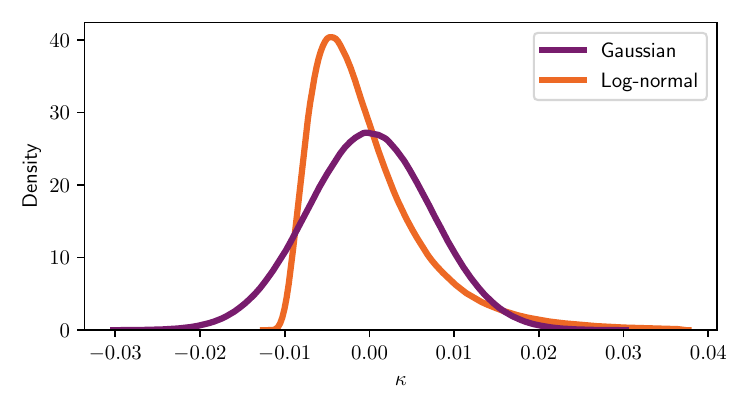}
    \vspace*{-0.5cm}
    \caption{Histogram showing the field values of a Gaussian and log-normal 
      realisation of the convergence field $\kappa$ for the same underlying
      power spectrum}
    \label{fig:kappa_histogram}
\end{figure}

The \texttt{Flask} software package was used to generate log-normal
maps~\citep{Xavier:2016elr}. The `shift parameter' that corresponds to the
minimum value of the convergence field required by \texttt{Flask} was set to
0.01214, following \cite{Hall:2022das}. The log-normal maps were generated at a
resolution of $\Nside = 1024$ and then downgraded to a resolution of $\Nside =
256$ as required to be processed through our estimators. A histogram showing the
distribution of $\kappa$ values for a Gaussian and log-normal realisation is
shown in Figure~\ref{fig:kappa_histogram}. This log-normal convergence field is
then propagated to a slightly modified shear field, which we can apply the
estimators to.

\begin{figure}
    \includegraphics[width=\columnwidth, trim={0cm 0cm 0cm 0cm}, clip]{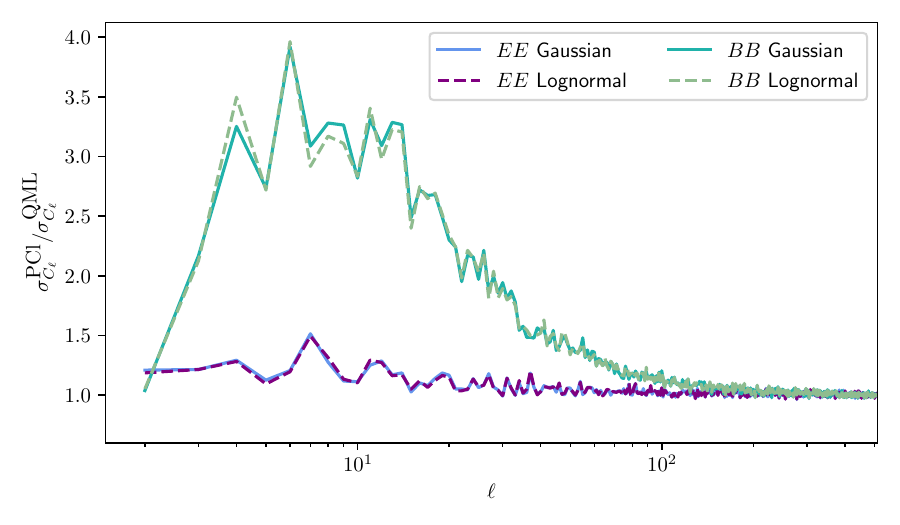}
    \vspace*{-0.5cm}
    \caption{Ratio of the power spectrum errors for the Pseudo-$\Cl$ method with
      respect to the QML estimator applied to both Gaussian and log-normal
      realisations. Here, we see that the results are indistinguishable between
      the two distributions. }
    \label{fig:Cl_std_ratio_nongaussian}
\end{figure}

Since the covariance of the QML estimator is no longer given by the inverse
Fisher matrix, we have to obtain estimates for the QML errors from an ensemble
of numerical realisations. The results of applying the QML and Pseudo-$\Cl$
estimators to an ensemble of $2\,500$ Gaussian and log-normal realisations is
shown in Figure~\ref{fig:Cl_std_ratio_nongaussian}. Here, we see that the ratio
of the power spectrum errors for the two distributions are virtually identical
which demonstrates that the relative behaviour of our estimators remains
unchanged even when applied to maps that the underlying likelihood does not
fully describe. Since the largest differences between the estimators for our
setup occur on the largest scales which is where the effects of the
non-Gaussianity is weak, this is not surprising.

\section{Conclusions}
\label{sec:Conclusions}

We have presented a new implementation of the optimal quadratic maximum
likelihood estimator that is the most efficient publicly available code of its
type. Using our new estimator, we have compared the statistical properties of
the expected power spectrum for forthcoming Stage-IV weak lensing surveys, using
realistic survey conditions, between our QML implementation and an existing
Pseudo-$\Cl$ code. We found that the sub-optimality of the Pseudo-$\Cl$
estimator resulted in marginally increased statistical errors for the $E$-mode
power spectra propagating to increased parameter contours when using Fisher
forecasting. In addition, we found a significant increase in the precision for
the $B$-mode power spectra when applying our QML estimator over the Pseudo-$\Cl$
method, which raises the hopes of being able to further constrain new $B$-mode
physics using forthcoming surveys. Our results show that the application of QML
methods to cosmic shear data provides a useful cross-check to existing methods,
and could have many interesting applications for the constraints of new $B$-mode
physics.

Our new estimator could be extended in numerous ways, for example the QML method
can be easily applied to spin-0 fields such as photometric galaxy clustering.
Since our estimator yield the best improvements on the largest physical scales,
scales at which primordial non-Gaussianity has the largest effects on the
observed signal, the use of our new estimator could enable tighter constraints
on primordial non-Gaussianity. In addition, photometric galaxy clustering data
can be combined with cosmic shear to form a combined $3 \times 2$-point
investigation which our estimator could be applied to. We leave these
applications of our estimator to future work. We also note that our estimator
could also be applied to thermal Sunyaev-Zeldovich data, which is a powerful
probe of cosmology at relatively low multipoles ($\ell \lesssim
10^{3}$)~\citep{Horowitz:2016dwk,Bolliet:2017lha}. This overlaps with the
multipole region where QML provides the best improvements over the Pseudo-$\Cl$
estimator, and so could provide sufficiently tighter constraints when applied to
these data-sets.

Since QML methods deal with data at the pixel-level, they are well suited for
dealing with contaminants and effects that can only be described accurately in
terms of pixels in the maps. One such problem is the effect of spatially varying
noise over survey area, which could arise from different seeing conditions
encountered as a telescope surveys the sky or the properties of the detector
evolving as data is taken or even from the slight varying of diffuse foreground
light levels which are aimed to be minimised for any cosmic shear survey but
may still nevertheless contaminate some pointings. These pixel-level effects
can easily be incorporated into the QML
estimator through an appropriate modification of the noise matrix
$\mathbfss{N}$, whereas the Pseudo-$\Cl$ method utilises Fourier transforms and
these pixel-level effects get diffused over a wide $\ell$ range and thus become
harder to model. This could further reduce the optimality of the Pseudo-$\Cl$
estimator. We leave a dedicated investigation of how such effects affect
the two estimators to future work.

To conclude, we have shown that the Pseudo-$\Cl$ estimator is close to optimal
on small scales for a simplified \textit{Euclid}-like weak lensing survey.
Despite this, the QML estimator is better suited for a variety of applications,
including $E$/$B$-mode separation, complex noise patterns, and complicated
survey geometries. With systematics expected to dominate the error budget of
upcoming surveys it is increasingly important to demonstrate the consistency of
results derived from different analysis pipelines - the fast, publicly available
implementation of the QML estimator that we have presented in this work
represents a significant step forward in this regard.

\section*{Data availability}

All data presented in this work has been generated by the authors. The code to
do so can be found on our GitHub repository located at \\
\href{https://github.com/AlexMaraio/WeakLensingQML}{\texttt{https://github.com/AlexMaraio/WeakLensingQML}
\faicon{github}}.

\section*{Acknowledgements}

AM would like to thank all members of Lensing Coffee at the IfA for many useful
conversations and invaluable support. AH thanks Uro\v s Seljak for useful
discussion. AH and AT are supported by a Science and Technology Facilities
Council (STFC) Consolidated Grant. For the purpose of open access, the author
has applied a Creative Commons Attribution (CC BY) licence to any Author
Accepted Manuscript version arising from this submission.



\section*{Appendices}

\begin{subappendices}

\section{Demonstration of unbiased estimators}
\label{app:unbiased_estimators}

\begin{figure}
    \centering
    \includegraphics[width=\columnwidth]{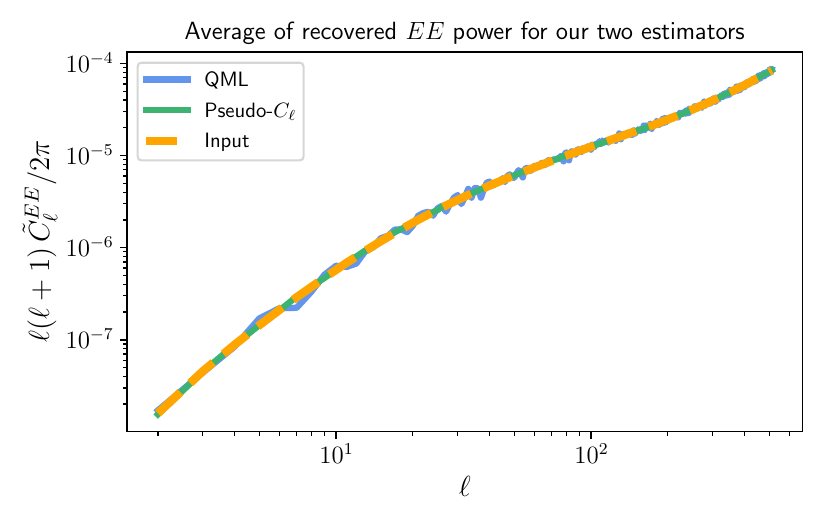}
    \vspace*{-0.55cm}
    \caption{Comparison of the average $\ClEE$ spectrum values for the QML and
      Pseudo-$\Cl$ estimators recovered from an ensemble of five thousand random
      realisations and the fiducial input spectrum.}
    \label{fig:Cl_EE_avg}
\end{figure}

\begin{figure}
    \centering
    \includegraphics[width=\columnwidth]{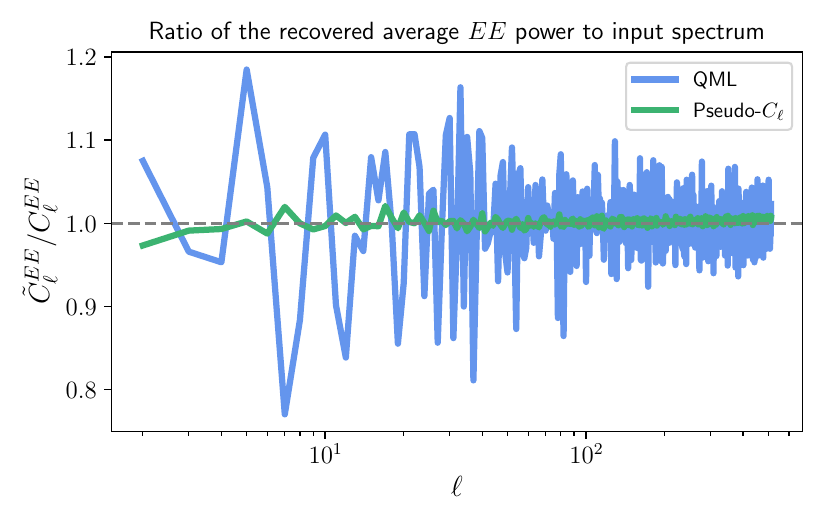}
    \vspace*{-0.55cm}
    \caption{Ratio of the average $\ClEE$ spectrum values for the QML and
      Pseudo-$\Cl$ estimators to the fiducial input spectrum. This shows that
      both estimators correctly recover the input spectrum on average as both
      estimator's averages exhibit roughly random scatter around the $y=1$ line
      (dashed grey line), and thus both estimators are unbiased estimators of
      the power spectrum.}
    \label{fig:Cl_EE_avg_ratio}
\end{figure}

\begin{figure}
    \centering
    \includegraphics[width=\columnwidth]{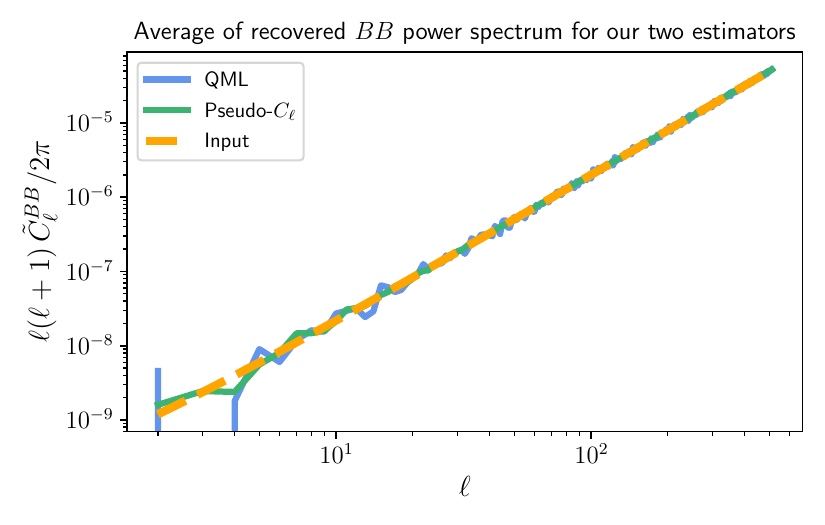}
    \vspace*{-0.55cm}
    \caption{Similar plot to Figure~\ref{fig:Cl_EE_avg}, but now for the 
      $BB$-spectrum.}
    \label{fig:Cl_BB_avg}
\end{figure}

\begin{figure}
    \centering
    \includegraphics[width=\columnwidth]{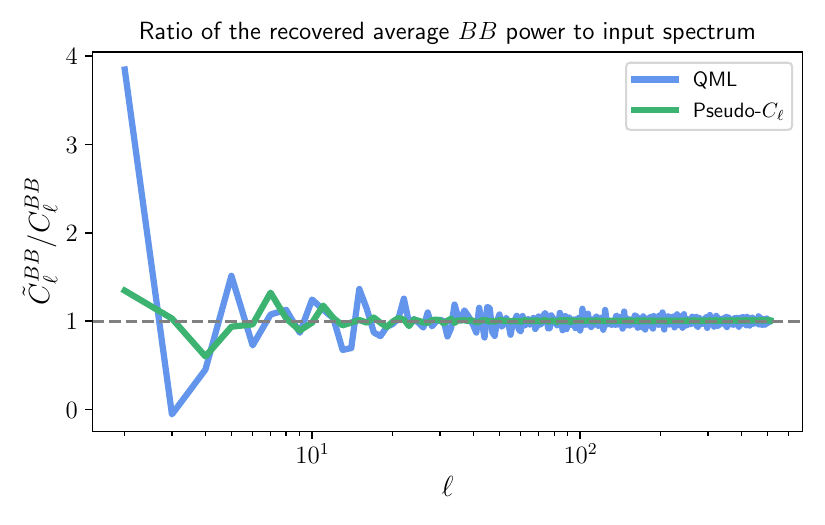}
    \vspace*{-0.55cm}
    \caption{Similar plot to Figure~\ref{fig:Cl_EE_avg_ratio}, but again now 
      for the $BB$-spectrum}
    \label{fig:Cl_BB_avg_ratio}
\end{figure}

The core feature of any power spectrum estimator is that the mean of
the recovered spectrum matches the expected values for a known input spectrum.
Here, we demonstrate that our new QML implementation correctly recovers the mean
of our input spectrum and that it is also consistent with the Pseudo-$\Cl$ mean.
Hence, any differences in their variances will not be due to a difference in
scaling between the two estimators. Figures \ref{fig:Cl_EE_avg} and
\ref{fig:Cl_BB_avg} show the average $EE$- and $BB$-spectra for an ensemble of
five thousand random realisations of the same input spectra, with the ratios of
the average recovered spectrum to the input spectrum shown in
Figures~\ref{fig:Cl_EE_avg_ratio} and~\ref{fig:Cl_BB_avg_ratio}. Here, we see
that both estimators accurately recover the input values over the entire $\ell$
range and thus our new QML estimator is unbiased and the differences in variance
between it and the Pseudo-$\Cl$ estimator is intrinsic to the method. However,
we note that Figures~\ref{fig:Cl_EE_avg_ratio} and~\ref{fig:Cl_BB_avg_ratio}
show that the average values produced for the QML estimator are significantly
noisier than those for the Pseudo-$\Cl$ estimator. This is a direct result of
the use of a numerical estimate of the Fisher matrix instead of its exact
analytic calculation. Through Figures~\ref{fig:Fisher_ratio}
and~\ref{fig:Numeric_to_analytic_Fisher_ratio_grid}, we have shown that our
numerical estimation method recovers the analytic calculation of the Fisher
matrix on average, with the level of noise determined by the number of maps that
have been averaged over (see Equation~\ref{eqn:FisherMatrixEstimation}). Hence,
the accuracy of the multipole estimates can be increased by simply computing the
Fisher matrix with an increased number of maps.  

\section{Ratio of numeric to analytic Fisher}
\label{app:Full_cl_fisher_ratio}

\begin{figure}
    \centering
    \includegraphics[width=0.85\textwidth]{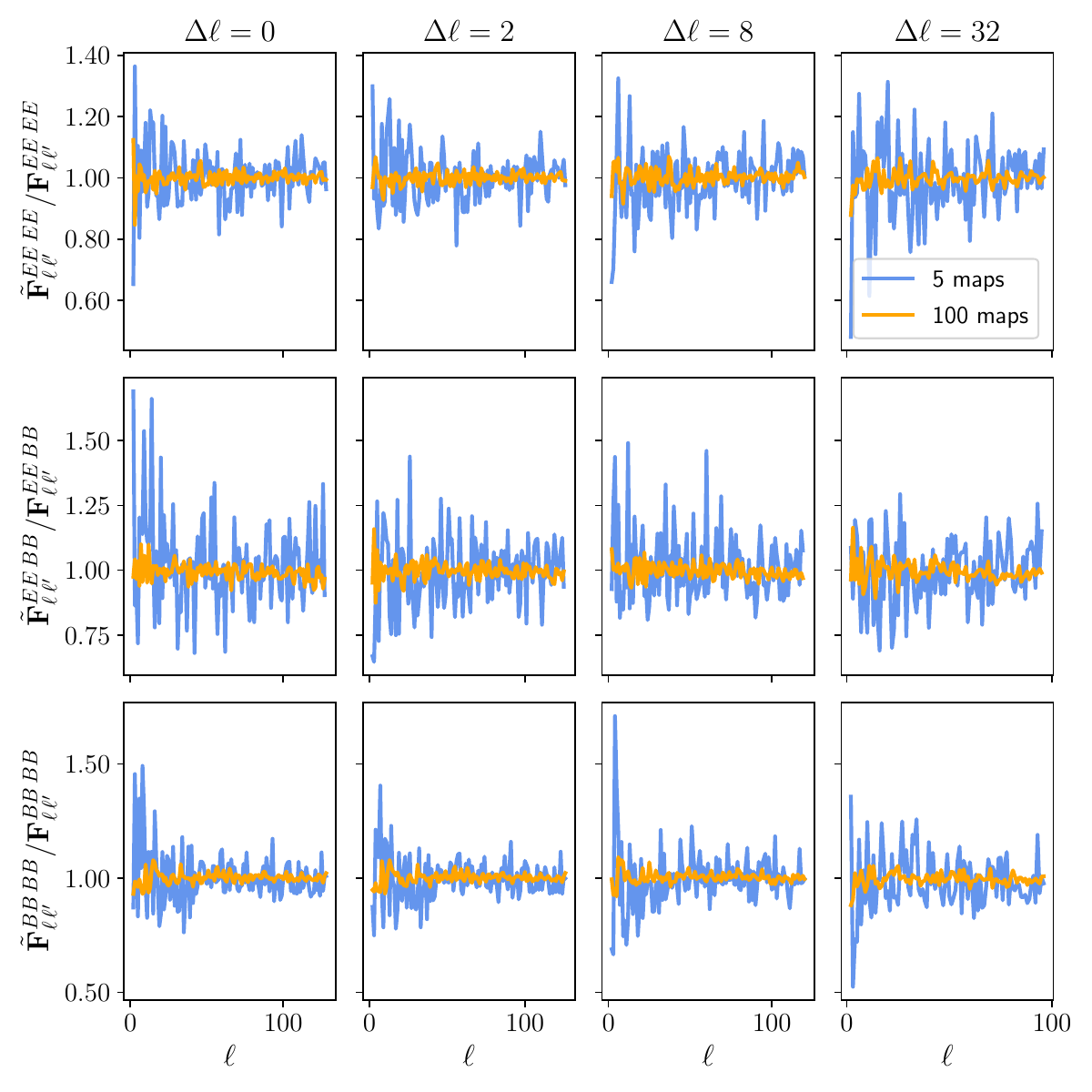}
    \caption{Ratio of our numerically-derived $\Cl$-Fisher matrix to the
    analytic result for a map resolution of $\Nside = 64$, presented for the
    cases where we average over five and one hundred maps. Here, we plot the
    $EE$-$EE$, $EE$-$BB$, and $BB$-$BB$ components separately, with varying
    off-sets from the diagonal in the different columns. We see good agreement
    between our estimator and existing results for all combination of spectra
    and off-sets, and so deduce that our numerical estimate is consistent
    with the analytic result.}
    \label{fig:Numeric_to_analytic_Fisher_ratio_grid}
\end{figure}

Figure \ref{fig:Numeric_to_analytic_Fisher_ratio_grid} shows the ratio of our
numerically computed $\Cl$ Fisher matrix to that computed using analytic methods
for a number of different off-set values from the diagonal. Here, we see that
all curves simply exhibit random scatter around unity which shows that our
numerical estimates of the Fisher matrix is an unbiased estimate of the true
values. The residual noise in the Fisher matrix gives rise to negligible 
differences in parameter confidence contours.

\clearpage
\section{Sensitivity to apodisation}
\label{app:Apodising_mask}

\begin{figure}
    \includegraphics[width=\columnwidth]{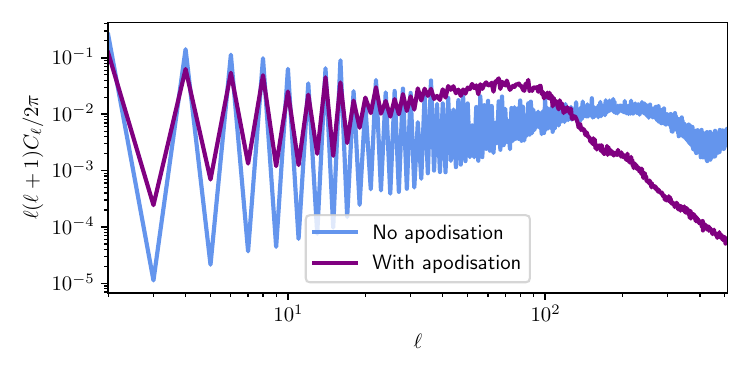}
    \vspace*{-0.5cm}
    \caption{Power spectrum of our mask, with stars included, for cases with and
      without apodisation applied.}
    \label{fig:Mask_power_spectrum_apodisation}
\end{figure}

\begin{figure}
    \includegraphics[width=\columnwidth]{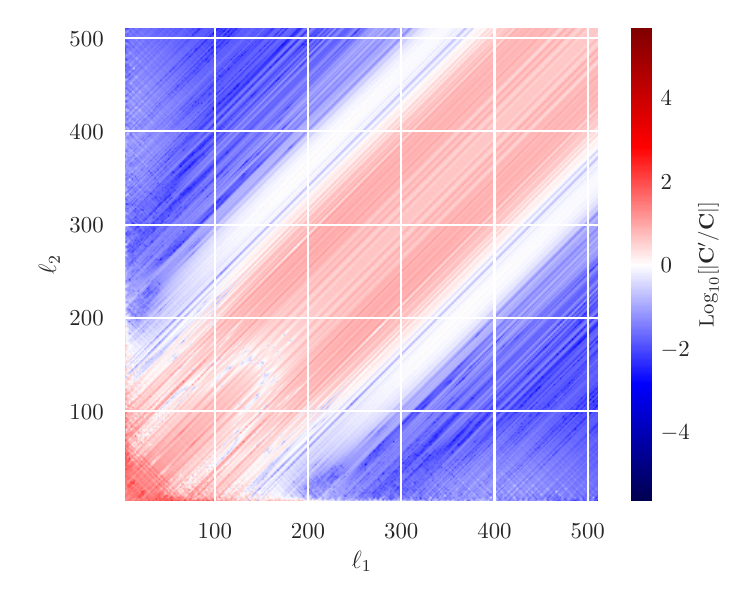}
    \vspace*{-0.5cm}
    \caption{Ratio of the analytic Pseudo-$\Cl$ covariance matrix for the
      $\ClEE$ power spectrum for the cases with ($\mathbfss{C}^{\prime}$) and
      without ($\mathbfss{C}$) mask apodisation applied.}
    \label{fig:PCl_cov_apo_ratio}
\end{figure}

Previously, we have discussed how apodisation of the mask is not required for
QML methods whereas there are certain advantages to doing so for the
Pseudo-$\Cl$ method. This is because apodisation reduces the effects of sharp
edges that may be present in the mask. To investigate the effects of
apodisation, we have applied a $2^{\circ}$ apodisation using the
$\mathcal{C}^{2}$ scheme as described in~\cite{Alonso:2018jzx} to our mask,
including stars. We note that for this apodisation scale and scheme, the sky
area reduces from $\fsky = 33 \, \%$ to $\fsky = 22 \, \%$.
Figure~\ref{fig:Mask_power_spectrum_apodisation} shows the power spectrum of our
mask with and without apodisation applied. Here, we see that the effect of
apodisation is to vastly reduce the small-scale power of the mask. The effect of
this suppression of small-scale power results in the reduction in long-range
correlations in the covariance matrix, as can be shown from
Equation~\ref{eqn:exact_pcl_covariance}, and thus the computation and inversion
of the mixing matrix should be more accurate when apodisation is applied. The
ratio of the analytic Pseudo-$\Cl$ covariance matrix for the $\ClEE$ spectrum
for the cases of with and without apodisation is shown in
Figure~\ref{fig:PCl_cov_apo_ratio}. This shows that for values along and close to
the diagonal, the loss of sky area causes significant increases in the variances
in the power spectrum. For mode-pairs that are highly separated we see
a notable decrease in their covariances, which once again can be seen from
Equation~\ref{eqn:exact_pcl_covariance}.

\begin{figure}
    \includegraphics[width=\columnwidth]{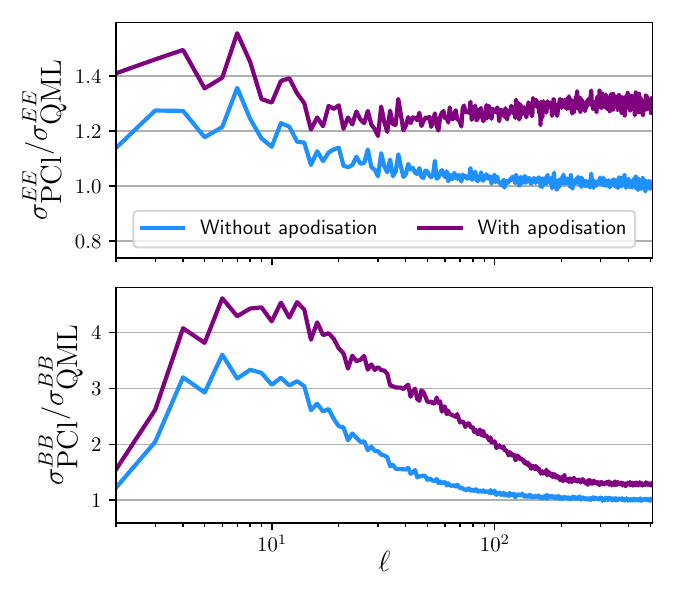}
    \caption{Ratio of $\Cl$ errors of the Pseudo-Cl method with respect to
    our QML estimator for the cases with and without apodisation.}
    \label{fig:Cl_err_ratio_with_apo}
\end{figure}

Figure~\ref{fig:Cl_err_ratio_with_apo} shows the ratio of the errors for the
Pseudo-$\Cl$ (the square-root of the diagonal of the covariance matrix) with
respect to our QML estimator for the case of with and without apodisation. Here,
we see that the effect of apodisation is to increase the errors of the
Pseudo-$\Cl$ method - which is a direct result in the loss of sky area that
apodisation produces.

\begin{figure}
    \includegraphics[width=\columnwidth, trim={0cm 0cm 0cm 0cm}, clip]{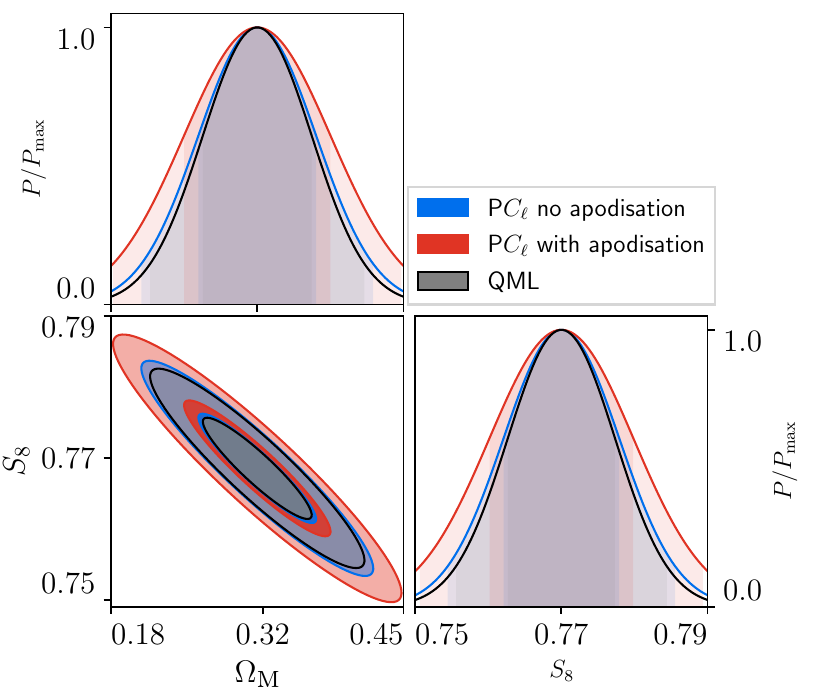}
    \vspace*{-0.5cm}
    \caption{Fisher parameter constraints comparison between QML and
      Pseudo-$\Cl$ where both estimators have a maximum multipole of $\lmax =
      512$ and for the case where apodisation has been applied for the
      Pseudo-$\Cl$ method. We see a large broadening for the Pseudo-$\Cl$
      contour with apodisation applied, which is consistent with the loss of
      sky area that apodisation results in.}
    \label{fig:Parameter_constraints_apodisation}
\end{figure}

The covariance matrix for the case where we have applied apodisation can then be
propagated into parameter constraints, which is shown in
Figure~\ref{fig:Parameter_constraints_apodisation}. Here, we see that the direct
loss of sky area associated with apodisation results in broadened parameter
contours which is not offset by the decrease in long-range correlations that
apodisation suppresses in the covariance matrix.

\end{subappendices}

\clearpage

\begin{savequote}[75mm]
  I foresee all sorts of unforeseen problems

  Such as?
  
  If I could foresee them, they wouldn't be unforeseen
  
  \qauthor{---Yes, Prime Minister}
\end{savequote}
\chapter[Mitigating baryon feedback bias in cosmic shear through a theoretical error covariance in the matter power spectrum]{Mitigating baryon feedback \\ bias in cosmic shear through a theoretical error covariance in \\ the matter power spectrum}
\label{chp:baryonic_effects}
\vspace*{-1cm}
\begin{mypaper}
  This chapter was published in \textit{Monthly Notices of the Royal 
    Astronomical Society} as Maraio, Hall, and Taylor (2025)~\citep{Maraio:2024xjz}.
\end{mypaper}
\vspace*{0.5cm}

\noindent
\begin{mytext}
    \textbf{Outline.} 
  Forthcoming cosmic shear surveys will make precise measurements of the matter
  density field down to very small scales, scales which are dominated by
  baryon feedback. The modelling of baryon feedback is crucial to ensure
  unbiased cosmological parameter constraints; the most efficient approach
  is to use analytic models, but these are limited by how well they can capture the physics of baryon feedback.
  We investigate the fitting and residual errors of various baryon feedback 
  models to a suite of hydrodynamic simulations, and
  propagate these to cosmological parameter constraints for cosmic shear. We
  present an alternative formalism to binary scale-cuts through the
  use of a theoretical error covariance, which is a well-motivated alternative
  using errors in the power spectrum modelling itself. We depart from previous
  works by modelling baryonic feedback errors directly in the matter power 
  spectrum, which is the
  natural basis to do so and thus preserves information in the lensing kernels.
  When including angular multipoles up to $\lmax = 5000$, and assuming \textit{Euclid}-like
  survey properties, we find that even multi-parameter models of baryon feedback can
  introduce significant levels of bias. In contrast, our theoretical error 
  reduces the bias in $\Omegam$ and $\Seight$ to acceptable levels, with only a modest increase in 
  parameter variances. The theoretical error approach bypasses the need to directly
  determine the per-bin $\lmax$ values, as it naturally suppresses the biassing 
  small-scale information. We also present a detailed study of how flexible 
  \HMCode-2020, a widely-used non-linear and baryonic feedback model, is at
  fitting a range of hydrodynamical simulations.
\end{mytext}

\section{Introduction}
\label{sec:paper2_Introduction}

Cosmic shear is the coherent distortion in the apparent shapes of galaxies due 
to the matter distribution of the large-scale structure of the universe 
\citep{Bartelmann:1999yn,Bartelmann:2010fz,Kilbinger:2014cea}. These 
distortions are sensitive to the total matter inhomogeneities along the 
line-of-sight, and thus are a powerful probe of the non-luminous dark matter, which
ordinarily cannot be directly observed using telescopes.
Cosmic shear is also sensitive to the detailed physics of baryons, particularly on small
scales, within the Universe (see Section~\ref{sec:baryon_feedback_in_cls}).
By making accurate measurements, 
along with robust theoretical modelling, cosmic shear is able to provide
detailed knowledge about the physics and geometry of the Universe.

Cosmic shear surveys have already placed tight constraints on the fundamental physics
and properties of our Universe, especially on the growth of structure parameter
$\Seight$, defined as $\Seight \equiv \sigmaeight \sqrt{\Omegam / 0.3}$.
The results from existing surveys have been well-studied, with data
coming from the Kilo-Degree Survey (KiDS-1000) \citep{Heymans:2020gsg,KiDS:2020suj,Li:2023azi},
the Dark Energy Survey (DES-Y3) \citep{DES:2021wwk,DES:2021bvc,DES:2021vln,DES:2022qpf},
and Hyper Suprime-Cam (HSC-Y3) \citep{Li:2023tui,Dalal:2023olq}.

Over the next decade, an unprecedented amount of high-quality cosmic shear data
will be released. This will come from the recently launched \textit{Euclid} 
space telescope \citep{Euclid:2011zbd,Euclid:2024yrr}, the Legacy Survey of Space and Time
(LSST) at the \textit{Rubin} observatory \citep{LSSTDarkEnergyScience:2012kar}, 
and the \textit{Roman} space telescope \citep{Spergel:2015sza}. Since the
quality and quantity of cosmic shear data that is expected to be produced is
so vast, the accuracy of the theoretical modelling is required to be as equally
precise. 

While the majority of the matter in the Universe is dark matter, which only
interacts gravitationally, the baryons in the Universe, while appearing to be
lighter in total mass, have an equally large affect on the dynamics of the Universe --
particularly on small scales. Baryons are responsible for the heating and
cooling of gas, the creation and demise of stars, and play an important part in
the feedback from active galactic nuclei (AGN) and supernovae, both of which can
have a considerable impact on the matter power spectrum over a wide range of
scales (see Figure~\ref{fig:cl_baryon_ratio_3param} showing suppressions upwards
of $20\,\%$ around angular scales of $\ell \sim 1000$, scales which have already 
been well-measured by current Stage-III surveys and will be measured with
increased precision using Stage-IV surveys)~\citep{vanDaalen:2011xb}.

The modelling of baryon feedback and its effects on cosmic shear cosmology
has been discussed extensively in the 
literature, with very many different methods and implementations proposed
to mitigate its effects. These include Schneider~\&~Teyssier (2015)~[\citenum{Schneider:2015wta}],
Giri \& Schneider (2021)~[\citenum{Giri:2021qin}], Aric\`o et~al. (2021)~[\citenum{Arico:2020lhq}],
Huang et~al. (2019)~[\citenum{Huang:2018wpy}], Salcido et~al. (2023)~[\citenum{Salcido:2023etz}]
and Mead et~al. (2020)~[\citenum{Mead:2020vgs}], among many others. Cosmic shear is a measurement
in angular-space due to the necessity of using coarse photometric redshift
estimates, and so this makes the measurements of cosmic shear sensitive to
high-$k$ wavenumbers throughout any $\ell$ values in the angular power spectrum.  
It is at high $k$ where feedback effects become relevant.

A popular approach to mitigating baryonic effects in cosmic shear 
analyses is to introduce scale-cuts into the data-vector. Here, physically 
smaller scale elements in the data vector beyond a cut-off, either in
real-space $\theta_{\textrm{min}}$ or in Fourier-space $\ell_{\textrm{max}}$,
are completely discarded even if high signal-to-noise observations have been 
taken. These scale-cuts are often dependent on the redshift of the source 
galaxies, as for further away redshift bins the same physical scale is given 
by smaller $\theta$ or larger $\ell$. One issue that arises with such scale-cuts
is that it by definition is a \textit{hard cut} on the data. For example, 
if it was chosen that $\ell_{\textrm{max}} = 2000$, then $\ell = 2000$ would
be included in an analysis whereas the $\ell = 2001$ mode would be excluded\footnote{\textit{I'm sorry Dave, I'm afraid I can't let you include that mode in an analysis} --- HAL 9000.}\!,
even though these modes will be highly coupled and have very similar
theoretical uncertainties.
The use of these binary scale cuts have been discussed heavily in previous
cosmic shear results, most recently in the Dark Energy Survey's Year 3 
results (real-space) \citep{DES:2021rex,DES:2021wwk},
the Kilo-Degree Survey fourth data release (KiDS-1000, in real-space) \citep{Li:2023azi},
and Hyper Suprime-Cam's Year 3 results
(harmonic-space) \citep{Dalal:2023olq}. 

Since the use and location of a hard scale cut could be considered somewhat
unphysical, in the sense that we have two consecutive data points one with
little error and one with infinite error, we investigate and present results for
an alternative: the use of a theoretical error covariance which acts as a soft
scale cut that is informed by our inability to correctly model baryonic feedback
down to arbitrarily small scales. This has been explored in the literature 
previously, with the theory being originally presented in Baldauf et~al. (2016)~[\citenum{Baldauf:2016sjb}],
expanded upon in Sprenger et~al. (2018)~[\citenum{Sprenger:2018tdb}],
and applied to mock cosmic shear analyses in
Moreira et~al.~[\citenum{Moreira:2021imm}], among others~\cite{Pellejero-Ibanez:2022efv}.
Baryonic feedback effects directly impact
the matter power spectrum, which is then integrated over to produce effects in
the lensing angular power spectrum. Thus, when we model our theoretical
uncertainties coming from baryonic feedback, it is most natural to do so in
the matter power spectrum and propagate these uncertainties to the angular
power spectrum. Moreira et~al. (2021)~[\citenum{Moreira:2021imm}] investigated theoretical uncertainties
with respect to the \textit{angular} power spectrum, and so we are revisiting this
formalism but extending it to the underlying matter power spectrum. Since
baryonic feedback contaminates only the matter power spectrum, not the lensing
kernels associated with cosmic shear, isolating the errors associated with
the matter power spectrum then propagating these to the angular power spectrum
is a sensible alternative approach. The theoretical uncertainties approach is
similar to analytic marginalisation of small-scale physics, which also
results in an additional term to the covariance matrix~\citep{Kitching:2010ab}.

This Chapter is structured as follows: in Section~\ref{sec:paper2_modelling_baryons} we
outline the need for analytic baryon feedback models and the use of hydrodynamical
simulations, Section~\ref{sec:paper2_Methodology} discusses our methodology for
constructing our theoretical error covariance, in Section~\ref{sec:paper2_results}
we present our results for cosmological parameter constraints using a selection
of baryon feedback models, and Section~\ref{sec:paper2_diss_and_conc} summarises
our findings.

\section{Modelling baryonic feedback in the matter power spectrum}
\label{sec:paper2_modelling_baryons}

\begin{sidewaysfigure}[tp]
  \centering
  \includegraphics[width=\columnwidth]{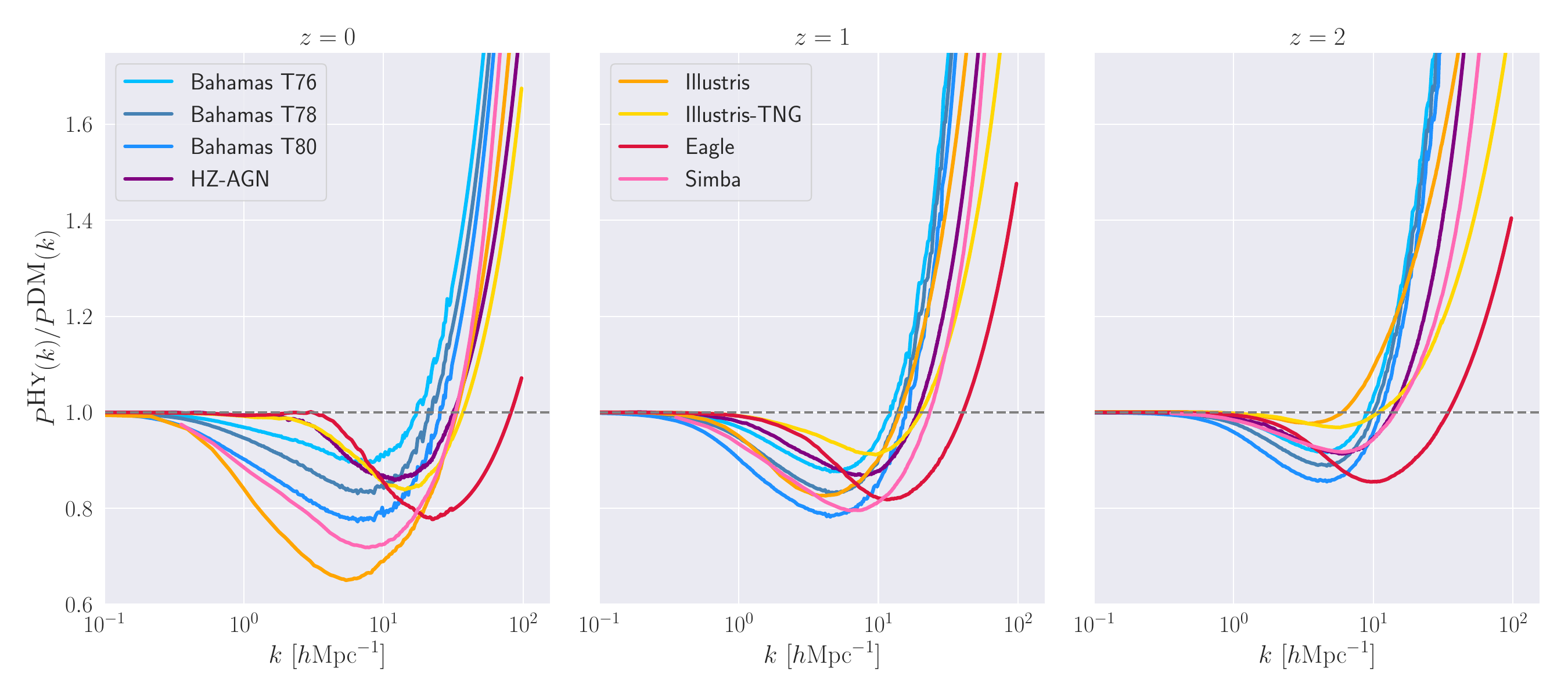}
  \vspace*{-0.25cm}
  \caption{Ratios of the matter power spectrum with baryonic feedback in to
    their dark-matter-only counterparts at redshifts 0, 1, and 2, for various
    hydrodynamical simulations, as given in Table~\ref{tbl:hydrosims}.
    Here, `\Bahamas~T$XY$' corresponds to the same underlying \Bahamas
    hydro-sim, just run at different $\Delta T_{\textrm{heat}}$ values, which
    corresponds to the amplitude of AGN feedback within their simulation~\citep{McCarthy:2016mry}.
    Note that \Simba was run on a smaller box-size with respect to the
    other simulations (as shown in Table~\ref{tbl:hydrosims}), and thus
    we do not have access to data below a certain $k$ for \Simba, which is much
    larger than the smallest $k$ modes available for the other hydro-sims.
    All data here (plus data for intermediate redshifts) has been taken from
    existing public repositories, see the 
    \hyperref[sec:data_avail]{data availability} section for more information.
    }
  \label{fig:My_Pk_ratio}
\end{sidewaysfigure}

Accurate evaluations of the non-linear dark-matter-only matter power spectrum is
of extreme importance to any cosmic shear analysis \citep{Schneider:2015yka}. 
There are numerous ways to accurately evaluate this. For example, many models
descend from the original halo model presented in the early 2000s
\citep{Seljak:2000gq,Peacock:2000MNRAS3181144P,Cooray:2002dia,Smith:2002dz}:
the \Halofit model presented in Ref.~\cite{Takahashi:2012em}, and \HMCode models
that originate with \HMCode-2015~\citep{Mead:2015yca}, then \HMCode-2016~\citep{Mead:2016zqy},
and most recently \HMCode-2020~\citep{Mead:2020vgs}. All these models serve to
provide accurate predictions for the non-linear dark-matter-only matter power
spectrum as a function of cosmological parameters.

We now wish to go one step further and include the effects of baryonic feedback
physics on the matter power spectrum into our numerical models. We are motivated
to develop numerical descriptions of baryon feedback because it allows for easy
comparison between theoretical models and observational data, thus allowing the 
data to place constraints on parameters of the model, or even in model selection.
Many of these parameter constraints or model selection analyses using Stage-IV
cosmic shear survey data will come from Markov chain Monte Carlo (MCMC) analyses~\citep{Boruah:2024tkq}.
These MCMC analyses depend heavily on the evaluation speed for each sample,
since very many samples are needed for accurate results ($\sim\!\!10^5$-$10^6$ samples required).
Thus, the speed of simple numerical models of baryonic feedback are desirable
to avoid excessive run-times. While analytic
tools are much sought-after, their limited accuracy in matching a 
wide-range of baryon feedback physics limits their use in current cosmological
surveys as to not bias parameter constraints.

The best tests of baryon physics come from fully hydrodynamical 
cosmological simulations (hydro-sims). These simulations aim to capture the physics of
dark matter clustering and galaxy formation and evolution using extremely large
volumes, with a box side length of $L \gtrsim 100 \, \textrm{Mpc}/h$ (in
comoving units), and across a wide range of redshifts. In addition,
hydrodynamical simulations aim to implement modelling of astrophysical 
processes which are significantly more complex than gravitational attraction,
and may include star formation, stellar feedback, chemical evolution of the
interstellar medium, active galactic nuclei (AGN) growth and feedback, among
others \citep{vanDaalen:2011xb}. For a recent review of hydro-sims, 
see Refs.~\cite{Vogelsberger:2019ynw} and~\cite{Crain:2023xap}.

The development and extraction of results from
hydro-sims are an incredibly active area of research with very many groups 
developing their own implementations of hydrodynamical effects within 
simulations. Recent works include the \textsc{Flamingo} simulation~\citep{Schaye:2023jqv},
\textsc{MillenniumTNG}~\citep{Hernandez-Aguayo:2022xcl}, with past works
including \textsc{Simba}~\citep{Dave:2019yyq}, \textsc{Illustris}~\citep{Nelson:2015dga},
\textsc{Illustris-TNG}~\citep{Springel:2017tpz}, \textsc{Horizon-AGN}~\citep{Chisari:2018prw},
\textsc{Bahamas}~\citep{McCarthy:2016mry}, Eagle~\citep{Schaye:2014tpa}, and
\textsc{Cosmo-OWLS}~\citep{Brun:2013yva} simulations.

These simulations mainly differ in their implementations of sub-grid physics.
Hydro-sims require the use of sub-grid models since the the scales of many
important physical processes that make up baryonic feedback (such as gas cooling,
star formation and stellar feedback, supermassive black hole formation and 
evolution, and AGN feedback), occur on scales smaller
than their ordinary particle mesh grid~\cite{Crain:2023xap}. Hence, to
model these physical processes, hydro-sims need to infer their existence 
in between nodes of the simulation, hence the name of sub-grid models. For
example, the scales over which AGN inject energy into the interstellar medium
occurs on scales which cannot be numerically resolved -- and thus hydro-sims
must approximate their macroscopic effects thorough sub-grid 
models~\cite{Crain:2023xap}. AGN feedback can be divided into two modes, each
with their own sub-grid implementation: a quasar mode which is associated with
the electromagnetic radiation emitted from the accreted matter, and a radio
mode which is formed from the highly collimated jets of relativistic matter. 
Since AGN activity impact a wide range of scales, matching the sub-grid nature
of the AGN themselves to the macroscopic impact on nearby nodes of the hydro-sim
is a challenging problem. Furthermore, supernovae explosions inject huge
quantities of energy into their surrounding interstellar medium, which is far
smaller than the simulated grid size. Thus, a sub-grid model for supernovae
feedback must be implemented, which can inject its energy and momentum either
thermally or kinematically~\cite{Vogelsberger:2019ynw}. 
There are suggestions that these sub-grid physics differences are connected to
the gas fraction in galaxy groups and clusters~\citep{Salcido:2024qrt}. 

Hydrodynamical simulations are extremely computationally expensive due to the need for
large volumes, to reduce cosmic variance and capture clustering on cosmological 
scales, and the need to model small-scale behaviour, which directly impacts the
larger-scale clustering. Thus, these simulations need sufficient resolution else
the baryonic feedback physics which we wish to capture will simply be washed out
on larger scales and also sufficient volume. This presents a challenging
dynamic range problem, hence the very many different simulations that 
implement these range of physics.

To isolate the effect of baryonic physics on the matter power spectrum when
compared to the dark-matter-only non-linear power spectrum, we use the baryon
response function, $R(k, z)$, defined as
\begin{align}
  R(k, z) \equiv \frac{\Phykz}{\Pdmkz},
  \label{eqn:baryonic_pk_ratio}
\end{align}
which isolates the effects of baryon physics from the general non-linear
spectrum, and is especially useful for reducing the effects of cosmic variance
on the power spectrum when using hydro-simulations. This ratio is approximately
insensitive to the cosmological parameters that $\Phy$ and $\Pdm$ were
generated using, provided that they are the same cosmology~\citep{vanDaalen:2019pst,Elbers:2024dad}.
This makes the baryon response function especially useful when trying to compare predictions between different
hydro-sims run at different cosmologies. Since every hydro-sim has a different 
implementation of their astrophysical processes, the range of curves for
$R(k, z)$ can be extremely broad. Figure~\ref{fig:My_Pk_ratio} plots this
ratio for a variety of different hydro-sims for three redshifts. This shows the
general behaviour of $R(k, z)$ that has been well explored previously~\citep{vanDaalen:2011xb}:
there is a dip in expected power on non-linear scales 
($1 \, h\Mpc^{-1} \lesssim  k \lesssim 10 \, h\Mpc^{-1}$ at $z=0$ )
due to baryon feedback processes, such as AGN, jets, and supernovae,
expelling matter reducing clustering, while the dramatic
increase on highly non-linear scales ($ k \gtrsim 10 \, h\Mpc^{-1}$) is due to
additional clustering from the ability of baryons to undergo radiative cooling,
increasing small-scale clustering, and from star formation within the simulations~\citep{Chisari:2019tus}.
Also shown in Figure~\ref{fig:My_Pk_ratio} is the level of scatter in the prediction of $R(k, z)$
for different hydro-sims, with the depth and location of the suppression and
location of the upturn highly simulation dependant with very little common 
consensus between these simulations. Ref.~\cite{vanDaalen:2019pst} has shown that
much of this scatter in the predictions for the suppression in the simulations can be
explained by how the amplitude of the suppression is strongly correlated with
the mean baryon fraction in haloes of mass~$\sim \!\! 10^{14} M_{\odot}$~\citep{Salcido:2024qrt}.

It should be noted that Figure~\ref{fig:My_Pk_ratio} shows that baryonic
feedback is also highly redshift dependent, both in the amplitude and
scale-dependence of the suppression, on a per-hydro-sim basis (with some
general scatter in this redshift scaling too). We see that the largest 
suppression of power comes from redshift zero, with suppression gradually
decreasing with increasing redshift. While it is true that the
star formation rate within galaxies and the accretion rate of matter into the
supermassive black holes, and thus the AGN luminosity, peaks around redshift 
$\sim2-4$~\cite{Fanidakis_2011}, this does not necessarily mean that the suppression
in the matter power spectrum should peak at this redshift. This is because
gravitational clustering is a time-evolutionary process, and so the suppression
at redshift $2$ can only \textit{grow} with respect to the dark-matter-only model
over the next ten billion years to redshift zero. 

Due to the large scatter in the predictions of the hydro-sims, we take an 
agnostic approach to their results: we assume that each of these simulations
are equally trustworthy, and thus any of these predictions for their $R(k, z)$
must be equally reliable. We must also assume that the true
description of our Universe lies somewhere in this ensemble in order to correctly
match observational data.
Hence, any baryon physics model must be able to recreate all of these curves
in order for us to sufficiently trust that it could potentially capture the real
physics of our Universe. This level of scatter in the hydro-sims has lead to the development of many
baryonic physics models, especially those which aim to match the simulations
through neural-network emulator methods, such as the \Bacco emulator presented
in Ref.~\cite{Arico:2020lhq}. \Bacco claims accuracy 1-2$\%$ for scales
$1 \, h\Mpc^{-1} <  k < 5 \, h\Mpc^{-1}$ and redshifts $0 < z < 1.5$, and again
finds that the most important parameter that controls baryon feedback physics
is the gas fraction per halo mass. Furthermore, \BCEmu~\citep{Giri:2021qin}
implements the `baryonification model' prescription~\citep{Schneider:2018pfw}
which claims percent-level accuracy for scales below $k \sim 10 \, h\Mpc^{-1}$
at redshifts $z < 2$.

\begin{figure}[t]
  \centering
  \includegraphics[width=\columnwidth]{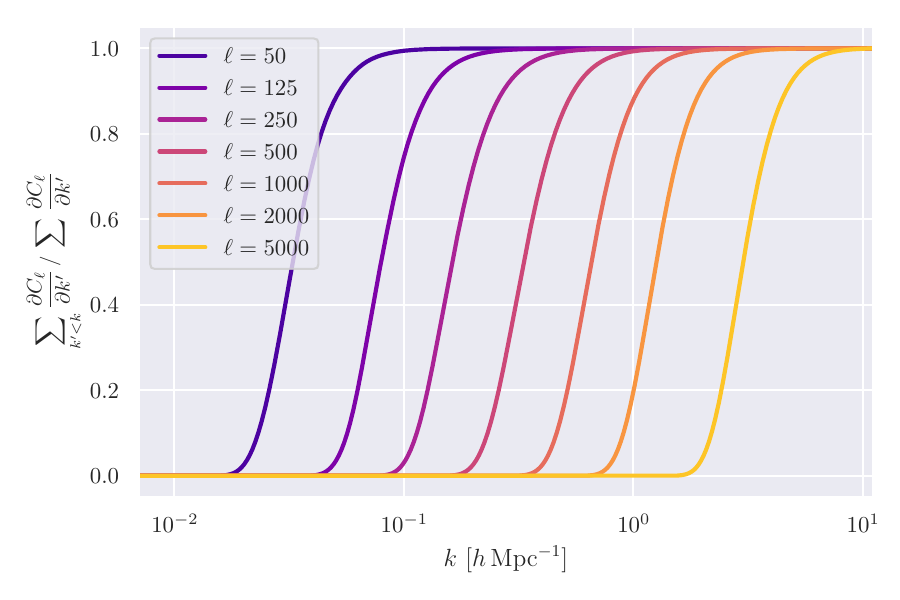}
  \vspace*{-0.25cm}
  \caption{Cumulative sum of the derivate of the cosmic shear power angular
    power spectrum coefficients $\Cl$, at certain $\ell$ modes, with respect to
    the wavenumber $k$ in the Limber integral (Equation~\ref{eqn:cosmic_shear_powspec}).
    The sum of the derivatives for each multipole is normalised to unity for
    easy comparison between different modes.
    This includes the contributions from the lensing kernels and the matter power spectrum. Derivatives were
    taken for a Gaussian source redshift distribution located at $\bar{z} = 0.33$
    and width $\sigma_{z} = 0.15$. }
  \label{fig:Cl_deriv_cumsum}
\end{figure}

We are motivated to find correct descriptions of baryon feedback physics due to
their large impact on results obtained from cosmic shear surveys. Future
Stage-IV cosmic shear surveys hope to probe extremely small scales, up to a
maximum multipole of $\lmax \sim 5000$ \citep{Euclid:2011zbd}, corresponding to
an approximate physical scale of around $1.2\,$Mpc, $4.1\,$Mpc, and $6.5\,$Mpc
at redshifts 0.25, 1, and 2, respectively. These small
angular scales probe the highly non-linear regime in the matter power spectrum,
as shown in Figure~\ref{fig:Cl_deriv_cumsum}. Here, we plot the cumulative sum
of the derivate of the angular power spectrum coefficients $\Cl$ at certain
$\ell$ values as a function of $k$. This shows the relative contribution for
each $k$ mode to each $\Cl$ value. Hence, we find that if we hope to probe
angular scales up to $\ell = 5000$, then we need to have a strong understanding of
the matter power spectrum up to scales $k \sim 10 \, h\Mpc^{-1}$ -- including
the details of baryonic physics. If these scales are not correctly modelled,
then this could induce catastrophic biases in a cosmic shear analysis~\citep{Semboloni:2011fe,Huang:2018wpy}, 
particularly for the constraints on dark energy~\citep{Copeland:2017hzu}.
The alternative is to reduce the maximum multipole $\lmax$ in our analyses,
though this discards potentially useful cosmological information.

This motivates us to test the accuracy of baryonic feedback models up to these
small scales, and assess how different treatments of baryonic physics impact
constraints obtained from forthcoming cosmic shear surveys.

\subsection{Theoretical error formalism}
\label{sec:paper2_theory_error_formalism}

Since we do not have models of baryonic physics that completely match
our complete set of hydrodynamical simulations, there will always be some
residual error for any given model of baryon feedback physics as a function of
scale and redshift. Rather than treating our models as a perfectly known 
quantity and limiting our scales of interest, we can directly incorporate these
known errors through the use of the `theoretical uncertainty' formalism 
first presented in Ref.~\cite{Baldauf:2016sjb}. We now have a theoretical error
data-vector $\mathbfit{e}$, which quantifies the difference between the true, underlying
physical model and what our approximative methods calculate, which is bound
by an envelope $\mathbfit{E}$. This theoretical error has its associated
covariance matrix, $\mathbfss{C}^{\textsc{e}}$, which can be written as a
function of the amplitude of the error, $\mathbfit{E}$, and a correlation matrix
$\boldsymbol{\rho}$ as
\begin{align}
  \mathbfss{C}^{\textsc{e}} = \mathbfit{E} \, \boldsymbol{\rho} \, \mathbfit{E}^{\textsc{t}}.
\end{align}
In the Gaussian approximation, we can simply model the inclusion of the 
theoretical error on the likelihood as simply the addition of our theoretical
error covariance, $\mathbfss{C}^{\textsc{e}}$, to our existing data covariance
matrix (which includes the contributions from cosmic variance),
$\mathbfss{C}^{\textsc{d}}$, to give
\begin{align}
  \mathbfss{C}^{\textsc{tot}} = \mathbfss{C}^{\textsc{d}} + \mathbfss{C}^{\textsc{e}}.
  \label{eqn:cov_addition}
\end{align}
The theoretical error formalism can be, in principle, applied to any effect
that changes summary statistics that is not yet perfectly modelled.
This approach of marginalising over the errors in the model was first applied
to baryonic physics effects for cosmic shear surveys in Ref.~\cite{Moreira:2021imm},
where they modelled the errors and correlations directly in $\ell$-space.
They constructed several versions of the error envelope function $E(\ell)$,
all featuring the similar term of the error ratio of best-fit $\Cl$ values
to the values generated from hydro-sims. They also assumed a Gaussian-like
function in $(\ell - \ell')^2$ for their correlation matrix, with a characteristic
correlation length~$L$ determining the widths of the correlation matrix. 

Since baryonic effects directly change the matter power spectrum, which is then
propagated to cosmic shear summary statistics, we are motivated to investigate
the theoretical errors in $k$-space for the matter power spectrum and then
propagate those to $\ell$-space. There is no theoretical error associated with
the lensing kernels, so by isolating the theoretical uncertainties to the matter
power spectrum only, we might hope to preserve information from geometry this way.
The task now becomes one to find an appropriate
theoretical error envelope function $\mathbfit{E}$ and correlation 
matrix $\boldsymbol{\rho}$ in $k$-space.

\section{Methodology}
\label{sec:paper2_Methodology}

\subsection{Modelling forthcoming cosmic shear surveys}

We use the standard prescription for the cosmic shear power 
spectrum~\citep{Bartelmann:1999yn,Bartelmann:2010fz,Kilbinger:2014cea} where
the power spectrum is, for tomographic redshift bins labelled by $a$ and $b$, is
given as
\begin{align}
  \Cl^{ab} = \frac{9}{4} \frac{\Omegam^2 H_0^4}{c^4} \int_{0}^{\chi_{\textrm{max}}}
  \!\! 
  \d \chi \! \frac{g_{a}(\chi) \, g_{b}(\chi)}{a^2(\chi)} \, P \! \left(k=\frac{\ell}{\chi}, \, z=z(\chi) \right),
  \label{eqn:cosmic_shear_powspec}
\end{align}
where $a(\chi)$ is the scale factor, $P$ is the non-linear matter power spectrum,
and $g(\chi)$ is the lensing kernel given as
\begin{equation}
  g_{a}(\chi) = \int_{\chi}^{\chi_{h}} \!\! \d \chi' \, n_{a}(\chi') \frac{f_{K}(\chi' - \chi)}{f_{K}(\chi')},
\end{equation}
where $n(\chi)$ is the probability density of source galaxies as a function of
comoving distance. To evaluate the non-linear matter power spectrum with baryonic
feedback, we used \HMCode-2020 with $\Tagn = 7.8$.

Since every galaxy has an intrinsic ellipticity, this introduces a shape
noise term into the power spectrum with a flat value $N_{\ell}$ given as\
\begin{align}
  N_{\ell}^{ab} = \frac{\sigma_{\epsilon}^2}{\bar{n}} \, \delta^{ab}, 
\end{align}
where $\sigma_{\epsilon}$ is the standard deviation of the intrinsic galaxy 
ellipticity dispersion per component, $\bar{n}$ is the expected number of
observed galaxies per steradian, and $\delta^{ab}$ is the Kronecker-$\delta$
symbol. We assume \textit{Euclid}-like values where it is expected that 
30 galaxies per square arcminute will be observed, but we divide these into five
photometric redshift bins, giving 
$\bar{n} = 6\,\mathrm{gals / arcmin}^{2}$~\citep{Euclid:2011zbd}. 
We take $\sigma_{\epsilon} = 0.21$. 

We use five Gaussian redshift bins with means located at 
$\bar{z} = \{0.33, 0.66, 1.0, 1.33, 1.66\}$ all with standard deviation of
$\sigma_z = 0.15$. This gives us 15 unique power spectrum combinations and
120 unique covariance matrix blocks.

We model the Gaussian covariance matrix as the four-point function~\citep{Euclid:2019clj},
given by
\begin{align}
  \textrm{Cov}[\Cl^{ab}, \Clp^{cd}] = 
  \frac{\Cl^{ac} \Cl^{bd} + \Cl^{ad} \Cl^{bc}}{(2 \ell + 1) \, \fsky} \delta_{\ell \ell'},
  \label{eqn:Gaussian_Cl_cov}
\end{align} 
where $\fsky$ is the fraction of sky observed by the cosmic shear survey. 
We take a \textit{Euclid}-like value of $\fsky = 0.35$.

To reduce the dimensionality of our power spectrum and covariance matrices, we
bin the power spectrum. We use linear binning of five bins up to $\ell = 100$,
and then use logarithmic binning with twenty bins from $\ell = 100$ to 
$\ell = 5\,000$. We use an $\ell$-mode weight of $\ell(\ell+1)$ when binning.

\subsection[Constructing our $k$-space theoretical covariance]{\boldmath Constructing our $k$-space theoretical covariance}

As outlined in Section~\ref{sec:paper2_theory_error_formalism}, the theoretical error
formalism requires us to quantify an error envelope function $\mathbfit{E}$ and
a correlation matrix $\boldsymbol{\rho}$, and since our uncertainties from
baryonic physics are best specified in $k$-space, we will first construct our
$k$-space covariance matrix. We will first discuss the construction of our
envelope function $\mathbfit{E}$.

There is no absolute choice for the functional form of $\mathbfit{E}$.
\cite{Moreira:2021imm} investigated many choices for the form of $\mathbfit{E}$,
with the `mirror' envelope being a good fit to the data. Here, we follow their
method and take the maximum deviation for each $k$-mode and redshift across a
suite of hydrodynamical simulations to the best-fitting values of a baryon 
feedback model (see Section~\ref{sec:paper2_fitting_hmcode_to_hydrosims}).
This gives the maximum relative difference in the baryonic
response function (Equation~\ref{eqn:baryonic_pk_ratio}) as
\begin{align}
  \Delta R(k, z) = \max_{k, \, z} \left\lvert \frac{R^{\textsc{Hydro}}(k, z)}{R^{\textsc{Model}}(k, z)} - 1 \right\rvert.
  \label{eqn:max_deviation}
\end{align} 
We then turn this into an amplitude by multiplying by a fiducial 
dark-matter-only power spectrum,
\begin{align}
  E(k, z) = P^{\textsc{dm}}(k, z) \, \Delta R(k, z).
\end{align}
where $\Pdmkz$ was evaluated using the \HMCode-2020 DM-only model.

With our amplitude function specified, we now wish to construct our $k$-space
correlation matrix. Here, we are considering the correlations between two
pairs of points $(k_1, z_1)$ and $(k_2, z_2)$. We follow previous works that
have employed the theoretical uncertainties approach (for example 
Refs.~\cite{Baldauf:2016sjb,Sprenger:2018tdb,Chudaykin:2019ock,Moreira:2021imm})
by choosing a factorisable Gaussian correlation
matrix of the form
\begin{align}
  \rho[(k_1, z_1), \, (k_2, z_2)] = 
  \exp \left[ -\frac{\log(k_1 / k_2)^2}{\sigma_{\log k}^2} \right] 
  \exp \left[ -\frac{(z_1 - z_2)^2}{\sigma_{z}^2} \right],
\end{align}
where $\sigma_{\log k}$ and $\sigma_z$ are the characteristic correlation scales in
$\log k$- and redshift-space, respectively. In our fiducial analyses, we used
values of $\sigma_{\log k} = 0.25$ and $\sigma_z = 0.25$. These values were motivated
from Figure~\ref{fig:My_Pk_ratio} where we see that the baryonic response
function $R(k, z)$ is correlated on scales of approximately a quarter of a
decade in log-$k$ space with approximately the same coupling in redshift-space.
We discuss the effects of changing $\sigma_{\log k}$ and $\sigma_z$ in Section~\ref{sec:paper2_sigma_kz}.

We are motivated to use factorisable
Gaussians since we know that baryon feedback has a relatively local effect on
the matter power spectrum, that is neighbouring wavenumbers and redshifts 
behave similarly and thus should have highly correlated covariances, whereas
vastly different wavenumbers and redshifts have very different evolutionary 
physics, and thus should be less correlated. Furthermore, we are motivated to 
use the logarithmic differences in $k$-space since our wavenumbers span many
orders of magnitude (see Figure~\ref{fig:My_Pk_ratio}), and thus an ordinary
difference might not properly reflect this. 

Combining our envelope function and correlation matrix, we find that our
complete $k$- and redshift-space covariance matrix is given by 
\begin{multline}
  \textrm{Cov}[(k_1, z_1), (k_2, z_2)] = \\
  P^{\textsc{dm}}(k_1, z_1) \, \Delta R(k_1, z_1) \,\,  P^{\textsc{dm}}(k_2, z_2) \, \Delta R(k_2, z_2) \, \times \\
  \exp \left[ -\frac{\log(k_1 / k_2)^2}{\sigma_{\log k}^2} \right] \times \exp \left[ -\frac{(z_1 - z_2)^2}{\sigma_{z}^2} \right].
  \label{eqn:k_space_cov}
\end{multline}

We note that by working directly with errors in the matter power spectrum, which is the
underlying quantity that we have limited knowledge in modelling (not the 
angular power spectrum values), this should capture the full phenomenology of 
the errors in baryonic feedback effects on the matter power spectrum. This 
contrasts with the work of \cite{Moreira:2021imm} which also aims to mitigate
baryonic physics effects through the use of a theoretical uncertainty, though
they quantify their theoretical uncertainty through the differences in
the angular power spectrum values.

\subsection[Propagating covariances to $\ell$-space]{\boldmath Propagating covariances to $\ell$-space}

With our $k$- and redshift-space covariance matrix specified, we can now 
propagate this into an $\ell$-space covariance matrix, which can then be added
to the data covariance to give our overall covariance matrix.

Equation~\ref{eqn:cosmic_shear_powspec} gave the cosmic shear power spectrum
as the Limber integral of the matter power spectrum weighted by the geometrical
factors.  Hence, if we now want to propagate our uncertainties in the $P(k)$ into an
additional covariance matrix for the $\Cl$ values, we can integrate this again, 
which yields
\begin{multline}
  \textrm{Cov}[C_{\ell_1}^{ab}, \, C_{\ell_2}^{cd}] =
  \left[ \frac{9}{4} \frac{\Omegam^2 H_0^4}{c^4} \right]^2 \times \\
  \int_{0}^{\chi_{\textrm{max}}} \!\! \int_{0}^{\chi_{\textrm{max}}}  \!\! \d \chi_1 \d \chi_2 \,
  \frac{g_a(\chi_1) \, g_b(\chi_1)}{a^2(\chi_1)} \, \frac{g_c(\chi_2) \, g_d(\chi_2)}{a^2(\chi_2)} \,\,
  \textrm{Cov}[(k_1, z_1), \, (k_2, z_2)],
  \label{eqn:cov_dbl_integral}
\end{multline}
where $k_1 = \frac{\ell_1}{\chi_1}$ and $k_2 = \frac{\ell_2}{\chi_2}$. 
We are motivated to use Limber's approximation here to simplify the double
integral since the low $\ell$ region in which the approximation is imprecise 
will have a negligible contribution to the total covariance.

\subsection{Numerical evaluation of the matter power spectrum with baryon 
feedback}
\label{sec:num_eval_pk_hmcode}

Our construction of the $(k, z)$-space covariance matrix can be applied to any
model of baryon feedback, with models ranging from purely analytical methods
using the Zel'dovich approximation~\citep{Mohammed:2014lja}, to semi-analytic
models (e.g. \HMCode \cite{Mead:2020vgs}), to purely numerical emulation (e.g.
\Bacco \cite{Arico:2020lhq} and \BCEmu \cite{Schneider:2018pfw}).

We have chosen \HMCode-2020 as our model of choice for the matter power spectrum
and baryonic feedback response since they claim that their dark-matter-only spectrum
has RMS errors of less than $5$ per-cent, and that their 
baryonic feedback response is accurate to within the $1$ per-cent level for redshifts 
$z < 1$ and scales ($k < 20 \, h\textrm{Mpc}^{-1}$), over a range of cosmologies~\citep{Mead:2020vgs}.
Thus, \HMCode-2020 is a natural choice for computing the
matter power spectrum for use in cosmic shear analyses, which is confirmed
by its use in recent major cosmic shear results such as the joint \KiDS--\DES
analysis \citep{Kilo-DegreeSurvey:2023gfr}.

\HMCode-2020 comes in two distinct flavours for modelling baryon physics:
\begin{galitemize}
  \item A full six-parameter description with free-parameters of the halo 
    concentration, $B$, the stellar mass fraction in haloes, $f_\star$, and the
    halo mass break scale, $\Mb$, along their redshift evolution counterparts.
    The redshift evolution for each physical parameter $X$ is given by 
    the fixed form of $X(z) = X_0 \times 10^{z \, X_z}$. 
  \item A one-parameter version where the feedback temperature $\Tagn$
    encapsulates the combined baryon physics of the full model. The 
    one-parameter model was constructed by fitting a linear relationship to 
    the six-parameter version using the \Bahamas simulations. Since the 
    one-parameter model was specifically constructed to the \Bahamas
    simulations, there is no guarantee of its accuracy to other hydrodynamical
    simulators or even the baryon physics of our own Universe.
  \item We also test a three-parameter version, which is analogous to the 
    six-parameter model but with fixed redshift evolution. This is a useful
    test as there arises significant degeneracies between the amplitude and
    redshift evolution of each parameter, and thus by fixing its evolution we
    can test if this extra degree of freedom is necessary.  
\end{galitemize} 

While a more general description of baryon feedback physics should have more freedom
to better match different physical models, it comes at the cost of additional
nuisance parameters which needs to be included and marginalised over in any
analysis. This could slow down the convergence of analysis pipelines, and in
the case of very many parameter models, unnecessarily decreasing constraints
on  the core cosmological parameters due to the need to excessively marginalise
over these baryonic nuisance parameters. Hence, while we are focusing on the
inclusion of theoretical uncertainties into our analysis pipeline, we also
look at how going from a one- to three- to six-parameter baryonic feedback
model changes our results.

\subsection{Chosen hydrodynamical simulations}

We make use of the `power spectrum library' presented 
in Ref.~\cite{vanDaalen:2019pst}\footnote{\url{https://powerlib.strw.leidenuniv.nl}}
and from the \Camels Project~\citep{CAMELS:2023wqd}\footnote{\url{https://camels.readthedocs.io/}}
to build up a suite of six hydrodynamical simulations (with three flavours of
the Bahamas simulations) as shown in Table~\ref{tbl:hydrosims}. This gives us a
wide range of baryon feedback models against which we can benchmark our analytic 
models to.



\begin{table}[t]
    
  \centering
  \caption{The six hydrodynamical simulations used in this work to benchmark
    and compare models of baryon feedback physics to.
    }
  
  \label{tbl:hydrosims}
  
  \addtolength{\tabcolsep}{-0.2em}

  \begin{tabular}{lrrrrr}
    \toprule

    Hydro-sim & \multicolumn{1}{p{1.5cm}}{\raggedleft Box-size \\ (Mpc)} & \multicolumn{1}{p{2cm}}{\raggedleft Number of\\ particles} & \multicolumn{1}{p{2.75cm}}{\raggedleft Baryonic mass \\ resolution $[M_{\odot}]$} & Cosmology & Reference\\
    
    \midrule
    Bahamas           & $400 / h$ & $2 \times 1024^{3}$ & $7.66 \times 10^{8} \, h^{-1}$ &  \textit{WMAP} 9 & \cite{McCarthy:2016mry} \\

    Horizon-AGN       & $100 / h$ & $1024^{3}$ & $8.3\times 10^{7} $ & \textit{WMAP} 7 & \cite{Chisari:2018prw} \\

    Illustris         & 106.5     & $1820^{3}$ & $1.6 \times 10^{6}$ & \textit{WMAP} 7 & \cite{Nelson:2015dga} \\

    Illustris-TNG 100 & $75 / h$  & $2 \times 1820^{3}$ & $9.44 \times 10^{5} \, h^{-1}$ &\textit{Planck} 2015 & \cite{Springel:2017tpz}\\

    Eagle             & 100       & $1504^{3}$ & $1.81 \times 10^{6}$ &  \textit{Planck} 2013 & \cite{Schaye:2014tpa} \\
    
    Simba             & $25 / h$  & $1024^{3}$ & $ 2.85 \times 10^{5}$ & \textit{Planck} 2013 & \cite{Dave:2019yyq} \\
    
    \bottomrule
  \end{tabular}
\end{table}

\subsection{Fitting HMCode to hydrodynamical simulations}
\label{sec:paper2_fitting_hmcode_to_hydrosims}

Armed with our three baryon feedback models and a suite of hydrodynamical 
simulations, we can now go about performing a best-fit analysis of our models
to the hydro-sims in order to quantify each model's maximum deviation as a 
function of wavenumber and redshift (Equation~\ref{eqn:max_deviation}). To do 
so, we performed a maximum likelihood fit where we simultaneously fitted 
across redshifts $z \leq 2$ with flat weighting in redshifts,
and wavenumbers over the range 
$0.01 \, h\textrm{Mpc}^{-1} \leq k \leq 20 \, h\textrm{Mpc}^{-1}$
with a $k^2$ weighting on each wavenumber to roughly approximate the cosmic 
variance contribution. We are free to choose an arbitrary $k$-mode weighting 
since our envelope function is arbitrary, and so there is no correct choice
for either a $k$- or $z$-mode weight here. We experimented with flat, and a
pure cosmic variance weight of $k^3$, and settled on our $k^2$ weighting
as somewhere in between. Note that, since we use equal spacing in $\log(k)$,
we find $\Delta \log k = \left[\Delta k\right] / k$ when summing over $k$.
This gives our loss function for each model and hydro-sim as
\begin{align}
  \mathcal{L} = \sum_{k, \, z} 
  k^2 \left(R^{\textsc{Hydro}}(k, z) - R^{\textsc{HMcode}}(k, z) \right) ^ 2.
  \label{eqn:chi_sq}
\end{align}
The \HMCode power spectra were generated at the same cosmology at which the
hydro-sims were run at, with the minimisation routine just varying the
astrophysical baryonic parameters. The \texttt{Minuit} optimiser, a robust
optimiser as part of the \Cosmosis\footnote{\url{https://cosmosis.readthedocs.io/}}
analysis framework~\citep{Zuntz:2014csq}, was used to maximise the fit
for the baryonic parameters. Since we there is freedom in the form of the 
weight function for each $(k, z)$-mode, different choices for the analytic form 
of $\mathcal{L}$ effectively adds little to that freedom.

\subsection[Fitting the $\Cl$ values]{\boldmath Fitting the $\Cl$ values}

To obtain estimates for the biases in cosmological parameters due to baryon
feedback, many MCMC analyses were run. We used a custom \Cosmosis pipeline
using the \texttt{PolyChord} nested sampler~\citep{Handley:2015fda,Handley:2015vkr}
with parameters \texttt{LivePoints = 1000}, \texttt{num\_repeats = 60}, 
\texttt{boost \_posteriors = 10}, and \texttt{tolerance = 0.001} for all analyses.
We sampled only over $\Omegac$ and $\As$ for our cosmological parameters, giving
results in terms of $\Omegam$, $\sigmaeight$, and $\Seight$, since cosmic shear
surveys are most sensitive to these cosmological parameters. Thus,
any induced bias will be greatest in these parameters.
In addition, we also
sample over our one, three, and six baryonic feedback parameters, depending
on the model, with wide priors, as shown in Table~\ref{tbl:baryon_priors}, on these nuisance parameters.

\begin{table}
  \caption{Uniform priors used for the \HMCode baryonic feedback models.
  $\Tagn$ is used only in the one-parameter models, with the three-parameter
  model sampling over just the individual amplitude parameters ($X_0$), and the
  six-parameter model including the redshift dependence too ($X_z$). See
  Figure~\ref{fig:HMCode_baryons} for the affect of these astrophysical 
  parameters on the matter power spectrum, its description for a full explanation
  of their physical interpretation, and Figure~\ref{fig:cl_baryon_ratio_3param} for their affect
  on the cosmic shear angular power spectrum. We give a brief reprise of their
  description here: $B$ is the halo concentration parameter which accounts for
  gas being expelled from halo centres through AGN activity, $f_\star$ representing
  the contribution of stellar objects at the centre of haloes, and $M_\textrm{b}$
  changing the value of the characteristic mass that haloes are susceptible to
  gas loss from AGN activity~\cite{Mead:2020vgs}. 
  }
  
  \label{tbl:baryon_priors}
  
  \centering

  \setlength\tabcolsep{0.55cm}
  \begin{tabular}{lr}
    \toprule

    Parameter & Uniform prior \\
    
    \midrule
    
    $\Tagn$ & $[4.0, \, 12.0]$ \\

    $B_0$ & $[0.25, \, 7.0]$ \\
    $B_z$ & $[-0.5, \, 0.5]$ \\

    $f_{\star, 0}$ & $[0.0, \, 5.0]$ \\
    $f_{\star, z}$ & $[-5.0, \,  5.0]$ \\
    
    $M_{\textrm{b}, 0}$ & $[5.0, \, 20.0]$ \\
    $M_{\textrm{b}, z}$ & $[-2.5, \, 2.5]$ \\
    
    \bottomrule
  \end{tabular}
\end{table}

We ran analyses with just the ordinary Gaussian covariance matrix for our cosmic
shear fields (Equation~\ref{eqn:Gaussian_Cl_cov}), and with the additional
theoretical error covariances (Equation~\ref{eqn:cov_dbl_integral}) added to
the Gaussian covariance (Equation~\ref{eqn:cov_addition}).

A Gaussian likelihood was used, which has been shown to be 
a good approximation to the full Wishart distribution on the 
cut-sky~\citep{Upham:2020klf,Hall:2022das} under the
assumption of Gaussian-only terms in the covariance matrix, which we are 
employing.

We generate the input cosmic shear $\Cl$ values for each hydro-sim by taking
each hydro-sim's baryon response function $R$ and multiplying it by \HMCode's
dark-matter only matter power spectrum at the hydro-sim's cosmology, giving
\begin{align}
  \Phykz = R(k, z) \, \times \, P^{\textsc{HMcode-DM}}(k, z).
  \label{eqn:baryon_response_func}
\end{align}
Thus, we are isolating the effects of baryon feedback in the $\Cl$ values, and
not studying other effects, for example the details of the non-linear matter
power spectrum.

\section{Results}
\label{sec:paper2_results}

\subsection{Results of fitting HMCode to hydro-sims}

\begin{figure}[t]
  \centering
  \includegraphics[width=0.85\columnwidth]{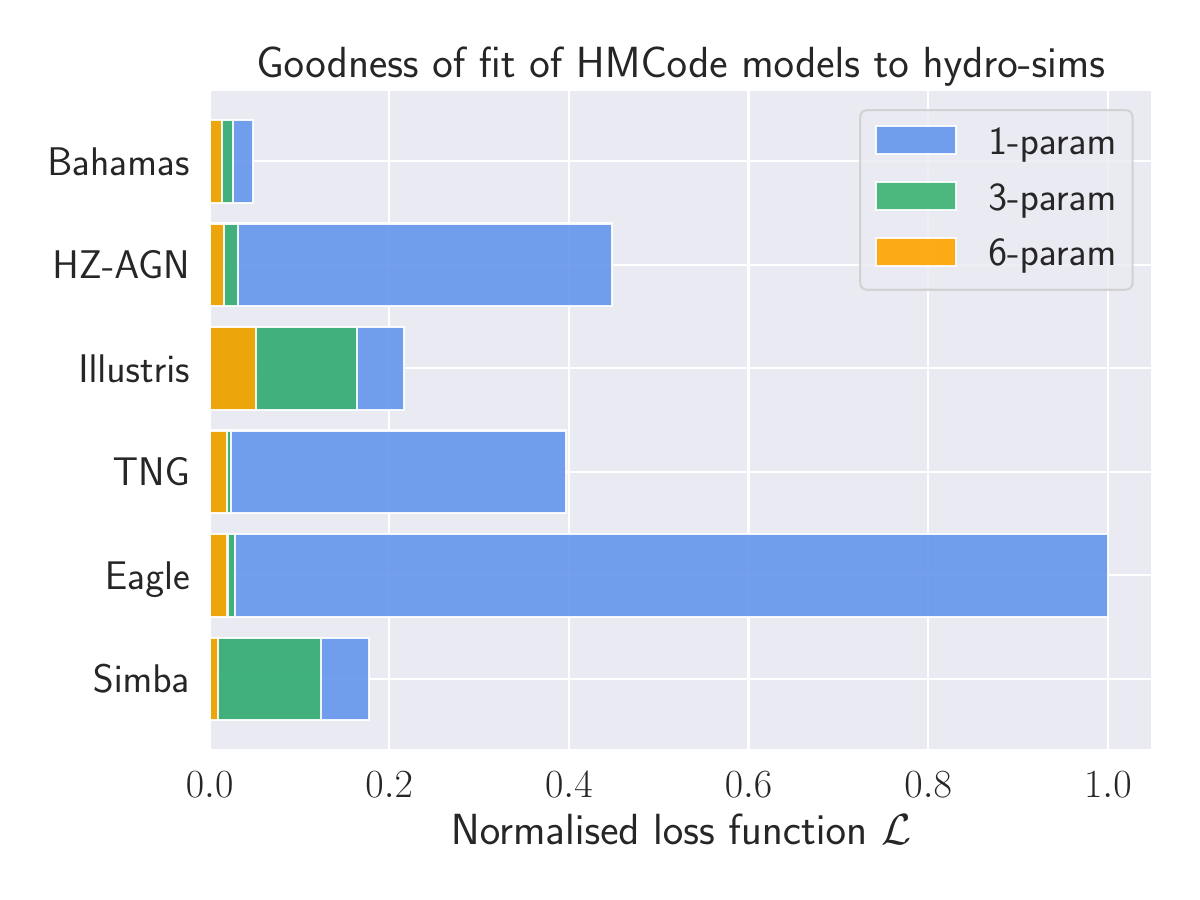}
  \vspace*{-0.5cm}
  \caption{Goodness of fit statistics for the \HMCode 1-, 3-, and 6-parameter
    models for our suite of hydro-simulations. We plot the reduced loss function
    $\mathcal{L}$ values (Equation~\ref{eqn:chi_sq} divided by the total number of data-points
    across our redshift and wavenumber ranges) since each hydro-sim has a
    different number of redshift bins and wavenumbers that the power spectra
    were evaluated at, normalised to the one-parameter model for the Eagle hydro-sim.  
    Here, we see the inflexibility of the 1-parameter model results in
    significant deviations across all simulations (except \Bahamas which it was
    constructed to fit well), which indicates a poor fit to the data. Extending
    the model to 3- and then 6-parameters further increases the goodness
    of fit to simulations, since we open up addition degrees of freedom within
    the model. We note that the feedback within \Illustris is quite extreme, and
    thus produces a degraded fit even with the six-parameter model. 
    }
  \label{fig:HMcode_Pk_bestfit}
\end{figure}

The first task was to fit our three \HMCode baryon physics models to the suite
of hydro-sims. Figure~\ref{fig:HMcode_Pk_bestfit} shows the goodness of fit
statistic for each hydro-sim for our 1-, 3-, and 6-parameter model. We see that 
the 1-parameter model generally provides a poor fit to our hydro-sims due to its
inability to match the general baryon feedback models. When we extend the
generality to 3- and 6-parameters, the goodness of fit increases which matches
our intuition that a more general model with higher degrees of freedom should
do a better job at matching arbitrary data sets. 
It should be noted that
the \HMCode 1-parameter model was specifically designed to closely match the
\Bahamas simulations, and so it is no surprise that it has a better fit to 
\Bahamas than other simulations. 

\begin{sidewaysfigure}
  \centering
  \includegraphics[width=\columnwidth]{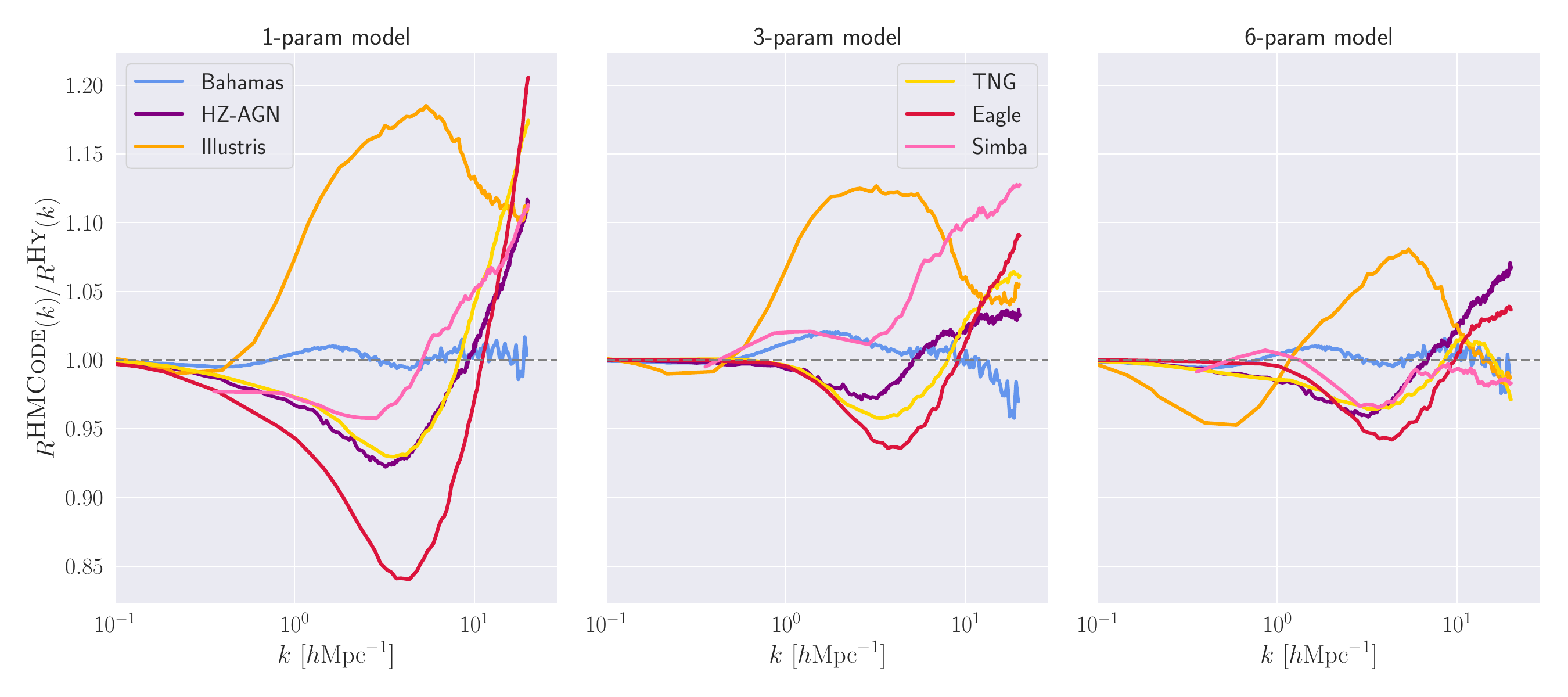}
  \vspace*{-0.25cm}
  \caption{Ratios of the best-fitting \HMCode one-, three-, and six-parameter 
    models to the underlying hydro-sims baryonic response function $R$ at redshift $z=0$.
    We see that \HMCode tends to fit the data better, as given by the smaller
    amplitude, as we increase the generality of the model, which is to be expected.
    Note that we fitted across all redshifts of each of the hydro-sims for 
    $z \leq 2$, where here we are just plotting the $z=0$ slice. 
    Similar curves are found at higher redshifts.}
  \label{fig:R_HMCode_hydros}
\end{sidewaysfigure}

While our goodness of fit values provide a valuable insight into how well our
three models fit the simulation data overall, we can look at the ratios for
the \HMCode predictions to the hydro-sims results to see how our fit changes
with scales. Figure~\ref{fig:R_HMCode_hydros} plots the ratio of the best-fit \HMCode
model to the hydro-sims baryonic response function for our three $n$-parameter
models as a function of scale at redshift $z=0$. 
We see that the more general three- and six-parameter models are
better able to match the hydro-sims than the one-parameter model, which is to be 
expected from more general models, and is shown by a reduced amplitude in the
relative differences. We also see the extreme nature of the \Illustris 
simulation, where the six-parameter model can only poorly match its feedback. 
This may question the validity of the \HMCode-2020 feedback model since
the more extreme nature of the feedback as found in \Illustris as been shown 
to more accurately match the feedback present in our Universe using 
Sunyaev-Zeldovich measurements~\cite{ACT:2024vsj} and strongly disfavours
simulations with lower feedback amplitudes~\cite{Bigwood:2025ism}. We leave
the task of reconstructing these fits with other, perhaps more general, models
of baryonic feedback that can fit larger-amplitude simulations to future work.

\subsection{Constructing the envelope}

Now that we know how well each of our feedback models match our suit of
hydro-sims, we can combine these to form our `envelope function' $\Delta R(k, z)$
introduced in Equation~\ref{eqn:max_deviation}. $\Delta R$ encodes the maximum
deviation of each baryon feedback model to all hydro-sims as a function of
wavenumber and redshift. This acts as a standard deviation to our theoretical
error as each baryon feedback model has an inherent uncertainty and we have 
quantified this with the envelope of the residual discrepancies between the
hydro-sims and the best-fitting $R(k, z)$ for each simulation.

\begin{sidewaysfigure}[tp]
  \centering
  \includegraphics[width=\columnwidth]{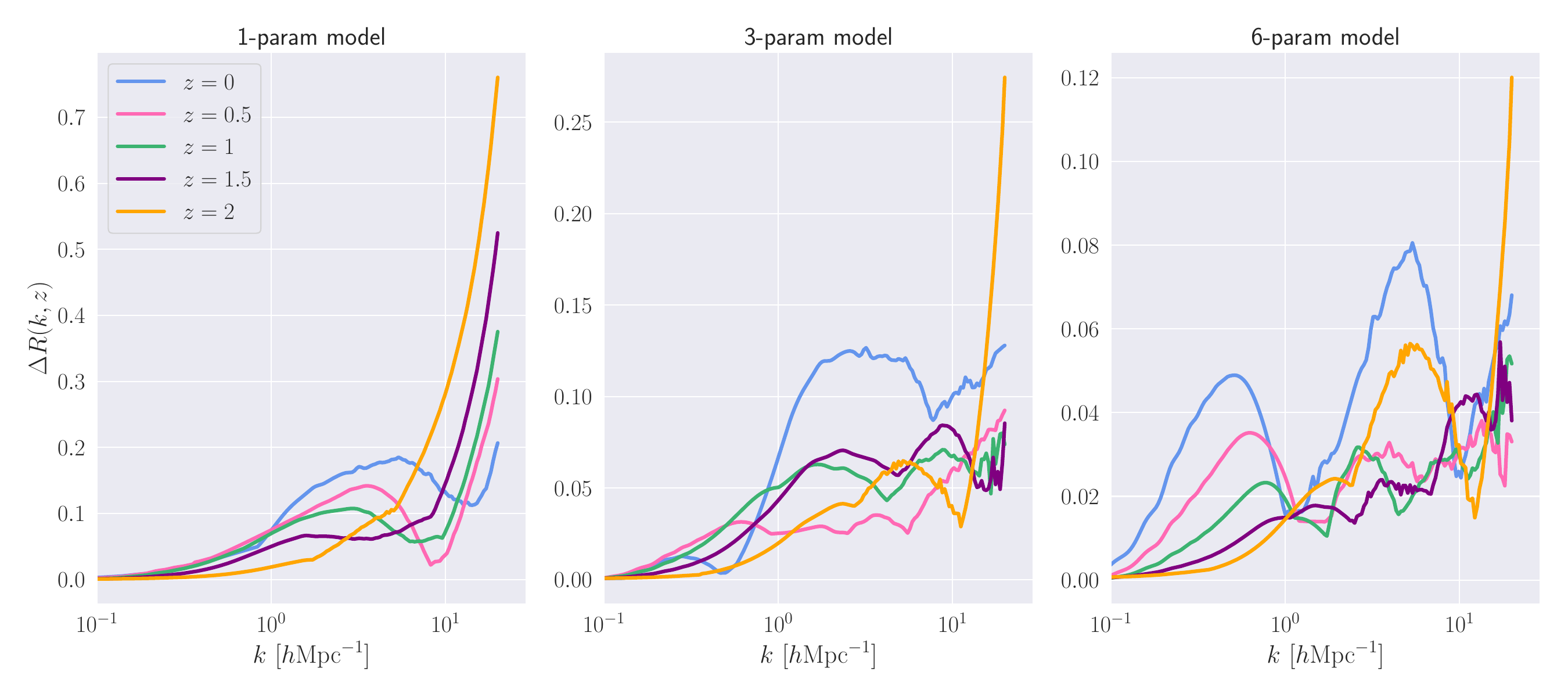}
  \vspace*{-0.25cm}
  \caption{Our envelope function $\Delta R(k, z)$ encoding the maximum deviation of
    our three \HMCode models to the hydro-sims as a function of wavenumber $k$
    plotted for select redshifts. We see that the amplitude of our envelope
    decreases as we go from one- to three- to six-parameters (as noted by the
    decreasing values in each of the panel's individual $y$-axes), and so we assign less
    theoretical uncertainties to those models that better match the data. The
    apparent noisy behaviour of these curves is due to numerical noise in the
    hydro-sims, but also from our choice of envelope function, which uses
    the absolute value of the error ratio. 
    }
  \label{fig:Pk_envelope}
\end{sidewaysfigure}

We plot $\Delta R(k, z)$ in Figure~\ref{fig:Pk_envelope}, which shows our envelope function as a
function of wavenumber $k$ for select redshift values. We see that, in general,
the errors increase with $k$ which corroborates our understanding that our
theoretical models do worse the smaller scales we probe. We also see that
higher redshifts tend to produce a better fit across all of the models which can
be understood through Figure~\ref{fig:My_Pk_ratio}: we see that at higher
redshifts, the suppression due to baryons is reduced, smoothing our response 
function $R$, and also moving the suppression to smaller wavenumbers. This acts
to allow our baryonic feedback models to better match the hydro-sims, though
our errors do increase to very large values on at high $k$ at high redshifts.

We note that the six parameter model has a distinctive double-hump feature,
which is a direct result of the poor fit to \Illustris on scales
$10^{-1} h \textrm{Mpc}^{-1} < k < 1 h \textrm{Mpc}^{-1}$, as shown in 
Figure~\ref{fig:R_HMCode_hydros}. One could remove \Illustris when constructing
the maximum deviation envelope, and such this double-hump would no longer appear. 

\subsection[Constructing the $\ell$-space theoretical uncertainty covariance matrix]{Constructing the $\bm \ell$-space theoretical uncertainty covariance matrix}

\begin{figure}[tp]
  \centering
  \includegraphics[width=0.95\columnwidth,trim={0.0cm 0.0cm 2.0cm 0.0cm},clip]{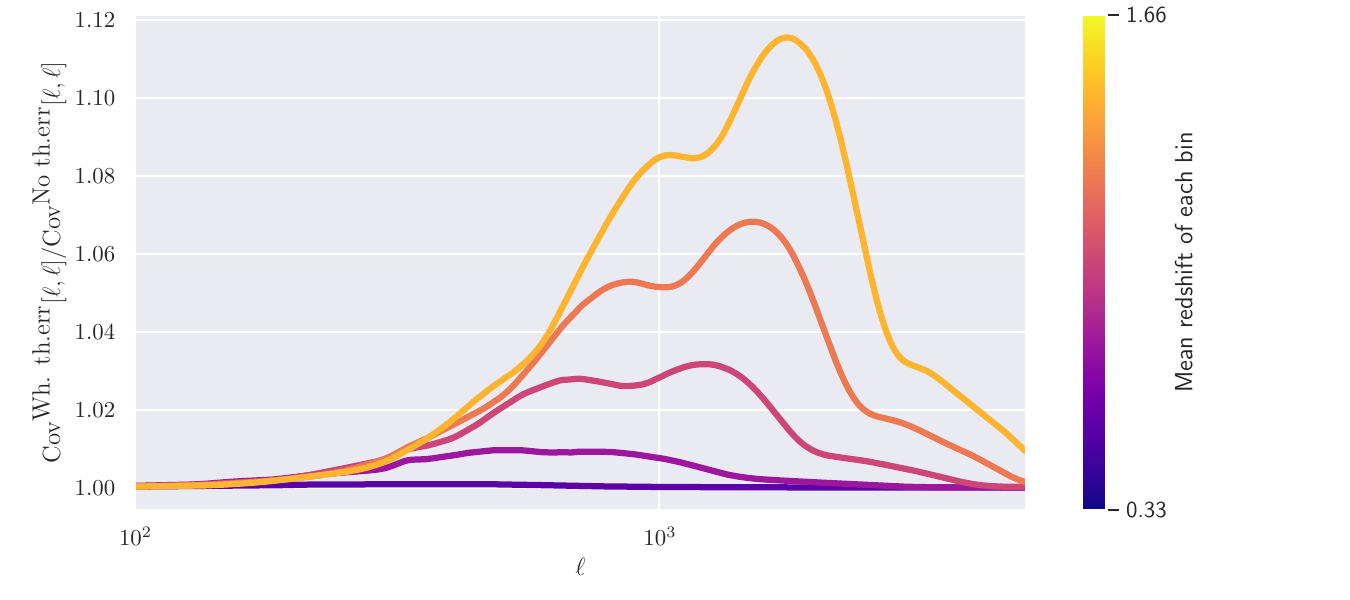}
  \vspace*{-0.25cm}
  \caption{Ratio of the diagonals of auto-spectra elements of our $\Cl$ covariance matrix with 
    our additional theoretical error to that without theoretical error. Here,
    we see that the amplitude of the additional theoretical error strongly
    depends on the redshift bins of the auto-spectra considered, with higher redshift bins
    containing more theoretical error than closer bins. We note that there is
    significant support on the off-diagonals for our theoretical covariance, and
    thus the total covariance with theoretical error is much larger than just
    the ratios showed here. 
    }

  \label{fig:Cl_cov_diag}
\end{figure}

Our numerical envelope in $(k, z)$-space can then be doubly integrated (Equation~\ref{eqn:cov_dbl_integral})
to give us our $\ell$-space covariance matrix. Figure~\ref{fig:Cl_cov_diag} plots
the ratio of the block diagonals of the covariance matrix with our additional
theoretical error to that without theoretical error. We see that for $\ell$ modes
below $\ell \simeq 200$ there is no effect of our theoretical error covariance,
since these $\ell$ modes are unaffected by baryon feedback physics. Above this,
we see that out theoretical error covariance acts to increase the total covariance,
thus suppressing $\ell$ modes here and down-weighting them in our likelihood
analyses. We note that this is strongly dependent on the combination of spectra
considered, with low redshift bins having a smaller error than those at high
redshift (which is a direct consequence of our baryon feedback modelling being
less accurate at high redshift). We see that, because we are considering the
covariances of the auto-spectra only, that the ratios tend to unity at
high-$\ell$ which is due to the inclusion of shape noise (which dominates the
auto-spectra for large $\ell$ modes) in the overall covariance matrix.

\begin{figure}[tp]
  \centering
  \includegraphics[width=\columnwidth,trim={0.0cm 0.0cm 0.0cm 0.0cm},clip]{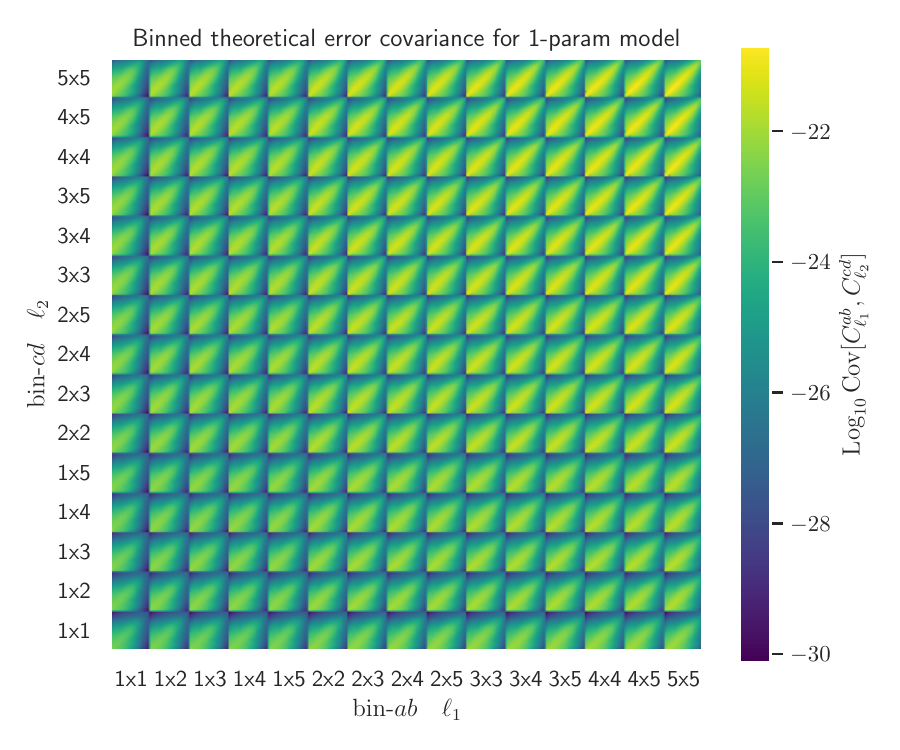}
  
  \vspace*{-0.25cm}

  \caption{Plot of the binned theoretical error matrix $\mathbfss{C}^{\textsc{e}}$
    for the one-parameter model across all fifteen redshift bin combinations and
    twenty five binned band-powers. Here, we see that the amplitude of our
    theoretical error matrix generally grows as we consider higher redshift bins
    (e.g. the 5x5$\times$5x5 sub-block is significantly brighter, and thus
    of larger magnitude, than the 1x1$\times$1x1 sub-block). We also see that
    for each sub-block, the theoretical error is greatest along the 
    $\ell_1 = \ell_2$ diagonal, though there is significant support along the
    off-diagonal terms, which serves to increase the overall effect of our
    theoretical uncertainty modelling.
     }

  \label{fig:Cl_cov_mat}
\end{figure}

We plot the total theoretical uncertainty matrix, $\mathbfss{C}^{\textsc{e}}$
in Figure~\ref{fig:Cl_cov_mat}. Here, we see that the amplitude of the full
matrix reflects what is shown in the diagonals only of Figure~\ref{fig:Cl_cov_diag},
that the amplitude of the theoretical uncertainties generally grow as we consider 
further away redshift bin combinations (going from bottom-left to top-right).
We also see that while each sub-block of the full matrix peaks along its diagonal,
it has significant support with comparable values for many off-diagonal entries,
which acts to boost the effects of the theoretical uncertainties over what
the diagonal values can provide.

\subsection{Parameter constraints and biases}

Using our three different baryon feedback models, we can run our MCMC pipeline
to estimate the biases in our cosmological parameters, the total matter density $\Omegam$
and the lensing amplitude $\Seight$, with and without our additional theoretical
error. We take each hydrodynamical simulations' baryon feedback response function
as the ground truth input values into our cosmic shear pipeline 
(Equation~\ref{eqn:baryon_response_func}).
We run our analyses for each baryon feedback model against each hydro-sim,
with and without the additional theoretical error to
determine the biases on the cosmological parameters due to the effects of
baryon feedback.

\subsubsection{Binary scale-cuts}
\begin{figure}[t]
  \centering

  \includegraphics[width=\columnwidth]{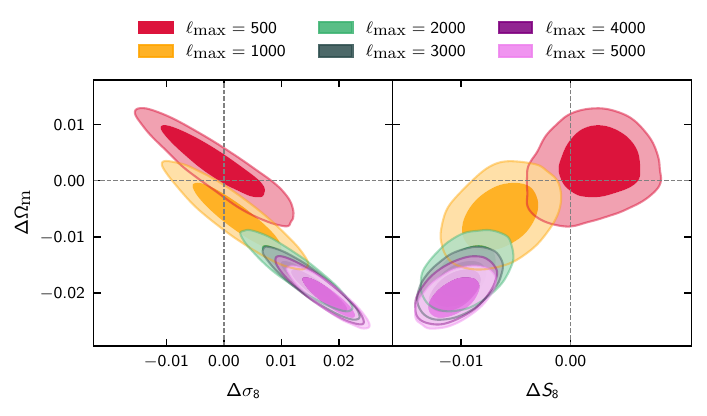}
  
  \vspace*{-0.25cm}

  \caption{2D $\Omegam$-$\sigmaeight$ and $\Omegam$-$\Seight$ offset contours
    for the one-parameter baryon feedback model for the \Simba hydro-sim for
    a range of maximum $\ell$-modes considered in the analysis, without the
    inclusion of our additional theoretical uncertainty covariance. As we increase
    $\lmax$, we are increasing the number of modes that are
    contaminated by baryon feedback, and thus bias our cosmological parameters
    away from the fiducial value. It is interesting to note that that even for
    the $\lmax = 1000$ analysis, which is a considerably conservative scale-cut
    even for Stage-III surveys (e.g. HSC-Y3 \cite{Dalal:2023olq}), our results
    are biased by more than $2\sigma$ in all cosmological parameters for our
    Stage-IV-like survey when using \Simba as the ground-truth. This motivates
    the development and inclusion of a theoretical uncertainty covariance as to
    retain the maximum information possible from the smallest scales, without
    incurring the significant biases seen in this Figure.
    }

  \label{fig:simba_scale_cuts}
\end{figure}

We first look at the value of the biases in $\Omegam$ and $\Seight$
as a function of maximum multipole 
when using a traditional binary scale-cuts approach. Figure~\ref{fig:simba_scale_cuts}
plots the $\Omegam$, $\sigmaeight$, and $\Seight$ offset contours for the Simba hydro-sim
when analysed using the one-parameter model for a range of maximum $\ell$-modes
allowed in the analysis. We see that only the $\lmax = 500$ analysis produces
results that are consistent with the input parameters to within $1\sigma$, with
even the $\lmax = 1000$ analysis producing biased results. Each subsequent
analysis where we increase the maximum $\ell$-mode serves to both increase the
raw bias in the cosmological parameters (since the high $\ell$-modes capture 
more of the impact from baryon feedback) and reduce the area of the contours (since
we are now including the more constraining higher $\ell$-modes), which vastly
increases the relative bias in the cosmological parameters.

Hence, if we were to introduce a binary scale-cut for our data that keeps the
$\Delta \Omegam - \Delta \Seight$ to be consistent within $1\sigma$, a
cut slightly larger than $\lmax = 500$ might be made for our idealised
Stage-IV like cosmic shear survey. However, to avoid making these analysis 
decisions ourselves, which have little physical motivation behind them and
apply simultaneously to all hydro-sims in our ensemble, we can turn to our
theoretical error modelling instead of making binary scale-cuts. 

\subsubsection{Including a theoretical error covariance}

Tables~\ref{tab:Omega_m_biases} and~\ref{tab:S8_biases} summarises our results
for the biases in $\Omegam$ and $\Seight$, respectively. Here, we present the
amplitude of the bias, that is the difference in the mean of our MCMC chains
to the fiducial value, and the relative bias, which is the difference divided by
the reported standard deviation from the MCMC chains for each parameter, with
and without the effects of our additional theoretical error. The baryonic
feedback parameters for each model are marginalised over when we present
results for $\Omegam$, $\sigmaeight$ and $\Seight$.

\subsubsection{One-parameter model}

In general, the one-parameter model without additional theoretical 
error produces a significant biases in the recovered parameters, with at least a
3$\sigma$ offset in either $\Omegam$ or $\Seight$ for all hydro-sims excluding
\Bahamas (recall that the one-parameter model was constructed by fitting to
\Bahamas data only). The results of the very significant biases in the remaining
five hydro-sims shows that the one-parameter model of baryon feedback for a
\textit{Euclid}-like  survey up to $\lmax = 5\,000$ is completely impractical if
we wish to recover unbiased results from the analyses of cosmic shear data. 

When we introduce our additional theoretical error into the analysis, we see
that biases in both cosmological parameters fall dramatically across all 
hydro-sims for our one-parameter model. We see a reduction in both the raw
offset values, and a significant decrease in the relative bias in terms of each
parameter's standard deviation. This demonstrates our theoretical uncertainties
modelling is correctly identifying the scales in which baryonic feedback are
not correctly modelled by the one-parameter model, down-weighting them in
our analyses, and thus resulting in significantly less biased cosmological
parameter constraints. However, we still see significant ($>\!1\sigma$)
biases in many hydro-sims even when using theoretical error. This is due to the
one-parameter model being inadequate when considering a \textit{Euclid}-like
weak lensing survey.

\subsubsection{Three- and six-parameter models}

As expected, the more general three- and six-parameter models result in smaller
absolute biases, which reflects their ability to better match more general
baryon feedback scenarios, but at the cost of decreased precision due to the
need to marginalise over more parameters. Hence, we see that the relative
biases show a strong decrease when going to our many-parameter versions, which
is an effect of decreased absolute bias and increased uncertainties.

\begin{figure}[tp]
  \centering
  \includegraphics[width=\columnwidth]{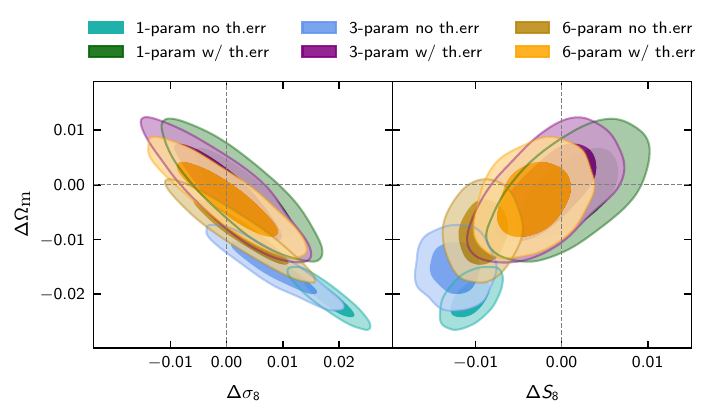}
  
  \vspace*{-0.25cm}

  \caption{2D $\Omegam$-$\sigmaeight$ and $\Omegam$-$\Seight$ offset contours
    for the three baryon feedback models with and without our theoretical
    error for the \Simba hydro-sim. We can clearly see that the one-parameter
    model without theoretical error (light grey contour) produces a tight 
    constraint on our cosmological parameters that is extremely
    far from the correct value. When we add our theoretical error to the data
    covariance (dark green contour), we find that the one-parameter model 
    \textit{is} able to recover the correct values, with an appropriate
    increase in the size of the contour. We see that the three- and 
    six-parameter models, while more flexible, are unable to correctly
    recover unbiased results without the addition of our theoretical
    uncertainties.
    }

  \label{fig:Simba_2D_contour}
\end{figure}

Figure~\ref{fig:Simba_2D_contour} plots the 2D $\Omegam$-$\sigmaeight$ and 
$\Omegam$-$\Seight$ offset contours for our three baryon feedback models with
and without our theoretical error added for the Simba hydro-sim. This highlights
the degeneracies that exist between these parameters, finding the usual `lensing
banana' that is the natural degeneracy between $\Omegam$ and $\sigmaeight$. 
These contours gradually move to the origin as we increase the number of
parameters in the baryon feedback model, and increase in size as we are
marginalising over more parameters.

\begin{sidewaysfigure}[tp]
  \centering
  \includegraphics[width=\columnwidth]{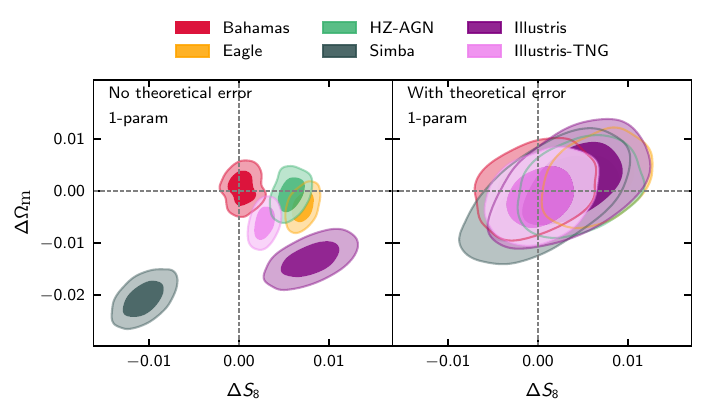}
  
  \vspace*{-0.25cm}

  \caption{2D $\Omegam$-$\Seight$ offset contours for the one-parameter
    baryon feedback model without and with the addition of our theoretical
    error for our range of hydro-sims considered. We can see that without
    theoretical error, the one-parameter model produces a large degree of scatter
    for all hydro-sims, with only \Bahamas being consistent with the input
    cosmology. 
    With the introduction of theoretical error, we see that the contours 
    increase and all except \Eagle become unbiased at the $2 \sigma$ level.
    }

  \label{fig:2D_one_param}
\end{sidewaysfigure}

Figure~\ref{fig:2D_one_param} plots the 2D $\Omegam$-$\sigmaeight$ offset 
contours for the one-parameter model for our suite of hydro-sims for the case of
with and without our theoretical error. We see that without the theoretical
error the recovered contours are highly constraining, which is a result of
marginalising over a single baryon feedback parameter only, with a large degree
of scatter between the hydro-sims. It is interesting to note that this scatter
appears to be roughly along the degeneracy direction, though it appears random
which quadrant each hydro-sim falls into.

\begin{sidewaysfigure}[tp]
  \centering
  \includegraphics[width=\columnwidth]{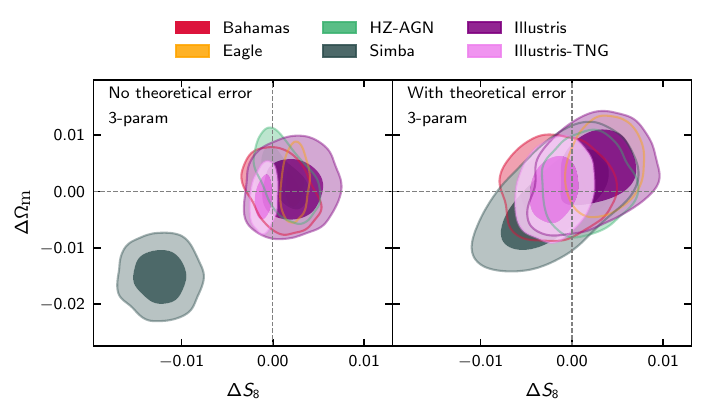}
  
  \vspace*{-0.25cm}

  \caption{2D $\Omegam$-$\Seight$ offset contours for the three-parameter
    baryon feedback model without and with the addition of our theoretical
    error for our range of hydro-sims considered. When compared to the less
    flexible one-parameter model, we see that the three parameter model without
    additional theoretical error is better able to recover the input cosmology
    for a wider range of simulations, though \Eagle, \TNG, and \Simba remain
    biased to at least the $2 \sigma$ level. We note the general increase in
    the contour's area going from the one- to three-parameter model which is
    associated with marginalising over more baryon feedback parameters.
    We see that with theoretical error, all contours are unbiased at the 
    $2\sigma$ level. 
    }

  \label{fig:2D_three_param}
\end{sidewaysfigure}

\begin{sidewaysfigure}[tp]
  \centering
  \includegraphics[width=\columnwidth]{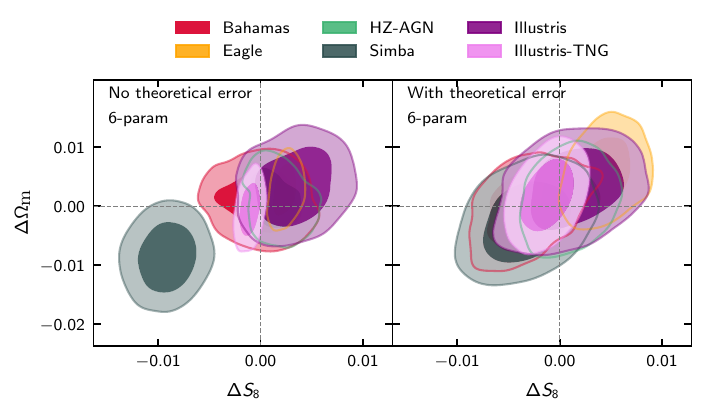}
  
  \vspace*{-0.25cm}

  \caption{2D $\Omegam$-$\Seight$ offset contours for the six-parameter
    baryon feedback model without and with the addition of our theoretical
    error for our range of hydro-sims considered. Again, we see an extension
    of the general trend in the contours without additional theoretical error:
    an increase in the contours with them moving towards the origin. We see that
    even with a six-parameter model, it is not sufficient to recover unbiased
    constraints from the \Simba simulation.
    }

  \label{fig:2D_six_param}
\end{sidewaysfigure}

\begin{sidewaystable}
  \centering
  \caption{Table summarising our results for the bias in $\Omegam$ as found from
    our MCMC analyses using our suite of six hydro-sims and three baryon feedback
    models, without and with our additional theoretical error modelling added
    to the covariance matrix when going to $\lmax = 5000$.
    We see that without theoretical error, the 
    one-parameter model generally produces significant biases ($> \! 3 \sigma$) 
    from the true value, with only \Bahamas giving small biases in $\Omegam$
    and $\Seight$.
    These biases reduce to less than $1 \sigma$ with the inclusion of our 
    theoretical error in the one-parameter model, highlighting the effectiveness
    of our theoretical error modelling. We see that the three- and six-parameter
    result in less bias in $\Omegam$, though some hydro-sims still give large
    biases for the six-parameter model (i.e. \Simba and \Eagle). The inclusion
    of our theoretical error covariance into the three- and six-parameter models
    further reduces the bias.
    }
  \label{tab:Omega_m_biases}
  
  \begin{tabular}{lrrrr}
    \toprule
    Hydro-sim and model
    & \quad $100 \Delta \Omegam$ No th. err. &  \quad $\Delta \Omegam / \sigma_{\Omegam}$  No th. err. 
    & \quad $100 \Delta \Omegam$ With th. err. & \quad  $\Delta \Omegam / \sigma_{\Omegam}$ With th. err. \\

    \midrule
    
    \Bahamas 1-param & 0.057 & 0.258 & 0.038 & 0.098 \\ 
    \Bahamas 3-param & 0.012 & 0.038 & 0.049 & 0.127 \\ 
    \Bahamas 6-param & 0.100 & 0.276 & -0.085 & -0.206 \\[0.5em]

    \Eagle 1-param & -0.306 & -1.535 & 0.224 & 0.587 \\
    \Eagle 3-param & 0.149 & 0.520 & 0.433 & 1.186 \\
    \Eagle 6-param & 0.274 & 0.962 & 0.573 & 1.459 \\[0.5em]

    \HZAGN 1-param & -0.057 & -0.255 & 0.079 & 0.199 \\
    \HZAGN 3-param & 0.224 & 0.631 & 0.123 & 0.331 \\
    \HZAGN 6-param & 0.109 & 0.325 & 0.105 & 0.269 \\[0.5em]

    \Illustris 1-param & -1.317 & -5.690 & 0.128 & 0.249 \\
    \Illustris 3-param & 0.048 & 0.127 & 0.373 & 0.828 \\
    \Illustris 6-param & 0.328 & 0.761 & 0.295 & 0.652 \\[0.5em]

    \Simba 1-param & -2.071 & -8.828 & -0.158 & -0.293 \\
    \Simba 3-param & -1.513 & -4.723 & -0.155 & -0.276 \\
    \Simba 6-param & -0.858 & -2.206 & -0.257 & -0.567 \\[0.5em]

    \TNG 1-param & -0.613 & -2.885 & -0.107 & -0.277 \\
    \TNG 3-param & -0.110 & -0.424 & 0.035 & 0.091 \\
    \TNG 6-param & -0.055 & -0.185 & 0.181 & 0.453 \\[0.5em]
    
    \bottomrule
  \end{tabular}
\end{sidewaystable}  

\begin{sidewaystable}
  \centering
  \caption{Table similar to Table~\ref{tab:Omega_m_biases} but now for the bias in $\Seight$ as found from
    our MCMC analyses. We see that without theoretical error, the 
    one-parameter model results in at least a 3-\!$\sigma$ bias from the true value
    for all hydro-sims except \Bahamas, which reduces when going to a multi-parameter model.
    Our theoretical error is able to significantly reduce the bias in the 
    one-parameter model, though we see a multi-parameter model is essential to
    ensure unbiased constraints across all hydro-sims.
    }
  \label{tab:S8_biases}
  
  \begin{tabular}{lrrrr}
    \toprule
    Hydro-sim and model 
    & \quad $100 \Delta \Seight$ No th. err. & \quad $\Delta \Seight / \sigma_{\Seight}$  No th. err. 
    & \quad $100 \Delta \Seight$ With th. err. & \quad $\Delta \Seight / \sigma_{\Seight}$ With th. err. \\

    \midrule
    
    \Bahamas 1-param & 0.038 & 0.386 & -0.028 & -0.101 \\
    \Bahamas 3-param & 0.083 & 0.492 & -0.210 & -0.819 \\
    \Bahamas 6-param & 0.010 & 0.042 & -0.360 & -1.468 \\[0.5em]

    \Eagle 1-param & 0.709 & 8.802 & 0.660 & 2.579 \\
    \Eagle 3-param & 0.240 & 3.742 & 0.360 & 2.039 \\
    \Eagle 6-param & 0.240 & 3.207 & 0.441 & 2.443 \\[0.5em]

    \HZAGN 1-param & 0.574 & 6.002 & 0.464 & 1.616 \\
    \HZAGN 3-param & 0.137 & 0.865 & 0.158 & 0.759 \\
    \HZAGN 6-param & 0.165 & 1.070 & 0.089 & 0.448 \\[0.5em]

    \Illustris 1-param & 0.796 & 3.710 & 0.377 & 1.024 \\
    \Illustris 3-param & 0.208 & 0.950 & 0.253 & 0.867 \\
    \Illustris 6-param & 0.325 & 1.289 & 0.135 & 0.435 \\[0.5em]

    \Simba 1-param & -1.059 & -7.096 & 0.081 & 0.207 \\
    \Simba 3-param & -1.235 & -6.469 & -0.180 & -0.480 \\
    \Simba 6-param & -0.916 & -4.903 & -0.328 & -1.156 \\[0.5em]

    \TNG 1-param & 0.281 & 3.701 & 0.033 & 0.130 \\
    \TNG 3-param & -0.100 & -1.602 & -0.193 & -1.083 \\
    \TNG 6-param & -0.106 & -1.628 & -0.122 & -0.715 \\[0.5em]

    \bottomrule
  \end{tabular}
\end{sidewaystable}  

\clearpage
\subsubsection{Dependency on $\sigma_k$ and $\sigma_z$}
\label{sec:paper2_sigma_kz}

In our propagated covariance matrix (Equation~\ref{eqn:k_space_cov}) we introduced
two coupling scales, $\sigma_k$ and $\sigma_z$, which were both set at $0.25$
in our fiducial analyses. Here, we investigate the effects of changing their
values, and the effects it has on parameter constraints. We note that in our
fiducial analyses, the baryon feedback physics of the \Simba simulation is strong
enough that even the six-parameter model of baryon feedback produces significantly
biased cosmological parameter constraints (Figure~\ref{fig:2D_six_param}).
Thus, \Simba is a good testing ground to see how the contours react to the
changing of these values.

In general, as $\sigma_k$ and $\sigma_z$ increase we are increasing the number 
of wavenumber and redshift modes that contribute to a given $(\ell_1, \, \ell_2)$
pair, respectively. This increases the size our $\ell$-space theoretical error
covariance, further suppressing the high-$\ell$ modes that are contaminated
by baryonic feedback. However, in the limit that $\sigma_k$ and $\sigma_z$ 
tend to infinity, then both Gaussian terms tend to unity and the double integral
becomes factorisable in $(k, \, k')$-space and thus is equivalent to some
rank-1 update of the form $\mathbfit{x} \, \mathbfit{x}^{\textsc{t}}$, 
where $\mathbfit{x}$ is some vector in $\ell$-space. Since this is a rank-1
addition to our data covariance matrix, this is equivalent to marginalisation
over a single parameter. However, since we know that single parameter models
for baryonic feedback do not correctly model the wide range of behaviour of
baryonic feedback effects seen in the hydro-sims, it would be incorrect
to take this limit for the values of $\sigma_k$ and $\sigma_z$.

The regime in which our theoretical error covariance reduces to a rank-1 matrix
will be determined by the maximum $k$-space difference between different
$\ell$-modes. For example, Figure~\ref{fig:Cl_deriv_cumsum} shows that there is
an approximate two decade difference in log-$k$ between our smallest and
largest $\ell$-mode. Thus, if one were to choose a value for $\sigma_k$ that
easily encompassed these values within one standard deviation, then we would
expect our theoretical error covariance to collapse to the rank-1 limit.
Since our values chosen for $\sigma_k$ are much less than two decades, we 
expect that our values to not fall within this limit.

\begin{figure}[tp]
  \includegraphics[width=\columnwidth]{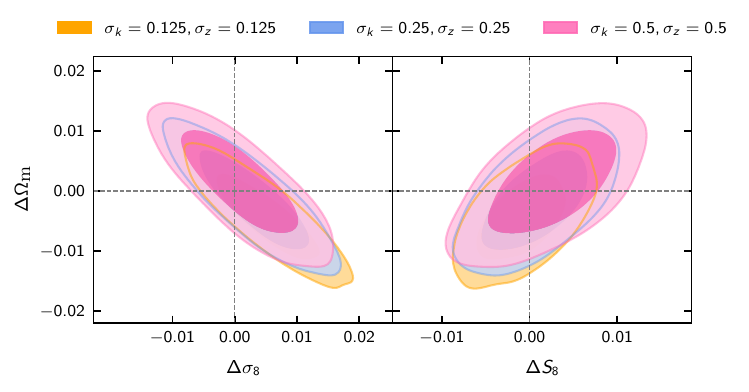}
  
  \vspace*{-0.25cm}

  \caption{2D $\Omegam$-$\sigmaeight$ and $\Omegam$-$\Seight$ contours for 
  the Simba hydro-sim using the one-parameter model with theoretical error
  where we are changing the values of $\sigma_k$ and $\sigma_z$. Here, we see
  that as we increase their values, the effects of the theoretical error
  covariance become larger - increasing the size and decreasing the biases 
  in the contours.
  }

  \label{fig:2D_sigma_kz}
\end{figure}

In Figure \ref{fig:2D_sigma_kz}, we plot the 2D contours for three different
combinations of $\sigma_k$ and $\sigma_z$. Larger values of
these parameters allows for longer-scale correlations in wavenumber and redshift,
and thus produce a larger amplitude for the theoretical error covariance
matrix. Figure~\ref{fig:2D_sigma_kz} shows that these larger values suppress
high-$\ell$ modes more than for smaller values, which is seen by the larger
contours and a reduction in the bias from baryon feedback. While on the range
of values that we tested larger values increase the effectiveness of the
theoretical error covariance, larger values also increases the correlation
length between different wavenumbers and redshifts in the covariance matrix
which decreases the adaptability of our theoretical error covariance to match
more general models of baryon feedback physics. In the limit that
these values go to infinity, we would be saying that all wavenumbers and redshifts
are $100\,\%$ correlated, which is unphysical since we know that baryonic effects
in the power spectrum are quite local at high $k$. Equally, in the limit that
these values go to zero, we would be maximally destroying information through
the covariance matrix by assuming that each wavenumber and redshift has an
independent error which are uncorrelated between similar modes. Ultimately,
there is less than a $1\sigma$ shift in contours plotted in 
Figure~\ref{fig:2D_sigma_kz}, and so our implementation of the theoretical
error approach is broadly insensitive to the values of $\sigma_{k}$ and
$\sigma_{z}$.

We investigate the effects of changing these coupling parameters in our three-
and six-parameter models in Appendix~\ref{app:sigma_kz}, which shows a similar
trend to the one-parameter model in that larger values produce increased 
contours with less baryonic bias in them.

\clearpage
\subsubsection{Growth-structure split}

The information in the cosmic shear power spectrum (Equation~\ref{eqn:cosmic_shear_powspec})
comes from two pieces: the geometrical factors of the lensing kernels, and
in the matter power spectrum, both depending on the matter density $\Omegam$.
It can be advantageous to decouple this matter
density in two: a factor that describes the geometry of the Universe through
the computation of the comoving distances in the lensing kernel, $\Omegamgeom$,
and the growth of structure in the universe that is used in the computation of
the matter power spectrum, $\Omegam$~\citep{Matilla:2017rmu}. This 
growth-structure split of the matter density has been applied to existing cosmic
shear data-sets, including the Dark Energy Survey~\citep{DES:2020iqt,Zhong:2023how}
and the Kilo-Degree Survey~\citep{Ruiz-Zapatero:2021rzl}.

We are motivated to apply this growth-structure split to our approach of
modelling the theoretical uncertainties at the matter power spectrum level. 
By quantifying the errors directly in $P(k)$, and then propagating to the
angular power spectrum, we are hopefully preserving information in the lensing
kernels. Our lensing kernels have no theoretical uncertainties associated with
them, and so we hope to preserve information about $\Omegamgeom$
even with our scale cuts present.

\begin{figure}[tp]
  \includegraphics[width=\columnwidth]{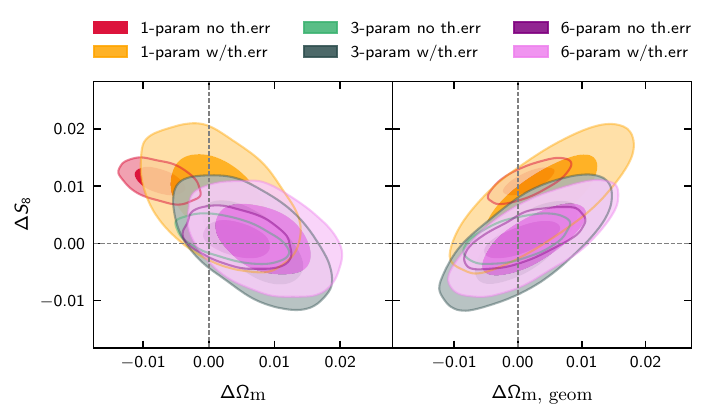}
  
  \vspace*{-0.25cm}

  \caption{2D $\Omegam$-$\Seight$ and $\Omegamgeom$-$\Seight$ contours for 
    different baryon feedback models without and with the addition of our theoretical
    error with the \Eagle hydro-sim. We see that without theoretical error, the 
    one-parameter model produces results that are biased in $\Seight$ and $\Omegam$,
    but can identify that the information from the lensing kernels is not affected by
    baryonic feedback, and thus produces unbiased estimates for $\Omegamgeom$.
  }

  \label{fig:2D_omegam}
\end{figure}

Thus, we can repeat a sub-set of our MCMC analyses to investigate how the
inclusion of our theoretical error changes the constraints on $\Omegam$ and
$\Omegamgeom$. This is presented in Figure~\ref{fig:2D_omegam}. 
Here, we see that without our theoretical uncertainties, 
the inflexible one-parameter model leads to a bias in $\Omegam$
but remains unbiased in the geometrical term $\Omegamgeom$. 
This tell us that while our data-vector may still be contaminated with baryonic
feedback from the matter power spectrum, information from the lensing kernels
can still be extracted from cosmic shear data. When we introduce our theoretical
error terms to the covariance matrix, we down-weight baryonic contaminated
modes, and so become consistent in $\Omegam$ and $\Omegamgeom$
at the cost of increased contour sizes.

\clearpage
\section{Discussion and conclusions}
\label{sec:paper2_diss_and_conc}

We have presented results for parameter constraints using various hydrodynamical
simulations as ground truths analysed with three different baryon feedback models,
with and without an additional theoretical error covariance. We have seen that
constraints on cosmological parameters are significantly biased for a forthcoming
Stage-IV like cosmic shear survey using as low as a maximum $\ell$-mode of
$\lmax = 1000$ when analysing the hydrodynamical simulation's data using
a single parameter for baryon feedback. This is consistent with previous
results in the literature, for example in Ref.~\cite{DES:2020daw} finding a 
harmonic-space scale cut of $\lmax \sim 500$ was needed for the case of
DES-Y3 mock data against a model without baryonic feedback in. 

Hence, we are motivated to more accurately
model the errors associated in the matter power spectrum arising from baryon
feedback over the more traditional binary scale-cuts approach, resulting in
our theoretical error covariance. This theoretical error was then applied to
our model of baryon feedback across our suite of hydrodynamical simulations,
finding that our our parameters biases decreased with smaller absolute and
relative off-sets when going to the same maximum multipole in our analysis. 

We note that some hydro-sims still yield a significant ($> \!\! 1\sigma$) bias
in cosmological parameters when using our theoretical error covariance
for the single-parameter model, which indicates that such a basic model will be
unsuitable for application to Stage-IV cosmic shear survey data.
While the multi-parameter models tended to do better in producing unbiased
results, our theoretical error covariance was still needed to ensure that all
hydro-sim results were unbiased when going to $\lmax = 5000$.

The theoretical error formalism is a general method for quantifying the known
errors of a method to realisations of the data. This method can be applied to a
wide-range of modelling problems within cosmology and astrophysics, where we
have applied it to baryonic feedback within the matter power spectrum. To that
end, the theoretical error is not a fixed quantity. If next generation
hydrodynamical simulations release that feature, for example, improved subgrid models
of physical processes that we know we are currently lacking~\citep{Crain:2023xap},
then it would make sense to replace
the older simulations in our suite with these new releases. This would alter
our error envelope function, improving or reducing our degree of trust in our
numerical baryonic feedback models. Alternatively, as new baryonic feedback
models are developed which are more actuate to the hydro-sims, the need 
for a theoretical error covariance becomes diminished. Though these new models
will have to come from advancements in the analytic or semi-analytic modelling
of baryonic feedback itself as numerical run-times from the convergence of MCMC
chains would be significantly affected if we were to use more-general models 
that contain many more free-parameters than the six used here.

With multi-parameter models of baryonic feedback, the effectiveness of
external priors (that is, information on baryonic feedback not from cosmic shear
alone) become increasingly powerful. In recent years, observational constraints
on baryonic feedback have come from the thermal Sunyaev-Zeldovich effect in
CMB observations~\citep{Troster:2021gsz}, and diffuse x-ray
backgrounds~\citep{Ferreira:2023syi}. With ever increasing precision data
taken on our Universe across the entire electromagnetic spectrum, the power of
external priors and cross-correlation with Stage-IV cosmic shear surveys
for baryonic feedback physics constraints will be immense.

The Holy Grail of
baryonic feedback models will, of course, be a model that can fit all hydro-sims
with no free parameters.\footnote{This is not a requirement on the baryonic
feedback models, per se, but on the requirements on the ensemble of hydrodynamical
simulations to converge to a single prediction for the evolution of $\Phykz$. 
With no scatter in the hydro-sims, deriving a fitting formulae for baryonic
feedback physics becomes trivial.} However, while we are making progress towards this goal,
for example \Bacco~\citep{Arico:2020lhq}, a theoretical error covariance is 
highly valuable until then. We expect the formalism that we have presented
and validated here will be of significant value to forthcoming Stage-IV
weak lensing surveys.

\section*{Acknowledgements}

We acknowledge HPC resources from the IRIS computing consortium.
AH acknowledges support from a Royal Society University Research Fellowship.
For the purpose of open access, the author
has applied a Creative Commons Attribution (CC BY) licence to any Author
Accepted Manuscript version arising from this submission.

\phantomsection
\section*{Data availability}
\label{sec:data_avail}

The hydrodynamical simulation data used in this article were taken from the 
`power spectrum library'~\citep{vanDaalen:2019pst}\footnote{\url{https://powerlib.strw.leidenuniv.nl}}
and from the \Camels Project~\citep{CAMELS:2023wqd}\footnote{\url{https://camels.readthedocs.io/}},
which include the \Bahamas~\citep{McCarthy:2016mry}, \HZAGN~\citep{Chisari:2018prw},
\Illustris~\citep{Nelson:2015dga}, \TNG~\citep{Springel:2017tpz}, \Eagle~\citep{Schaye:2014tpa},
and \Simba~\citep{Dave:2019yyq} simulations. We thank their authors for making 
their data publicly available. No new data was generated in support of this 
research.

\section*{Appendices}
\begin{subappendices}

\section{Dependency of the covariance on the coupling parameters}
\label{app:sigma_kz}

\begin{figure}[tp]
  \includegraphics[width=\columnwidth]{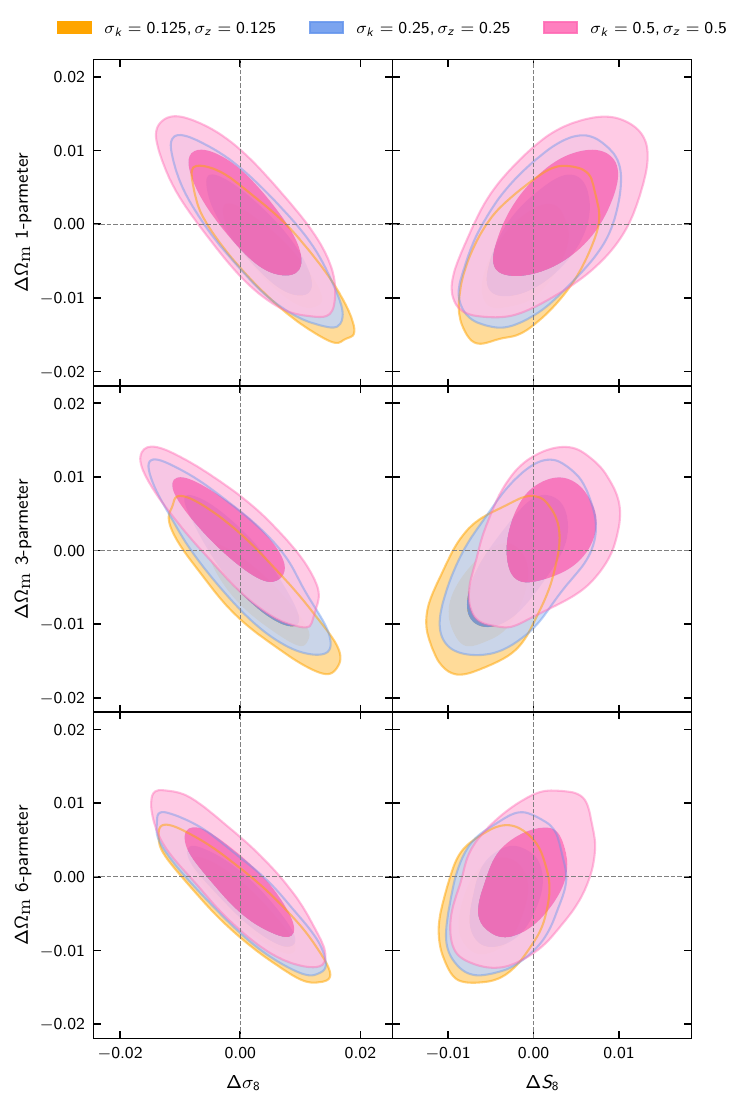}
  
  \vspace*{-0.25cm}

  \caption{$\Omegam$-$\sigmaeight$ and $\Omegam$-$\Seight$ contours for 
  the \Simba hydro-sim using the one-, three-, and six-parameter models (descending rows)
  with theoretical error for different values of $\sigma_k$ and $\sigma_z$ (different coloured contours).
  We see that increasing these values produces a stronger effect for the
  theoretical covariance, increasing the suppression of small-scale modes and
  thus reducing the bias from baryonic feedback.  
  }

  \label{fig:2D_sigma_kz_136param}
\end{figure}

\begin{figure}[tp]
  \includegraphics[width=\columnwidth]{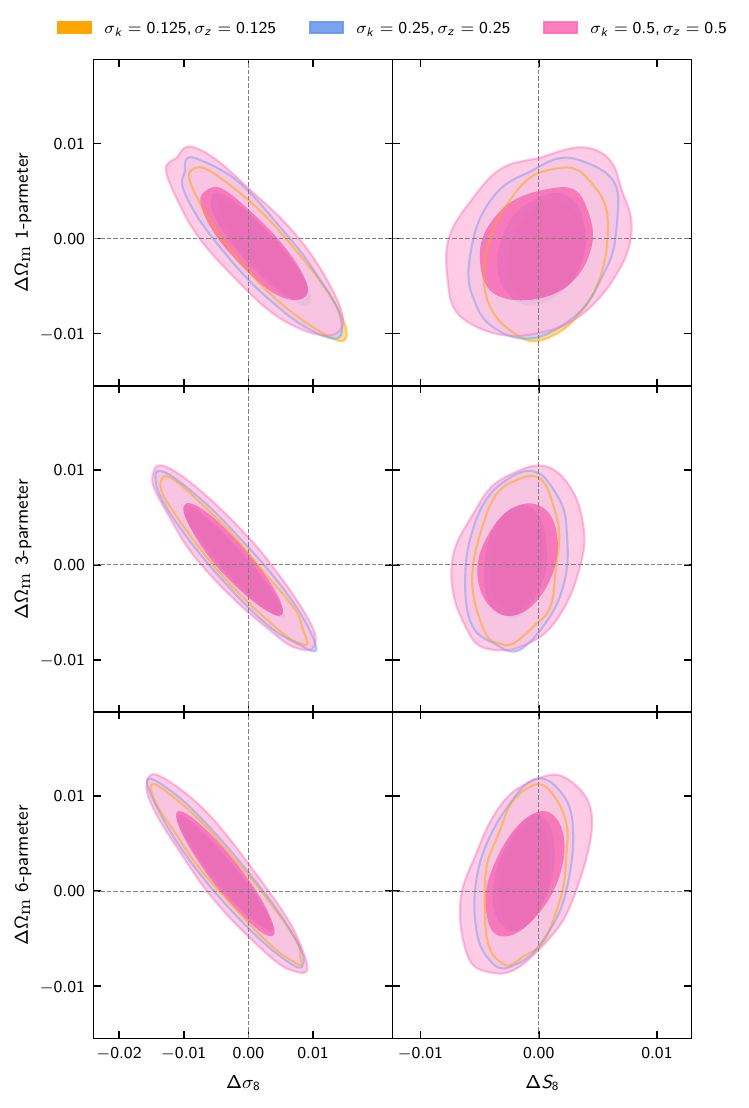}
  
  \vspace*{-0.25cm}

  \caption{Similar plot to Figure~\ref{fig:2D_sigma_kz_136param}, but for the
    \TNG hydro-sim instead of \Simba. Since \TNG can be well fitted by each of our \HMCode
    models with theoretical error, the effect of changing $\sigma_k$ and 
    $\sigma_z$ is just to increase or decrease the size of the contours,
    without effecting the bias in these parameters. 
  }

  \label{fig:2D_sigma_kz_136param_TNG}
\end{figure}

In Section~\ref{sec:paper2_sigma_kz} we looked at the one-parameter model and its
dependency on the coupling parameters $\sigma_k$ and $\sigma_z$. Here, we
present a comparison between all three baryon feedback models and investigate
their dependency on the coupling parameters.
Figure~\ref{fig:2D_sigma_kz_136param} uses \Simba as the ground-truth, which
features quite significant baryonic feedback, and
Figure~\ref{fig:2D_sigma_kz_136param_TNG} using \TNG, which has less extreme
feedback. When using \Simba as the ground-truth, we see a that there is a strong
link between the coupling parameter's values and the cosmological parameter 
biases for the one-parameter model, whereas the more general three- and 
six-parameter models are approximately insensitive to $\sigma_k$ and $\sigma_z$
due to their ability to better fit the data (thus have less significant
theoretical uncertainties associated with these models). We also see that,
when using \TNG as the ground-truth, since all three feedback models can 
adequately fit the data, increasing $\sigma_k$ and $\sigma_z$ serves to
slightly broaden the contours only, with no significant effect on the parameter
means.
\clearpage

\end{subappendices}

\clearpage

\begin{savequote}[65mm]
  Quest\textinterrobang
  
  I'm already on a quest!
  
  \qauthor{---Shrek}

  Hold on, I'm not done,

  One more time, with feeling!

  \qauthor{---Bon Jovi}
\end{savequote}
\chapter[Mitigating baryon feedback bias in cosmic shear through development of new techniques to optimise binary cuts]{Mitigating baryon feedback \\ bias in cosmic shear through development of new techniques \\ to optimise binary cuts}
\label{chp:binary_cuts}
\vspace*{-1cm}

\begin{mytext}
	\textbf{Outline.} In this chapter, I present an overview of existing
    methods for the derivation of binary scale cuts that mitigate baryonic
    feedback bias in a Stage-III survey. I then show how if we applied these
    methods to Stage-IV-like mock data, then headline constraints on $\Seight$
    would be no better than those from existing surveys. Using this as motivation,
    I go on to extend existing methods by explicitly incorporating baryonic
    feedback models, and show how we can extract much more information while
    keeping bias at acceptable levels. While specialised to baryonic feedback
    in cosmic shear only, the methods presented here are easily applicable
    to $3 \! \times \! 2$pt analyses, and to the mitigation of other effects,
    such as intrinsic alignments and non-linear galaxy clustering.
\end{mytext}

\section{Introduction \& motivation}

In this thesis, we have already seen how the modelling of baryonic physics
in the matter power spectrum is a hugely complex task, both in terms of numerical implementations
in hydrodynamical simulations and analytical descriptions in our Boltzmann code
solvers. These baryonic physical processes have no pure analytical expressions,
and thus we are forced to rely on semi-analytical and numerical methods.
As with any method, we aim to benchmark these methods to the available data to
quantify their accuracy and errors over the physical domain that these models
are used in. Since we are concerned with baryonic feedback, our data
comes from numerical hydrodynamical simulations which aim to capture 
`the Universe in a box'. This led to the use of the theoretical uncertainties 
approach, which was successfully used to mitigate baryonic feedback in 
Chapter~\ref{chp:baryonic_effects} of this thesis.  

Our application of the theoretical uncertainties approach was
limited to the cosmic shear angular power spectrum, $\Cl^{ab}$. However, there exist 
other estimators for cosmic shear they have been used in existing analyses,
namely the two-point angular correlation functions ($\xi_{\pm}^{ab}$),
band powers ($\mathcal{B}^{ab}_{n}$), and COSEBIs ($E_n^{ab}$). The theoretical
expectation value of these new observables are simply integrals of the 
underlying angular power spectrum with different weighting functions, though they
do measure different fundamental properties of the cosmic shear maps and thus are
distinct observables from the angular power spectrum.
Hence, one could propagate our $\Cl$ theoretical
uncertainty matrix into these bases, just as we propagated our $k$-space
covariance into $\ell$-space. However, these further integrations could be a
problem of considerable numerical complexity. Therefore, the investigation
of baryonic feedback biases for these additional observables would still require
the use of binary cuts, and so they maintain a place within our cosmic
shear analysis toolkits. 

Furthermore, our investigation of the theoretical
uncertainty covariance only looked at the cosmic shear spectrum. This is
just one prong of the three-headed beast that is the `$3\! \times\! 2$pt'
analyses, which include galaxy clustering and galaxy-galaxy lensing. The
$3\!\times\!2$pt data-vector will be the foundation of cosmological analyses 
using Stage-IV galaxy surveys, since it provides maximal constraining power. Thus, while
the theoretical uncertainties approach has been applied to mitigate non-linear 
galaxy bias in the $2 \! \times \! 2$pt data-vector (which lacks the pure 
shear-shear signal) separately~\cite{Aires:2024wze}, it may not be trivial to
combine these approaches in one consistent package.

Additionally, there exists the Bernardeau-Nishimichi-Taruya
(BNT) transformation \cite{Bernardeau:2013rda}, which aims to isolate the
redshift information from each source galaxy bin though the use of a linear
mixing matrix. This isolation in redshift-space forms a much tighter
$k \Leftrightarrow \ell$ relation in the matter power spectrum, and thus if one were to apply a $\kmax$ to
an analysis of a given model, then this would result in more appropriate
$\lmax$ values than in the ordinary Limber approximation~\cite{Euclid:2024yrr}.
It has been proposed to use the BNT transform in all \textit{Euclid} weak lensing
cosmological analyses, and it forms part of the core \textit{Euclid} \texttt{CLOE} 
code~\cite{Euclid:2024yrr}. 

Perhaps the strongest reason to investigate binary scale cuts is that they have
been used widely in existing Stage-III survey 
analyses~\cite{DES:2020daw,DES:2021vln,Kohlinger:2017sxk,KiDS:2021opn,HSC:2018mrq,Miyatake:2023njf},
and thus form the baseline default choice for Stage-IV surveys.

Hence, we are suitably motivated to investigate the properties of binary scale
cuts for future, Stage-IV cosmic shear surveys. The enhanced statistical
precision that the forthcoming surveys have require us to investigate the
properties of all statistical methods that will be used on the next-generation 
of data, which include the application and derivation of scale cuts. We will
again focus on the impact of baryonic feedback as the source of our biases
that we will be mitigating through scale cuts, though our methods are extendable 
to any source of bias that varies as a function of scale.

\section{The Dark Energy Survey's scale cuts approach}

We are motivated to look at existing Stage-III approaches to scale cuts to
inform us about our approach to the future
Stage-IV surveys. The approach that DES took to mitigate the effects
of baryonic physics is detailed in Ref~\cite{DES:2021vln}. Here,
they quantify baryonic effects in the cosmic shear observables ($\xi_{\pm}(\vartheta)$
for DES) through the use of the $\chi^2$ statistic. First, they generate
a synthetic data-vector without any baryonic feedback $\xi_{\pm, \textrm{base}}^{ab}$
and then one with baryonic `contamination', $\xi_{\pm, \textrm{baryon}}^{ab}$.
To generate this contaminated data-vector, they take the OWLS hydrodynamical
simulation's matter power spectrum, since that represents an extreme
scenario of AGN feedback. It is important to note that these
data-vectors are fixed quantities, and do not change as a function 
of scale-cuts.

To determine the scale cuts from these two data-vectors, they
construct a $\chi^2$ difference between them, where they
only focus on a specific combination of spectra, $ab$. This gives the
per-bin $\chi^2$ criterion as
\begin{align}
    \left(\xi_{\pm, \textrm{baryon}}^{ab} - \xi_{\pm, \textrm{base}}^{ab} \right)
    \, \mathbf{C}^{-1}_{abab} \,
    \left(\xi_{\pm, \textrm{baryon}}^{ab} - \xi_{\pm, \textrm{base}}^{ab} \right)^{\textsc{t}}
    \leq \frac{\chi^2_{\textrm{crit}}}{N},
    \label{eqn:chi_sq_crit}
\end{align}
where $\chi^2_{\textrm{crit}}$ is the targeted $\chi^2$ value that is split
between $N$ spectra, and $\mathbf{C}^{-1}_{abab}$ is the sub-block
corresponding to the spectra $ab$ of the inverse covariance matrix and has
shape $N[\vartheta^{ab}_{\pm}] \times N[\vartheta^{ab}_{\pm}]$ where $N[\vartheta^{ab}_{\pm}]$ is
the number of angular modes in bin combination $ab$ for either the $\xi_{+}$ or $\xi_{-}$ function.
Note that in the DES-Y3 approach, they treat $\xi_{+}$ and $\xi_{-}$ as 
\textit{separate} spectra. Thus, they find different values for 
$\vartheta_{\textrm{min}}$ for $\xi^{ab}_{+}$ and $\xi^{ab}_{-}$ for the same
bin combination $ab$. Hence, while they may have four tomographic bins, which
would give ten unique combinations for the angular power spectrum, they have
$N=20$ since there is an additional factor of two coming from the two shear
correlation functions. They describe an `iterative procedure' was used to obtain
the value of $\chisqcrit < 0.5$~\cite{DES:2021vln}. The values of
$\vartheta_{\textrm{min}}^{ab}$ would then be adjusted
to fit the criteria above, taking the minimum value that would satisfy it.

While this is certainly a way to optimise your scale cuts such that
it mitigates baryonic feedback in a data-vector, it suffers from
several key issues:
\begin{galitemize}
    \item The $\chi^2$ statistic is in data-space. While this tells us
        how statistically significant the difference between the
        two data-vectors are, it does not tell us anything about how
        these differences in the data-vectors affect our cosmological
        inference from these data-vectors. Naturally, a larger
        $\chi^2$ would correlate with larger differences in
        cosmological parameters, but this depends on the
        parameter's response to the observables --- which is
        different for each observable. Hence, a determination of the
        relation between $\chisqcrit$ and the level of bias induced for each
        combination of observables considered needs to be computed
        for each survey and is not necessarily easily comparable between
        different surveys and data releases.
    
    \item Equation~\ref{eqn:chi_sq_crit} works bin-by-bin, in
        that while it still contains some information from other bins
        through the inverse covariance matrix, it only looks at
        the scale cuts on a bin-by-bin basis. Thus, it does
        not include contributions from the cross-correlations
        of the spectra in the inverse covariance matrix that
        would otherwise contribute to the $\chi^2$ (and thus
        final likelihood). Let us demonstrate with a simple
        $2 \times 2$ example. Consider a data-vector of length
        two: a mean vector $\vec{\mu} = (\mu_A, \, \mu_B)$ and a realisation
        vector $\vec{d} = (d_A, \, d_B)$ giving the vector $\vec{\delta} \equiv \vec{d} - \vec{\mu} = (A, B)$.
        It has an associated covariance matrix $\mathbf{C}$ given by
        $\mathbf{C} = 
        \begin{pmatrix}
            C_{AA}, \, C_{AB} \\
            C_{AB}, \, C_{BB}
        \end{pmatrix}$
        and inverse covariance matrix\footnote{I've used $D$ to label the elements of the inverse
        covariance matrix as to not confuse between the elements of the non-inverted
        covariance matrix. Thus $D_{AB} \equiv (\mathbf{C}^{-1})_{AB}$.} $\mathbf{C}^{-1} = 
        \begin{pmatrix}
            D_{AA}, \, D_{AB} \\
            D_{AB}, \, D_{BB}
        \end{pmatrix}$. The full $\chi^2$ of this is given by
        \begin{align}
            \chi^2 &= \vec{\delta} \, \mathbf{C}^{-1} \, \vec{\delta}^{\textsc{t}} \\
                   &= \left(A, B\right) \, \begin{pmatrix}
                    D_{AA}, \, D_{AB} \\
                    D_{AB}, \, D_{BB}
                    \end{pmatrix}
                    \begin{pmatrix}
                        A \\
                        B
                    \end{pmatrix} \\
                   &= A^2 D_{AA} + 2 AB D_{AB} + B^2 D_{BB},
        \end{align}
        which clearly has contributions from the cross-correlations
        in the (inverse) covariance matrix. If we were to repeat this
        exercise, but instead isolating the $A$ and $B$ components separately
        and computing a `total' $\chi^2$ from them, we find
        \begin{align}
            \chi^2_{A} &= A^2 D_{AA} \\
            \chi^2_{B} &= B^2 D_{BB} \\
            \therefore \chi^2_{\textrm{sum}} &= \chi^2_{A} + \chi^2_{B} = A^2 D_{AA} + B^2 D_{BB} \neq \chi^2. \label{eqn:chi_sq_sum_neq}
        \end{align}
        Thus, the sum of our individual $\chi^2$  is not necessarily the same as the 
        total $\chi^2$ computed when including the cross-correlations.

        Since these cross-correlations carry cosmological information, and are included when
        computing the likelihood for cosmological inference, it is important to keep
        them when determining the scale cuts to be used in any analysis. Thus, a holistic
        approach where we include the entire covariance matrix should be used when determining
        any scale cuts. 

    \item The DES approach aims to mitigate against \textit{any and all} baryonic feedback. This is
        because they take their base data-vector to have no baryonic feedback in at all. While this
        approach may have certainly worked in the lower-precision Stage-III era, it's not
        particularly suitable to apply to the forthcoming Stage-IV era experiments. This is
        because baryonic feedback affects a large range of scales, and thus mitigating
        against all effects throws away significant amounts of data. Additionally, we
        have models of baryonic feedback, some of which are extremely accurate down to
        small scales and high redshifts, so it's not a case of having no appropriate models at all.
        Thus, if we were to apply even a basic model of baryonic feedback physics then
        we would find that the scale cuts allow the inclusion of smaller scales, thus
        increasing the constraining power of our data-vector. This does, however, require
        that we marginalise over any new baryonic feedback parameters included in our
        models. It is hoped that the increase in precision coming from the inclusion
        of much more constraining angular modes 
        mitigates against any information lost from marginalisation.
\end{galitemize}

Do note that while DES used $\xi_{\pm}$ as their summary statistics, we are
using the power spectrum $\Cl$ values instead. Thus, when investigating our
scale cuts, we are interested in the maximum multipole to include in our
analyses, $\lmax$, which is equivalent to their quest for $\vartheta_{\textrm{min}}$.

\section[Detailed look at the per-bin $\chi^2$]{Detailed look at the per-bin \boldmath$\rchi^2$}

As explained above, the scale cuts used in the DES-Y3 analysis were derived from optimising
Equation~\ref{eqn:chi_sq_crit}, which worked bin-by-bin. We then showed
in Equation~\ref{eqn:chi_sq_sum_neq} that these individual $\chi^2$ values do
not necessarily reflect the total $\chi^2$ of the entire data-vector.  We can
now investigate the properties of how going bin-by-bin is reflected in the $\chi^2$
values. 

To set up our two fiducial data-vectors, we computed our base baryon-free
set of $\Cl$ values using the \texttt{mead2020} model as included in CAMB as
part of the Core Cosmological Library (CCL). For our baryon-contaminated
data-vector, the \texttt{mead2020\_feedback} one-parameter model was used
with $\TAGN = 8.0$. To be broadly equivalent to the DES-Y3 setup, four
Gaussian photometric redshift bins were created, with means $\bar{z} = \{0.4, 0.8, 1.2, 1.6\}$
with widths $\sigma_{z} = 0.2$. With four bins, we get ten different 
unique spectra: $\bins{1}{1}, \, \bins{1}{2}, \, \bins{1}{3}, \, \bins{1}{4}, \, 
\bins{2}{2}, \, \bins{2}{3}, \, \bins{2}{4}, \, \bins{3}{3}, \, \bins{3}{4}, \, \bins{4}{4}$.
Thus, we have ten different $\chi^2$ curves to investigate as a function of
$\lmax$.

\begin{figure}[t]
    \centering
    \includegraphics[width=\linewidth]{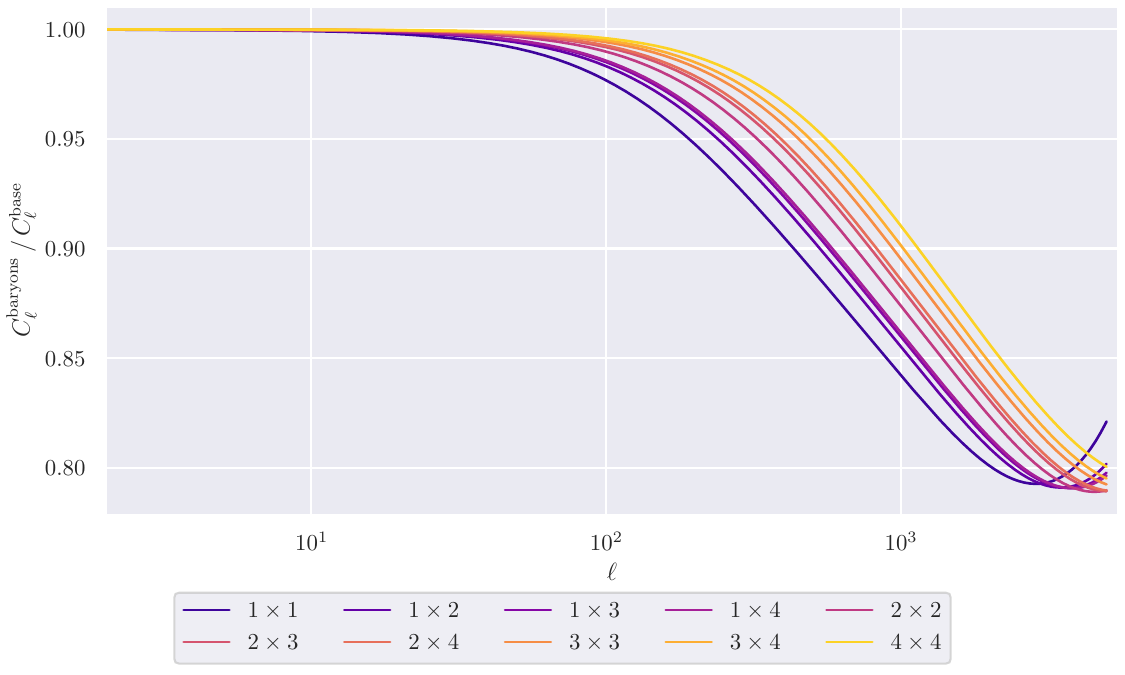}
    \caption{Ratio of our $\Cl$ data-vectors with baryonic feedback to that of
        no baryonic feedback. We can see that AGN and supernovae activity
        causes a large suppression in the power spectrum that effects a
        large range of $\ell$ values. We also see that baryonic physics causes
        the suppression to move to lager $\ell$ values for higher redshift bins. 
    }
    \label{fig:cl_baryon_ratio}
\end{figure}

First, let us look at how including baryonic feedback in our $\Cl$
data vector changes their values. This is shown in Figure~\ref{fig:cl_baryon_ratio}.
Here, we see that the $\Cl$ values are the same on the very largest scales,
however as soon as we cross about $\ell \gtrsim 20$, then the AGN and supernovae
activity within our baryonic feedback model starts suppressing the $\Cl$ values.
This suppression grows as we go to smaller and smaller scales, eventually
reaching about a $20\, \%$ decrement. We also note that this suppression profile
is bin-dependent, with the lowest redshift bins deviating from the base 
data-vector at a much lower $\ell$ value than the higher redshift bins. Thus,
na\"ievely, one may assume that the lower redshift bins would have a larger
$\chi^2$ difference between the base data-vector and our baryon contaminated
data-vector based on the above plot. However, it is important to note that
due to the lager cosmic shear signal from higher redshift bins, the overall
signal-to-noise of these high redshift bins is much larger than the lower
redshift bins. Hence, when we look at our $\chi^2$ differences per-bin, we
find that the higher redshift bin differences are more statistically significant.

\begin{figure}[t]
    \centering
    \includegraphics[width=\linewidth]{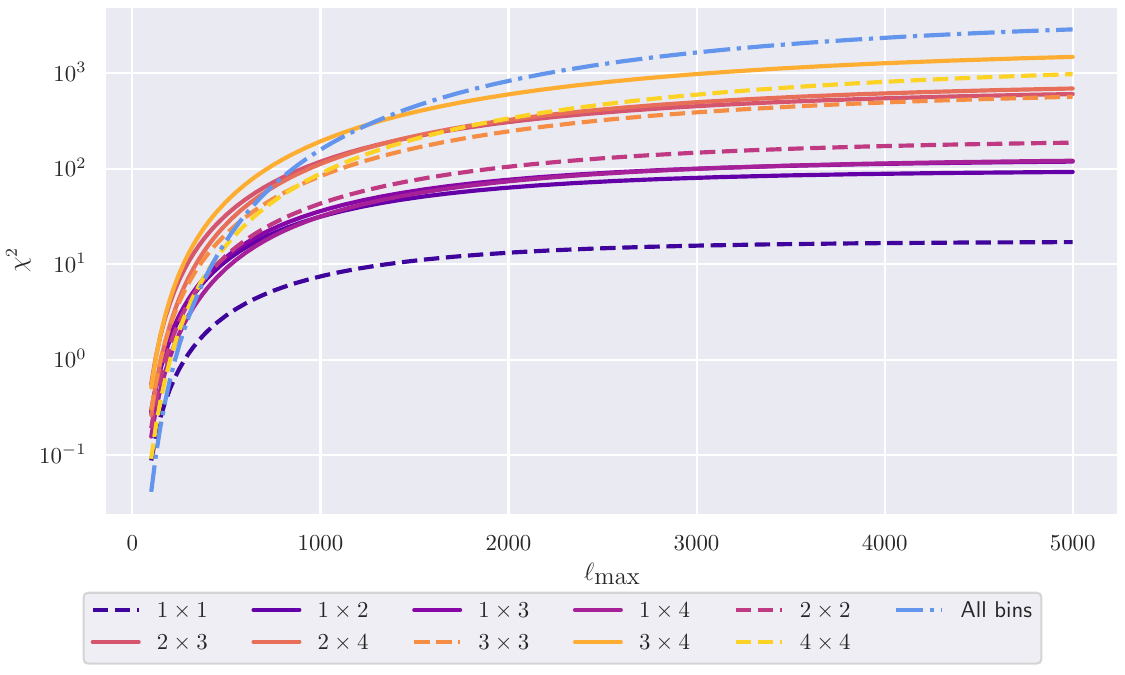}
    \caption{Plot of the different $\chi^2$ values as a function of $\lmax$
        for each of our ten different spectra, in solid and dashed curves, 
        along with the total $\chi^2$ for the entire data-vector in the dot-dashed
        blue curve. We plot the auto-spectra with dashed lines to make it
        immediately obvious that, due to the inclusion of shape noise in the
        auto-spectra only, these curves have a lower $\chi^2$ compared to
        their cross-spectra counterparts. 
        }
    \label{fig:cl_chi_sq_diff}
\end{figure}

Figure~\ref{fig:cl_chi_sq_diff} plots the $\chi^2$ differences obtained
using the left-hand side of Equation~\ref{eqn:chi_sq_crit} for each
spectra (that is the angular power spectrum for each unique auto- and
cross-correlation combination of our four bins), as a function of maximum
multipole. Here, we see that the larger
signal-to-noise of the higher redshift bins causes the $\chi^2$ differences 
to be much larger than for the lower redshift bins. This results in a much
lower $\lmax$ for the higher redshift bins, which is exactly what was found 
by the Dark Energy Survey when producing their scale cuts.\footnote{Though, 
of course, DES worked in correlation functions so they found a larger 
$\vartheta_{\textrm{min}}$ for the higher redshift bins.}
It is also important to note that we plot the $\chi^2$ values of the entire
data-vector as a function of $\lmax$ in the dot-dashed blue line. This shows
that the overall $\chi^2$ is not simply the sum of the individual $\chi^2$
values, and that it has a slightly different behaviour to that of the
individual spectra.

We also note that the $\chi^2$ values in Figure~\ref{fig:cl_chi_sq_diff} are 
quite large, since we are applying Stage-IV survey specifications to our
fiducial data-vectors. Thus, if we were to apply the DES-Y3 cut of
$\chi^2_{\textrm{crit}} = 0.5$, split equally between the ten bin pairs, then
this would result in extraordinarily aggressive scale cuts.

\section{Insufficiencies of the DES-Y3 approach in the Stage-III era}

While the DES-Y3 approach to scale cuts may appear to be a sensible approach to
mitigate baryonic feedback in Stage-III data, we can show that their scale cuts
did not effectively remove contamination from their data-vectors. Since their
$\chi^2$ analysis aimed to reduce baryonic feedback in their data-vectors
to `acceptable levels', which was defined as a shift in the parameter's mean
of no greater than $0.3\sigma$~\cite{DES:2021vln}, if we were to analyse these data-vectors using a model with
and without baryonic feedback we should find the same results for cosmological
parameters. 

To this end, we have performed a re-analysis of the joint DES-Y3 and KiDS-1000
cosmic shear analysis~\cite{KIDS_DES:2023gfr}, which presents the tightest 
constraints using cosmic shear Stage-III data yet. Hence, this is a good
testing ground to see of changing baryonic feedback models and its effects
$\Seight$ and $\Omegam$ values inferred from the data. 

Results were obtained using the DES-Y3+KiDS-1000 example pipeline as provided with
\textsc{Cosmosis}, which was altered to use either a `zero-parameter' model,
which is the \texttt{mead2020} model without baryonic feedback, and the one- and 
three-parameter \texttt{mead2020\_ feedback} models. It should be noted that the
one-parameter model was used in their fiducial analysis, though they also
investigated neglecting baryonic feedback though a dark-matter-only 
matter power spectrum. The \texttt{PolyChord} sampler was used, matching
the joint analysis, with the same precision settings too. In addition to the
astrophysical baryonic feedback parameters already mentioned, our analysis
matched that presented in Ref.~\cite{KIDS_DES:2023gfr} by sampling over six
cosmological parameters ($\Omegab h^2$, $\Omegac h^2$, $h$, $n_\textrm{s}$,
$\Seight$, $\Sigma m_{\nu}$), four intrinsic alignment, and thirteen
redshift calibration nuisance parameters.  

\begin{figure}[tp]
    \centering
    \includegraphics[width=\linewidth]{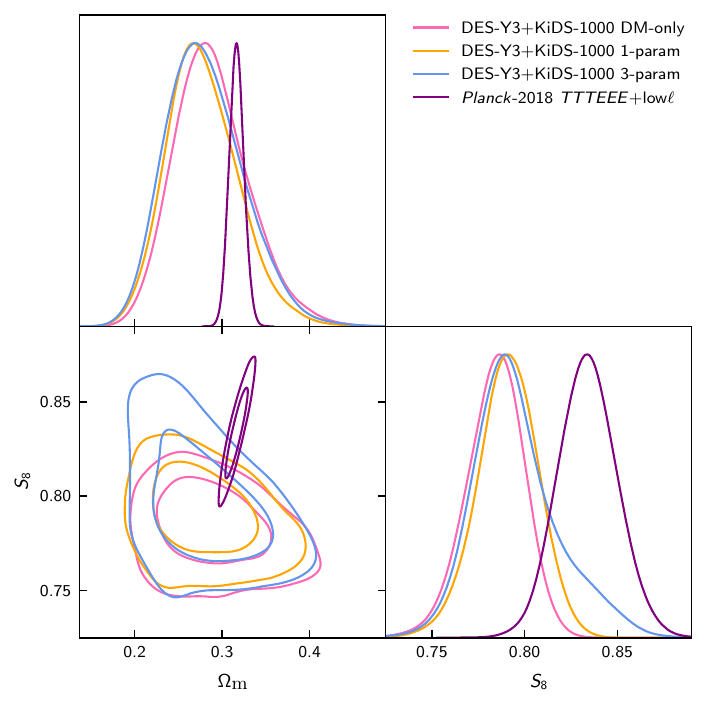}
    \caption{Triangle plot highlighting the distribution of the key cosmic shear statistics,
        $\Omegam$, $\sigmaeight$, and $\Seight$, for the DES-Y3+KiDS-1000
        data-vector using three different models of baryonic feedback, with the
        \textit{Planck} 2018 results present for comparison~\cite{Planck:2018vyg}. We see that the
        changing of the baryonic feedback model does have an impact on the
        distribution of the cosmological parameters, particularly for $\Seight$
        -- the most well constrained parameter using cosmic shear. We see that 
        is a significant tail out to large $\Seight$ values when considering the
        three-parameter model, that isn't present in either the zero- or
        one-parameter models. The exact values for the mean and $1\sigma$ errors
        one-dimensional marginalised values are presented in Table~\ref{tab:DES_KiDS_cosmo_values}.\\
        }
    \label{fig:2D_KiDS_DES_baryons}
\end{figure}

Figure~\ref{fig:2D_KiDS_DES_baryons} presents the triangle plot for the
distribution of $\Omegam$, $\sigmaeight$ and $\Seight$ for our re-analysis
of the DES-Y3+KiDS-1000 data-vector, along with the results from \textit{Planck}
2018 for comparison. Table~\ref{tab:DES_KiDS_cosmo_values} also shows the
one-dimensional marginalised means and $1\sigma$ errors for the three
cosmological parameters. We see that while the peaks in the one-dimensional
distributions are extremely close between all three baryonic feedback models,
there is a significant tail out to larger $\sigmaeight$ and $\Seight$ values
in the one- and three-parameter models when compared to the dark-matter only
analysis. We expect that the inclusion of one and three additional nuisance
parameters would act to slightly broaden the marginalised contours of our cosmological
parameters, but in an overall scaling of the $1\sigma$ errors. However, what we
see is that there is a clear one-sided broadening to the parameter's 
distributions in favour of larger $\sigmaeight$ and $\Seight$ values. Hence,
we can say that the one- and three-parameter models are extracting additional
information from the data-vector that the dark-matter-only model cannot, and
thus their scale cuts were not totally sensitive to all baryonic feedback.

To reiterate, the scale cuts used in Ref.~\cite{KIDS_DES:2023gfr} for the
DES-Y3 data-vector were derived using the DES-Y3 methodology previously
described (Equation~\ref{eqn:chi_sq_crit}, and presented in Ref.~\cite{DES:2021vln}).
For the KiDS-1000 part of the data-vector, they mitigate baryonic feedback
by raising the minimum angular scale in the COSEBI integral from
$\vartheta_{min} = 0.5\,$arcmin to $\vartheta_{min} = 2.0\,$arcmin and the
use of the one-parameter baryonic feedback model. The KiDS team do not 
explicitly do a DES-like per-bin $\chi^2$ analysis to determine per-bin
biases, though it should be noted that DES-Y3 contributes a significant part
of the overall signal-to-noise of the joint analysis.

This extension of $\Seight$ out to larger values when changing the baryonic
feedback model was also seen in Ref.~\cite{Garcia-Garcia:2024gzy}, which
jointly analysed DES-Y3, KiDS-1000, and HSC-Y1 data using the \texttt{Baccoemu}
baryonification
emulator~\cite{Angulo:2020vky,Arico:2021izc,Arico:2020lhq,Zennaro:2021bwy,Pellejero-Ibanez:2022efv},
and taking the existing data-vectors down to significantly smaller scales (an
$\lmax$ of 4500).
\texttt{Baccoemu} features seven free parameters in its baryonic feedback
model, and so should be even more flexible than our three-parameter model. 
It should be noted that Ref.~\cite{Garcia-Garcia:2024gzy} found an anomalously
low value of $\Omegam$ of $\Omegam = 0.212 ^{+0.017} _{-0.032}$, which is in
tension to \textit{Planck} at the $\sim \! 3\sigma$ level, and lower than each 
of the survey's own individual analyses\footnote{Naturally, it may cause alarms
that in trying to solve the $\sim\!2\sigma$ tension in $\Seight$ one has caused
a $\sim\!3\sigma$ tension in $\Omegam$ which was previously not under any
tensions. Ref.~\cite{Garcia-Garcia:2024gzy} do not find their results 
particularly alarming, since there could be many systematic or astrophysical
effects, such as intrinsic alignments, which show up at such large
multipoles using existing Stage-III survey data. Cosmic shear analyses using
Stage-IV data, which were designed to probe these small-scales with as little
systematic effects possible, on these small-scales will be the key to unravelling
the $\Seight$ (and this $\Omegam$) tension.}\!.
 Therefore, we are currently in a 
situation where analysing existing data-sets with different numerical algorithms 
(the baryonic feedback models) and analysis choices (the chosen scale cuts)
yield different results for the key cosmic shear observables ($\Omegam$ and $\Seight$),
which are in tension with \textit{Planck} at some level or another. If we
are to use the Stage-III surveys as a springboard into the Stage-IV era, then
we should quantify and acknowledge these short-comings before applying the same
inadequate methods to much more precise data in an actual cosmological analysis.

\begin{table}[t]
    \centering
    \caption{Results for the mean and $1\sigma$ errors for the 
        DES-Y3+KiDS-1000 analysis using three different models of baryonic
        feedback, along with results from \textit{Planck} 2018 `TT,TE,EE+lowE'~\cite{Planck:2018vyg}
        analysis for comparison. These values are in excellent agreement with the values
        presented in Ref.~\cite{KIDS_DES:2023gfr} for the DM-only and
        one-parameter models. We note that the mean of the $\Seight$
        distributions systematically increases as we go from DM-only, to
        one- and then three-parameter models. Though these values are still
        within a $1 \sigma$ uncertainty of each other, the mean of $\Seight$
        distributions shifts by larger than the stated $0.3\sigma$ tolerance.
        }
    \label{tab:DES_KiDS_cosmo_values}
    \bgroup
    {\renewcommand{\arraystretch}{1.5}
    \setlength\tabcolsep{0.45cm}
    \begin{tabular}{r|ccc}
        \toprule
        Survey \& modelling      & $\Seight$               &  $\sigmaeight$             & $\Omegam$                     \\

        \midrule
        
        DES+KiDS DM-only    & $0.784^{+0.016}_{-0.014}$    & $0.807 \pm 0.065$          & $0.288^{+0.036}_{-0.048}$     \\

        DES+KiDS 1-param    & $0.791^{+0.017}_{-0.015}$    & $0.828 \pm 0.068$          & $0.278^{+0.035}_{-0.045}$     \\

        DES+KiDS 3-param    & $0.797^{+0.017}_{-0.026}$    & $0.830^{+0.067}_{-0.088}$  & $0.281^{+0.037}_{-0.051}$     \\

        \textit{Planck}     & $0.834 \pm 0.016$            & $0.8120 \pm 0.0073$        & $0.3166 \pm 0.0084$           \\

        \bottomrule
    \end{tabular}        
    \egroup
    }
\end{table}

\section{Insufficiencies of the DES-Y3 approach in the Stage-IV era}

We have seen that the offsets between baryonic feedback models is relatively
small for the DES\nobreakdash-Y3 +KiDS\nobreakdash-1000 data vector when using the DES-Y3 approach to
determining scale cuts from baryonic physics, though non-zero.
While this automatically suggests that these approaches might be insufficient
for Stage-IV surveys, we can perform an investigation into just how
insufficient these methods might be with future data-sets.

\subsection{Stage-IV cosmic shear surveys}

To properly understand if the DES-Y3 approach of determining scale cuts through
Equation~\ref{eqn:chi_sq_crit} is sufficient in the Stage-IV era, 
we can generate sets of scale cuts using an adapted version of this method
and see how they compare between Stage-III and Stage-IV surveys. 
For comparison, we will use three mock surveys: DES-Y1-like, DES-Y3-like, and
a \textit{Euclid}-DR3-like (the final data release from the planned six-year survey)
surveys. The properties of the cosmic shear surveys can
be summarised into their total observed galaxy density, $\ngal$, and their sky
fraction observed, $\fsky$. The values for our three mock surveys are specified
in Table~\ref{tab:survey_properties}. Of course, there are many more differences
between Stage-III and Stage-IV surveys beyond these two summary statistics,
such as the accuracy of the shear measurements, precision on the photometric
redshift estimates of each galaxy, number of tomographic redshift bins used in
an analysis, and the maximum redshift at which source
galaxies have been observed to (changing the $n(z)$ distributions). However,
as $\ngal$ and $\fsky$ directly feed into the data-space covariance matrix, 
changing their values have a direct effect on the impact of baryonic feedback
in cosmic shear surveys. Thus, by just changing these two values, we can test
the reaction of our scale cuts to different survey specifications.

Through having increased resolution detectors and higher sensitivity down
to fainter magnitude detectors, Stage-IV surveys will observe many more galaxies 
than their Stage-III predecessors, thus increasing $\ngal$. By being able to
average many more galaxies ellipticities in each pixel, this deceases the 
contributions of shape noise in the power spectrum, thereby increasing the
size of the signal-dominated region in $\ell$-space and allowing more 
information to be extracted from the power spectra.

By surveying more of the sky to produce a larger $\fsky$, more pairs of pixels
can be included for each $\ell$ mode. This deceases the $\ell$-space covariance 
matrix for every $\ell$ mode. Thus, surveys with larger $\ngal$ and $\fsky$ are
able to meaningfully probe higher $\ell$ modes and extract more information
from every available $\ell$ mode, thus are highly sensitive to baryonic physics. 

\begin{table}[t]
    \centering
    \caption{Assumed survey specifications used in our analyses. We have
        condensed the properties of a Stage-III and Stage-IV survey into their
        total observed galaxy density, $\ngal$, and the observed sky fraction,
        $\fsky$. 
        All three surveys assume $\sigma_\epsilon = 0.26$. 
        }
    \label{tab:survey_properties}
    \bgroup
    \setlength\tabcolsep{0.45cm}
    \begin{tabular}{lrrr}
        \toprule
        Survey & Total $\ngal$& Survey area [$\deg^2$] & $\fsky$ \\
        \midrule
        
        DES-Year1 & 6 & 1,321 & $3.2\%$ \\
        DES-Year3 & 6 & 5,000 & $12.1\%$ \\
        \textit{Euclid}-DR3 & 30 & 14,500 & $35.0\%$ \\
        
        \bottomrule
    \end{tabular}        
    \egroup
\end{table}

Table~\ref{tab:survey_properties} highlights just how large the increase in
precision in the data that forthcoming Stage-IV surveys will have over existing 
Stage-III results. With a nearly order of magnitude increase in the galaxy
density and a trebling of the sky area, Stage-IV surveys will both probe to
higher $\ell$ modes, and have them more tightly constrained than Stage-III.
Hence, with this dramatic increase in the precision on the data, existing
methods to mitigate baryonic feedback physics may not be sufficient,
which is what we will explore now.  

To simplify our analyses to a `proof-of-concept' rather than a more rigorous
detailed analysis, we restrict ourselves to three tomographic bins for both our
Stage-III and Stage-IV mock surveys. This restricts the number of power
spectra to six, and thus we only need to derive six scale cuts using our
methods.

\subsection{Extending the DES-Y3 approach to power spectra}

The Dark Energy Survey chose to use the correlation functions 
($\xi_{\pm}^{ab}(\vartheta)$) as their cosmic shear summary statistic, 
which is a perfectly valid choice. However, it can be considered
slightly more natural to work in the power spectrum ($C_{\ell}^{ab}$), and thus
we can formulate an equivalent equation to Equation~\ref{eqn:chi_sq_crit}
but now working in $\ell$-space. For a single bin\footnote{\textit{Give me a bin, Vasili. One bin only, pleash} --- Captain Ramius.}
combination $ab$, this gives the $\chi^2$ criterion as
\begin{align}
    \left(C_{\ell, \textrm{baryon}}^{ab} - C_{\ell, \textrm{base}}^{ab} \right)
    \, \mathbf{C}^{-1}_{abab} \,
    \left(C_{\ell, \textrm{baryon}}^{ab} - C_{\ell, \textrm{base}}^{ab} \right)^{\textsc{t}}
    \leq \frac{\chi^2_{\textrm{crit}}}{N}.
    \label{eqn:chi_sq_crit_ell_space}
\end{align}
For our $\ell$-space covariance matrix, we assume the standard Gaussian
form of
\begin{align}
    \textrm{Cov}\left[C^{ab}_{\ell}, \, C^{cd}_{\ell'}\right] = \frac{\delta_{\ell \ell'}}{(2 \ell + 1) \, \fsky}
    \left(C^{ac}_{\ell} \, C^{bd}_{\ell} + C^{ad}_{\ell} \, C^{bc}_{\ell} \right),
    \label{eqn:gaussian_cl_cov_2}
\end{align} 
where we can now directly see how a larger value of $\fsky$ reduces the
covariance between $\ell$-modes. Note that the $\Cl$ values in 
Equation~\ref{eqn:gaussian_cl_cov_2} refer to the \textit{total} power spectrum,
that of the cosmological signal plus noise,
\begin{align}
    C_{\ell}^{ab} = C_{\ell, \textrm{signal}}^{ab} + N_{\ell}^{ab}.
\end{align}
Here, $N_{\ell}^{ab}$ is the noise power spectrum in bin $ab$ and is given by
\begin{align}
    N_{\ell}^{ab} = \frac{\sigma_{\epsilon}^2}{\bar{n}} \, \delta_{ab},
\end{align}
where $\bar{n}$ is the average surface galaxy density of that bin, and the
Kronecker-$\delta$ ensures that only auto-spectra contain shape noise. Thus,
observing more galaxies which, in turn, increases $\bar{n}$ and decreases the
amplitude of the noise spectra. We use $\chisqcrit = 1.5$ which gives a
per-bin value of $\chisqcrit = 0.25$ in all analyses. This was chosen through an
iterative scheme where a range of $\chisqcrit$ values were trialled and the
resulting scale cuts and parameter contours were compared against each other. 
Note that this is a factor of three larger than the value of $\chisqcrit = 0.5$
as used in the DES-Y3 analysis, which is a result of us using simplified
survey configurations and a simplified covariance matrix. 

Note that we also use the same value of $\chisqcrit$ for all surveys in our
analyses, since we have assumed that the mapping between $\chi^2$ and parameter
biases are independent on the survey specification. For a realistic analysis,
it may be more appropriate to determine the value of $\chisqcrit$ which lead to
acceptable levels of bias on a per-survey basis. It would be worth investigating
how survey specifications change the $\chi^2$ to parameter biases mapping,
especially since a survey such as \textit{Euclid} will go from a DR1, to DR2, and
finally a DR3 analysis and so the allowable $\chisqcrit$ value may well change
between data releases.

\subsection{Scale cuts}

With our $\chi^2$ criterion in $\ell$-space constructed, we can apply it
for our mock DES-Y1, DES-Y3, and \textit{Euclid}-DR3 surveys and explore
how the scale-cuts behave for our surveys. To do so, we need to generate
sets of $C_{\ell, \textrm{base}}^{ab}$ and $C_{\ell, \textrm{baryon}}^{ab}$.
For the base data-vector, this is simple as we can use the dark-matter-only 
\texttt{mead2020} model as part of HMCode-2020. For the baryon contaminated
data-vector, DES-Y3 used the OverWhelmingly Large Simulations (OWLS)~\cite{vanDaalen:2011xb,Schaye:2009bt}
which, they claim, features an `extreme' implementation of baryonic feedback
compared to other hydrodynamical simulations in the literature. However,
since its release in 2011, there have been numerous other releases of hydrodynamical
simulations which claim that they feature more realistic implementations of
baryonic feedback\footnote{Generally, the matter power spectrum is of little
importance to the benchmarking (and thus determination of what `realistic' means)
since thus far there has been little constraint on the shape and amplitude of
baryonic feedback in the matter power spectrum (hopefully Stage-IV surveys
will change this!). Hence, the use of other physical observables which are
much better constrained have driven the development of the hydro-sims which aim
to match these observables to ever increasing accuracy. Examples
include the galaxy stellar mass function, the hot gas fraction within haloes, and
the star formation rate~\cite{Bigwood:2025ism,Jones:2024MNRAS5351293J}.}\!.
For our baryonic contaminated data-vectors, we use an ensemble of five 
hydro-sims: \Bahamas, \HZAGN, \Illustris, \TNG, and \Eagle. Thus,
when we compute Equation~\ref{eqn:chi_sq_crit_ell_space}, we average over our
five hydro-sims to compute an average $\chi^2$ value which can be compared
against our critical value $\chisqcrit$.

We are motivated to average over an ensemble of hydrodynamical simulations,
which brings added numerical complexity and increased run-times, since each
hydro-sim has its own baryonic feedback profile (see Figure~\ref{fig:My_Pk_ratio}) and so have different errors
for each $\ell$-mode and redshift bin pair. Hence, one hydro-sim could produce
large errors on intermediate $\ell$ modes, while fitting the high-$\ell$ region
successfully while another hydro-sim features the opposite. Thus, if we are to
adequately derive scale cuts which successfully mitigate baryonic feedback, then
we wish to include information from all available hydro-sims. Note that we
average here instead of taking the maximum since the average is more numerically
stable when repeatedly fitting our models to the data. We again take an agnostic
approach to the hydro-sims and provide each simulation with an equal weight
in the average.

\begin{table}[t]
    \centering
    \caption{Results for our scale cuts (that is the maximum possible $\ell$ 
        multipole that satisfies the criterion) using the DES-Y3 ($\chi^2$, no baryons)
        approach for our three mock surveys. We clearly see the increase in
        sky areas serves to decrease the allowable $\lmax$ from DES-Y1 to DES-Y3,
        and then the dramatic decrease the $\lmax$'s our \textit{Euclid}-like survey.
        This also clearly shows the decrease in the $\lmax$ as we go to
        higher redshift bins, and the impact of shape noise on the auto-spectra.
        An $\lmax$ of $50$ was the lower bound of the optimisation. 
        }
    \label{tab:scale_cuts_chi_sq_no_baryons}

    \bgroup
    \setlength\tabcolsep{0.75cm}
    \begin{tabular}{lrrr}
        \toprule
        Bin & DES-Y1 & DES-Y3 & \textit{Euclid}-DR3 \\
        \midrule
        
        $\bins{1}{1}$ &   5000 &    870 &     226 \\
        $\bins{1}{2}$ &    898 &    408 &      80 \\
        $\bins{1}{3}$ &    888 &    440 &     124 \\
        $\bins{2}{2}$ &    851 &    390 &      91 \\
        $\bins{2}{3}$ &    506 &    183 &      73 \\
        $\bins{3}{3}$ &    695 &    319 &     122 \\

        \bottomrule
    \end{tabular}        
    \egroup
\end{table}

The results for our six bin-dependent scale cuts using the $\chi^2$ method
mitigating against all baryonic feedback for our three mock surveys are
presented in Table~\ref{tab:scale_cuts_chi_sq_no_baryons}. Here, we see that the
higher signal-to-noise of the higher redshift bins act to decrease the allowable
$\lmax$ for the given $\chisqcrit$ value. We also see that, due to the inclusion
of shape noise in the auto-spectra only, these auto-spectra generally have
higher allowable $\lmax$ values over their comparable cross-spectra. We also
see the dramatic impact that the larger $\ngal$ and $\fsky$ values of
our \textit{Euclid}-DR3-like survey has on our derived scale cuts. Not even
being able to reach an $\lmax$ of one hundred for various bins before the
impact of baryon feedback becomes significant shows that extremely aggressive
scale cuts are needed if we were to eliminate all baryonic feedback from 
our Stage-IV data-vectors.

\subsection{Illustration of data-loss from aggressive scale-cuts}

\begin{figure}[tp]
    \centering
    \includegraphics[width=0.975\linewidth]{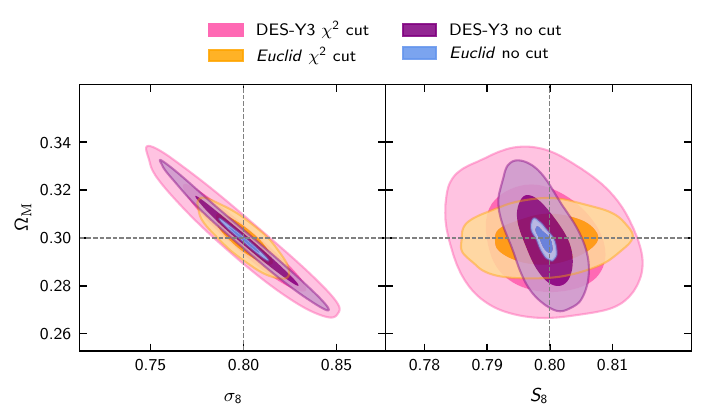}
    \caption{Two-dimensional $\Omegam$-$\sigmaeight$ and 
        $\Omegam$-$\Seight$ contours for our DES-Y3-like and 
        \textit{Euclid}-DR3-like surveys using the $\chi^2$ derived scale cuts
        (pink and orange) and no scale cuts (taking a maximum multipole of
        $\lmax = 5000$ for all bins, in purple and blue) for comparison.
        \textit{Left panel}: For our DES-Y3 survey, we see imposing scale cuts
            derived from the $\chi^2$ method does not significantly effect
            the one-dimensional parameter constraints on either $\Omegam$
            or $\sigmaeight$. However, it drastically reduces the width of the
            $\Omegam$-$\sigmaeight$ degeneracy, which is probed by $\Seight$.
            For our \textit{Euclid}-DR3-like survey, even with aggressively
            small scale cuts, we still see improvements in $\Omegam$ and 
            $\sigmaeight$ over the DES-Y3 results. 
        \textit{Right panel}: Plotting $\Seight$, which characterises the width
            of the $\Omegam$-$\sigmaeight$ degeneracy, we now see that when we
            impose the $\chi^2$ criterion for the scale cuts, this acts to
            optimise the $\Seight$ distribution regardless of survey specifications.
            Hence, even with dramatically improved quality of data from
            future Stage-IV surveys, our $\Seight$ constraints would be no
            better off than our existing Stage-III results when imposing no
            baryonic feedback in the data-vector. Since a key goal of Stage-IV
            experiments is to probe the $\Seight$ tension, having an $\Seight$
            distribution no tighter than DES-Y3 would be a disappointing outcome.
        }
    \label{fig:2D_contours_DESY3_Euclid_no_baryons}
\end{figure}

To illustrate how much cosmological information is lost from the extremely
aggressive scale cuts for our \textit{Euclid}-DR3-like survey, presented in
Table~\ref{tab:scale_cuts_chi_sq_no_baryons}, we can perform a simple 
Monte Carlo Markov chain (MCMC) analysis. Using a fiducial noise-free
uncontaminated data-vector, we can obtain cosmological parameter contours
from our DES-Y3-like and \textit{Euclid}-DR3-like surveys using both no
cuts (all scales down to $\lmax = 5000$ for all bins), and the $\chi^2$
cuts from Table~\ref{tab:scale_cuts_chi_sq_no_baryons}. This is
shown in Figure~\ref{fig:2D_contours_DESY3_Euclid_no_baryons}, plotting
the two-dimensional contours for $\Omegam$-$\sigmaeight$ and $\Omegam$-$\Seight$.

Here, we see that our scale cuts dramatically reduce the amount of cosmological
information from our data-vectors, as shown by the increased size of the
contours with respect to the no-cut curves. It is interesting to see that 
the one-dimensional distribution of $\Seight$ when we impose the $\chi^2$-derived
scale cuts is extremely similar between our DES-Y3 and \text{Euclid}-like surveys. 
This suggests that, since $\Seight$ is the most well constrained parameter
and is essentially the amplitude of the lensing signal~\cite{Hall:2021qjk},
the $\chi^2$ cuts are indirectly placing limits on $\Seight$. Hence, if we are
to use Stage-IV surveys to probe the $\Seight$ tension, then we need to come up
with alternate scale cut strategies to optimise our $\Seight$ distributions.

\subsection{Verifying the scale cuts}

Table~\ref{tab:scale_cuts_chi_sq_no_baryons} presents the sets of scale cuts
for our three mock surveys that aimed to eliminate any baryonic contamination
from the data-vectors. But how can we be sure that these scale cuts effectively
eliminate baryonic feedback? To do so, we can use one of our hydrodynamical
simulations' data-vector that features a large amplitude of baryonic feedback,
and repeat the same exercise of Figure~\ref{fig:2D_contours_DESY3_Euclid_no_baryons}
where we aim to recover cosmological parameter contours from our now
contaminated data-vector. Since we are using a dark-matter-only matter power
spectrum in our analysis, we \textit{expect} that our results will be biased away
from the ground truth unless we have adequately mitigated against baryonic
physics through our scale cuts. Hence, this is a good test of the method.

\begin{figure}[tp]
    \centering
    \includegraphics[width=0.975\linewidth]{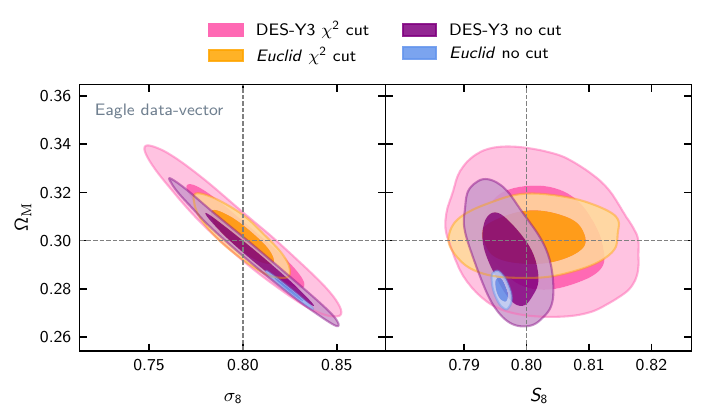}
    \caption{Two-dimensional $\Omegam$-$\sigmaeight$ and 
        $\Omegam$-$\Seight$ contours for our DES-Y3-like and 
        \textit{Euclid}-DR3-like surveys using $\chi^2$-derived and no
        scale cuts, using the \Eagle hydro-sim featuring baryonic feedback
        as the ground truth, and recovered using a dark-matter-only model.
        We see that without scale cuts, our two surveys are highly biased,
        which is to be expected when a dark-matter-only model is used. When
        we impose our scale cuts to eliminate baryonic feedback from our
        data-vector, we find that our contours are centred correctly, albeit
        with extremely compromised precision on $\Omegam$, $\sigmaeight$,
        and $\Seight$.
        }
    \label{fig:2D_contours_DESY3_Euclid_Eagle}
\end{figure}

Figure~\ref{fig:2D_contours_DESY3_Euclid_Eagle} presents the two-dimensional 
$\Omegam$-$\sigmaeight$ and $\Omegam$-$\Seight$ contours using the same scale
cuts and methods as Figure~\ref{fig:2D_contours_DESY3_Euclid_no_baryons}, but
now using the \Eagle hydrodynamical simulation as the ground truth. Here, we
see that when we apply our scale cuts, even using a dark-matter-only matter
power spectrum we correctly recover the ground truth cosmological parameters.
This shows that our $\chi^2$ cuts did effectively remove enough scales to
not be impacted by baryonic physics in this situation. For comparison, we plot
contours for our DES-Y3 and \textit{Euclid}-DR3 surveys without scale cuts, both
of which we expect biases in due to our dark-matter-only modelling. We see that
the DES-Y3 results are biased at the level of around $1 \sigma$, whereas
the results for our \textit{Euclid}-like survey are extremely biased at 
$>\!5\sigma$.

Hence, the balancing act becomes one of trying to preserve as much information
as possible to extract as much cosmological information from the data vectors
as possible without becoming susceptible to biases from baryonic
physics.

\subsection[Comparing to a $k$-space cut]{Comparing to a \bm$k$-space cut}

It is interesting to note that Table~\ref{tab:scale_cuts_chi_sq_no_baryons}
shows that as we increase the mean redshift of each spectra, the allowable
$\lmax$ decreases. This is in contrast to other previous analyses for
cosmic shear scale cuts which were motivated by a cut in wavenumber-space
($k$-cuts), for example the DES-Y1 approach~\cite{DES:2020daw}.

\begin{figure}[tp]
    \centering
    \includegraphics[width=0.975\linewidth]{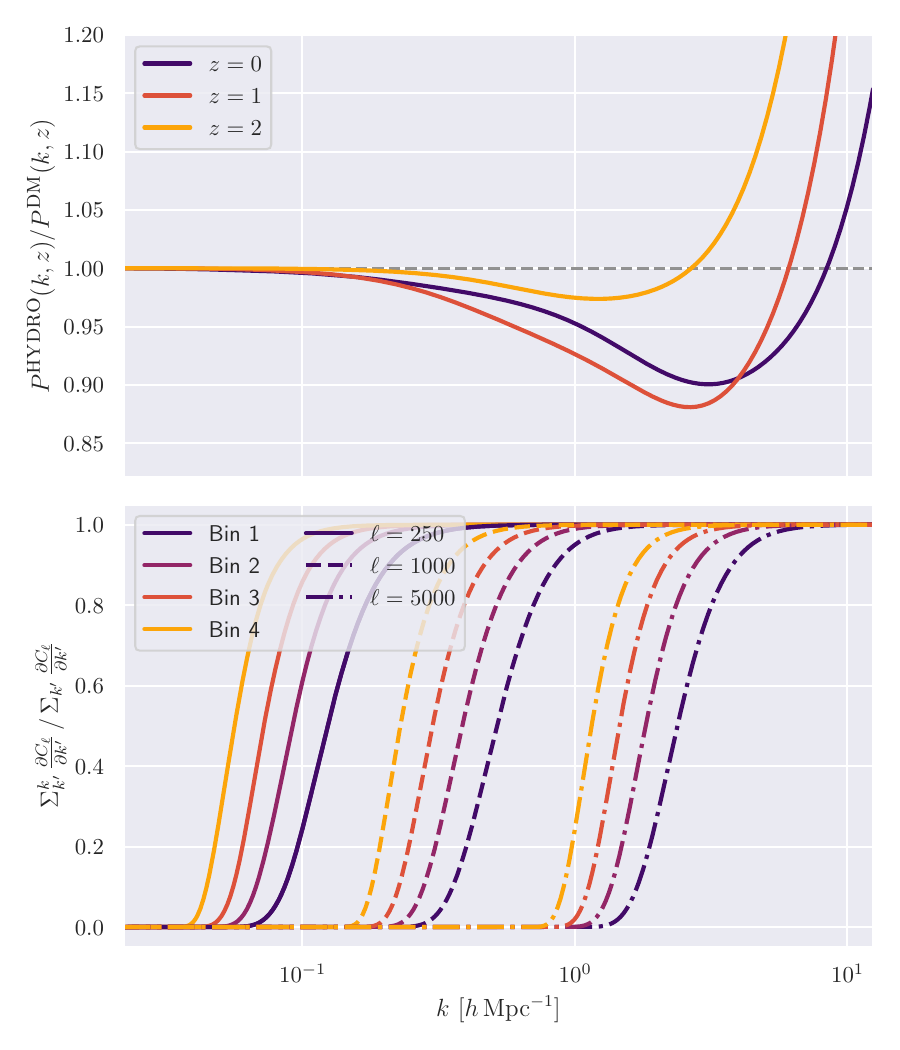}
    \caption{\textit{Top panel}: Ratio of the matter power spectrum with 
        baryonic feedback in to the dark-matter-only power spectrum as
        a function of wavenumber $k$ for three fixed redshifts 
        $z = 0, 1, 2$. We see that baryonic feedback is highly redshift
        dependent, and so a simple redshift-independent cut in $k$-space
        is insufficient to accurately model it.
        \textit{Bottom panel}: Normalised cumulative sum of the derivative of the
        $\Cl$ values with respect to $k$ for four different Gaussian redshift
        bins with centres $\bar{z} = \{0.33, 0.66, 1.00, 1.50\}$ and for
        three fixed $\ell$ values. We see that for fixed $\ell$, the
        higher redshift bins have contributions from smaller $k$ values,
        which would indicate that these bins have less contributions
        from baryonic feedback, however the top panel shows that baryonic
        feedback is highly redshift dependent and so this relation isn't
        as simple.
        }
    \label{fig:d_cl_d_k_cumsum}
\end{figure}

$k$-cuts are often motivated by the desire to eliminate physical scales
that feature baryonic feedback in the matter power spectrum from our
cosmic shear summary statistics. When looking at the Limber integral for
the $\Cl$ values using the matter power spectrum, of
\begin{align}
    C_{\ell}^{ab} = \frac{9 \Omegam^2 H_0^4}{4}
    \int_{0}^{\chi_\textsc{h}} \!\! \d \chi \, \, 
    \frac{g_a(\chi) \, g_b(\chi)}{a^2(\chi)} \, 
    P_{\delta} \left( k = \frac{\ell}{\chi}, \, z=z(\chi)\right),
\end{align}
we see that for fixed wavenumber $k$, larger comoving distances (as is the case
for higher redshift bins) produces larger angular multipole values.  
Thus, if one chooses a na\"ive scale cut in $k$-space only, then one
would find that for fixed $k_{\textrm{max}}$ higher redshift bins give larger
allowable multipoles $\lmax$ -- which is opposite of the results from using the
$\chi^2$ method. While it is true that for fixed a $\ell$ mode, the corresponding
wavenumber values are smaller for higher redshift bins (shown in Figure~\ref{fig:d_cl_d_k_cumsum}),
which would indicate that baryonic physics plays a smaller role in the $\Cl$ values, the dramatic
increase in signal-to-noise of these higher redshifts bins places tighter
constraints on the allowable errors than for low redshift bins.

\section[Extending the $\chi^2$ method]{Extending the \bm$\chi^2$ method }

It is clear that if we were to continue to apply the DES-Y3 approach of
reducing impact of baryonic feedback on our data-vectors to acceptable levels,
then headline constraints from forthcoming Stage-IV surveys would only be
marginally better than existing Stage-III surveys -- despite the 
overwhelmingly large increase in the precision on the data. Hence,
we must come up with alternative methods, or adaptations of existing
methods, if we are to truly unleash the constraining power of Stage-IV
surveys. The major limitation with the DES-Y3 approach is that we compare
the hydrodynamical simulations with dark-matter-only models. This naturally
results in harsh scale cuts from the poor modelling. While modelling baryonic
feedback is inherently a difficult process, we do have many existing models 
in the literature -- each with their own error profiles. 
Thus, we can extend the DES-Y3 approach by now
taking the difference between the hydro-sims and the results from baryonic
feedback models. By taking the same approach of computing the average $\chi^2$
to our ensemble of hydro-sims, we can now derive sets of physically motivated
binary scale cuts for each baryonic feedback model for each mock survey.

\subsection{Fitting baryonic feedback models to hydro-sims}

Now that we are using models of baryonic feedback, our $\chi^2$ criterion
becomes
\begin{align}
    \left(C_{\ell, \textrm{baryon}}^{ab} - C_{\ell, \textrm{model}}^{ab} \right)
    \, \mathbf{C}^{-1}_{abab} \,
    \left(C_{\ell, \textrm{baryon}}^{ab} - C_{\ell, \textrm{model}}^{ab} \right)^{\textsc{t}}
    \leq \frac{\chi^2_{\textrm{crit}}}{N},
    \label{eqn:chi_sq_crit_ell_space_baryons}
\end{align}
where $C_{\ell, \textrm{model}}^{ab}$ are the $\Cl$ values using a particular
model. The question now becomes how to find these model $\Cl$ values for a given
hydro-sim, which gives us the error in these values to be used in our $\chi^2$
criterion. To do so, we used a maximum-likelihood optimiser to find the 
best-fitting $\Cl$ values for a given model. The baryonic parameters,
along with cosmological parameters $\As$ and $\Omegac$, were varied by the
optimiser to find the best-fit values.

\begin{figure}[tp]
    \centering
    \includegraphics[width=\linewidth]{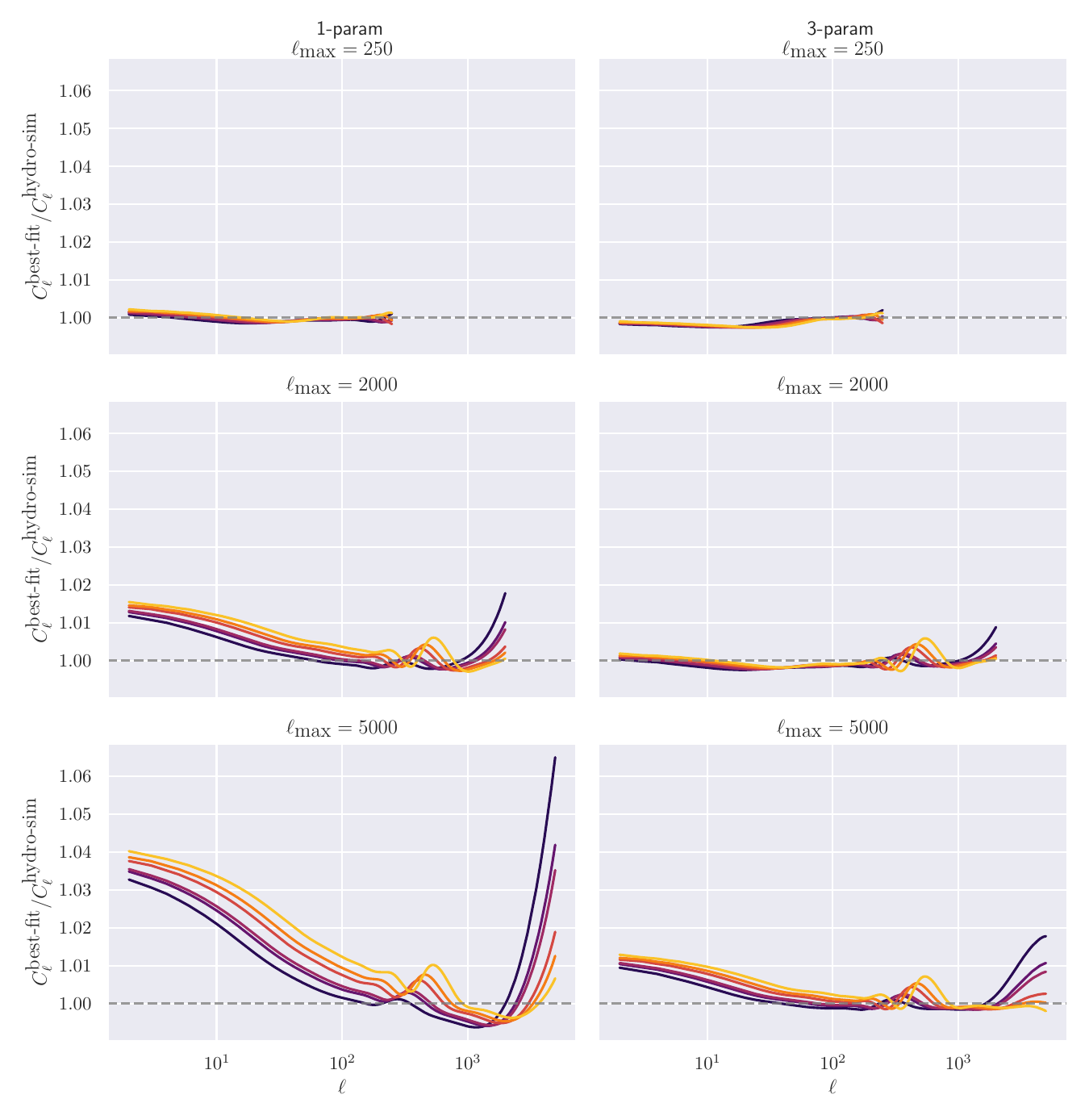}
    \caption{Plot of the ratio of the best-fitting $\Cl$ values
        of our one- and three-parameter models to that of the baryonic feedback
        from the \TNG hydrodynamical simulation. We see that when 
        limiting the fit to an $\lmax$ of $250$, the errors are much less than 
        $1\, \%$ across the entire $\ell$ range. However, when we extend the
        fit down to an $\lmax$ of $2000$ and then $5000$, these previously
        well-fitted values now have significantly larger errors in the fit,
        since the optimiser is optimising the global likelihood that sacrifices
        the fit of low $\ell$ values. The curves are coloured to approximate their
        redshift, with the lighter yellow curves for higher redshift bins. We
        see that at high $\ell$, where the signal-to-noise is largest, the
        higher redshift bins tend to have a better fit than for lower redshift
        bins, owing to the larger signal-to-noise of these higher redshift bins.
        This also shows how the more flexible three-parameter model produces
        significantly better fits when baryonic feedback becomes important at 
        $\lmax$'s of 2000 and 5000. Thus, when determining 
        $(\vec{C}_{\ell}^{\textrm{truth}} - \vec{C}_{\ell}^{\textrm{model}})$,
        this is an $\lmax$ dependent quantify.
        }
    \label{fig:Cl_best_fit_TNG}
\end{figure}

Since we are now fitting our models to the hydro-sims, it is important to note
that these fits become $\lmax$ dependent. This is because, for an insufficient
model of baryonic feedback, the optimiser has to sacrifice accuracy at lower
multipoles when trying to fit down to higher multipoles (since those have a
larger signal-to-noise than larger angular scales) to optimise the maximisation
of the global likelihood. This problem is illustrated
in Figure~\ref{fig:Cl_best_fit_TNG}, which plots the ratio of the
best-fitting $\Cl$ values for our one-parameter and three-parameter models 
against the ground-truth values for the \TNG hydrodynamical simulation,
for three values of $\lmax$. We see that for an $\lmax$ of $250$, there is
very little baryonic contamination within our data-vectors, and so achieve 
greater than percent-level accuracy with our fits for both baryonic feedback
models. When we extend down to an $\lmax$ of 2000 and then 5000, we significantly
increase the effects of baryonic contamination within our data-vectors, and so
become sensitive to the accuracy of our chosen model of baryonic feedback. In
order to optimise the global likelihood, which is weighted towards larger $\ell$
values from their higher signal-to-noise, the optimiser prefers to alter the
values of $\As$ and $\Omegac$ such that a better fit at high $\ell$ can be
obtained at the expense of low $\ell$ accuracy. This problem becomes particularly
acute for the one-parameter model at an $\lmax$ of 5000, since the model
simply does not fit the data well.

It should be noted that for very high $\ell$ multipoles, shape noise dominates 
the covariance matrix and so the contribution to the $\chi^2$ from these
modes is reduced. Hence, the model can have excursions out to very large
relative amplitudes without changing the $\chi^2$ by significant amounts.

We refer the reader back to Sections~\ref{sec:modelling_baryonic_feedback} 
and~\ref{sec:num_eval_pk_hmcode}, and Table~\ref{tbl:baryon_priors} 
for an introduction, discussion, and reminder
about the relevant astrophysical baryonic feedback parameters that feature in
the HMCode-2020 model. We also refer back to Table~\ref{tbl:hydrosims} presenting
the specifications of the hydrodynamical simulations chosen in this work.

It is clear looking at Figure~\ref{fig:Cl_best_fit_TNG} that if were one to
go about computing a $\chi^2$ for an $\lmax$ of 250 for all three fits, then
one would arrive at three very different values. Hence, when we wish to go 
about obtaining an estimate for
$\left(C_{\ell, \textrm{baryon}}^{ab} - C_{\ell, \textrm{model}}^{ab} \right)$
in our determination of the scale cuts (Equation~\ref{eqn:chi_sq_crit_ell_space_baryons})
then this itself is an $\lmax$ dependent quantity. This dependence should be
accounted for when optimising the scale cuts using any criterion.

To be explicit, the steps that we used to obtain scale cuts using our
modified $\chi^2$ method were as follows:
\begin{galitemize}
    \item On a grid ranging from $\lmax = 150$ to $\lmax = 5000$ with
        $\ell = 25$ spacing, fit the one- and three-parameter HMCode to each
        of the five hydro-sims in our ensemble. We use $\As$ and $\Omegam$
        along with the baryonic parameters as the free-parameters in the fit.
        For the angular power spectrum and its covariance matrix, we
        use five linearly-spaced bins from $\ell = 2$ to $\ell = 100$, and
        then twenty logarithmically-spaced bins from $\ell = 101$ to $\ell = \lmax$.
        We use Equation~\ref{eqn:gaussian_cl_cov} to determine the $\ell$-space
        covariance matrix, which is used in the optimisation of the Gaussian
        likelihood.
    
    \item With the fits to the hydro-sims obtained, we can go about evaluating
        the $\chi^2$ criterion 
        to determine the scale cuts for each bin pair $ab$. Since we are only
        performing a one-dimensional optimisation, the \texttt{GridMax} sampler
        worked well for our use-case. 

    \item For each $\lmax$ considered by the \texttt{GridMax} sampler, we round
        up to the nearest multiple of 25, which allows us to load in the
        previously calculated best-fits of each hydro-sim. We then take the
        $ab$ sub-block of the data vector for each hydro-sim and the $abab$
        sub-block of the inverse covariance matrix calculated for the given 
        $\lmax$ to compute the $\chi^2$ for each hydro-sim. We then take the
        maximum $\chi^2$ value from the set of hydro-sims, which is then
        compared against $\chisqcrit$ to form our likelihood. 

    \item The \texttt{GridMax} sampler then optimises the maximum multipole
        such that the likelihood, and thus criterion, is maximised.

\end{galitemize}

\subsection[Results for our modified $\chi^2$ method]{Results for our modified \bm$\chi^2$ method}

With our best-fit analysis performed using our two baryon feedback models on
our suite of hydrodynamical simulations, we can use our `modified $\chi^2$'
criterion to determine scale cuts. These are shown in
Table~\ref{tab:scale_cuts_chi_sq_wh_baryons}. We see the general trends that
emerged in the no-baryons case continue for the case where baryonic feedback
is included in the theoretical models: that the extra constraining power
of \textit{Euclid} over DES results in much lower scale cuts across the board,
which can be partially mitigated by using a three-parameter model over
a single-parameter model. 

Therefore, using our set of derived scale cuts, we now want to see what impact
these have on the mitigation of baryonic feedback using contaminated data-vectors.

\begin{table}[t]
    \centering
    \caption{Results for our scale cuts using our `modified $\chi^2$'
        approach for our two mock surveys and two baryonic feedback models.
        We again see that the increased precision going from our mock DES-Y3
        to mock \textit{Euclid}-DR3 survey lowers the allowable maximum 
        multipole across the bin ranges. We also see that the increased
        flexibility of the three-parameter model over the one-parameter model 
        reduces errors in the $\Cl$ fits, and thus increases the maximum multipole
        for a given error budget. Though we still see extremely low values of
        $\lmax$ in the $\bins{2}{3}$ bins as a result of using the $\chi^2$
        criterion for our \textit{Euclid}-like survey.
        }
    \label{tab:scale_cuts_chi_sq_wh_baryons}

    \bgroup
    \setlength\tabcolsep{0.35cm}
    \begin{tabular}{lrrrr}
        \toprule
        Bin & DES-Y3 1-param & \textit{Euclid} 1-param & DES-Y3 3-param & \textit{Euclid} 3-param \\
        \midrule
        
        $\bins{1}{1}$ &     5000 &      5000 &      5000 &      5000 \\
        $\bins{1}{2}$ &     2897 &       923 &      5000 &      2897 \\
        $\bins{1}{3}$ &     4497 &      1501 &      5000 &      4497 \\
        $\bins{2}{2}$ &     4798 &       485 &      5000 &       531 \\
        $\bins{2}{3}$ &     3703 &       200 &      4892 &       210 \\
        $\bins{3}{3}$ &     4892 &       337 &      5000 &       489 \\

        \bottomrule
    \end{tabular}        
    \egroup
\end{table}

\subsubsection{MCMC results using scale cuts}

\begin{figure}[t]
    \centering
    \includegraphics[width=\linewidth]{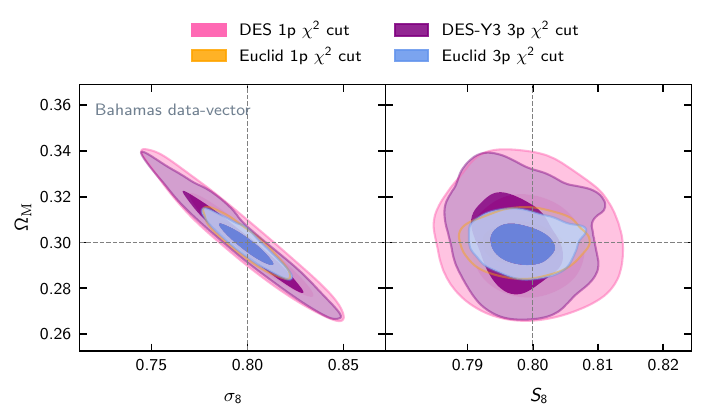}
    \vspace*{-1cm}
    \caption{Two-dimensional $\Omegam$-$\sigmaeight$ and 
        $\Omegam$-$\Seight$ contours for our DES-Y3-like and 
        \textit{Euclid}-DR3-like surveys using the $\chi^2$ with a baryonic
        feedback model derived scale cuts for the \Bahamas hydro-sim as input.
        First, we see that there is no baryonic feedback bias present in any
        of the analyses presented (though that is to be expected from the
        \Bahamas hydro-sim since the one-parameter HMCode model was explicitly
        trained on \Bahamas only). Secondly, we see that the increased constraining 
        power of \textit{Euclid} over DES-Y3 can be utilised to further 
        constrain $\Seight$ over what our mock DES-Y3 can achieve. Thirdly, we
        see that the more relaxed scale cuts from using the three-parameter
        over the one-parameter model allow us to extract more information from
        our data vectors (the purple and blue contours being smaller than their
        pink and yellow counterparts) without inducing biases for the \Bahamas
        data-vector. 
    }

    \label{fig:2D_contours_chi_sq_wh_baryons_Bahamas}
\end{figure}

\begin{figure}[t]
    \centering
    \includegraphics[width=\linewidth]{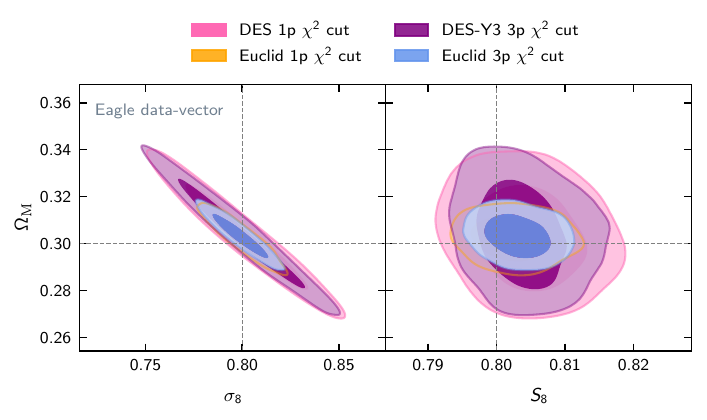}
    \vspace*{-1cm}
    \caption{Two-dimensional $\Omegam$-$\sigmaeight$ and 
        $\Omegam$-$\Seight$ contours for our DES-Y3-like and 
        \textit{Euclid}-DR3-like surveys using the $\chi^2$ with a baryonic
        feedback model derived scale cuts for the \Eagle hydro-sim as input.
        As \Eagle features more significant baryonic feedback over \Bahamas, 
        it provides a good testing ground for our scale cuts methods. What we
        see is broadly similar to the results for \Bahamas presented in 
        Figure~\ref{fig:2D_contours_chi_sq_wh_baryons_Bahamas}: we are now
        able to take advantage of the additional information in the 
        \textit{Euclid} covariance with smaller parameter constraints for all
        three cosmological parameters; and that the three-parameter model can
        still deliver unbiased results with smaller parameter covariances
        over the one-parameter model. This is because, while the three-parameter
        model can access higher-order multipoles, the increase in constraining
        power is somewhat counteracted by the two additional nuisance parameters
        that need marginalising over to present these cosmological contours.
    }

    \label{fig:2D_contours_chi_sq_wh_baryons_Eagle}
\end{figure}

To obtain estimates of how our cosmic shear analyses react to the scale cuts
imposed in Table~\ref{tab:scale_cuts_chi_sq_wh_baryons}, we can repeat our
MCMC analyses where we use various hydrodynamical simulations as the input
power spectra, and investigate the associated biases in cosmological parameters
from them. Figure~\ref{fig:2D_contours_chi_sq_wh_baryons_Bahamas} plots the
2D parameter contours using our modified $\chi^2$ method for the \Bahamas
hydro-sim as the ground truth, with Figure~\ref{fig:2D_contours_chi_sq_wh_baryons_Eagle}
using \Eagle as the ground truth.

In these figures, we see that through using a baryonic feedback method and 
deriving scale cuts using these methods, we arrive at results that allow our
\textit{Euclid}-like survey to harness its increased statistical power over
our DES-Y3-like survey to better constrain all three cosmological parameters
(which was not true for the no-baryons case). This certainly gives some 
reassurances behind the motivation for Stage-IV surveys! Secondly, 
Table~\ref{tab:scale_cuts_chi_sq_wh_baryons} showed that the scale cuts from
our three-parameter model allowed the probing of higher-order multipoles than
those allowed from the one-parameter model. While these new multipoles should
bring additional constraining power, this is somewhat mitigated by sampling
over two additional nuisance parameters. However, what we see in 
Figures~\ref{fig:2D_contours_chi_sq_wh_baryons_Bahamas}
and~\ref{fig:2D_contours_chi_sq_wh_baryons_Eagle} is that there is a net
decrease in the parameter's covariance for both hydro-sims when going from
one to three parameters. This provides strong motivation for us to use more
general models of baryonic feedback which allow us to probe smaller-scales
than simpler models, even if this comes at the expense of additional
nuisance parameters that need marginalising over. The extra information from
accessing the higher-order multipoles more than compensates for the additional
parameters.

\section{Scale cuts in parameter-space}

As we have described, the $\chi^2$ values operate in data-space. Whilst this is
a natural choice, especially since it does not require any transformations
of our data-vector or its covariance matrix, it may not properly reflect the 
biases in cosmological constraints that these data-vectors are used to obtain.
Thus, we wish to transform these quantities into parameter-space to obtain
more physical scale cuts. 

\subsection{The Fisher matrix and figures of merit \& bias}

The two quantities that we are interested in are
the figure of merit (FoM) and the figure of bias (FoB). Both of these are
derived from the Fisher matrix, $\mathbf{F}_{\alpha \beta}$, defined as~\cite{Euclid:2019clj}
\begin{align}
    \mathbf{F}_{\alpha \beta} =
    \frac{\partial \vec{C}_{\ell}}{\partial \vartheta_{\alpha}}
    \mathbf{C}^{-1}
    \frac{\partial \vec{C}_{\ell}}{\partial \vartheta_{\beta}}.
\end{align}
Thus, the Fisher matrix encodes both the covariance of our underlying
data-vector, and the response of our data-vector to the parameters that
we are interested in constraining through our analyses. From the Fisher
matrix, we can construct the figure of merit, which encodes how well
our parameters are constrained, with larger values giving increased
constraining power. To calculate the constraining power of two specific
parameters $\alpha \& \beta$, we need to use the marginalised Fisher
matrix $\tilde{\mathbf{F}}_{\alpha \beta}$, which is computed by inverting
the full Fisher matrix, isolating the rows and columns that correspond to our
parameters only, and then re-inverting the matrix. This gives our figure of
merit as
\begin{align}
    \textrm{FoM}_{\alpha \beta} = \sqrt{\det \left[\tilde{\mathbf{F}}_{\alpha \beta}\right]}.
\end{align}

From our Fisher matrix, we can also compute the figure of bias, which encodes
how a mis-specification in the data-vector can lead to biases in the
cosmological parameters. 
The bias on parameter $\vartheta_{\alpha}$, $\delta \vartheta_{\alpha}$, is
given by~\cite{Amara:2007as}
\begin{align}
    \delta \vartheta_{\alpha} = \left[\mathbf{F}^{-1}\right]_{\alpha \beta}
    \left( \vec{C}_{\ell}^{\textrm{true}} - \vec{C}_{\ell}^{\textrm{model}}  \right)
    \mathbf{C}^{-1} \, \frac{\partial \vec{C}_{\ell}}{\partial \vartheta_{\beta}},
    \label{eqn:delta_theta_alpha}
\end{align}
where $\vec{C}_{\ell}^{\textrm{true}}$ is the underlying data-vector that we wish
to fit our model to, and $\vec{C}_{\ell}^{\textrm{model}}$ is the resulting 
best-fit data-vector using our model. The figure of bias can then be computed
from the combination of individual parameter biases through~\cite{Gordon:2024jaj}
\begin{align}
    \textrm{FoB} = \sqrt{\overrightarrow{\delta \vartheta}_{\alpha} \, \tilde{\mathbf{F}}_{\alpha \beta} \, \overrightarrow{ \delta \vartheta}_{\beta}},
\end{align}
thus providing an overall summary statistic for how an insufficient model leads
to biased parameters.

It is important to note that while the figures of merit and bias may appear
to have similar properties, they behave significantly different when considering
different parametrisations. This is because the figure of merit is unitful. That is,
since it is inversely proportional to the area of the 2D contour, a reparametrisation 
of the form $\vartheta_{\alpha} \rightarrow \vartheta_{\alpha'} = 10 \vartheta_{\alpha}$,
would lead to each entry in the Fisher matrix being one hundred times smaller. 
Thus, a figure of merit with the new parametrisation is ten times smaller ($\textrm{FoM}_{\alpha \beta}
\rightarrow \textrm{FoM}_{\alpha' \beta} = \frac{1}{10} \textrm{FoM}_{\alpha \beta}$)
than in the old parametrisation. Though, of course, the physical interpretation
of either parametrisation would lead to the same conclusions, and so we
mostly care about differences in the figure of merit, rather than its
absolute value.

Conversely, the figure of bias \textit{is} unitless and so simple reparametrisations
do not affect its value. Hence, it is always the absolute values of the FoB that
inform us about parameter biases. Since the Fisher matrix, and thus figure of bias,
were derived under Gaussian conditions, we can use the results for the multivariate
$\chi^2$ distribution to find critical values of the FoB which give certain
levels of $\sigma$ offsets. Table~\ref{tab:chi_sq_crit_vals} gives these critical
values as a function of the $\sigma$ offset for one, two, and four degrees
of freedom models. Hence, if we were to target no greater than a $1 \sigma$
offset for a two-parameter model, then we could accept a figure of bias no
greater than $\sqrt{2.30}$. This number is then independent of any parametrisations
that we might have, and only depends on our desired precision and number of
free parameters in our model.

\begin{table}[t]
    \centering
    \caption{Values of $\Delta \chi^2$ / FoB$^2$ as a function
        of confidence level and degrees of freedom, taken from Ref.~\cite{NumRec:2007}.
        These can be calculated from the inverse cumulative distribution 
        function of the $\chi^2$ distribution. 
        }
    \label{tab:chi_sq_crit_vals}
    \bgroup
    \setlength\tabcolsep{0.35cm}
    \begin{tabular}{ccccc}
        \toprule
        $\sigma$ offset \quad & $p$ \quad & $ \nu = 1$ \quad & $\nu = 2$ \quad & $\nu = 4$ \quad\\
        \midrule
        
        $1 \sigma$ & $68.3 \, \%$ & 1.00 & 2.30 & 4.72 \\
        $2 \sigma$ & $95.4 \, \%$ & 4.00 & 6.17 & 9.70 \\
        $3 \sigma$ & $99.7 \, \%$ & 9.00 & 11.8 & 16.3 \\
        
        \bottomrule
    \end{tabular}        
    \egroup
\end{table}

\subsection{Using the figure of bias for scale cuts}

With the figure of bias as our statistic of choice, we can now go about trying
to use it to obtain a set of scale cuts.
Since we have an ensemble of hydro-sims, each with their own baryonic feedback,
we want to guard against \textit{any and all} of them with our figure of bias
cuts. Thus, when determining the scale-cuts, we optimise against the maximum
of the set,
\begin{align}
    \max \left[\textrm{FoB}_{\textrm{sims}} (\vec{\ell}_{\textrm{max}})  \right] \leq \textrm{FoB}_{\textrm{crit}},
\end{align}
where $\textrm{FoB}_{\textrm{crit}}$ is a free-parameter of the method. We chose
to set $\textrm{FoB}_{\textrm{crit}} = 0.5$, since that provided a good
compromise between constraining power and biases, and ensured that no bias 
would ever be larger than one sigma.

\section{Scale cuts using the figure of bias statistic}

\subsection{One-dimensional optimisations}

Thus far, we have constructed our scale cuts using our `no baryons' and `with
baryons' $\chi^2$ methods for a single bin at a time. The advantage of this
approach is that there is a unique solution for each bin, and that the problem
becomes $N \times 1$D optimisations, rather than a $1 \times N$D problem. Hence, we wish to
continue this approach of using one-dimensional optimisations, but using
the figure of bias as our critical statistic. 

\begin{table}[t]
    \centering
    \caption{Results for scale cuts using the one-dimensional optimisations
        using the figure of bias as the critical statistic. We see the same
        general trends that emerged when using the $\chi^2$ statistic, that
        the enhanced statistical precision of \textit{Euclid} demands tighter
        scale cuts to accept the same level of bias as our DES-Y3-like survey,
        and that the better-fitting three-parameter model yields looser cuts.
    }

    \label{tab:scale_cuts_FoB_1D}

    \bgroup
    \setlength\tabcolsep{0.35cm}
    \begin{tabular}{lrrrr}
        \toprule
        Bin & DES-Y3 1-param & \textit{Euclid} 1-param & DES-Y3 3-param & \textit{Euclid} 3-param \\
        \midrule
        
        $\bins{1}{1}$ &     5000 &  2859 &  5000 &  5000  \\
        $\bins{1}{2}$ &     5000 &  199  &  4850 &  3399 \\
        $\bins{1}{3}$ &     5000 &  349  &  4897 &  2130 \\
        $\bins{2}{2}$ &     4649 &  226  &  4850 &  3399 \\
        $\bins{2}{3}$ &     1663 &  249  &  1830 &  3148 \\
        $\bins{3}{3}$ &     2150 &  240  &  4649 &  1337 \\

        \bottomrule
    \end{tabular}        
    \egroup

\end{table}
Table~\ref{tab:scale_cuts_FoB_1D} shows the set of derived scale cuts using
the figure of bias as our critical statistic, for our set of six one-dimensional
optimisations. Here, we see the same general trends that emerged when using the
$\chi^2$ as the critical statistic. Since we are applying the FoB criterion
on each bin pair individually, and some of the bin pairs are not very 
constraining, we can go to very high $\lmax$ in these bins without inducing
significant bias. For all analyses, we applied a global maximum of $\lmax = 5000$,
which is the maximum multipole that the cosmic shear signal can be accurately
measured down to for any \textit{Euclid} analyses~\cite{Euclid:2024xqh}. 

\begin{figure}[tp]
    \centering
    \includegraphics[width=\linewidth]{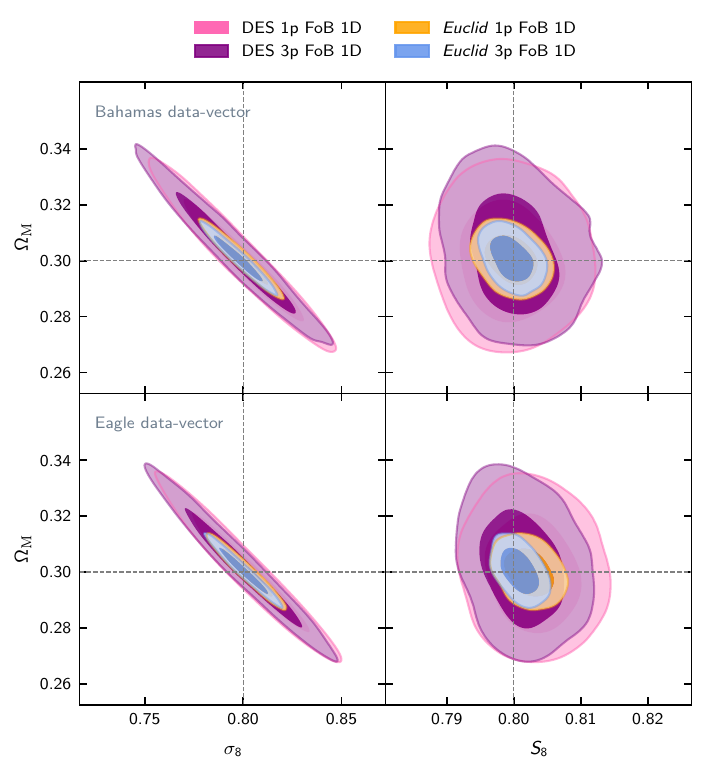}
    \vspace*{-1cm}
    \caption{Two-dimensional $\Omegam$-$\sigmaeight$ and 
        $\Omegam$-$\Seight$ contours for our DES-Y3-like and 
        \textit{Euclid}-DR3-like surveys using the figure of bias as our
        critical statistic with one-dimensional optimisations for the \Bahamas
        and \Eagle hydro-sims as input data-vectors. We see that all FoB-derived
        scale cuts yield unbiased results for our three cosmological parameters,
        which is verification that the figure of bias can indeed be used to 
        construct scale cuts which correctly mitigate baryonic feedback. We
        also see that, for our \textit{Euclid}-like survey, the use of
        the three-parameter model leads to smaller contours over the one-parameter
        model, since it is able to access higher-order multipoles without
        biasing our results. 
    }

    \label{fig:2D_contours_FoB_1D_Bahamas_Eagle}
\end{figure}

Using our sets of derived scale cuts, we can now repeat the same MCMC analyses
shown in Figures~\ref{fig:2D_contours_chi_sq_wh_baryons_Bahamas} 
and~\ref{fig:2D_contours_chi_sq_wh_baryons_Eagle} to determine what the
effects of these scale cuts are on our parameter covariances and biases. This is
presented in Figure~\ref{fig:2D_contours_FoB_1D_Bahamas_Eagle}, which shows that
all sets of scale cuts derived from our figure of bias statistic correctly 
produce unbiased contours, verifying our use of the figure of bias as our
criterion. We also see that for our \textit{Euclid}-like survey, we find
smaller parameter contours for the three-parameter model over the one-parameter
model. This is due to the three-parameter model being able to access higher-order
multipoles without being susceptible to significant biases.

\subsection[Comparison between 1D $\chi^2$ and FoB scale cuts]{Comparison between 1D \bm$\chi^2$ and FoB scale cuts}

We are now in a position where we have developed sets of scale cuts that have
been derived from two statistics: the $\chi^2$ using baryonic feedback models,
and the figure of bias. We now wish to compare the performance of these sets
of scale cuts to see if there are any differences in the results of these
methods, which could tell us if using either data- or parameter-space lead to
more optimal cuts.

\begin{figure}[tp]
    \centering
    \includegraphics[width=\linewidth]{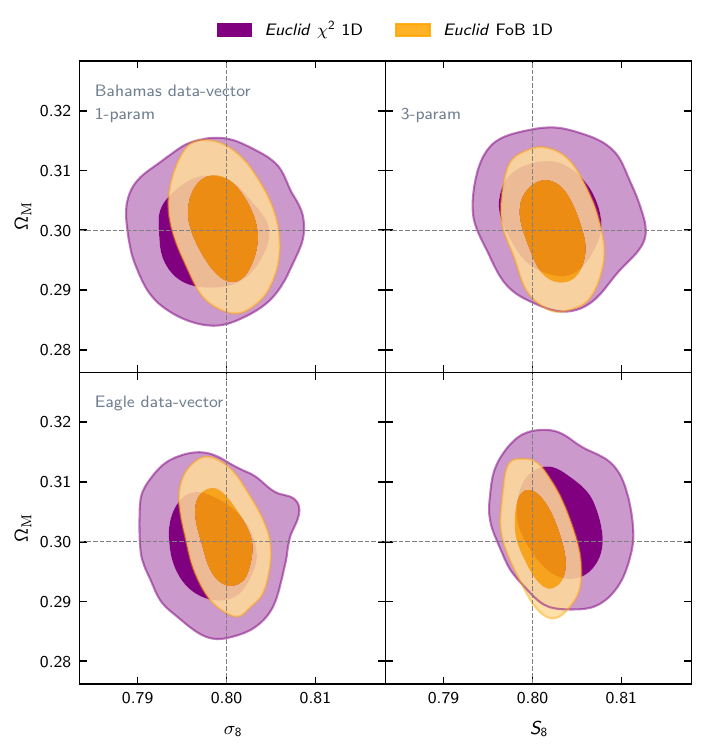}
    \vspace*{-1cm}
    \caption{Two-dimensional $\Omegam$-$\sigmaeight$ and 
        $\Omegam$-$\Seight$ contours for our \textit{Euclid}-DR3-like survey
        comparing the results obtained using both the $\rchi^2$ (purple) and
        figure of bias (orange) derived scale cuts. The rows of the plot denote
        the \Bahamas and \Eagle hydro-sims as input, with the columns plotting the
        one- and three-parameter models, respectively. We see systematically
        smaller contours for the figure of bias derived cuts, which suggests that
        this statistic is more optimal than optimising using the raw 
        data-space $\rchi^2$ values. Here, `1D' refers to use optimising each
        bin combination individually, and not optimising the global $\rchi^2$
        or FoB for all bin combinations simultaneously.
    }

    \label{fig:2D_contours_chi_sq_and_FoB_1D_Bahamas_Eagle}
\end{figure}

Figure~\ref{fig:2D_contours_chi_sq_and_FoB_1D_Bahamas_Eagle} plots the
parameter contours derived from using the $\rchi^2$ and figure of bias statistics
for our \textit{Euclid}-like survey. Here, we see that the contours derived
from the figure of bias criterion are systematically tighter than their $\rchi^2$
counterparts. This suggests that the figure of bias is correctly identifying
and excluding the modes that actually lead to biases in the cosmological 
parameters, not just the raw differences in the data-vectors that the $\rchi^2$
is computed from. This provides significant verification that the figure
of bias can be used for the optimal computation of scale cuts for Stage-IV cosmic
shear surveys.

\subsection{Six-dimensional optimisation}

Thus far, we have only been doing scale cut optimisation on a bin-by-bin basis.
This matches what has been done in the literature previously, and considerably
reduces the numerical complexity of our algorithms to determine the scale cuts.
However, we know that the scale cuts used in any analysis are not totally
independent, since valuable information comes from the cross-correlations
between the different spectra. Hence, when we optimise against either a 
$\rchi^2$ or FoB criterion, then one should account for these cross-correlations
in the optimisations. Thus, the optimisation algorithm vary all of the scale
cuts together, and so for three tomographic bins, we have six scale cuts to
optimise.

We use the one-dimensional optimisations as starting positions for the
six-dimensional optimisation. This is because, while we are now including the
effects of cross-correlations, if each bin satisfies the FoB criterion itself,
then it is hoped that the overall figure of bias will be close to the
desired value. This means that the optimisation algorithm hopefully converges
much faster now that we are starting it from an informed position, rather than
them to random or non-sensible values.  

With our starting position given by our one-dimensional optimisations using the
figure of bias as our critical statistic, we can move up to exploring the full
six-dimensional parameter space and optimising them together. However, as has
been found from experimenting with the optimisation algorithm, the six-dimensional
figure of bias likelihood surface is far from simple, and so traditional
optimisation methods often failed to correctly converge. Hence, a different
optimisation technique was employed, which was the \texttt{GridMax} sampler
as part of \textsc{Cosmosis}. This optimises the likelihood by probing each
dimension individually, while keeping all over variables fixed. This aims to
`spiral in' to an optimal solution, and was found to work much better for our
complex likelihood surface than other optimisers (such as \texttt{Minuit} and
\texttt{maxlike} as provided with \textsc{Cosmosis}). 

One subtlety to using the \texttt{GridMax} sampler is the order in which it
probes each dimension in. Since it keeps all other variables fixed as it probes
along each axis, the response of the likelihood to the current dimension 
strongly depends on the other scale cut values. To investigate this effect,
numerous chains of the \texttt{GridMax} sampler were run, with different
orderings of the optimisation. For example, one chain was run with the
fiducial ordering of $(\bins{1}{1}, \, \bins{1}{2}, \, \bins{1}{3}, \, \bins{2}{2}, \, \bins{2}{3}, \, \bins{3}{3})$,
whereas another with its inverse ordering 
$(\bins{3}{3}, \, \bins{2}{3}, \, \bins{2}{2}, \, \bins{1}{3}, \, \bins{1}{2}, \, \bins{1}{1})$.
Both chains give the same figure of bias, but very different results for each
bin's scale cuts. 

Tables~\ref{tab:fob_cuts_1param} and~\ref{tab:fob_cuts_3param} feature ten
different sets of scale cuts that were the result of running the \texttt{GridMax}
sampler in different order of the bin combinations.
Even though the value for the figure of bias is the same for each chain (or at
least close to be within a small numerical tolerance), the value for the
figure of merit and $\rchi^2$ vary quite significantly between each of
the runs. We see that there is a loose connection between higher allowable
multipoles in the $\bins{2}{3}$ and $\bins{3}{3}$ spectra and increased
constraining power, via the figure of merit. This suggests that the `bias budget'
should be used up in the most constraining spectra, which then necessitates 
tighter cuts in the less constraining lower redshift spectra. However, it is
clear that by using just the figure of bias, no optimisation routine can give
unique solutions, so either additional terms in the likelihood could be included
(such as a small weight to the $\rchi^2$ or figure of merit), or another
optimisation method is needed to be found. 

\begin{table}[tp]
    \centering
    \caption{Scale cuts (that is the maximum possible $\ell$ multipole)
        for the one-parameter model obtained using our 6-dimensional optimisation
        approach using the figure of bias as the critical statistic. We note
        that there is not a single unique solution which satisfies our
        criterion, but instead a whole family of solutions. We see that, in general,
        the optimisation routine has chosen to set one bin's $\lmax$ to be quite
        low (of the order a few hundred), with the remaining five bins having a
        much larger maximum multipole, with some approaching our hard maximum
        of 5000. We have sorted the results by their figure of merit for 
        $(\Seight, \Omegam)$. We see that the most constraining set of scale cuts,
        that is the set with the largest FoM, is one in which the smallest $\lmax$
        is still in the thousands. This suggests that the most-optimal set of
        cuts are where the $\lmax$ values are all about the same order, and
        not having an over order of magnitude difference between smallest
        and largest multipoles. 
        }
    \label{tab:fob_cuts_1param}

    \bgroup
    \setlength\tabcolsep{0.45cm}
    \begin{tabular}{rrrrrrrrrr}
        \toprule
        
        $\bins{1}{1}$ & $\bins{1}{2}$ & $\bins{1}{3}$ & $\bins{2}{2}$ & $\bins{2}{3}$ & $\bins{3}{3}$ & $\chi^2$ & FoM \\
        
        \midrule

        5000 &        1681 &        2957 &        5000 &        4744 &           405 &            3.74 &      59000 \\
        2859 &        4234 &        5000 &         150 &        5000 &          1426 &            3.69 &      67300 \\
        4301 &        4489 &         673 &         606 &        4946 &          2575 &            4.10 &      68700 \\
        1936 &        3978 &        5000 &        5000 &        5000 &          1426 &            4.44 &      69300 \\
         915 &        4059 &        5000 &        5000 &        5000 &          1574 &            4.35 &      69900 \\
        4489 &        5000 &        5000 &        3723 &        1426 &          3468 &            4.51 &      69900 \\
        4744 &        5000 &        5000 &        5000 &        5000 &          1439 &            4.02 &      70100 \\
        3978 &        5000 &        5000 &        3723 &        5000 &          2447 &            4.54 &      73000 \\
         277 &        5000 &        5000 &        5000 &        4489 &          3213 &            4.29 &      73400 \\
        2192 &        5000 &        5000 &        3723 &        5000 &          4489 &            4.41 &      74800 \\

        \bottomrule
    \end{tabular}        
    \egroup
\end{table}

\begin{table}[tp]
    \centering
    \caption{Scale cuts (that is the maximum possible $\ell$ multipole)
        for the three-parameter model obtained using our 6-dimensional optimisation
        approach using the figure of bias as the critical statistic. We note
        that there is not a single unique solution which satisfies our
        criterion, but instead a whole family of solutions. 
        }
    \label{tab:fob_cuts_3param}

    \bgroup
    \setlength\tabcolsep{0.45cm}
    \begin{tabular}{rrrrrrrrr}
        \toprule
        
        $\bins{1}{1}$ & $\bins{1}{2}$ & $\bins{1}{3}$ & $\bins{2}{2}$ & $\bins{2}{3}$ & $\bins{3}{3}$ & $\chi^2$ & FoM \\
        
        \midrule

        4234 &        1224 &         432 &        3481 &         512 &        1171 &  1.25 &  33100 \\
        741  &        1305 &         432 &        3508 &        3468 &        1681 &  1.39 &  36400 \\
        2192 &        1251 &         660 &        3213 &        2957 &        2702 &  1.57 &  39200 \\
        2575 &        1412 &         888 &        3468 &         660 &        3723 &  1.45 &  39600 \\
        2931 &        1650 &         485 &        3428 &         888 &        3871 &  1.59 &  39900 \\
        2192 &        1802 &        2447 &        2702 &        3213 &        2259 &  1.55 &  40300 \\
        2957 &        1681 &        1426 &        3428 &        3656 &        2286 &  1.40 &  40300 \\
        915  &        1426 &        2702 &        1171 &        2957 &        4234 &  1.52 &  40400 \\
        5000 &        1936 &        4489 &        3468 &        3468 &        2339 &  1.62 &  42400 \\
        5000 &        4072 &        4744 &        4919 &        3817 &        2393 &  2.17 &  43400 \\
        
        \bottomrule
    \end{tabular}        
    \egroup
\end{table}

There are many different choices of scale cuts which keep the overall bias due 
to mis-modelled baryon feedback within acceptable levels. These differ in the
tightness of the resulting constraints on cosmological parameters, and on the
overall goodness of fit of the best-fit model.

\section{Discussion and conclusions}

In this chapter, we have presented several alternative methods for the derivation
of binary scale cuts for cosmic shear surveys. We found that if we were to use
DES-Y3-inspired methods where we mitigated against baryonic feedback using a
dark-matter-only fiducial model, then the headline constraints on $\Seight$
from a \textit{Euclid}-like survey will be no better than those existing
in the literature today. 

This provided excellent motivation into investigating
alternative approaches, the first of which was the `modified $\rchi^2$' method,
which modified the $\rchi^2$ through the inclusion of a baryonic feedback model
into our fiducial model. This added the complication of fitting baryonic
feedback models to the hydro-sims, where now these fits are
dependent on the maximum multipole included in the fit. With this challenge
overcome, new sets of scale cuts were obtained for our DES-Y3-like and 
\textit{Euclid}-like surveys. These showed that higher-order multipoles
could be included in an analysis without inducing statistically significant
bias, which is expected from going from a dark-matter-only model to one
including baryonic feedback. These scale cuts were then benchmarked against
various hydrodynamical simulations, finding that they do indeed guard against
baryonic feedback bias. 

Since we measure parameter biases in, well, parameter space (duh!), it makes 
sense that our scale cuts could be constructed in this basis too. This provided
the motivation for constructing our set of scale cuts that optimised against
the value of the figure of bias rather than the $\rchi^2$. This required the
computation of the derivative of the data-vector with respect to our
cosmological and astrophysical parameters. With these computed, sets of
scale cuts could be derived from the figure of bias criterion. These were found
to mitigate baryonic feedback, and result in smaller contours than their 
$\rchi^2$ counterparts. This indicates that by constructing scale cuts in
parameter-space, we are more able to isolate scales which contaminate our
cosmological parameters with less reduction in the constraining power over
the $\rchi^2$. 

We also extended this method to look at optimising our scale cuts together, in
a single six-dimensional optimisation. This was significantly computationally
challenging, since the likelihood surface of the scale cuts was far from
trivial or well-behaved. This required the use of a non-standard optimiser,
the \texttt{GridMax} sampler. We then found that there are no unique solutions
to our figure of bias criterion, where different chains that were optimised in
different orders have the same FoB but different individual scale-cuts and FoM
values. This highlights the challenges for deriving sets of scale cuts
for future weak lensing surveys that feature very many tomographic redshift bins. 

These results presented here are entirely new results in the field, the degeneracy
and non-uniqueness of scale cuts has not yet been thoroughly explored in the 
literature. Furthermore, surveys tend to adopt just one approach to scale
cuts without much scrutiny of alternative approaches or statistics, or do not
use scale cuts at all. I have shown that using different methods and statistics
can lead to the derivation of different sets of scale cuts, with interesting
effects on the derived cosmological parameter constraints.

\subsection{Outlook}

While we have focused on mitigating the
effects of baryonic feedback physics on our cosmic shear summary statistics,
the methods presented here using either the $\rchi^2$ or figure of bias statistics
provide an excellent basis for extending to other systematic effects which
introduce scale-dependent bias into observables. This includes the properties
of the non-linear matter power spectrum, the effect of intrinsic alignments, 
and non-linear bias in galaxy clustering. Each of these physical phenomena and
their effects on forthcoming weak lensing galaxy surveys needs to be quantified
and mitigated against to ensure that final results are unbiased. The methods
that we have presented for determining binary cuts can easily be extended to
include contributions from all of these effects, or even `unknown unknown'
effects that may be unravelled with increased precision data.

It is also clear that this work is highly idealised scenario for any cosmic shear
survey: sampling over only two cosmological parameters and neglecting all over
systematic effects apart from baryonic feedback. We also used only three redshift
bins for our analyses, a far cry from the expected thirteen bins that the full
\textit{Euclid} analysis could use~\cite{Euclid:2021osj}. This gives us only six spectra to compute
the scale cuts from, not ninety-one! This will make the determination of the
scale cuts used in their analyses though a rigorous and robust method more
important than ever. Hence, a more thorough approach to determining the
baryonic biases using the full complexities of a Stage-IV cosmic shear survey
is needed before any scale cuts can be accurately determined for that survey.


\vspace*{-5cm}
\begin{savequote}[65mm]
  This was a triumph, 
  
  I'm making a note here: `huge success'
  \qauthor{---GLaDOS}
\end{savequote}
\chapter{Conclusions}
\label{chp:conclusions}

\begin{mytext}
	\textbf{Outline.} In this chapter, I conclude the results featured in
    my thesis, aim to place these results within the wider cosmological context, 
    and offer some thoughts for the future. Hooray, the thesis is nearly over!
\end{mytext}


\section{Our place in the Universe}

Throughout my thesis, we have seen the great power that forthcoming
weak lensing galaxy surveys will have on our ability to measure, constrain, and
provide physical insight into our Universe. But as Spider-Man once warned us,
`With great power comes great responsibility (not to fuck it up!)'. As I
write, \textit{Euclid} is sat at L$_{2}$ and taking data that is just marvellous
(I mean, just look at Figure~\ref{fig:Euclid_ERO}!). The technical teams have done a
tremendous job to build such a fabulous instrument, and so the onus is very much
on us theorists and data analysts to interpret, analyse, and extract the most
amount of information possible from the data. This is far from a straightforward
process, I have focused on two key areas of research: minimising the errors
associated with power spectrum estimation, and mitigating baryonic feedback
bias that is induced in our cosmic shear observables in a way that still leaves
us with the most amount of information possible from the raw data. It is hoped
that the development of these accurate and precise statistical methods will shed
light on some of our greatest cosmological mysteries.

While the success of Einstein's general relativity, the $\Lambda$CDM model,
and the hot Big Bang model all combine to form the truly marvellous modern
cosmological model, see, for example, the brightness of Type~Ia supernovae over a
large range of redshifts in Figure~\ref{fig:DES_SNe}, the black-body spectrum of
the CMB in Figure~\ref{fig:COBE_CMB}, the temperature anisotropies of the CMB in 
Figure~\ref{fig:Planck_Cl}, the peak in the clustering
of galaxies around the BAO scale in Figure~\ref{fig:DESI_BAO}, and the deflection
of light from the gravitational field of the Sun in Figure~\ref{fig:Eddington1919},
we don't really know what \textit{actually} makes up $95 \, \%$ of our Universe! 
While we may give attractive names to the mysterious components of our Universe,
`dark matter' and `dark energy', we have yet to narrow down the exact
physical characteristics of these phenomena since their discovery in the 
1970's~\cite{Rubin:1980ApJ471R} and 1990's~\cite{Perlmutter:1998np,Riess:1998cb},
respectively. 

Though we have extremely strong observational evidence of the gravitational
interaction of our dark fields --- that is dark matter is some cold, collisionless matter
that interacts only through the gravitational field~\cite{Davis:1985ApJ292},
and dark energy is a fundamental property of space-time that acts as a
repulsive gravitational force, if dark energy is the cosmological constant $\Lambda$
--- we don't have satisfying theoretical models
of these physical phenomena. Many well-motivated theories to explain
dark matter have emerged in the literature, ranging from weakly-interacting
massive particles (WIMPs) to primordial black holes, have been suggested to
explain the existence of dark matter~\cite{Arbey:2021gdg}. The theoretical
motivation behind dark energy is no better; if one computes the vacuum energy
density using quantum field theory, then you arrive at an answer that is
about 120 orders of magnitude larger than the upper limit on the value of $\Lambda$ from
our cosmological observations. This has been dubbed `probably the worst theoretical
prediction in the history of physics!'~\cite{Hobson:2006se}.

It's ironic that the most familiar component of our Universe, that of ordinary
baryonic matter that you and I are made of, is almost always too complex to
model on cosmological scales and so we need to develop methods to mitigate
their mischievous effects on our observables (Chapters~\ref{chp:baryonic_effects}
and~\ref{chp:binary_cuts}). It would certainty be easier to perform a cosmic
shear survey in a universe without baryons since we wouldn't have to marginalise
over their properties, though we wouldn't be there to observe it ---
nor would there be any light from distant galaxies to measure cosmic shear with,
save for the faint cosmic microwave background afterglow from the Big Bang.

\textit{Euclid} will soon be joined by its NASA cousin the \textit{Roman}
space telescope~\cite{Spergel:2015arXiv150303757S},
and ground-based cousin the \textit{Legacy Survey of Space
and Time} (LSST) at the \textit{Rubin} observatory~\cite{LSST:2008ijt},
and has already joined its older cousin the ground-based 
\textit{Dark Energy Spectroscopic Instrument} (DESI)~\cite{DESI:2016fyo}.
These make up the four great Stage-IV galaxy survey observatories and will undoubtedly make measurements
that make unveiling the mysterious properties of dark matter and dark energy
a distinct probability over the next decade. A measurement of $w_0 \neq -1$
or $w_a \neq 0$ would be as significant as the measurements that proved 
$\Omega_{\Lambda} > 0$ and the existence of dark energy in the first place!\footnote{
    DESI's data release one has already shown slight preference for $w_0 > -1$
    and $w_a < 0$ at a minimum of the $2.5 \, \sigma$ level~\cite{DESI:2024mwx}.
    While exciting, these results are the \textit{first} hints at a departure from the
    $\Lambda$CDM model, and so additional, more constraining data from future
    DESI releases and other Stage-IV surveys is needed before we can truly
    say that dark energy cannot be described by the cosmological constant. Let's
    not forget that the BICEP2 collaboration announced a $7.0 \, \sigma$
    detection of non-zero inflationary $B$-modes~\cite{BICEP2:2014owc}, before
    \textit{Planck} found that their signal was entirely consistent with
    the emissions from cosmic dust~\cite{Planck:2014dmk}. 
}

Thus, while we may not know the detailed physical properties of $95 \, \%$
of our Universe, and the $5 \, \%$ that we can explain is too hard to 
accurately model, it certainly won't stop us cosmologists from trying to observe
our Universe to exquisite precision in the hopes of finally understanding it all.
Our new surveys could well be the `\textit{Start of Something New}' for cosmology.

\section{Summary of work and results}

The core theme of my thesis has been to develop statistical methods that 
allow for the extraction of the most amount of information possible from
cosmic shear survey data. First, the development of
a new implementation of the quadratic maximum likelihood estimator allows us
to measure the cosmic shear power spectrum with significantly reduced error bars
when compared to the standard Pseudo-$\Cl$ method. We see that by using QML
methods, we are able to reduce the errors on the largest angular scales in 
the $E$-modes by around $20 \, \%$, which is equivalent to an increase in the
sky area of around $40 \, \%$. Given that \textit{Euclid} is already at the
limit of what an all-sky weak lensing galaxy survey can achieve, an effective
increase of $40 \, \%$ in the survey area by using an alternative analysis
method is not to be passed over! Combine this with the over order of magnitude
decrease in the error bars for the $B$-modes, we conclude that analysing
future cosmic shear survey data with QML methods would be a worthwhile 
endeavour that may just help unearth something surprising about our Universe.

The monumental decrease in runtime and RAM usage of my new method over
existing QML methods (Figure~\ref{fig:QML_code_comparison}), which permits
the execution of the QML estimator to far higher multiples than was previously
possible, will allow the application of QML methods to more physical 
problems and far larger data-sets than existing methods could cope with.
It would not be out of the realm of possibility to think that such methods,
which measure the $B$-modes to far greater precision than Pseudo-$\Cl$ methods,
could be instrumental in forming a non-zero detection of $B$-modes in either
the cosmic shear signal or in the cosmic microwave background polarisation.

An accurate determination of the observationally-measured cosmic shear power
spectrum is useless without equally accurate and precise theoretical modelling
of the power spectrum, taking into account all of the cosmological and
astrophysical effects that contribute to cosmic shear. I have investigated one
of the largest sources of said astrophysical effects, which is the impact of
baryonic feedback that exists on small-scales within our Universe on cosmic 
shear. I have shown that using an incorrect, or non-suitable, analytic 
description of baryonic feedback can induce significant bias in the deduced
cosmological parameters from a Stage-IV survey 
(Figures~\ref{fig:simba_scale_cuts} and~\ref{fig:2D_one_param}). Using this
as motivation, I extended the existing theoretical uncertainties approach
to quantify the errors associated with three different implementations of the
\HMCode-2020 baryonic feedback model. I based my analysis though fitting the
baryonic feedback response function of our three analytic models compared to
the predictions for an ensemble of six hydrodynamical simulations, which is the
natural basis to perform such comparison as it is the underlying physical
quantity, and is a new result in the field. From this, I then propagated this 
into an additional term in the covariance matrix for the angular power spectrum.

Using this additional theoretical error covariance, I showed how using even a
basic prescription of baryonic feedback applied to a hydrodynamical simulation
featuring extreme baryonic feedback allows the recovered cosmological
parameters to be consistent with their true values (Figures~\ref{fig:2D_one_param},
\ref{fig:2D_three_param}, and~\ref{fig:2D_six_param}). This paves the way for
using alternative methods to traditional binary scale cuts which still mitigate
baryonic feedback contamination, while attempting to keep as much information
content as possible. 

Having said that, binary scale cuts have featured extensively in previous
Stage-III surveys, and so are prime candidates to be used in the analyses of
our Stage-IV surveys. Thus, Chapter~\ref{chp:binary_cuts} was dedicated to
investigating the properties of existing binary cuts methods on Stage-IV surveys,
finding that if we apply existing methods in the literature to the new data,
then our cosmological constraints will be no better than existing results. 
This drove the development and analysis of alternative methods to derive sets
of binary cuts, where we are now explicitly including the properties of 
baryonic feedback models into the analysis. I compared the use of two statistics,
the $\chi^2$ and figure of bias, to derive sets of scale cuts for our next
generation of surveys, finding that there are no unique solutions to this 
problem.

\section{Future work}

Throughout the development of my new implementation of the QML method in
Chapter~\ref{chp:QML_estimator}, I focused solely on the cosmic shear signal
coming from a single redshift bin. Thus, a natural extension to my work is to
investigate the properties of the performance of the QML method with respect
to the Pseudo-$\Cl$ method for the case where we include galaxy-galaxy lensing
and galaxy position-position information, forming the $3 \times 2$pt data-vector
which Stage-IV surveys can take full advantage of. The additional information,
present from both the new cosmological probes and cross-correlations with
the increased number of redshift bins, could uncover new properties of the
QML estimator when compared to the Pseudo-$\Cl$ method.

The next generation of CMB experiments, the Simons Observatory and CMB-S4,
aim to measure the polarisation of the microwave background to unprecedented
accuracy. Since the polarisation of the CMB is a spin-2 field and thus
completely analogous to our cosmic shear field, the application of my QML
method to CMB polarisation would be rather straightforward. Furthermore, since
one of the main science goals of these next generation CMB experiments is to
place stringent limits on the detection of $B$-mode polarisation. If the 
dramatic decrease in the $B$-mode error bars found in my cosmic shear analysis
were to propagate over to the CMB, then these QML methods would form an essential
role in the detection of $B$-modes.  

Throughout the development of my implementation of the theoretical uncertainties
method to the matter power spectrum, the use of our ensemble of hydrodynamical
simulations has underpinned my entire numerical pipeline. Just as we are
experiencing a rapid growth in the quality of observations, the hydrodynamical
simulation community has continued to make great advancements with new simulations
that feature larger volumes, increased baryonic particle mass resolution, total
particle counts, and increased accuracy of sub-grid physics. Simulations such
as \textsc{Flamingo}~\cite{Schaye:2023jqv}, \textsc{Camels}~\cite{CAMELS:2020cof,CAMELS:2023wqd}
and \textsc{Flares}~\cite{Lovell:2021MNRAS5002127L} have all arisen as the
latest state-of-the-art hydro-sims. Hence, a straightforward extension of
Chapter~\ref{chp:baryonic_effects} would be to include these new simulations in
our ensemble, perhaps replacing some of the older simulations such as
\Bahamas and \Illustris, and seeing how the theoretical error envelope and
our \HMCode baryonic feedback models react to these new inclusions. 

For simplicity, the application of my theoretical error covariance was limited 
to analyses where we sampled only over the two main cosmological parameters
for cosmic shear, $\Omegam$ and $\sigmaeight$, and using five redshift bins.
This was done to speed up the convergence of the MCMC chains,
and to get estimates for the worst-case scenario
for a \textit{Euclid}-like survey. However, the full analysis of \textit{Euclid}
data is expected to include sampling over seven cosmological
parameters, using thirteen tomographic redshift bins which necessitates the
inclusion of well over thirty nuisance parameters for the photometric
sample alone~\cite{Euclid:2024yrr}. This takes the number of spectra from
fifteen in my analysis to a whopping ninety one! Thus, an important result
would be to first quantify the level of bias associated with using inadequate
methods of baryonic feedback when using hydro-sims as the ground truth, and then
investigate the mitigation of these biases though either the inclusion of the
theoretical error covariance, or application of binary cuts methods. 

At present, the current preliminary forecasts for the final \textit{Euclid} survey
data have used the one-parameter model of \HMCode-2020 with 32
logarithmic-spaced bins ranging from $\ell \subseteq [10, \, 5000]$. Based on
work presented in this thesis, these are certainly \textit{optimistic}
targets, though I have no doubt that thorough and extensive validation tests
will be performed by the supremely talented \textit{Euclideans} that I have had
the pleasure to work with in my PhD to verify these analysis choices.

\section{Outlook to the future}

The future of the cosmological dark sectors have never looked as bright as they
have at the time of writing this thesis. The rise of the four great Stage-IV
galaxy survey observatories, future CMB experiments, the James Webb Space
Telescope~\cite{Gardner:2006ky}, 
the \textit{Athena} X-ray telescope~\cite{Barcons:2012arXiv1207.2745B},
and the \textit{LISA} space-based gravitational wave observatory~\cite{LISA:2017pwj}
will all probe our Universe across the entire electromagnetic and gravitational
wave spectrum with immense precision.

While each probe will undoubtedly uncover great physical insights into our
Universe, just like the Avengers (or perhaps more fittingly, 
the \textit{Guardians of the Galaxy}) their combined power is far greater than
the sum of their parts. For example, while we have treated baryonic feedback
as a nuisance astrophysical effect that needs to be marginalised over in our
weak lensing cosmological analyses, the physics of baryonic feedback can be
directly probed though the kinematic Sunyaev-Zel'dovich (kSZ) effect in observations
of the CMB~\cite{Sunyaev:1980ARA&A18537S,ACT:2024vsj}, and gas fractions in galaxy clusters from
X-ray observations~\cite{Schneider:2021wds,Salcido:2024qrt}. The development
of these external data-sets will allow us to place tight constraints on the 
physics of baryonic feedback that is independent of our cosmic shear observations
and so provide valuable additional insights into both the development of analytic
baryonic feedback models, and the development of the hydrodynamical simulations.
For example, Ref.~\cite{Hadzhiyska:2023wae} found that observations of the
kSZ effect strongly disfavoured the more modern \TNG hydrodynamical simulation,
instead favouring the more extreme feedback as featured in the original
\Illustris simulation. Thus, by using and combining different cosmological 
probes, we can gather huge quantities of data that tests the fundamental 
properties of our Universe.

\begin{figure}[t]
  \centering
  \includegraphics[width=\columnwidth]{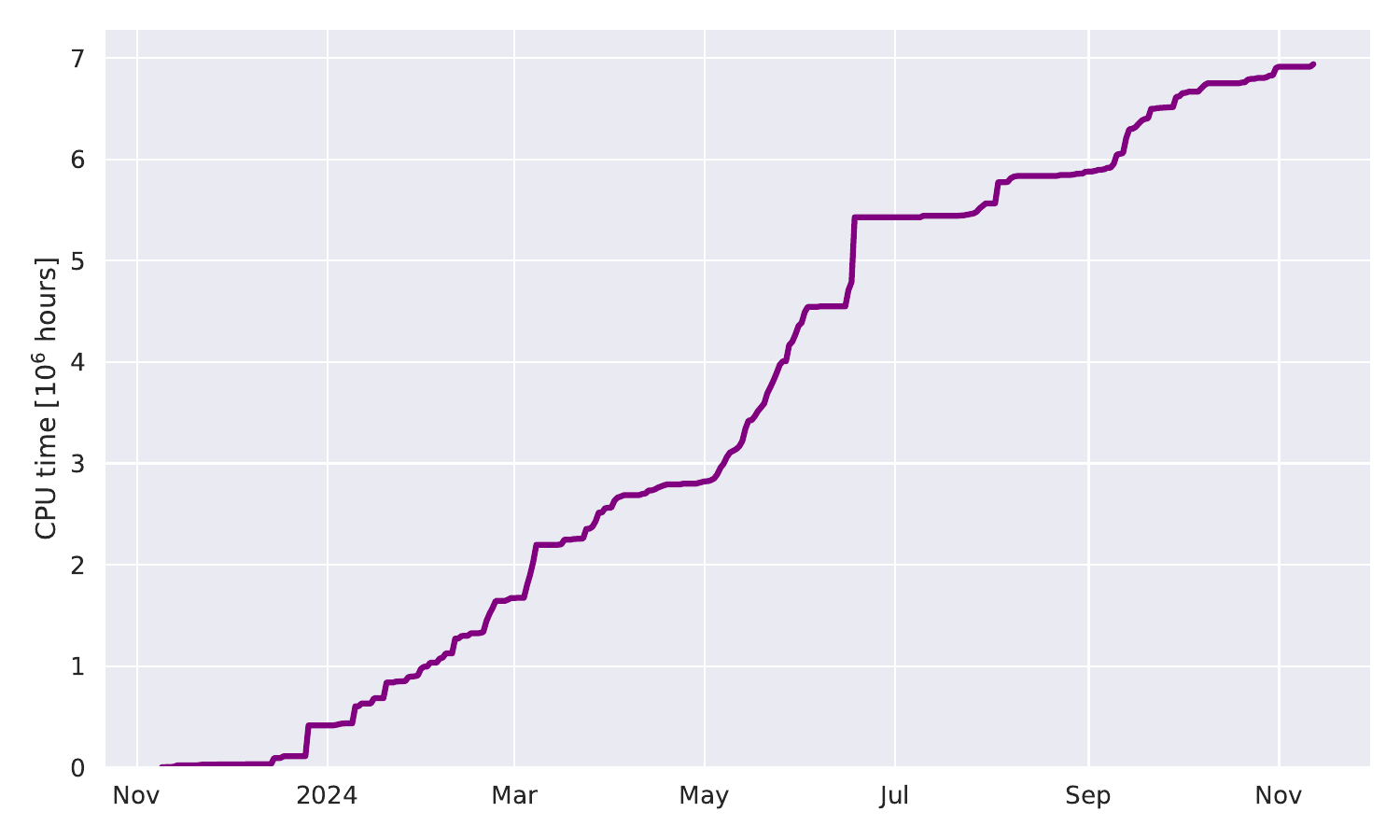}
  
  \vspace*{-0.25cm}

  \caption{Cumulative CPU hours used on the \textit{Euclid}:UK high-performance
    computing cluster during the last year of my PhD. It shows that in a single
    year of just one cosmology PhD, my analysis methods required nearly
    $7 \times 10^6$ CPU hours. 
    }

  \label{fig:cum_sum_cpu_hours}
\end{figure}

While the availability and exploitation of such precision data is sure to get
any cosmologist buzzing, it is important to recognise that we cannot just
blindly increase the size of our data-sets without restriction. To demonstrate 
this, Figure~\ref{fig:cum_sum_cpu_hours} shows the cumulative computing time
in CPU hours\footnote{A CPU hour is defined as one thread of a processor
working for a wall-clock time of an hour. For example, a 64 core job running for
two hours would use 128 CPU hours.}
that I used in the last year of my PhD to perform the analyses presented in
Chapters~\ref{chp:baryonic_effects} and~\ref{chp:binary_cuts}. This shows an
impressive total of nearly $7 \times 10^6$ CPU hours, just shy of $800$ 
CPU years. This computing usage has consumed an approximate $36\,000\,$kWh
of energy which, using the UK's electricity average carbon intensity of
$150\,\textrm{gCO}_2 \, \textrm{kWh}^{-1}$ in 2024~\cite{UKGrid}, gives carbon dioxide
emissions of around $5400\,\textrm{kgCO}_2$, or about ten transatlantic
flights from London to New York~\cite{Kwan:2014}. It should be noted that
the UK's electricity carbon intensity has fallen dramatically over the last
decade, as it averaged $500\,\textrm{gCO}_2 \, \textrm{kWh}^{-1}$ in 2012, and
hopefully will continue to further fall. Sourcing clean, renewable energy to
power our supercomputers is essential if we want the next generations of
astronomers and cosmologists to enjoy the same planet that we currently inhabit. 
Or we could abandon Python and switch to \textsc{Fortran} and \Cpp~\cite{PortegiesZwart:2020pdu}. 

\begin{figure}[t]
  \centering
  \includegraphics[width=\columnwidth]{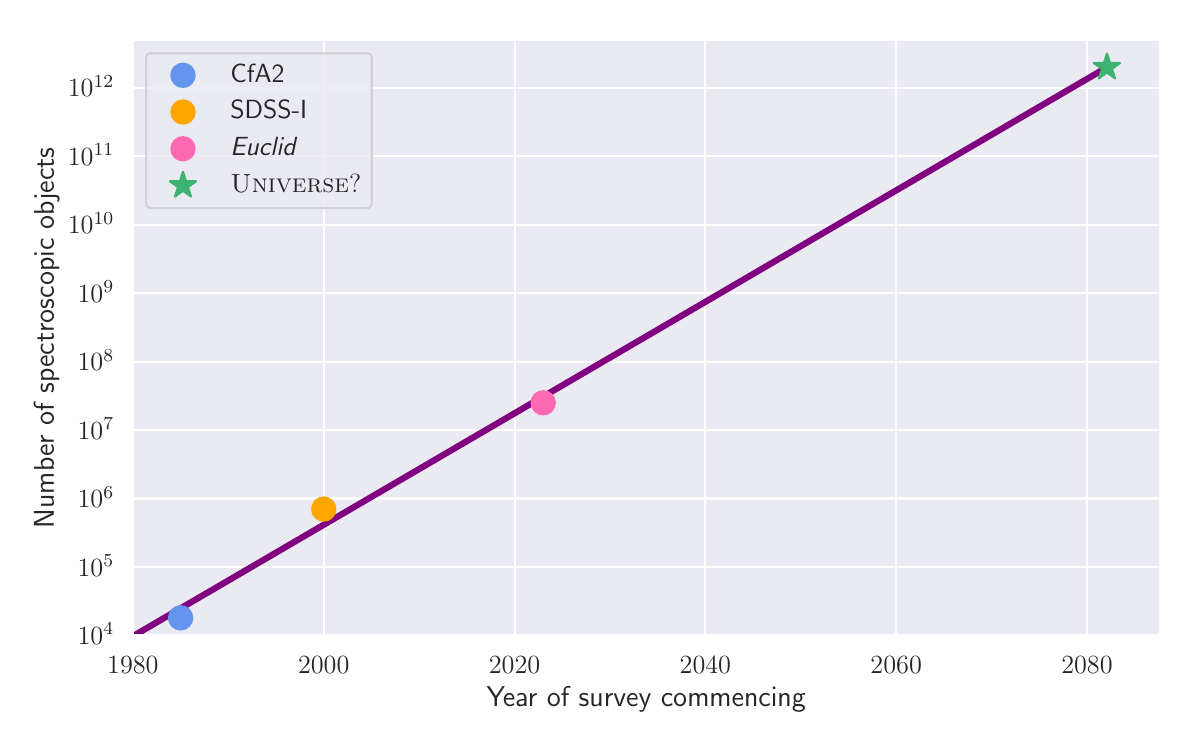}
  
  \vspace*{-0.25cm}

  \caption{Evolution of the number of galaxies in selected spectroscopic surveys
    as a function of their year of inauguration, ranging from the Center for
    Astrophysics survey 2 (CfA2)~\cite{Geller:1989Sci897G}, to the Sloan Digital
    Sky Survey (SDSS-I)~\cite{SDSS:2000hjo}, and finally to the \textit{Euclid}
    spectroscopic survey~\cite{Euclid:2024yrr}. If we fit a power-law to these
    three data-points, and then extend to include all galaxies in the
    observable Universe, of $2\times10^{12}$, then the \textsc{Universe} survey
    could become reality around 2080. I hope to live to see that survey!
    }

  \label{fig:THE_galaxy_survey}
\end{figure}

To end on a slightly whimsical note, which is about par for my thesis,
we can investigate what it would take for us to complete \textit{the} galaxy 
survey --- that is observing \textit{all two trillion galaxies} in our observable Universe~\cite{Conselice:2016ApJ83083C}
spectroscopically. Using some optimistic assumptions, a 280 metre telescope operating
at the second Lagrange point, $L_{2}$, could perform a spectroscopic analysis of every galaxy in
the observable Universe over a ten-year survey~\cite{Rhodes:2020xwp}. Of course,
massive technological advances would be required to build and launch such a
massive telescope into space, but, just like dark energy, one should never
discount the accelerated growth of human ingenuity! 

This survey may well be
closer than we know it. Figure~\ref{fig:THE_galaxy_survey} plots the growth
of the number of galaxies in spectroscopic surveys over time, and so conclude
that this mythical survey could become reality around 2080.
To get ahead of the curve, I decree that this galaxy survey to end all surveys
shall be known as \textsc{Universe}: UNiversal Investigation of the
Vast Extragalactic Regions of Space-time Evolution\footnote{Created with
the brainstorming help of ChatGPT~\cite{OpenAI:2023ktj}.}. All I know is that
you could get some seriously tight contours on $\Omegam$ and $\sigmaeight$
by using every galaxy in the Universe, but I certainly wouldn't want to be
the one running the MCMC to analyse the data -- it could well take a universal age
to compute!

\clearpage

\appendix
\addcontentsline{toc}{chapter}{Appendicies}

\begin{savequote}
  My power spectra has doubled since we last met, Count

  \qauthor{---Anakin Skywalker}
  
  Good, twice the matter, double the shear
  \qauthor{---Count Dooku}
\end{savequote}

\chapter{Power spectrum of a simple mask}
\label{chp:appendix_A}

\begin{mytext}
    \textbf{Outline.}
    Appendix detailing the derivation for the power spectra
    of simple masks that vary in either the $\vartheta$ or $\phi$ directions,
    which have been used in my thesis.
\end{mytext}

\section[Simple mask in the $\vartheta$ plane]{Simple mask in the $\bm \vartheta$ plane}

Much of the first science project (Chapter~\ref{chp:QML_estimator}) was spent
investigating the properties of masks and how they impact the statistics of
fields that are recovered by deconvolving a mask. In the Pseudo-$\Cl$ method,
this deconvolution is 
sensitive to the power spectrum of the mask. Thus, if we are to understand
the mixing matrix well, then we can investigate the analytic properties of
the power spectrum of a mask first. 

\begin{figure}[t]
    \centering
    \includegraphics[width=0.975\linewidth,trim={0.25cm 1.5cm 0.25cm 0.7cm},clip]{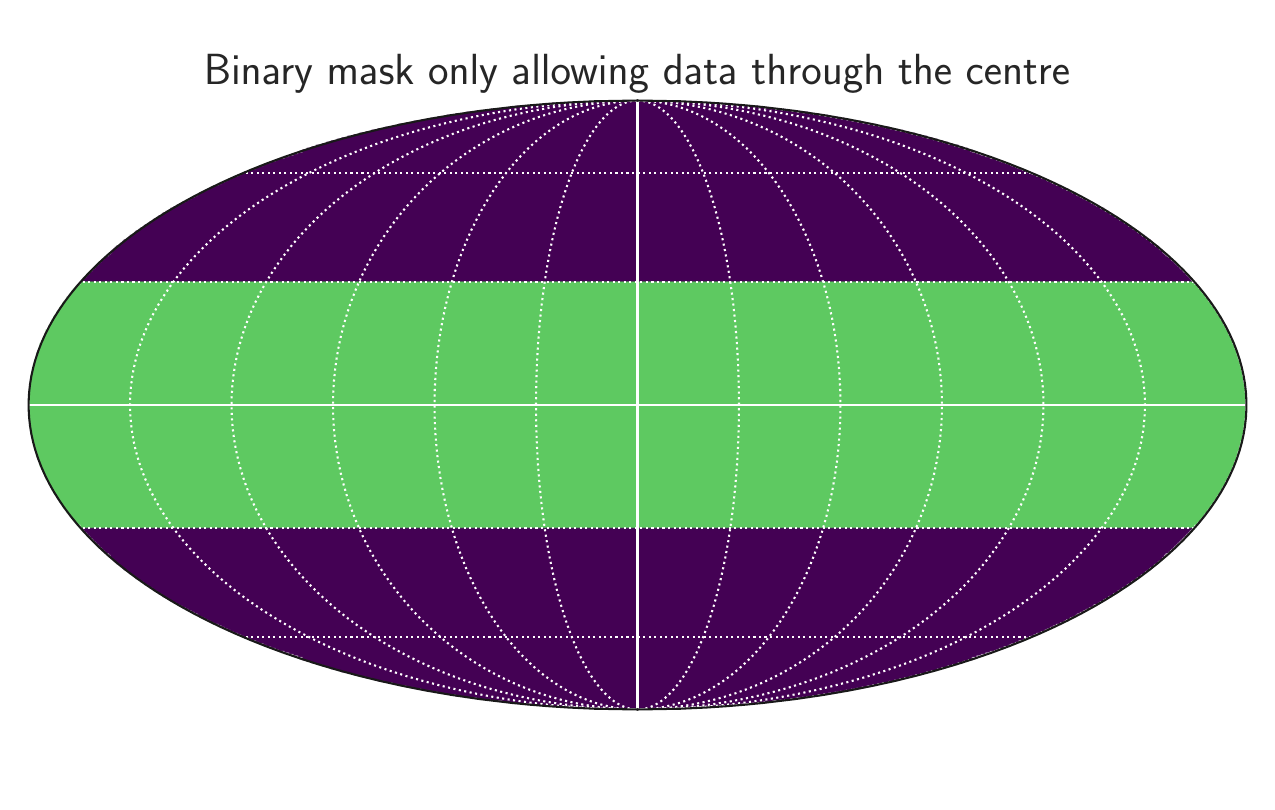}
    \caption{Simple binary mask that only allows data through within the
        central angular region $\frac{\pi}{3} \leq \vartheta \leq \frac{2\pi}{3}$,
        shown in green, with the purple region masking out data.
        }
    \label{fig:binary_theta_mask}
\end{figure}

We start out with a simple binary mask that only varies in the $\vartheta$
plane. For example, Figure~\ref{fig:binary_theta_mask} shows a mask that only allows
data through in a region between $\vartheta = \pi / 3$ and $\vartheta = 2\pi/3$,
with no $\phi$ dependence. To find its power spectrum, we first need to evaluate
its spherical harmonic transform into its corresponding $\alm$ values. Recalling
Equation~\ref{eqn:alm_def} for the definition of the $\alm$ values for a function
on the sphere, we have
\begin{align}
    \alm = \int \!\! \d \Omega(\hat{n}) \,\,  f(\vartheta, \phi) \, \Ylm^*(\vartheta, \phi).
\end{align}
If our mask is only a function of $\vartheta$, then we can write our mask simply as
\begin{align}
    f(\vartheta, \phi) = f(\vartheta).
\end{align}
The definition of the spherical harmonics, $\Ylm$, are given as
\begin{align}
    Y_{\ell m}(\vartheta, \phi) = \sqrt{\frac{(2\ell + 1)}{4 \pi} \, \frac{(\ell - m)!}{(\ell + m)!}}
    \, P_{\ell m}(\cos \vartheta) \, e^{i m \phi},
\end{align}
where $P_{\ell m}(x)$ are the associated Legendre polynomials. Recalling the
differential element on the sphere can be written as 
$\d \Omega = \sin \vartheta \, \d \vartheta \, \d \phi$, we find we can split
the $\alm$ integral into
\begin{align}
    \alm = \sqrt{\frac{(2\ell + 1)}{4 \pi} \, \frac{(\ell - m)!}{(\ell + m)!}} \,
    \int_{0}^{\pi} \! \d \vartheta \, \sin \vartheta \, f(\vartheta) P_{\ell m} (\cos \theta)
    \int_{0}^{2\pi} \! \d \phi \, e^{-i m \phi}.
    \label{eqn:curr_int_alm}
\end{align}
The $\phi$ integral can be done immediately, yielding a Kronecker-$\delta$ since
it is only non-zero when $m$ is zero, giving
\begin{align}
    \int_{0}^{2\pi} \!\! \d \phi e^{-im\phi} = 2\pi \delta_{m}.
\end{align}
Since this forces the $m$ to be zero to give non-zero $\alm$ values, we can
substitute in $m=0$ into our integral of Equation~\ref{eqn:curr_int_alm}, we find
the $\alm$ values can be written as
\begin{align}
    \alm = \delta_{m} \sqrt{ \pi (2\ell + 1)}
    \int_{0}^{\pi} \! \d \vartheta \, \sin \vartheta \, f(\vartheta) P_{\ell} (\cos \vartheta),
\end{align}
where $P_{\ell}(\cos \vartheta)$ are the regular Legendre polynomials. Using our
definition of our masking function $f(\vartheta)$ that it is only non-zero within
a certain range of $\vartheta$ values of $[A, B]$, we find our integral becomes
\begin{align}
    \alm = \delta_{m} \sqrt{ \pi (2\ell + 1)}
    \int_{A}^{B} \!\!\! \d \vartheta \, \sin \vartheta \, P_{\ell} (\cos \theta).
\end{align}
This is now an integral of known Legendre polynomials, which has the standard
result of
\begin{align}
    \int_{A}^{B} \!\!\! \d \vartheta \, \sin \vartheta \, P_{\ell} (\cos \theta) =
    \frac{-P_{\ell - 1}(\cos A) + P_{\ell + 1}(\cos A) + P_{\ell - 1}(\cos B) - P_{\ell + 1}(\cos B)}{2 \ell + 1}.
\end{align}
This gives the $\alm$ values of our mask as
\begin{align}
    \alm = \delta_m \sqrt{\frac{\pi}{2 \ell + 1}} \,
    \left[ -P_{\ell - 1}(\cos A) + P_{\ell + 1}(\cos A) + P_{\ell - 1}(\cos B) - P_{\ell + 1}(\cos B)\right].
\end{align}
This, transforming these $\alm$ values into our power spectrum coefficients, 
$\Cl$, we find that the mask's power spectrum to be given by
\begin{align}
    \Cl = \frac{\pi}{(2 \ell + 1)^2} \,
    \left[ -P_{\ell - 1}(\cos A) + P_{\ell + 1}(\cos A) + P_{\ell - 1}(\cos B) - P_{\ell + 1}(\cos B)\right]^2.
\end{align}

If we now assume that our mask is symmetric around $\vartheta = \pi / 2$, then
we can use the parity of the Legendre polynomials~\cite{Boas:913305}, of
\begin{align}
    P_{\ell}(-x) = (-1)^{\ell} \, P_{\ell}(x),
\end{align}
we can write the power spectrum as
\begin{align}
    \Cl = \frac{\pi}{(2 \ell + 1)^2} \,
    \left[
        \left\{ -1 + \left(-1\right)^{\ell - 1} \right\} P_{\ell - 1}(\cos A) +
        \left\{ 1 - \left(-1\right)^{\ell - 1} \right\} P_{\ell + 1}(\cos A)
    \right]^2.
    \label{eqn:mask_pow_spec}
\end{align}
Thus we see that for odd $\ell$ values, the two terms in curly brackets cancel
out and are left with zero for the $\Cl$ values. This simplifies our power
spectrum for even $\ell$ as
\begin{align}
    \Cl = \frac{4 \pi}{\left(2 \ell + 1\right)^2} \left[
        P_{\ell + 1} \left(\cos A \right) - P_{\ell - 1} \left(\cos A \right)
    \right]^2.
    \label{eqn:simple_mask_pow_spec}
\end{align}

\subsubsection[Limit in large $\ell$]{Limit in large $\bm \ell$}

We can further simplify our equation for the power spectrum of our simple mask
by taking the limit $\ell \gg 1$. Since most of our cosmological information 
in the power spectrum comes from high-$\ell$ modes, we are motivated to 
understand the behaviour of the mask's power spectrum in this limit. Since
our power spectrum is a function of Legendre polynomials, we can take the
large-$\ell$ asymptotic limit of them, finding
\begin{align}
    P_{\ell} \left(\cos \vartheta \right) \simeq \frac{2}{\sqrt{2 \pi \ell \, \sin \vartheta}}
    \cos \left(\left[\ell + \frac{1}{2}\right] \vartheta - \frac{\pi}{4} \right),
\end{align}
which is correct up to order $\mathcal{O}(\ell^{-3/2})$ corrections. Applying
this limit, we find our mask's power spectrum, for even $\ell$, to be
given by
\begin{align}
    \Cl = \frac{4 \pi}{\left(2 \ell + 1\right)^2} \left[
        \frac{2}{\sqrt{\sqrt{3} \pi (\ell + 1)}} 
        \cos \left(\frac{\pi \ell}{3}\right) -
        \frac{2}{\sqrt{\sqrt{3} \pi (\ell - 1)}} 
        \cos \left(\frac{\pi \ell - 2 \pi}{3}\right)
    \right]^2
    \label{eqn:mask_pow_spec_cos}
\end{align}
Where we have used $\sin \pi / 3 = \sin 2 \pi / 3 = \sqrt{3} / 2$.
The cosine terms will cause some oscillations in the $\Cl$ values that depend on
their $\ell$ value, however as they are bounded between $[-1, \,+1]$, we can simply label
their values as $\alpha$ and $\beta$, as they will only contribute to $\mathcal{O}(1)$
corrections, and no impact the $\ell$ scaling of the $\Cl$ values. Therefore, this
gives us
\begin{align}
    \Cl = \frac{\pi}{(2 \ell + 1)^2} \left[ \frac{2}{\sqrt{\sqrt{3}\pi (\ell -1)}} \alpha
    + \frac{2}{\sqrt{\sqrt{3}\pi (\ell + 1)}} \beta \right]^2
\end{align}
If we take $\ell \gg 1$, we find the $\Cl$ values as approximately
\begin{align}
    \Cl &\simeq \frac{\pi}{4 \ell^2} \left[\frac{2}{\sqrt{\sqrt{3}\pi \ell}} (\alpha + \beta)\right]^2\!\!, \\
    \Cl &\propto \frac{1}{\ell^3}. \label{eqn:mask_pow_spec_ell}
\end{align}
Thus we find that the $\Cl$ values go as $1 / \ell^3$, and thus when we plot 
$\ell (\ell + 1) \Cl$ we find their values scale as $1 / \ell$.

Additionally, we note that for our simple mask, the cosine terms in
Equation~\ref{eqn:mask_pow_spec_cos} scale as $\ell \pi / 3$, and thus repeat
with a period of $\ell \rightarrow \ell' = \ell + 6$. Hence, when we plot the
$\Cl$ values of our mask, we can plot them starting at $\ell = 2$, $\ell = 4$,
and $\ell = 6$, and take each starting point up in multiples of six. This
cancels out the effects of the oscillations by plotting them separately, and
thus we are able to discern the overall $1 / \ell$ scaling behaviour more
clearly. 

\begin{figure}[t]
    \centering
    \includegraphics[width=0.975\linewidth,trim={0.25cm 0.25cm 0.25cm 0.7cm},clip]{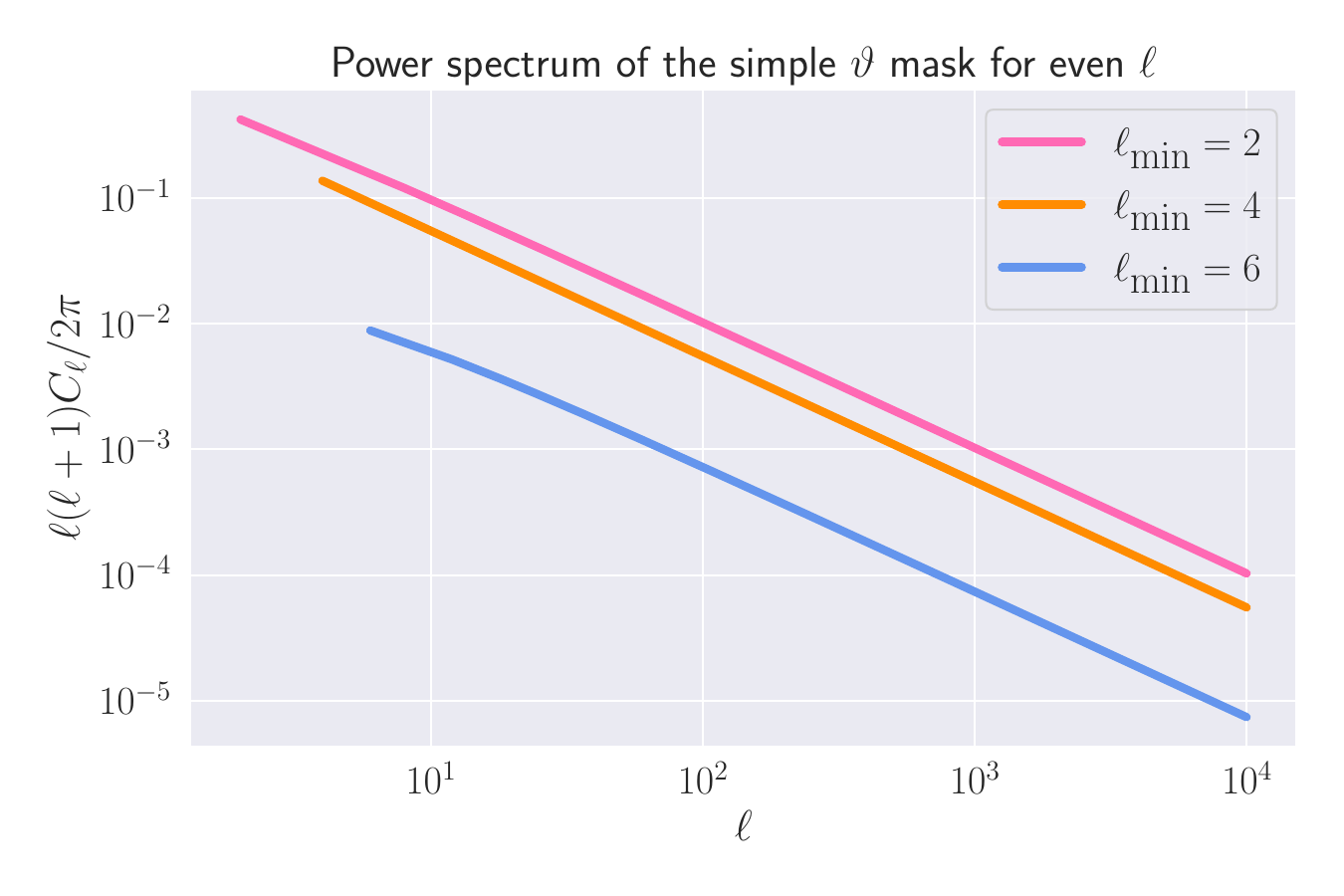}
    \caption{The power spectrum of our simple binary mask shown in 
        Figure~\ref{fig:binary_theta_mask} and computed in Equation~\ref{eqn:simple_mask_pow_spec}.
        We start from our three starting $\ell$ positions as demonstrated in 
        Equation~\ref{eqn:mask_pow_spec_cos}, increasing in steps of six. We
        note the clear $1 / \ell$ behaviour which is shown in Equation~\ref{eqn:mask_pow_spec_ell}.
        These $\Cl$ values computed theoretically are in perfect agreement
        to their numerical counterpart computed using \healpy.
        }
    \label{fig:binary_theta_mask_powspec}
\end{figure}

We plot the power spectrum of our simple mask in Figure~\ref{fig:binary_theta_mask_powspec}

\section{The convolution of power from masking}

When we derived the properties of the Pseudo-$\Cl$ estimator in 
Section~\ref{sec:Review_of_pseudocl}, the power spectrum of the mask was key
to understanding the mixing matrix, which quantifies how different modes on the
unmasked-sky couple together when we apply a mask. We can now demonstrate this
mode-coupling visually, by investigating how different masks convolve the
power spectrum in slightly different ways.

Using a fiducial power spectrum where we have power only for a single multipole,
say $\ell = 50$, we can generate a map using this input spectrum, apply the mask,
and then compute its masked spectrum. By using only a single active multipole
in the original spectrum, this is completely analogous to taking a
one-dimensional slice of the mixing matrix $\Mmat$ at constant $\ell'$~\cite{Hivon:2001jp}.

\begin{figure}[t]
    \centering
    \includegraphics[width=0.975\linewidth,trim={0.25cm 0.25cm 0.25cm 0.7cm},clip]{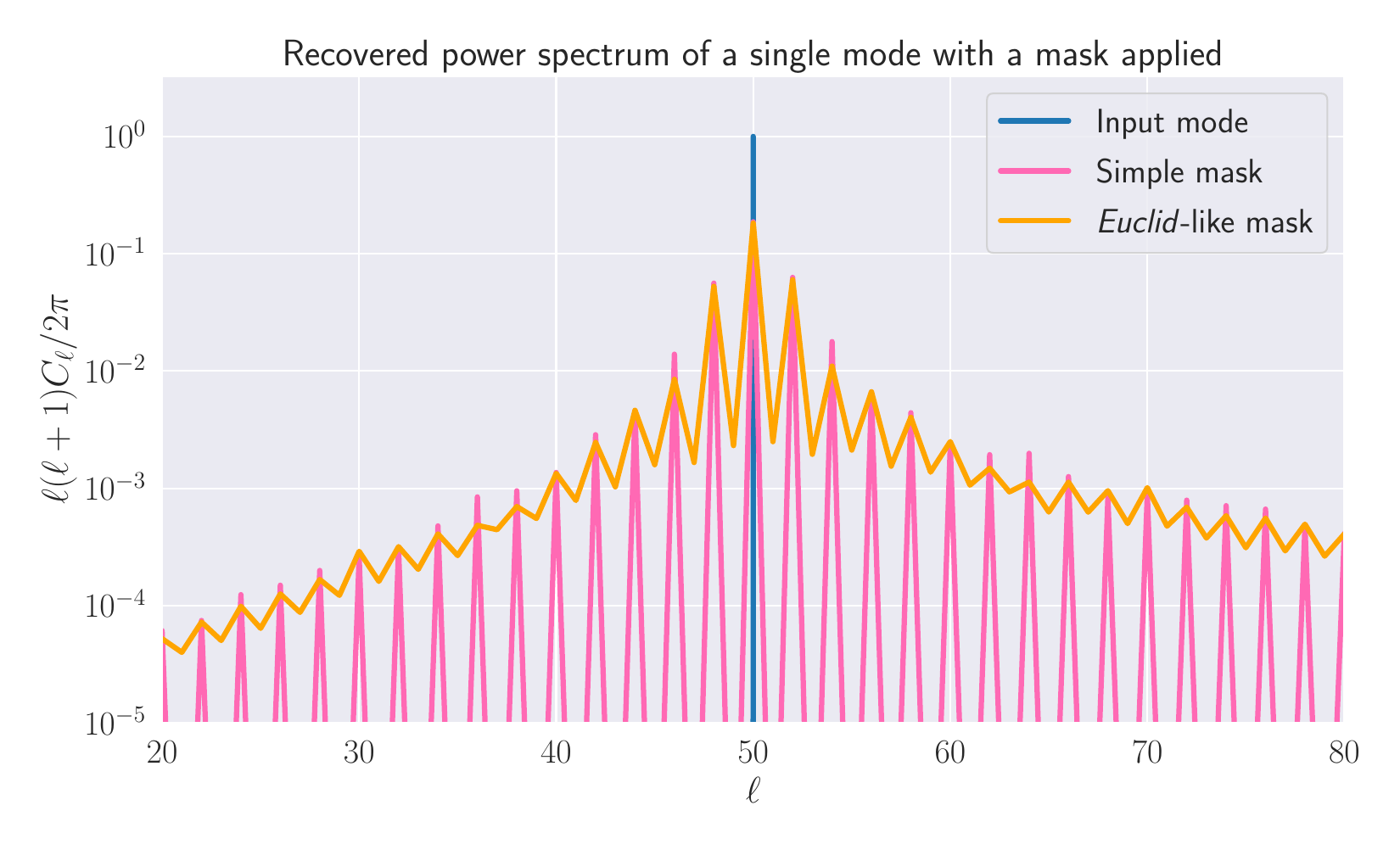}
    \caption{One-dimensional slices of the Pseudo-$\Cl$ mixing matrix $\Mmat$,
        with $\ell'=50$. We see strong support for even $\ell$ increments, which
        stem from how both mask's have approximate $\vartheta$ symmetry, and
        so their power spectra are suppressed for odd $\ell$ multipoles. 
        The additional features in the \textit{Euclid}-like mask serve to
        increase the coupling for odd $\ell$ intervals over our simple mask. 
        We note that the convolved power produces an asymmetric masked spectrum,
        with an increased tail out to larger $\ell$ values. Hence, when one
        deconvolves the mixing matrix, higher-order multipoles should be computed
        and included in the inversion, and then cut down post-inversion. 
        }
    \label{fig:injected_power}
\end{figure}

Figure~\ref{fig:injected_power} plots the recovered spectra using two masks:
a simple mask featuring two cuts in the galactic and ecliptic plane (approximating
the necessary cuts to block the Milky Way and our Solar System, respectively);
and a final-mission \textit{Euclid}-like space-based survey, which includes masking
out of discrete objects such as the Magellanic Clouds and bright stars. We see
that the additional objects in our \textit{Euclid}-like mask induce additional
coupling for odd-$\ell$ increments over our simple mask, though is still 
dominated by even-$\ell$ contributions. 

\clearpage

\begin{savequote}[80mm]
  The neutrinos have mutated,
  
  and they're heating up the planet!

  \qauthor{---2012}

  Oh, I can spend the rest of my life
  
  having this conversation. Now, please,
  
  please try to understand before one of us dies.
  \qauthor{---Fawlty Towers}
\end{savequote}

\chapter[Dark energy, massive neutrinos, and their impact on \texorpdfstring{\\}{} cosmology]{Dark energy, massive neutrinos,\\and their impact on cosmology}
\label{chp:appendix_B}
\vspace*{-1cm}
\begin{mytext}
    \textbf{Outline.}
    Appendix highlighting the impact of the `dark sector', that is dark energy
    and massive neutrinos, on the growth of structure within our Universe,
    their impact on cosmic shear, and how we can observe these physical 
    phenomena.  
\end{mytext}

\section{Dark energy}

The driving force behind this thesis has been to maximise the information content from
weak lensing galaxy surveys, such that we can obtain as much
constraining power on the physics and contents of our Universe as possible.   
Of particular importance has been to constrain the properties of dark energy,
which is some cosmological fluid with a negative equation of state such that
it provides the accelerated expansion that has been observed in the late-time Universe. 
The simplest explanation of dark energy is that it is the effect of the
cosmological constant $\Lambda$ that can be included in the Einstein equations
(Equation~\ref{eqn:Einstein_fild_eqn}), and thus $\Lambda$ is some fundamental
property of spacetime which does not evolve with time nor is positional 
dependant. While we are free to include $\Lambda$ in the Einstein equations,
which gives us the accelerated expansion as observed in the late-time Universe,
including $\Lambda$ as the source of dark energy is not entirely theoretically 
satisfying since many unanswered questions arise from its inclusion, such as why
does $\Lambda$ has the value that it has, and does it link in anyway to the
hypothesised inflationary epoch in the very early Universe? 

Hence, there have been huge efforts in cosmology over the past two decades since
the first detections of the accelerated expansion to narrow down the properties
of dark energy, and to provide well-motivated theoretical explanations for its
existence. The primary goal of these observational searches have been to
determine if there is any time evolution in the value of $\Lambda$, and thus it
wouldn't be a true cosmological constant. This motivates the reparametrisation of
the dark energy equation of state ($w_{\Lambda}$) into the time-dependant
form of~\cite{Linder:2002et,Linder:2005in,Chevallier:2000qy}
\begin{align}
	w_{\Lambda}(a) = w_0 + w_a \times \left(1 - a\right),
	\label{eqn:w0_wa_app}
\end{align}
where $w_0$ is the value today and $w_a$ encodes the scale factor dependence.

In keeping with the rest of this thesis, we shall denote $\Lambda$ to mean
any dark energy fluid, even if it is not a true cosmological constant.

\begin{figure}[t]
    \centering
    \includegraphics[width=\linewidth,trim={0.0cm 0.0cm 0.0cm 0.0cm},clip]{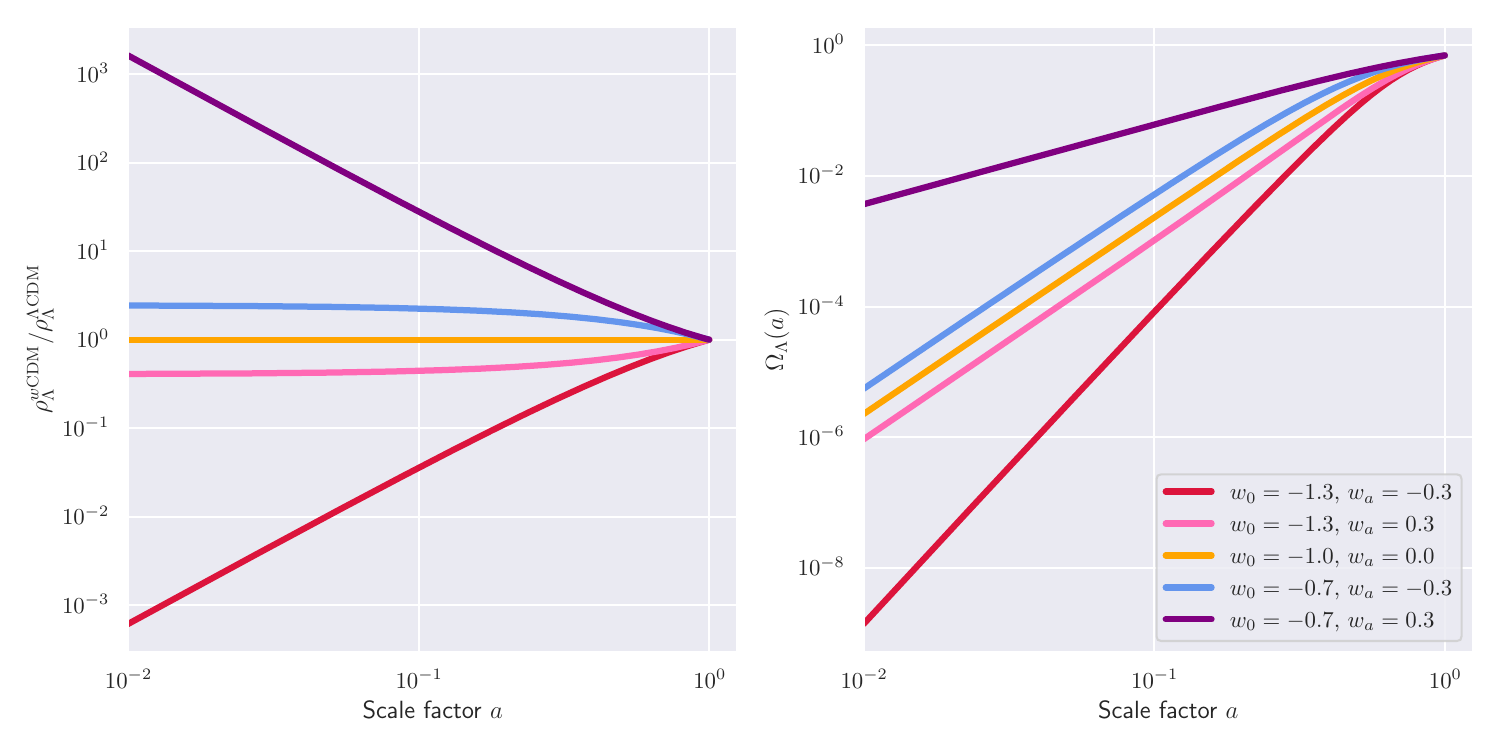}
    \caption{The physical density of time-evolving dark energy models relative
        to that of a cosmological constant (left panel) and dark energy density
        parameters (right panel) for five combinations of $w_0$ and $w_a$ values,
        where the orange curve corresponds to the cosmological constant $\Lambda$.
        We see that more positive values for the equation of state give higher
        dark energy densities over the Universe's evolution, and more
        negative values giving smaller dark energy densities. We see that the
        densities for the pink and blue curves track the cosmological constant
        at earlier times, since their equation of states tend to that of the
        cosmological constant, $w_{\Lambda} \rightarrow -1$ as $a \rightarrow 0$,
        only differing in the late-time Universe ($a \rightarrow 1$). The
        purple and red curves do not have such asymptotic trends of their 
        equation of state, and thus exhibit large deviances in their densities
        in the early Universe. Note that we require all models to have the same
        dark energy density today.
        }
    \label{fig:dark_energy_densities}
\end{figure}

Figure~\ref{fig:dark_energy_densities} plots the relative physical density
and density parameter of dark energy for five different evolutions of the
equation of state. Here, we see that more positive values for the equation
of state's evolution gives larger dark energy densities in the past. In
particular, for models whose equations of state do not tend to $-1$ as 
$a \rightarrow 0$ give large deviances from the prediction of the
cosmological constant $\Lambda$. Hence, any cosmological observations that probe
the evolution of the dark energy or total energy densities over cosmic time (such
as through the evolution of the Hubble rate, $H(z)$) will be sensitive to
the values of $w_0$ and $w_a$. Though, as Figure~\ref{fig:dark_energy_densities}
shows, the varying of $w_0$ and $w_a$ values only slightly modifies the
dark energy density, particularly out to redshift $z \sim 2$ (which is the limit
that most cosmological surveys are sensitive out to) and for the case that
$w_{\Lambda} \rightarrow -1$ as $a \rightarrow 0$. Hence, high-quality precision
measurements are needed over a wide range of redshifts to place meaningful
constraints on $w_0$ and $w_a$.

With this in mind, cosmological surveys over the last two decades have been designed
from the ground up to place tight constraints on $w_0$ and $w_a$. These have
primarily arisen from the mid-2000s Dark Energy Task Force~\cite{Albrecht:2006um}
which discussed and recommended the best approaches for placing tight constraints on $w_0$
and $w_a$. As we will see, dark energy both modifies the distance-redshift
relation through a modification of the background dynamics of the Universe,
and the suppression in the growth of structure thanks to the additional expansion.
Since weak lensing is sensitive to both effects, it is a prime candidate to
measure the evolution of dark energy, as we shall see in this Appendix.

\subsection{The distance-redshift relation}

One of the major impacts of dark energy is that it affects the cosmological
distance-redshift relation. After all, it was through the use of this relation
that the existence of dark energy was discovered! Hence, any observable that
uses this relation will also be sensitive to the amount and evolutionary
history of the dark energy fluid. For example, if distant galaxies are receding
away from us at a greater rate, then a given redshift will actually correspond 
to a closer comoving distance (via Hubble's law) compared to smaller receding
rates. Therefore, dark energy has a crucial role in the cosmological
distance-redshift relation and so it is worth investigating how evolving
dark energy models modify this relation. 

In a universe that features evolving dark energy, such that its equation of
state follows Equation~\ref{eqn:w0_wa_app}, then the distance-redshift relation
is given by~\cite{Tsutsui:2008cu}
\begin{align}
    \rchi(z) = \frac{1}{H_0} \,\int_0^{z} \!\! \d z' \left[\Omegam (1+z')^3
    + \Omega_{\Lambda} (1 + z')^{3(1 + w_0 + w_a)} e^{-3w_a \frac{z'}{1+z'}}\right]^{-\frac{1}{2}}.
    \label{eqn:app_dist_redshift}
\end{align}
Hence, we can see that for any universe where $w_0 \neq -1$ and $w_a \neq 0$, 
there is a redshift dependence to the dark energy density, and so constraining
the distance-redshift relation will allow us to directly probe these parameters.

\begin{figure}[t]
    \centering
    \includegraphics[width=\linewidth,trim={0.0cm 0.0cm 0.0cm 0.0cm},clip]{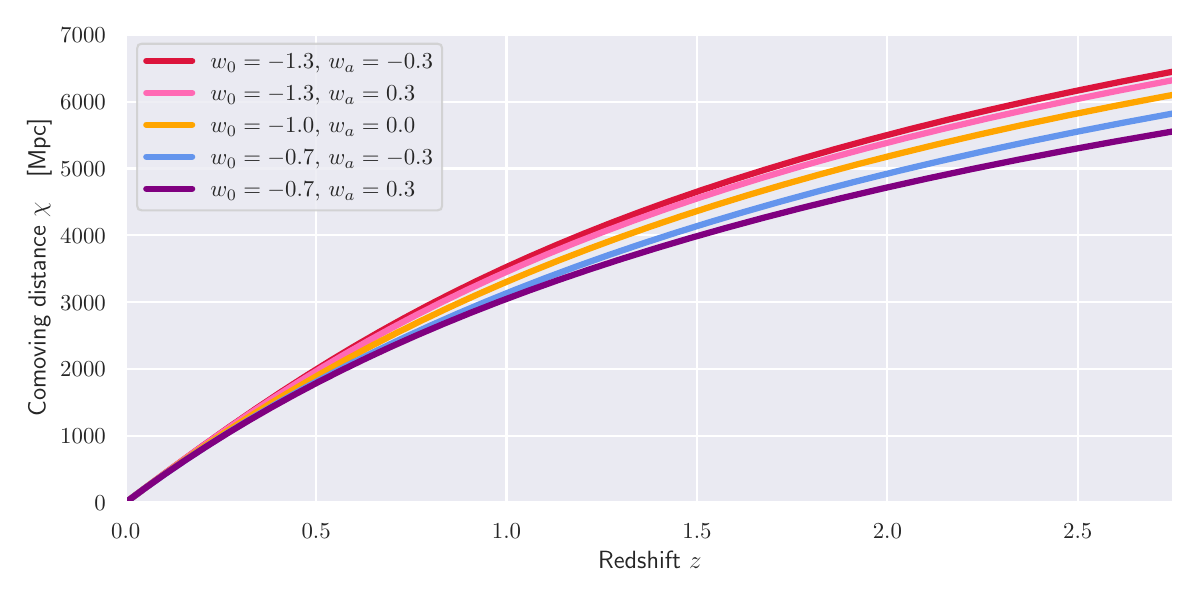}
    \caption{The distance-redshift relation for five different combination of
        $(w_0, w_a)$ values, where the orange curve represents that of the cosmological
        constant $\Lambda$. We see that for more positive values of the equation
        of state (such as the blue and purple curves), which represents an
        increase in the historic dark energy density over the evolution of the
        universe, yield a smaller comoving distance for a given redshift when
        compared to the cosmological constant, and a larger distance for more
        negative equations of state (pink and red curves). Constraining the
        distance-redshift relation to higher redshifts, where the difference
        between dark energy models is greatest, will help in determining
        the values of $w_0$ and $w_a$ of our Universe.
        }
    \label{fig:app_distance_redshifts}
\end{figure}

Figure~\ref{fig:app_distance_redshifts} plots the distance-redshift relation for
five different combinations of $w_0$ and $w_a$, where the orange curve 
corresponds to the cosmological constant $\Lambda$. We see that for universes
where $w > -1$, which corresponds to a greater dark energy density in the
universe's past, a given redshift is actually closer to us. If we use the
low-redshift approximation of Hubble's law ($z = H_0 d$), then if we have
increased accelerated expansion from dark energy (a higher $H_0$) then a 
source must be closer (smaller distance $d$) for a given redshift $z$, and
vice-versa. This distance-redshift relation is utilised extensively in weak
lensing, specifically in the lensing kernels and the redshift to evaluate the
matter power spectrum at, and so the impact of the changing distance-redshift
relation on weak lensing will be investigated next.

\subsection{Dark energy in weak lensing}

Since this thesis has focused on weak lensing, and specifically the power
spectrum of weak lensing as our observable of interest, it is natural to
enquire how evolving dark energy models change our observables --- and thus if
cosmic shear can discern between models of dark energy. First, let us recall the
definition of the cosmic shear angular power spectrum 
(Equation~\ref{eqn:converg_power_spec}), which was the Limber integral of
the matter power spectrum $P(k, z)$ weighted by the lensing kernels $g_a(\rchi)$,
and given as  
\begin{align}
    C_{\ell}^{ab} = \frac{9 \Omegam^2 H_0^4}{4}
    \int_{0}^{\chimax} \!\! \d \rchi \, \, 
    \frac{g_a(\rchi) \, g_b(\rchi)}{a^2(\rchi)} \,\,
    P \!\left( k = \frac{\ell}{\fk(\rchi)}, \, z=z(\rchi)\right).
\end{align}
Hence, information about the physics and properties of our Universe
come from the combination of these two functions. Dark energy impacts both,
since it alters the background dynamics and growth of structure in the Universe,
and so information about dark energy in the angular power spectrum will
come from the interplay between these two functions. We shall see how evolving
dark energy impacts these functions next\dots

\subsection{The lensing kernel}

The lensing kernel informs us about the efficiency of the intervening
large-scale structure of the Universe at producing lensing distributions in the
images of background galaxies. For a given perturbation in the distribution of
matter, it produces the maximum distortion in the shapes of galaxies when it is
half-way between the source and observer. Mathematically, they are written as
an integral over the comoving distance as (Equation~\ref{eqn:lensing_kernels})
\begin{align}
    g(\chi) = \int_{\chi}^{\chimax} \!\! \d \chi' \, n(\chi') \, 
    \frac{\fk(\chi' - \chi)}{\fk(\chi')},
    \label{eqn:lens_kern_app}
\end{align} 
where $n(\chi)$ is the number density of source galaxies that were imaged.
Normally, we assign the position of galaxies that are required in
Equation~\ref{eqn:lens_kern_app} thorough their photometric redshift $z$, since
this is the direct observable. Thus, we require the use of the distance-redshift
relation to transform their
observed redshifts to comoving coordinate $\rchi$. This is done through the
relationship $n(\rchi) \d \rchi = n(z) \d z$, hence $n(\rchi) = n(z) \frac{\d z}{\d \rchi}$.
Since this derivative is linked to the expansion history of the Universe
through Equation~\ref{eqn:app_dist_redshift}, the $z$--$\chi$ mapping is
sensitive to the amount and properties of dark energy, and thus our lensing 
kernels are sensitive to this too.

\begin{sidewaysfigure}
    \centering  
    \includegraphics[width=\columnwidth]{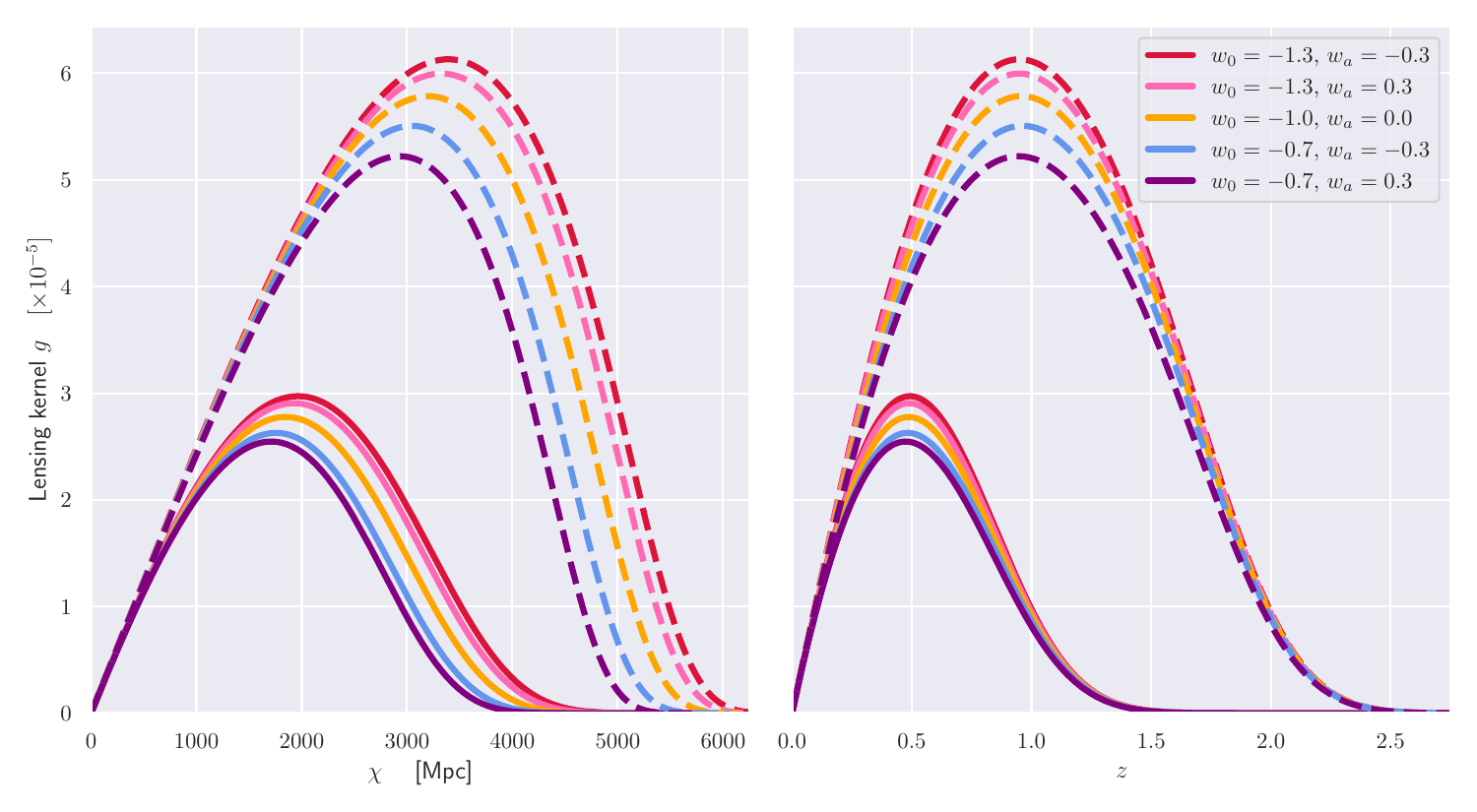}
    \vspace*{-0.25cm}
    \caption{Lensing kernels as a function of comoving distance $\rchi$
        (left panel) and redshift (right panel) for five combinations of
        $(w_0, w_a)$ values --- where the orange curve corresponds to that of the
        cosmological constant $\Lambda$, and for two different Gaussian redshift
        bins centred at $\bar{z} = 1$ (solid curves) and $\bar{z} = 2$ (dashed
        curves). We see that changing $w_0$ and $w_a$ has a direct impact on the
        shape and amplitude of the lensing kernels, though their response in 
        distance- and redshift-space is slightly different, as discussed in 
        the text.
        }
    \label{fig:app_lensing_kernels}
\end{sidewaysfigure}

Figure~\ref{fig:app_lensing_kernels} plots the lensing kernel as a function
of comoving distance $\rchi$ and redshift $z$ for five different combinations
of $w_0$ and $w_a$ values. Since we measure the source galaxy positions through
their redshift, the lensing kernels do not shift in redshift-space since
the redshift observations are anchored observables and irrespective of
$w_0$ and $w_a$ values. However, we still see that the amplitude of the lensing
kernels decrease in redshift-space for models with increased historic dark
energy density, since the photons from distant galaxies travel less physical
distance and so decreased opportunities to encounter lensing distortions. When
we look at the lensing kernels in comoving distance-space, then we see the
direct effects of dark energy changing the distance-redshift relation. As now
curves with an increased historic dark energy density correspond to a decreased
comoving distance for fixed redshift, and so the lensing kernels loose support
at much smaller comoving distances.

\clearpage
\subsection{The matter power spectrum}

The second key ingredient to the cosmic shear angular power spectrum is the
matter power spectrum, which is evaluated over a range of scales and redshifts.
Since dark energy serves to accelerate the expansion of the Universe, it makes
sense that an increased dark energy density hinders the ordinary gravitational
attraction of matter, and thus the growth of structure is suppressed. This
decreases the amplitude of the matter power spectrum, and thus decreases the
amplitude of our lensing power spectrum. Just as we have done so for the
distance-redshift relation and the lensing kernels, we would like to produce
predictions for the matter power spectrum for a variety of dark energy evolutions
and see the impact it has on the power spectrum. However, one subtlety here is
the choice of amplitude normalisation, we can either use the amplitude of the
primordial perturbations $\As$ (which is more suitable for early-Universe
measurements, such as the CMB), or directly normalise the matter power
spectrum through $\sigmaeight$ (which is more suitable for late-Universe
measurements, such as galaxy clustering and weak lensing). By anchoring the
amplitude of the perturbations to either the early or late Universe, the 
evolution behaviour of the matter power spectrum ratios with redshift changes
quite significantly.

\begin{figure}[tp]
    \centering
    \includegraphics[width=\linewidth,trim={0.0cm 0.0cm 0.0cm 0.0cm},clip]{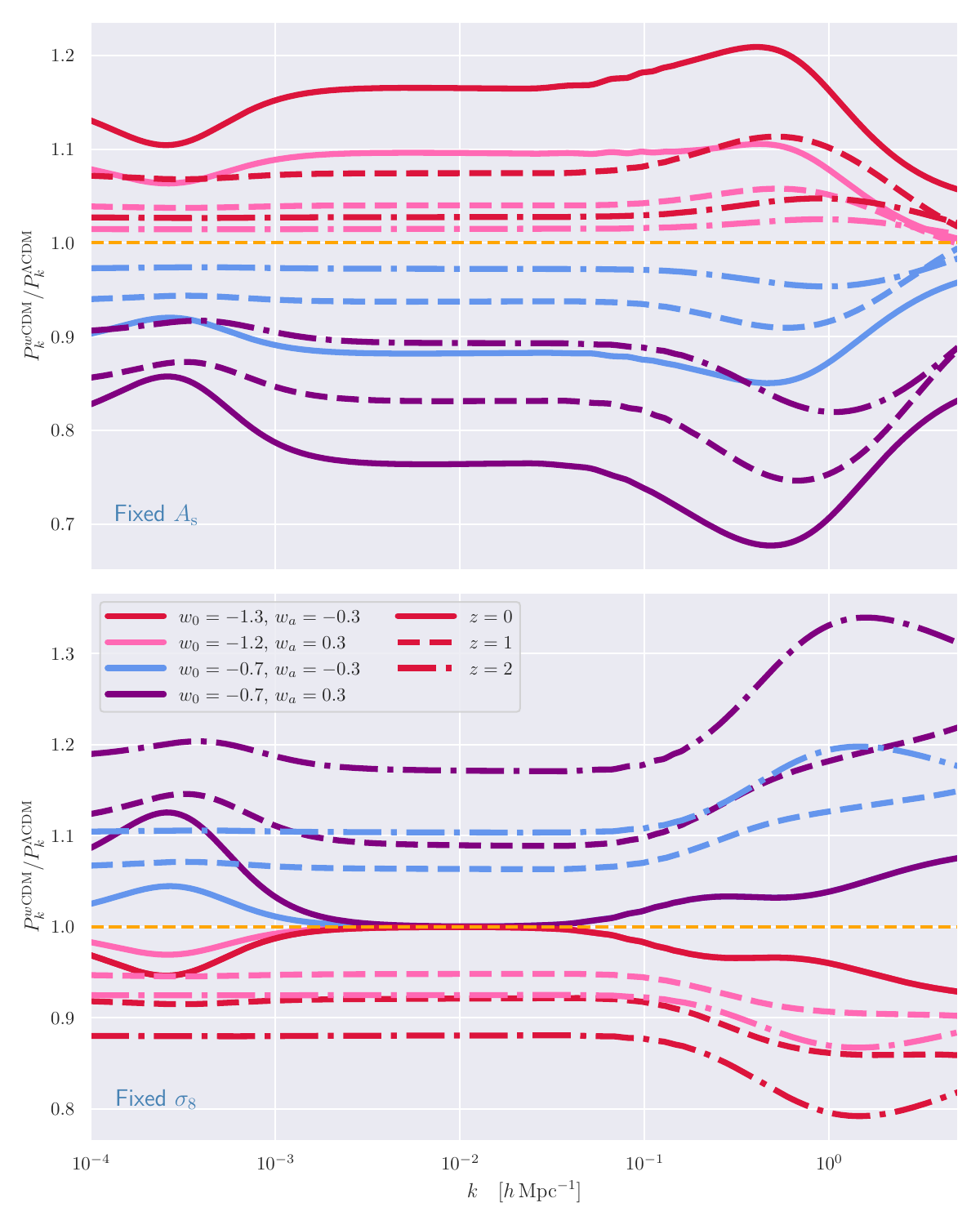}
    \caption{Ratio of the matter power spectrum for four models of time-evolving dark
        energy densities to the matter power spectrum for the cosmological 
        constant, where we have fixed $\As$ in the upper panel and $\sigmaeight$
        in the lower panel, and for three redshifts. Here, we see the
        difference between anchoring the amplitude of the density perturbations
        in either the early ($\As$) or late ($\sigmaeight$) universe. For the
        case of anchoring $\As$, we see that the differences in the $P(k)$
        grow with redshift, whereas anchoring $\sigmaeight$ we are fixing the
        amplitude at $z=0$ on linear scales where the background dynamics predict
        large deviances at higher redshifts. Hence, when using models of
        time-evolving dark energy we should be careful which amplitude we fix.  
        }
    \label{fig:app_mat_pow_ratios}
\end{figure}

Figure~\ref{fig:app_mat_pow_ratios} plots the ratio of the matter power spectrum
for four models of evolving dark energy to that of the cosmological constant
at three redshifts, using either fixed $\As$ values (top panel) or fixed
$\sigmaeight$ values (bottom panel). The choice of normalisation dramatically
changes the evolution of the matter power spectrum with redshift:
\begin{galitemize}
    \item \textbf{Fixed $\bm\As$}. With the amplitude of the primordial
        perturbations fixed, the amplitude of the perturbations that we see
        in the late-time Universe are a direct result of the evolution of
        the initial conditions (Section~\ref{sec:perturbation_evolution}).
        Hence, the growth of structure is subject to the background dynamics
        of the universe. In a universe with an increased dark energy density,
        the growth of structure is suppressed since the additional dark energy
        serves to increase the expansion rate, reducing the effect of
        gravitational collapse. This is why in Figure~\ref{fig:app_mat_pow_ratios},
        the blue and purple curves (which represent an increased historic dark
        energy density) feature suppression in their matter power spectra
        with respect to that of the cosmological constant, and that this
        suppression grows as structures evolve to the present time ($z=0$). In
        contrast, the pink and red curves (which represent a decreased historic
        dark energy density) feature an increase in their matter power spectra,
        and this increase grows as we evolve to redshift zero. 
    
    \item \textbf{Fixed $\bm\sigmaeight$}. By fixing $\sigmaeight$ we are
        fixing the amplitude of the linear matter power spectrum at redshift 
        zero. This is why all four dynamical dark energy models converge at
        unity around $k \sim 10^{-2}\,h$Mpc$^{-1}$. If we are fixing
        the amplitude today and are in a universe where the perturbations evolve
        slower than for the cosmological constant (as is the case for increased
        dark energy), then the amplitude of the perturbations must have been
        larger in the past, since they evolve more slowly. Hence, we now see
        that the blue and purple curves predict an excess in power on all scales
        at higher redshifts, which is in constant to when we fixed $\As$. The
        reverse effect happens for models with decreased dark energy density
        (pink and red curves), as now the perturbations can evolve quicker and 
        thus have a smaller amplitude at higher redshifts.
\end{galitemize}

Since the choice of normalisation can have quite a dramatic effect on the 
evolution of the matter power spectrum (whether we evolve forwards, using $\As$,
or backwards, using $\sigmaeight$) we should be extremely careful in comparing
results for $\As$ and $\sigmaeight$ in models of time-evolving dark energy.

\subsection[The $\As$--$\sigmaeight$ relation]{The $\bm\As$--$\bm\sigmaeight$ relation}

\begin{figure}[tp]
    \centering
    \includegraphics[width=\linewidth,trim={0.0cm 0.0cm 0.0cm 0.0cm},clip]{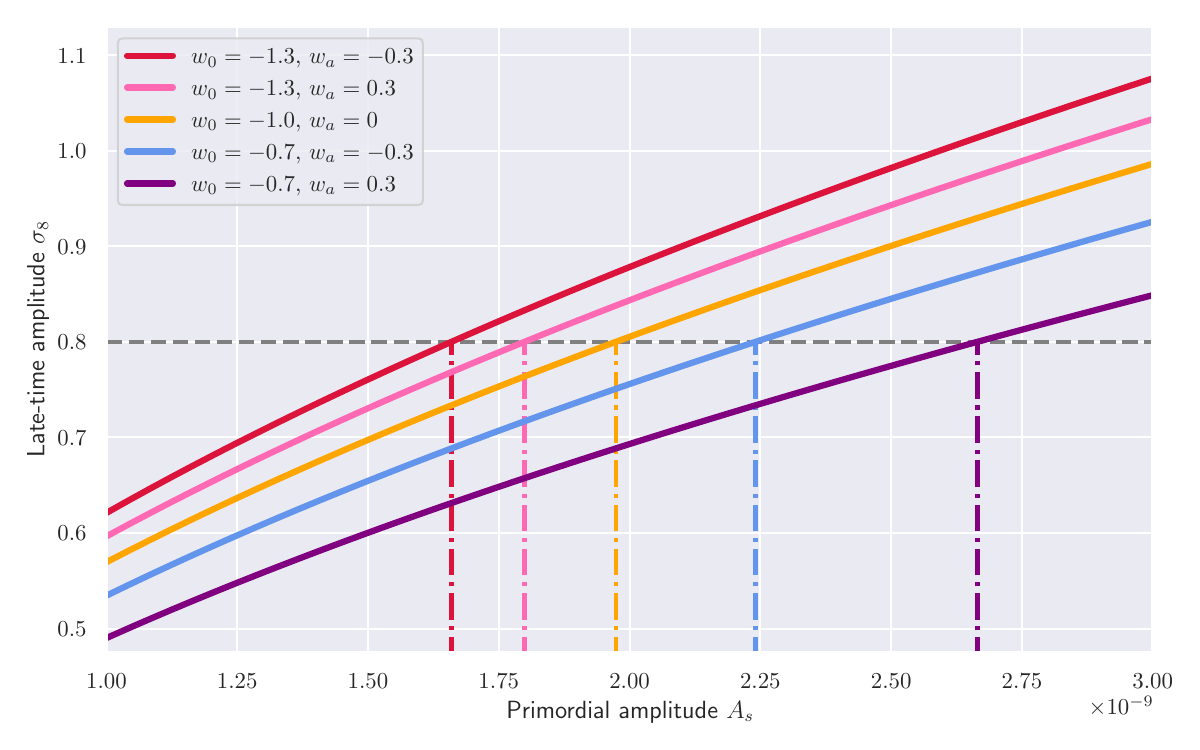}
    \caption{The relationship between the late-time amplitude $\sigmaeight$
        to the primordial amplitude $\As$ for five different dark energy 
        evolutionary models, where the orange curve corresponds to the
        cosmological constant. While we see that each curve follows the basic
        trend, that $\sigmaeight \propto \sqrt{\As}$, the normalisation is
        sensitive to the dark energy dynamics. Models which feature a larger
        dark energy densities require a larger primordial amplitude $\As$
        for a given value of $\sigmaeight$ observed today, as shown in the
        dot-dashed lines. Hence, constraining $\As$ and $\sigmaeight$
        individually would allow us to constrain the time-evolution of dark energy. 
        }
    \label{fig:app_As_s8_relation}
\end{figure}

For an otherwise fixed cosmology, there exists a one-to-one mapping between
$\As$ and $\sigmaeight$, given by $\As \propto \sigmaeight^2$. While this
general relationship is still true for any universe with a time-evolving
dark energy field, the exact constant of proportionality changes when compared
to the cosmological constant model of dark energy. Figure~\ref{fig:app_As_s8_relation}
plots this relationship between the primordial and late-time amplitudes for our
five models of dynamical dark energy. We see that for a fixed value of 
$\sigmaeight$ that is observed today, models that feature a decreased historic
dark energy (such as the pink and red curves) require a lower value of the
primordial amplitude when compared to that of the cosmological constant. 
This shows that the growth of structure is much more efficient in these 
models which feature decreased dark energy densities. 

\begin{figure}[t]
    \centering
    \includegraphics[width=\linewidth,trim={0.0cm 0.0cm 0.0cm 0.0cm},clip]{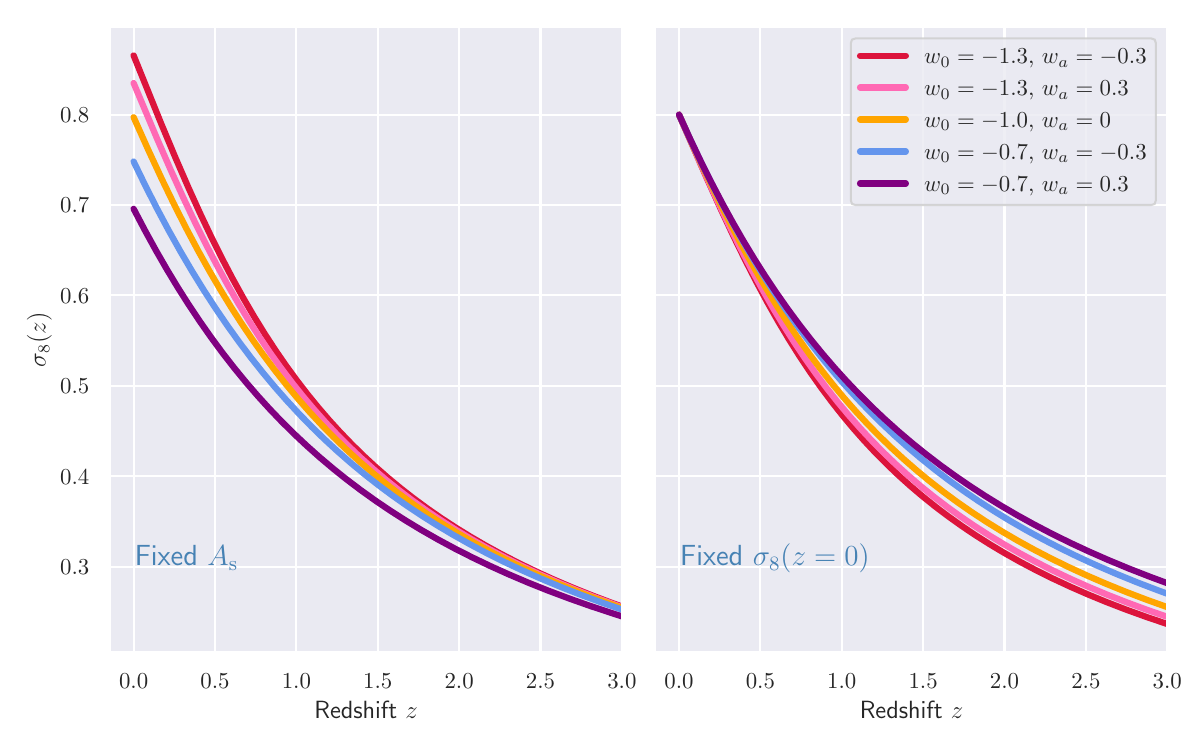}
    \caption{The evolution in the late-time matter perturbation parameter 
        $\sigmaeight$ as a function of redshift for five different dark energy
        evolutionary models, where we have either fixed the value of the
        primordial perturbation amplitude (left panel) or the value of 
        $\sigmaeight$ observed today (right panel). 
        }
    \label{fig:app_s8_z_relation}
\end{figure}

To demonstrate this
further, we can plot the evolution of the the late-time matter perturbation 
amplitude $\sigmaeight$ as a function of redshift, i.e. $\sigmaeight(z)$, 
where we independently fix either $\As$ or $\sigmaeight$. This is shown in
Figure~\ref{fig:app_s8_z_relation} where we see that when $\As$ is fixed, 
the different background dynamics of the dark energy models lead to suppressed
or enhanced clustering strength compared to that of the cosmological constant 
and so give different values for $\sigmaeight$ at $z=0$. However, when we fix
the value of $\sigmaeight$ at $z=0$, we require different primordial
amplitudes since the structure formation efficiency has been altered. Since
cosmic shear is sensitive to the growth of structure over the Universe's
evolution, it should be able to discern models of evolving dark energy and
their $w_0$ and $w_a$ parameters.

\subsection{Cosmic shear angular power spectrum}

Now that we know that dark energy effects both the weak lensing kernels and
the matter power spectrum, we can quantify the effects of dynamical dark
energy models on the cosmic shear angular power spectrum. Since we found that
by choosing to fix either $\As$ or $\sigmaeight$ had a dramatic effect on the
evolution of the matter power spectrum, we can predict that this choice will 
also have an impact on the cosmic shear power spectrum too. Hence, we can fix
them individually and see how the $\Cl$ values react in models of dynamical
dark energy.

\begin{figure}[tp]
    \centering
    \includegraphics[width=\linewidth,trim={0.0cm 0.0cm 0.0cm 0.0cm},clip]{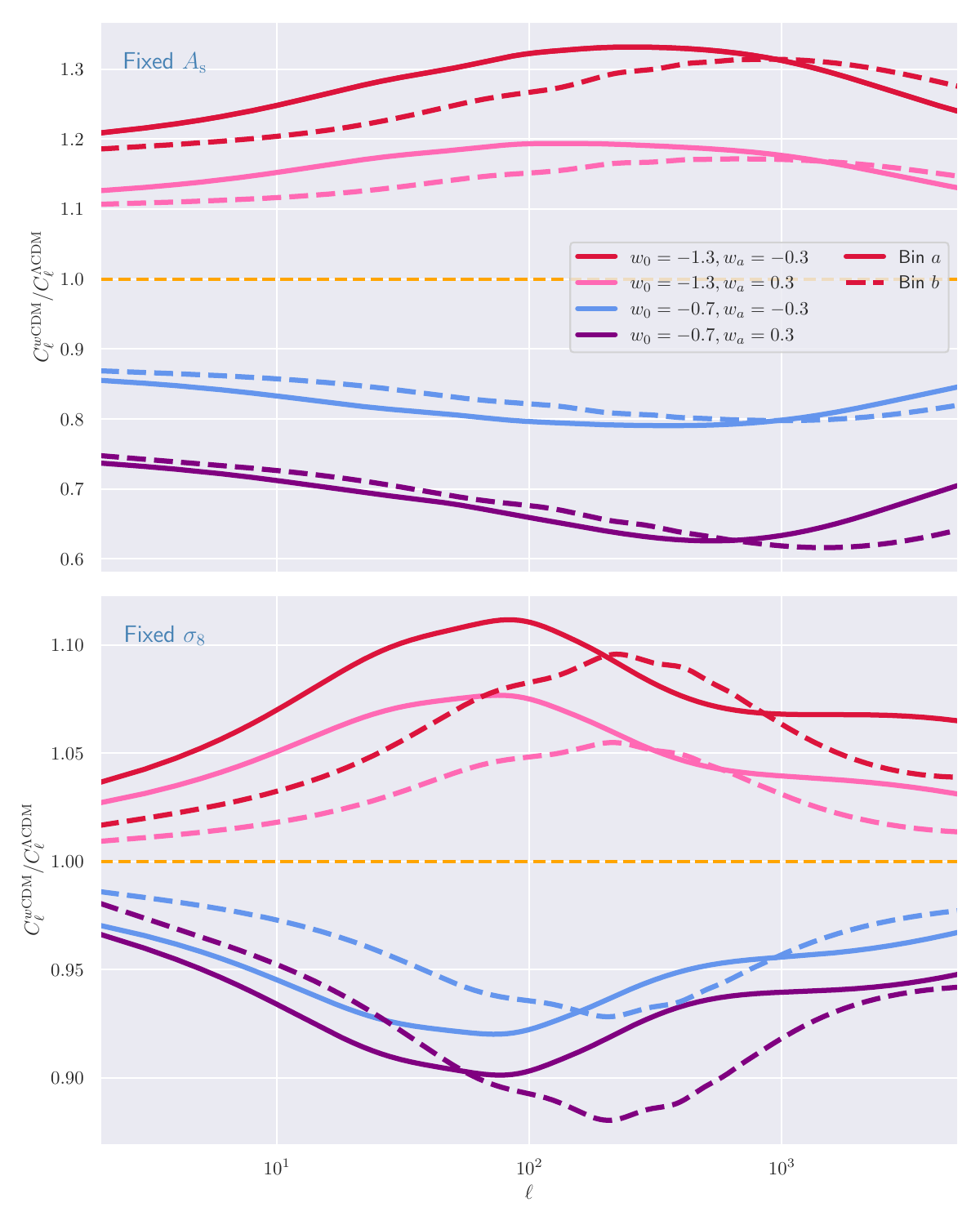}
    \caption{Ratio of the cosmic shear angular power spectrum for four models of
        time-evolving dark energy to that of the cosmological constant,
        where we have fixed $\As$ in the upper panel and $\sigmaeight$ in the
        lower panel, and for two Gaussian redshift bins centred at 
        $\bar{z} = 0.5$ (solid curves) and $\bar{z} = 2.0$ (dashed curves).
        When fixing $\As$, we see almost scale-invariant growth or suppression
        in the angular power spectrum for models with decreased or increased
        dark energy densities, respectively. This is a result of how the
        lensing kernels and matter power spectrum both scale together with
        the dark energy density when fixing $\As$.
        However, when we fix $\sigmaeight$, the lensing
        kernels and matter power spectrum react to changing dark energy densities
        in opposite ways. Hence, we see an increased $\ell$-dependence to
        the ratios and that the size of the ratios decreased when compared to
        the fixed $\As$ ratios, since the two contributions somewhat cancel
        each other out. 
        }
    \label{fig:app_cl_ratios}
\end{figure}

Figure~\ref{fig:app_cl_ratios} plots the ratios of the cosmic shear angular 
power spectrum for four models of dynamical dark energy to that of the
cosmological constant for two Gaussian redshift bins. For the case where
we fix $\As$, we see an almost scale-invariant enhancement or suppression in 
the angular power spectrum for models with a decreased or increased historic
dark energy density, respectively. This arises due to the lensing kernels and
matter power spectrum both scaling the same way with variable dark energy 
density using fixed $\As$. However, when we change to fixing $\sigmaeight$ 
instead, we see that the ratios become much more $\ell$-dependant and that
the amplitude of the enhancement or suppression becomes much smaller. This is
because now the lensing kernels and matter power spectrum act in opposite
directions to the effects of evolving dark energy, and so give rise to the
curves seen. We also see that the ratios also diverge with increasing redshift 
(the centre of bin $a$ is closer than that of bin $b$), which make sense given 
that dark energy is a late-time effect, and so lower redshift sources will be 
more susceptible to variable dark energy than higher redshift sources. 
Note that time-evolving dark energy also affects the redshift that the matter power
spectrum is evaluated over in the $\Cl$ Limber integral, since it is evaluated
at $z = z(\rchi)$ --- thus dependent on the distance-redshift relation ---
which further alters the $\Cl$ values and their distribution across redshift bins.

\clearpage
\section{Neutrinos}

In addition to placing tight constraints on the evolution of dark energy, 
forthcoming Stage-IV cosmic shear surveys are prime candidates to place
place tight constraints on the properties of neutrino, through their impact on
cosmological observables. Neutrinos, whose name were dubbed by Enrico Fermi
meaning \textit{little neutral one} in humorous and grammatically incorrect
Italian~\cite{AMALDI19841}, are neutral leptonic elementary particles that
interact via the weak nuclear and gravitational forces. The Standard Model
of particle physics predicts that neutrinos come in three flavours: the electron
neutrino, the muon neutrino, and the tauon neutrino --- in analogy to the charged
leptonic particles --- and this has been experimentally confirmed by accelerator
experiments~\cite{Mele:2015etc}. 

Neutrino oscillation experiments are sensitive to the squared mass differences
in the individual neutrino masses, that is $\Delta m^2_{ab} = m_a^2 - m_b^2$,
and have shown that at least two of the neutrino flavours are 
massive~\cite{ParticleDataGroup:2022pth}. Cosmological observations are sensitive
to the total neutrino mass, $\mnu$, and so provide an excellent compliment to the
oscillation experiments in constraining the neutrino masses. Cosmology is
also sensitive to the effective number of neutrinos, $\Neff$, which is given by
$\Neff = 3.044$ for the Standard Model~\cite{Bennett:2020zkv}.

Massive neutrinos impact cosmology in two distinct ways since their small,
but non-zero, mass means that they act as relativistic particles in the early
universe, and then transition to being non-relativistic in the later 
universe --- the only such particles to undergo such transition~\cite{DESI:2024mwx}.
When neutrinos are relativistic in the early universe, they free-stream
over large distances. This suppresses the gravitational clustering of matter,
and thus causes a suppression in the matter power spectrum on small scales. 
The amplitude of suppression is given by approximately $P^{\mnu}(k) / P^{0m_{\nu}}(k)
= 1 - 8f_{\nu}$, where $f_\nu = \Omega_{\nu, 0} / \Omega_{\mathrm{m}, 0}$ is the 
neutrino mass fraction~\cite{Lesgourgues:2013sjj}. When neutrinos become 
non-relativistic, they act the same way as baryons and dark matter do, and hence
affect the background expansion rate of the universe and slightly alter observables
such as the distance-redshift relation. Therefore, cosmology (and in particular
the Stage-IV surveys) are prime candidates for measuring neutrino 
properties~\cite{Euclid:2024imf}.

\begin{figure}[tp]
    \centering
    \includegraphics[width=\linewidth,trim={0.0cm 0.0cm 0.0cm 0.0cm},clip]{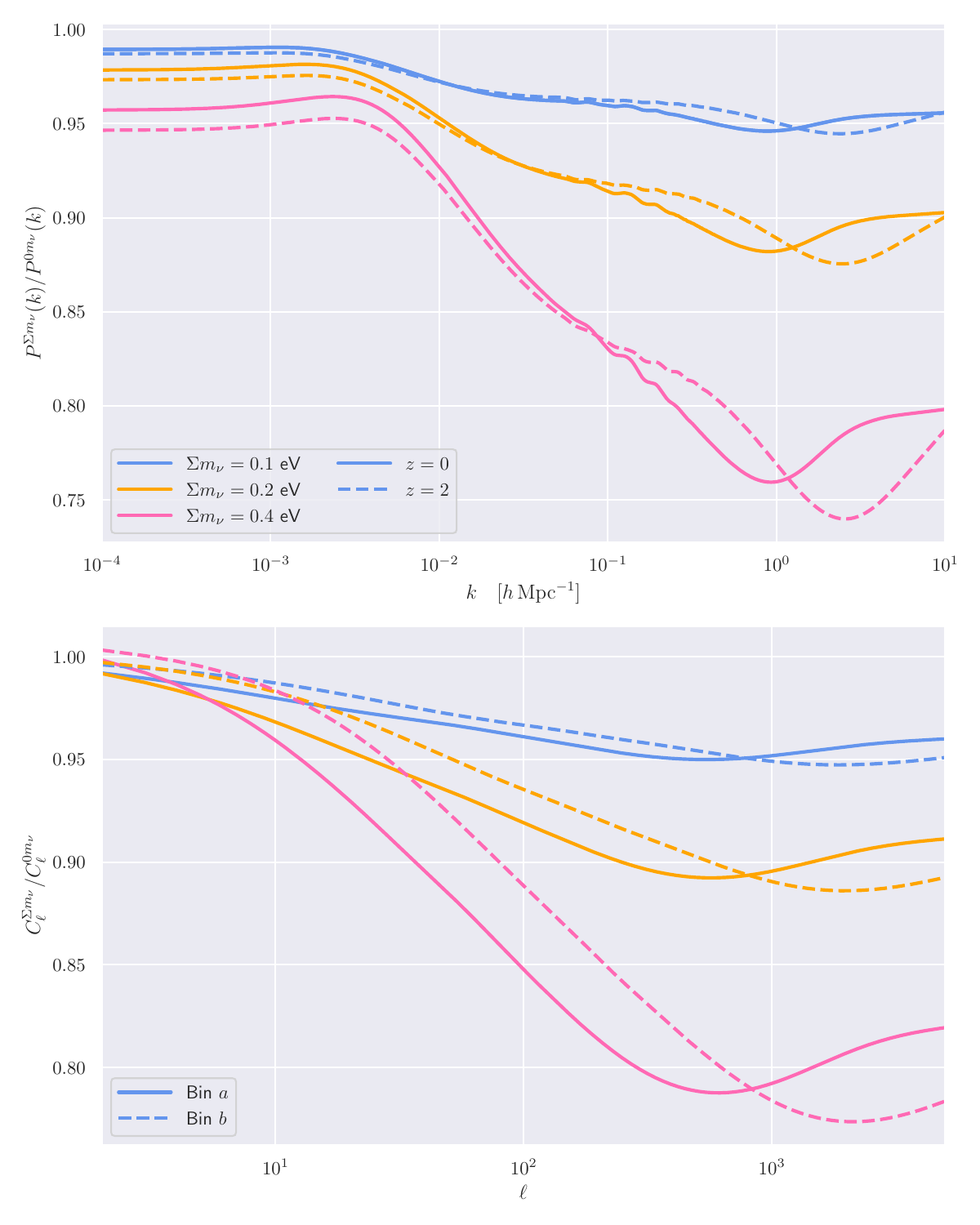}
    \caption{Ratio of the matter power spectrum (top panel) and cosmic shear
        angular power spectrum (bottom panel) for three models of non-zero
        neutrino masses with respect to that of zero neutrino masses, for two 
        redshifts (in the $P(k)$) and for two Gaussian redshift bins centred at 
        $\bar{z} = 0.5$ (solid curves) and $\bar{z} = 2.0$ (dashed curves)
        for the $\Cl$ values. We see that increasing the neutrino masses 
        directly increases the suppression in the clustering on small scales.
        Hence, accurately measuring the cosmic shear angular power spectrum
        down to small scales will enable tight constraints on the neutrino 
        masses. Ratios computed for an otherwise fixed cosmology for a
        flat $\Lambda$CDM Universe that is compatible with observations.
        }
    \label{fig:app_neutrino_ratios}
\end{figure}

Figure~\ref{fig:app_neutrino_ratios} plots the ratios of the matter power 
spectrum (top panel) and cosmic shear angular power spectrum (bottom panel)
for three cosmologies with different non-zero neutrino masses with respect to
that of massless neutrinos. We see the confirmation of the linear approximation,
that the matter power spectrum is suppressed on small scales due to neutrino
free-streaming, and that the amplitude of suppression scales with the neutrino
masses. This suppression in the matter power spectrum then directly feeds into
a suppression in the cosmic shear angular power spectrum, though now this is
convolved with the effects of neutrino masses in the lensing kernels thanks
to the slightly altered background dynamics.

\section{Summary}

We have seen that the properties of time-evolving dark energy and massive 
neutrinos both affect the cosmic shear angular power spectrum in their own ways,
with particular impact on small angular scales. Hence, an accurate and precise
determination of the experimental power spectrum from observations, coupled
with robust theoretical modelling of physical phenomena such as baryonic feedback
and the determination of associated scale-cuts will ensure that the
constraints on evolving dark energy and neutrinos are unbiased and
precise as possible. Both of which have been discussed heavily in this thesis,
and showing how the application of techniques developed here can be used to
maximise information from weak lensing analyses. {\hfill \textit{Tara procrastinators!}}


\clearpage
\titleformat{\chapter}[hang]{\filleft\Huge\bfseries}{}{0pt}{\Huge\seriffontsemi}[\vspace{0ex}\rule{\textwidth}{1.15pt}\vspace*{1.25ex}]
\bibliographystyle{jcap}

\begin{savequote}[65mm]
  If I have seen further,

  it is by standing on the shoulders of giants.

  \qauthor{---Isaac Newton}
\end{savequote}

\phantomsection
\bibliography{references/cosmo_refs,references/lensing_refs,references/QML_refs,references/theoretical_errors_refs,references/conclusion_refs,references/binary_cuts_refs}

\providecommand{\href}[2]{#2}\begingroup\raggedright\begin{thebibliography}{100}

\bibitem{Maraio:2022ywi}
A.~Maraio, A.~Hall and A.~Taylor, \emph{{Testing quadratic maximum likelihood estimators for forthcoming Stage-IV weak lensing surveys}}, \href{https://doi.org/10.1093/mnras/stad426}{\emph{Mon. Not. Roy. Astron. Soc.} {\bfseries 520} (2023) 4836} [\href{https://arxiv.org/abs/2207.10412}{{\ttfamily 2207.10412}}].

\bibitem{Maraio:2024xjz}
A.~Maraio, A.~Hall and A.~Taylor, \emph{{Mitigating baryon feedback bias in cosmic shear through a theoretical error covariance in the matter power spectrum}}, \href{https://doi.org/10.1093/mnras/staf113}{\emph{Mon. Not. Roy. Astron. Soc.} {\bfseries 537} (2025) 1749} [\href{https://arxiv.org/abs/2410.12500}{{\ttfamily 2410.12500}}].

\bibitem{Numpy:2020array}
C.~R. Harris, K.~J. Millman, S.~J. van~der Walt, R.~Gommers, P.~Virtanen, D.~Cournapeau et~al., \emph{Array programming with {NumPy}}, \href{https://doi.org/10.1038/s41586-020-2649-2}{\emph{Nature} {\bfseries 585} (2020) 357} [\href{https://arxiv.org/abs/2006.10256}{{\ttfamily 2006.10256}}].

\bibitem{SciPy:2020}
P.~Virtanen, R.~Gommers, T.~E. Oliphant, M.~Haberland, T.~Reddy, D.~Cournapeau et~al., \emph{{{SciPy} 1.0: Fundamental Algorithms for Scientific Computing in Python}}, \href{https://doi.org/10.1038/s41592-019-0686-2}{\emph{Nature Methods} {\bfseries 17} (2020) 261} [\href{https://arxiv.org/abs/1907.10121}{{\ttfamily 1907.10121}}].

\bibitem{Pandas:2010}
W.~{M}c{K}inney, \emph{{D}ata {S}tructures for {S}tatistical {C}omputing in {P}ython},  in \emph{{P}roceedings of the 9th {P}ython in {S}cience {C}onference}, S.~Van~der {W}alt and J.~{M}illman, eds., pp.~56 -- 61, 2010, \href{https://doi.org/10.25080/Majora-92bf1922-00a}{DOI}.

\bibitem{Matplotlib:2007}
J.~D. Hunter, \emph{{Matplotlib: A 2D Graphics Environment}}, \href{https://doi.org/10.1109/MCSE.2007.55}{\emph{Computing in Science \& Engineering} {\bfseries 9} (2007) 90}.

\bibitem{Seaborn:2021}
M.~L. Waskom, \emph{seaborn: statistical data visualization}, \href{https://doi.org/10.21105/joss.03021}{\emph{Journal of Open Source Software} {\bfseries 6} (2021) 3021}.

\bibitem{Lewis:2002ah}
A.~Lewis and S.~Bridle, \emph{{Cosmological parameters from CMB and other data: A Monte Carlo approach}}, \href{https://doi.org/10.1103/PhysRevD.66.103511}{\emph{Phys.\ Rev.\ D} {\bfseries 66} (2002) 103511} [\href{https://arxiv.org/abs/astro-ph/0205436}{{\ttfamily astro-ph/0205436}}].

\bibitem{Lewis:1999bs}
A.~Lewis, A.~Challinor and A.~Lasenby, \emph{{Efficient computation of CMB anisotropies in closed FRW models}}, \href{https://doi.org/10.1086/309179}{\emph{Astrophys.\ J.} {\bfseries 538} (2000) 473} [\href{https://arxiv.org/abs/astro-ph/9911177}{{\ttfamily astro-ph/9911177}}].

\bibitem{Howlett:2012mh}
C.~Howlett, A.~Lewis, A.~Hall and A.~Challinor, \emph{{CMB power spectrum parameter degeneracies in the era of precision cosmology}}, \href{https://doi.org/10.1088/1475-7516/2012/04/027}{\emph{JCAP} {\bfseries 04} (2012) 027} [\href{https://arxiv.org/abs/1201.3654}{{\ttfamily 1201.3654}}].

\bibitem{Lewis:2019xzd}
A.~Lewis, \emph{{GetDist: a Python package for analysing Monte Carlo samples}},  \href{https://arxiv.org/abs/1910.13970}{{\ttfamily 1910.13970}}.

\bibitem{LSSTDarkEnergyScience:2018yem}
{\scshape LSST Dark Energy Science} collaboration, \emph{{Core Cosmology Library: Precision Cosmological Predictions for LSST}}, \href{https://doi.org/10.3847/1538-4365/ab1658}{\emph{Astrophys. J. Suppl.} {\bfseries 242} (2019) 2} [\href{https://arxiv.org/abs/1812.05995}{{\ttfamily 1812.05995}}].

\bibitem{Xavier:2016elr}
H.~S. Xavier, F.~B. Abdalla and B.~Joachimi, \emph{{Improving lognormal models for cosmological fields}}, \href{https://doi.org/10.1093/mnras/stw874}{\emph{Mon. Not. Roy. Astron. Soc.} {\bfseries 459} (2016) 3693} [\href{https://arxiv.org/abs/1602.08503}{{\ttfamily 1602.08503}}].

\bibitem{Tessore:2023zyk}
N.~Tessore, A.~Loureiro, B.~Joachimi, M.~von Wietersheim-Kramsta and N.~Jeffrey, \emph{{GLASS: Generator for Large Scale Structure}}, \href{https://doi.org/10.21105/astro.2302.01942}{\emph{The Open Journal of Astrophysics} {\bfseries 6} (2023) } [\href{https://arxiv.org/abs/2302.01942}{{\ttfamily 2302.01942}}].

\bibitem{Gorski:2004by}
K.~M. Gorski, E.~Hivon, A.~J. Banday, B.~D. Wandelt, F.~K. Hansen, M.~Reinecke et~al., \emph{{HEALPix - A Framework for high resolution discretization, and fast analysis of data distributed on the sphere}}, \href{https://doi.org/10.1086/427976}{\emph{Astrophys. J.} {\bfseries 622} (2005) 759} [\href{https://arxiv.org/abs/astro-ph/0409513}{{\ttfamily astro-ph/0409513}}].

\bibitem{Zonca2019}
A.~Zonca, L.~Singer, D.~Lenz, M.~Reinecke, C.~Rosset, E.~Hivon et~al., \emph{{healpy: equal area pixelization and spherical harmonics transforms for data on the sphere in Python}}, \href{https://doi.org/10.21105/joss.01298}{\emph{Journal of Open Source Software} {\bfseries 4} (2019) 1298}.

\bibitem{Alonso:2018jzx}
{\scshape LSST Dark Energy Science} collaboration, \emph{{A unified pseudo-$C_\ell$ framework}}, \href{https://doi.org/10.1093/mnras/stz093}{\emph{Mon. Not. Roy. Astron. Soc.} {\bfseries 484} (2019) 4127} [\href{https://arxiv.org/abs/1809.09603}{{\ttfamily 1809.09603}}].

\bibitem{eigenweb}
G.~Guennebaud et~al., ``\textit{Eigen v3}.'' \url{http://eigen.tuxfamily.org}, 2010.

\bibitem{Zuntz:2014csq}
J.~Zuntz, M.~Paterno, E.~Jennings, D.~Rudd, A.~Manzotti, S.~Dodelson et~al., \emph{{CosmoSIS: modular cosmological parameter estimation}}, \href{https://doi.org/10.1016/j.ascom.2015.05.005}{\emph{Astron. Comput.} {\bfseries 12} (2015) 45} [\href{https://arxiv.org/abs/1409.3409}{{\ttfamily 1409.3409}}].

\bibitem{Newton:1687eqk}
I.~Newton, \emph{{Philosophi\ae{} Naturalis Principia Mathematica}}. England, 1687.

\bibitem{Einstein:1916vd}
A.~Einstein, \emph{{The Foundation of the General Theory of Relativity}}, \href{https://doi.org/10.1002/andp.200590044, 10.1002/andp.19163540702}{\emph{Annalen Phys.} {\bfseries 49} (1916) 769}.

\bibitem{Einstein:1917ce}
A.~Einstein, \emph{{Cosmological Considerations in the General Theory of Relativity}}, {\emph{Sitzungsber. Preuss. Akad. Wiss. Berlin (Math. Phys. )} {\bfseries 1917} (1917) 142}.

\bibitem{ORaifeartaigh:2017uct}
C.~O'Raifeartaigh, M.~O'Keeffe, W.~Nahm and S.~Mitton, \emph{{Einstein\textquoteright{}s 1917 static model of the universe: a centennial review}}, \href{https://doi.org/10.1140/epjh/e2017-80002-5}{\emph{Eur. Phys. J. H} {\bfseries 42} (2017) 431} [\href{https://arxiv.org/abs/1701.07261}{{\ttfamily 1701.07261}}].

\bibitem{Slipher:1917Obs40304S}
V.~M. {Slipher}, \emph{{Radial velocity observations of spiral nebulae}}, {\emph{The Observatory} {\bfseries 40} (1917) 304}.

\bibitem{Weinberg:1993ftmmW}
S.~{Weinberg}, \emph{{The first three minutes: A modern view of the origin of the Universe}}. Basic Books, New York, 1993.

\bibitem{Lemaitre:1931MNRAS91490L}
G.~{Lema{\^\i}tre}, \emph{{The expanding universe}}, \href{https://doi.org/10.1093/mnras/91.5.490}{\emph{Mon. Not. Roy. Astron. Soc.} {\bfseries 91} (1931) 490}.

\bibitem{Leavitt:1907AnHar6087L}
H.~S. {Leavitt}, \emph{{1777 variables in the Magellanic Clouds}}, {\emph{Annals of Harvard College Observatory} {\bfseries 60} (1907) 87}.

\bibitem{Leavitt:1912HarCi1731L}
H.~S. {Leavitt} and E.~C. {Pickering}, \emph{{Periods of 25 Variable Stars in the Small Magellanic Cloud.}}, {\emph{Harvard College Observatory Circular} {\bfseries 173} (1912) 1}.

\bibitem{Hubble168}
E.~Hubble, \emph{A relation between distance and radial velocity among extra-galactic nebulae}, \href{https://doi.org/10.1073/pnas.15.3.168}{\emph{Proc. Natl. Acad. Sci. U.S.A.} {\bfseries 15} (1929) 168}.

\bibitem{Lemaitre:1931MNRAS483L}
G.~{Lema{\^\i}tre}, \emph{{A homogeneous universe of constant mass and increasing radius accounting for the radial velocity of extra-galactic nebulae}}, \href{https://doi.org/10.1093/mnras/91.5.483}{\emph{Mon. Not. Roy. Astron. Soc.} {\bfseries 91} (1931) 483}.

\bibitem{Dyson:1920abc}
F.~W. Dyson, A.~S. Eddington and C.~Davidson, \emph{{A determination of the deflection of light by the sun's gravitational field, from observations made at the total eclipse of May 29, 1919}}, \href{https://doi.org/10.1098/rsta.1920.0009}{\emph{Phys. Trans. Roy. Soc. London} {\bfseries 220} (1920) 291}.

\bibitem{Bartelmann:1999yn}
M.~Bartelmann and P.~Schneider, \emph{{Weak gravitational lensing}}, \href{https://doi.org/10.1016/S0370-1573(00)00082-X}{\emph{Phys. Rept.} {\bfseries 340} (2001) 291} [\href{https://arxiv.org/abs/astro-ph/9912508}{{\ttfamily astro-ph/9912508}}].

\bibitem{Kilbinger:2014cea}
M.~Kilbinger, \emph{{Cosmology with cosmic shear observations: a review}}, \href{https://doi.org/10.1088/0034-4885/78/8/086901}{\emph{Rept. Prog. Phys.} {\bfseries 78} (2015) 086901} [\href{https://arxiv.org/abs/1411.0115}{{\ttfamily 1411.0115}}].

\bibitem{Zwicky:1933AcHPh6110Z}
F.~{Zwicky}, \emph{{Die Rotverschiebung von extragalaktischen Nebeln}}, {\emph{Helvetica Physica Acta} {\bfseries 6} (1933) 110}.

\bibitem{Zwicky:2009GReGr41207Z}
F.~{Zwicky}, \emph{{Republication of: The redshift of extragalactic nebulae}}, \href{https://doi.org/10.1007/s10714-008-0707-4}{\emph{General Relativity and Gravitation} {\bfseries 41} (2009) 207}.

\bibitem{Rubin:1980ApJ471R}
V.~C. {Rubin}, J.~{Ford}, W.~K. and N.~{Thonnard}, \emph{{Rotational properties of 21 SC galaxies with a large range of luminosities and radii, from NGC 4605 (R=4kpc) to UGC 2885 (R=122kpc)}}, \href{https://doi.org/10.1086/158003}{\emph{Astrophys.\ J.} {\bfseries 238} (1980) 471}.

\bibitem{Perlmutter:1998np}
{\scshape Supernova Cosmology Project} collaboration, \emph{{Measurements of $\Omega$ and $\Lambda$ from 42 high redshift supernovae}}, \href{https://doi.org/10.1086/307221}{\emph{Astrophys. J.} {\bfseries 517} (1999) 565} [\href{https://arxiv.org/abs/astro-ph/9812133}{{\ttfamily astro-ph/9812133}}].

\bibitem{Riess:1998cb}
{\scshape Supernova Search Team} collaboration, \emph{{Observational evidence from supernovae for an accelerating universe and a cosmological constant}}, \href{https://doi.org/10.1086/300499}{\emph{Astron. J.} {\bfseries 116} (1998) 1009} [\href{https://arxiv.org/abs/astro-ph/9805201}{{\ttfamily astro-ph/9805201}}].

\bibitem{Lahav:2022poa}
O.~Lahav and A.~R. Liddle, \emph{{The Cosmological Parameters (2021)}},  \href{https://arxiv.org/abs/2201.08666}{{\ttfamily 2201.08666}}.

\bibitem{PlanckCollaboration:2018eyx}
{\scshape Planck} collaboration, \emph{{Planck 2018 results. VI. Cosmological parameters}}, \href{https://doi.org/10.1051/0004-6361/201833910}{\emph{Astron. Astrophys.} {\bfseries 641} (2020) A6} [\href{https://arxiv.org/abs/1807.06209}{{\ttfamily 1807.06209}}].

\bibitem{Schneider:1992bmb}
P.~Schneider, J.~Ehlers and E.~E. Falco, \emph{{Gravitational Lenses}}, Astronomy and Astrophysics Library. Springer, Berlin, 1992, \href{https://doi.org/10.1007/978-3-662-03758-4}{DOI}.

\bibitem{Kaiser:1992ps}
N.~Kaiser and G.~Squires, \emph{{Mapping the dark matter with weak gravitational lensing}}, \href{https://doi.org/10.1086/172297}{\emph{Astrophys. J.} {\bfseries 404} (1993) 441}.

\bibitem{Kaiser:2000if}
N.~Kaiser, G.~Wilson and G.~A. Luppino, \emph{{Large scale cosmic shear measurements}},  \href{https://arxiv.org/abs/astro-ph/0003338}{{\ttfamily astro-ph/0003338}}.

\bibitem{Bacon:2000sy}
D.~J. Bacon, A.~R. Refregier and R.~S. Ellis, \emph{{Detection of weak gravitational lensing by large-scale structure}}, \href{https://doi.org/10.1046/j.1365-8711.2000.03851.x}{\emph{Mon. Not. Roy. Astron. Soc.} {\bfseries 318} (2000) 625} [\href{https://arxiv.org/abs/astro-ph/0003008}{{\ttfamily astro-ph/0003008}}].

\bibitem{vanWaerbeke:2000rm}
L.~van Waerbeke et~al., \emph{{Detection of correlated galaxy ellipticities on CFHT data: First evidence for gravitational lensing by large scale structures}}, {\emph{Astron. Astrophys.} {\bfseries 358} (2000) 30} [\href{https://arxiv.org/abs/astro-ph/0002500}{{\ttfamily astro-ph/0002500}}].

\bibitem{Wittman:2000tc}
D.~M. Wittman, J.~A. Tyson, D.~Kirkman, I.~Dell'Antonio and G.~Bernstein, \emph{{Detection of weak gravitational lensing distortions of distant galaxies by cosmic dark matter at large scales}}, \href{https://doi.org/10.1038/35012001}{\emph{Nature} {\bfseries 405} (2000) 143} [\href{https://arxiv.org/abs/astro-ph/0003014}{{\ttfamily astro-ph/0003014}}].

\bibitem{Hobson:2006se}
M.~P. Hobson, G.~P. Efstathiou and A.~N. Lasenby, \emph{{General relativity: An introduction for physicists}}. Cambridge Univ. Pr., Cambridge, 2006, \href{https://doi.org/10.1017/CBO9780511790904}{DOI}.

\bibitem{Schwarzschild:1916Abh}
K.~{Schwarzschild}, \emph{{On the Gravitational Field of a Mass Point According to Einstein's Theory}}, \href{https://doi.org/10.48550/arXiv.physics/9905030}{\emph{Sitzungsber. Preuss. Akad. Wiss. Berlin (Math. Phys.)} {\bfseries 1916} (1916) 189} [\href{https://arxiv.org/abs/physics/9905030}{{\ttfamily physics/9905030}}].

\bibitem{COBE:1993ij}
J.~C. Mather et~al., \emph{{Measurement of the Cosmic Microwave Background spectrum by the COBE FIRAS instrument}}, \href{https://doi.org/10.1086/173574}{\emph{Astrophys.\ J.} {\bfseries 420} (1994) 439}.

\bibitem{Planck:2018nkj}
{\scshape Planck} collaboration, \emph{{Planck 2018 results. I. Overview and the cosmological legacy of Planck}}, \href{https://doi.org/10.1051/0004-6361/201833880}{\emph{Astron. Astrophys.} {\bfseries 641} (2020) A1} [\href{https://arxiv.org/abs/1807.06205}{{\ttfamily 1807.06205}}].

\bibitem{Griffiths_2023}
D.~J. Griffiths, \emph{Introduction to Electrodynamics}. Cambridge University Press, 5~ed., 2023, \href{https://doi.org/10.1017/9781009397735}{DOI}.

\bibitem{Huterer:1998qv}
D.~Huterer and M.~S. Turner, \emph{{Prospects for probing the dark energy via supernova distance measurements}}, \href{https://doi.org/10.1103/PhysRevD.60.081301}{\emph{Phys. Rev. D} {\bfseries 60} (1999) 081301} [\href{https://arxiv.org/abs/astro-ph/9808133}{{\ttfamily astro-ph/9808133}}].

\bibitem{Linder:2002et}
E.~V. Linder, \emph{{Exploring the expansion history of the universe}}, \href{https://doi.org/10.1103/PhysRevLett.90.091301}{\emph{Phys. Rev. Lett.} {\bfseries 90} (2003) 091301} [\href{https://arxiv.org/abs/astro-ph/0208512}{{\ttfamily astro-ph/0208512}}].

\bibitem{Linder:2005in}
E.~V. Linder, \emph{{Cosmic growth history and expansion history}}, \href{https://doi.org/10.1103/PhysRevD.72.043529}{\emph{Phys. Rev. D} {\bfseries 72} (2005) 043529} [\href{https://arxiv.org/abs/astro-ph/0507263}{{\ttfamily astro-ph/0507263}}].

\bibitem{DESI:2024mwx}
{\scshape DESI} collaboration, \emph{{DESI 2024 VI: Cosmological constraints from the measurements of baryon acoustic oscillations}}, \href{https://doi.org/10.1088/1475-7516/2025/02/021}{\emph{JCAP} {\bfseries 02} (2025) 021} [\href{https://arxiv.org/abs/2404.03002}{{\ttfamily 2404.03002}}].

\bibitem{Schneider:2015eaci}
P.~{Schneider}, \emph{{Extragalactic Astronomy and Cosmology: An Introduction}}. Springer Berlin, Heidelberg, 2nd~ed., 2015, \href{https://doi.org/10.1007/978-3-642-54083-7}{DOI}.

\bibitem{Gamow:1948pob}
G.~Gamow, \emph{{The Evolution of the Universe}}, \href{https://doi.org/10.1038/162680a0}{\emph{Nature} {\bfseries 162} (1948) 680}.

\bibitem{ParticleDataGroup:2024cfk}
{\scshape Particle Data Group} collaboration, \emph{{Review of particle physics}}, \href{https://doi.org/10.1103/PhysRevD.110.030001}{\emph{Phys. Rev. D} {\bfseries 110} (2024) 030001}.

\bibitem{Alpher:PhysRev73803}
R.~A. Alpher, H.~Bethe and G.~Gamow, \emph{The origin of chemical elements}, \href{https://doi.org/10.1103/PhysRev.73.803}{\emph{Phys. Rev.} {\bfseries 73} (1948) 803}.

\bibitem{Peebles:1965ApJ}
R.~H. Dicke, P.~J.~E. Peebles, P.~G. Roll and D.~T. Wilkinson, \emph{Cosmic black-body radiation}, \href{https://doi.org/10.1086/148306}{\emph{Astrophys.\ J.} {\bfseries 142} (1965) 414}.

\bibitem{Liddle:2015int}
A.~Liddle, \emph{An Introduction to Modern Cosmology}. Wiley, Chichester, 3rd~ed., 2015.

\bibitem{Guth:1980zm}
A.~H. Guth, \emph{{The Inflationary Universe: A Possible Solution to the Horizon and Flatness Problems}}, \href{https://doi.org/10.1103/PhysRevD.23.347}{\emph{Phys. Rev.} {\bfseries D23} (1981) 347}.

\bibitem{Linde:1981mu}
A.~D. Linde, \emph{{A New Inflationary Universe Scenario: A Possible Solution of the Horizon, Flatness, Homogeneity, Isotropy and Primordial Monopole Problems}}, \href{https://doi.org/10.1016/0370-2693(82)91219-9}{\emph{Phys. Lett.} {\bfseries 108B} (1982) 389}.

\bibitem{Milton:2006cp}
K.~A. Milton, \emph{{Theoretical and experimental status of magnetic monopoles}}, \href{https://doi.org/10.1088/0034-4885/69/6/R02}{\emph{Rept. Prog. Phys.} {\bfseries 69} (2006) 1637} [\href{https://arxiv.org/abs/hep-ex/0602040}{{\ttfamily hep-ex/0602040}}].

\bibitem{ParticleDataGroup:2022pth}
{\scshape Particle Data Group} collaboration, \emph{{Review of Particle Physics}}, \href{https://doi.org/10.1093/ptep/ptac097}{\emph{PTEP} {\bfseries 2022} (2022) 083C01}.

\bibitem{Linde:1983psb}
A.~D. Linde, \emph{{Primordial Inflation Without Primordial Monopoles}}, \href{https://doi.org/10.1016/0370-2693(83)90316-7}{\emph{Phys. Lett. B} {\bfseries 132} (1983) 317}.

\bibitem{Liddle:2003as}
A.~R. Liddle and S.~M. Leach, \emph{{How long before the end of inflation were observable perturbations produced?}}, \href{https://doi.org/10.1103/PhysRevD.68.103503}{\emph{Phys. Rev.} {\bfseries D68} (2003) 103503} [\href{https://arxiv.org/abs/astro-ph/0305263}{{\ttfamily astro-ph/0305263}}].

\bibitem{Dodelson:2003vq}
S.~Dodelson and L.~Hui, \emph{{A Horizon ratio bound for inflationary fluctuations}}, \href{https://doi.org/10.1103/PhysRevLett.91.131301}{\emph{Phys. Rev. Lett.} {\bfseries 91} (2003) 131301} [\href{https://arxiv.org/abs/astro-ph/0305113}{{\ttfamily astro-ph/0305113}}].

\bibitem{Ryden:1970vsj}
B.~Ryden, \emph{{Introduction to cosmology}}. Cambridge University Press, Cambridge, 2nd~ed., 2016, \href{https://doi.org/10.1017/9781316651087}{DOI}.

\bibitem{Marzouk:2021tsz}
K.~Marzouk, A.~Maraio and D.~Seery, \emph{{Non-Gaussianity in D3-brane inflation}}, \href{https://doi.org/10.1088/1475-7516/2022/02/013}{\emph{JCAP} {\bfseries 02} (2022) 013} [\href{https://arxiv.org/abs/2105.03637}{{\ttfamily 2105.03637}}].

\bibitem{Torrado:2017qtr}
J.~Torrado, C.~T. Byrnes, R.~J. Hardwick, V.~Vennin and D.~Wands, \emph{{Measuring the duration of inflation with the curvaton}}, \href{https://doi.org/10.1103/PhysRevD.98.063525}{\emph{Phys. Rev.} {\bfseries D98} (2018) 063525} [\href{https://arxiv.org/abs/1712.05364}{{\ttfamily 1712.05364}}].

\bibitem{Peacock_1998}
J.~A. Peacock, \emph{Cosmological Physics}. Cambridge University Press, 1998.

\bibitem{Earnshaw:2024zbv}
Z.~Earnshaw, \emph{{Measurement of the one lepton final state of $t\bar{t}Z$ with $Z \to \nu\bar{\nu}$ with the ATLAS detector}}, PhD thesis, University of Sussex, 2024.
\newblock \href{https://doi.org/https://hdl.handle.net/10779/uos.28045418.v1}{DOI}.

\bibitem{Kofman:1994rk}
L.~Kofman, A.~D. Linde and A.~A. Starobinsky, \emph{{Reheating after inflation}}, \href{https://doi.org/10.1103/PhysRevLett.73.3195}{\emph{Phys. Rev. Lett.} {\bfseries 73} (1994) 3195} [\href{https://arxiv.org/abs/hep-th/9405187}{{\ttfamily hep-th/9405187}}].

\bibitem{Lyth:2009zz}
D.~H. Lyth and A.~R. Liddle, \emph{{The Primordial Density Perturbation: Cosmology, Inflation and the Origin Of Structure}}. Cambridge Univ. Pr., Cambridge, 2009.

\bibitem{Liddle:2000cg}
A.~R. Liddle and D.~H. Lyth, \emph{{Cosmological inflation and large scale structure}}. Cambridge Univ. Pr., Cambridge, 2000, \href{https://doi.org/10.1017/CBO9781139175180}{DOI}.

\bibitem{griffiths2008}
D.~Griffiths, \emph{Introduction to Elementary Particles}. Wiley, 2nd~ed., 2008, \href{https://doi.org/10.1002/9783527618460}{DOI}.

\bibitem{Higgs:1964pj}
P.~W. Higgs, \emph{{Broken Symmetries and the Masses of Gauge Bosons}}, \href{https://doi.org/10.1103/PhysRevLett.13.508}{\emph{Phys. Rev. Lett.} {\bfseries 13} (1964) 508}.

\bibitem{Englert:1964et}
F.~Englert and R.~Brout, \emph{{Broken Symmetry and the Mass of Gauge Vector Mesons}}, \href{https://doi.org/10.1103/PhysRevLett.13.321}{\emph{Phys. Rev. Lett.} {\bfseries 13} (1964) 321}.

\bibitem{Djouadi:2024has}
A.~Djouadi and J.~I. Illana, \emph{{Steven Weinberg and Higgs physics}}, \href{https://doi.org/10.1016/j.nuclphysb.2024.116541}{\emph{Nucl. Phys. B} {\bfseries 1004} (2024) 116541} [\href{https://arxiv.org/abs/2401.07838}{{\ttfamily 2401.07838}}].

\bibitem{ATLAS:2012tfa}
{\scshape ATLAS} collaboration, \emph{{Observation of a new particle in the search for the Standard Model Higgs boson with the ATLAS detector at the LHC}}, \href{https://doi.org/10.1016/j.physletb.2012.08.020}{\emph{Phys. Lett.} {\bfseries B716} (2012) 1} [\href{https://arxiv.org/abs/1207.7214}{{\ttfamily 1207.7214}}].

\bibitem{CMS:2012xdj}
{\scshape CMS} collaboration, \emph{{Observation of a New Boson at a Mass of 125 GeV with the CMS Experiment at the LHC}}, \href{https://doi.org/10.1016/j.physletb.2012.08.021}{\emph{Phys. Lett.} {\bfseries B716} (2012) 30} [\href{https://arxiv.org/abs/1207.7235}{{\ttfamily 1207.7235}}].

\bibitem{Weinberg:1995mt}
S.~Weinberg, \emph{{The Quantum theory of fields. Vol. 1: Foundations}}. Cambridge Univ. Pr., Cambridge, 2005, \href{https://doi.org/10.1017/CBO9781139644167}{DOI}.

\bibitem{Dodelson:2021ft}
S.~Dodelson and F.~Schmidt, \emph{{Modern Cosmology}}. Academic Press, Amsterdam, 2nd~ed., 2021, \href{https://doi.org/https://doi.org/10.1016/C2017-0-01943-2}{DOI}.

\bibitem{Byrnes:2014pja}
C.~T. Byrnes, \emph{{Lecture notes on non-Gaussianity}}, \href{https://doi.org/10.1007/978-3-319-44769-8_3}{\emph{Astrophys. Space Sci. Proc.} {\bfseries 45} (2016) 135} [\href{https://arxiv.org/abs/1411.7002}{{\ttfamily 1411.7002}}].

\bibitem{Wands:2010af}
D.~Wands, \emph{{Local non-Gaussianity from inflation}}, \href{https://doi.org/10.1088/0264-9381/27/12/124002}{\emph{Class. Quant. Grav.} {\bfseries 27} (2010) 124002} [\href{https://arxiv.org/abs/1004.0818}{{\ttfamily 1004.0818}}].

\bibitem{Lyth:2005fi}
D.~H. Lyth and Y.~Rodriguez, \emph{{The Inflationary prediction for primordial non-Gaussianity}}, \href{https://doi.org/10.1103/PhysRevLett.95.121302}{\emph{Phys. Rev. Lett.} {\bfseries 95} (2005) 121302} [\href{https://arxiv.org/abs/astro-ph/0504045}{{\ttfamily astro-ph/0504045}}].

\bibitem{Bartolo:2004if}
N.~Bartolo, E.~Komatsu, S.~Matarrese and A.~Riotto, \emph{{Non-Gaussianity from inflation: Theory and observations}}, \href{https://doi.org/10.1016/j.physrep.2004.08.022}{\emph{Phys. Rept.} {\bfseries 402} (2004) 103} [\href{https://arxiv.org/abs/astro-ph/0406398}{{\ttfamily astro-ph/0406398}}].

\bibitem{Meszaros:1974tb}
P.~Meszaros, \emph{{The behaviour of point masses in an expanding cosmological substratum}}, {\emph{Astron. Astrophys.} {\bfseries 37} (1974) 225}.

\bibitem{Amendola:2015ksp}
L.~Amendola and S.~Tsujikawa, \emph{{Dark Energy}: {Theory and Observations}}. Cambridge University Press, 1, 2015, \href{https://doi.org/10.1017/CBO9780511750823}{DOI}.

\bibitem{Schaefer:2018jwu}
B.~M. Schaefer, \emph{{Formation of the First Black holes: Formation and evolution of the cosmic large-scale structure}},  \href{https://arxiv.org/abs/1807.06269}{{\ttfamily 1807.06269}}.

\bibitem{Prat:2025ucy}
J.~Prat and D.~Bacon, \emph{{Weak Gravitational Lensing}},  \href{https://arxiv.org/abs/2501.07938}{{\ttfamily 2501.07938}}.

\bibitem{Tkachev:2017fdu}
I.~Tkachev, \emph{{Cosmology and Dark Matter}},  in \emph{{2016 European School of High-Energy Physics}}, vol.~5, pp.~259--294, 2017, [\href{https://arxiv.org/abs/1802.02414}{{\ttfamily 1802.02414}}], \href{https://doi.org/10.23730/CYRSP-2017-005.259}{DOI}.

\bibitem{Bernardeau:2001qr}
F.~Bernardeau, S.~Colombi, E.~Gaztanaga and R.~Scoccimarro, \emph{{Large scale structure of the universe and cosmological perturbation theory}}, \href{https://doi.org/10.1016/S0370-1573(02)00135-7}{\emph{Phys. Rept.} {\bfseries 367} (2002) 1} [\href{https://arxiv.org/abs/astro-ph/0112551}{{\ttfamily astro-ph/0112551}}].

\bibitem{Bahr-Kalus:2023ebd}
B.~Bahr-Kalus, D.~Parkinson and E.-M. Mueller, \emph{{Measurement of the matter-radiation equality scale using the extended baryon oscillation spectroscopic survey quasar sample}}, \href{https://doi.org/10.1093/mnras/stad1867}{\emph{Mon. Not. Roy. Astron. Soc.} {\bfseries 524} (2023) 2463} [\href{https://arxiv.org/abs/2302.07484}{{\ttfamily 2302.07484}}].

\bibitem{Peebles:1974aaa}
P.~J.~E. {Peebles}, \emph{{The Gravitational-Instability Picture and the Nature of the Distribution of Galaxies}}, \href{https://doi.org/10.1086/181462}{\emph{Astrophys. J.} {\bfseries 189} (1974) L51}.

\bibitem{Peebles:1980abc}
P.~J.~E. Peebles, \emph{The Large-Scale Structure of the Universe}. Princeton University Press, Princeton, 1980, \href{https://doi.org/10.23943/princeton/9780691209838.001.0001}{DOI}.

\bibitem{Seljak:2000abc}
U.~Seljak, \emph{{Analytic model for galaxy and dark matter clustering}}, \href{https://doi.org/10.1046/j.1365-8711.2000.03715.x}{\emph{Mon. Not. Roy. Astron. Soc.} {\bfseries 318} (2000) 203} [\href{https://arxiv.org/abs/astro-ph/0001493}{{\ttfamily astro-ph/0001493}}].

\bibitem{Peacock:2000qk}
J.~A. Peacock and R.~E. Smith, \emph{{Halo occupation numbers and galaxy bias}}, \href{https://doi.org/10.1046/j.1365-8711.2000.03779.x}{\emph{Mon. Not. Roy. Astron. Soc.} {\bfseries 318} (2000) 1144} [\href{https://arxiv.org/abs/astro-ph/0005010}{{\ttfamily astro-ph/0005010}}].

\bibitem{Ma:2000ik}
C.-P. Ma and J.~N. Fry, \emph{{Deriving the nonlinear cosmological power spectrum and bispectrum from analytic dark matter halo profiles and mass functions}}, \href{https://doi.org/10.1086/317146}{\emph{Astrophys. J.} {\bfseries 543} (2000) 503} [\href{https://arxiv.org/abs/astro-ph/0003343}{{\ttfamily astro-ph/0003343}}].

\bibitem{Asgari:2023mej}
M.~Asgari, A.~J. Mead and C.~Heymans, \emph{{The halo model for cosmology: a pedagogical review}}, \href{https://doi.org/10.21105/astro.2303.08752}{\emph{The Open Journal of Astrophysics} {\bfseries 6} (2023) } [\href{https://arxiv.org/abs/2303.08752}{{\ttfamily 2303.08752}}].

\bibitem{Smith:2002dz}
{\scshape VIRGO Consortium} collaboration, \emph{{Stable clustering, the halo model and nonlinear cosmological power spectra}}, \href{https://doi.org/10.1046/j.1365-8711.2003.06503.x}{\emph{Mon. Not. Roy. Astron. Soc.} {\bfseries 341} (2003) 1311} [\href{https://arxiv.org/abs/astro-ph/0207664}{{\ttfamily astro-ph/0207664}}].

\bibitem{Takahashi:2012em}
R.~Takahashi, M.~Sato, T.~Nishimichi, A.~Taruya and M.~Oguri, \emph{{Revising the Halofit Model for the Nonlinear Matter Power Spectrum}}, \href{https://doi.org/10.1088/0004-637X/761/2/152}{\emph{Astrophys. J.} {\bfseries 761} (2012) 152} [\href{https://arxiv.org/abs/1208.2701}{{\ttfamily 1208.2701}}].

\bibitem{Mead:2015yca}
A.~Mead, J.~Peacock, C.~Heymans, S.~Joudaki and A.~Heavens, \emph{{An accurate halo model for fitting non-linear cosmological power spectra and baryonic feedback models}}, \href{https://doi.org/10.1093/mnras/stv2036}{\emph{Mon. Not. Roy. Astron. Soc.} {\bfseries 454} (2015) 1958} [\href{https://arxiv.org/abs/1505.07833}{{\ttfamily 1505.07833}}].

\bibitem{Mead:2016zqy}
A.~Mead, C.~Heymans, L.~Lombriser, J.~Peacock, O.~Steele and H.~Winther, \emph{{Accurate halo-model matter power spectra with dark energy, massive neutrinos and modified gravitational forces}}, \href{https://doi.org/10.1093/mnras/stw681}{\emph{Mon. Not. Roy. Astron. Soc.} {\bfseries 459} (2016) 1468} [\href{https://arxiv.org/abs/1602.02154}{{\ttfamily 1602.02154}}].

\bibitem{Mead:2020vgs}
A.~Mead, S.~Brieden, T.~Tr\"oster and C.~Heymans, \emph{{HMcode-2020: Improved modelling of non-linear cosmological power spectra with baryonic feedback}}, \href{https://doi.org/10.1093/mnras/stab082}{\emph{Mon. Not. Roy. Astron. Soc.} {\bfseries 502} (2021) 1401} [\href{https://arxiv.org/abs/2009.01858}{{\ttfamily 2009.01858}}].

\bibitem{EHT:2024a}
{\scshape The Event Horizon Telescope} collaboration, \emph{{First Sagittarius A\!* Event Horizon Telescope Results. VII. Polarization of the Ring}}, \href{https://doi.org/10.3847/2041-8213/ad2df0}{\emph{The Astrophysical Journal Letters} {\bfseries 964} (2024) L25}.

\bibitem{EHT:2024b}
{\scshape The Event Horizon Telescope} collaboration, \emph{{First Sagittarius A\!* Event Horizon Telescope Results. VIII. Physical Interpretation of the Polarized Ring}}, \href{https://doi.org/10.3847/2041-8213/ad2df1}{\emph{The Astrophysical Journal Letters} {\bfseries 964} (2024) L26}.

\bibitem{Balick:1974ApJ}
B.~{Balick} and R.~L. {Brown}, \emph{{Intense sub-arcsecond structure in the galactic center}}, \href{https://doi.org/10.1086/153242}{\emph{Astrophys. J.} {\bfseries 194} (1974) 265}.

\bibitem{Hawking:1988qt}
S.~W. Hawking, \emph{{A Brief History of Time}: {From the Big Bang to Black Holes}}. Bantam Press, London, 1988.

\bibitem{Bambi:2019xzp}
C.~Bambi, \emph{{Astrophysical Black Holes: A Review}}, \href{https://doi.org/10.22323/1.362.0028}{\emph{Proc. of Sci.} {\bfseries 362} (2020) 028} [\href{https://arxiv.org/abs/1906.03871}{{\ttfamily 1906.03871}}].

\bibitem{LIGOScientific:2016aoc}
{\scshape LIGO Scientific, Virgo} collaboration, \emph{{Observation of Gravitational Waves from a Binary Black Hole Merger}}, \href{https://doi.org/10.1103/PhysRevLett.116.061102}{\emph{Phys. Rev. Lett.} {\bfseries 116} (2016) 061102} [\href{https://arxiv.org/abs/1602.03837}{{\ttfamily 1602.03837}}].

\bibitem{KAGRA:2021vkt}
{\scshape KAGRA, Virgo, LIGO Scientific} collaboration, \emph{{GWTC-3: Compact Binary Coalescences Observed by LIGO and Virgo during the Second Part of the Third Observing Run}}, \href{https://doi.org/10.1103/PhysRevX.13.041039}{\emph{Phys. Rev. X} {\bfseries 13} (2023) 041039} [\href{https://arxiv.org/abs/2111.03606}{{\ttfamily 2111.03606}}].

\bibitem{GRAVITY:2023avo}
{\scshape GRAVITY} collaboration, \emph{{Polarimetry and astrometry of NIR flares as event horizon scale, dynamical probes for the mass of Sgr A\!*}}, \href{https://doi.org/10.1051/0004-6361/202347416}{\emph{Astron. Astrophys.} {\bfseries 677} (2023) L10} [\href{https://arxiv.org/abs/2307.11821}{{\ttfamily 2307.11821}}].

\bibitem{EventHorizonTelescope:2019ggy}
{\scshape Event Horizon Telescope} collaboration, \emph{{First M87 Event Horizon Telescope Results. VI. The Shadow and Mass of the Central Black Hole}}, \href{https://doi.org/10.3847/2041-8213/ab1141}{\emph{Astrophys. J. Lett.} {\bfseries 875} (2019) L6} [\href{https://arxiv.org/abs/1906.11243}{{\ttfamily 1906.11243}}].

\bibitem{Wu_2015}
X.-B. Wu, F.~Wang, X.~Fan, W.~Yi, W.~Zuo, F.~Bian et~al., \emph{An ultraluminous quasar with a twelve-billion-solar-mass black hole at redshift 6.30}, \href{https://doi.org/10.1038/nature14241}{\emph{Nature} {\bfseries 518} (2015) 512} [\href{https://arxiv.org/abs/1502.07418}{{\ttfamily 1502.07418}}].

\bibitem{Sorce:2022sgz}
{\scshape Euclid} collaboration, \emph{{Euclid Legacy Science prospects}}, \href{https://doi.org/10.22323/1.414.0096}{\emph{PoS} {\bfseries ICHEP2022} (2022) 096} [\href{https://arxiv.org/abs/2211.11709}{{\ttfamily 2211.11709}}].

\bibitem{Bosman:2024NatAs81054B}
S.~E.~I. {Bosman}, J.~{{\'A}lvarez-M{\'a}rquez}, L.~{Colina}, F.~{Walter}, A.~{Alonso-Herrero}, M.~J. {Ward} et~al., \emph{{A mature quasar at cosmic dawn revealed by JWST rest-frame infrared spectroscopy}}, \href{https://doi.org/10.1038/s41550-024-02273-0}{\emph{Nature Astronomy} {\bfseries 8} (2024) 1054} [\href{https://arxiv.org/abs/2307.14414}{{\ttfamily 2307.14414}}].

\bibitem{Yang:2023pnw}
J.~Yang et~al., \emph{{A SPectroscopic Survey of Biased Halos in the Reionization Era (ASPIRE): A First Look at the Rest-frame Optical Spectra of z \ensuremath{>} 6.5 Quasars Using JWST}}, \href{https://doi.org/10.3847/2041-8213/acc9c8}{\emph{Astrophys. J. Lett.} {\bfseries 951} (2023) L5} [\href{https://arxiv.org/abs/2304.09888}{{\ttfamily 2304.09888}}].

\bibitem{Padovani:2017zpf}
P.~Padovani et~al., \emph{{Active galactic nuclei: what\textquoteright{}s in a name?}}, \href{https://doi.org/10.1007/s00159-017-0102-9}{\emph{Astron. Astrophys. Rev.} {\bfseries 25} (2017) 2} [\href{https://arxiv.org/abs/1707.07134}{{\ttfamily 1707.07134}}].

\bibitem{Gow:2024yba}
A.~D. Gow, P.~Clark and D.~Rycanowski, \emph{{I'm in AGNi: A new standard for AGN pluralisation}},  \href{https://arxiv.org/abs/2403.20302}{{\ttfamily 2403.20302}}.

\bibitem{Levine:2006ApJ57L}
R.~{Levine} and N.~Y. {Gnedin}, \emph{{Active Galactic Nucleus Outflows and the Matter Power Spectrum}}, \href{https://doi.org/10.1086/508370}{\emph{The Astrophysical Journal Letters} {\bfseries 649} (2006) L57} [\href{https://arxiv.org/abs/astro-ph/0604308}{{\ttfamily astro-ph/0604308}}].

\bibitem{vanDaalen:2011xb}
M.~P. van Daalen, J.~Schaye, C.~M. Booth and C.~D. Vecchia, \emph{{The effects of galaxy formation on the matter power spectrum: A challenge for precision cosmology}}, \href{https://doi.org/10.1111/j.1365-2966.2011.18981.x}{\emph{Mon. Not. Roy. Astron. Soc.} {\bfseries 415} (2011) 3649} [\href{https://arxiv.org/abs/1104.1174}{{\ttfamily 1104.1174}}].

\bibitem{McNamara:2007ww}
B.~R. McNamara and P.~E.~J. Nulsen, \emph{{Heating Hot Atmospheres with Active Galactic Nuclei}}, \href{https://doi.org/10.1146/annurev.astro.45.051806.110625}{\emph{Ann. Rev. Astron. Astrophys.} {\bfseries 45} (2007) 117} [\href{https://arxiv.org/abs/0709.2152}{{\ttfamily 0709.2152}}].

\bibitem{Fabian:2012xr}
A.~C. Fabian, \emph{{Observational Evidence of AGN Feedback}}, \href{https://doi.org/10.1146/annurev-astro-081811-125521}{\emph{Ann. Rev. Astron. Astrophys.} {\bfseries 50} (2012) 455} [\href{https://arxiv.org/abs/1204.4114}{{\ttfamily 1204.4114}}].

\bibitem{Croton:2005hbr}
D.~J. Croton, V.~Springel, S.~D.~M. White, G.~De~Lucia, C.~S. Frenk, L.~Gao et~al., \emph{{The Many lives of AGN: Cooling flows, black holes and the luminosities and colours of galaxies}}, \href{https://doi.org/10.1111/j.1365-2966.2006.09994.x}{\emph{Mon. Not. Roy. Astron. Soc.} {\bfseries 365} (2006) 11} [\href{https://arxiv.org/abs/astro-ph/0602065}{{\ttfamily astro-ph/0602065}}].

\bibitem{Sunyaev:1980ARA&A18537S}
R.~A. {Sunyaev} and I.~B. {Zeldovich}, \emph{{Microwave background radiation as a probe of the contemporary structure and history of the universe}}, \href{https://doi.org/10.1146/annurev.aa.18.090180.002541}{\emph{Annual Review of Astron and Astrophys} {\bfseries 18} (1980) 537}.

\bibitem{LeBrun:2015sgq}
A.~M.~C. Le~Brun, I.~G. McCarthy and J.-B. Melin, \emph{{Testing Sunyaev\textendash{}Zel'dovich measurements of the hot gas content of dark matter haloes using synthetic skies}}, \href{https://doi.org/10.1093/mnras/stv1172}{\emph{Mon. Not. Roy. Astron. Soc.} {\bfseries 451} (2015) 3868} [\href{https://arxiv.org/abs/1501.05666}{{\ttfamily 1501.05666}}].

\bibitem{Amodeo:2020mmu}
S.~Amodeo et~al., \emph{{Atacama Cosmology Telescope: Modeling the gas thermodynamics in BOSS CMASS galaxies from kinematic and thermal Sunyaev-Zel\textquoteright{}dovich measurements}}, \href{https://doi.org/10.1103/PhysRevD.103.063514}{\emph{Phys. Rev. D} {\bfseries 103} (2021) 063514} [\href{https://arxiv.org/abs/2009.05558}{{\ttfamily 2009.05558}}].

\bibitem{Mohammed:2014lja}
I.~Mohammed and U.~Seljak, \emph{{Analytic model for the matter power spectrum, its covariance matrix, and baryonic effects}}, \href{https://doi.org/10.1093/mnras/stu1972}{\emph{Mon. Not. Roy. Astron. Soc.} {\bfseries 445} (2014) 3382} [\href{https://arxiv.org/abs/1407.0060}{{\ttfamily 1407.0060}}].

\bibitem{Li:stu1571}
S.~Li, A.~Frank and E.~G. Blackman, \emph{{Triggered star formation and its consequences}}, \href{https://doi.org/10.1093/mnras/stu1571}{\emph{Mon. Not. Roy. Astron. Soc.} {\bfseries 444} (2014) 2884} [\href{https://arxiv.org/abs/1404.5924}{{\ttfamily 1404.5924}}].

\bibitem{Gallart:2021ApJ909192G}
C.~{Gallart}, M.~{Monelli}, T.~{Ruiz-Lara}, A.~{Calamida}, S.~{Cassisi}, M.~{Cignoni} et~al., \emph{{The Star Formation History of Eridanus II: On the Role of Supernova Feedback in the Quenching of Ultrafaint Dwarf Galaxies}}, \href{https://doi.org/10.3847/1538-4357/abddbe}{\emph{Astrophy. J.} {\bfseries 909} (2021) 192} [\href{https://arxiv.org/abs/2101.04464}{{\ttfamily 2101.04464}}].

\bibitem{Dashyan:2020qnw}
G.~Dashyan and Y.~Dubois, \emph{{Cosmic ray feedback from supernovae in dwarf galaxies}}, \href{https://doi.org/10.1051/0004-6361/201936339}{\emph{Astron. Astrophys.} {\bfseries 638} (2020) A123} [\href{https://arxiv.org/abs/2003.09900}{{\ttfamily 2003.09900}}].

\bibitem{Strait:2023ApJ949L23S}
V.~{Strait}, G.~{Brammer}, A.~{Muzzin}, G.~{Desprez}, Y.~{Asada}, R.~{Abraham} et~al., \emph{{An Extremely Compact, Low-mass Galaxy on its Way to Quiescence at z = 5.2}}, \href{https://doi.org/10.3847/2041-8213/acd457}{\emph{Astrophys. J. Lett.} {\bfseries 949} (2023) L23} [\href{https://arxiv.org/abs/2303.11349}{{\ttfamily 2303.11349}}].

\bibitem{Jones:2024MNRAS5351293J}
E.~{Jones}, B.~{Smith}, R.~{Dav{\'e}}, D.~{Narayanan} and Q.~{Li}, \emph{{SIMBA-EOR: early galaxy formation in the SIMBA simulation including a new sub-grid interstellar medium model}}, \href{https://doi.org/10.1093/mnras/stae2445}{\emph{Mon. Not. Roy. Astron. Soc.} {\bfseries 535} (2024) 1293} [\href{https://arxiv.org/abs/2402.06728}{{\ttfamily 2402.06728}}].

\bibitem{NFW:1997ApJ493N}
J.~F. {Navarro}, C.~S. {Frenk} and S.~D.~M. {White}, \emph{{A Universal Density Profile from Hierarchical Clustering}}, \href{https://doi.org/10.1086/304888}{\emph{Astrophys. J.} {\bfseries 490} (1997) 493} [\href{https://arxiv.org/abs/astro-ph/9611107}{{\ttfamily astro-ph/9611107}}].

\bibitem{Cappellaro:2022xen}
E.~Cappellaro, \emph{{Supernovae and their cosmological implications}}, \href{https://doi.org/10.1007/s40766-022-00034-1}{\emph{Riv. Nuovo Cim.} {\bfseries 45} (2022) 549}.

\bibitem{Zhao:2006ChJAA}
F.~Y. {Zhao}, R.~G. {Strom} and S.~Y. {Jiang}, \emph{{The Guest Star of AD185 must have been a Supernova}}, \href{https://doi.org/10.1088/1009-9271/6/5/17}{\emph{Chinese J. Astron. Astrophys.} {\bfseries 6} (2006) 635}.

\bibitem{Branch:1992rv}
D.~Branch and G.~A. Tammann, \emph{{Type ia supernovae as standard candles}}, \href{https://doi.org/10.1146/annurev.aa.30.090192.002043}{\emph{Ann. Rev. Astron. Astrophys.} {\bfseries 30} (1992) 359}.

\bibitem{DES:2018ioi}
{\scshape DES} collaboration, \emph{{First Cosmology Results Using Type Ia Supernovae From the Dark Energy Survey: Analysis, Systematic Uncertainties, and Validation}}, \href{https://doi.org/10.3847/1538-4357/ab08a0}{\emph{Astrophys. J.} {\bfseries 874} (2019) 150} [\href{https://arxiv.org/abs/1811.02377}{{\ttfamily 1811.02377}}].

\bibitem{DES:2018paw}
{\scshape DES} collaboration, \emph{{First Cosmology Results using Type Ia Supernovae from the Dark Energy Survey: Constraints on Cosmological Parameters}}, \href{https://doi.org/10.3847/2041-8213/ab04fa}{\emph{Astrophys. J. Lett.} {\bfseries 872} (2019) L30} [\href{https://arxiv.org/abs/1811.02374}{{\ttfamily 1811.02374}}].

\bibitem{Hicken:2009ApJ331H}
M.~{Hicken}, P.~{Challis}, S.~{Jha}, R.~P. {Kirshner}, T.~{Matheson}, M.~{Modjaz} et~al., \emph{{CfA3: 185 Type Ia Supernova Light Curves from the CfA}}, \href{https://doi.org/10.1088/0004-637X/700/1/331}{\emph{Astrophys. J.} {\bfseries 700} (2009) 331} [\href{https://arxiv.org/abs/0901.4787}{{\ttfamily 0901.4787}}].

\bibitem{Hicken:2012ApJS12H}
M.~{Hicken}, P.~{Challis}, R.~P. {Kirshner}, A.~{Rest}, C.~E. {Cramer}, W.~M. {Wood-Vasey} et~al., \emph{{CfA4: Light Curves for 94 Type Ia Supernovae}}, \href{https://doi.org/10.1088/0067-0049/200/2/12}{\emph{Astrophys. J. S.} {\bfseries 200} (2012) 12} [\href{https://arxiv.org/abs/1205.4493}{{\ttfamily 1205.4493}}].

\bibitem{Contreras:2010AJ519C}
C.~{Contreras}, M.~{Hamuy}, M.~M. {Phillips}, G.~{Folatelli}, N.~B. {Suntzeff}, S.~E. {Persson} et~al., \emph{{The Carnegie Supernova Project: First Photometry Data Release of Low-Redshift Type Ia Supernovae}}, \href{https://doi.org/10.1088/0004-6256/139/2/519}{\emph{Astronomical J.} {\bfseries 139} (2010) 519} [\href{https://arxiv.org/abs/0910.3330}{{\ttfamily 0910.3330}}].

\bibitem{Stritzinger:2011AJ156S}
M.~D. {Stritzinger}, M.~M. {Phillips}, L.~N. {Boldt}, C.~{Burns}, A.~{Campillay}, C.~{Contreras} et~al., \emph{{The Carnegie Supernova Project: Second Photometry Data Release of Low-redshift Type Ia Supernovae}}, \href{https://doi.org/10.1088/0004-6256/142/5/156}{\emph{Astronomical J.} {\bfseries 142} (2011) 156} [\href{https://arxiv.org/abs/1108.3108}{{\ttfamily 1108.3108}}].

\bibitem{COBE:1996fi}
D.~J. {Fixsen}, E.~S. {Cheng}, J.~M. {Gales}, J.~C. {Mather}, R.~A. {Shafer} and E.~L. {Wright}, \emph{{The Cosmic Microwave Background Spectrum from the Full COBE FIRAS Data Set}}, \href{https://doi.org/10.1086/178173}{\emph{Astrophys.\ J.} {\bfseries 473} (1996) 576} [\href{https://arxiv.org/abs/astro-ph/9605054}{{\ttfamily astro-ph/9605054}}].

\bibitem{Penzias:1965wn}
A.~A. Penzias and R.~W. Wilson, \emph{{A Measurement of excess antenna temperature at 4080-Mc/s}}, \href{https://doi.org/10.1086/148307}{\emph{Astrophys.\ J.} {\bfseries 142} (1965) 419}.

\bibitem{Planck:2018yye}
{\scshape Planck} collaboration, \emph{{Planck 2018 results. IV. Diffuse component separation}}, \href{https://doi.org/10.1051/0004-6361/201833881}{\emph{Astron. Astrophys.} {\bfseries 641} (2020) A4} [\href{https://arxiv.org/abs/1807.06208}{{\ttfamily 1807.06208}}].

\bibitem{Simons:2018sbj}
{\scshape Simons Observatory} collaboration, \emph{{The Simons Observatory: Science goals and forecasts}}, \href{https://doi.org/10.1088/1475-7516/2019/02/056}{\emph{JCAP} {\bfseries 02} (2019) 056} [\href{https://arxiv.org/abs/1808.07445}{{\ttfamily 1808.07445}}].

\bibitem{Simons:2019BAAS147L}
A.~{Lee}, M.~H. {Abitbol}, S.~{Adachi}, P.~{Ade}, J.~{Aguirre}, Z.~{Ahmed} et~al., \emph{{The Simons Observatory}},  in \emph{Bulletin of the American Astronomical Society}, vol.~51, p.~147, Sept., 2019, [\href{https://arxiv.org/abs/1907.08284}{{\ttfamily 1907.08284}}], \href{https://doi.org/10.48550/arXiv.1907.08284}{DOI}.

\bibitem{CMB-S4:2019BAAS}
J.~{Carlstrom}, K.~{Abazajian}, G.~{Addison}, P.~{Adshead}, Z.~{Ahmed}, S.~W. {Allen} et~al., \emph{{CMB-S4}},  in \emph{Bulletin of the American Astronomical Society}, vol.~51, p.~209, Sept., 2019, [\href{https://arxiv.org/abs/1908.01062}{{\ttfamily 1908.01062}}], \href{https://doi.org/10.48550/arXiv.1908.01062}{DOI}.

\bibitem{CMB-S4:2022ght}
{\scshape CMB-S4} collaboration, \emph{{Snowmass 2021 CMB-S4 White Paper}},  \href{https://arxiv.org/abs/2203.08024}{{\ttfamily 2203.08024}}.

\bibitem{Fergusson:2006pr}
J.~R. Fergusson and E.~P.~S. Shellard, \emph{{Primordial non-Gaussianity and the CMB bispectrum}}, \href{https://doi.org/10.1103/PhysRevD.76.083523}{\emph{Phys. Rev. D} {\bfseries 76} (2007) 083523} [\href{https://arxiv.org/abs/astro-ph/0612713}{{\ttfamily astro-ph/0612713}}].

\bibitem{DESI:2023bgx}
{\scshape DESI} collaboration, \emph{{First detection of the BAO signal from early DESI data}}, \href{https://doi.org/10.1093/mnras/stad2618}{\emph{Mon. Not. Roy. Astron. Soc.} {\bfseries 525} (2023) 5406} [\href{https://arxiv.org/abs/2304.08427}{{\ttfamily 2304.08427}}].

\bibitem{Coles:2019abc}
P.~Coles, \emph{{Einstein, Eddington and the 1919 eclipse}}, \href{https://doi.org/10.1038/d41586-019-01172-z}{\emph{Nature} {\bfseries 568} (2019) 306}.

\bibitem{Smail:1994sx}
I.~Smail, R.~S. Ellis and M.~J. Fitchett, \emph{{Gravitational lensing of distant field galaxies by rich clusters: I. - faint galaxy redshift distributions}}, \href{https://doi.org/10.1093/mnras/270.2.245}{\emph{Mon. Not. Roy. Astron. Soc.} {\bfseries 270} (1994) 245} [\href{https://arxiv.org/abs/astro-ph/9402048}{{\ttfamily astro-ph/9402048}}].

\bibitem{Euclid:2019clj}
{\scshape Euclid} collaboration, \emph{{Euclid preparation: VII. Forecast validation for Euclid cosmological probes}}, \href{https://doi.org/10.1051/0004-6361/202038071}{\emph{Astron. Astrophys.} {\bfseries 642} (2020) A191} [\href{https://arxiv.org/abs/1910.09273}{{\ttfamily 1910.09273}}].

\bibitem{Euclid:2011zbd}
{\scshape EUCLID} collaboration, \emph{{Euclid Definition Study Report}},  \href{https://arxiv.org/abs/1110.3193}{{\ttfamily 1110.3193}}.

\bibitem{Hubble:1926ApJ}
E.~P. {Hubble}, \emph{{Extragalactic nebulae}}, \href{https://doi.org/10.1086/143018}{\emph{Astrophys. J.} {\bfseries 64} (1926) 321}.

\bibitem{Hu:1999ek}
W.~Hu, \emph{{Power spectrum tomography with weak lensing}}, \href{https://doi.org/10.1086/312210}{\emph{Astrophys. J. Lett.} {\bfseries 522} (1999) L21} [\href{https://arxiv.org/abs/astro-ph/9904153}{{\ttfamily astro-ph/9904153}}].

\bibitem{Baker:2025NatAs141B}
W.~M. {Baker}, S.~{Tacchella}, B.~D. {Johnson}, E.~{Nelson}, K.~A. {Suess}, F.~{D'Eugenio} et~al., \emph{{A core in a star-forming disc as evidence of inside-out growth in the early Universe}}, \href{https://doi.org/10.1038/s41550-024-02384-8}{\emph{Nature Astronomy} {\bfseries 9} (2025) 141} [\href{https://arxiv.org/abs/2306.02472}{{\ttfamily 2306.02472}}].

\bibitem{NIRSpec:2022A&A80J}
P.~{Jakobsen}, P.~{Ferruit}, C.~{Alves de Oliveira}, S.~{Arribas}, G.~{Bagnasco}, R.~{Barho} et~al., \emph{{The Near-Infrared Spectrograph (NIRSpec) on the James Webb Space Telescope. I. Overview of the instrument and its capabilities}}, \href{https://doi.org/10.1051/0004-6361/202142663}{\emph{Astron. Astrophys.} {\bfseries 661} (2022) A80} [\href{https://arxiv.org/abs/2202.03305}{{\ttfamily 2202.03305}}].

\bibitem{Bolzonella:2000js}
M.~{Bolzonella}, J.~M. {Miralles} and R.~{Pell{\'o}}, \emph{{Photometric redshifts based on standard SED fitting procedures}}, \href{https://doi.org/10.48550/arXiv.astro-ph/0003380}{\emph{Astron. Astrophys.} {\bfseries 363} (2000) 476} [\href{https://arxiv.org/abs/astro-ph/0003380}{{\ttfamily astro-ph/0003380}}].

\bibitem{Euclid:2024yrr}
{\scshape Euclid} collaboration, \emph{{Euclid. I. Overview of the Euclid mission}},  \href{https://arxiv.org/abs/2405.13491}{{\ttfamily 2405.13491}}.

\bibitem{LSST:2008ijt}
{\scshape LSST} collaboration, \emph{{LSST: from Science Drivers to Reference Design and Anticipated Data Products}}, \href{https://doi.org/10.3847/1538-4357/ab042c}{\emph{Astrophys. J.} {\bfseries 873} (2019) 111} [\href{https://arxiv.org/abs/0805.2366}{{\ttfamily 0805.2366}}].

\bibitem{Carnall:2018MNRAS4804379C}
A.~C. {Carnall}, R.~J. {McLure}, J.~S. {Dunlop} and R.~{Dav{\'e}}, \emph{{Inferring the star formation histories of massive quiescent galaxies with BAGPIPES: evidence for multiple quenching mechanisms}}, \href{https://doi.org/10.1093/mnras/sty2169}{\emph{Mon. Not. Roy. Astron. Soc.} {\bfseries 480} (2018) 4379} [\href{https://arxiv.org/abs/1712.04452}{{\ttfamily 1712.04452}}].

\bibitem{Dalal:2023olq}
R.~Dalal et~al., \emph{{Hyper Suprime-Cam Year 3 results: Cosmology from cosmic shear power spectra}}, \href{https://doi.org/10.1103/PhysRevD.108.123519}{\emph{Phys. Rev. D} {\bfseries 108} (2023) 123519} [\href{https://arxiv.org/abs/2304.00701}{{\ttfamily 2304.00701}}].

\bibitem{Euclid:2020gbk}
{\scshape Euclid} collaboration, \emph{{Euclid preparation. X. The $Euclid$ photometric-redshift challenge}}, \href{https://doi.org/10.1051/0004-6361/202039403}{\emph{Astron. Astrophys.} {\bfseries 644} (2020) A31} [\href{https://arxiv.org/abs/2009.12112}{{\ttfamily 2009.12112}}].

\bibitem{Euclid:2022oea}
{\scshape Euclid} collaboration, \emph{{Euclid: Calibrating photometric redshifts with spectroscopic cross-correlations}}, \href{https://doi.org/10.1051/0004-6361/202244795}{\emph{Astron. Astrophys.} {\bfseries 670} (2023) A149} [\href{https://arxiv.org/abs/2208.10503}{{\ttfamily 2208.10503}}].

\bibitem{Euclid:2022vkk}
{\scshape Euclid} collaboration, \emph{{Euclid preparation. XVIII. The NISP photometric system}}, \href{https://doi.org/10.1051/0004-6361/202142897}{\emph{Astron. Astrophys.} {\bfseries 662} (2022) A92} [\href{https://arxiv.org/abs/2203.01650}{{\ttfamily 2203.01650}}].

\bibitem{Mehrem:2009ip}
R.~Mehrem, \emph{The plane wave expansion, infinite integrals and identities involving spherical bessel functions}, \href{https://doi.org/10.1016/j.amc.2010.12.004}{\emph{Appl. Math. Comput.} {\bfseries 217} (2011) 5360 } [\href{https://arxiv.org/abs/0909.0494}{{\ttfamily 0909.0494}}].

\bibitem{Limber:1954zz}
D.~Limber, \emph{{The Analysis of Counts of the Extragalactic Nebulae in Terms of a Fluctuating Density Field. II}}, \href{https://doi.org/10.1086/145870}{\emph{Astrophys. J.} {\bfseries 119} (1954) 655}.

\bibitem{Gebhardt:2017chz}
H.~S. Grasshorn~Gebhardt and D.~Jeong, \emph{{Fast and accurate computation of projected two-point functions}}, \href{https://doi.org/10.1103/PhysRevD.97.023504}{\emph{Phys. Rev.} {\bfseries D97} (2018) 023504} [\href{https://arxiv.org/abs/1709.02401}{{\ttfamily 1709.02401}}].

\bibitem{LoVerde:2008re}
M.~LoVerde and N.~Afshordi, \emph{{Extended Limber Approximation}}, \href{https://doi.org/10.1103/PhysRevD.78.123506}{\emph{Phys. Rev. D} {\bfseries 78} (2008) 123506} [\href{https://arxiv.org/abs/0809.5112}{{\ttfamily 0809.5112}}].

\bibitem{Helmholtz+1858+25+55}
H.~Helmholtz, \emph{Über integrale der hydrodynamischen gleichungen, welche den wirbelbewegungen entsprechen.}, \href{https://doi.org/doi:10.1515/crll.1858.55.25}{\emph{Journal für die reine und angewandte Mathematik} {\bfseries 1858} (1858) 25}.

\bibitem{Euclid:2025nwk}
{\scshape Euclid} collaboration, \emph{{Euclid: Optimising tomographic redshift binning for 3$\times$2pt power spectrum constraints on dark energy}},  \href{https://arxiv.org/abs/2501.07559}{{\ttfamily 2501.07559}}.

\bibitem{Hall:2021qjk}
A.~Hall, \emph{{Cosmology from weak lensing alone and implications for the Hubble tension}}, \href{https://doi.org/10.1093/mnras/stab1563}{\emph{Mon. Not. Roy. Astron. Soc.} {\bfseries 505} (2021) 4935} [\href{https://arxiv.org/abs/2104.12880}{{\ttfamily 2104.12880}}].

\bibitem{Copeland:2019bho}
D.~Copeland, A.~Taylor and A.~Hall, \emph{{Towards determining the neutrino mass hierarchy: weak lensing and galaxy clustering forecasts with baryons and intrinsic alignments}}, \href{https://doi.org/10.1093/mnras/staa314}{\emph{Mon. Not. Roy. Astron. Soc.} {\bfseries 493} (2020) 1640} [\href{https://arxiv.org/abs/1905.08754}{{\ttfamily 1905.08754}}].

\bibitem{Heymans:2013fya}
C.~Heymans et~al., \emph{{CFHTLenS tomographic weak lensing cosmological parameter constraints: Mitigating the impact of intrinsic galaxy alignments}}, \href{https://doi.org/10.1093/mnras/stt601}{\emph{Mon. Not. Roy. Astron. Soc.} {\bfseries 432} (2013) 2433} [\href{https://arxiv.org/abs/1303.1808}{{\ttfamily 1303.1808}}].

\bibitem{Lemos:2017arq}
P.~Lemos, A.~Challinor and G.~Efstathiou, \emph{{The effect of Limber and flat-sky approximations on galaxy weak lensing}}, \href{https://doi.org/10.1088/1475-7516/2017/05/014}{\emph{JCAP} {\bfseries 05} (2017) 014} [\href{https://arxiv.org/abs/1704.01054}{{\ttfamily 1704.01054}}].

\bibitem{Schneider:2010pm}
P.~Schneider, T.~Eifler and E.~Krause, \emph{{COSEBIs: Extracting the full E-/B-mode information from cosmic shear correlation functions}}, \href{https://doi.org/10.1051/0004-6361/201014235}{\emph{Astron. Astrophys.} {\bfseries 520} (2010) A116} [\href{https://arxiv.org/abs/1002.2136}{{\ttfamily 1002.2136}}].

\bibitem{Asgari:2012ir}
M.~Asgari, P.~Schneider and P.~Simon, \emph{{Cosmic Shear Tomography and Efficient Data Compression using COSEBIs}}, \href{https://doi.org/10.1051/0004-6361/201218828}{\emph{Astron. Astrophys.} {\bfseries 542} (2012) A122} [\href{https://arxiv.org/abs/1201.2669}{{\ttfamily 1201.2669}}].

\bibitem{Percival:2006ss}
W.~J. Percival and M.~L. Brown, \emph{{Likelihood methods for the combined analysis of CMB temperature and polarisation power spectra}}, \href{https://doi.org/10.1111/j.1365-2966.2006.10910.x}{\emph{Mon. Not. Roy. Astron. Soc.} {\bfseries 372} (2006) 1104} [\href{https://arxiv.org/abs/astro-ph/0604547}{{\ttfamily astro-ph/0604547}}].

\bibitem{Upham:2020klf}
R.~E. Upham, M.~L. Brown and L.~Whittaker, \emph{{Sufficiency of a Gaussian power spectrum likelihood for accurate cosmology from upcoming weak lensing surveys}}, \href{https://doi.org/10.1093/mnras/stab522}{\emph{Mon. Not. Roy. Astron. Soc.} {\bfseries 503} (2021) 1999} [\href{https://arxiv.org/abs/2012.06267}{{\ttfamily 2012.06267}}].

\bibitem{Trotta:2017wnx}
R.~Trotta, \emph{{Bayesian Methods in Cosmology}},  \href{https://arxiv.org/abs/1701.01467}{{\ttfamily 1701.01467}}.

\bibitem{Jeffreys:1946RSPSA}
H.~{Jeffreys}, \emph{{An Invariant Form for the Prior Probability in Estimation Problems}}, \href{https://doi.org/10.1098/rspa.1946.0056}{\emph{Proceedings of the Royal Society of London Series A} {\bfseries 186} (1946) 453}.

\bibitem{Hadzhiyska:2023wae}
B.~Hadzhiyska, K.~Wolz, S.~Azzoni, D.~Alonso, C.~Garc\'\i{}a-Garc\'\i{}a, J.~Ruiz-Zapatero et~al., \emph{{Cosmology with 6 parameters in the Stage-IV era: efficient marginalisation over nuisance parameters}}, \href{https://doi.org/10.21105/astro.2301.11895}{\emph{The Open Journal of Astrophysics} {\bfseries 6} (2023) 23} [\href{https://arxiv.org/abs/2301.11895}{{\ttfamily 2301.11895}}].

\bibitem{Heavens:2009nx}
A.~Heavens, \emph{{Statistical techniques in cosmology}},  \href{https://arxiv.org/abs/0906.0664}{{\ttfamily 0906.0664}}.

\bibitem{Skilling:BA127}
J.~Skilling, \emph{{Nested sampling for general Bayesian computation}}, \href{https://doi.org/10.1214/06-BA127}{\emph{Bayesian Analysis} {\bfseries 1} (2006) 833 }.

\bibitem{Ashton:2022grj}
G.~Ashton et~al., \emph{{Nested sampling for physical scientists}}, \href{https://doi.org/10.1038/s43586-022-00121-x}{\emph{Nature} {\bfseries 2} (2022) } [\href{https://arxiv.org/abs/2205.15570}{{\ttfamily 2205.15570}}].

\bibitem{Feroz:2007kg}
F.~Feroz and M.~P. Hobson, \emph{{Multimodal nested sampling: an efficient and robust alternative to MCMC methods for astronomical data analysis}}, \href{https://doi.org/10.1111/j.1365-2966.2007.12353.x}{\emph{Mon. Not. Roy. Astron. Soc.} {\bfseries 384} (2008) 449} [\href{https://arxiv.org/abs/0704.3704}{{\ttfamily 0704.3704}}].

\bibitem{Feroz:2008xx}
F.~Feroz, M.~P. Hobson and M.~Bridges, \emph{{MultiNest: an efficient and robust Bayesian inference tool for cosmology and particle physics}}, \href{https://doi.org/10.1111/j.1365-2966.2009.14548.x}{\emph{Mon. Not. Roy. Astron. Soc.} {\bfseries 398} (2009) 1601} [\href{https://arxiv.org/abs/0809.3437}{{\ttfamily 0809.3437}}].

\bibitem{Feroz:2013hea}
F.~Feroz, M.~P. Hobson, E.~Cameron and A.~N. Pettitt, \emph{{Importance Nested Sampling and the MultiNest Algorithm}}, \href{https://doi.org/10.21105/astro.1306.2144}{\emph{Open J. Astrophys.} {\bfseries 2} (2019) 10} [\href{https://arxiv.org/abs/1306.2144}{{\ttfamily 1306.2144}}].

\bibitem{Handley:2015fda}
W.~J. Handley, M.~P. Hobson and A.~N. Lasenby, \emph{{PolyChord: nested sampling for cosmology}}, \href{https://doi.org/10.1093/mnrasl/slv047}{\emph{Mon. Not. Roy. Astron. Soc.} {\bfseries 450} (2015) L61} [\href{https://arxiv.org/abs/1502.01856}{{\ttfamily 1502.01856}}].

\bibitem{Handley:2015vkr}
W.~J. Handley, M.~P. Hobson and A.~N. Lasenby, \emph{{PolyChord: next-generation nested sampling}}, \href{https://doi.org/10.1093/mnras/stv1911}{\emph{Mon. Not. Roy. Astron. Soc.} {\bfseries 453} (2015) 4385} [\href{https://arxiv.org/abs/1506.00171}{{\ttfamily 1506.00171}}].

\bibitem{Tegmark:1997rp}
M.~Tegmark, \emph{{Measuring cosmological parameters with galaxy surveys}}, \href{https://doi.org/10.1103/PhysRevLett.79.3806}{\emph{Phys. Rev. Lett.} {\bfseries 79} (1997) 3806} [\href{https://arxiv.org/abs/astro-ph/9706198}{{\ttfamily astro-ph/9706198}}].

\bibitem{Tegmark:1996bz}
M.~Tegmark, A.~Taylor and A.~Heavens, \emph{{Karhunen-Loeve eigenvalue problems in cosmology: How should we tackle large data sets?}}, \href{https://doi.org/10.1086/303939}{\emph{Astrophys. J.} {\bfseries 480} (1997) 22} [\href{https://arxiv.org/abs/astro-ph/9603021}{{\ttfamily astro-ph/9603021}}].

\bibitem{Amara:2007as}
A.~Amara and A.~Refregier, \emph{{Systematic Bias in Cosmic Shear: Beyond the Fisher Matrix}}, \href{https://doi.org/10.1111/j.1365-2966.2008.13880.x}{\emph{Mon. Not. Roy. Astron. Soc.} {\bfseries 391} (2008) 228} [\href{https://arxiv.org/abs/0710.5171}{{\ttfamily 0710.5171}}].

\bibitem{Gordon:2024jaj}
J.~Gordon, B.~F. de~Aguiar, J.~a. Rebou\c{c}as, G.~Brando, F.~Falciano, V.~Miranda et~al., \emph{{Modeling nonlinear scales with the comoving Lagrangian acceleration method: Preparing for LSST Y1}}, \href{https://doi.org/10.1103/PhysRevD.110.083529}{\emph{Phys. Rev. D} {\bfseries 110} (2024) 083529} [\href{https://arxiv.org/abs/2404.12344}{{\ttfamily 2404.12344}}].

\bibitem{Euclid:2021ilj}
{\scshape Euclid} collaboration, \emph{{Euclid: Covariance of weak lensing pseudo-C\ensuremath{\ell} estimates - Calculation, comparison to simulations, and dependence on survey geometry}}, \href{https://doi.org/10.1051/0004-6361/202142908}{\emph{Astron. Astrophys.} {\bfseries 660} (2022) A114} [\href{https://arxiv.org/abs/2112.07341}{{\ttfamily 2112.07341}}].

\bibitem{Euclid:2023ove}
{\scshape Euclid} collaboration, \emph{{Euclid preparation - LII. Forecast impact of super-sample covariance on 3\texttimes{}2pt analysis with Euclid}}, \href{https://doi.org/10.1051/0004-6361/202348389}{\emph{Astron. Astrophys.} {\bfseries 691} (2024) A318} [\href{https://arxiv.org/abs/2310.15731}{{\ttfamily 2310.15731}}].

\bibitem{Riess:2021jrx}
A.~G. Riess et~al., \emph{{A Comprehensive Measurement of the Local Value of the Hubble Constant with 1 km s$^{-1}$ Mpc$^{-1}$ Uncertainty from the Hubble Space Telescope and the SH0ES Team}}, \href{https://doi.org/10.3847/2041-8213/ac5c5b}{\emph{Astrophys. J. Lett.} {\bfseries 934} (2022) L7} [\href{https://arxiv.org/abs/2112.04510}{{\ttfamily 2112.04510}}].

\bibitem{Abdalla:2022yfr}
E.~Abdalla et~al., \emph{{Cosmology intertwined: A review of the particle physics, astrophysics, and cosmology associated with the cosmological tensions and anomalies}}, \href{https://doi.org/10.1016/j.jheap.2022.04.002}{\emph{JHEAp} {\bfseries 34} (2022) 49} [\href{https://arxiv.org/abs/2203.06142}{{\ttfamily 2203.06142}}].

\bibitem{Kilo-DegreeSurvey:2023gfr}
{\scshape Kilo-Degree Survey, Dark Energy Survey} collaboration, \emph{{DES Y3 + KiDS-1000: Consistent cosmology combining cosmic shear surveys}}, \href{https://doi.org/10.21105/astro.2305.17173}{\emph{Open J. Astrophys.} {\bfseries 6} (2023) 2305.17173} [\href{https://arxiv.org/abs/2305.17173}{{\ttfamily 2305.17173}}].

\bibitem{Albrecht:2006um}
A.~Albrecht et~al., \emph{{Report of the Dark Energy Task Force}},  \href{https://arxiv.org/abs/astro-ph/0609591}{{\ttfamily astro-ph/0609591}}.

\bibitem{Bartelmann:2010fz}
M.~Bartelmann, \emph{{Gravitational Lensing}}, \href{https://doi.org/10.1088/0264-9381/27/23/233001}{\emph{Class. Quant. Grav.} {\bfseries 27} (2010) 233001} [\href{https://arxiv.org/abs/1010.3829}{{\ttfamily 1010.3829}}].

\bibitem{LSSTDarkEnergyScience:2012kar}
{\scshape LSST Dark Energy Science} collaboration, \emph{{Large Synoptic Survey Telescope: Dark Energy Science Collaboration}},  \href{https://arxiv.org/abs/1211.0310}{{\ttfamily 1211.0310}}.

\bibitem{Spergel:2015sza}
D.~Spergel et~al., \emph{{Wide-Field InfrarRed Survey Telescope-Astrophysics Focused Telescope Assets WFIRST-AFTA 2015 Report}},  \href{https://arxiv.org/abs/1503.03757}{{\ttfamily 1503.03757}}.

\bibitem{Heymans:2020gsg}
C.~Heymans et~al., \emph{{KiDS-1000 Cosmology: Multi-probe weak gravitational lensing and spectroscopic galaxy clustering constraints}}, \href{https://doi.org/10.1051/0004-6361/202039063}{\emph{Astron. Astrophys.} {\bfseries 646} (2021) A140} [\href{https://arxiv.org/abs/2007.15632}{{\ttfamily 2007.15632}}].

\bibitem{DES:2022qpf}
{\scshape DES} collaboration, \emph{{Dark energy survey year 3 results: cosmological constraints from the analysis of cosmic shear in harmonic space}}, \href{https://doi.org/10.1093/mnras/stac1826}{\emph{Mon. Not. Roy. Astron. Soc.} {\bfseries 515} (2022) 1942} [\href{https://arxiv.org/abs/2203.07128}{{\ttfamily 2203.07128}}].

\bibitem{Hamana:2019etx}
T.~Hamana et~al., \emph{{Cosmological constraints from cosmic shear two-point correlation functions with HSC survey first-year data}}, \href{https://doi.org/10.1093/pasj/psz138}{\emph{Publ. Astron. Soc. Jap.} {\bfseries 72} (2020) 16} [\href{https://arxiv.org/abs/1906.06041}{{\ttfamily 1906.06041}}].

\bibitem{Kaiser:1991qi}
N.~Kaiser, \emph{{Weak gravitational lensing of distant galaxies}}, \href{https://doi.org/10.1086/171151}{\emph{Astrophys. J.} {\bfseries 388} (1992) 272}.

\bibitem{Schneider:2002jd}
P.~Schneider, L.~van Waerbeke, M.~Kilbinger and Y.~Mellier, \emph{{Analysis of two-point statistics of cosmic shear: I. estimators and covariances}}, \href{https://doi.org/10.1051/0004-6361:20021341}{\emph{Astron. Astrophys.} {\bfseries 396} (2002) 1} [\href{https://arxiv.org/abs/astro-ph/0206182}{{\ttfamily astro-ph/0206182}}].

\bibitem{KiDS:2020suj}
{\scshape KiDS} collaboration, \emph{{KiDS-1000 Cosmology: Cosmic shear constraints and comparison between two point statistics}}, \href{https://doi.org/10.1051/0004-6361/202039070}{\emph{Astron. Astrophys.} {\bfseries 645} (2021) A104} [\href{https://arxiv.org/abs/2007.15633}{{\ttfamily 2007.15633}}].

\bibitem{Hu:2000ax}
W.~Hu and M.~J. White, \emph{{Power spectra estimation for weak lensing}}, \href{https://doi.org/10.1086/321380}{\emph{Astrophys. J.} {\bfseries 554} (2001) 67} [\href{https://arxiv.org/abs/astro-ph/0010352}{{\ttfamily astro-ph/0010352}}].

\bibitem{Brown:2004jn}
M.~L. Brown, P.~G. Castro and A.~N. Taylor, \emph{{CMB temperature and polarisation pseudo-C(l) estimators and covariances}}, \href{https://doi.org/10.1111/j.1365-2966.2005.09111.x}{\emph{Mon. Not. Roy. Astron. Soc.} {\bfseries 360} (2005) 1262} [\href{https://arxiv.org/abs/astro-ph/0410394}{{\ttfamily astro-ph/0410394}}].

\bibitem{Hikage:2010sq}
C.~Hikage, M.~Takada, T.~Hamana and D.~Spergel, \emph{{Shear Power Spectrum Reconstruction using Pseudo-Spectrum Method}}, \href{https://doi.org/10.1111/j.1365-2966.2010.17886.x}{\emph{Mon. Not. Roy. Astron. Soc.} {\bfseries 412} (2011) 65} [\href{https://arxiv.org/abs/1004.3542}{{\ttfamily 1004.3542}}].

\bibitem{Hivon:2001jp}
E.~Hivon, K.~M. Gorski, C.~B. Netterfield, B.~P. Crill, S.~Prunet and F.~Hansen, \emph{{Master of the cosmic microwave background anisotropy power spectrum: a fast method for statistical analysis of large and complex cosmic microwave background data sets}}, \href{https://doi.org/10.1086/338126}{\emph{Astrophys. J.} {\bfseries 567} (2002) 2} [\href{https://arxiv.org/abs/astro-ph/0105302}{{\ttfamily astro-ph/0105302}}].

\bibitem{Tegmark:1996qt}
M.~Tegmark, \emph{{How to measure CMB power spectra without losing information}}, \href{https://doi.org/10.1103/PhysRevD.55.5895}{\emph{Phys. Rev. D} {\bfseries 55} (1997) 5895} [\href{https://arxiv.org/abs/astro-ph/9611174}{{\ttfamily astro-ph/9611174}}].

\bibitem{Tegmark:2001zv}
M.~Tegmark and A.~de~Oliveira-Costa, \emph{{How to measure CMB polarization power spectra without losing information}}, \href{https://doi.org/10.1103/PhysRevD.64.063001}{\emph{Phys. Rev. D} {\bfseries 64} (2001) 063001} [\href{https://arxiv.org/abs/astro-ph/0012120}{{\ttfamily astro-ph/0012120}}].

\bibitem{Szapudi:2000xj}
I.~Szapudi, S.~Prunet, D.~Pogosyan, A.~S. Szalay and J.~R. Bond, \emph{{Fast CMB analyses via correlation functions}}, \href{https://doi.org/10.1086/319105}{\emph{{Astrophys. J.}} {\bfseries 548} (2000) L115} [\href{https://arxiv.org/abs/astro-ph/0010256}{{\ttfamily astro-ph/0010256}}].

\bibitem{Chon:2003gx}
G.~Chon, A.~Challinor, S.~Prunet, E.~Hivon and I.~Szapudi, \emph{{Fast estimation of polarization power spectra using correlation functions}}, \href{https://doi.org/10.1111/j.1365-2966.2004.07737.x}{\emph{Mon. Not. Roy. Astron. Soc.} {\bfseries 350} (2004) 914} [\href{https://arxiv.org/abs/astro-ph/0303414}{{\ttfamily astro-ph/0303414}}].

\bibitem{HSC:2018mrq}
{\scshape HSC} collaboration, \emph{{Cosmology from cosmic shear power spectra with Subaru Hyper Suprime-Cam first-year data}}, \href{https://doi.org/10.1093/pasj/psz010}{\emph{Publ. Astron. Soc. Jap.} {\bfseries 71} (2019) 43} [\href{https://arxiv.org/abs/1809.09148}{{\ttfamily 1809.09148}}].

\bibitem{Nicola:2020lhi}
A.~Nicola, C.~Garc\'\i{}a-Garc\'\i{}a, D.~Alonso, J.~Dunkley, P.~G. Ferreira, A.~Slosar et~al., \emph{{Cosmic shear power spectra in practice}}, \href{https://doi.org/10.1088/1475-7516/2021/03/067}{\emph{JCAP} {\bfseries 03} (2021) 067} [\href{https://arxiv.org/abs/2010.09717}{{\ttfamily 2010.09717}}].

\bibitem{Garcia-Garcia:2019bku}
C.~Garc\'\i{}a-Garc\'\i{}a, D.~Alonso and E.~Bellini, \emph{{Disconnected pseudo-$C_\ell$ covariances for projected large-scale structure data}}, \href{https://doi.org/10.1088/1475-7516/2019/11/043}{\emph{JCAP} {\bfseries 11} (2019) 043} [\href{https://arxiv.org/abs/1906.11765}{{\ttfamily 1906.11765}}].

\bibitem{KiDS:2021opn}
{\scshape KiDS, Euclid} collaboration, \emph{{KiDS and Euclid: Cosmological implications of a pseudo angular power spectrum analysis of KiDS-1000 cosmic shear tomography}}, \href{https://doi.org/10.1051/0004-6361/202142481}{\emph{Astron. Astrophys.} {\bfseries 665} (2022) A56} [\href{https://arxiv.org/abs/2110.06947}{{\ttfamily 2110.06947}}].

\bibitem{Efstathiou:2003dj}
G.~Efstathiou, \emph{{Myths and truths concerning estimation of power spectra}}, \href{https://doi.org/10.1111/j.1365-2966.2004.07530.x}{\emph{Mon. Not. Roy. Astron. Soc.} {\bfseries 349} (2004) 603} [\href{https://arxiv.org/abs/astro-ph/0307515}{{\ttfamily astro-ph/0307515}}].

\bibitem{Schneider:2001af}
P.~Schneider, L.~Van~Waerbeke and Y.~Mellier, \emph{{B-modes in cosmic shear from source redshift clustering}}, \href{https://doi.org/10.1051/0004-6361:20020626}{\emph{Astron. Astrophys.} {\bfseries 389} (2002) 729} [\href{https://arxiv.org/abs/astro-ph/0112441}{{\ttfamily astro-ph/0112441}}].

\bibitem{Heavens:2000ad}
A.~Heavens, A.~Refregier and C.~Heymans, \emph{{Intrinsic correlation of galaxy shapes: Implications for weak lensing measurements}}, \href{https://doi.org/10.1046/j.1365-8711.2000.03907.x}{\emph{Mon. Not. Roy. Astron. Soc.} {\bfseries 319} (2000) 649} [\href{https://arxiv.org/abs/astro-ph/0005269}{{\ttfamily astro-ph/0005269}}].

\bibitem{Thomas:2016xhb}
D.~B. Thomas, L.~Whittaker, S.~Camera and M.~L. Brown, \emph{{Estimating the weak-lensing rotation signal in radio cosmic shear surveys}}, \href{https://doi.org/10.1093/mnras/stx1468}{\emph{Mon. Not. Roy. Astron. Soc.} {\bfseries 470} (2017) 3131} [\href{https://arxiv.org/abs/1612.01533}{{\ttfamily 1612.01533}}].

\bibitem{Kohlinger:2015tza}
F.~K\"ohlinger, M.~Viola, W.~Valkenburg, B.~Joachimi, H.~Hoekstra and K.~Kuijken, \emph{{A direct measurement of tomographic lensing power spectra from CFHTLenS}}, \href{https://doi.org/10.1093/mnras/stv2762}{\emph{Mon. Not. Roy. Astron. Soc.} {\bfseries 456} (2016) 1508} [\href{https://arxiv.org/abs/1509.04071}{{\ttfamily 1509.04071}}].

\bibitem{Kohlinger:2017sxk}
F.~K\"ohlinger et~al., \emph{{KiDS-450: The tomographic weak lensing power spectrum and constraints on cosmological parameters}}, \href{https://doi.org/10.1093/mnras/stx1820}{\emph{Mon. Not. Roy. Astron. Soc.} {\bfseries 471} (2017) 4412} [\href{https://arxiv.org/abs/1706.02892}{{\ttfamily 1706.02892}}].

\bibitem{vanUitert:2017ieu}
E.~van Uitert et~al., \emph{{KiDS+GAMA: cosmology constraints from a joint analysis of cosmic shear, galaxy\textendash{}galaxy lensing, and angular clustering}}, \href{https://doi.org/10.1093/mnras/sty551}{\emph{Mon. Not. Roy. Astron. Soc.} {\bfseries 476} (2018) 4662} [\href{https://arxiv.org/abs/1706.05004}{{\ttfamily 1706.05004}}].

\bibitem{SDSS:2011gwu}
{\scshape SDSS} collaboration, \emph{{The SDSS Coadd: Cosmic Shear Measurement}}, \href{https://doi.org/10.1088/0004-637X/761/1/15}{\emph{Astrophys. J.} {\bfseries 761} (2012) 15} [\href{https://arxiv.org/abs/1111.6622}{{\ttfamily 1111.6622}}].

\bibitem{Efstathiou:2006eb}
G.~Efstathiou, \emph{{Hybrid estimation of cmb polarization power spectra}}, \href{https://doi.org/10.1111/j.1365-2966.2006.10486.x}{\emph{Mon. Not. Roy. Astron. Soc.} {\bfseries 370} (2006) 343} [\href{https://arxiv.org/abs/astro-ph/0601107}{{\ttfamily astro-ph/0601107}}].

\bibitem{Horowitz:2018tbe}
B.~Horowitz, U.~Seljak and G.~Aslanyan, \emph{{Efficient Optimal Reconstruction of Linear Fields and Band-powers from Cosmological Data}}, \href{https://doi.org/10.1088/1475-7516/2019/10/035}{\emph{JCAP} {\bfseries 10} (2019) 035} [\href{https://arxiv.org/abs/1810.00503}{{\ttfamily 1810.00503}}].

\bibitem{Seljak:2017rmr}
U.~Seljak, G.~Aslanyan, Y.~Feng and C.~Modi, \emph{{Towards optimal extraction of cosmological information from nonlinear data}}, \href{https://doi.org/10.1088/1475-7516/2017/12/009}{\emph{JCAP} {\bfseries 12} (2017) 009} [\href{https://arxiv.org/abs/1706.06645}{{\ttfamily 1706.06645}}].

\bibitem{Oh:1998sr}
S.~P. Oh, D.~N. Spergel and G.~Hinshaw, \emph{{An Efficient technique to determine the power spectrum from cosmic microwave background sky maps}}, \href{https://doi.org/10.1086/306629}{\emph{Astrophys. J.} {\bfseries 510} (1999) 551} [\href{https://arxiv.org/abs/astro-ph/9805339}{{\ttfamily astro-ph/9805339}}].

\bibitem{Elsner:2012fe}
F.~Elsner and B.~D. Wandelt, \emph{{Efficient Wiener filtering without preconditioning}}, \href{https://doi.org/10.1051/0004-6361/201220586}{\emph{Astron. Astrophys.} {\bfseries 549} (2013) A111} [\href{https://arxiv.org/abs/1210.4931}{{\ttfamily 1210.4931}}].

\bibitem{Bunn:2016lxi}
E.~F. Bunn and B.~Wandelt, \emph{{Pure E and B polarization maps via Wiener filtering}}, \href{https://doi.org/10.1103/PhysRevD.96.043523}{\emph{Phys. Rev. D} {\bfseries 96} (2017) 043523} [\href{https://arxiv.org/abs/1610.03345}{{\ttfamily 1610.03345}}].

\bibitem{Ramanah:2018enp}
D.~K. Ramanah, G.~Lavaux and B.~D. Wandelt, \emph{{Optimal and fast $\mathcal {E}/\mathcal {B}$ separation with a dual messenger field}}, \href{https://doi.org/10.1093/mnras/sty341}{\emph{Mon. Not. Roy. Astron. Soc.} {\bfseries 476} (2018) 2825} [\href{https://arxiv.org/abs/1801.05358}{{\ttfamily 1801.05358}}].

\bibitem{Estrada:2021hdo}
N.~Estrada, B.~R. Granett and L.~Guzzo, \emph{{Cosmology behind the mask: constraining the parameters of \ensuremath{\Lambda}CDM with the unmasked galaxy density field from VIPERS}}, \href{https://doi.org/10.1093/mnras/stac515}{\emph{Mon. Not. Roy. Astron. Soc.} {\bfseries 512} (2022) 2817} [\href{https://arxiv.org/abs/2108.01926}{{\ttfamily 2108.01926}}].

\bibitem{Philcox:2020vbm}
O.~H.~E. Philcox, \emph{{Cosmology without window functions: Quadratic estimators for the galaxy power spectrum}}, \href{https://doi.org/10.1103/PhysRevD.103.103504}{\emph{Phys. Rev. D} {\bfseries 103} (2021) 103504} [\href{https://arxiv.org/abs/2012.09389}{{\ttfamily 2012.09389}}].

\bibitem{Bilbao-Ahedo:2017uuk}
J.~D. Bilbao-Ahedo, R.~B. Barreiro, D.~Herranz, P.~Vielva and E.~Mart\'\i{}nez-Gonz\'alez, \emph{{On the regularity of the covariance matrix of a discretized scalar field on the sphere}}, \href{https://doi.org/10.1088/1475-7516/2017/02/022}{\emph{JCAP} {\bfseries 02} (2017) 022} [\href{https://arxiv.org/abs/1701.06617}{{\ttfamily 1701.06617}}].

\bibitem{Bond:1998zw}
J.~R. Bond, A.~H. Jaffe and L.~Knox, \emph{{Estimating the power spectrum of the cosmic microwave background}}, \href{https://doi.org/10.1103/PhysRevD.57.2117}{\emph{Phys. Rev. D} {\bfseries 57} (1998) 2117} [\href{https://arxiv.org/abs/astro-ph/9708203}{{\ttfamily astro-ph/9708203}}].

\bibitem{Bilbao-Ahedo:2021jhn}
J.~D. Bilbao-Ahedo, R.~B. Barreiro, P.~Vielva, E.~Mart\'\i{}nez-Gonz\'alez and D.~Herranz, \emph{{ECLIPSE: a fast Quadratic Maximum Likelihood estimator for CMB intensity and polarization power spectra}}, \href{https://doi.org/10.1088/1475-7516/2021/07/034}{\emph{JCAP} {\bfseries 07} (2021) 034} [\href{https://arxiv.org/abs/2104.08528}{{\ttfamily 2104.08528}}].

\bibitem{Seljak:1997ep}
U.~Seljak, \emph{{Weak lensing reconstruction and power spectrum estimation: minimum variance methods}}, \href{https://doi.org/10.1086/306225}{\emph{Astrophys. J.} {\bfseries 506} (1998) 64} [\href{https://arxiv.org/abs/astro-ph/9711124}{{\ttfamily astro-ph/9711124}}].

\bibitem{Pen:2003mu}
U.-L. Pen, \emph{{Fast power spectrum estimation}}, \href{https://doi.org/10.1046/j.1365-2966.2003.07118.x}{\emph{Mon. Not. Roy. Astron. Soc.} {\bfseries 346} (2003) 619} [\href{https://arxiv.org/abs/astro-ph/0304513}{{\ttfamily astro-ph/0304513}}].

\bibitem{McDonald:2018mfm}
P.~McDonald, \emph{{Renormalization group computation of likelihood functions for cosmological data sets}}, \href{https://doi.org/10.1103/PhysRevD.99.043538}{\emph{Phys. Rev. D} {\bfseries 99} (2019) 043538} [\href{https://arxiv.org/abs/1810.08454}{{\ttfamily 1810.08454}}].

\bibitem{McDonald:2019efe}
P.~McDonald, \emph{{Improved renormalization group computation of likelihood functions for cosmological data sets}}, \href{https://doi.org/10.1103/PhysRevD.100.043511}{\emph{Phys. Rev. D} {\bfseries 100} (2019) 043511} [\href{https://arxiv.org/abs/1906.09127}{{\ttfamily 1906.09127}}].

\bibitem{Leistedt:2013gfa}
B.~Leistedt, H.~V. Peiris, D.~J. Mortlock, A.~Benoit-L\'evy and A.~Pontzen, \emph{{Estimating the large-scale angular power spectrum in the presence of systematics: a case study of Sloan Digital Sky Survey quasars}}, \href{https://doi.org/10.1093/mnras/stt1359}{\emph{Mon. Not. Roy. Astron. Soc.} {\bfseries 435} (2013) 1857} [\href{https://arxiv.org/abs/1306.0005}{{\ttfamily 1306.0005}}].

\bibitem{Mortlock:2000zw}
D.~J. Mortlock, A.~D. Challinor and M.~P. Hobson, \emph{{Analysis of cosmic microwave background data on an incomplete sky}}, \href{https://doi.org/10.1046/j.1365-8711.2002.05085.x}{\emph{Mon. Not. Roy. Astron. Soc.} {\bfseries 330} (2002) 405} [\href{https://arxiv.org/abs/astro-ph/0008083}{{\ttfamily astro-ph/0008083}}].

\bibitem{Hu:2000ee}
W.~Hu, \emph{{Weak lensing of the CMB: A harmonic approach}}, \href{https://doi.org/10.1103/PhysRevD.62.043007}{\emph{Phys. Rev. D} {\bfseries 62} (2000) 043007} [\href{https://arxiv.org/abs/astro-ph/0001303}{{\ttfamily astro-ph/0001303}}].

\bibitem{Martinet:2020mqm}
N.~Martinet, J.~Harnois-D\'eraps, E.~Jullo and P.~Schneider, \emph{{Probing dark energy with tomographic weak-lensing aperture mass statistics}}, \href{https://doi.org/10.1051/0004-6361/202039679}{\emph{Astron. Astrophys.} {\bfseries 646} (2021) A62} [\href{https://arxiv.org/abs/2010.07376}{{\ttfamily 2010.07376}}].

\bibitem{Euclid:2021icp}
{\scshape Euclid} collaboration, \emph{{Euclid preparation: I. The Euclid Wide Survey}}, \href{https://doi.org/10.1051/0004-6361/202141938}{\emph{Astron. Astrophys.} {\bfseries 662} (2022) A112} [\href{https://arxiv.org/abs/2108.01201}{{\ttfamily 2108.01201}}].

\bibitem{Szego1975orthogonal}
G.~Szeg{\"{o}}, \emph{Orthogonal Polynomials}, American Math. Soc: Colloquium publ. American Mathematical Society, 1975.

\bibitem{Vanneste:2018azc}
S.~Vanneste, S.~Henrot-Versill\'e, T.~Louis and M.~Tristram, \emph{{Quadratic estimator for CMB cross-correlation}}, \href{https://doi.org/10.1103/PhysRevD.98.103526}{\emph{Phys. Rev. D} {\bfseries 98} (2018) 103526} [\href{https://arxiv.org/abs/1807.02484}{{\ttfamily 1807.02484}}].

\bibitem{Lewis:2001hp}
A.~Lewis, A.~Challinor and N.~Turok, \emph{{Analysis of CMB polarization on an incomplete sky}}, \href{https://doi.org/10.1103/PhysRevD.65.023505}{\emph{Phys. Rev. D} {\bfseries 65} (2002) 023505} [\href{https://arxiv.org/abs/astro-ph/0106536}{{\ttfamily astro-ph/0106536}}].

\bibitem{Smith:2005gi}
K.~M. Smith, \emph{{Pseudo-$C_\ell$ estimators which do not mix E and B modes}}, \href{https://doi.org/10.1103/PhysRevD.74.083002}{\emph{Phys. Rev. D} {\bfseries 74} (2006) 083002} [\href{https://arxiv.org/abs/astro-ph/0511629}{{\ttfamily astro-ph/0511629}}].

\bibitem{Grain:2009wq}
J.~Grain, M.~Tristram and R.~Stompor, \emph{{Polarized CMB spectrum estimation using the pure pseudo cross-spectrum approach}}, \href{https://doi.org/10.1103/PhysRevD.79.123515}{\emph{Phys. Rev. D} {\bfseries 79} (2009) 123515} [\href{https://arxiv.org/abs/0903.2350}{{\ttfamily 0903.2350}}].

\bibitem{Hall:2022das}
A.~Hall and A.~Taylor, \emph{{Non-Gaussian likelihood of weak lensing power spectra}}, \href{https://doi.org/10.1103/PhysRevD.105.123527}{\emph{Phys. Rev. D} {\bfseries 105} (2022) 123527} [\href{https://arxiv.org/abs/2202.04095}{{\ttfamily 2202.04095}}].

\bibitem{Taruya:2002vy}
A.~Taruya, M.~Takada, T.~Hamana, I.~Kayo and T.~Futamase, \emph{{Lognormal property of weak-lensing fields}}, \href{https://doi.org/10.1086/340048}{\emph{Astrophys. J.} {\bfseries 571} (2002) 638} [\href{https://arxiv.org/abs/astro-ph/0202090}{{\ttfamily astro-ph/0202090}}].

\bibitem{Hilbert:2011xq}
S.~Hilbert, J.~Hartlap and P.~Schneider, \emph{{Cosmic-shear covariance: The log-normal approximation}}, \href{https://doi.org/10.1051/0004-6361/201117294}{\emph{Astron. Astrophys.} {\bfseries 536} (2011) A85} [\href{https://arxiv.org/abs/1105.3980}{{\ttfamily 1105.3980}}].

\bibitem{Horowitz:2016dwk}
B.~Horowitz and U.~Seljak, \emph{{Cosmological constraints from thermal Sunyaev\textendash{}Zeldovich power spectrum revisited}}, \href{https://doi.org/10.1093/mnras/stx766}{\emph{Mon. Not. Roy. Astron. Soc.} {\bfseries 469} (2017) 394} [\href{https://arxiv.org/abs/1609.01850}{{\ttfamily 1609.01850}}].

\bibitem{Bolliet:2017lha}
B.~Bolliet, B.~Comis, E.~Komatsu and J.~F. Mac\'\i{}as-P\'erez, \emph{{Dark energy constraints from the thermal Sunyaev\textendash{}Zeldovich power spectrum}}, \href{https://doi.org/10.1093/mnras/sty823}{\emph{Mon. Not. Roy. Astron. Soc.} {\bfseries 477} (2018) 4957} [\href{https://arxiv.org/abs/1712.00788}{{\ttfamily 1712.00788}}].

\bibitem{Li:2023azi}
S.-S. Li et~al., \emph{{KiDS-1000: Cosmology with improved cosmic shear measurements}}, \href{https://doi.org/10.1051/0004-6361/202347236}{\emph{Astron. Astrophys.} {\bfseries 679} (2023) A133} [\href{https://arxiv.org/abs/2306.11124}{{\ttfamily 2306.11124}}].

\bibitem{DES:2021wwk}
{\scshape DES} collaboration, \emph{{Dark Energy Survey Year 3 results: Cosmological constraints from galaxy clustering and weak lensing}}, \href{https://doi.org/10.1103/PhysRevD.105.023520}{\emph{Phys. Rev. D} {\bfseries 105} (2022) 023520} [\href{https://arxiv.org/abs/2105.13549}{{\ttfamily 2105.13549}}].

\bibitem{DES:2021bvc}
{\scshape DES} collaboration, \emph{{Dark Energy Survey Year 3 results: Cosmology from cosmic shear and robustness to data calibration}}, \href{https://doi.org/10.1103/PhysRevD.105.023514}{\emph{Phys. Rev. D} {\bfseries 105} (2022) 023514} [\href{https://arxiv.org/abs/2105.13543}{{\ttfamily 2105.13543}}].

\bibitem{DES:2021vln}
{\scshape DES} collaboration, \emph{{Dark Energy Survey Year 3 results: Cosmology from cosmic shear and robustness to modeling uncertainty}}, \href{https://doi.org/10.1103/PhysRevD.105.023515}{\emph{Phys. Rev. D} {\bfseries 105} (2022) 023515} [\href{https://arxiv.org/abs/2105.13544}{{\ttfamily 2105.13544}}].

\bibitem{Li:2023tui}
X.~Li et~al., \emph{{Hyper Suprime-Cam Year 3 results: Cosmology from cosmic shear two-point correlation functions}}, \href{https://doi.org/10.1103/PhysRevD.108.123518}{\emph{Phys. Rev. D} {\bfseries 108} (2023) 123518} [\href{https://arxiv.org/abs/2304.00702}{{\ttfamily 2304.00702}}].

\bibitem{Schneider:2015wta}
A.~Schneider and R.~Teyssier, \emph{{A new method to quantify the effects of baryons on the matter power spectrum}}, \href{https://doi.org/10.1088/1475-7516/2015/12/049}{\emph{JCAP} {\bfseries 12} (2015) 049} [\href{https://arxiv.org/abs/1510.06034}{{\ttfamily 1510.06034}}].

\bibitem{Giri:2021qin}
S.~K. Giri and A.~Schneider, \emph{{Emulation of baryonic effects on the matter power spectrum and constraints from galaxy cluster data}}, \href{https://doi.org/10.1088/1475-7516/2021/12/046}{\emph{JCAP} {\bfseries 12} (2021) 046} [\href{https://arxiv.org/abs/2108.08863}{{\ttfamily 2108.08863}}].

\bibitem{Arico:2020lhq}
G.~Aric\`o, R.~E. Angulo, S.~Contreras, L.~Ondaro-Mallea, M.~Pellejero-Iba\~nez and M.~Zennaro, \emph{{The BACCO simulation project: a baryonification emulator with neural networks}}, \href{https://doi.org/10.1093/mnras/stab1911}{\emph{Mon. Not. Roy. Astron. Soc.} {\bfseries 506} (2021) 4070} [\href{https://arxiv.org/abs/2011.15018}{{\ttfamily 2011.15018}}].

\bibitem{Huang:2018wpy}
H.-J. Huang, T.~Eifler, R.~Mandelbaum and S.~Dodelson, \emph{{Modelling baryonic physics in future weak lensing surveys}}, \href{https://doi.org/10.1093/mnras/stz1714}{\emph{Mon. Not. Roy. Astron. Soc.} {\bfseries 488} (2019) 1652} [\href{https://arxiv.org/abs/1809.01146}{{\ttfamily 1809.01146}}].

\bibitem{Salcido:2023etz}
J.~Salcido, I.~G. McCarthy, J.~Kwan, A.~Upadhye and A.~S. Font, \emph{{SP(k) \textendash{} a hydrodynamical simulation-based model for the impact of baryon physics on the non-linear matter power spectrum}}, \href{https://doi.org/10.1093/mnras/stad1474}{\emph{Mon. Not. Roy. Astron. Soc.} {\bfseries 523} (2023) 2247} [\href{https://arxiv.org/abs/2305.09710}{{\ttfamily 2305.09710}}].

\bibitem{DES:2021rex}
{\scshape DES} collaboration, \emph{{Dark Energy Survey Year 3 Results: Multi-Probe Modeling Strategy and Validation}},  \href{https://arxiv.org/abs/2105.13548}{{\ttfamily 2105.13548}}.

\bibitem{Baldauf:2016sjb}
T.~Baldauf, M.~Mirbabayi, M.~Simonovi\'c and M.~Zaldarriaga, \emph{{LSS constraints with controlled theoretical uncertainties}},  \href{https://arxiv.org/abs/1602.00674}{{\ttfamily 1602.00674}}.

\bibitem{Sprenger:2018tdb}
T.~Sprenger, M.~Archidiacono, T.~Brinckmann, S.~Clesse and J.~Lesgourgues, \emph{{Cosmology in the era of Euclid and the Square Kilometre Array}}, \href{https://doi.org/10.1088/1475-7516/2019/02/047}{\emph{JCAP} {\bfseries 02} (2019) 047} [\href{https://arxiv.org/abs/1801.08331}{{\ttfamily 1801.08331}}].

\bibitem{Moreira:2021imm}
M.~G. Moreira, F.~Andrade-Oliveira, X.~Fang, H.-J. Huang, E.~Krause, V.~Miranda et~al., \emph{{Mitigating baryonic effects with a theoretical error covariance}}, \href{https://doi.org/10.1093/mnras/stab2481}{\emph{Mon. Not. Roy. Astron. Soc.} {\bfseries 507} (2021) 5592} [\href{https://arxiv.org/abs/2104.01397}{{\ttfamily 2104.01397}}].

\bibitem{Pellejero-Ibanez:2022efv}
M.~Pellejero-Ibanez, R.~E. Angulo, M.~Zennaro, J.~Stuecker, S.~Contreras, G.~Arico et~al., \emph{{The bacco simulation project: bacco hybrid Lagrangian bias expansion model in redshift space}}, \href{https://doi.org/10.1093/mnras/stad368}{\emph{Mon. Not. Roy. Astron. Soc.} {\bfseries 520} (2023) 3725} [\href{https://arxiv.org/abs/2207.06437}{{\ttfamily 2207.06437}}].

\bibitem{Kitching:2010ab}
T.~D. Kitching and A.~N. Taylor, \emph{{On Mitigation of the Uncertainty in Nonlinear Matter Clustering for Cosmic Shear Tomography}}, \href{https://doi.org/10.1111/j.1365-2966.2011.18772.x}{\emph{Mon. Not. Roy. Astron. Soc.} {\bfseries 416} (2011) 1717} [\href{https://arxiv.org/abs/1012.3479}{{\ttfamily 1012.3479}}].

\bibitem{McCarthy:2016mry}
I.~G. McCarthy, J.~Schaye, S.~Bird and A.~M.~C. Le~Brun, \emph{{The BAHAMAS project: Calibrated hydrodynamical simulations for large-scale structure cosmology}}, \href{https://doi.org/10.1093/mnras/stw2792}{\emph{Mon. Not. Roy. Astron. Soc.} {\bfseries 465} (2017) 2936} [\href{https://arxiv.org/abs/1603.02702}{{\ttfamily 1603.02702}}].

\bibitem{Schneider:2015yka}
A.~Schneider, R.~Teyssier, D.~Potter, J.~Stadel, J.~Onions, D.~S. Reed et~al., \emph{{Matter power spectrum and the challenge of percent accuracy}}, \href{https://doi.org/10.1088/1475-7516/2016/04/047}{\emph{JCAP} {\bfseries 04} (2016) 047} [\href{https://arxiv.org/abs/1503.05920}{{\ttfamily 1503.05920}}].

\bibitem{Seljak:2000gq}
U.~Seljak, \emph{{Analytic model for galaxy and dark matter clustering}}, \href{https://doi.org/10.1046/j.1365-8711.2000.03715.x}{\emph{Mon. Not. Roy. Astron. Soc.} {\bfseries 318} (2000) 203} [\href{https://arxiv.org/abs/astro-ph/0001493}{{\ttfamily astro-ph/0001493}}].

\bibitem{Peacock:2000MNRAS3181144P}
J.~A. {Peacock} and R.~E. {Smith}, \emph{{Halo occupation numbers and galaxy bias}}, \href{https://doi.org/10.1046/j.1365-8711.2000.03779.x}{\emph{{Mon. Not. Roy. Astron. Soc.}} {\bfseries 318} (2000) 1144} [\href{https://arxiv.org/abs/astro-ph/0005010}{{\ttfamily astro-ph/0005010}}].

\bibitem{Cooray:2002dia}
A.~Cooray and R.~K. Sheth, \emph{{Halo Models of Large Scale Structure}}, \href{https://doi.org/10.1016/S0370-1573(02)00276-4}{\emph{Phys. Rept.} {\bfseries 372} (2002) 1} [\href{https://arxiv.org/abs/astro-ph/0206508}{{\ttfamily astro-ph/0206508}}].

\bibitem{Boruah:2024tkq}
{\scshape LSST Dark Energy Science} collaboration, \emph{{Machine Learning LSST 3x2pt analyses -- forecasting the impact of systematics on cosmological constraints using neural networks}},  \href{https://arxiv.org/abs/2403.11797}{{\ttfamily 2403.11797}}.

\bibitem{Vogelsberger:2019ynw}
M.~Vogelsberger, F.~Marinacci, P.~Torrey and E.~Puchwein, \emph{{Cosmological Simulations of Galaxy Formation}}, \href{https://doi.org/10.1038/s42254-019-0127-2}{\emph{Nature Rev. Phys.} {\bfseries 2} (2020) 42} [\href{https://arxiv.org/abs/1909.07976}{{\ttfamily 1909.07976}}].

\bibitem{Crain:2023xap}
R.~A. Crain and F.~van~de Voort, \emph{{Hydrodynamical simulations of the galaxy population: enduring successes and outstanding challenges}}, \href{https://doi.org/10.1146/annurev-astro-041923-043618}{\emph{Ann. Rev. Astron. Astrophys.} {\bfseries 61} (2023) 473} [\href{https://arxiv.org/abs/2309.17075}{{\ttfamily 2309.17075}}].

\bibitem{Schaye:2023jqv}
J.~Schaye et~al., \emph{{The FLAMINGO project: cosmological hydrodynamical simulations for large-scale structure and galaxy cluster surveys}}, \href{https://doi.org/10.1093/mnras/stad2419}{\emph{Mon. Not. Roy. Astron. Soc.} {\bfseries 526} (2023) 4978} [\href{https://arxiv.org/abs/2306.04024}{{\ttfamily 2306.04024}}].

\bibitem{Hernandez-Aguayo:2022xcl}
C.~Hern\'andez-Aguayo et~al., \emph{{The MillenniumTNG Project: high-precision predictions for matter clustering and halo statistics}}, \href{https://doi.org/10.1093/mnras/stad1657}{\emph{Mon. Not. Roy. Astron. Soc.} {\bfseries 524} (2023) 2556} [\href{https://arxiv.org/abs/2210.10059}{{\ttfamily 2210.10059}}].

\bibitem{Dave:2019yyq}
R.~Dav\'e, D.~Angl\'es-Alc\'azar, D.~Narayanan, Q.~Li, M.~H. Rafieferantsoa and S.~Appleby, \emph{{Simba: Cosmological Simulations with Black Hole Growth and Feedback}}, \href{https://doi.org/10.1093/mnras/stz937}{\emph{Mon. Not. Roy. Astron. Soc.} {\bfseries 486} (2019) 2827} [\href{https://arxiv.org/abs/1901.10203}{{\ttfamily 1901.10203}}].

\bibitem{Nelson:2015dga}
D.~Nelson et~al., \emph{{The Illustris Simulation: Public Data Release}}, \href{https://doi.org/10.1016/j.ascom.2015.09.003}{\emph{Astron. Comput.} {\bfseries 13} (2015) 12} [\href{https://arxiv.org/abs/1504.00362}{{\ttfamily 1504.00362}}].

\bibitem{Springel:2017tpz}
V.~Springel et~al., \emph{{First results from the IllustrisTNG simulations: matter and galaxy clustering}}, \href{https://doi.org/10.1093/mnras/stx3304}{\emph{Mon. Not. Roy. Astron. Soc.} {\bfseries 475} (2018) 676} [\href{https://arxiv.org/abs/1707.03397}{{\ttfamily 1707.03397}}].

\bibitem{Chisari:2018prw}
N.~E. Chisari, M.~L.~A. Richardson, J.~Devriendt, Y.~Dubois, A.~Schneider, A.~L. Brun, M.~C. et~al., \emph{{The impact of baryons on the matter power spectrum from the Horizon-AGN cosmological hydrodynamical simulation}}, \href{https://doi.org/10.1093/mnras/sty2093}{\emph{Mon. Not. Roy. Astron. Soc.} {\bfseries 480} (2018) 3962} [\href{https://arxiv.org/abs/1801.08559}{{\ttfamily 1801.08559}}].

\bibitem{Schaye:2014tpa}
J.~Schaye et~al., \emph{{The EAGLE project: Simulating the evolution and assembly of galaxies and their environments}}, \href{https://doi.org/10.1093/mnras/stu2058}{\emph{Mon. Not. Roy. Astron. Soc.} {\bfseries 446} (2015) 521} [\href{https://arxiv.org/abs/1407.7040}{{\ttfamily 1407.7040}}].

\bibitem{Brun:2013yva}
A.~M. C.~L. Brun, I.~G. McCarthy, J.~Schaye and T.~J. Ponman, \emph{{Towards a realistic population of simulated galaxy groups and clusters}}, \href{https://doi.org/10.1093/mnras/stu608}{\emph{Mon. Not. Roy. Astron. Soc.} {\bfseries 441} (2014) 1270} [\href{https://arxiv.org/abs/1312.5462}{{\ttfamily 1312.5462}}].

\bibitem{Salcido:2024qrt}
J.~Salcido and I.~G. McCarthy, \emph{{Implications of feedback solutions to the $S_8$ tension for the baryon fractions of galaxy groups and clusters}},  \href{https://arxiv.org/abs/2409.05716}{{\ttfamily 2409.05716}}.

\bibitem{vanDaalen:2019pst}
M.~P. van Daalen, I.~G. McCarthy and J.~Schaye, \emph{{Exploring the effects of galaxy formation on matter clustering through a library of simulation power spectra}}, \href{https://doi.org/10.1093/mnras/stz3199}{\emph{Mon. Not. Roy. Astron. Soc.} {\bfseries 491} (2020) 2424} [\href{https://arxiv.org/abs/1906.00968}{{\ttfamily 1906.00968}}].

\bibitem{Elbers:2024dad}
W.~Elbers et~al., \emph{{The FLAMINGO project: the coupling between baryonic feedback and cosmology in light of the $S_8$ tension}}, \href{https://doi.org/10.1093/mnras/staf093}{\emph{Mon. Not. Roy. Astron. Soc.} {\bfseries 537} (2025) 2160} [\href{https://arxiv.org/abs/2403.12967}{{\ttfamily 2403.12967}}].

\bibitem{Chisari:2019tus}
N.~E. Chisari et~al., \emph{{Modelling baryonic feedback for survey cosmology}}, \href{https://doi.org/10.21105/astro.1905.06082}{\emph{Open J. Astrophys.} {\bfseries 2} (2019) 4} [\href{https://arxiv.org/abs/1905.06082}{{\ttfamily 1905.06082}}].

\bibitem{Fanidakis_2011}
N.~Fanidakis, C.~M. Baugh, A.~J. Benson, R.~G. Bower, S.~Cole, C.~Done et~al., \emph{{The evolution of active galactic nuclei across cosmic time: what is downsizing?}}, \href{https://doi.org/10.1111/j.1365-2966.2011.19931.x}{\emph{Mon. Not. Roy. Astron. Soc.} {\bfseries 419} (2011) 2797} [\href{https://arxiv.org/abs/1011.5222}{{\ttfamily 1011.5222}}].

\bibitem{Schneider:2018pfw}
A.~Schneider, R.~Teyssier, J.~Stadel, N.~E. Chisari, A.~M.~C. Le~Brun, A.~Amara et~al., \emph{{Quantifying baryon effects on the matter power spectrum and the weak lensing shear correlation}}, \href{https://doi.org/10.1088/1475-7516/2019/03/020}{\emph{JCAP} {\bfseries 03} (2019) 020} [\href{https://arxiv.org/abs/1810.08629}{{\ttfamily 1810.08629}}].

\bibitem{Semboloni:2011fe}
E.~Semboloni, H.~Hoekstra, J.~Schaye, M.~P. van Daalen and I.~J. McCarthy, \emph{{Quantifying the effect of baryon physics on weak lensing tomography}}, \href{https://doi.org/10.1111/j.1365-2966.2011.19385.x}{\emph{Mon. Not. Roy. Astron. Soc.} {\bfseries 417} (2011) 2020} [\href{https://arxiv.org/abs/1105.1075}{{\ttfamily 1105.1075}}].

\bibitem{Copeland:2017hzu}
D.~Copeland, A.~Taylor and A.~Hall, \emph{{The impact of baryons on the sensitivity of dark energy measurements}}, \href{https://doi.org/10.1093/mnras/sty2001}{\emph{Mon. Not. Roy. Astron. Soc.} {\bfseries 480} (2018) 2247} [\href{https://arxiv.org/abs/1712.07112}{{\ttfamily 1712.07112}}].

\bibitem{Chudaykin:2019ock}
A.~Chudaykin and M.~M. Ivanov, \emph{{Measuring neutrino masses with large-scale structure: Euclid forecast with controlled theoretical error}}, \href{https://doi.org/10.1088/1475-7516/2019/11/034}{\emph{JCAP} {\bfseries 11} (2019) 034} [\href{https://arxiv.org/abs/1907.06666}{{\ttfamily 1907.06666}}].

\bibitem{CAMELS:2023wqd}
{\scshape CAMELS} collaboration, \emph{{The CAMELS Project: Expanding the Galaxy Formation Model Space with New ASTRID and 28-parameter TNG and SIMBA Suites}}, \href{https://doi.org/10.3847/1538-4357/ad022a}{\emph{Astrophys. J.} {\bfseries 959} (2023) 136} [\href{https://arxiv.org/abs/2304.02096}{{\ttfamily 2304.02096}}].

\bibitem{ACT:2024vsj}
{\scshape ACT, DESI} collaboration, \emph{{Evidence for large baryonic feedback at low and intermediate redshifts from kinematic Sunyaev-Zel'dovich observations with ACT and DESI photometric galaxies}},  \href{https://arxiv.org/abs/2407.07152}{{\ttfamily 2407.07152}}.

\bibitem{Bigwood:2025ism}
L.~Bigwood, M.~A. Bourne, V.~Irsic, A.~Amon and D.~Sijacki, \emph{{The case for large-scale AGN feedback in galaxy formation simulations: insights from XFABLE}},  \href{https://arxiv.org/abs/2501.16983}{{\ttfamily 2501.16983}}.

\bibitem{Matilla:2017rmu}
J.~M.~Z. Matilla, Z.~Haiman, A.~Petri and T.~Namikawa, \emph{{Geometry and growth contributions to cosmic shear observables}}, \href{https://doi.org/10.1103/PhysRevD.96.023513}{\emph{Phys. Rev. D} {\bfseries 96} (2017) 023513} [\href{https://arxiv.org/abs/1706.05133}{{\ttfamily 1706.05133}}].

\bibitem{DES:2020iqt}
{\scshape DES} collaboration, \emph{{DES Y1 results: Splitting growth and geometry to test $\Lambda$CDM}}, \href{https://doi.org/10.1103/PhysRevD.103.023528}{\emph{Phys. Rev. D} {\bfseries 103} (2021) 023528} [\href{https://arxiv.org/abs/2010.05924}{{\ttfamily 2010.05924}}].

\bibitem{Zhong:2023how}
K.~Zhong, E.~Saraivanov, V.~Miranda, J.~Xu, T.~Eifler and E.~Krause, \emph{{Growth and geometry split in light of the DES-Y3 survey}}, \href{https://doi.org/10.1103/PhysRevD.107.123529}{\emph{Phys. Rev. D} {\bfseries 107} (2023) 123529} [\href{https://arxiv.org/abs/2301.03694}{{\ttfamily 2301.03694}}].

\bibitem{Ruiz-Zapatero:2021rzl}
J.~Ruiz-Zapatero et~al., \emph{{Geometry versus growth - Internal consistency of the flat $\Lambda$CDM model with KiDS-1000}}, \href{https://doi.org/10.1051/0004-6361/202141350}{\emph{Astron. Astrophys.} {\bfseries 655} (2021) A11} [\href{https://arxiv.org/abs/2105.09545}{{\ttfamily 2105.09545}}].

\bibitem{DES:2020daw}
{\scshape DES} collaboration, \emph{{Consistency of cosmic shear analyses in harmonic and real space}}, \href{https://doi.org/10.1093/mnras/stab661}{\emph{Mon. Not. Roy. Astron. Soc.} {\bfseries 503} (2021) 3796} [\href{https://arxiv.org/abs/2011.06469}{{\ttfamily 2011.06469}}].

\bibitem{Troster:2021gsz}
T.~Tr\"oster et~al., \emph{{Joint constraints on cosmology and the impact of baryon feedback: Combining KiDS-1000 lensing with the thermal Sunyaev\textendash{}Zeldovich effect from Planck and ACT}}, \href{https://doi.org/10.1051/0004-6361/202142197}{\emph{Astron. Astrophys.} {\bfseries 660} (2022) A27} [\href{https://arxiv.org/abs/2109.04458}{{\ttfamily 2109.04458}}].

\bibitem{Ferreira:2023syi}
T.~Ferreira, D.~Alonso, C.~Garcia-Garcia and N.~E. Chisari, \emph{{X-Ray\textendash{}Cosmic-Shear Cross-Correlations: First Detection and Constraints on Baryonic Effects}}, \href{https://doi.org/10.1103/PhysRevLett.133.051001}{\emph{Phys. Rev. Lett.} {\bfseries 133} (2024) 051001} [\href{https://arxiv.org/abs/2309.11129}{{\ttfamily 2309.11129}}].

\bibitem{Aires:2024wze}
A.~Aires, N.~Kokron, R.~Rosenfeld, F.~Andrade-Oliveira and V.~Miranda, \emph{{Mitigation of nonlinear galaxy bias with a theoretical-error likelihood}},  \href{https://arxiv.org/abs/2410.08930}{{\ttfamily 2410.08930}}.

\bibitem{Bernardeau:2013rda}
F.~Bernardeau, T.~Nishimichi and A.~Taruya, \emph{{Cosmic shear full nulling: sorting out dynamics, geometry and systematics}}, \href{https://doi.org/10.1093/mnras/stu1861}{\emph{Mon. Not. Roy. Astron. Soc.} {\bfseries 445} (2014) 1526} [\href{https://arxiv.org/abs/1312.0430}{{\ttfamily 1312.0430}}].

\bibitem{Miyatake:2023njf}
H.~Miyatake et~al., \emph{{Hyper Suprime-Cam Year 3 results: Cosmology from galaxy clustering and weak lensing with HSC and SDSS using the emulator based halo model}}, \href{https://doi.org/10.1103/PhysRevD.108.123517}{\emph{Phys. Rev. D} {\bfseries 108} (2023) 123517} [\href{https://arxiv.org/abs/2304.00704}{{\ttfamily 2304.00704}}].

\bibitem{KIDS_DES:2023gfr}
{\scshape Kilo-Degree Survey, DES} collaboration, \emph{{DES Y3 + KiDS-1000: Consistent cosmology combining cosmic shear surveys}}, \href{https://doi.org/10.21105/astro.2305.17173}{\emph{Open J. Astrophys.} {\bfseries 6} (2023) 2305.17173} [\href{https://arxiv.org/abs/2305.17173}{{\ttfamily 2305.17173}}].

\bibitem{Garcia-Garcia:2024gzy}
C.~Garc\'\i{}a-Garc\'\i{}a, M.~Zennaro, G.~Aric\`o, D.~Alonso and R.~E. Angulo, \emph{{Cosmic shear with small scales: DES-Y3, KiDS-1000 and HSC-DR1}}, \href{https://doi.org/10.1088/1475-7516/2024/08/024}{\emph{JCAP} {\bfseries 08} (2024) 024} [\href{https://arxiv.org/abs/2403.13794}{{\ttfamily 2403.13794}}].

\bibitem{Angulo:2020vky}
R.~E. Angulo, M.~Zennaro, S.~Contreras, G.~Aric\`o, M.~Pellejero-Iba\~nez and J.~St\"ucker, \emph{{The BACCO simulation project: exploiting the full power of large-scale structure for cosmology}}, \href{https://doi.org/10.1093/mnras/stab2018}{\emph{Mon. Not. Roy. Astron. Soc.} {\bfseries 507} (2021) 5869} [\href{https://arxiv.org/abs/2004.06245}{{\ttfamily 2004.06245}}].

\bibitem{Arico:2021izc}
G.~Aric\`o, R.~E. Angulo and M.~Zennaro, \emph{{Accelerating Large-Scale-Structure data analyses by emulating Boltzmann solvers and Lagrangian Perturbation Theory}},  \href{https://arxiv.org/abs/2104.14568}{{\ttfamily 2104.14568}}.

\bibitem{Zennaro:2021bwy}
M.~Zennaro, R.~E. Angulo, M.~Pellejero-Ib\'a\~nez, J.~St\"ucker, S.~Contreras and G.~Aric\`o, \emph{{The BACCO simulation project: biased tracers in real space}}, \href{https://doi.org/10.1093/mnras/stad2008}{\emph{Mon. Not. Roy. Astron. Soc.} {\bfseries 524} (2023) 2407} [\href{https://arxiv.org/abs/2101.12187}{{\ttfamily 2101.12187}}].

\bibitem{Planck:2018vyg}
{\scshape Planck} collaboration, \emph{{Planck 2018 results. VI. Cosmological parameters}}, \href{https://doi.org/10.1051/0004-6361/201833910}{\emph{Astron. Astrophys.} {\bfseries 641} (2020) A6} [\href{https://arxiv.org/abs/1807.06209}{{\ttfamily 1807.06209}}].

\bibitem{Schaye:2009bt}
J.~Schaye, C.~Dalla~Vecchia, C.~M. Booth, R.~P.~C. Wiersma, T.~Theuns, M.~R. Haas et~al., \emph{{The physics driving the cosmic star formation history}}, \href{https://doi.org/10.1111/j.1365-2966.2009.16029.x}{\emph{Mon. Not. Roy. Astron. Soc.} {\bfseries 402} (2010) 1536} [\href{https://arxiv.org/abs/0909.5196}{{\ttfamily 0909.5196}}].

\bibitem{NumRec:2007}
W.~Press, S.~Teukolsky, W.~Vetterling and B.~Flannery, \emph{Numerical Recipes in FORTRAN: The Art of Scientific Computing}. Cambridge University Press, 3rd~ed., 2007.

\bibitem{Euclid:2024xqh}
{\scshape Euclid} collaboration, \emph{{Euclid preparation: LIX. Angular power spectra from discrete observations}}, \href{https://doi.org/10.1051/0004-6361/202452018}{\emph{Astron. Astrophys.} {\bfseries 694} (2025) A141} [\href{https://arxiv.org/abs/2408.16903}{{\ttfamily 2408.16903}}].

\bibitem{Euclid:2021osj}
{\scshape Euclid} collaboration, \emph{{Euclid preparation. XII. Optimizing the photometric sample of the Euclid survey for galaxy clustering and galaxy-galaxy lensing analyses}}, \href{https://doi.org/10.1051/0004-6361/202141061}{\emph{Astron. Astrophys.} {\bfseries 655} (2021) A44} [\href{https://arxiv.org/abs/2104.05698}{{\ttfamily 2104.05698}}].

\bibitem{Davis:1985ApJ292}
M.~{Davis}, G.~{Efstathiou}, C.~S. {Frenk} and S.~D.~M. {White}, \emph{{The evolution of large-scale structure in a universe dominated by cold dark matter}}, \href{https://doi.org/10.1086/163168}{\emph{Astrophys.\ J.} {\bfseries 292} (1985) 371}.

\bibitem{Arbey:2021gdg}
A.~Arbey and F.~Mahmoudi, \emph{{Dark matter and the early Universe: a review}}, \href{https://doi.org/10.1016/j.ppnp.2021.103865}{\emph{Prog. Part. Nucl. Phys.} {\bfseries 119} (2021) 103865} [\href{https://arxiv.org/abs/2104.11488}{{\ttfamily 2104.11488}}].

\bibitem{Spergel:2015arXiv150303757S}
D.~Spergel et~al., \emph{{Wide-Field InfrarRed Survey Telescope-Astrophysics Focused Telescope Assets WFIRST-AFTA 2015 Report}},  \href{https://arxiv.org/abs/1503.03757}{{\ttfamily 1503.03757}}.

\bibitem{DESI:2016fyo}
{\scshape DESI} collaboration, \emph{{The DESI Experiment Part I: Science,Targeting, and Survey Design}},  \href{https://arxiv.org/abs/1611.00036}{{\ttfamily 1611.00036}}.

\bibitem{BICEP2:2014owc}
{\scshape BICEP2} collaboration, \emph{{Detection of $B$-Mode Polarization at Degree Angular Scales by BICEP2}}, \href{https://doi.org/10.1103/PhysRevLett.112.241101}{\emph{Phys. Rev. Lett.} {\bfseries 112} (2014) 241101} [\href{https://arxiv.org/abs/1403.3985}{{\ttfamily 1403.3985}}].

\bibitem{Planck:2014dmk}
{\scshape Planck} collaboration, \emph{{Planck intermediate results. XXX. The angular power spectrum of polarized dust emission at intermediate and high Galactic latitudes}}, \href{https://doi.org/10.1051/0004-6361/201425034}{\emph{Astron. Astrophys.} {\bfseries 586} (2016) A133} [\href{https://arxiv.org/abs/1409.5738}{{\ttfamily 1409.5738}}].

\bibitem{CAMELS:2020cof}
{\scshape CAMELS} collaboration, \emph{{The CAMELS project: Cosmology and Astrophysics with MachinE Learning Simulations}}, \href{https://doi.org/10.3847/1538-4357/abf7ba}{\emph{Astrophys. J.} {\bfseries 915} (2021) 71} [\href{https://arxiv.org/abs/2010.00619}{{\ttfamily 2010.00619}}].

\bibitem{Lovell:2021MNRAS5002127L}
C.~C. {Lovell}, A.~P. {Vijayan}, P.~A. {Thomas}, S.~M. {Wilkins}, D.~J. {Barnes}, D.~{Irodotou} et~al., \emph{{First Light And Reionization Epoch Simulations (FLARES) - I. Environmental dependence of high-redshift galaxy evolution}}, \href{https://doi.org/10.1093/mnras/staa3360}{\emph{Mon. Not. Roy. Astron. Soc.} {\bfseries 500} (2021) 2127} [\href{https://arxiv.org/abs/2004.07283}{{\ttfamily 2004.07283}}].

\bibitem{Gardner:2006ky}
J.~P. Gardner et~al., \emph{{The James Webb Space Telescope}}, \href{https://doi.org/10.1007/s11214-006-8315-7}{\emph{Space Sci. Rev.} {\bfseries 123} (2006) 485} [\href{https://arxiv.org/abs/astro-ph/0606175}{{\ttfamily astro-ph/0606175}}].

\bibitem{Barcons:2012arXiv1207.2745B}
X.~Barcons et~al., \emph{{Athena (Advanced Telescope for High ENergy Astrophysics) Assessment Study Report for ESA Cosmic Vision 2015-2025}},  \href{https://arxiv.org/abs/1207.2745}{{\ttfamily 1207.2745}}.

\bibitem{LISA:2017pwj}
{\scshape LISA} collaboration, \emph{{Laser Interferometer Space Antenna}},  \href{https://arxiv.org/abs/1702.00786}{{\ttfamily 1702.00786}}.

\bibitem{Schneider:2021wds}
A.~Schneider, S.~K. Giri, S.~Amodeo and A.~Refregier, \emph{{Constraining baryonic feedback and cosmology with weak-lensing, X-ray, and kinematic Sunyaev\textendash{}Zeldovich observations}}, \href{https://doi.org/10.1093/mnras/stac1493}{\emph{Mon. Not. Roy. Astron. Soc.} {\bfseries 514} (2022) 3802} [\href{https://arxiv.org/abs/2110.02228}{{\ttfamily 2110.02228}}].

\bibitem{UKGrid}
{UK National Energy System Operator}, \emph{Historic GB Generation Mix}. 2024.

\bibitem{Kwan:2014}
I.~Kwan and D.~Rutherford, \emph{Transatlantic Airline Fuel Efficiency Ranking, 2014}. {The International Council on Clean Transportation}, 2014.

\bibitem{PortegiesZwart:2020pdu}
S.~Portegies~Zwart, \emph{{The Ecological Impact of High-performance Computing in Astrophysics}}, \href{https://doi.org/10.1038/s41550-020-1208-y}{\emph{Nature Astron.} {\bfseries 4} (2020) 819} [\href{https://arxiv.org/abs/2009.11295}{{\ttfamily 2009.11295}}].

\bibitem{Geller:1989Sci897G}
M.~J. {Geller} and J.~P. {Huchra}, \emph{{Mapping the Universe}}, \href{https://doi.org/10.1126/science.246.4932.897}{\emph{Science} {\bfseries 246} (1989) 897}.

\bibitem{SDSS:2000hjo}
{\scshape SDSS} collaboration, \emph{{The Sloan Digital Sky Survey: Technical Summary}}, \href{https://doi.org/10.1086/301513}{\emph{Astron. J.} {\bfseries 120} (2000) 1579} [\href{https://arxiv.org/abs/astro-ph/0006396}{{\ttfamily astro-ph/0006396}}].

\bibitem{Conselice:2016ApJ83083C}
C.~J. {Conselice}, A.~{Wilkinson}, K.~{Duncan} and A.~{Mortlock}, \emph{{The Evolution of Galaxy Number Density at z < 8 and Its Implications}}, \href{https://doi.org/10.3847/0004-637X/830/2/83}{\emph{Astrophys.\ J.} {\bfseries 830} (2016) 83} [\href{https://arxiv.org/abs/1607.03909}{{\ttfamily 1607.03909}}].

\bibitem{Rhodes:2020xwp}
J.~Rhodes, E.~Huff, D.~Masters and A.~Nierenberg, \emph{{The End of Galaxy Surveys}}, \href{https://doi.org/10.3847/1538-3881/abbe86}{\emph{Astron. J.} {\bfseries 160} (2020) 261} [\href{https://arxiv.org/abs/2010.06064}{{\ttfamily 2010.06064}}].

\bibitem{OpenAI:2023ktj}
OpenAI et~al., \emph{{GPT-4 Technical Report}},  \href{https://arxiv.org/abs/2303.08774}{{\ttfamily 2303.08774}}.

\bibitem{Boas:913305}
M.~L. \mychar{B}oas{\footnote{\textit{How did you get a B in a red box?} --- Peter Thomas}}, \emph{{Mathematical Methods in the Physical Sciences}}. Wiley, Hoboken, USA, 3rd~ed., 2006.

\bibitem{Chevallier:2000qy}
M.~Chevallier and D.~Polarski, \emph{{Accelerating universes with scaling dark matter}}, \href{https://doi.org/10.1142/S0218271801000822}{\emph{Int. J. Mod. Phys. D} {\bfseries 10} (2001) 213} [\href{https://arxiv.org/abs/gr-qc/0009008}{{\ttfamily gr-qc/0009008}}].

\bibitem{Tsutsui:2008cu}
R.~Tsutsui, T.~Nakamura, D.~Yonetoku, T.~Murakami, S.~Tanabe, Y.~Kodama et~al., \emph{{Constraints on $w_0$ and $w_a$ of Dark Energy from High Redshift Gamma Ray Bursts}}, \href{https://doi.org/10.1111/j.1745-3933.2008.00604.x}{\emph{Mon. Not. Roy. Astron. Soc.} {\bfseries 394} (2009) L31} [\href{https://arxiv.org/abs/0807.2911}{{\ttfamily 0807.2911}}].

\bibitem{AMALDI19841}
E.~Amaldi, \emph{From the discovery of the neutron to the discovery of nuclear fission}, \href{https://doi.org/https://doi.org/10.1016/0370-1573(84)90214-X}{\emph{Physics Reports} {\bfseries 111} (1984) 1}.

\bibitem{Mele:2015etc}
S.~Mele, \emph{{The Measurement of the Number of Light Neutrino Species at LEP}}, \href{https://doi.org/10.1142/9789814644150_0004}{\emph{Adv. Ser. Direct. High Energy Phys.} {\bfseries 23} (2015) 89}.

\bibitem{Bennett:2020zkv}
J.~J. Bennett, G.~Buldgen, P.~F. De~Salas, M.~Drewes, S.~Gariazzo, S.~Pastor et~al., \emph{{Towards a precision calculation of $N_{\rm eff}$ in the Standard Model II: Neutrino decoupling in the presence of flavour oscillations and finite-temperature QED}}, \href{https://doi.org/10.1088/1475-7516/2021/04/073}{\emph{JCAP} {\bfseries 04} (2021) 073} [\href{https://arxiv.org/abs/2012.02726}{{\ttfamily 2012.02726}}].

\bibitem{Lesgourgues:2013sjj}
J.~Lesgourgues, G.~Mangano, G.~Miele and S.~Pastor, \emph{{Neutrino Cosmology}}. Cambridge University Press, Cambridge, 2013, \href{https://doi.org/10.1017/CBO9781139012874}{DOI}.

\bibitem{Euclid:2024imf}
{\scshape Euclid} collaboration, \emph{{Euclid preparation - LIV. Sensitivity to neutrino parameters}}, \href{https://doi.org/10.1051/0004-6361/202450859}{\emph{Astron. Astrophys.} {\bfseries 693} (2025) A58} [\href{https://arxiv.org/abs/2405.06047}{{\ttfamily 2405.06047}}].

\end{thebibliography}\endgroup

\clearpage

\vspace*{7.5cm}

\noindent\rule{\textwidth}{0.5pt}
\vspace*{0.25cm}

\noindent \textit{I always rip out the last page of a book. Then it doesn't have to end. I hate endings!}

\vspace*{-0.25cm}

\begin{flushright}
  {---The Doctor}
\end{flushright}

\vspace*{-0.25cm}

\noindent\rule{\textwidth}{0.5pt}

\end{document}